%% file: thesis.tex
\documentclass[12pt,oneside,letterpaper]{report}

\newcommand{\draftfinal}[2]{\ifdefined\draftversion#1\else#2\fi}

\newcommand{\finalonly}[1]{\draftfinal{}{#1}}

\newcommand{\thesistitle}{{\LARGE Locally Supersymmetric Effective Field Theories\\of Inflation}}
\newcommand{\thesisauthor}{{\large Hun Jang}}
\newcommand{\thesisadvisor}{  \href{https://as.nyu.edu/faculty/massimo-porrati.html}{\color{black}{{\large Prof. Massimo Porrati}}}}

\newcommand{\thesisdept}{Physics}
\newcommand{\gradmonth}{May}
\newcommand{\gradyear}{2022}

\RequirePackage[margin=1in, includefoot, letterpaper]{geometry}

\RequirePackage[prologue]{xcolor}
\definecolor[named]{ThesisBlue}{cmyk}{1,0.1,0,0.1}
\definecolor[named]{ThesisYellow}{cmyk}{0,0.16,1,0}
\definecolor[named]{ThesisOrange}{cmyk}{0,0.42,1,0.01}
\definecolor[named]{ThesisRed}{cmyk}{0,0.90,0.86,0}
\definecolor[named]{ThesisLightBlue}{cmyk}{0.49,0.01,0,0}
\definecolor[named]{ThesisGreen}{cmyk}{0.20,0,1,0.19}
\definecolor[named]{ThesisPurple}{cmyk}{0.55,1,0,0.15}
\definecolor[named]{ThesisDarkBlue}{cmyk}{1,0.58,0,0.21}

\definecolor{SchoolColor}{rgb}{0.3412, 0.0235, 0.5490} 
\definecolor{chaptergrey}{rgb}{0.2600, 0.0200, 0.4600} 
\definecolor{midgrey}{rgb}{0.4, 0.4, 0.4}

\usepackage{cancel}
\usepackage[utf8]{inputenc}
\usepackage{tocloft}

\usepackage{chemformula}
\usepackage{array}
\RequirePackage[labelfont={bf,sf,small,singlespacing},
                textfont={sf,small,singlespacing},
                margin=0pt,
                figurewithin=chapter,
                tablewithin=chapter]{caption}

\usepackage{fix-cm}
\RequirePackage[raggedright,sc]{titlesec}
\definecolor{gray75}{gray}{0.75}
\newcommand{\hsp}{\hspace{20pt}}

\titleformat{\chapter}[hang]
{\Huge\sc}
{\textcolor{SchoolColor}{\thechapter}\hsp\textcolor{gray75}{|}\hsp}
{0pt}{\Huge\sc\raggedright}

\usepackage{setspace}

\finalonly{
  \doublespacing 
}

\usepackage{lipsum}

\newcommand{\nocomments}{}

\newcommand{\notodos}{}

\newcommand{\fcomment}[2]{\ifdefined\nocomments{}\else\footnote{\textcolor{red}{#1:} #2}\fi}

\input{defs}

\usepackage{comment}

\usepackage{hyperref}
\hypersetup{colorlinks,
  linkcolor=ThesisDarkBlue,
  citecolor=ThesisPurple,
  urlcolor=ThesisDarkBlue,
  filecolor=ThesisDarkBlue}

\usepackage[
    backend=biber,
    style=numeric-comp,
    natbib=true,
    url=true, 
    doi=true,
    eprint=true,
    sorting=none]{biblatex}
\renewbibmacro{in:}{}

 \usepackage{lscape}
 
\begin{document}

\pagenumbering{roman}
%
\thispagestyle{empty}
%

\vspace*{25pt}
\begin{center}

  {\Large
    \begin{doublespace}
      {\textcolor{SchoolColor}{\textsc{\thesistitle}}}
    \end{doublespace}
  }
  \vspace{.7in}

  {\large by}
  \vspace{.7in}

  \thesisauthor
  \vfill

  \begin{doublespace}
    \textsc{
    {\large A dissertation submitted in partial fulfillment\\
    of the requirements for the degree of\\
    Doctor of Philosophy\\
    Department of \thesisdept\\
    New York University\\
    \gradmonth, \gradyear}}
  \end{doublespace}
\end{center}
\vfill

\noindent\makebox[\textwidth]{\hfill\makebox[2.5in]{\hrulefill}}\\
\makebox[\textwidth]{\hfill\makebox[2.5in]{\hfill\thesisadvisor}}

\newpage

\thispagestyle{empty}
\vspace*{25pt}
\begin{center}
  \scshape \noindent \small \copyright \  \small  \thesisauthor \\
  All rights reserved, \gradyear
\end{center}
\vspace*{0in}
\newpage

\cleardoublepage
\phantomsection
\chapter*{Dedication}
\addcontentsline{toc}{chapter}{Dedication}
This doctoral dissertation is dedicated to my beloved fiancee, {\it Dongju Lee}, who has loved me all the times, has encouraged me to pursue my dreams towards theoretical physicist, and has been a sustaining source of support and encouragement during my PhD journey at NYU Physics for five years from September 2017 to May 2022. I am truly thankful for having her in my life. Also, this work is dedicated to my beloved family, who all have backed me up.
\vfill
\newpage

\chapter*{Acknowledgements}
\addcontentsline{toc}{chapter}{Acknowledgments}

\input{chapters/acknowledge}

\newpage

\chapter*{Abstract}
\addcontentsline{toc}{chapter}{Abstract}

\input{abstract}

\newpage

\tableofcontents

\cleardoublepage
\phantomsection
\addcontentsline{toc}{chapter}{List of Figures}
\listoffigures
\newpage


\cleardoublepage
\phantomsection
\addcontentsline{toc}{chapter}{List of Appendices}
\listofappendices
\newpage

\pagenumbering{arabic} 




\chapter{Introduction} \label{ch1}
\input{chapters/1}

\chapter{Review 1: Inflationary Cosmology} \label{ch2}

\input{chapters/2}

\chapter{Review 2: Effective Field Theory} \label{ch3}

\input{chapters/3}

\chapter{Review 3: Supersymmetry} \label{ch4}
\input{chapters/4}

\chapter{Review 4: $4D$ $\mathcal{N}=1$ Supergravity in the Superconformal Formalism} \label{ch5}
\input{chapters/5}

\chapter{Review 5: Recent Developments of Inflationary Models in Supergravity}\label{ch5-1}

\input{chapters/5-1}

\chapter{Reconstruction of Liberated $\mathcal{N}=1$ Supergravity in the Superconformal Formalism} \label{ch6}

\input{chapters/6}

\chapter{Revisiting New K\"{a}hler-Invariant Fayet-Iliopoulos Terms in the Superconformal Formalism} \label{ch7}
\input{chapters/7}

\chapter{Relaxed Supergravity} \label{ch8}
\input{chapters/8}

\chapter{Inflationary Model 1: A minimal model of single-field and slow-roll inflation in liberated $\mathcal{N}=1$ supergravity} \label{ch9}
\input{chapters/9}

\chapter{Inflationary Model 2: Inflation, Gravity-Mediated Supersymmetry Breaking, and de Sitter Vacua in Supergravity with a Kähler-Invariant Fayet-Iliopoulos Term} \label{ch10}
\input{chapters/10}

\chapter{Inflationary Model 3: Single-Field Slow-Roll Inflation, Minimal Supersymmetric Standard Model (MSSM), and de Sitter Vacua from Minimal Supergravity with New Fayet-Iliopoulos Terms} \label{ch11}
\input{chapters/11}

\label{chp-conclusion}

\chapter{Conclusion} \label{ch12}
\input{chapters/12}
\label{chp-conclusion}



\appendix

\chapter{Spinor Algebras in General Dimensions}\appcaption{Spinor Algebras in General Dimensions}\label{Spinor_algebras}
\input{chapters/spinor_algebra}

\chapter{Superconformal Tensor Calculus}\appcaption{Superconformal Tensor Calculus of $\mathcal{N}=1$ Supergravity}\label{STC}

\input{chapters/SCC}

\chapter{Derivation of Fermion Masses in the Supergravity Model of Inflation in Ch.~\ref{ch10}}\appcaption{Derivation of Fermion Masses in the Supergravity Model of Inflation in Ch.~\ref{ch10}}\label{deri}

\input{chapters/deri}

\chapter{Scanning Nonrenormalizable Terms of New FI Terms}\appcaption{Scanning Nonrenormalizable Terms of New FI Terms}\label{alpha_cal}

\input{chapters/alpha_cal}




\cleardoublepage
\phantomsection


\addcontentsline{toc}{chapter}{Bibliography}
\printbibliography

\end{document}

%% file: chapters/acknowledge.tex
I wish to acknowledge two generous financial supports from New York University (NYU), which fully covered the five years of my PhD study at Department of Physics in NYU, under which this doctoral research has been carried: {\it Henry Mitchell MacCracken Fellowship} (for four years) funded by Graduate School of Arts and Sciences (GSAS), and {\it James Arthur Graduate Associate (JAGA) Fellowship} (for the last year) funded by Center for Cosmology and Particle Physics (CCPP) at NYU Physics. 

First and foremost, I am deeply grateful to my doctoral advisor, {\it Prof. Massimo Porrati}, for his patience, insightful feedback, and continuous support with full encouragement and enthusiasm during the nice five years of my PhD study at NYU. These all pushed me to sharpen my thinking and brought my work to a higher level. In particular, his expertise on supergravity and string theory was invaluable in formulating questions and ideas of my research. Also, I would like to express my sincere gratitude to my doctoral co-advisor, {\it Prof. Matthew Kleban}, for his sincere guidance and encouragement during my first three doctoral years. His expertise on inflation and string theory was invaluable as well in studying inflationary cosmology. 
My sincere thanks also goes to the other three members of my doctoral thesis committee, {\it Profs. Joshua Ruderman, Allen Mincer, and Neal Weiner}, for their patience and understanding that went into the production of this dissertation. I would like to offer my special thanks to {\it Prof. Ken Van Tilburg} for his participation in my oral qualifying examination.

I would also like to thank NYU Physics professors who led the classes of coursework I have taken for about three years, {\it Profs. Massimo Porrati (String Theory), Matthew Kleban (QFT1\&2), Gregory Gabadadze (QFT3), Joshua Ruderman (Early Universe), Yacine Ali-Haïmoud (General Relativity), L. Andrew Wray (Quantum Mechanics 1), David Grier (Statistical Physics), Andrei Gruzinov (Quantum Mechanics 2), Roman Scoccimarro (Classical Dynamics, Electromagnetism, Cosmology), Andrew MacFadyen (Computational Physics), Kyle Cranmer (Statistics and Data Science), and Tycho Sleator (Advanced Experimental Physics)}. Through all these classes, I was able to improve knowledge of what is related to my doctoral research. Moreover, I would like to specially thank {\it Prof. Allen Mincer}, who strongly inspired me. I was able to learn his philosophy of teaching and how to communicate with students by serving as a Teaching Assistant (TA) to assist his classes. Also, I would like to thank {\it Prof. Andre Adler}, who passed away in May 2020 and was left as an ideal teacher in my mind, for his advice, support, and lessons during my teaching experiences as a TA with him at NYU Physics. I also thank Undergraduate Lab Manager {\it Michael Salvati} for his sincere help during the laboratory sessions I led. I would like to thank Program Administrators {\it Bill LePage} and {\it Evette Ma} at the Physics Department for their administrative support and kindly consideration during my PhD journey at NYU Physics. 

I would like to specially thank my doctoral roommates, {\it Ekapob Kulchoakrungsun} and {\it Yucheng Zhang}. My doctoral journey along with them for about five years at the Stuytown in Manhattan has been very nice and memorable. I would like to also thank the other doctoral friends I have met at NYU Physics, {\it Jack Donahue, Milad Noorikuhani, Kaizhe Wang, Shiyuan Hu, Xingyang Yu, Austin McDowell, Lauren Altman, Kate Storey-Fisher, Samuele Crespi, Marek Narozniak, Alexis Clavijo, Xuyao Hu, {\it Nanoom Lee}, {\it Dr. Jayme Kim}, Xucheng Gan, Joan Manuel La Madrid, Conghuan Luo, Shahrzad Zare, and others}. In addition, I would like to thank many Korean friends I have ever met at NYU for highly enjoyable moments with them. Lastly, I would like to thank my old friends in South Korea who gave me moral support and encouragement during my PhD study, {\it Seungwoo Lee, Jong-inn Im, Hyojin Kim, Kyung Ju Lee, Yeseal Yim, Geunwoong Jeon, Hosu Park, Unyong Eom, Joobyoung Kim, Kyungoh Kwon}, and {\it Jihyang Kim}.

%% file: abstract.tex
Supergravity, a locally supersymmetric gauge theory, may provide to describe new physics beyond the Standard Model (BSM), such as slow-roll inflation, the cosmological constant, and dark sectors. In this sense, cosmological applications of supergravity can be the arena for probing outcomes of supergravity. It is also attractive that supergravity can appear as a low-energy effective theory of superstrings, a possible candidate of quantum gravity. Nevertheless, it is not trivial to build inflationary models in supergravity due to difficulties arising mainly from the extremely constrained form of supergravity scalar potentials, complicated structure of interaction terms, and excessive scalar degrees of freedom. These obstructions generally make it challenging to contrive a desirable inflationary trajectory, perform the moduli stabilization to obtain the stable de-Sitter phase, and make extra scalars to be much heavier than the Hubble scale to get single field inflation. Besides, supergravity predicts many non-renormalizable interactions. It thus arises as effective field theory (EFT) which can be valid only up to typical energies $E$ below its ultraviolet cutoff scale $\Lambda_{cut}$, and up to some accuracy of $(E/\Lambda_{cut})^n$ that we desire. We note that these non-renormalizable terms may affect physics during and/or after inflation. 

From such points of view, it is very important in supergravity to find the method for relaxing the scalar potentials, and flexible scalar field dynamics (particularly for inflaton), and examine self-consistency at the quantum level. In this thesis, therefore, we construct locally supersymmetric effective field theories of inflation by taking into account recently-proposed reformulations of $\mathcal{N}=1$ supergravity that can enlarge the space of scalar potentials. The reformulations involve {\it liberated $\mathcal{N}=1$ supergravity} and {\it new K\"{a}hler-invariant Fayet-Iliopoulos terms which do not require gauging R-symmetry}. Then, we build minimal supergravity models of single-field slow-roll inflation and de Sitter vacua in the KKLT string background in the reformulated supergravities. At the same time, we identify possible constraints on the cutoff for their self-consistency as EFT by inspecting the suppression of non-renormalizable terms within the superconformal formalism, which is very convenient to systematically manage the non-renormalizable terms.

%% file: chapters/1.tex
Cosmological inflation and its standard $\Lambda$CDM cosmology of the Early Universe are backed by the latest experimental data on the Cosmic Microwave Background (CMB) given in the Planck 2018 results \cite{Planck2018_review, Planck2018_cosmo_para,Planck2018_infl}. This data strongly supports the existence of the so-called {\it cold dark matter} (CDM) and {\it cosmological constant} ($\Lambda$), and a primordial accelerated expansion of the universe, considered to be led by a hypothetical particle called {\it inflaton} particularly in form of {\it single-field slow-roll inflation}\footnote{This will be reviewed in Ch. 2 in this thesis.}. In particular, inflation is considered as a very essential epoch of the early universe since it is a key for resolving several cosmological issues like ``the Horizon problem'' \cite{Baumann_cosmo}. It also has long been mysterious where theoretical origin of inflation comes from. Unfortunately, cosmological observations cannot theoretically be predicted solely by the Standard Model (SM) that led to the successful discovery of Higgs scalar boson in 2012 \cite{Higgs_discover}. Therefore, it is crucial to have new theories for physics beyond the Standard Mdoel (BSM) in order to tackle such problems of cosmological phenomena. 

Supersymmetry (SUSY) may be a good candidate of the BSM physics \cite{BSM_SUSY}. SUSY proposes a pairing of bosons and fermions in which fields of different spins but with the same mass belong to an irreducible representation called ``a superfield'' of the supersymmetry algebras (See Ref.~\cite{Quevedo_SUSY} for an introduction to SUSY). This means that SUSY can provide extra degrees of freedom for BSM physics in a consistent way with a theoretical robustness, so that there may exist other elementary particles for explaining the BSM sectors as superpartners of the SM bosons and fermions. 


In principle, SUSY may be realized either globally or locally. Another attracting feature of SUSY is a locally supersymmetric field theory called {\it Supergravity} (SUGRA) that can even carry the SM fundamental forces, and even gravitational interaction, thanks to its local symmetry called {\it super-Poincar\'{e} group}. It predicts a spin-3/2 fermionic superpartner of the spin-2 bosonic field {\it graviton}, which is called {\it gravitino} (See a self-contained supergravity textbook in Ref.~\cite{Superconformal_Freedman} (or Refs.~\cite{Superspace_DallAgata,SUGRAprimer}) for a detailed review of superconformal (or superspace) formalism of supergravity). 


Regretfully, experimental evidence of SUSY has never been seen until now \cite{No_SUSY}. Nevertheless, this may not be the end of the SUSY story. This is because supersymmetry may be spontaneously broken at some scale of energy, as in the case of the spontaneously symmetry breaking of the electroweak interaction by the stabilization of Higgs scalar field around the true vacuum. We call such a scale as supersymmetry breaking scale $M_S$. This implies that particles that belong to the same superfield may have different masses, implying that superpartners of the SM particles may be to be detectable. Hence, supergravity can still be a viable option for BSM physics as long as supersymmetry breaking is considered, and it can be employed at the quantum level as a so-called {\it Effective Field Theory (EFT)}, which will be explained more in the following. 

In quantum field theory (QFT), one conventionally investigates dynamical phenomena of elementary particles by computing ``convergent'' scattering (S-matrix) amplitudes (and even quantum loop integrals) of various interactions among their quanta. When a amplitude can be made with only the finite number of counter-terms, we call such theory as renormalizable. Such renormalizable interactions are also called ``relevant or marginally relevant.'' On the other hand, ``irrelevant'' couplings in a quantum field theory may push the theory to be ill-behaved at high energies due to divergences from quantum loop corrections to the amplitude constructed by such couplings. This type of QFT is called non-renormalizable. 

Notwithstanding this problematic feature, there is still a chance for non-renormalizable theory to be {\it effective} for our practical use in real world. Non-renormalizable theory can typically be considered as a theory that can be valid only up to energies below some scale $\Lambda_{cut}$ called ``{\it cutoff}.'' One then treats the divergences from the quantum loop corrections by allowing ignorance beyond this scale and dealing with the cutoff $\Lambda_{cut}$ as the actual ultraviolet cutoff on any momentum integrals in the theory. In this way, one can obtain convergent results which are valid to some accuracy of $(E/\Lambda_{cut})^n$ with some order $n$. This type of non-renormalizable theory is called as {\it Effective Field Theory} mentioned above. This will be one of the core notions in this thesis.   

Normally, a quantized theory of gravity suffers from the UV divergences of quantum loop integrals. In pure gravity, the one-loop S-matrix is finite, but the divergences first appear generically at the two-loop level. In gravity coupled to matter, the situation gets worse. The first divergences appear at the one-loop level of two graviton exchanges. For these reasons, quantization of gravity is a most difficult and ongoing issue. 


On the contrary, in supergravity, remarkable cancellation of divergences in loop integrals may take place thanks to the {\it supertrace} structure \cite{One_loop}. If supersymmetry is perfectly preserved, then the supertrace will wash out the divergences because of the same mass eigenvalues in a supermultiplet. However, in broken supersymmetry, the problem is not trivial because the mass gap of the superfields in the supermultiplet will diverge. Hence, in this sense, I comment that future investigations of supergravity divergences of quantum loop effects in broken supersymmetry (for this research as well) should be performed, but I will not deal with this issue in this thesis since it is out of its scope.

Including these issues, a quantum theory of supergravity remains one of the big puzzling questions in theoretical high energy physics. Superstring theory\footnote{In superstring theory, every elementary particle can be described by a vibrating mode of a fundamental supersymmetric string.} improved this point of view, which is a self-consistent realization of quantum gravity. This is because supergravity emerges as its low-energy effective field theory. Specifically, at energies below Planck scale $M_{pl}\equiv \sqrt{\frac{\hbar c}{8\pi G}}$, the five ten-dimensional superstring theories\footnote{Moreover, the eleven-dimensional M-theory that incorporates the five superstring models through T and S dualities can produce a 11D $\mathcal{N}=1$ supergravity.} (`Type I, $SO(32)$, $E_8\times E_8$' and `Type II-A \& B') can be approximated into 10D `$\mathcal{N}=1$' and `$\mathcal{N}=2$' supergravities\footnote{Supergravity can also be obtained by locally supersymmetric theory independently.} respectively (see Ref. \cite{Taylor} for a review of supergravity and string vacua). Much below $M_{pl}$, it is demanded for a supergravity theory to reduce to the four-dimensional (4D) $\mathcal{N}=1$ minimal supersymmetric standard model (MSSM) \cite{MSSMWorkingGroup} for phenomenological reasons\footnote{The $\mathcal{N}>1$ supergravities are less viable phenomenologically because Yukawa couplings are proportional to a gauge coupling and because matter fermions belong to either real or pseudo-real representations of gauge groups.}. In this sense, $\mathcal{N}=1$ supergravity is the intermediate link between the low energy MSSM and its ultraviolet (UV) completion done by superstring theory, thus pushing 4D $\mathcal{N}=1$ supergravity to be a strong candidate for a realistic low-energy effective description of superstring theory. 

Supergravity has recently been of great interests as a natural arena to explore inflationary cosmology \cite{Ferrara_SUGRA40}. This is because supergravity can offer us theoretical clues not only about inflation but also about the subsequent epochs of the primitive soup of elementary particles in the expanding universe. In fact, realization of an inflationary plateau by single-field inflaton and de Sitter (dS) vacua in supergravity or string theory is rather challenging \cite{Yamaguchi}. This is basically due to two main problems: one is the existence of extra scalar modes, and the other is the highly-constrained form of the so-called D and F-term scalar potentials (together with possibly quantum one-loop effective potentials) predicted by the theories. The so-called $\eta$ problem\footnote{The $\eta$ problem can be avoided by no-scale F-term scalar potential \cite{Ellis_no_scale}.} \cite{eta_problem} is a consequence of such difficulty in building inflatioanry models in the context of supergravity.

In general, it is inevitably required for one to uplift a supersymmetric Anti de Sitter (AdS) vacuum in standard 4D $\mathcal{N}=1$ supergravity to some de Sitter (dS) one (taking cancellation between terms of the potential to obtain a positive-definite but very small cosmological constant) by utilizing mechanisms that go beyond the standard supergravity. To reach single-field slow-roll inflation, it is also required for a theory to keep all extra scalars but inflaton much heavier than the Hubble scale during inflation, which is typically demanding to realize as well. 

The Kachru-Kallosh-Linde-Trivedi (KKLT) mechanism \cite{KKLT} is a prototype model to resolve the dS issue. The underlying background of KKLT model, say KKLT string background, is a no-scale 4D $\mathcal{N}=1$ supergravity with a K\"{a}hler potential for its volume modulus and with a superpotential that is induced by some possible string-theoretical nonperturbative corrections\footnote{The correactions are obtained from either Euclidean D3 branes in type IIB compactifications or gaugino condensation due to D7 branes.}. The superpotential does not allow the volume modulus to transform under a R-symmetry. Specifically, KKLT proposed a mechanism for uplifting the supersymmetric Anti-de Sitter (AdS) vacuuum to a dS vacuum by adding anti-D3 brane contribution to the superpotential in the supergravity background, making a ``bump'' which eliminates an inflationary plateau along the volume modulus in the scalar potential. As an improved modification of KKLT, Kachru, Kallosh, Linde, Maldacena, McAllister and Trivedi (KKLMMT) \cite{KKLMMT} suggested another mechanism by taking into account a contribution arising from the anti-D3 tension in a highly warped compactifications. Nevertheless, appropriate realization of inflation and moduli stabilizations has been left as an ongoing issue.

In this thesis, I will search how supergravity scalar potentials can be relaxed in a manner that allows flexible scalar field dynamics. This will be done by introducing new models of the four-dimensional $\mathcal{N}=1$ supergravity like liberated supergravity and new K\"{a}hler-invariant Fayet-Iliopoulos terms without gauging R-symmetry. Then, I will focus on how to build minimal supergravity models of the single-field slow-roll inflation and de Sitter vacua at once together with appropriate scalar stabilization around the minimum of the potential. Here, because string-theoretical motivation appears, I will consider KKLT K\"{a}hler potential and superpotentials. Moreover, I determine maximum energy scale for which such models can be valid as effective field theory. To do this, I inspect non-renormalizable interactions involved in the proposed models using the superconformal tensor calculus in order to systematically identify possible constraints on the cutoff in the efficient manner. Further investigations of quantum loop effects will be left for the future.

This thesis consists of twelve chapters including the introduction. In Ch.~\ref{ch2}, I review the basics of inflationary cosmology that I desire to study in the supergravity language in a self-contained manner. This is to figure out an overall cosmological picture of various epochs of the early universe before, during, and after inflation. In Ch.~\ref{ch3}, I explain a conceptual overview of effective field theory, which is one of the core notions carried in this research. In Ch.~\ref{ch4}, I give a brief introduction to supersymmetry, which is an underlying idea of supergravity. In Ch.~\ref{ch5}, together with appendixes \ref{Spinor_algebras} (which is about spinor algebras in general dimensions) and \ref{STC} (which is about superconformal tensor calculus), I elucidate the supergravity language used in this work by reviewing in some detail the fundamental techniques of the superconformal formalism of $\mathcal{N}=1$ supergravity. In Ch.~\ref{ch5-1}, I review the recent developments on reformed supergravities, and their inflationary models.

From Ch.~\ref{ch6} through Ch.~\ref{ch11}, I present original studies I carried out during my doctorate. In particular, over the chapters from \ref{ch6} to \ref{ch8}, I focus on the supergravity framework for investigating inflationary cosmology. Then, from Ch.~\ref{ch9} through Ch.~\ref{ch11}, I explain some proposals of minimal supergravity models of cosmological inflation. Specifically, in Ch.~\ref{ch6}, I reconstruct the equivalent action of liberated $\mathcal{N}=1$ supergravity in the ``superconformal'' formalism, which was originally constructed by Farakos et al. in the ``superspace'' formalism. In Ch.~\ref{ch7}, I revisit the superconformal actions of new Fayet-Iliopoulos (FI) terms and find some core constraints on the new FI terms. In Ch.~\ref{ch8}, I propose a new class of $\mathcal{N}=1$ supergravity called ``relaxed supergravity'' which can enlarge the space of scalar potentials. In Ch.~\ref{ch9}, I show how to build a toy model of inflation in the liberated supergravity by suggesting a special phase transition of the supersymmetry breaking scale from Planck to the electroweak scale (i.e. TeV). In Ch.~\ref{ch10}, I offer a supergravity model of inflation based on the KKLT-type supergravity with K\"{a}hler-invariant FI terms. In this model, I show that inflation may take place in a hidden sector and supersymmetry may be broken at high scale via gravity mediation, leading to soft supersymmetry breaking terms in an observable sector. In Ch.~\ref{ch11}, I study an improved supergravity model of inflation and minimal supersymmetric standard model (MSSM). In this model, single-field slow-roll inflation can be clearly realized thanks to production of extra scalars but inflaton that can be sufficiently heavier than the Hubble scale during inflation. I will also analyze the mass spectra of the visible particles in such models. Lastly, in Ch.~\ref{ch12}, I conclude this thesis and discuss some possible future developments that may arise from this research.

%% file: chapters/2.tex
In this chapter, we review inflationary cosmology in a concise but self-contained manner. This chapter is based on Ch. 1 and Ch. 2 of the lecture in Ref.~\cite{Cosmology}.

\section{Geometry of Maximally-Symmetric Spatial Slice of Spacetime}
Following the ``mostly-minus'' sign convention of metric ($\eta_{\mu\nu}=(+1,-1,-1,-1)$), the geometry of spacetime manifold of universe is represented by its line element 
\begin{eqnarray}
ds^2 = dt^2 - dl^2 \qquad  \textrm{where} \quad dl^2 \equiv \gamma_{ij}dx^idx^j,
\end{eqnarray}
where we define ``{\bf physical coordinates}'' $x^i$, ``{\bf physical distance}'' $dl^2$, and ``{\bf physical metric}'' $\gamma_{ij}$.

The three-dimensional spatial line element $dl^2$ with the metric $\gamma_{ij}$ is {\bf homogeneous} if it is invariant under the spatial translation $x'^i = x^i + a^i$ for some constant shift $a^i$, i.e. $dx'^i = dx^i$. The $dl^2$ is {\bf isotropic} if it is invariant under the spatial rotations $x'^i=R^i_j x^j$ for some rotation group $\{R^i_j\}$, i.e. $dx'^i =R^i_j dx^j$. If these two symmetries (i.e. homogeneity and isotropy) are imposed on a metric, we call this metric as ``{\bf maximally symmetric}.'' In particular, the manifolds with the maximally symmetric spatial metric $dl^2$ are given by three types of space, such as 3D-Euclidean $E^3$, 3D-Sphere $S^3$, and 3D-Hyperboloid $H^3$. By embedding a 3 dimensional manifold ($\Vec{x}$) into 4 dimensional Euclidean space ($\Vec{x},u$), we find the induced metrics of the spaces like $S^3$ and $H^3$. Let $dl^2=d\Vec{x}^2 \pm du^2$ be the 4D metric (where $+$ for $S^3$, $-$ for $H^3$). Then, from {\bf slice constraints} $\Vec{x}^2 \pm u^2 = \pm a^2$ for some constant ``$a$'' as a {\bf scale factor}, by solving $d(\Vec{x}^2 \pm u^2)=0$, we find $du = \mp \frac{1}{u} \Vec{x}\cdot d\Vec{x}$, so that $du^2=\frac{(\Vec{x}\cdot d\Vec{x})^2}{u^2}=\frac{(\Vec{x}\cdot d\Vec{x})^2}{a^2 \mp \Vec{x}^2}$. Notice that the fourth component $u$ is now fixed and expressed in terms of the coordinates $(\Vec{x})$. Hence, we have $dl^2 = d\Vec{x}^2 \pm \frac{(\Vec{x}\cdot d\Vec{x})^2}{a^2 \mp \Vec{x}^2}$. Taking a redefinition $x \longrightarrow ax$, it reduces to $dl^2 =a^2 \Big( d\Vec{x}^2 \pm \frac{(\Vec{x}\cdot d\Vec{x})^2}{1 \mp \Vec{x}^2} \Big) \equiv a^2 \Big( d\Vec{x}^2 +k \frac{(\Vec{x}\cdot d\Vec{x})^2}{1 -k \Vec{x}^2} \Big)$ where $k=0$ for $E^3$, $k=+1$ for $S^3$, and $k=-1$ for $H^3$. In spherical coordinates, $d\Vec{x}^2 = dr^2 +r^2 d\Omega^2$, $d\Omega^2 = d\theta^2 + \sin^2 \theta d\phi^2$, and $\Vec{x}\cdot d\Vec{x} = rdr + r^2 \hat{r}\cdot d\hat{r} = rdr$. Therefore, putting these into the line element, we obtain the maximally symmetric spatial metric represented by
\begin{eqnarray}
dl^2 = a^2 \left( \Big(\frac{dr}{\sqrt{1-kr^2}}\Big)^2 + r^2 d\Omega^2\right) = a^2 \Big( d\chi^2 + r^2(\chi) d\Omega^2 \Big) \equiv a^2 dl_c^2  \quad \textrm{where} \quad d\chi \equiv \frac{dr}{\sqrt{1-kr^2}},
\end{eqnarray}
\begin{figure}[t!]
    \centering
    \includegraphics[width=15cm]{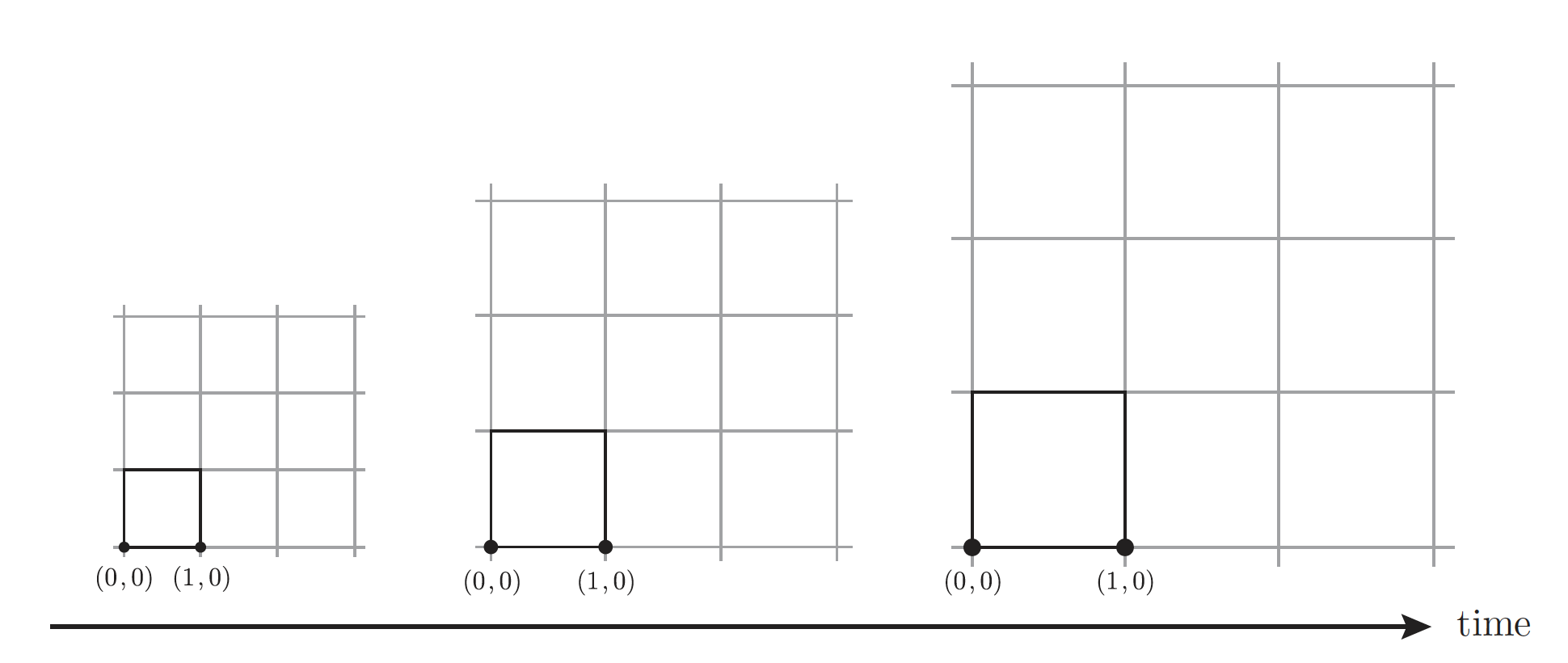}
    \caption{Evolution of comoving and physical coordinates in the expanding universe. The grid is a chart of comoving coordinates. As the universe expands, the comoving distance between the two points $(0,0)$ and $(1,0)$ remains the same, but the corresponding physical distance increases. Figure taken from \cite{Baumann_cosmo}.}
    \label{coordinates}
\end{figure}
where we defined ``{\bf comoving coordinates}'' $(\chi,\Omega(\theta,\phi))$,  ``{\bf comoving distance\footnote{Refer to Fig. \ref{coordinates}}.}'' $dl_c^2$, and ``{\bf comoving metric}'' $\textrm{diag}(1,r^2(\chi))$, and $r=r(\chi)$ is called ``{\bf comoving metric distance}''. {\it This line element means the metric of the spatial slice with a fixed scale factor $a$ and spatial curvature sign $k$ as being embedded in the 4D Euclidean space}. Moreover, {\it we can see this comoving line element $dl_c^2$ as the line element of a cylindrical coordinate system whose chart is identified with $(\rho=r(\chi),\varphi= \Omega, z=\chi)$ (with $dl_c^2 |_{\textrm{fixed}~\rho} = \underbrace{d\rho^2}_{=0} + \rho^2 d\varphi^2 + dz^2$) when $\rho=r(\chi)$ is fixed}.

Metric distance $d_m\equiv r(\chi)$ is calculated as follows: from the definition
\begin{eqnarray}
d\chi \equiv \frac{dr}{\sqrt{1-kr^2}} \implies  d_m \equiv r(\chi)=
\begin{cases}
E^3~(k=0):  \chi \\
S^3~(k>0):  \frac{1}{\sqrt{k}}\sin(\sqrt{k}\chi)\\
H^3~(k<0):  \frac{1}{\sqrt{|k|}}\sinh(\sqrt{|k|}\chi)\\
\end{cases}
\end{eqnarray}

\section{Friedmann–Lemaître–Robertson–Walker (FLRW) Geometry of Expanding Universe}

The so-called ``{\bf Friedmann–Lemaître–Robertson–Walker (FLRW) metric}'' is defined by the maximally-symmetric spacetime geometry consisting of time line element $dt^2$ and maximally-symmetric spatial line element $dl^2$ where the scale factor $a$ now depends on time variable $t$. Thus, the FLRW metric is given by 
\begin{eqnarray}
ds^2 = dt^2 - a^2(t) \Big( d\chi^2 + r^2(\chi) d\Omega^2 \Big).
\end{eqnarray}
Note that the physical distance $dl^2$ must be different according to the value of the scale factor $a(t)$, while the comoving distance $dl_c^2$ must be the same for any $a(t)$. 

Let us consider one ``{\it physical observer $O_P$ (who uses the chart of physical coordinates)}'' and one ``{\it comoving observer $O_C$ (who uses the chart of comoving coordinates)}'', and some dynamical object in space. For comoving distance $r_c$ and scale factor $a(t)$, {\bf physical velocity} $v_{phy}$ that the physical observer $O_P$ measures is given by the time derivative of physical distance $r_{phy}(t)=a(t)r_c(t)$
\begin{eqnarray}
    v_{phy} \equiv \frac{dr_{phy}}{dt} =\underbrace{\frac{da}{dt}r_c}_{\equiv~ v_{H}} + \underbrace{a \frac{dr_c}{dt}}_{\equiv~ v_{pec}} = v_H + v_{pec} . 
\end{eqnarray}
Notice that physical velocity $-c\leq v_{phy} < \infty$ of the object that $O_P$ measures is equal to the sum of {\bf Hubble flow or recession velocity} $0 \leq v_H < \infty$ (= velocity of a comoving virtual circle of radius $r_c$ that the comoving observer $O_C$ measures), and {\bf peculiar velocity} $-c\leq v_{pec} \leq c$ (= velocity of the dynamical object in comoving coordinates that the comoving observer $O_C$ measures). Also, we note that while $v_{pec}$ obeys the principle of constant speed of light by relativity, the physical velocity does not follow it anymore. This is basically due to the fact that the recession velocity is the velocity of not an object but expanding space itself. Plus, remark that $v>0$ is away from the observer, while $v<0$ is approaching to the observer.

In particular, the Hubble flow velocity is observed by the physical observer $O_P$ as follows: defining $H(t) \equiv \frac{\dot{a}(t)}{a(t)}$ which is called ``{\bf Hubble parameter},'' the physical observer measures the following in physical coordinates
\begin{eqnarray}
v_{H} \equiv \underbrace{ \frac{da}{dt}r_c = \dot{a} r_c = Har_c }_{\textrm{measured by}~O_C} = \underbrace{Hr_{phy}}_{\textrm{measured by}~O_P}
\end{eqnarray}
Especially, if a dynamical object is fixed at the comoving grids (i.e. $\frac{dr_c}{dt}=0$) but apart from the observer by $r_{phy}$, then its physical velocity is given solely by the Hubble flow velocity $v_H = H r_{phy}$. Also, the peculiar velocity can be measured by the physical observer in the way that 
\begin{eqnarray}
v_{pec} = \underbrace{a\frac{dr_c}{dt}}_{\textrm{measured by}~O_C} = \underbrace{v_{phy} - v_H}_{\textrm{measured by}~ O_P}.
\end{eqnarray}

Next, there is an useful concept called ``{\bf comoving Hubble horizon (or Hubble radius)}'' defined as {\it the radius of a comoving virtual circle whose recession velocity is luminal $c=1$}, so that 
\begin{eqnarray}
v_H = Har_c  \overset{!}{=} c=1 \implies r_c = (aH)^{-1}.
\end{eqnarray}
For example, if some object exists inside of the Hubble radius, then its recession velocity is sub-luminal. On the contrary, if some object exists outside of the Hubble radius, then its recession velocity is super-luminal. 

Here is an important remark. If the comoving distance between two objects is either equal to or greater than the Hubble radius, then they can never communicate with each other by exchanging photons since the physical velocity of the photons they emit is measured as $v_{phy} \geq 0$ by the two objects. This means that the photons are always propagating ``away'' from the other object in practice. Hence, {\it the comoving Hubble horizon or Hubble radius can also be defined as a maximal comoving distance only inside of which particles are able to do ``physical interaction'' between them in the causally-connected way.} For instance, thermal equilibrium of particles can take place only if all of them are inside of the same comoving Hubble horizon. This property will be related to the so-called ``Horizon problem'' about the {\it uniform} Cosmic Microwave Background (CMB) radiations.

Meanwhile, we can reexpress the spacetime metric by introducing a {\bf conformal or comoving time} ``$\tau$'' given by $\tau \equiv t / a(t)$. By inserting this into the metric, we find 
\begin{eqnarray}
ds^2 =a^2(\tau) \underbrace{[d\tau^2 - (d\chi^2 + r^2(\chi)d\Omega^2)]}_{\textrm{static/comoving FLRW metric}}
\end{eqnarray}
where $a(\tau)$ is now a time-dependent scale factor. This metric form is convenient for studying the propagation of light.

Next, let us consider propagation of light in the expanding universe. In this case, it is convenient to use the conformal coordinates. Assume that we look at a fixed solid angle $\Omega$. Then, the light travels in the radial direction $d\chi$. The corresponding metric is given by $ds^2 = a^2(\tau)[d\tau^2-d\chi^2]$. Since photons travel along the null geodesic (i.e. $ds^2=0$), we have the relation $d\tau^2=d\chi^2$, so that $d\chi=\pm d\tau$. Then, we have lightcone now. For the observer at a conformal time $\tau$ at the center of the lightcone, the $+$ corresponds to null-geodesic outgoing from the observer, while $-$ corresponds to null-geodesic incoming to the observer. Then, we can compute the comoving distance $\Delta \chi$ between two times $\tau_A,\tau_B$ (or $t_A,t_B$):
\begin{eqnarray}
\Delta \chi = \chi(\tau_B) - \chi(\tau_A) = \int_{\tau_A}^{\tau_B} d\tau = \int_{t_A}^{t_B}  \frac{dt}{a(t)} = \int_{a_A}^{a_B} \frac{da}{a\dot{a}} = \int_{\ln a_A}^{\ln a_B} \frac{d\ln a}{\dot{a}} = \int_{\ln a_A}^{\ln a_B} \frac{1}{aH}d\ln a. 
\end{eqnarray}
where $t_i (\tau_i)$ is some past (conformal) time, and $t_f (\tau_f)$ is some future (conformal) time. Notice that comoving distance depends on the evolution of Hubble radius $(aH)^{-1}$.

\begin{itemize}
    \item {\bf Causal influence:} Only comoving particles whose worldlines intersect the past lighcone of an observer at $p$ can {\it send} a signal to the observer at $p$.  Only comoving particles whose worldlines intersect the future lighcone of an observer at $p$ can {\it receive} a signal from the observer at $p$.
    \item {\bf (comoving) particle horizon} ``$\chi_{ph}(\tau)$'' is defined by the maximal comoving distance where the observer at $\tau$ can {\it receive} the null-geodesic signals from the past events at $\tau_i$. 
    \begin{eqnarray}
    \chi_{ph}(\tau) \equiv \int_{\tau_i}^{\tau} d\tau = \int_{t_i}^{t}  \frac{dt}{a(t)} = \int_{\ln a_i}^{\ln a} \frac{1}{aH}d\ln a
    \end{eqnarray}
    In particular, this particle horizon is related to the horizon problem of standard Big Bang cosmology. The case of $a_i=0$ is called ``{\bf Big Bang singularity}.'' Plus, we also call particle horizon as ``{\bf causal contact or patch}'' of the observer.
    \item  {\bf (comoving) event horizon} ``$\chi_{eh}(\tau)$'' is defined by the maximal comoving distance where the observer at $\tau$ can {\it send} the null-geodesic signals to the future events at $\tau_f$.
    \begin{eqnarray}
    \chi_{eh}(\tau) \equiv \int_{\tau}^{\tau_f} d\tau = \int_{t}^{t_f}  \frac{dt}{a(t)} = \int_{\ln a}^{\ln a_f} \frac{1}{aH}d\ln a.
    \end{eqnarray}
    \item Every observer has his own particle and event horizons.
\end{itemize}

\section{Geodesic Motion and Redshift in FLRW Spacetime}

In general, Tthe geodesic equation is given by
\begin{eqnarray}
\frac{d^2x^{\mu}}{d\tau} + \Gamma_{\alpha\beta}^{\mu} \frac{dx^{\alpha}}{d\tau}\frac{dx^{\beta}}{d\tau} =0  \Longleftrightarrow 
\begin{cases}
u^{\alpha} \nabla_{\alpha}u^{\mu} = 0 \quad \textrm{where} \quad \nabla_{\alpha}u^{\mu} \equiv \partial_{\alpha}u^{\mu} + \Gamma_{\alpha\beta}^{\mu}u^{\beta} \\
p^{\alpha} \nabla_{\alpha}p^{\mu} = 0 \quad \textrm{where} \quad \nabla_{\alpha}p^{\mu} \equiv \partial_{\alpha}p^{\mu} + \Gamma_{\alpha\beta}^{\mu}p^{\beta} 
\end{cases}
\end{eqnarray}
where $u^{\mu} \equiv dx^{\mu}/d\tau$ is the 4-velocity; $p^{\mu} \equiv mu^{\mu}$ is the 4-momentum, and $\Gamma_{\alpha\beta}^{\mu} \equiv \frac{1}{2}g^{\mu\lambda}(\partial_{\alpha}g_{\lambda\beta}+\partial_{\beta}g_{\lambda\alpha}-\partial_{\lambda}g_{\alpha\beta})$ is Christoffel symbol.

The FLRW metric is $g_{\mu\nu} = \textrm{diag}(1,-a^2\gamma_{ij})$ and its inverse is  $g^{\mu\nu} = \textrm{diag}(1,-a^{-2}\gamma^{ij})$. The only non-vanishing Christoffel symbol components are 
\begin{eqnarray}
\Gamma_{ij}^0 = a\dot{a}\gamma_{ij}, \quad \Gamma_{0j}^i = \frac{\dot{a}}{a}\delta_j^i, \quad \Gamma_{jk}^i = \frac{1}{2} \gamma^{il}(\partial_j \gamma_{lk} + \partial_{k} \gamma_{lj} -\partial_l \gamma_{jk}).\label{FLRW_Gammas}
\end{eqnarray}
In fact, due to homogeneity ($p^{\mu}(x')=p^{\mu}(x+a) \approx p^{\mu}(x)+a^i\partial_i p^{\mu} \overset{!}{=} p^{\mu}(x)$, so that $\partial_i p^{\mu}=0$), the geodesic equation of any massless/massive particle in the FLRW spacetime reduces to
\begin{eqnarray}
p^0 \frac{dp^{\mu}}{dt} = - (2\Gamma_{0j}^{\mu}p^0 + \Gamma_{ij}^{\mu}p^i)p^j.
\end{eqnarray}
For $\mu=0$ component, we have 
\begin{eqnarray}
E\frac{dE}{dt} = -a\dot{a}\gamma_{ij}p^ip^j = -a\dot{a} \Big(-\frac{1}{a^2}g_{ij}\Big) = \frac{\dot{a}}{a} g_{ij}p^ip^j = \frac{\dot{a}}{a} |\Vec{p}|^2. \label{Geodesic}
\end{eqnarray}
From the Einstein relation $E^2 = |\Vec{p}|^2 + m^2$, 
\begin{eqnarray}
E \frac{dE}{dt} = |\Vec{p}|\frac{d|\Vec{p}|}{dt} \label{Einstein}
\end{eqnarray}
By combining \eqref{Geodesic} and \eqref{Einstein}, we obtain ``{\bf 3-momentum decay in FLRW spacetime}'' given by
\begin{eqnarray}
\frac{d|\Vec{p}|}{|\Vec{p}|} = - \frac{da}{a} \implies |\Vec{p}| \propto \frac{1}{a}  \implies
\begin{cases}
|\Vec{p}| = E \propto \frac{1}{a} \\
|\Vec{p}| = \gamma m |\Vec{v}| \propto \frac{1}{a}, \label{three_momentum_decay}
\end{cases}
\end{eqnarray}
where $\Vec{v}$ is 3-velocity as ``peculiar velocity,'' and $\gamma$ is the Lorentz factor. The result means that the physical 3-momentum $|\Vec{p}|$ of any massless/massive particle decays with the expansion of the universe, so that the peculiar velocity becomes small as well. Due to this, the physical velocity will be close to the Hubble flow speed (i.e. recession velocity $v_H$). 

Remarkably, combining the result of ``3-momentum decay'' \eqref{three_momentum_decay} and de Broglie wave ($\lambda = h/|\Vec{p}|$) deduces ``{\bf red-shifting of wave}''
\begin{eqnarray}
\lambda \propto a \quad  \Longleftrightarrow \quad  \frac{\lambda_f}{\lambda_i} = \frac{a_f}{a_i}.
\end{eqnarray}
In particular, let us consider that light with wavelength $\lambda_1$ is emitted from a distant source at $t=t_1$. If an observer detects the ``redshifted'' wavelength $\lambda_0$ of the light at time $t_0$, then a redshifted parameter $z$ is defined by
\begin{eqnarray}
z \equiv \frac{\lambda_0-\lambda_1}{\lambda_1} = \frac{\lambda_0}{\lambda_1} -1 \geq 0 \quad \implies \quad  z = \frac{a_0}{a_1}-1 \quad \textrm{for red-shifting in FLRW metric},\label{z1}
\end{eqnarray}
where we usually set $a(t_0)=1$. Notice that any scale factor $a(t_1)$ at a given time $t_1$ can be represented in terms of the redshift parameter $z$, i.e.
\begin{eqnarray}
a(t_1) = (1+z)^{-1},
\end{eqnarray}
which means that as $z$ gets large, $a(t_1)$ gets small and we get close to the past. Moreover, by Taylor-expanding $a(t_1)$, we get 
\begin{eqnarray}
a(t_1) \approx a(t_0) + \frac{da(t_0)}{dt}(t_1-t_0) = a_0 + H(t_0) a(t_0) (-d_p) = a(t_0) [1-H_0d_p].\label{z2}
\end{eqnarray}
Combining \eqref{z1} and \eqref{z2}, up to the first order in $t$, we obtain
\begin{eqnarray}
z \approx H_0 d_p,
\end{eqnarray}
where $H_0 \equiv H(t_0)$ is Hubble parameter at ``today'' and $d_p \equiv c(t_0-t_1)$ for $c=1$ is physical distance.

\section{Einstein Field Equations in FLRW Metric: 1st and 2nd Fridemannn, and Continuity Equations}

The dynamics of the universe is determined by the Einstein field equation 
\begin{eqnarray}
G_{\mu\nu} = 8\pi G T_{\mu\nu},
\end{eqnarray}
where $G_{\mu\nu}\equiv R_{\mu\nu} -\frac{1}{2}R g_{\mu\nu}$ is the Einstein tensor (where $R_{\mu\nu}$ is Ricci tensor and $R$ is Ricci scalar), $T_{\mu\nu}$ is the energy-momentum tensor, and $G$ is the gravitational constant. For the FLRW spacetime, we require homogeneity and isotropy. Thus, isotropy under $x'^i=R_j^ix^j$ requires the average of 3-vector in time to vanish: $\left<v^i\right>=0 \implies v^i=0$. Particularly, for isotropy around $x^i=0$, average of a rank-2 3-tensor $T^{ij}$ can be proportional to identity tensor, i.e. $\left<T^{ij}\right> \propto \delta^{ij} \approx g^{ij}$. Homogeneity under $x'^i=x^i+a^i$ requires a scalar to be only a function of time in order for its average to vanish: $\left<S\right>=0 \implies S=S(t) $. Due to these, we must consider that 
\begin{eqnarray}
&& \textrm{By isotropy:}\qquad  \left<T_{i0}\right>=0 \implies T_{i0}=0,\\
&& \textrm{By isotropy around ~x=0:}\qquad  \left<T_{ij}\right> = \left<Cg_{ij}\right>=\left<C(-a^2\delta_{ij})\right> \propto \delta_{ij} \nonumber\\
&& \qquad \qquad \qquad \qquad\qquad \qquad  \implies \left<C\right>=0 \implies C=-P(t),\\
&& \textrm{By homogeneity:}\qquad  \left<T_{00}\right>=0  \implies T_{00} = \rho(t),\label{homo_isotropy_condition}
\end{eqnarray}
where $\rho$ is interpreted as {\it energy density} and $P(t)$ is interpreted as {\it pressure}. Therefore, the energy-momentum tensor is given by
\begin{eqnarray}
T_{\mu\nu} = 
\begin{pmatrix}
\rho(t) & 0 \\
0 & -P(t)g_{ij}\\
\end{pmatrix}
\implies
T^{\mu}_{~\nu} = g^{\mu\lambda}T_{\lambda\nu}= \textrm{diag}(\rho(t),-P(t),-P(t),-P(t))~ \textrm{in comoving frame},\nonumber\\{} \label{stress_tensor_in_comoving}
\end{eqnarray}
where $g^{\mu\lambda} = \textrm{diag}(1,g^{mn})$. In fact, this is the stress-tensor of a ``perfect fluid'' seen by a comoving observer. $\rho(t)$ and $P(t)$ are measured for a fluid at rest in the comoving coordinates. The relative 4-velocity of the perfect fluid is $u^{\mu}  =(1,\Vec{0})^T,~u_{\nu}=(1,\Vec{0})$, which gives a matrix $u^{\mu}u_{\nu}= \textrm{diag}(1,0_{3\times 3})$. When rewriting the stress tensor as
\begin{eqnarray}
T^{\mu}_{~\nu} =  \textrm{diag}(\rho(t),-P(t),-P(t),-P(t)) = (\rho(t)+P(t)) \cdot \textrm{diag}(1,0_{3\times 3}) -P(t) \cdot   \textrm{diag}(1,1_{3\times 3}),
\end{eqnarray}
by plugging the 4-velocity back to the tensor, we can easily find the general form of the stress tensor which is valid in any coordinate system 
\begin{eqnarray}
T^{\mu}_{~\nu} = (\rho(t)+P(t))u^{\mu}u_{\nu} -P(t) \delta_{\nu}^{\mu} \quad \textrm{in any reference frame},
\end{eqnarray}
where $u^{\mu} = dx^{\mu}/d\tau$ is defined by the relative 4-velocity between the fluid and observer. 

The continuity equation is then given by
\begin{eqnarray}
\nabla_{\mu}T^{\mu}_{~\nu} = \partial_{\mu}T^{\mu}_{~\nu} + \Gamma_{\mu\lambda}^{\mu} T^{\lambda}_{~\nu} - \Gamma_{\mu\nu}^{\lambda} T^{\mu}_{~\lambda} = 0.\label{continuity_eq}
\end{eqnarray}
In comoving frame, by plugging $\Gamma_{0j}^{i}=\frac{\dot{a}}{a}\delta_{j}^{i}$ and Eq.~\eqref{stress_tensor_in_comoving} into Eq.~\eqref{continuity_eq} for $\nu=0$, we find a {\bf continuity equation of the stress tensor in FLRW metric}
\begin{eqnarray}
\nabla_{\mu}T^{\mu}_{~0} = \partial_0T^0_{~0} + \Gamma_{i0}^i T^0_{~0} - \Gamma_{i0}^{j} T^i_{~j} =0 \implies \frac{d\rho}{dt} + 3\frac{\dot{a}}{a}(\rho+P) =0.
\end{eqnarray}
Moreover, we may represent the pressure $P(t)$ in terms of the energy density $\rho(t)$ using the {\bf equation of state}
\begin{eqnarray}
\omega \equiv \frac{P}{\rho},
\end{eqnarray}
where $\omega$ is a constant. Hence, we obtain ``{\bf energy density evolution}'' given by
\begin{eqnarray}
&& \frac{d\rho}{dt} + 3(1+\omega)\frac{\dot{a}}{a}\rho =0 \implies \rho(t) = A [a(t)]^{-3(1+\omega)} \Longleftrightarrow \rho(t) a(t)^{3(1+\omega)} = A = \textrm{constant} \nonumber\\
&&  \rho(t)   \propto 
\begin{cases}
a^{-3} \quad ~\textrm{for matter}~\omega = 0 \\
a^{-4} \quad ~\textrm{for radiation}~\omega = 1/3 \\
a^0\quad ~\textrm{for vacuum/dark energy}~\omega = -1. \label{energy_density_evolution}
\end{cases}
.
\end{eqnarray}

Next, let us compute the components of Ricci tensor in the FLRW metric. We find that by isotropy, the 3-vectors vanish, i.e.
\begin{eqnarray}
R_{i0} = R_{0i} = 0.
\end{eqnarray}
Using the Christoffel symbol components \eqref{FLRW_Gammas} (here, i.e. $\Gamma_{0j}^{i}=\frac{\dot{a}}{a}\delta_{j}^{i}$), we get 
\begin{eqnarray}
R_{00} = -\partial_0 \Gamma_{0l}^l - \Gamma_{0l}^r \Gamma_{r0}^l = -3\frac{\ddot{a}}{a}.
\end{eqnarray}
In addition, taking the limit of $\gamma_{ij}$ around $\vec{x}=0$ as
\begin{eqnarray}
\gamma_{ij} \equiv \delta_{ij} + k \bigg(\frac{x_ix_j}{1-k x_kx^k}\bigg) \approx \delta_{ij} + k (x_ix_j) \implies \gamma^{jk} \approx \delta^{jk} - k(x^jx^k),
\end{eqnarray}
we find $\Gamma_{jk}^i \approx k x^i \delta_{jk}$ and thus 
\begin{eqnarray}
R_{ij}(x) \approx (a\ddot{a}+2\dot{a}^2 + 2k + kx^2)\delta_{ij}-k^2 x_ix_j.
\end{eqnarray}
At $x^i=0$, it reduces to
\begin{eqnarray}
R_{ij}(x=0) = (a\ddot{a}+2\dot{a}^2 + 2k)\delta_{ij} =
-\frac{1}{a^2}(a\ddot{a}+2\dot{a}^2 + 2k)g_{ij}(x=0),
\end{eqnarray}
where we used $\gamma_{ij}(x=0) \approx \delta_{ij}$ and $g_{ij}= -a^2\gamma_{ij}$. Since the spatial dependence of the Ricci tensor comes from that of the metric tensor $g_{ij}$, $R_{ij}(x=0)$ holds for general $x$ as well, so that
\begin{eqnarray}
R_{ij}(x) = - \Big(\frac{\ddot{a}}{a}+2 \left(\frac{\dot{a}}{a}\right)^2+2\frac{k}{a^2}\Big) g_{ij}(x).
\end{eqnarray}
Then, it is straightforward to compute the Ricci scalar $R$ 
\begin{eqnarray}
R = g^{\mu\nu}R_{\mu\nu} = g^{00}R_{00} + g^{ij}R_{ij} = -6 \Big(\frac{\ddot{a}}{a}+ \left(\frac{\dot{a}}{a}\right)^2+\frac{k}{a^2}\Big).
\end{eqnarray}
Therefore, by inserting the values of Ricci tensor components and Ricci scalar into the Einstein tensor ($G_{\mu\nu}\equiv R_{\mu\nu} -\frac{1}{2}Rg_{\mu\nu}$), we find all of its components as follows:
\begin{eqnarray}
G^0_{~0} = 3 \Big(\left(\frac{\dot{a}}{a}\right)^2+\frac{k}{a^2}\Big), \quad G^i_{~j} = \Big(2\frac{\ddot{a}}{a}+ \left(\frac{\dot{a}}{a}\right)^2+\frac{k}{a^2}\Big)\delta_{j}^i, \quad G^{i}_{~0}=G^0_{~j}=0.
\end{eqnarray}
Finally, we are ready to derive the Einstein field equations in the FLRW metric. The so-called ``{\bf 1st Fridemannn equation}'' is obtained from $00$-component Einstein equation:
\begin{eqnarray}
G^{0}_{~0} = 8\pi G T^0_{~0} \implies H^2 = \left(\frac{\dot{a}}{a}\right)^2 = \frac{8\pi G}{3}\rho - \frac{k}{a^2}.
\end{eqnarray}
Plus, the so-called ``{\bf 2nd Fridemannn equation}'' is obtained from $ij$-component Einstein equation:
\begin{eqnarray}
G^{i}_{~j} = 8\pi G T^i_{~j} \implies  \frac{\ddot{a}}{a} = -\frac{4\pi G}{3}(\rho+3P).
\end{eqnarray}

\section{Dimensionless Density Parameter, Cosmic Energy Budget, and $\Lambda$CDM}

The so-called ``{\bf critical density}'' $\rho_c$ is defined by the {\it total energy density at today} ($t_0$): $\rho_c(t_0) \equiv \sum_I\rho_I(t_0)$ for some species $I=r,m,\Lambda$ (i.e. radiation, matter, dark energy, respectively). In fact, this critical density can be obtained by the 1st Fridemannn equation
\begin{eqnarray}
H^2 = \frac{8\pi G}{3}\rho -\frac{k}{a^2} \approx \frac{8\pi G}{3}\rho \implies \rho(t=t_0) = \frac{3H(t_0)^2}{8\pi G} \equiv \rho_c(t_0),\label{critical_density}
\end{eqnarray}
where $\rho(t)$ is the total energy density at $t$, and we used $|k| \ll 1$. Then, we define ``{\bf dimensionless density parameter}'' $\Omega_I$ for a certain specie $I$ as {\it a ratio of the $I$'s density at today $t_0$ to the critical density $\rho_c$ at today $t_0$}, i.e.
\begin{eqnarray}
\Omega_I \equiv \frac{\rho_I(t_0)}{\rho_c(t_0)} = \frac{\rho_I(t_0)}{\sum_I\rho_I(t_0)}.\label{density_parameter}
\end{eqnarray}
Taking this definition, we can find alternative of the 1st Fridemannn equation in terms of the dimensionless density parameters. Using Eq. \eqref{critical_density}, we reach
\begin{eqnarray}
H^2 = \frac{\rho(t)}{\rho_c(t_0)} H_0^2 - \frac{k}{a(t)^2}= \sum_I\frac{\rho_I(t)}{\rho_c(t_0)} H_0^2 - \frac{k}{a(t)^2}.
\end{eqnarray}
Since we have seen that $\rho(t) \propto a(t)^{-3(1+\omega)} \implies  \rho(t)a(t)^{3(1+\omega)} = \textrm{constant}$, defining $\rho_I \propto a(t)^{-3(1+\omega_I)}$, we have
\begin{eqnarray}
\rho_I(t) =\rho_I(t_0) \left(\frac{a(t_0)}{a(t)}\right)^{3(1+\omega_I)} ,
\end{eqnarray}
which gives rise to
\begin{eqnarray}
H^2(t) =  H_0^2 \bigg[ \sum_I \Omega_I \left(\frac{a(t_0)}{a(t)}\right)^{3(1+\omega_I)} + \Omega_k \left(\frac{a(t_0)}{a(t)}\right)^2\bigg] \equiv H_0^2 \sum_{\alpha=I,k} \Omega_{\alpha}\left(\frac{a(t_0)}{a(t)}\right)^{3(1+\omega_{\alpha})} ,
\end{eqnarray}
where we also introduced a new definition of the $k$, i.e. $\Omega_k \equiv -\frac{k}{(a_0H_0)^2}$. Then, defining $a(t_0)=1$ for today, it reduces to the final form of the alternative of the 1st Fridemannn equation
\begin{eqnarray}
\frac{H^2(t)}{H_0^2} = \sum_{\alpha=I,k} \Omega_{\alpha} a(t)^{-3(1+\omega_{\alpha})} = \Omega_r a^{-4} + \Omega_m a^{-3} + \Omega_k a^{-2} + \Omega_{\Lambda}.\label{Alter_Fridemannn}
\end{eqnarray}
The cosmological observation predicts the following ``{\bf cosmic energy budget}'' that 
\begin{eqnarray}
|\Omega_k| \leq 0.01, \qquad \Omega_r \approx 9.4 \times 10^{-5}, \qquad \Omega_m \approx 0.32, \qquad \Omega_{\Lambda} \approx 0.68,
\end{eqnarray}
where the smallness of the first budget $\Omega_k$ is called ``{\bf Flatness Problem}.'' In particular, it is remarkable that the matter contribution to the parameter in fact does not come from solely the ordinary matter (i.e. baryons mostly) since it was measured as $\Omega_b \approx 0.05$. Hence, it is inevitable to assume that there exists a novel type of matter called ``{\bf Dark Matter\footnote{Dark matter does not do electromagnetic interaction since it has no its gauge charges. In this sense, it is called ``dark.''} (DM)}'' for compensating the difference in the density parameter of matter. We thus consider the sum $\Omega_m = \Omega_{b} + \Omega_{CDM}$, and obtain $\Omega_{CDM}\approx 0.27$, which is the case of {\bf Cold\footnote{This means that the velocity is ``non-relativistic.''} Dark Matter (CDM)}. 

Next, by taking advantage of the alternative of the 1st Fridemannn equation, we are going to investigate {\it the behavior of scale factor $a(t)$ in (conformal) time} according to the species. In fact, the different scalings of the energy densities of radiation ($\rho \propto a^{-4}$), matter ($\rho \propto a^{-3}$), and dark energy ($\rho \propto a^{0}$) imply that the universe was dominated by a single component for most of the history of our universe. Given a single species $I$ only, the Fridemannn equation reduces to
\begin{eqnarray}
H^2 = H_0^2 \Omega_I a^{-3(1+\omega_I)} \Longleftrightarrow a^{-1+\frac{3}{2}(1+\omega_I)} da = H_0 \sqrt{\Omega_I} dt. 
\end{eqnarray}
Then, the corresponding solutions to this are given as follows: when $\omega_I \neq -1$,
\begin{eqnarray}
a(t) = \bigg[ \frac{3(1+\omega_I)}{2}\sqrt{\Omega_I} H_0 (t-t_0) + 1 \bigg]^{\frac{2}{3(1+\omega_I)}} = \bigg[ \frac{3(1+\omega_I)}{2}\sqrt{\Omega_I} H_0t \bigg]^{\frac{2}{3(1+\omega_I)}} \propto t^{\frac{2}{3(1+\omega_I)}},
\end{eqnarray}
where we used $a(t=0)=0$ at which Big Bang occurs, and $t_0 = \frac{2}{3(1+\omega_I)\sqrt{\Omega_I}H_0}$. When $\omega_I = -1$, we find $a(t) = e^{H_0\sqrt{\Omega_I}(t-t_0)} \propto e^{H_0\sqrt{\Omega_I}t} \approx e^{Ht}$. Therefore, we have
\begin{eqnarray}
a(t) \propto \begin{cases}
t^{1/2} \textrm{  for radiation-dominated (RD)}, \quad i.e.\quad \omega_r=1/3 \\
t^{2/3} \textrm{  for matter-dominated (MD)}, \quad i.e. \quad \omega_m=0 \\
e^{Ht} \textrm{  for $\Lambda$-dominated ($\Lambda$D)}, \quad i.e.  \quad\omega_{\Lambda}=-1 \\
\end{cases}.
\end{eqnarray}
After Big Bang at $t=0$, expansion of the universe was first dominated by radiation, and then matter, and has been being dominated by dark energy until now. It is also possible to re-express the above equation in conformal time $\tau$. Using $dt = ad\tau$, we find 
\begin{eqnarray}
a^{-2+\frac{3}{2}(1+\omega_I)}da = H_0\sqrt{\Omega_I} d\tau.
\end{eqnarray}
The solutions to this are given by
\begin{eqnarray}
a(\tau) = \bigg[ \frac{(1+3\omega_I)}{2}H_0 \sqrt{\Omega_I} \tau \bigg]^{\frac{2}{(1+3\omega_I)}} \propto \tau^{\frac{2}{(1+3\omega_I)}}.\label{Comove_Evolution_scale_factor}
\end{eqnarray}
Thus, 
\begin{eqnarray}
a(\tau) \propto \begin{cases}
\tau^{2} \textrm{  for radiation-dominated (RD)}, \quad i.e.\quad \omega_r=1/3 \\
\tau \textrm{  for matter-dominated (MD)}, \quad i.e. \quad \omega_m=0 \\
-\tau^{-1} \textrm{  for $\Lambda$-dominated ($\Lambda$D)}, \quad i.e.  \quad\omega_{\Lambda}=-1 \\
\end{cases}.
\end{eqnarray}
The evolution of the universe that is made by the cosmic energy budget we have seen is called ``{\bf $\Lambda$CDM standard cosmology}.''

\section{Horizon Problem and Inflation as Its Solution}

In the previous section, we have seen that the alternative form of the 1st Fridemannn equation is given by Eq.~\eqref{Alter_Fridemannn}. Now, we are going to see how Hubble radius $(aH)^{-1}$ evolves as the universe expands. Let us consider a single component species with $\omega$. Then, the corresponding dimensionless density parameter can be set to unity (i.e. $\Omega=1$), and thus the Fridemannn equation gives the evolution of Hubble radius 
\begin{eqnarray}
\frac{H^2}{H_0^2} = a^{-3(1+\omega)} \implies (aH)^{-1} = H_0^{-1}a^{(1+3\omega)/2}. \label{Hubble_radius_evolution}
\end{eqnarray}
We observe that since all the matter sources satisfy the ``{\bf strong energy condition (SEC), i.e. $1+3\omega >0$},'' the corresponding Hubble radius increases as the universe expands. Next, by inserting the evolution of the scale factor in Eq.~\eqref{Comove_Evolution_scale_factor} for the single component species with $\Omega=1$, i.e. $a(\tau) = \bigg[ \frac{(1+3\omega_I)}{2}H_0 \tau \bigg]^{\frac{2}{(1+3\omega_I)}}$, into Eq.~\eqref{Hubble_radius_evolution}, we obtain the evolution of Hubble radius in comoving time
\begin{eqnarray}
(aH)^{-1}  =  \frac{(1+3\omega)}{2} \tau \implies (aH)^{-1} \propto \begin{cases}
\qquad\tau \qquad \textrm{for Radiation-dominated ($\omega_r=1/3$)} \\
\qquad \tau/2 \qquad \textrm{for Matter-dominated ($\omega_m=0$)} \\
\qquad - \tau \qquad \textrm{for Dark energy-dominated ($\omega_{\Lambda}=-1$)} \\
\end{cases},
\end{eqnarray}
which also implies that the particle horizon $\chi_{ph}(\tau)$ is given by
\begin{eqnarray}
\chi_{ph}(\tau) = \tau = \frac{2}{(1+3\omega)}(aH)^{-1}.
\end{eqnarray}
In the meantime, from $a(\tau) = \bigg[ \frac{(1+3\omega_I)}{2}H_0 \tau \bigg]^{\frac{2}{(1+3\omega_I)}}$, we find 
\begin{eqnarray}
\tau = \frac{2H_0^{-1}}{(1+3\omega)}a^{(1+3\omega)/2}.
\end{eqnarray}
Notice that {\it when $(1+3\omega)>0$ by the strong energy condition, the initial conformal time $\tau_i$ at the Big Bang singularity (i.e. $a(\tau_i)\equiv 0$) becomes zero, which also means that the size of the particle horizon at the singularity in our past lightcone is zero}. This {\it vanishing particle horizon at the singularity} means that past events at the Big Bang singularity has never existed that were able to send signals to us at present. However, this result seriously contradicts our observation of {\bf cosmic microwave background (CMB)} radiation!

Regarding CMB, about 380000 years ago after the standard ``Hot'' Big Bang, the universe had cooled down sufficiently to allow ``{\bf Recombination}'' process where {\it hydrogen atoms were formed and photons were decoupled from the primordial plasma of the hydrogen atoms}. We call such photons created from the decoupling during the recombination process as ``{\bf CMB radiation}'' since we have observed them in form of CMB. In the meantime, according to the observation, the CMB radiation is almost perfectly isotropic with anisotropies in the CMB temperature being much smaller than the average of the CMB temperature $ T_{CMB} \sim 10^{-4} \textrm{eV} \sim 2.725 K$, i.e. $\delta T \ll T_{CMB}$. 

The uniform temperature of the CMB radiation implies that they must be in thermal equilibrium in causal contact in the same particle horizon in the past. Therefore, the discovery of the uniform CMB radiation supports that there did exist the past events that were able to send the CMB photons to us at present, and thus particle horizon at the Big Bang singularity must be ``spacious'' enough to include all the possible causal contacts/patches\footnote{Here, the  ``{\it causal contacts/patches}'' are defined by particle horizons of signals that an observer has received.} of the CMB photons that we have seen! 

\begin{figure}[t!]
    \centering
    \includegraphics[width=15cm]{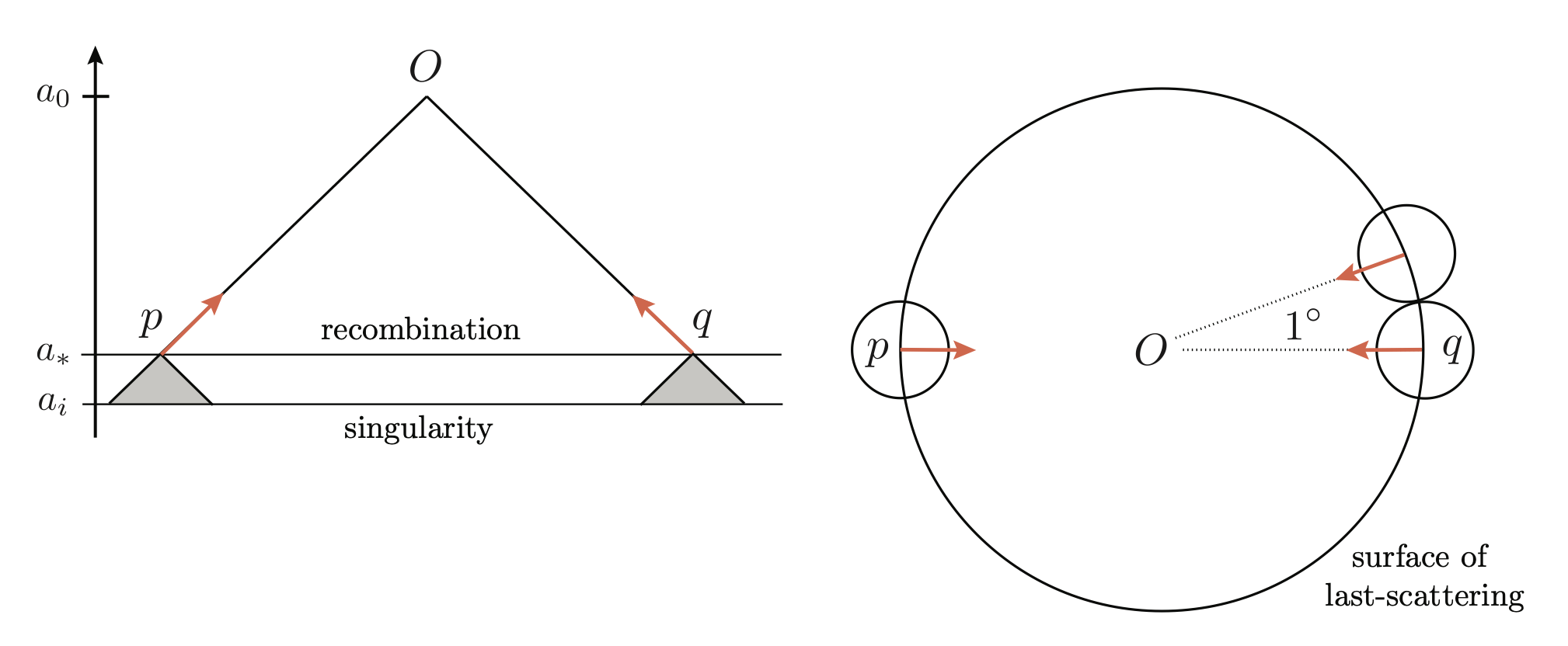}
        \includegraphics[width=15cm]{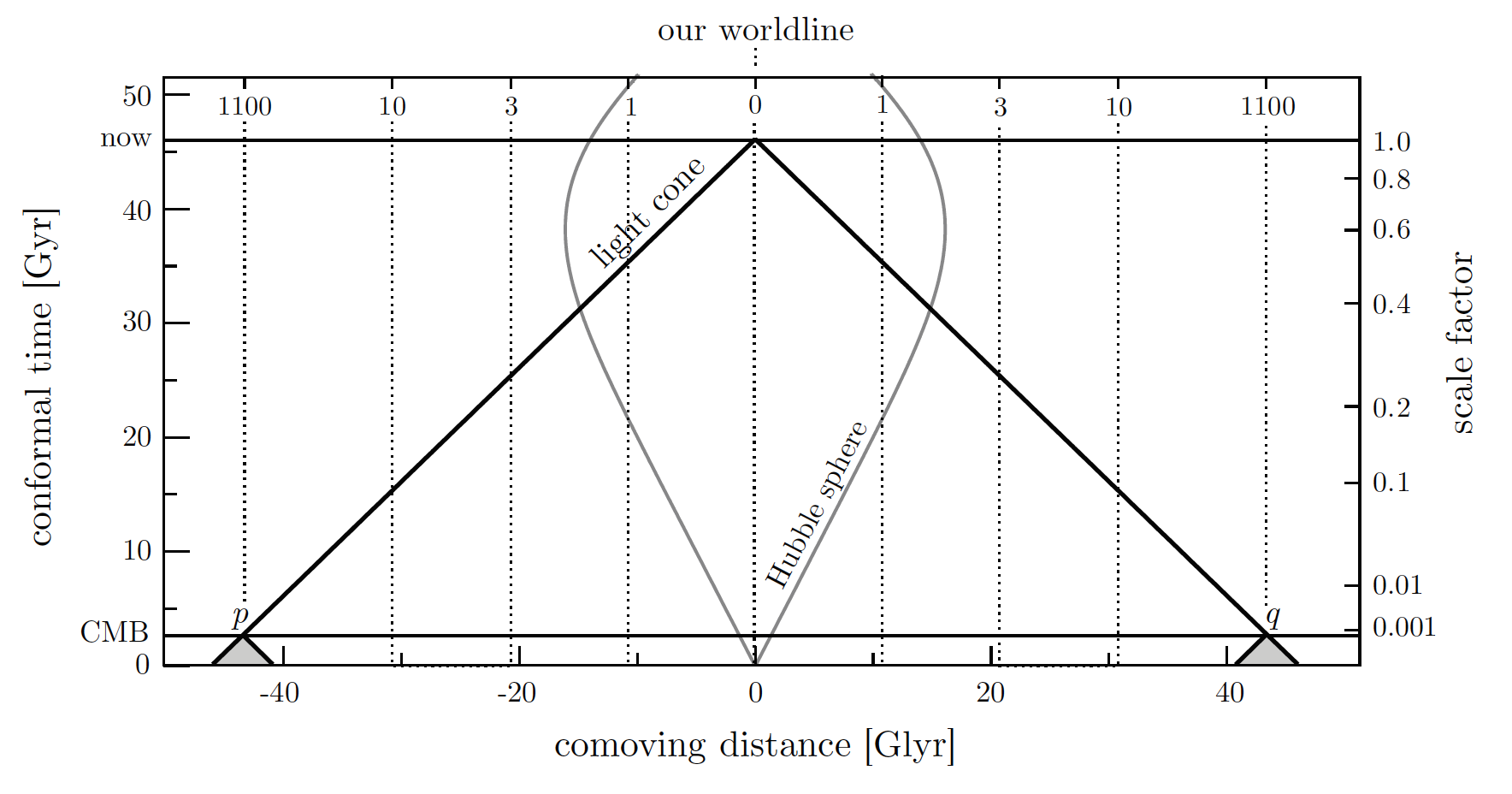}
    \caption{(Horizon problem) At the last-scattering surface of recombination, any two points $p$ and $q$ that are apart from each other by more than one degree seem to be not in causal contact and outside of Hubble radius since their past lightcones do not overlap before the singularity and the Hubble radius at singularity is zero, so that they seem causally-disconnected. However, {\it uniform} CMB temperature has been measured with very small anisotropies over the sky, implying that CMB radiations should have had thermal equilibrium in some causal contact in the past. The upper figure is taken from \cite{Baumann}, and the lower figure is taken from \cite{Baumann_cosmo}.}
    \label{Horizon_problem}
\end{figure}

Now we face a critical problem due to the discordance between the {\it vanishing particle horizon} (obtained from the ordinary sources satisfying the strong energy condition) and {\it requirement of non-vanishing particle horizon (which is predicted by our observation of the uniform CMB radiation)} at the Big Bang singularity in the past. This discordance issue is called ``{\bf Horizon Problem}'' in Fig.~\ref{Horizon_problem} in cosmology.

Certainly, given the result of the vanishing particle horizon, the CMB photons may seem to be sent from causally-disconnected patches of particle horizon. Hence, for our observation of the uniform CMB temperature to be valid, there must be ``long enough'' conformal time for the causally-disconnected patches to be causally-connected inside of {\it the common single non-vanishing particle horizon at the singularity in the past}. This is a starting point to resolve the Horizon problem.

\begin{figure}[t!]
    \centering
    \includegraphics[width=15cm]{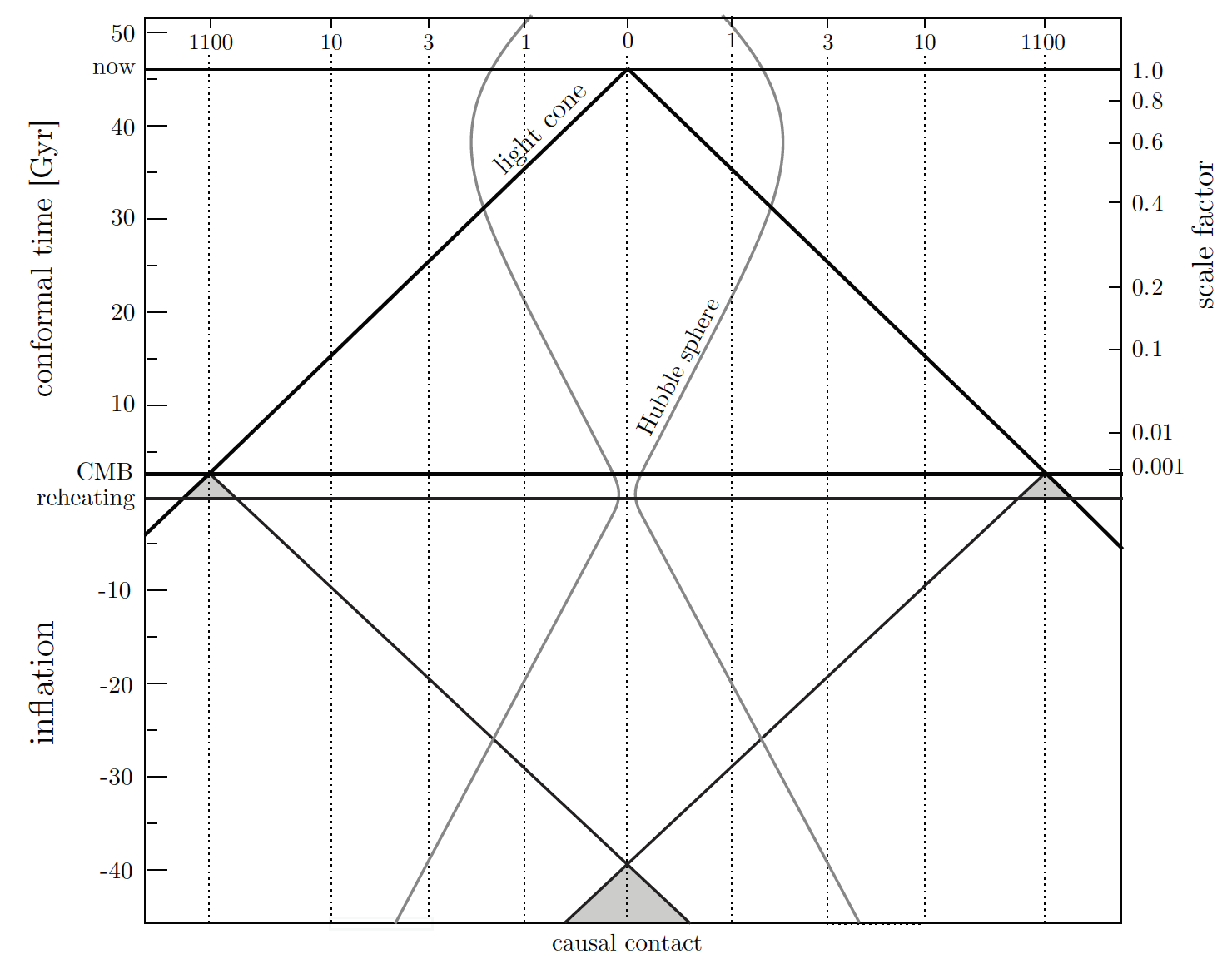}
    \caption{(Inflation as a key solution to the Horizon problem) An era of a so-called inflation is introduced to push the singularity to the far past, so that the CMB radiations could be in causal contact inside of the Hubble radius at some moment before singularity. Figure taken from \cite{Baumann_cosmo}.}
    \label{pushing_singularity}
\end{figure}

In respect of this, we need to move the initial singularity $\tau_i$ to the further far point in the past to obtain a past lightcone where the common causal contact can be established. This is illustrated in Fig.~\ref{pushing_singularity}. To do this, we push the singularity to negative conformal time to have such an enough time 
\begin{eqnarray}
\tau_i = \frac{2H_0^{-1}}{(1+3\omega)}a_i^{(1+3\omega)/2} \implies \lim_{a_i\rightarrow 0} \tau_i= -\infty.
\end{eqnarray}
This type of $\tau_i$ can only exist when there is a {\bf SEC-violating fluid} such that
\begin{eqnarray}
1+3\omega < 0.\label{SEC-violating}
\end{eqnarray}
Now we ask: how can we get such SEC-violation giving the proper singularity? In fact, the simple answer to this is to conjecture that there was a phase of decreasing Hubble radius in the early universe. Let us recall the evolution of Hubble radius \eqref{Hubble_radius_evolution}, and take its time derivative:
\begin{eqnarray}
(aH)^{-1} = H_0^{-1}a^{(1+3\omega)/2} \implies \frac{d}{dt}(aH)^{-1} = H_0^{-1}\frac{(1+3\omega)}{2}a^{3\omega/2}.
\end{eqnarray}
Then, since we are interested in the SEC-violation \eqref{SEC-violating} (i.e. $(1+3\omega)<0$), we conclude that we need to conjecture that a {\bf shrinking Hubble sphere or radius} existed in the past
\begin{eqnarray}
\frac{d}{dt} (aH)^{-1} < 0,
\end{eqnarray}
which exactly enables us to have much conformal time in the negative direction beyond $\tau=0$. This is because the initial singularity of $a_i=0$ is now placed at $\tau_i = -\infty$! Moreover, the shrinking Hubble sphere condition means ``{\bf acceleration of the scale factor of space}'' or the so-called ``{\bf Inflation}'' because 
\begin{eqnarray}
\frac{d}{dt}(aH)^{-1} = \frac{d}{dt} (a\frac{\dot{a}}{a})^{-1} = \frac{d}{dt}(\dot{a}^{-1}) = - \dot{a}^{-2} \ddot{a} <0 \implies \ddot{a} > 0.
\end{eqnarray}
That is, it is required that accelerated expansion of the universe occurred in the past. This is why we call inflation as a period of acceleration.

Here, we are finally able to define a sufficient time interval between $\tau=0$ and a certain instant $\tau_I<0$ (such that $0 > \tau_I > \tau_i$) at which the common causal contact or particle horizon of the CMB photons could exist in the past lightcone of an observer in the past thanks to the shrinking Hubble sphere. This special time interval $\Delta \tau = 0-\tau_I$ during which inflation has lasted is called ``{\bf Era of Inflation}.'' Hence, $\tau=0$ is no longer the initial singularity but instead it becomes only ``a critical point of phase transition from inflation before $\tau=0$ to the standard Hot Big Bang after $\tau=0$,'' which is called ``{\bf Reheating}''.

Equivalently, the shirinking Hubble sphere condition deduces that 
\begin{eqnarray}
\frac{d}{dt}(aH)^{-1} &=& -(aH)^{-2} \frac{d}{dt}(aH) = -(aH)^{-2}[\dot{a}H+a\dot{H}]\nonumber\\
&=& -\frac{a}{(aH)^2}[H^2+\dot{H}]=-\frac{1}{a}\Big(1+\frac{\dot{H}}{H^2}\Big) = -\frac{1}{a}(1-\varepsilon) < 0 \implies \varepsilon < 1 
\end{eqnarray}
where we define ``{\bf 1st slow-roll parameter}''
\begin{eqnarray}
\varepsilon \equiv - \frac{\dot{H}}{H^2} = -\frac{d\ln H}{H dt} < 1.
\end{eqnarray}
In fact, this additionally implies the ``{\bf almost-constant Hubble parameter}'' during inflation
\begin{eqnarray}
\varepsilon < 1 \implies \textrm{For perfect inflation, } \varepsilon=0 \implies \dot{H} = 0 \ll 1 \implies H = \textrm{constant in time}.\label{Constant_Hubble}
\end{eqnarray}

Meanwhile, recalling the definition of Hubble parameter, we find the solution for the scale factor 
\begin{eqnarray}
H \equiv \frac{\dot{a}}{a} = \frac{1}{a}\frac{da}{dt} = \frac{d\ln a}{dt} \implies a(t_f) = a(t_i) \exp\Big(\int_{t_i}^{t_f}H(t)dt\Big).
\end{eqnarray}
Now we assume that the Hubble parameter is almost constant during $\Delta t = t_f -t_i$. Then, we have
\begin{eqnarray}
a(t_f) = e^{H\Delta t} \equiv e^N a(t_i), \qquad N \equiv H\Delta t,
\end{eqnarray}
where $N$ is called ``{\bf The number of e-folds of inflation}.'' In particular, it takes $t_H \equiv \Delta t = H^{-1}$ for the scale factor to get $e \approx 2.72$ times for $N=1$ (one e-fold). We call such $t_H$ as ``{\bf Hubble expansion time}.''

The next question here is how much change of the comoving particle horizon $\Delta \chi$ can be large as the number of e-folds $N$ increases. To do this, let us introduce a {\bf time-independent characteristic scale} ``$\lambda$'' which is defined as {\it the characteristic comoving distance and, at the same time, the increasing unit of particle horizon as $N$ increases} such that:
\begin{eqnarray}
\Delta\chi = N\lambda \Longleftrightarrow \lambda \equiv \frac{\Delta \chi}{N} = \frac{\Delta\chi}{H\Delta t} = \frac{a}{da/dt}\frac{d\chi}{dt}= a \frac{d\chi}{da} \Longleftrightarrow ad\chi = \lambda da, \label{chracteristic_scale_lambda}
\end{eqnarray}
where we see that the given comoving particle horizon is $N$-times of the characteristic scale $\lambda$. Moreover, from the definition of Hubble parameter, we can find the time element
\begin{eqnarray}
H = \frac{\dot{a}}{a}= \frac{1}{a} \frac{da}{dt} \implies dt = (aH)^{-1} da. \label{time_element}
\end{eqnarray}
Note that we can consider the {\bf Hubble horizon (or radius)} $(aH)^{-1}(t)$ as {\it the comoving distance over which particles can travel for one Hubble (expansion) time $H^{-1}(t)$ and interact with each other at a given moment ``$t$'' within the next Hubble time}. 

By putting \eqref{chracteristic_scale_lambda} and \eqref{time_element} into the FLRW metric (i.e. $ds^2 = dt^2 - a^2 d\chi^2$), we obtain its equivalent form and can determine {\bf causality} of the scale $\lambda$ with respect to the Hubble radius $(aH)^{-1}$:
\begin{eqnarray}
&& ds^2 = dt^2 - a^2 d\chi^2 = [(aH)^{-2}-\lambda^2]da^2 
\nonumber\\
&& \implies 
\begin{cases}
\textrm{Subhorizon: } (aH)^{-1} > \lambda \implies ds^2 >0~(\textrm{ timelike = causally-connected })\\
\textrm{On-horizon: } (aH)^{-1} = \lambda \implies ds^2 =0~(\textrm{ null-like = causally-connected } )\\
\textrm{Superhorizon: } (aH)^{-1} < \lambda \implies ds^2 <0~(\textrm{ spacelike = causally-disconnected } )\\
\end{cases}
\end{eqnarray}
That is, consider that two particles are separated by a comoving distance $\lambda$ at a given moment ``$t_i$.'' Then, the Hubble radius ``$(aH)^{-1}(t_i)$'' at the time can be used for judging whether they can interact (i.e. start to send a signal at the given time $t_i$ and receive it at some later time $t_f$) with each other at the given time $t_i$ within one Hubble time (i.e. $\Delta t = t_f-t_i \leq  t_H=H^{-1}$). That is, if $\lambda > (aH)^{-1}(t)$, then they are causally-disconnected and thus cannot interact with each other within one Hubble time. On the contrary, the particle horizon $\chi_{ph}$ can be used for judging whether they were able to interact with each other at a given past moment $t_{i'}$. That is, if $\lambda > \chi_{ph}$ at some past time, then they have never been able to interact with each other.

Now we are ready to investigate how inflation of the early universe in the past can wash out the Horizon problem. Since for the observer at today $t_0$, the Hubble horizon around the observer is $(a_0H_0)^{-1}$, the characteristic scale $\lambda$ must be less than the present Hubble horizon for the initial particle horizon $\lambda$ to be causally-connected to the observer at $t_0$: i.e. $0 \leq \lambda \leq (a_0H_0)^{-1}$. We observe that $(a_0H_0)^{-1}$ is the maximum of $\lambda$. This means that the initial particle horizon $\lambda$ must be either less than or equal to $(a_0H_0)^{-1}$, which is the comoving size of the observable universe. Next, let us imagine that we go to the starting point of inflation. Then, for the observer at the inflation-starting time $t_I$, the Hubble sphere is given by $(a_IH_I)^{-1}$. For any $\lambda$ of the regions $\lambda \leq (a_0H_0)^{-1}$ to be causally-connected to the observer at $t_I$, it is needed to impose that $ \lambda (\leq (a_0H_0)^{-1}) \leq (a_IH_I)^{-1}$. Therefore, we find the sufficient condition for inflation 
\begin{eqnarray}
(a_0H_0)^{-1} \leq (a_IH_I)^{-1}.\label{HR_inflation_condition}
\end{eqnarray}

\begin{figure}[t!]
    \centering
    \includegraphics[width=15cm]{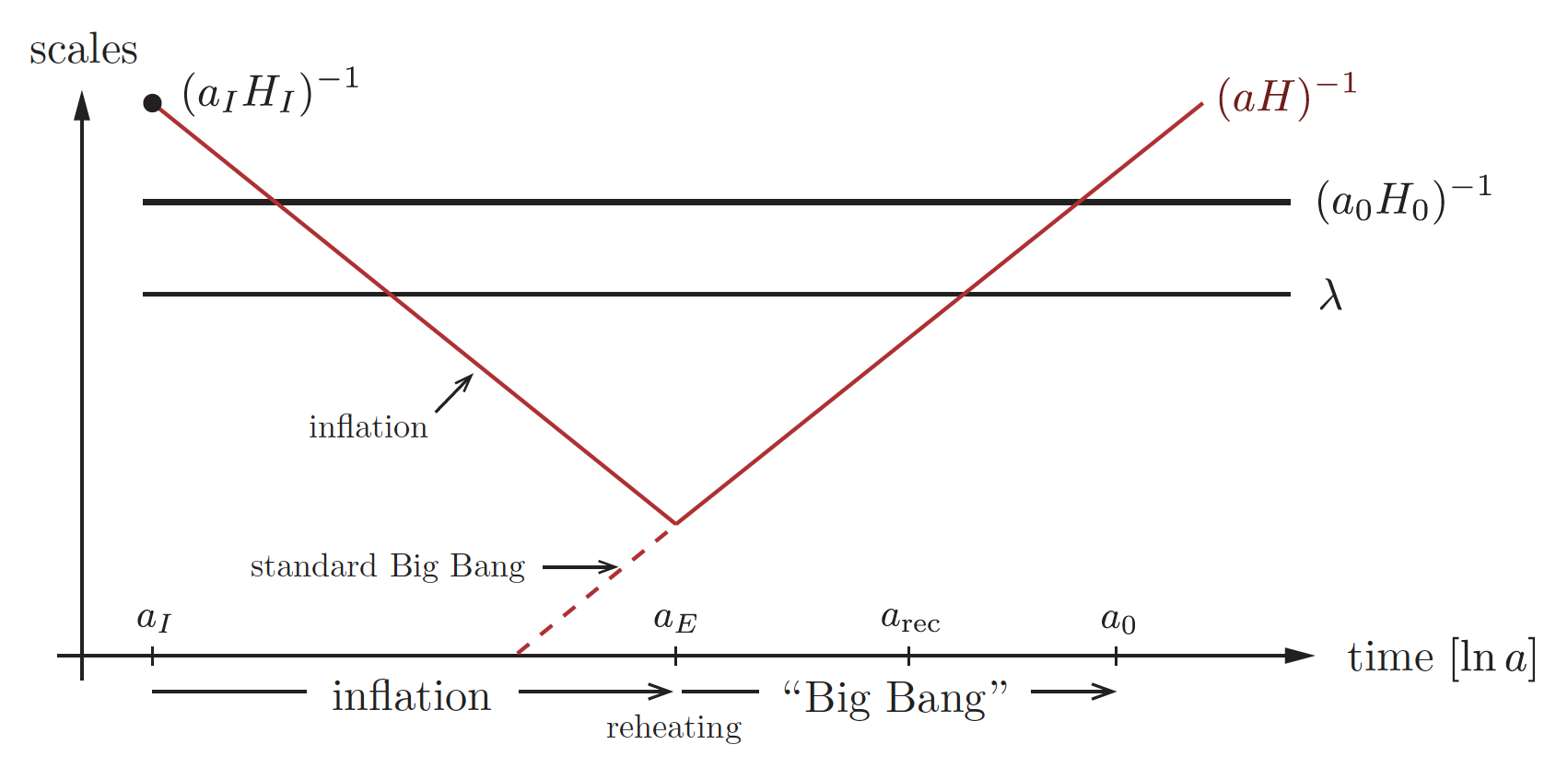}
    \caption{Hubble horizon (or radius) hierarchy for inflation to be able to resolve the Horizon problem in a consistent way. Figure taken from \cite{Baumann_cosmo}.}
    \label{inflation_duration}
\end{figure}
Here we comment that since normally particle horizon $\chi_{pl}=\lambda+\Delta \chi = (1+N)\lambda$ (except for the initial particle horizon $\lambda$ at $t_I$) can be larger than the Hubble horizon $(aH)^{-1}$ all the times, it is more conservative to use the inflation Hubble horizon $(a_IH_I)^{-1}$ as a means of judging the horizon problem. 

{\bf We claim that if the entire observable universe (i.e. $\lambda \lesssim (a_0H_0)^{-1}$) was within the comoving Hubble horizon at the beginning of inflation (i.e. $\lambda \lesssim (a_0H_0)^{-1} \lesssim (a_IH_I)^{-1}$ as shown in Fig.~\ref{inflation_duration}), then there is no Horizon problem.}

\section{Conditions for Inflation and Its Duration}

In this section, we investigate the conditions for inflation in detail. How long should inflation have lasted? The answer to this question will give us how to satisfy the first inflation condition given by Eq. \eqref{HR_inflation_condition}. To estimate the duration of inflation, let us assume that at the end of inflation at $t=t_E$, the universe was radiation-dominated. Hence, recalling Eq.~\eqref{Alter_Fridemannn} and putting $\Omega=1$ and $\omega_r = 1/3$ into Eq.~\eqref{Alter_Fridemannn}, we can consider the following single-component radiation-dominated evolution of $H$: 
\begin{eqnarray}
H = H_0 \left(\frac{a_0}{a}\right)^{2}.
\end{eqnarray}
Setting $t=t_E$, we find 
\begin{eqnarray}
\frac{H_E}{H_0} = \frac{a_0^2}{a_E^2} \implies \frac{a_EH_E}{a_0H_0} = \frac{a_0}{a_E}.
\end{eqnarray}
From the energy density evolution in Eq.~\eqref{energy_density_evolution}, we have seen that the radiation-dominated case corresponds to $\rho \sim a^{-4}$. In addition, we know that energy density of radiation is proportional to the fourth power of temperature $T$ according to {\bf Stefan-Boltzman Law} (i.e. $\rho(T) \propto T^4$. This leads to the fact that $a^{-1} \propto T$. Hence, the above relation can reduce to 
\begin{eqnarray}
\frac{a_EH_E}{a_0H_0} = \frac{a_0}{a_E} = \frac{T_0}{T_E}.
\end{eqnarray}
In particular, we have observation of $T_0$ and $T_E$, which are $T_0 \sim 10^{-4}\textrm{~eV}$ (which is today's CMB temperature) and $T_E \sim 10^{15} \textrm{~GeV}=10^{24}\textrm{~eV}$ at the end of inflation. Therefore, we find \begin{eqnarray}
\frac{a_EH_E}{a_0H_0} = \frac{a_0}{a_E} = \frac{T_0}{T_E} = 10^{-28} \implies (a_0H_0)^{-1} = 10^{28} (a_EH_E)^{-1}.
\end{eqnarray}
By inserting this result into the condition \eqref{HR_inflation_condition}, we get 
\begin{eqnarray}
(a_0H_0)^{-1} = 10^{28} (a_EH_E)^{-1} \leq (a_IH_I)^{-1}.
\end{eqnarray}
Since we have seen that $H \sim \textrm{constant in time}$ from Eq.~\eqref{Constant_Hubble} during inflation, we have $H_I \approx H_E$, and thus the inequality leads to the condition on the number of e-folds for inflation
\begin{eqnarray}
 \frac{a_E}{a_I} \geq 10^{28} \implies \ln \left( \frac{a_E}{a_I}\right) = N \geq 28\ln 10 \sim 64 \implies N > 64,
\end{eqnarray}
where we used $a_E = e^N e_I$. Also, from the result of $(a_0H_0)^{-1} \sim 10^{28}(a_EH_E)^{-1}$, we get $a_E = e^{64}a_0$. This tells us that inflation must long last over the number of e-folds greater than 64! 

In fact, there are other equivalent forms of the condition for inflation:
\begin{itemize}
    \item 1. Shrinking Hubble Sphere: $\dfrac{d}{dt}(aH)^{-1} <0$.
    \item 2. Accelerated Expansion: $\ddot{a} >0$.
    \item 3. Slowly-varying Hubble parameter: $\varepsilon = -\dfrac{\dot{H}}{H^2} = -\dfrac{d\ln H}{dN} < 1$. This condition means that the fractional change of the Hubble parameter (i.e. $d\ln H = dH/H$) is slowly-varying over the period of one e-fold $N=1$.
    \item 4. Quasi-de Sitter expansion: $\varepsilon \approx 0 \ll 1 \implies H \approx \textrm{constant in time}$. This means that 
    \begin{eqnarray}
     H = \frac{\dot{a}}{a} \implies a(t) \propto e^{Ht}.
    \end{eqnarray}
    \item 5. Negative pressure: $\omega < -1/3$. By differentiating the 1st Fridemann equation with $k=0$ (spatially-flat) with respect to time and using the continuity equation $\dot{\rho}=-3H(\rho+P)$, we get $\dot{H}= -4\pi G(\rho+P)$. Then, by adding $H^2$ to this, we get 
    \begin{eqnarray}
     \dot{H} + H^2 = - \frac{H^2}{2}\Big(1+\frac{3P}{\rho}\Big) \implies -\frac{\dot{H}}{H^2}-1 = \frac{1}{2}\Big(1+\frac{3P}{\rho}\Big) \implies \varepsilon -1 = \frac{1}{2} + \frac{3}{2} \frac{P}{\rho},
    \end{eqnarray}
    which gives 
    \begin{eqnarray}
     \varepsilon = \frac{3}{2}\Big(1+\frac{P}{\rho}\Big) = \frac{3}{2}(1+\omega) < 1 \implies \omega < -1/3.
    \end{eqnarray}
    As we can see, inflation requires ``negative pressure'' (or equivalently a violation of the strong energy condition). 
    \item 6. Constant density: $\bigg|\dfrac{d\ln \rho}{d\ln a}\bigg|= 2\varepsilon < 1$. From the continuity equation, we have 
    \begin{eqnarray}
     \frac{d\rho}{dt} = -3H\rho(1+\omega) \implies \frac{d\rho}{\rho} = -3\frac{\dot{a}}{a}(1+\omega)dt \implies d\ln\rho = -3(1+\omega)d\ln a.
    \end{eqnarray}
    Thus, we have 
    \begin{eqnarray}
     \frac{d\ln\rho}{d\ln a} = -3(1+\omega) = -2\varepsilon \implies \bigg| \frac{d\ln\rho}{d\ln a}\bigg| = 2\varepsilon < 1.
    \end{eqnarray}
    For small $\varepsilon$, $\ln\rho$ will be constant in $\ln a$. Thus, the energy density is thus nearly constant during expansion. However, conventional matter sources all dilute with expansion; for example, $\rho_r \propto a^{-4}$ and $\rho_m \propto a^{-3}$. This means that we need to explore a new matter source that is beyond the conventional matter form, which will be identified with a real scalar quantum field called ``{\bf Inflaton}''
\end{itemize}

We have seen that in a given FLRW spacetime with the Hubble parameter $H$, cosmological inflation can take place if and only if $\varepsilon < 1$ holds. Thus, it is needed to have a sufficiently long time for $\varepsilon$ to be small enough during inflation. To acquire this, we have to consider $\varepsilon$ to remain small for a large number of e-folds of inflation, e.g. over 60 e-folds. In this respect, we introduce a new parameter $\eta$ where the information of slowly-varying $\varepsilon$ over the e-folds is encoded. This $\eta$ parameter is defined as a ratio of fractional change of $\varepsilon$ to change of the number of e-folds, so that
\begin{eqnarray}
 \eta \equiv \frac{d\ln \varepsilon}{dN} = \frac{d\ln \varepsilon}{Hdt}= \frac{\dot{\varepsilon}}{H\varepsilon}   \Longleftrightarrow \varepsilon_f = \varepsilon_i e^{\eta \Delta N}.
\end{eqnarray}
Since we are interested in $\varepsilon_{i},\varepsilon_f\ll 1$, it is required to impose $\eta \Delta N \rightarrow 0$. This means that for any $\Delta N$, $\eta \rightarrow 0$, which is equivalent to 
\begin{eqnarray}
 |\eta |\ll 1.
\end{eqnarray}
In summary, we call the two variables $\varepsilon$ and $\eta$ as ``{\bf Hubble slow-roll parameters}'':
\begin{eqnarray}
 \varepsilon \equiv -\frac{\dot{H}}{H^2} = -\frac{d\ln H}{dN}  < 1, \qquad \eta \equiv \frac{d\ln\varepsilon}{dN}=\frac{\dot{\varepsilon}}{H\varepsilon}\quad \textrm{with} \quad |\eta|<1.
\end{eqnarray}

\section{Scalar Field Dynamics of ``Inflaton'': Slow-Roll Inflation}

In this section, we discuss a new matter form given by a real scalar field $\phi(t,\vec{x})$ called ``{\bf Inflaton}'' and its scalar field dynamics. Let us assume that there exists a real scalar field $\phi$ and this has kinetic and potential energies given by
\begin{eqnarray}
 \mathcal{L} = \frac{1}{2} g^{\mu\nu}\partial_{\mu}\phi \partial_{\nu}\phi - V(\phi).
\end{eqnarray}
If the energy-momentum tensor of the inflaton dominates the universe, it sources the evolution of the FLRW backgroud. Hence, it is desirable to know under which conditions the inflaton can lead to the accelerated expansion, i.e. inflation. The corresponding energy-momentum tensor is given by
\begin{eqnarray}
 T^{\mu}_{~\nu}=g^{\mu\lambda}T_{\lambda\nu} = \partial^{\mu}\phi\partial_{\nu}\phi -\delta^{\mu}_{\nu} \Big(\frac{1}{2} g^{\alpha\beta}\partial_{\alpha}\phi \partial_{\beta}\phi - V(\phi)\Big),
\end{eqnarray}
where we used $T_{\lambda\nu} \equiv \frac{\partial \mathcal{L}}{\partial \partial_{\lambda}\phi}\partial_{\nu} -g_{\lambda\nu}\mathcal{L}$. In particular, due to the homogeneity and isometry of the FLRW spacetime \eqref{homo_isotropy_condition}, we impose that the inflaton field is given by only a function of time: $\phi(t)$. Then, from Eq.~\eqref{stress_tensor_in_comoving} in the comoving frame, the components of the corresponding energy-momentum tensor are given by
\begin{eqnarray}
T^0_{~0} = \frac{1}{2}\dot{\phi}^2 + V(\phi) = \rho_{\phi}, \qquad T^i_{~j} = \delta^i_{j} \Big( -\frac{1}{2}\dot{\phi}^2 + V(\phi) \Big) = -P_{\phi}\delta^i_{j}  
\end{eqnarray}
or equivalently
\begin{eqnarray}
\rho_{\phi} = \frac{1}{2}\dot{\phi}^2 + V(\phi), \qquad P_{\phi} = \frac{1}{2}\dot{\phi}^2 -V(\phi).
\end{eqnarray}
Then, the negative pressure condition is 
\begin{eqnarray}
\omega < -1/3 \Longleftrightarrow \omega = \frac{P_{\phi}}{\rho_{\phi}} = \frac{\frac{1}{2}\dot{\phi}^2 -V(\phi)}{\frac{1}{2}\dot{\phi}^2 + V(\phi)} < -\frac{1}{3} \implies \frac{1}{2}\dot{\phi}^2 < V(\phi).
\end{eqnarray}

Next, let us study the evolution of the inflaton field $\phi(t)$. Using the relation $8\pi G = M_P^{-2}$ (where $M_{P} \equiv \sqrt{\frac{\hbar c}{8\pi G}} =1/ \sqrt{8\pi G}$ is ``{\bf reduced Planck constant}'' for $\hbar=c=1$), we can re-express the 1st Fridemann equation with spatially-flat condition $k=0$
\begin{eqnarray}
H^2 = \frac{\rho}{3M_P^2}
\end{eqnarray}
and we get 
\begin{eqnarray}
H^2 = \frac{\rho_{\phi}}{3M_P^2} = \frac{1}{3M_P^2}\Big(\frac{\dot{\phi}^2}{2}+V(\phi)\Big).\label{1st_eq_inflaton}
\end{eqnarray}
From the continuity equation $\dot{\rho}_{\phi}=-3H(\rho_{\phi}+P_{\phi})$, the left-hand side of this is given by $\dot{\phi}\ddot{\phi}+\dot{\phi}V'(\phi)$ where $V'(\phi)\equiv dV(\phi)/d\phi$, while the right-hand side of the equation is given by $-3H\dot{\phi}^2$. Therefore, we obtain the equation of motion for the inflaton from the continuity equation as follows:
\begin{eqnarray}
\ddot{\phi} + 3H\dot{\phi} + \frac{dV(\phi)}{d\phi} = 0.\label{2nd_eq_inflaton}
\end{eqnarray}
Notice that this is Klein-Gordon equation for the inflaton with a friction term $3H\dot{\phi}$ caused by the expansion of the universe. 

Next, let us compute the slow-roll parameters $\varepsilon$ and $\eta$. By taking the time derivative of the 1st Fridemann equation and using the continuity equation, we obtain
\begin{eqnarray}
&& H^2 = \frac{\rho_{\phi}}{3M_P^2} \implies 2H\dot{H}=\frac{\dot{\rho}_{\phi}}{3M_P^2} = \frac{1}{3M_P^2}[-3H(\rho_{\phi}+P_{\phi})] = -\frac{H}{M_{P}^2}(\rho_{\phi}+P_{\phi}) \nonumber\\
&& \implies \dot{H} = -\frac{1}{2M_P^2}(\rho_{\phi}+P_{\phi}) = -\frac{\dot{\phi}^2}{2M_P^2},
\end{eqnarray}
and thus the first slow-roll parameter is found to be
\begin{eqnarray}
\varepsilon = -\frac{\dot{H}}{H^2} = \frac{\frac{1}{2}\dot{\phi}^2}{M_P^2H^2}.
\label{epsilon}
\end{eqnarray}
To get the shrinking Hubble sphere, we need to impose $\varepsilon <1$, which leads to
\begin{eqnarray}
\frac{1}{2}\dot{\phi}^2 < M_P^2H^2 < 3M_P^2H^2=\rho_{\phi} \implies \frac{1}{2}\dot{\phi}^2\ll V(\phi).
\end{eqnarray}
\begin{figure}[t!]
    \centering
    \includegraphics[width=10cm]{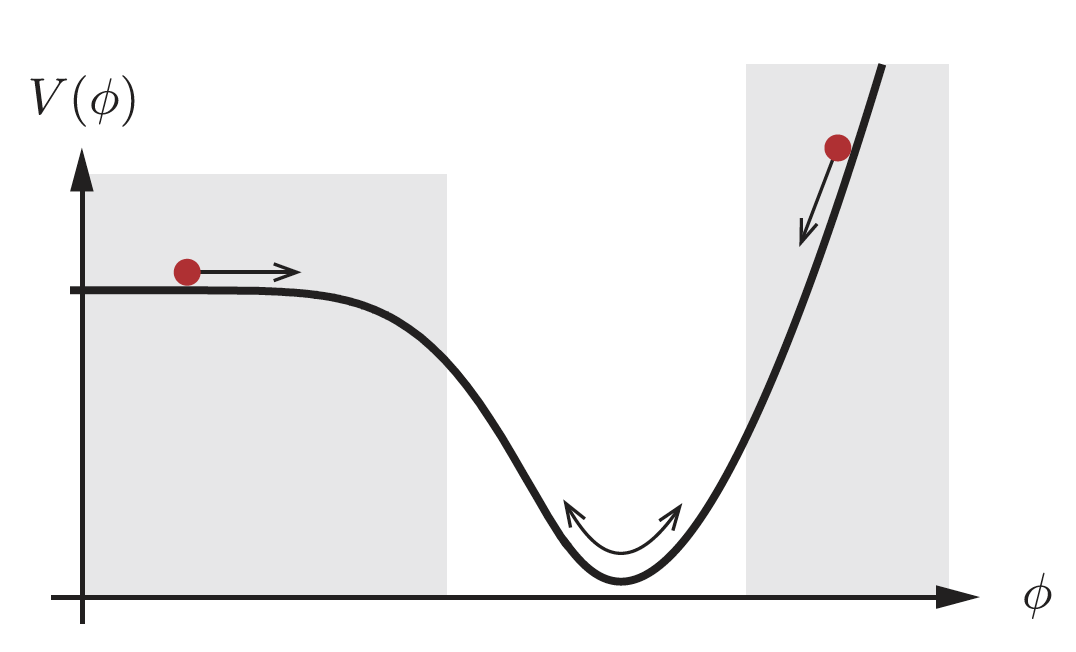}
    \caption{Example of a slow-roll inflationary potential $V(\phi)$ of inflaton field $\phi$. In the shaded regions, inflation may take place. Figure taken from \cite{Baumann_cosmo}.}
    \label{slow-roll}
\end{figure}
We note that {\it the smallness of $\varepsilon$ requires kinetic energy of the inflaton to be very smaller than it total or potential energy}! That is to say, it is necessary for the inflaton to roll down very slowly along the potential. In this sense, we call this situation for inflation as ``{\bf Slow-Roll Inflation},'' which is illustrated in Fig.~\ref{slow-roll}. Furthermore, we have talked about the persistence of small $\varepsilon$ during inflation. This is reflected in another slow-roll parameter $\eta$. Let us calculate this in the following. From Eq.~\eqref{epsilon}, by taking its time derivative, we obtain
\begin{eqnarray}
\dot{\varepsilon} = \frac{1}{M_P^2H^2}(\dot{\phi}\ddot{\phi}-\dot{\phi}^2\dot{H}).
\end{eqnarray}
Thus, the $eta$ parameter is found to be
\begin{eqnarray}
\eta = \frac{\dot{\varepsilon}}{\varepsilon H} = \frac{1}{M_P^2H^2}(\dot{\phi}\ddot{\phi}-\dot{\phi}^2\dot{H}) \bigg( \frac{\dot{\phi}^2}{2M_P^2H}\bigg)^{-1} = 2\bigg(\frac{\ddot{\phi}}{\dot{\phi}H}-\frac{\dot{H}}{H}\bigg) \equiv 2(\varepsilon-\delta),
\end{eqnarray}
where we define $\delta \equiv - \ddot{\phi}/(\dot{\phi}H)$. For $\varepsilon$ to remain small for a long time (i.e. to satisfy $ \{\varepsilon,|\eta|\} \ll 1$), we must take the so-called ``{\bf Slow-roll approximation condition}'' that
\begin{eqnarray}
\{\varepsilon,|\delta|\} \ll 1,
\end{eqnarray}
where inflation can occurs and persists. 

In fact, taking the slow-roll approximation condition, we are able to simplify the 1st Fridemann equation \eqref{1st_eq_inflaton}, and the equation of motion for the inflaton \eqref{2nd_eq_inflaton}.  First, from $\frac{\dot{\phi}^2}{2} \ll V(\phi)$ which is equivalent to $\varepsilon \ll 1$, the 1st Fridemann equation reduces to
\begin{eqnarray}
H^2 \approx \frac{V(\phi)}{3M_{P}^2}.\label{SRA_eq1}
\end{eqnarray}
Remark that the Hubble expansion is determined by the scalar potential $V(\phi)$ of the inflaton! Plus, from the condition $|\delta|=|\ddot{\phi}/(\dot{\phi}H)|\ll 1$, we can ignore the term of $\ddot{\phi}$ in the equation of motion for the inflaton, so that 
\begin{eqnarray}
\ddot{\phi} + 3H\dot{\phi} + \frac{dV(\phi)}{d\phi} \approx 3H\dot{\phi} + \frac{dV(\phi)}{d\phi} = 0 \implies \dot{\phi} \approx -\frac{V'(\phi)}{3H}.\label{SRA_eq2}
\end{eqnarray}

Finally, we are ready to compute the slow-roll parameters in terms of the potential $V(\phi)$. Using the results of Eqs. \eqref{SRA_eq1} and \eqref{SRA_eq2}, we find 
\begin{eqnarray}
\varepsilon = \frac{\dot{\phi}^2}{2M_P^2H^2} = \frac{1}{2M_P^2}\frac{V'^2}{9H^2} \frac{1}{H^2} = \frac{1}{2M_P^2}\frac{V'^2}{9} \frac{9M_P^4}{V^2}  \implies \varepsilon = \frac{M_P^2}{2}\bigg(\frac{V'}{V}\bigg)^2 \equiv \varepsilon_V,
\end{eqnarray}
where we defined a new parameter denoted by $\varepsilon_V$ since this is now determined by the potential $V(\phi)$. Furthermore, taking the time derivative of the approximated equation of motion in Eq. \eqref{SRA_eq2}, we find 
\begin{eqnarray}
3(\dot{H}\dot{\phi}+H\ddot{\phi}) \approx -\dot{\phi}V'' \implies 3\bigg(\frac{\dot{H}}{H^2}+\frac{\ddot\phi}{\dot{\phi}H}\bigg) \approx - \frac{V''}{H^2} \implies 3(\varepsilon+\delta) \approx \frac{V''}{H^2} \approx 3M_P^2\frac{V''}{V},
\end{eqnarray}
where $\delta \equiv  -\ddot{\phi}/(\dot{\phi}H)$ and which gives rise to
\begin{eqnarray}
\varepsilon + \delta \approx \frac{V''}{3H^2} \approx M_P^2 \bigg(\frac{V''}{V}\bigg) \equiv \eta_V.\label{etaV_definition}
\end{eqnarray}
Since $\eta = 2(\varepsilon-\delta)$ and $\eta_V \approx \varepsilon +\delta$, we find $\eta \approx 2(2\varepsilon-\eta_V) \approx 2(2\varepsilon_V -\eta_V)$. Hence, considering $\eta \ll 1$ for slow-roll inflation, we can find the following ``{\bf Slow-roll inflation and inflation-end/no-inflation conditions}''
\begin{eqnarray}
\begin{cases}
\textrm{~~Slow-Roll Inflation:}\quad \{\varepsilon,|\eta|,\varepsilon_V,|\eta_V|,\delta\} \ll 1 \\
\textrm{~~Inflation-End/No-Inflation:}\quad \{\varepsilon,|\eta|,\varepsilon_V,|\eta_V|,\delta \} \gtrsim 1 
\end{cases}\label{Slowroll_No_Inflation_conditions}
\end{eqnarray}
and we emphasize that a convenient way of judging whether a given scalar potential $V(\phi)$ of an inflaton can lead to slow-roll inflation is to calculate the ``{\bf potential slow-roll parameters}''  $\varepsilon_V,\eta_V$, which are given by
\begin{eqnarray}
\varepsilon_V \equiv \frac{M_P^2}{2}\left(\frac{V'(\phi)}{V(\phi)}\right)^2 \approx \varepsilon \ll 1,\qquad |\eta_V| \equiv M_P^2 \left|\frac{V''(\phi)}{V(\phi)}\right|\approx \bigg|\frac{V''(\phi)}{3H^2}\bigg| \approx |\varepsilon+\delta| \ll 1,
\end{eqnarray}
which implies that the inflaton mass $m_{\phi}$ must be much lighter than the Hubble scale $H$ during inflation, i.e.
\begin{eqnarray}
|\eta_V| \ll 1 \quad \Longleftrightarrow \quad  m_{\phi} \ll H \quad \textrm{during inflation}.
\end{eqnarray}
Also, we note that as shown in \eqref{Slowroll_No_Inflation_conditions}, {\it when $\varepsilon_V \approx \eta_V \approx 1$, the inflation can end up. Thus, through this condition, it is possible to compute the field value of $\phi$, say $\phi_E$, at which the inflation can stop}.

The last step we have to do is to compute the number of e-folds $N$ for the time interval from $t_I$ (i.e. at the beginning of inflation) to $t_E$ (i.e. at the end of inflation) in terms of the potential $V(\phi)$. The total e-folding number $N$ for the whole duration of inflation is given by
\begin{eqnarray}
N(\phi_I,\phi_E) &\equiv& \int_{t_I}^{t_E} H(t) dt = \int_{\phi(t_I)}^{\phi(t_E)} \frac{H}{\dot{\phi}}d\phi =\int_{\phi(t_I)}^{\phi(t_E)} \frac{1}{\sqrt{2\varepsilon}} \frac{|d\phi|}{M_P} \nonumber\\
&\approx& \int_{\phi(t_I)}^{\phi(t_E)} \frac{1}{\sqrt{2\varepsilon_V}}\frac{|d\phi|}{M_P} = \int_{\phi_I}^{\phi_E} \frac{1}{M_P^2} \frac{V(\phi)}{V'(\phi)} |d\phi|,
\end{eqnarray}
where we note that $|d\phi|=-d\phi$ since $d\phi < 0$ when $\phi_I < \phi_E$, while $|d\phi|=d\phi$ since $d\phi > 0$ when $\phi_I > \phi_E$. Here, we point out that while $\phi_E$ can be computed from the inflation-end condition (i.e. $\varepsilon_V \approx \eta_V \approx 1$), the initial field value of $\phi_I$ at which the inflation starts cannot be determined exactly but estimated as a possible starting point of inflation by imposing the necessary number of e-folds for inflation. 

\section{``Reheating'' After Inflation: Scalar Field Oscillations, Inflaton Decay, and Thermalization}

During inflation, most of the energy density in the universe is in the form of the inflaton potential $V(\phi)$. Then, when the potential steepens, inflation ends and the inflaton begins to pick up its kinetic energy, and subsequently oscillates around the minimum of the potential. This is the first process after inflation called ``{\bf Scalar Field Oscillation}.'' Then, via  ``{\bf Inflaton Decay},'' the energy of inflaton has to be transferred to those of standard model (SM) particles, so that the SM particles can be produced in the universe. The last process is ``{\bf Thermalization}'' of the primordial plasma of produced particles. After the inflaton is completely frozen, the SM particles produced by the inflaton decay can interact with each other, create other particles through their interactions, and the primordial plasma of all the particles will eventually reach thermal equilibrium with some characteristic temperature $T_{rh}$ called ``{\bf Reheating Temperature}''. In particular, the process of the three steps are called ``{\bf Reheating}.'' 
~\\
~~\\
\noindent {\large \it Scalar Field Oscillation:} When the slow-roll parameters $\varepsilon_V,\eta_V$ reach around $1$, the inflation ends. After inflation, then, the inflaton field $\phi$ begins to oscillate around the minimum of its potential $V(\phi)$. About this minimum, it is possible to assume that the potential can be approximated to a quadratic potential $V(\phi) \approx \frac{1}{2}m^2_{\phi}\phi^2$ corresponding to the mass term of inflaton, where the amplitude of $\phi$ is small. The inflaton $\phi(t)$ can still be considered as a homogeneous field depending only on the time $t$. Then, from the definition of $\eta_V$ in Eq.~\eqref{etaV_definition}, we have 
\begin{eqnarray}
 \varepsilon + \delta \approx \frac{V''}{3H^2} \approx M_P^2 \frac{V''}{V} \equiv \eta_V \implies \eta_V \approx \frac{V''}{3H^2} = \frac{1}{3}\frac{m_{\phi}^2}{H^2}.
\end{eqnarray}
Since $\eta_V \gtrsim 1$ after inflation, we get the following result 
\begin{eqnarray}
\frac{m_{\phi}^2}{H^2} \gtrsim \mathcal{O}(1).
\end{eqnarray}
This means that the oscillation period $T_{\textrm{osc}}= m_{\phi}^{-1}$ of the inflaton around the potential minimum must be very shorter than the expansion time $t_H = H^{-1}$. Thus, the mass of inflaton when inflation is not active must be heavier than the Hubble scale $H$: $H^{-1} \gg m^{-1}_{\phi} \quad \Longleftrightarrow \quad m_{\phi} \gg H$. In this mass limit $m_{\phi} \gg H$, it is possible to neglect the friction term $3H\dot{\phi}$ in the equation of motion for the inflaton, so that $\ddot{\phi} \approx -m^2_{\phi}\phi$, whose solution is found to be the conventional oscillating one $\phi = A\sin(m_{\phi}t)$. Next, let us consider the continuity equation (i.e. $\dot{\rho}_{\phi} = -3H(\rho_{\phi}+P_{\phi})$). Then, we obtain
\begin{eqnarray}
\dot{\rho}_{\phi} + 3H\rho_{\phi} = -3HP_{\phi} \implies  \dot{\rho}_{\phi} + 3H\rho_{\phi} = -3H(\frac{\dot{\phi}^2}{2}-\frac{1}{2}m_{\phi}^2\phi^2) = \frac{3H}{2}(m_{\phi}^2\phi^2 -\dot{\phi}^2).
\end{eqnarray}
Taking the time average over one period of oscillation $T_{osc}$ (i.e. $\left<f(t)\right>_T \equiv \frac{1}{T}\int^{+T/2}_{-T/2}f(t)dt$), we find 
\begin{eqnarray}
\left<\dot{\rho}_{\phi} + 3H\rho_{\phi}\right>_T = \frac{3H}{2}(m_{\phi}^2\left<\phi^2\right>_T -\left<\dot{\phi}^2\right>_T) = \frac{3HA^2m_{\phi}^2}{2}( \left<\sin^2(m_{\phi}t)\right>_T-\left<\cos^2(m_{\phi}t)\right>_T) = 0,\nonumber\\{}
\end{eqnarray}
where we used the solution $\phi = A\sin(m_{\phi}t)$ and the fact that $\left<\cos^2(m_{\phi}t)\right>_T=\left<\sin^2(m_{\phi}t)\right>_T=1/2$. The result ``$\left<\dot{\rho}_{\phi} + 3H\rho_{\phi}\right>_T=0$'' means that the energy density decays as 
\begin{eqnarray}
\left<\rho_{\phi}(t_f)\right>_T= \left<\rho_{\phi}(t_i)\right>_T e^{-3\int_{t_i}^{t_f} H dt} =\left<\rho_{\phi}(t_i)\right>_T e^{-3N} = \left<\rho_{\phi}(t_i)\right>_T\Big(\frac{a(t_f)}{a(t_i)}\Big)^{-3} \propto a(t_f)^{-3},\label{decay_of_rho}
\end{eqnarray}
where we used $a(t_f)=e^N a(t_i)$. We observe that as the universe expands, decaying behavior of the energy density of the inflaton after inflation is the same as that of the conventional ``pressureless ($\omega=P/\rho=0$)'' matter! Also, the result \eqref{decay_of_rho} means that the amplitude of oscillation of the inflaton decreases as the universe expands since 
\begin{eqnarray}
\left<\rho_{\phi}\right>_T = \left< \frac{1}{2}\dot{\phi}^2+\frac{1}{2}m_{\phi}^2\phi^2 \right>_T= \frac{1}{2}m_{\phi}^2A^2\left<(\cos^2(m_{\phi}t)+\sin^2(m_{\phi}t))\right>_T =\frac{1}{2}m_{\phi}^2A^2 \implies  A^2 \propto a^{-3}.\nonumber\\{}
\end{eqnarray}
\noindent {\large \it Inflaton Decay:} There must be interaction terms of the inflaton field coupling to the standard model fields in their Lagrangians because the universe must be filled with matter. The energy stored in the inflaton field must then be transferred to the ordinary particles of matter. If the infaton can decay into bosons via a mechanism of ``parametric resonance'' sourced by bose condensation effects, then the decay may be very rapid, which is called ``{\bf Preheating}.'' If the inflaton can only decay into fermions, then the decay is slow. In this case, the energy density follows: $\dot{\rho}_{\phi} + 3H\rho_{\phi} = -\Gamma_{\phi}\rho_{\phi}$ where $\Gamma_{\phi}$ parameterizes the inflaton decay rate.
~\\
~~\\
\noindent {\large \it Thermalization:} After the inflaton decay, particles of matter have to be produced, and then they will form a primordial plasma. Then, in this plasma, the particles reach thermal equilibrium with some temperature $T_{rh}$ called ``{\bf Reheating Temperature}'' via their possible interactions with creating and annihilating other particles. The reheating temperature is determined by the energy density $\rho_{rh}$ at the end of the reheating epoch. Because the universe is cooled down as it expands, $\rho_{rh}$ is less than the energy density of the inflaton $\rho_{\phi,E}$ at the end of inflation. IF the reheating takes long, then it may be possible that $\rho_{rh} \ll \rho_{\phi,E}$. Some particles (e.g. gravitino) never reach thermal equilibrium when their interactions are very weak. Moreover, as long as momentum of a particle is much higher than its mass, the energy density of the particle behaves like radiation. After thermalization of the plasma, the era of the {\bf standard Hot Big Bang} begins.

%% file: chapters/3.tex
In this chapter, we briefly review conceptual aspects of effective field theory (EFT). This chapter is based on the lecture of EFT in Ref. \cite{EFT}.

\section{What is Effective Field Theory (EFT)? Why EFT?}

In high energy physics, we pursue quantum field theory that is valid at {\it all energies} from infrared (IR) to ultraviolet (UV) scales, and contains {\it full physical information} (such as dynamical field degrees of freedom, Lagrangians, symmetries, partition function, etc). In this sense, we call such a theory as ``{\bf UV theory},'' while we call the theory that is valid only at {\it low energies} and lost the information of the UV physics as ``{\bf IR theory},'' so that this includes partial physical information. In fact, in real world, we face particular physical systems that are in the region of particular energies (i.e. not all energies). In this situation, it is sufficient to have the practical theory that can be valid only at {\it certain energies of our interest} and contains {\it partial physical information of our interest}. We call this practical theory as ``{\bf Effective Field Theory (EFT)}''. Usually, EFT is obtained by a low-energy approximation (also called ``IR limit'') of a UV theory, so that EFT is considered as IR theory.

Why do we use the effective Lagrangian? There are some motivations to work with $\mathcal{L}_{EFT}$ rather than the fundamental one $\mathcal{L}_{UV}$. The first is {\bf simplicity}; that is, we can handle finitely-many terms. The second is {\bf calculability}; that is, we can resum the large logarithms arising from loop-integral calculations into a renormalization gorup flow of the EFT parameters. The last is {\bf agnosticity}; that is, it is hard to know the fundamental UV theory.

\section{Locality and Perturbativity in Field Theory}

In the following, I describe the conceptual aspects of EFT that are necessary to understand the research of this dissertation. Let us consider a physical system whose dynamics is governed by a UV Lagrangian $\mathcal{L}_{UV}(\phi,H)$ for some light fields $\{\phi\}$ that we wish to treat, and heavy fields $\{H\}$ that we cannot access or wish to ignore. Then, the UV theory is fully characterized by the following partition function 
\begin{eqnarray}
Z_{UV}[J_{\phi},J_{H}] = \int [D\phi][DH] \exp \Big( i \underbrace{ \int d^dx (\mathcal{L}_{UV}(\phi,H)+J_{\phi}\phi + J_H H)}_{=S_{UV}[\phi,H]}\Big),
\end{eqnarray}
where $J_{\phi,H}$ are the external currents as source corresponding to some fields $\phi,H$ respectively. Every $n$-point correlator (i.e. S-matrix scattering amplitude), the main field-theoretical observable, of the relevant fields can be obtained by differentiating the partition function $Z$ with respect to the current $J$. In EFT, however, we are interested in the correlators of $\phi$'s only through the EFT partition function $Z_{EFT}[J_{\phi}]$ represented by
\begin{eqnarray}
Z_{EFT}[J_{\phi}] = \int [D\phi]\exp \Big( i \underbrace{\int d^dx (\mathcal{L}_{EFT}(\phi)+J_{\phi}\phi)}_{=S_{EFT}[\phi]}\Big),
\end{eqnarray} 
which can be obtained by integrating out\footnote{See Ref. \cite{EFT} for a review of how to integrate out the heavy field modes, or Ref.~
\cite{FPR_Integrating_out_functional_method} for one recent work about a simplified method of how to integrate out heavy modes in the functional formalism, or see Ref.~\cite{Dittmaier} for another work on the integrating-out method.} the ``heavy'' field degrees of freedom $H$ in the path integral. Here, from the UV partition function, we can define the EFT one as follows: 
\begin{eqnarray}
Z_{EFT}[J_{\phi}] \equiv Z_{UV}[J_{\phi},J_H=0].
\end{eqnarray}

Now it is essential to understand what locality and perturbaility are in field theory:

{\bf Locality (of Lagrangian in field theory) ---} {\it Analytic} mathematical object like function or functional $F(x)$ is called ``{\it local at a fixed local point of evaluation}'' if it is {\it finitely Taylor-expandable\footnote{{\bf Taylor expansion} is a mathematical way of representing {\it any ``analytic'' function} as a {\it local expansion} in form of {\it an infinite sum of polynomials}. In principle, non-local function can be written only by an infinite sum of polynomials if Taylor-expanded. We call a function as finitely Taylor-expandable when its Taylor expansion can be approximated by {\it a finite sum of polynomials} up to some order of interest if the expansion parameter can be very small. In this case, we are blind to the information of the higher-order terms beyond the order of interest since a perturbative expansion lost them.} in polynomials\footnote{Polynomial means that the exponent of a term is given only by non-negative integer. Non-polynomial is not polynomial, i.e. fractional/negative exponent like $f(x)=\sqrt{x}$ or $x^{-1}$.}} {\it at the fixed local point $x$ where we want to evaluate the function $F(x)$} in an open covering of some local coordinate chart $X$ ($x\in \{X\}$). If something is not in the case, it is called ``{\it non-local}'' because it will be {\it infinitely Taylor-expandable in polynomials} when the non-local object is evaluated at a single local point of interest in the end. 

For example, let us assume a ``non-contact'' interaction term of two fields, $\mathcal{L}(x,y)= \phi_1(x) \phi_2(y)$, depending on two local points $x,y$. We express $y=x+\Delta x$ for some interval $\Delta x$. Then, at some single point $x$, we have 
\begin{eqnarray}
\mathcal{L}(x,y)|_{\textrm{evaluated at $x$}}&=& \{  \phi_1(x) \phi_2(x+\Delta x)\}|_{\textrm{evaluated at $x$}} \nonumber\\
&=& \underbrace{ \phi_1(x) \sum_{n=0}^{\infty} \frac{\phi^{(n)}_2(x)}{n!} (\Delta x)^n}_{\textrm{evaluated at $x$}} = \phi_1(x) \Big( e^{\Delta x \cdot \frac{\partial}{\partial x}}\phi_2(x)\Big),
\end{eqnarray}
where we Taylor-expanded $\phi_2(x+\Delta x)$ at the single local point $x$ of our interest. We observe that $\mathcal{L}(x,x+\Delta x)$ is non-local. This is because while $\phi_1(x)$ remains the same after evaluation at $x$, $\phi_2(x+\Delta x)$ becomes an ``infinitely Taylor-expandable'' function at the evaluation point $x$. Especially, we can see that after the last equality, the function $\mathcal{L}$ can also be represented by the product of ``{\it nonlinear differential operator}'' (i.e. $e^{\Delta x \cdot \frac{\partial}{\partial x}}$) and $\phi_2(x)$! Also, we see that when $\Delta x\rightarrow 0$, the function $\mathcal{L}(x)$ becomes local in fields as expected. Conclusively, if a function depends on multiple points of evaluation (i.e. literally, non-local) or includes nonlinear differential operator (with respect to the evaluation variable) at a single evaluation point, then the function is non-local.

The effective Lagrangian $\mathcal{L}_{EFT}$ is a {\it non-local}\footnote{Here, in EFT, {\bf Non-local} means non-polynomial in the fields and their derivatives.} object. As above, after Taylor expanding at a single local point in the space of fields and their derivatives, the effective Lagrangian $\mathcal{L}_{EFT}$ can also be represented by a local Lagrangian, but with infinitely-many interaction terms in principle. Fortunately, however, there is a particular situation where the non-local effective Lagrangian $\mathcal{L}_{EFT}$ can be properly approximated to a ``local'' effective Lagrangian with finitely-many terms\footnote{This is because it is impossible to us to treat infinitely-many terms in real world.} only if the non-local Lagrangian can be {\it perturbative}:

{\bf Perturbativity ---} An analytic (local or non-local) object is called ``{\it perturbative}'' if it can be {\it local (i.e. finitely Taylor-expandable in polynomials at a single local point) up to certain order of accuracy we wish when a very small expansion (also called ``perturbation'') parameter $\epsilon \ll 1$} is given.

For example, $\mathcal{L}_{EFT} \supset \phi^2(x) \square \phi^2(x)$ is local in fields and their derivatives, while $\mathcal{L}_{EFT} \supset \phi^2(x)(\square + M^2)^{-1}\phi^2(x)$ is non-local because at the evaluation point $x$, we face the situation that  $(\square + M^2)^{-1} = M^{-2} - M^{-4}\square + \cdots$. But if we are able to take a low-energy (or long-distance) limit of $|p| \ll M$ (or $L=|p|^{-1} \gg M^{-1}$) when its 4-momentum is much smaller than the mass scale $M$, then we can express the non-local Lagrangian as a local Lagrangian with finitely-many terms up to some order using the local expansion $(\square + M^2)^{-1} \approx M^{-2} - M^{-4}\square + \cdots$, so that $\mathcal{L}_{EFT} \supset \phi^2M^{-2}\phi^2 - \phi^2M^{-4}\square\phi^2 + \cdots$. In this case of the low-energy limit, we say that the non-local Lagrangian is perturbative. However, remember that we are blind to physics at high energy (or short-distance) scale $|p| \gg M$ (or $L=|p|^{-1} \ll M^{-1}$). 

{\bf We can only treat and analyze finitely-many terms in practice, so that we necessitate local Lagrangians by taking advantage of perturbativity. That is, perturbativity is a strategy of approximating non-local object to be local up to some accuracy.}

\section{Scaling}

We have seen that the effective Lagrangian is non-local, and thus this has infinitely-many interaction terms in general ``before imposing the low-energy limit.'' Certainly, this is not favored in physics because the infinitely-many terms are beyond our control and knowledge! Hence, it is inevitable for one to be able to treat the non-local effective Lagrangian $\mathcal{L}_{EFT}$ as a {\it perturbative expansion}. Once such perturbativity is given to $\mathcal{L}_{EFT}$, it is then needed to have a special way of organizing the calculations in a consistent expansion and single out the most relevant contributions, which is called ``{\bf Power counting rule}.'' 

To achieve this, let us talk about ``scaling property'' of Lagrangian and action. Let us consider a 4-dimensional local relativistic EFT Lagrangian $\mathcal{L}_{EFT}$ of mass dimension 4 (or action $S_{EFT}$ of mass dimension zero):
\begin{eqnarray}
S_{EFT} \equiv \int d^4x\mathcal{L}_{EFT}(\phi) = \int d^4x \bigg[ (\partial_{\mu} \phi)^2 - m^2\phi^2 -\kappa \mu \phi^3 - \lambda \phi^4 - \sum_{n+d>4}^{\textrm{Finite}~N} \frac{c_{n,d}}{\Lambda^{n+d-4}} \phi^{n-1} \partial^d_{\mu}\phi\bigg], 
\end{eqnarray}
where $\mathcal{O}^{(n+d)}=\phi^{n-1} \partial^d_{\mu}\phi$ are dimension-($n+d$) operators; $c_{n,d}$ are dimensionaless Wilson coefficients; $\Lambda$ is a (dimension-1) cutoff scale of the EFT; $\partial$ is the (dimension-1) derivative operator with respect to the spacetime coordinate $x^{\mu}$, and $n,d$ are positive and non-negative integers, respectively. Again, this Lagrangian is perturbative. Next, we consider re-scaling by going from old $x$ to new coordinates $x'$, which is given by
\begin{eqnarray}
x_{\mu} \longrightarrow \xi x'_{\mu},\quad i.e.\quad   x_{\mu} = \xi x'_{\mu} \Longleftrightarrow x'_{\mu} = x_{\mu}/\xi,
\end{eqnarray}
where $\xi$ is a re-scaling dimensionless parameter. The point is that $\xi \longrightarrow 0$ means ``zooming in\footnote{For instance, we can see microorganisms (of ``micro meter scales'' in reality) as that of ``centimeter scales'' on our eyes by zooming in (i.e. taking $\xi \longrightarrow 0$ or sizing up $x'$ equivalently) through microscope.}'' on short-distance scales or high energies in the new reference frame, while $\xi \longrightarrow \infty$ means ``zooming out'' on long-distance scales or low energies in the new reference frame. In particular, we will be interested in the low-energy limit since we wish to ignore or are not able to know something. 

By investigating how the effective Lagrangian $\mathcal{L}_{EFT}$ changes as the re-scaling parameter increases (i.e. when taking the low-energy limit $\xi \longrightarrow \infty$ or sizing down $x'$ equivalently), we can figure out the importance of various relevant terms as we go to lower energies away from the UV theory underlying EFT. When taking the rescaling $x = \xi x'$, the first kinetic action changes to $\int d^4x(\partial \phi)^2 =\int d^4 x' \xi^4 \cdot \xi^{-2}(\partial' \phi)^2=\int d^4 x' \xi^2(\partial' \phi)^2$. However, this is not canonically normalized. Hence, we need to take additional rescaling for the field $\phi$ in the way 
\begin{eqnarray}
\phi= \xi^{-1} \phi',
\end{eqnarray}
so that the kinetic action can become canonical as $\int d^4x(\partial \phi)^2 =\int d^4 x' (\partial' \phi')^2$. Then, considering both re-scalings, we find 
\begin{eqnarray}
S_{EFT}= \int d^4x' \bigg[ (\partial_{\mu} \phi')^2 - m^2\xi^2\phi'^2 -\kappa (\xi\mu)\phi'^3 - \lambda \phi'^4 - \sum_{n+d>4}^{\textrm{Finite}~N} \frac{c_{n,d}}{(\xi\Lambda)^{n+d-4}} \phi'^{n-1} \partial'^d_{\mu}\phi'\bigg]. 
\end{eqnarray}
We observe that as $\xi \longrightarrow \infty$ (low-energy limit), the $\phi^2$ and $\phi$ terms grows. We call these ``{\it growing}'' terms as ``{\bf relevant}'' interaction term. We see that the $\phi^4$ term remains constant. The ``{\it constant}'' terms are called ``{\bf marginally relevant}'' interaction term. On the contrary, inside the finite sum, we note that the terms are suppressed with the power of $\xi^{-D}$ where we call $D = n+d-4 >0$ as ``{\bf scaling dimension of $\xi$}'' that uniquely determines the scaling behavior of the interaction term. We call such ``{\it suppressed}'' terms as ``{\bf irrelevant}'' interaction term. The interaction term becomes more irrelevant since more suppressed as the scaling dimension of $\xi$ (i.e. $D$) gets larger.

\section{EFT expansion}\label{EFT_expansion}

In $d$-dimensional spacetime, the effetive Lagrangian $\mathcal{L}_{EFT}$ that includes all the effective operators can be represented by an infinite sum of local, gauge-invariant, Lorentz-invariant, mass dimension-$\delta$ operators $\mathcal{O}_i^{(\delta)}(x)$ with their mass dimension-($d-\delta$) expansion coefficients $C_i^{(d-\delta)}$ 
\begin{eqnarray}
\mathcal{L}_{EFT}(x) = \sum_{\delta \geq 0}^{\infty} \sum_i C_i^{(d-\delta)}\mathcal{O}_i^{(\delta)}(x),
\end{eqnarray}
where the Lagrangian's mass dimension is given by $[\mathcal{L}]=d$ since the action $S=\int d^dx \mathcal{L}$ is dimensionless and $[d^dx]=-d$. In particular, we derive the mass dimensions of fields through those of their canonical kinetic Lagrangians\footnote{For example, scalar $\phi$, spinor $\psi$, and vector $A_{\mu}$ fields have the corresponding mass dimensions $[\phi]=\frac{d-2}{2}$, $[\psi]=\frac{d-1}{2}$, and $[A_{\mu}]=\frac{d-2}{2}$, respectively, because $[\frac{1}{2}|\partial_{\mu} \phi|^2]=[\Bar{\psi}i\cancel{\partial}\psi]=[-\frac{1}{4}|F_{\mu\nu}|^2]=d$. Plus, when the vector is gauged, we have $[D_{\mu}]=[\partial_{\mu}]=[gA_{\mu}]=[g]+[A_{\mu}]=1$, so that the gauge coupling parameter may be dimensionful because $[g]=1-[A_{\mu}]=\frac{4-d}{2}$. Moreover, in $d=4-2\epsilon$ of dimensional regularization (DR), $[g]=\epsilon$. Thus, making the gauge coupling to be dimensionless, we take redefinition $g=\hat{g}\mu^{\epsilon}$ for some DR mass scale $\mu$ with mass dimension-1, i.e. $[\mu]=1$, and dimensionless coupling $\hat{g}$, i.e. $[\hat{g}]=0$.}. 
 
Furthermore, we can extract the mass dimension from the dimensionful coefficients. By introducing the so-called ``{\bf cutoff scale}'' $\Lambda$ with mass dimension-1, i.e. $[\Lambda]=1$, we re-define the coefficients as follows:
\begin{eqnarray}
C_i^{(d-\delta)} \equiv \frac{c_i^{\delta}}{\Lambda^{\delta-d}},
\end{eqnarray}
where $c_i^{\delta}$ is defined as the ``{\it dimensionless}'' parameter\footnote{Usually, this dimensionless coefficient will be of order of $\mathcal{O}(1)$.} that couples to an effective operator $\mathcal{O}_i^{(\delta)}(x)$ and called as ``{\bf Wilson (dimensionless) coefficient}.'' With this redefinition, we define ``{\bf EFT expansion}'' given by 
\begin{eqnarray}
\mathcal{L}_{EFT}(x) = \sum_{\delta \geq 0}^{\infty}\sum_i \frac{c_i^{\delta}}{\Lambda^{\delta-d}}\mathcal{O}_i^{(\delta)}(x) = \sum_{\delta \geq 0}^{\infty} \frac{1}{\Lambda^{\delta-d}}\mathcal{L}_{\delta} \qquad \textrm{where} \quad \mathcal{L}_{\delta} \equiv \sum_i c_i^{\delta} \mathcal{O}_i^{\delta}(x),
\end{eqnarray}
where $\mathcal{L}_{\delta}$ is called ``{\it dimension-$\delta$ effective interaction}.'' It is worth noting that the cutoff scale $\Lambda$ means the short-distance/high-energy scale at which new physics occurs.
Moreover, for later use, we shall call ``$\delta-d$'' as ``{\bf cutoff scaling dimension}\footnote{That is, it is equal to ``Effective-Operator Minus Spacetime (EOMS) dimensions.''}.''

\section{Power Counting Rule and Renormalizability}

In quantum field theory (QFT), Lagrangian is classified into two types as follows. The ``relevant'' Lagrangian with ``{\it negative cutoff scaling}'' (i.e. $\delta-d < 0$) is called ``{\bf super-renormalizable}''; the ``marginally-relevant'' Lagrangian with ``{\it zero cutoff scaling}'' is called ``{\bf marginally renormalizable},'' and the ``irrelevant'' Lagrangian with ``{\it positive cutoff scaling}'' (i.e. $\delta-d > 0$) is called ``{\bf non-renormalizable}.'' In short, {\it ``{\bf renormalizability}\footnote{See Ref.~\cite{Marko} for a brief review on renormalization in QFT, which gives an insightful overview of renormalization.}'' in QFT is applicable to the (marginally) relevant Lagrangians with ``non-positive cutoff scaling.''}

In particular, we call the quantum field theory that only includes renormalizable Lagrangians as ``{\bf renormalizable theory},'' while the theory that also includes non-renormalizable Lagrangians as ``{\bf non-renormalizable theory}.'' In fact, effective field theory is non-renormalizable theory.

In this section, I will explain how power counting rule is associated with renormalizability; how different renormalizable and effective field (or non-renormalizable) theories are, and why effective field theory is physically useful in spite of its non-renormalizability. 

Let us consider a dimensionless scattering amplitude $\mathcal{A}$ in $d$-dimensional spacetime. If one works at some typical momentum scale $p$, then a single insertion of a dimension-$\delta$ effective operator in the Feynman's scattering graph gives the following scaling behavior of the scattering amplitude
\begin{eqnarray}
\mathcal{A} \sim \left(\frac{p}{\Lambda}\right)^{\delta-d}.
\end{eqnarray}
We note that the ``{\bf amplitude scaling}'' in $(p/\Lambda)$ is equal to the cutoff scaling. Furthermore, when inserting a set of higher dimensional operators in any Feynman's scattering graph, the ampliture scales as
\begin{eqnarray}
\mathcal{A} \sim \left(\frac{p}{\Lambda}\right)^{n} \qquad \textrm{where} \quad n \equiv \sum_i (\delta_i -d),
\end{eqnarray}
where $i$ runs over all the inserted operators, and the amplitude scaling $n$ is given by the sum of cutoff scalings of all the inserted operators. We call the equation of amplitude scaling $n$ as ``{\bf Power counting formula}'' and $n$ means ``{\it the order of correction to amplitude}'' as well 
\begin{eqnarray}
n \equiv \sum_i (\delta_i -d).
\end{eqnarray}

{\it The power counting formula let us know how to organize the amplitude calculations}. For example, in ($d=4$)-dimensional spacetime, if one want to compute a scattering amplitude $\mathcal{A}$ to the leading order $n=0$, i.e. $(p/\Lambda)^0$, then one can only use the interactions with non-positive cutoff scaling, i.e. $\mathcal{L}_{\delta-d \leq 0}$, such as $\mathcal{L}_2$, $\mathcal{L}_3$, and $\mathcal{L}_4$. For the correction $(p/\Lambda)^1$, one has $1=\sum_i (\delta_i-4)$ and thus need to consider a single insertion of dimension-5 interaction $\mathcal{L}_5$. For the correction $(p/\Lambda)^2$, one has $2=\sum_i (\delta_i-4)$ and thus need to consider single insertion of $\mathcal{L}_6$ or two insertions of $\mathcal{L}_5$, and so on. 

The {\bf Power counting rule} is to use the power counting formula for a fixed amplitude scaling $n$ of our interest in order to know which interactions $\mathcal{L}_{\delta}$'s should be inserted into the scattering graph of a correction in $(p/\Lambda)^n$ to the final amplitude. 

Now we are ready to compare the renormalizable and non-renormalizable (or effective field) theories. First, as an example, let us say that we consider some ``divergent'' amplitude with loops for a fixed $n$ made by inserting $m$-multiples of a single operator $\mathcal{L}_{\delta}$ into the graph. Then, we have $n=m(\delta-d)$. Since the graph is divergent, it is required to add a counterterm (CT) to make the Lagrangian renormalizable. Hence, if we assume that a new single operator $\mathcal{L}_{\delta'}$ is enough to create the counterterm, then we must have it such that $n= \delta' - d$. Therefore, by considering two power counting formulae, we can solve for $\delta'$ as follows: 
\begin{eqnarray}
m(\delta -d) = \delta' -d \implies \delta' = m(\delta-d)+d.
\end{eqnarray}
The response to this result will be different as follows:
\begin{itemize}
\item {\bf Non-Renormalizable Theory (e.g. Effective Field Theory)}: 
Since we now consider $\mathcal{L}_{\delta}$ as ``non-renormalizable'' operator, $\mathcal{L}_{\delta}$ has ``positive cutoff scaling,'' i.e. $\delta-d >0$. This leads to the condition that $\delta' > d$ or equivalently $\delta'-d >0$. {\it Notice that this condition is not upperbound on the dimension of the new operator $\mathcal{L}_{\delta'}$}. For example, for a single dimension-5 operator $\mathcal{L}_5$, in general, the loop graphs with `two insertions' of $\mathcal{L}_5$ (so, we consider $n=2(5-4)=2$ order correction to amplitude) are divergent. Thus, one need a counterterm to cancel out the divergence from the loops. That is, the counterterm should be given by the dimension-6 operator $\mathcal{L}_6$ because we have $\delta' = 2(5-4)+4 = 6$ where $m=2,\delta=5,d=4$ are input. Then, the renormalized Lagrangian must be written as
\begin{eqnarray}
\mathcal{L} = \mathcal{L}_4 + \frac{1}{\Lambda} \mathcal{L}_5  \longrightarrow \mathcal{L}_{ren} = \mathcal{L}_4 + \frac{1}{\Lambda} \mathcal{L}_5 + \frac{1}{\Lambda^2}\mathcal{L}_6^{(CT)}.
\end{eqnarray}
Notice that this is merely ``{\it renormalizable to the correction $(p/\Lambda)^{2}$}.'' Continuing this way, one can generate operators of arbitrarily higher dimensions $\delta' > \delta > d$ by inserting multiples of operators with $\delta - d>0$. The renormalization process of cancelling out the divergences continues infinitely. This means that we generate an infinite sum of non-renormalizable terms in the effective Lagrangian $\mathcal{L}_{EFT}=\sum_{\delta\geq0} \frac{\mathcal{L}_{\delta}}{\Lambda^{\delta-d}}$ in the end game.
    
\item {\bf Renormalizable Theory}: In renormalizable theory, since we have operators with non-positive cutoff scaling, i.e. $\mathcal{L}_{\delta\leq d}$, the possible new operators can be considered such that $\delta'=m(\delta-d)+d \leq d$. This is exactly the upperbound on the dimension of new operators $\mathcal{L}_{\delta'}$! For example, for the square-shape one-loop graph made by four insertions of the dimension-4 vertex operator of $e^-e^+A_{\mu}$ coupling, the corresponding counterterm can be made by the new operator $\mathcal{L}_{\delta'=4}$ since $\delta' = 4(4-4)+4 = 4 \leq d=4$ where $m=4,d=4,\delta=4$ are input. We note that {\it the new operators contributing to the counterterms\footnote{We do not need to add counterterms for the negative dimension operators like $\phi(x)^{-2}$ since there are no divergences of this type.} have already been included in the original renormalizable Lagrangian $\mathcal{L}_{\delta \leq d}$}. In conclusion, renormalizable terms are those with non-positive cutoff scailing $n \leq 0$. Divergences in such a QFT can be absorbed by local operators with $0 \leq \delta \leq d$.

\end{itemize}

{\bf Concluding remarks:}

\begin{itemize}
    \item Renormalizable theory is a special case of EFT where we take $\Lambda \longrightarrow \infty$. Thus, scattering amplitudes can be computed to arbitrary accuracy because of no corrections in $(p/\Lambda)^n$ to the amplitude.
    \item A theory with operators of dimension $\delta > d$ (i.e. ``over-spacetime-dimension'') is referred to as non-renormalizable theory because the infinitely-many higher dimensional operators are needed to renormalize the theory. 
    \item However, as long as one is interested in corrections of the maximal value of the amplitude scaling $n$ up to some accuracy, it is sufficient to consider finitely-many operators that can contribute to the amplitude of interest. This is the situation that non-local Lagrangian is approximated into the local one up to the desirable accuracy. This implies that non-renormalizable theory can be as good as renormalizable theory from the viewpoint of practical use of a theory!
    \item Again, EFT (= non-renormalizable theory) can have only finite number of effective operators if one wish to keep the corrections in $(p/\Lambda)^n$ only up to some maximal value of $n$.
    \item The $(p/\Lambda)$ suppression of a given graph at low energy $p \ll \Lambda$ implies that non-local effective Lagrangian can be well-approximated into a local Lagrangian as a finite sum of the operators:
\begin{eqnarray}
\mathcal{L}_{EFT}(x) = \sum_{\delta \geq 0}^{\infty} \frac{1}{\Lambda^{\delta-d}}\mathcal{L}_{\delta} \approx  \sum_{\delta \geq 0}^{\textrm{Finite N}} \frac{1}{\Lambda^{\delta-d}}\mathcal{L}_{\delta} \quad \textrm{when} \quad p \ll \Lambda.
\end{eqnarray}
According to this result, it is worth noticing that if we have some ``effective'' Lagrangian with a characteristic mass scale $M$ and dimensionless ``remnant'' Wilson coefficient $C^{\delta}$ generically assumed to be apart from $\mathcal{O}(1)$, i.e. in the following form
\begin{eqnarray}
\mathcal{L}_{EFT}(x) \supset \sum_{\delta \geq 0}^{\textrm{Finite N}} \frac{C^{\delta}}{M^{\delta-d}} \mathcal{L}_{\delta}, \label{EFT_alternative}
\end{eqnarray}
then for this effective Lagrangian to be ``local,'' it must be required that
\begin{eqnarray}
\frac{C^{\delta}}{M^{\delta-d}} \lesssim \frac{1}{\Lambda^{\delta-d}} \implies C^{\delta} \lesssim \left(\frac{M}{\Lambda}\right)^{\delta -d}. \label{EFT_constraint}
\end{eqnarray}
We observe that {\it the remnant Wilson coefficient $C^{\delta}$ must be constrained}. If $M \sim \Lambda$, then the coefficient $C^{\delta}$ becomes of order of $\mathcal{O}(1)$ as the usual one. This argument will be used throughout this dissertation.

\end{itemize}

%% file: chapters/4.tex
This chapter is based on Refs.~\cite{Quevedo_SUSY,Wess,Kirsten}.

For several decades, supersymmetry has been developed as a main framework that could explain new physics beyond the Standard Model (SM). Particularly, it is motivated by the fact that it can give us the answers to some problems about the gauge coupling unification, dark matter candidate (Lightest Supersymmetric Particle: LSP), and Higgs mass hierarchy. Unfortunately, however, we have never seen yet any experimental evidence of supersymmetry even though many experiments have been done. For this reason, it is considered that supersymmetry may be broken spontaneously at some point in the past as so does the electroweak interaction in the SM. Neverthless, supersymmetry has been deeply studied much due to its mathematically cogent structure, which may be expected to be realized in nature in a certain way. For example, taking advantage of the superalgebra, gravitation and other fundamental forces can be unified within supergravity formulation which is a locally supersymmetric field theory and even this can be used as a model of the inflationary cosmology. In addition to this, supersymmetry is a fundamental ingredient of string theory. 

In that sense, this review is devoted to its mathematical features and techniques in order for one to understand how supersymmetry is formed and what physical applications can be obtained in a somewhat rigorous way. From the viewpoint of this, we need to first look at the Coleman-Mandula theorem given by 
\begin{theorem}[Coleman-Mandula Theorem]
The most general Lie algebra of symmetries of the S-matrix contains the energy momentum generator $P_{\mu}$ and the Lorentz rotation generator $M_{\mu\nu}$, and a finite number of Lorentz invariant bosonic generators $B_l$ of some internal compact Lie group.
\end{theorem}

However, we can make this restrictions to be relaxed by taking the supersymmetry, which were proposed by Haag, {\L}opusza\'{n}ski, and Sohnius. Supersymmetry is defined as a graded Lie algebra that has additional algebraic system using anti-commutator in addition to the usual Lie algebra.

\section{Supersymmetry Algebras}

In this section, we wish to talk about how mathematics of supersymmetry is defined, and it can overcome the limitation given by the Coleman-Mandula theorem. 

\begin{definition}[Graded Algebra]
Let $L$ be the direct sum of N+1 ($N\ge1$) vector subspaces $L_k$ as follows:
\begin{eqnarray}
L = \bigoplus^N_{k=0} L_k. 
\end{eqnarray}
Then, the space $L$ satisfies a multiplication rule $\circ$ that is defined by a map $\circ: L\times L \rightarrow L$ such that $u_k \in L_k \implies u_j \circ u_k \in L_{j+k ~\textrm{mod(N+1)}}$. This is called $Z_{(N+1)}$-graded algebra because the group whose elements are the numbers of indices for the vector subspaces $L_k$ is isomorphic to a finite cyclic group $Z/(N+1)Z$ \footnote{e.g. $Z_2 \equiv Z/2Z = \{0,1\} = \{x \in Z | x = n~\textrm{mod}(2) \wedge n \in Z \}$ }with the addition operation modulo $(N+1)$. Such a property with the multiplication $\circ$ is called ``grading''.
\end{definition}

Now we utilize this property to define the so-called ``super-Poincar\'{e} algebra'' that is Poincare algebra equipped with the superalgebra that has structure of a $Z_2$-graded Lie algebra which can avoid the restrictions offered by the Coleman-Mandula theorem. First of all, we are going to see the defintion of the $Z_2$-graded Lie algebra as follows:
\begin{definition}[Superalgebra]
A superalgebra is a $Z_2$-graded Lie algebra defined by a $Z_2$-graded algebra whose multiplication is given by a Lie bracket such that:
(1) Supersymmetrization:
For all $x_i \in L_i,x_j \in L_j ~(i,j=0,1)$, $x_i \circ x_j = (-1)^{(i \times j)+1}x_j \circ x_i$. (2) Generalized Jacobi identity: \begin{eqnarray}
&&(-1)^{ik}(x_i \circ x_j) \circ  x_k +(-1)^{ji} (x_j \circ  x_k) \circ  x_i + (-1)^{kj}(x_k \circ x_i) \circ  x_j = 0.
\end{eqnarray}
\end{definition}

Then, we are finally ready to introduce the Super-Poincar\'{e}. 

\begin{definition}[Super-Poincar\'{e} algebra]
Super-Poincar\'{e} algebra is defined by a $Z_2$-graded Lie algebra \footnote{ The elements of $L_0$ and $L_1$ are considered as `bosonic' and `fermionic' variables, respectively.}  $L = L_0 \oplus L_1$ where $L_0$ is an union of the Poincare algebra of $SO(3,1)$ and a finite set of Lorentz scalar (internal) Lie algebras of compact Lie groups $G_l$ ($l=1,2,3,\cdots,M$, $M$ is the number of the internal Lie groups), and $L_1 \equiv \bigoplus^{\mathcal{N}}_{i=1} L_1^{(i)}$ whose $L_1^{(i)}$ is $Span\{Q_a^i\},a=1,2,3,4$ where $Q_a^i$ \footnote{The reason why we consider different $Q_a^i$s with the index `$i$' is that for all $a$ of the Majorana spinor $Q_a^i$ the multiplication $\circ$ between the elements of the internal Lie algebras in the $L_0$ and the $L_1$ must be closed back into the $L_1$, while the product between the elements of the internal Lie algebras and the Poincar\'{e} algebra is closed into the $L_0$ naturally. Therefore, in general, the indices of the elements in the two subspaces $L_0$ and $L_1$ have to be allowed to mix via structure constants of their commutation relation. That is why we can consdier a general $\mathcal{N}$-extended supersymmetry algebra.} are Majorana spinors and $i=~1,~2,~3,\cdots,~\mathcal{N}$ (this $\mathcal{N}$ is called `dimension of supersymmetry') such that 
\begin{itemize}
    \item $P_\mu \circ Q_a^i = [P_\mu, Q_a^i] = 0$
    \item $M_{\mu\nu} \circ Q_a^i = [M_{\mu\nu}, Q_a^i] = (M_{\mu\nu})_{ab}Q_b^i$ 
    \item $Q_a^i \circ Q_b^j = \{Q_a^i,Q_b^j\} = -2(\gamma^\mu C)_{ab}P_\mu$
    \item $\bar{Q}_a^i \circ \bar{Q}_b^j = \{\bar{Q}_a^i,\bar{Q}_b^j\} = -2(C \gamma^\mu )_{ab}P_\mu$
    \item $Q_a^i \circ \bar{Q}_b^j = \{Q_a^i,\bar{Q}_b^j\} = 2(\gamma^\mu )_{ab}P_\mu$
    \item $[B_l, Q_a^i]=f_l^{~ij}Q_a^j$
    \item $[B_l,P_\mu]=[B_l,M_{\mu\nu}]=0$
\end{itemize}
 where $P_\mu,M_{\mu\nu},B_l \in L_0$ and $Q_a^i \in L_1^{(i)}$; $\gamma^\mu$ are the gamma matrices; $C$ is the charge conjugation matrix\footnote{In any representation (Dirac, Weyl, Majorana), the following relations have to hold:
\begin{eqnarray}
 C \equiv i\gamma^2\gamma^0, \quad  C^2 = -1,\quad  C^{-1}=C^T=C^\dag=-C,\quad  ( \gamma^\mu )^T = C\gamma^\mu C,~ ( \gamma^5 )^T = -C\gamma^5 C.
\end{eqnarray}}, and $\bar{Q}_a = (Q^TC)_a, Q_a = (C\bar{Q}^T)_a$ (Majorana conjugate). Note that $[\cdot,\cdot]$ is anti-symmetric but $\{\cdot,\cdot\}$ is symmetric. However, we call them as commutator and anti-commutator respectively.
\end{definition}

Here, the commutation relations can be obtained by using the generalized Jacobi identity of three variables and their given algebras. For instance, the commutation relations $[P,Q]$, $[M,Q]$, and $\{Q,Q\}$ (we dropped the relevant indices for simplicity)  can be found by evaluating the Jacobi identities of the three elements ($P,P,Q$), ($M,M,Q$), and ($Q,Q,P$) respectively.  

The detailed derivations of the commutators are as follows. At first, consider the $P,P,Q$ set, and suppose that the supercharge has a commutation relation defined by  $ [P^\mu,Q_a^i] \equiv c(\gamma^\mu C)_{ab} Q_b^i \in L_1^{(i)}$, where $c(\gamma^\mu C)_{ab}$ are its structure constants and $c$ is a constant. Then, their Jacobi identity is 
\begin{eqnarray}
&& [[P^\mu,P^\nu],Q_a^i]+[[P^\nu,Q_a^i],P^\mu]+[[Q_a^i,P^\mu],P^\nu]=0 \nonumber\\
&& \implies 
c^2[\gamma^\nu C\gamma^\mu C-\gamma^\mu C\gamma^\nu C]_{ac}Q_c^i=0
\nonumber\\
&& \implies 
c^2[\gamma^\nu (\gamma^\mu)^T -\gamma^\mu (\gamma^\nu)^T]_{ac}Q_c^i=0 \nonumber\\
&& \implies c=0
\end{eqnarray}
Thus, it turns out that $[Q_b^i,P^\nu] = 0$. This is as expected because the supercharge does not depend on spacetime coordinate, which means it has spacetime translation invariance so that we consider global supersymmetry.  

Next, consider the $M,M,Q$ set, and define $[M^{\mu\nu},Q_a^i]\equiv (b^{\mu\nu})_{ab}Q_b^i$, where $(b^{\mu\nu})_{ab}$ are its structure constants. Applying the Jacobi identity to the set and after some tedious calculations, we find
\begin{eqnarray}
&& [[M^{\mu\nu},M^{\rho\sigma}],Q_a^i]+[[M^{\rho\sigma},Q_a^i],M^{\mu\nu}] +[[Q_a^i,M^{\mu\nu}],M^{\rho\sigma}]=0 \nonumber\\
&& \implies [b^{\mu\nu},b^{\rho\sigma}]_{ab} = -i(\eta^{\mu\rho}b^{\nu\sigma}-\eta^{\nu\rho}b^{\mu\sigma}-\eta^{\mu\sigma}b^{\nu\rho}+\eta^{\nu\sigma}b^{\mu\rho}),
\end{eqnarray}
which is exactly same as the commutation relation of $M^{\mu\nu}$s as representation of the Lorentz algebra for a spinor. Thus, we can conclude that $[M^{\mu\nu},Q_a^i]\equiv (M^{\mu\nu})_{ab}Q_b^i$.

Plus, let us consider a Jacobi identity of the $Q,Q,P$ set. Then, the identity is given by
\begin{eqnarray}
[\{Q^i_a,Q^j_b\},P^\mu]-\{[Q^j_b,P^\mu],Q^i_a\}+\{[P^\mu,Q^i_a],Q^j_b\}=0.
\end{eqnarray}
Since a supercharge and the spacetime translation operator already commute, the Jacobi identity implies that \begin{eqnarray}
[\{Q^i_a,Q^j_b\},P^\mu]=0.
\end{eqnarray}
This tells us that $\{Q^i_a,Q^j_b\}$ must have no a Lorentz generator $M^{\mu\nu}$ because this does not commute with the momentum. That is, the anti-commutator can be proportional to any operator that commutes with the spacetime translation generator $P^\mu$. 

Now, let us find the most general admissible form of the anti-commutator. Since the supercharges are anticommuting elements of a Clifford algebra, their commutation relations must be closed in the algebra as well. Thus, a result of a commutation relation can be represented by a linear combination of the basis elements of the Clifford algebra, i.e. the $\gamma$-matrices and their products.   

Since $(\gamma^\mu)^T=C \gamma^\mu C$ in any representation and $\{Q_a^i,Q_b^j\}$ is symmetric under the interchange $a,i \leftrightarrow b,j$, we have to consider the charge conjugation matrix $C$ and the definite symmetries. Hence, from the possible basis matrices of a general commutator of two anti-commuting variables, given by 
\begin{eqnarray}
C,~\gamma_5C, ~\gamma^\mu C, \gamma^\mu\gamma_5 C,~\gamma^\mu\gamma^\nu C,~\gamma^\mu\gamma^\nu\gamma_5 C, \cdots,
\end{eqnarray}
we can only choose $C, \gamma^\mu C, \gamma^\mu\gamma^5 C$. This is because there is no generators but $P_\mu,M_{\mu\nu}$ allowed by the Coleman-Mandula theorem; that is, there are no the corresponding tensorial generators for the $\gamma^\mu\gamma^5 C$ etc and $[\{Q_a^i,Q_b^j\},M_{\mu\nu}] \neq 0$. However, there can be bosonic generators that can be allowed to exist in the Coleman-Mandula theorem. Therefore, the commutator can be represented by
\begin{eqnarray}
\{Q_a^i,Q_b^j\} = (\gamma^\mu C)_{ab}m^{ij}P_\mu + C_{ab}V^{ij} + i(\gamma^5C)_{ab}Z^{ij},
\end{eqnarray}
where $m^{ij}=m^{ji},V^{ij}=-V^{ji},Z^{ij}=-Z^{ji}$ that are required to meet the interchange symmetries $a,i \leftrightarrow b,j$ and they are real matrices.

Moreover, since $\{Q_a^i,Q_b^j\}=0$ when $i\neq j$, we have to impose $m^{ij} \equiv a \delta^{ij}$ where $a$ is a constant, which will be determined by imposing that (when $\mathcal{N}=1$) $[\delta_1,\delta_2]$ using the above expression with the $a$ must be equal to $[\delta_1,\delta_2] = 2\left((\bar{\theta}_{2})_a (\gamma^\mu)_{ab} (\theta_{1})_b \right)P_\mu$ in the Majorana representation \footnote{If we represent the supercharges $Q,\bar{Q}$ and $\theta,\bar{\theta}$ in the Weyl representation, we need to evaluate the commutation relation $[\delta_1,\delta_2]$ in the same representation.}. Then, it turns out that $a = -2$, which give us the following:
\begin{eqnarray}
\{Q_a^i,Q_b^j\} = -2(\gamma^\mu C)_{ab}\delta^{ij}P_\mu + C_{ab}V^{ij} + i(\gamma^5C)_{ab}Z^{ij},
\end{eqnarray}
where we call the bosonic generators $V^{ij}$ and $Z^{ij}$ as ``Central charges'' because they commute with each other and even all the Poincar\'{e} and supercharge generators. Hence, the relation reduces to
\begin{eqnarray}
\{Q_a,Q_b\} = -2(\gamma^\mu C)_{ab}P_\mu. 
\end{eqnarray}

Also, by multiplying the charge conjugation matrix to this, we can get the following:
\begin{eqnarray}
\{Q_a,Q_b\}C_{bc}&=& Q_aQ_bC_{bc}+Q_bC_{bc}Q_a  \nonumber\\ 
&=& Q_a(Q^TC)_c+(Q^TC)_cQ_a \nonumber\\
&=& Q_a\bar{Q}_c+\bar{Q}_cQ_a = \{Q_a,\bar{Q}_b\},
\end{eqnarray}
and $-2(\gamma^\mu C^2)_{ac} P_\mu = 2(\gamma^\mu)_{ac} P_\mu$, together with $C^2=-1$ and $\bar{Q}=Q^TC$. This gives us 
\begin{eqnarray}
\{Q_a,\bar{Q}_b\} = 2(\gamma^\mu)_{ab}P_\mu.
\end{eqnarray}
Similarly, we can get 
\begin{eqnarray}
\{\bar{Q}_a,\bar{Q}_b\} = -2(C\gamma^\mu)_{ab}P_\mu.
\end{eqnarray}

So far, we have derived all the commutation relations that define the super-Poincar\'{e} algebras.

In the meantime, we are going to only consider $\mathcal{N}=1$ supersymmetry (also called `pure supersymmetry') so that we can drop the indices $i,j$ from now on.

\section{Some Remarks of Supersymmetry Algebras}

In this section, we now look into the representations of supersymmetry. 
First of all, we are going to see the three special properties of supermultiplets.
\begin{itemize}
    \item All component fields belonging to an irreducible representation of supersymmetry (a supermultiplet) have the identical mass.
    \item The energy $P^0$ in supersymmtric theory is always positive.
    \item A supermultiplet always has an equal number of boson and fermion degrees of freedom.
\end{itemize}

Let us now check these properties. The first statement comes from the following:
\begin{eqnarray}
[P^2,P^\mu]=[P^2,M^{\mu\nu}]=[P^2,Q_a^i]=0.
\end{eqnarray}

For the second statement, we can easily see this by evaluating 
\begin{eqnarray}
\sum_{i=1}^\mathcal{N}\textrm{Tr}[\{Q_a^i,\bar{Q}_a^i\}(\gamma^0)] 
&=& 2\mathcal{N}\textrm{Tr}[(\gamma^\mu\gamma^0)]P_\mu = 8\mathcal{N}P_0 \ge 0,
\nonumber\\
\end{eqnarray}
where we used $\textrm{Tr}[\gamma^\mu\gamma^\nu]=4\eta^{\mu\nu}$, and the trace is done over the spinor indices. Thus, it implies that $P_0 \ge 0$. 
To get the third one, let us introduce a fermionic operator defined by $(-1)^{N_f}$ in which $(-1)^{N_f} = +1$ for bosonic states and $(-1)^{N_f} = -1$ for fermionic states. This now anti-commute with the supercharge: $\{(-1)^{N_f},Q^i_a\}=0$.

Then, for any supermultiplet, by taking the trace over the supermultiplet indices, we have 
\begin{eqnarray}
\textrm{Tr}[(-1)^{N_f}\{Q_a^i,\bar{Q}_b^j\}]&=&\textrm{Tr}[(-1)^{N_f}Q_a^i\bar{Q}_b^j+(-1)^{N_f}\bar{Q}_b^jQ_a^i] \nonumber\\
&=& \textrm{Tr}[-Q_a^i(-1)^{N_f}\bar{Q}_b^j+Q_a^i(-1)^{N_f}\bar{Q}_b^j] \nonumber\\
&=& 0,
\end{eqnarray}
where the first term is obtained by using the commutator $\{(-1)^{N_f},Q^i_a\}=0$, while the second term is obtained by taking a permutation for the three operators inside the trace. 

By the way, since $\{Q_a,\bar{Q}_b\} = 2(\gamma^\mu)_{ab}P_\mu$, for $i=j$, we have 
\begin{eqnarray}
\textrm{Tr}[(-1)^{N_f}2(\gamma^\mu)_{ab}P_\mu]&=&2(\gamma^\mu)_{ab}P_\mu \textrm{Tr}[(-1)^{N_f}] = 0 
\implies \textrm{Tr}[(-1)^{N_f}] = 0.
\end{eqnarray}
This tells us that there must be the same number of the bosonic and fermionic degrees of freedom on a supermultiplet. 

In fact, by the theorem of Haag, {\L}opusza\'{n}ski, and Sohnius, it is known that the supersymmetry algebra is the most general superalgebra admissible as a symmetry of quantum field theory. Therefore, this can be considered as an extension to superalgebra of the Coleman-Mandula theorem.  

We omit putting the representations of supersymmetry algebras, which are explianed in detail in Ref. \cite{1,2,3}. The point is that for $\mathcal{N}=1$ global supersymmetry, there is a massless supermultiplet $|p^\mu,\lambda>,|p^\mu,\lambda+1/2)>$, while a massive supermultiplet is given by $2|m,j=y;p^\mu,j_3>,|m,j=y+1/2;p^\mu,j_3>,|m,j=y-1/2;p^\mu,j_3>$, where $y$ is superspin number that is a label of a clifford vacuum $|\Omega>=|m,j=y;p^\mu,j_3>$.

\section{Superspace Formalism and Supersymetry Transformations}

\subsection{Superspace}

In this section, we introduce a new concept called  ``superspace''. The coordinate $z$ on this space is called ``super-coordinates'' and defined by $z \equiv (x^\mu, \theta^\alpha, \bar{\theta}_{\dot{\alpha}})$. 

Here, we use the Weyl representation, instead of the Majorana for convenience. 

From the anti-commutator $\{Q_a,\bar{Q}_b\} = 2(\gamma^\mu)_{ab}P_\mu$, we can find 
\begin{eqnarray}
\{Q_\alpha,\bar{Q}_{\dot{\alpha}}\} = 2(\sigma^\mu)_{\alpha \dot{\alpha}}P_\mu.
\end{eqnarray}
Then, it gives us 
\begin{eqnarray}
[\theta Q, \bar{\theta}\bar{Q}] =[\theta^\alpha Q_\alpha, \bar{\theta}_{\dot{\alpha}}\bar{Q}^{\dot{\alpha}}] = 2\theta^\alpha(\sigma^\mu)_{\alpha\dot{\beta}}\bar{\theta}^{\dot{\beta}}P_\mu, 
\end{eqnarray}
where $\theta Q \equiv \theta^\alpha Q_\alpha, ~ (\theta Q)^\dag = (\theta^\alpha Q_\alpha)^\dag = (Q_\alpha)^\dag(\theta^\alpha)^\dag = \bar{Q}^{\dot{\alpha}}\bar{\theta}_{\dot{\alpha}}=- \bar{Q}_{\dot{\alpha}}\bar{\theta}^{\dot{\alpha}} =\bar{\theta}^{\dot{\alpha}} \bar{Q}_{\dot{\alpha}}\equiv \bar{\theta}\bar{Q}, ~\theta \sigma^m \bar{\theta} \equiv \theta^\alpha (\sigma^m)_{\alpha \dot{\alpha}} \bar{\theta}^{\dot{\alpha}}$.

Then now we are ready to introduce a general group element that generates a Super-Poincar\'{e} group defined by
\begin{eqnarray}
g(z,\omega) &=& g(x,\theta,\bar{\theta},\omega) \nonumber\\
&\equiv& \exp[i(-x^\mu P_\mu -\frac12 \omega^{\mu\nu}M_{\mu\nu} + \theta^\alpha Q_\alpha + \bar{\theta}_{\dot{\alpha}}\bar{Q}^{\dot{\alpha}} )],\nonumber\\&&
\end{eqnarray}
where $z$ is the super-coordinate and $\omega^{\mu\nu}$ is the corresponding parameter to the Lorentz generator.

For simplicity, we set $\omega^{\mu\nu} =0$, which is not of interest. Then, the group element is $g(z) \equiv \exp[i(x^\mu P_\mu + \theta^\alpha Q_\alpha + \bar{\theta}_{\dot{\alpha}}\bar{Q}^{\dot{\alpha}} )] \equiv \exp(iz^AG_A)$.

Now, let us consider a product of two group elements to see what happens:
\begin{eqnarray}
g(z_1)g(z_2)&=& \exp(-iz_1^AG_A)\exp(-iz_2^AG_A)= g(0,\epsilon,\bar{\epsilon})g(x,\theta,\bar{\theta})
\nonumber\\ 
&=&
\exp[-i(\epsilon^\alpha Q_\alpha + \bar{\epsilon}_{\dot{\alpha}}\bar{Q}^{\dot{\alpha}} )]\exp[-i(-x^\mu P_\mu + \theta^\alpha Q_\alpha + \bar{\theta}_{\dot{\alpha}}\bar{Q}^{\dot{\alpha}} )] 
\nonumber\\
&=&  \exp[-i(\epsilon^\alpha Q_\alpha + \bar{\epsilon}_{\dot{\alpha}}\bar{Q}^{\dot{\alpha}} -x^\mu P_\mu + \theta^\alpha Q_\alpha + \bar{\theta}_{\dot{\alpha}}\bar{Q}^{\dot{\alpha}} +\frac12 [\epsilon^\alpha Q_\alpha + \bar{\epsilon}_{\dot{\alpha}}\bar{Q}^{\dot{\alpha}},\theta^\alpha Q_\alpha + \bar{\theta}_{\dot{\alpha}}\bar{Q}^{\dot{\alpha}} ])] 
\nonumber\\
&=&\exp[-i( -x^\mu P_\mu + (\theta^\alpha+\epsilon^\alpha) Q_\alpha + (\bar{\theta}_{\dot{\alpha}}+\bar{\epsilon}_{\dot{\alpha}})\bar{Q}^{\dot{\alpha}} +\frac12 \left( [\epsilon^\alpha Q_\alpha, \bar{\theta}_{\dot{\alpha}}\bar{Q}^{\dot{\alpha}} ]+[ \bar{\epsilon}_{\dot{\alpha}}\bar{Q}^{\dot{\alpha}},\theta^\alpha Q_\alpha]\right) )] 
\nonumber\\
&=&\exp[-i(- x^\mu P_\mu + (\theta^\alpha+\epsilon^\alpha) Q_\alpha + (\bar{\theta}_{\dot{\alpha}}+\bar{\epsilon}_{\dot{\alpha}})\bar{Q}^{\dot{\alpha}} +\frac12 \left( 2\epsilon^\alpha(\sigma^\mu)_{\alpha\dot{\beta}}\bar{\theta}^{\dot{\beta}}P_\mu-2\theta^\alpha(\sigma^\mu)_{\alpha\dot{\beta}}\bar{\epsilon}^{\dot{\beta}}P_\mu\right) )] 
\nonumber\\
&=&\exp[-i(- (x^\mu-\epsilon^\alpha(\sigma^\mu)_{\alpha\dot{\beta}}\bar{\theta}^{\dot{\beta}}+\theta^\alpha(\sigma^\mu)_{\alpha\dot{\beta}}\bar{\epsilon}^{\dot{\beta}} ) P_\mu + (\theta^\alpha+\epsilon^\alpha) Q_\alpha + (\bar{\theta}_{\dot{\alpha}}+\bar{\epsilon}_{\dot{\alpha}})\bar{Q}^{\dot{\alpha}})] \nonumber\\
&=& g(x',\theta',\bar{\theta}') = \exp(iz'^AG_A) = \exp(i\Lambda^A_B z^B G_A), 
\end{eqnarray}
where we used $[\epsilon Q, \theta Q]=[\bar{\epsilon}\bar{Q}, \bar{\theta}\bar{Q}]=0$ and $z'^A=\Lambda^A_B z^B$. Hence, the group operation generates the following transformations 
\begin{eqnarray}
x^\mu \rightarrow x^\mu-\epsilon^\alpha(\sigma^\mu)_{\alpha\dot{\beta}}\bar{\theta}^{\dot{\beta}}+\theta^\alpha(\sigma^\mu)_{\alpha\dot{\beta}}\bar{\epsilon}^{\dot{\beta}},\quad \theta^\alpha \rightarrow  \theta^\alpha+\epsilon^\alpha, \quad \bar{\theta}_{\dot{\alpha}} \rightarrow \bar{\theta}_{\dot{\alpha}} + \bar{\epsilon}_{\dot{\alpha}}.
\end{eqnarray}

\subsection{General supersymmetry transformations of a Lorentz invariant superfield}

Meanwhile, let us consider a general function on the superspace $\Phi(z)$, which is called ``superfield'' and can be expanded in terms of component fields:
\begin{eqnarray}
\Phi(x,\theta,\bar{\theta}) 
&=& \varphi(x) + \theta \psi(x) + \bar{\theta} \bar{\chi}(x) + \theta\theta M(x) + \bar{\theta}\bar{\theta} N(x) \nonumber\\
&&+ (\theta \sigma^\mu \bar{\theta})V_\mu(x) + (\theta\theta)\bar{\theta} \bar{\lambda} + (\bar{\theta}\bar{\theta}) \theta \rho(x) + (\theta\theta)(\bar{\theta}\bar{\theta}) D(x), 
\end{eqnarray}
where $\varphi,M,N,D$ are scalar fields; $\psi,\chi,\lambda,\rho$ are spinor fields, and $V_\mu$ is a vector field. These are called ``component fields''.

Then, using the usual symmetry argument \footnote{ For a finite general symmetry transformation as a realization (or fundamental representation) of a symmetry group that acts on the ``Realization space'' $\mathcal{R}$ of a (local) field $\phi(z)$ (which is considered as an element in a `Function space', e.g. Hilbert space) given by $\mathcal{U}(\epsilon) \equiv e^{\delta(\epsilon)} = e^{\epsilon^A \mathcal{G}_A}$  where $\mathcal{G}_A$ is the symmetry realization generator and $\epsilon^A$ is the corresponding transformation parameter which is a conjugate of the generator, the transformation is given by $\phi'(z)=\mathcal{U}(\epsilon)\phi(z)=\phi(z')$, where $z'\equiv z'(\epsilon) = \Lambda(\epsilon)z $ and $\Lambda(\epsilon)$ is the coordinate symmetry transformation of the coordinates $z$ generated by the $\mathcal{U}(\epsilon)$. On the other hand, as an adjoint representation of a symmetry group that acts on the ``Adjoint representation space (i.e. the Lie algebra of the group)'' $R$ of a (local) field $\phi(z)$ (which is now considered as a quantum operator on the `Hilbert space'), the transformation is given by $\phi'(z)=U(\epsilon)\phi(z) U(\epsilon)^{-1}$, where $U(\epsilon) \equiv e^{-\epsilon^A G_A}$, $G_A$ is the adjoint representation symmetry generator, and $\epsilon^A$ is the same symmetry parameter. Therefore, from the two transformations, we find an equality such that $U(\epsilon)\phi(z) U(\epsilon)^{-1}=\mathcal{U}(\epsilon)\phi(z)=\phi(z')$, where $z'\equiv z'(\epsilon) = \Lambda(\epsilon)z $. Also, we can see that the $U(\epsilon)$ generates the $\Lambda(\epsilon)$ after all. This implies that $\delta(\epsilon)\phi =-\epsilon^A[G_A,\phi]=\epsilon^A\mathcal{G}_A \phi$.} for the supersymmetry transformation,
we have
\begin{eqnarray}
 e^{-i(-a^\mu P_\mu+\epsilon Q +\bar{\epsilon}\bar{Q})} \Phi(x^\mu, \theta^\alpha, \bar{\theta}_{\dot{\alpha}}) e^{+i(-a^\mu P_\mu+\epsilon Q +\bar{\epsilon}\bar{Q})}  \equiv \Phi(x'^\mu, \theta'^\alpha, \bar{\theta}'_{\dot{\alpha}}) = \Phi(x^\mu, \theta^\alpha, \bar{\theta}_{\dot{\alpha}}, a^\mu, \epsilon^\alpha, \bar{\epsilon}_{\dot{\alpha}}).
\end{eqnarray}
Then, by redefining the variables, we can get
\begin{eqnarray}
 e^{-i(-x^\mu P_\mu+\theta Q +\bar{\theta}\bar{Q})} \Phi(0, 0, 0) e^{+i(-x^\mu P_\mu+\theta Q +\bar{\theta}\bar{Q})} \equiv \Phi(x^\mu, \theta^\alpha, \bar{\theta}_{\dot{\alpha}}) = g(z_2) \Phi(0,0,0) g(z_2)^{-1}.
\end{eqnarray}
Thus, using this relation, we can reach the following 
\begin{eqnarray}
e^{-i(\epsilon Q +\bar{\epsilon}\bar{Q})} \Phi(x^\mu, \theta^\alpha, \bar{\theta}_{\dot{\alpha}})  e^{+i(\epsilon Q +\bar{\epsilon}\bar{Q})} &=& g(z_1)\Phi(x^\mu, \theta^\alpha, \bar{\theta}_{\dot{\alpha}}) g(z_1)^{-1} 
\nonumber\\&=&  g(z_1)g(z_2) \Phi(0,0,0) (g(z_1)g(z_2))^{-1} 
\nonumber\\&=&  g(z') \Phi(0,0,0) g(z')^{-1} 
\equiv \Phi(x'^\mu, \theta'^\alpha, \bar{\theta}'_{\dot{\alpha}}) \nonumber\\
&=&  \Phi(x^\mu-\epsilon^\alpha(\sigma^\mu)_{\alpha\dot{\beta}}\bar{\theta}^{\dot{\beta}}+\theta^\alpha(\sigma^\mu)_{\alpha\dot{\beta}}\bar{\epsilon}^{\dot{\beta}} ,\theta^\alpha+\epsilon^\alpha,  \bar{\theta}_{\dot{\alpha}}+\bar{\epsilon}_{\dot{\alpha}}) \nonumber\\&=&  e^{i(\epsilon \mathcal{Q} + \bar{\epsilon}\bar{\mathcal{Q}})}  \Phi(x^\mu, \theta^\alpha, \bar{\theta}_{\dot{\alpha}}),
\end{eqnarray}
where $\mathcal{Q}$ is a realization of the transformation acting on the function space of $\Phi$. 
Next, let us find the explicit form of the realization generators $\mathcal{Q}$. From the above equation, by expanding the transformed $\Phi$, we have
\begin{eqnarray}
&& \Phi(x^\mu-\epsilon^\alpha(\sigma^\mu)_{\alpha\dot{\beta}}\bar{\theta}^{\dot{\beta}}+\theta^\alpha(\sigma^\mu)_{\alpha\dot{\beta}}\bar{\epsilon}^{\dot{\beta}} ,\theta^\alpha+\epsilon^\alpha,  \bar{\theta}_{\dot{\alpha}}+\bar{\epsilon}_{\dot{\alpha}}) \nonumber\\
&& \approx \Phi(x,\theta,\bar{\theta}) + i(-\epsilon^\alpha(\sigma^\mu)_{\alpha\dot{\beta}}\bar{\theta}^{\dot{\beta}}+\theta^\alpha(\sigma^\mu)_{\alpha\dot{\beta}}\bar{\epsilon}^{\dot{\beta}})\partial_\mu \Phi \nonumber\\&& + \epsilon^\alpha \frac{\partial}{\partial \theta^\alpha} \Phi  + \bar{\epsilon}_{\dot{\alpha}} \frac{\partial}{\partial \bar{\theta}_{\dot{\alpha}}} \Phi + \cdots  
\nonumber\\&& = \Phi + i(\epsilon^\alpha \mathcal{Q}_\alpha + \bar{\epsilon}_{\dot{\alpha}}\bar{\mathcal{Q}}^{\dot{\alpha}})\Phi.
\end{eqnarray}
By using $\bar{\epsilon}_{\dot{\alpha}}\bar{\mathcal{Q}}^{\dot{\alpha}}=-\bar{\epsilon}^{\dot{\alpha}}\bar{\mathcal{Q}}_{\dot{\alpha}}=\bar{\mathcal{Q}}_{\dot{\alpha}}\bar{\epsilon}^{\dot{\alpha}}$; $\theta^\alpha(\sigma^\mu)_{\alpha\dot{\beta}}\bar{\epsilon}^{\dot{\beta}} = \theta^\beta(\sigma^\mu)_{\beta\dot{\alpha}}\bar{\epsilon}^{\dot{\alpha}}$; and $\bar{\epsilon}_{\dot{\beta}} \frac{\partial}{\partial \bar{\theta}_{\dot{\beta}}}=- \frac{\partial}{\partial \bar{\theta}_{\dot{\beta}}}\bar{\epsilon}_{\dot{\beta}}=\varepsilon^{\dot{\beta}\dot{\alpha}}\frac{\partial}{\partial \bar{\theta}^{\dot{\alpha}}}\bar{\epsilon}_{\dot{\beta}}=-\varepsilon^{\dot{\alpha}\dot{\beta}}\frac{\partial}{\partial \bar{\theta}^{\dot{\alpha}}}\bar{\epsilon}_{\dot{\beta}}$, and by identifying each term in both sides, we finally have the realizations of the supercharges as follows:
\begin{eqnarray}
\mathcal{Q}_\alpha &=& -i\frac{\partial}{\partial \theta^\alpha} - (\sigma^\mu)_{\alpha\dot{\beta}}\bar{\theta}^{\dot{\beta}}\partial_\mu, \\
\bar{\mathcal{Q}}_{\dot{\alpha}} &=& +i\frac{\partial}{\partial \bar{\theta}^{\dot{\alpha}}} + \theta^\beta(\sigma^\mu)_{\beta\dot{\alpha}}\partial_\mu, \\
\mathcal{P}_\mu &=& -i\partial_\mu.
\end{eqnarray}

After some lengthy calculations with the above realizations and the symmetry transformation relation given by  
\begin{eqnarray}
\delta(\epsilon) \Phi = i(\epsilon^\alpha \mathcal{Q}_\alpha+\bar{\epsilon}_{\dot{\alpha}}\bar{\mathcal{Q}}^{\dot{\alpha}})\Phi (= -i[\epsilon^\alpha Q_\alpha+\bar{\epsilon}_{\dot{\alpha}}\bar{Q}^{\dot{\alpha}},\Phi]),
\end{eqnarray}
we can get the explicit forms of the infinitesimal supersymmetry transformations of the component fields of the superfield $\Phi$ as follows:
\begin{eqnarray}
\delta_{\epsilon} \varphi &=& \epsilon \psi + \bar{\epsilon}\bar{\chi},\\
\delta_{\epsilon} \psi &=& 2\epsilon M + \sigma^\mu \bar{\epsilon}(i\partial_\mu \varphi + V_\mu),\\
\delta_{\epsilon} \bar{\chi} &=& 2\bar{\epsilon}N -\epsilon \sigma^\mu (i\partial_\mu \varphi - V_\mu),\\
\delta_{\epsilon} M &=& \bar{\epsilon}\bar{\lambda}-\frac {i}{2} \partial_\mu \psi \sigma^\mu \bar{\epsilon},\\
\delta_{\epsilon} N &=& \epsilon\rho+\frac {i}{2} \epsilon \sigma^\mu \partial_\mu \bar{\chi},\\
\delta_{\epsilon} V_\mu &=& \epsilon \sigma_\mu \bar{\lambda} + \rho \sigma_\mu \bar{\epsilon} + \frac{i}{2} (\partial^\nu \psi \sigma_\mu \bar{\sigma}_\nu \epsilon -\bar{\epsilon} \bar{\sigma}_\nu \sigma_\mu \partial^\nu \bar{\chi}),\nonumber\\ && \\
\delta_{\epsilon} \bar{\lambda} &=& 2\bar{\epsilon}D + \frac{i}{2} (\bar{\sigma}^\nu \sigma^\mu \bar{\epsilon})\partial_\mu V_\nu + i\bar{\sigma}^\mu \epsilon \partial_\mu M,\\
\delta_{\epsilon} \rho &=& 2\bar{\epsilon}D- \frac{i}{2} (\sigma^\nu \bar{\sigma}^\mu \epsilon)\partial_\mu V_\nu + i\sigma^\mu \bar{\epsilon} \partial_\mu N,\\
\delta_{\epsilon} D &=& \frac{i}{2}\partial_\mu (\epsilon \sigma^\mu \bar{\lambda}-\rho \sigma^\mu \bar{\epsilon}),
\end{eqnarray}
where the $\delta_{\epsilon}D$ is a total derivative in particular \footnote{$\theta$-derivatives to be added}.  

\subsection{Classification of superfields}

The following are superfields:
\begin{itemize}
    \item Linear combinations of superfields, $\Phi+\Psi$, etc.
    \item Products of superfields, $\Phi\Psi$, etc.
    \item Spacetime derivatives of superfields, $\partial _\mu \Phi$, etc.
    \item Super-covariant derivatives of superfields, $\mathcal{D}_{\alpha}\Phi,\bar{\mathcal{D}}_{\dot{\alpha}}\Phi$, etc. Note that $\partial_{\alpha}\Phi$ is not a superfield because $\delta_{\epsilon} (\partial_{\alpha}\Phi)=-i[\epsilon Q +\bar{\epsilon}\bar{Q},\partial_{\alpha}\Phi] = -i\partial_{\alpha}[\epsilon Q +\bar{\epsilon}\bar{Q},\Phi] = \partial_{\alpha} i(\epsilon\mathcal{Q}+\bar{\epsilon}\bar{\mathcal{Q}}) \Phi \neq i(\epsilon\mathcal{Q}+\bar{\epsilon}\bar{\mathcal{Q}}) (\partial_{\alpha}\Phi) $, i.e. $[\partial_{\alpha},\epsilon\mathcal{Q}+\bar{\epsilon}\bar{\mathcal{Q}}] \neq 0$. That is why we define a super-covariant derivative given by
    \begin{eqnarray}
    \mathcal{D}_{\alpha} &\equiv& \partial_{\alpha} + i(\sigma^\mu)_{\alpha\dot{\beta}}\bar{\theta}^{\dot{\beta}}\partial_\mu,\\
    \bar{\mathcal{D}}_{\dot{\alpha}} &\equiv& -\partial_{\alpha} - i\theta^\beta(\sigma^\mu)_{\beta\dot{\alpha}}\bar{\theta}^{\dot{\beta}}\partial_\mu,
    \end{eqnarray}
    which satisfy $[\mathcal{D}_{\alpha},\epsilon\mathcal{Q}+\bar{\epsilon}\bar{\mathcal{Q}}]=[ \bar{\mathcal{D}}_{\dot{\alpha}},\epsilon\mathcal{Q}+\bar{\epsilon}\bar{\mathcal{Q}}]=0$ and some anticommutation relations \footnote{$\{\mathcal{D}_{\alpha}, \mathcal{Q}_{\beta}\}=\{\mathcal{D}_{\alpha}, \bar{\mathcal{Q}}_{\dot{\beta}}\}=\{\bar{\mathcal{D}}_{\dot{\alpha}}, \mathcal{Q}_{\beta}\}=\{\bar{\mathcal{D}}_{\dot{\alpha}}, \bar{\mathcal{Q}}_{\dot{\beta}}\}=\{\mathcal{D}_{\alpha}, \mathcal{D}_{\beta}\}=\{\bar{\mathcal{D}}_{\dot{\alpha}}, \bar{\mathcal{D}}_{\dot{\beta}} \}=0$, but $\{\mathcal{D}_{\alpha},\bar{\mathcal{D}}_{\dot{\beta}}\} = -2i(\sigma^\mu)_{\alpha\dot{\beta}}\partial_\mu$.}.
    \item Constant scalar $\Phi=f$ can be superfield, but constant spinor $c$, in $\Phi = \theta c$, cannot because there exists $\delta_{\epsilon}\Phi = \epsilon c$.
\end{itemize}

In general, the most general superfield $\Phi$ is not an irreducible representation of supersymmetry. Thus, we can get rid of some degrees of freedom of the component fields by imposing possible constraints on a superfield. The possible constraints are as follows:
\begin{itemize}
    \item ``Chiral superfield'' $\Phi$ is a superfield that obeys a constraint $\bar{\mathcal{D}}_{\dot{\alpha}}\Phi=0$.
    \item ``Anti-Chiral superfield'' $\bar{\Phi}$ is a superfield that obeys a constraint $\mathcal{D}_{\alpha}\bar{\Phi}=0$, where $\bar{\Phi} \equiv \Phi^\dag$.
    \item ``Vector superfield'' $V$ is a superfield that obeys a constraint $V^\dag = V$.
    \item ``Linear superfield'' $L$ is a superfield that obeys two constaints $\mathcal{D}\mathcal{D}L =0$ and $L^\dag=L$.
\end{itemize}

\subsubsection{Chiral superfields}

To find a chiral superfield, let us consider $\Phi(y,\theta,\bar{\theta})$ with $y^\mu=x^\mu+i\theta \sigma^\mu \bar{\theta}$. Then, the constraint gives 
\begin{eqnarray}
\bar{\mathcal{D}}_{\dot{\alpha}}\Phi = -\bar{\partial}_{\dot{\alpha}} \Phi = 0,
\end{eqnarray}
which implies that the chiral superfield does not depend on the $\bar{\theta}$ in the new super-coordinate $(y,\theta,\theta)$.
Thus, it can be expanded as
\begin{eqnarray}
\Phi(y,\theta) = A(y)+\sqrt{2}\theta \psi(y) + \theta\theta F(y),
\end{eqnarray}
where we replaced $\theta$ in the general superfield into $\sqrt{2}\theta$ for convention. The chiral superfield can also be written by
\begin{eqnarray}
\Phi(x,\theta,\bar{\theta})= A + \sqrt{2} \theta\psi + \theta\theta F + i\theta\sigma^\mu\bar{\theta}\partial_\mu A -\frac{i}{\sqrt{2}}(\theta\theta)\partial_\mu \psi \sigma^\mu \bar{\theta} -\frac14(\theta\theta)(\bar{\theta}\bar{\theta})\partial_\mu\partial^\mu A,
\end{eqnarray}
which transform as
\begin{eqnarray}
\delta A &=& \sqrt{2}\epsilon\psi,\\
\delta \psi &=& i\sqrt{2} \sigma^\mu \bar{\epsilon} \partial_\mu A +\sqrt{2}\epsilon F,\\
\delta F &=& i\sqrt{2} \bar{\epsilon}\bar{\sigma}^\mu  \partial_\mu \psi.
\end{eqnarray}

Particularly, the product of chiral superfields is a chiral superfield, and any holomorphic function $f(\Phi)$ of a chiral superfield $\Phi$ is always a chiral superfield. However, $\Phi^\dag\Phi$ and $\Phi+\Phi^\dag$ are not chiral but real superfields. 

\subsubsection{Vector superfields}

Another constrained one is a real vector superfield. Let $V$ be a real vector superfield such that $V^\dag(x,\theta,\bar{\theta})=V(x,\theta,\bar{\theta})$. Then,
we have 
$\varphi^*=\varphi \equiv C,\psi = \chi,M = N^*,V_\mu = V_\mu^\dag,\lambda = \rho,D^* = D$ and
\begin{eqnarray}
V(x,\theta,\bar{\theta}) &=& C(x) + \theta \phi(x) + \bar{\theta} \bar{\phi}(x) + \theta\theta M(x) + \bar{\theta}\bar{\theta} M^*(x) \nonumber\\
&&+ (\theta \sigma^\mu \bar{\theta})V_\mu(x) + (\theta\theta)\bar{\theta} \bar{\lambda} + (\bar{\theta}\bar{\theta}) \theta \lambda(x) + (\theta\theta)(\bar{\theta}\bar{\theta}) D(x). 
\end{eqnarray}

By the way, a real superfield $\Phi+\Phi^\dag$ gives us 
\begin{eqnarray}
\Phi + \Phi^\dag &=& A+A^* +\sqrt{2}\theta\psi +\sqrt{2}\bar{\theta}\bar{\psi} +(\theta\theta)F+(\bar{\theta}\bar{\theta})F^* \nonumber\\&& 
+ i(\theta\sigma^\mu\bar{\theta})\partial_\mu [A-A^*] -\frac{i}{\sqrt{2}}(\theta\theta)\bar{\theta}\sigma^\mu \partial_\mu \psi \nonumber\\&& 
-\frac{i}{\sqrt{2}}(\theta\sigma^\mu\bar{\theta})\theta\sigma^\mu \partial_\mu \bar{\psi}-\frac14(\theta\theta)(\bar{\theta}\bar{\theta})\square [A+A^*].\nonumber\\&&
\end{eqnarray}
The interesting fact is that there is a term $i\partial_\mu [A+A^*]$ as the coefficient of $(\theta\sigma^\mu\bar{\theta})$ which is similar to the usual gauge transformation of a vector gauge field. Hence, now we are going to choose a special vector superfield by shifting two components of the $V$ as $\lambda \rightarrow \lambda -\frac{i}{2}\sigma^\mu \partial_\mu \bar{\phi}$ and $D \rightarrow D -\frac14 \square C$. Then, it turns out that
\begin{eqnarray}
V(x,\theta,\bar{\theta}) &=& C + \theta \phi + \bar{\theta} \bar{\phi} + \theta\theta M + \bar{\theta}\bar{\theta} M^* + (\theta \sigma^\mu \bar{\theta})V_\mu + (\theta\theta)\bar{\theta} [\bar{\lambda} -\frac{i}{2}\bar{\sigma}^\mu \partial_\mu \phi] \nonumber\\&&
+ (\bar{\theta}\bar{\theta}) \theta [\lambda -\frac{i}{2}\sigma^\mu \partial_\mu \bar{\phi}]+ (\theta\theta)(\bar{\theta}\bar{\theta}) [D -\frac14 \square C]. 
\end{eqnarray}
For this specially chosen vector superfield, we can construct a supersymmetric generalization of the gauge transformation:
\begin{eqnarray}
V \rightarrow V' = V -( \Phi + \Phi^\dag) = V -\frac12 i[\Lambda-\Lambda^\dag],
\end{eqnarray}
where $\Phi \equiv \frac12 i\Lambda$. The component transformations are obtained as
\begin{eqnarray}
C' &=& C + A + A^*,\\
\phi' &=& \sqrt{2} \psi,\\
M' &=& M+F,\\
V_\mu' &=& V_\mu + i\partial_\mu [A-A^*],\\
\lambda' &=& \lambda,\\
D' &=& D,
\end{eqnarray}
where we can see that the $\lambda,D$ are gauge invariant fields. 
In the meantime, there is a special gauge, which is called ``Wess-Zumino Gauge''. This gauge requires 
\begin{eqnarray}
V' \equiv V_{WZ},C' = M' = \phi'=0.
\end{eqnarray}
This leads to
\begin{eqnarray}
V_{WZ} = (\theta\sigma^\mu\bar{\theta})[V_\mu +i\partial_\mu (A-A^*)]+(\theta\theta)\bar{\theta} \bar{\lambda} + (\bar{\theta}\bar{\theta}) \theta \lambda+(\theta\theta)(\bar{\theta}\bar{\theta}) D,
\end{eqnarray}
where the Wess-Zumino gauge does not fix the gauge freedom because we dod not fix the imaginary part of the scalar component field $A$, which gives the gauge transformation of the vector component $V_\mu$. On the other hand, the Wess-Zumino gauge breaks the supersymmetry because the gauge conditions $\phi=M=0$ cannot be compatible with the supersymmetry transformations of the $\phi,M$ in the sense that there still exist their supersymmetry transformations given by $\delta_{\epsilon} \phi|_{WZ} = \sigma^\mu \bar{\epsilon}V_\mu$ and $\delta_{\epsilon} M |_{WZ}= \bar{\epsilon}\bar{\lambda}$. If $A$ is purely real, then we have
\begin{eqnarray}
V_{WZ} &=& (\theta\sigma^\mu\bar{\theta})V_\mu+(\theta\theta)\bar{\theta} \bar{\lambda} + (\bar{\theta}\bar{\theta}) \theta \lambda+(\theta\theta)(\bar{\theta}\bar{\theta}) D,\nonumber\\&&
\end{eqnarray}
which produces 
\begin{eqnarray}
V^2_{WZ} &=& \frac12 (\theta\theta)(\bar{\theta}\bar{\theta}) V_\mu V^\mu,~V^3_{WZ} = 0,\\
e^{V_{WZ}} &=& 1+V+\frac12 V^2 =  1 + (\theta\sigma^\mu\bar{\theta})V_\mu+(\theta\theta)\bar{\theta} \bar{\lambda} 
+ (\bar{\theta}\bar{\theta}) \theta \lambda+(\theta\theta)(\bar{\theta}\bar{\theta}) [D +\frac14 (\theta\theta)(\bar{\theta}\bar{\theta})]
\end{eqnarray}

There is another supersymmetric superfield in terms of a general vector superfield $V$. This is called ``supersymmetric field strength'' $\mathcal{W}_A$ defined by
\begin{eqnarray}
\mathcal{W}_{\alpha} &\equiv& -\frac14 (\bar{\mathcal{D}}\bar{\mathcal{D}})\mathcal{D}_{\alpha} V(x,\theta,\bar{\theta}),\\
\bar{\mathcal{W}}_{\dot{\alpha}} &\equiv& -\frac14 (\mathcal{D}\mathcal{D})\bar{\mathcal{D}}_{\dot{\alpha}} V(x,\theta,\bar{\theta}),
\end{eqnarray}
where satisfy $\mathcal{D}_{\alpha} \bar{\mathcal{W}}_{\dot{\alpha}} = \bar{\mathcal{D}}_{\dot{\alpha}} \mathcal{W}_{\alpha} = 0,~ \mathcal{D}^{\alpha}\mathcal{W}_{\alpha} = \bar{\mathcal{D}}_{\dot{\alpha}} \bar{\mathcal{W}}^{\dot{\alpha}}$ and they are gauge invariant, $\mathcal{W}_{\alpha}'=\mathcal{W}_{\alpha},~\bar{\mathcal{W}}_{\dot{\alpha}}'=\bar{\mathcal{W}}_{\dot{\alpha}}$.

If the vector superfield contains non-abelian vector components, then its supersymmetric field strength is given by
\begin{eqnarray}
\mathcal{W}_{\alpha} &\equiv& -\frac14 (\bar{\mathcal{D}}\bar{\mathcal{D}})e^{-V}\mathcal{D}_{\alpha} e^V,\\
\bar{\mathcal{W}}_{\dot{\alpha}} &\equiv& -\frac14 (\mathcal{D}\mathcal{D})e^{-V}\bar{\mathcal{D}}_{\dot{\alpha}} e^V,
\end{eqnarray}
where they transform under $V' = V + \Phi +\Phi^\dag$ as 
\begin{eqnarray}
\mathcal{W}_{\alpha}' = e^{\Phi^\dag} \mathcal{W}_{\alpha} e^{\Phi}. 
\end{eqnarray}

\section{Supersymmetric Gauge Theory}

To implement the Standard Model (SM), let us consider $\mathcal{N}=1$ global supersymmetry, which is only a realistic possibility for describing the SM. 

First of all, since we want to get a supersymmetric Lagrangian which is `invariant up to the total derivative' under the supersymmetry transformations, we have to find out which one can be such a supersymmetric one.  

We observe that for a general scalar superfield, in the $\theta^2\bar{\theta}^2D$-term, the D-term transforms as the total derivative, while for a chiral superfield, in the $\theta^2 F$-term, the F-term transforms as the total derivative. Therefore, we can expect that the supersymmetric Lagrangian must be obtained only from the D-term of certain scalar superfields and the F-term of certain chiral superfields.  

In particular, we can construct the most general supersymmetric Lagrangian in terms of chiral superfields
\begin{eqnarray}
\mathcal{L} = [K(\Phi^\dag, \Phi)] |_D +[ W(\Phi)+\bar{W}(\bar{\Phi})]|_F ,
\end{eqnarray}
where $K$ is called ``K\"{a}hler-potential'' and $W$ is called ``Super-potential''.

A simple model for this is called ``Wess-Zumino model'', which is given by
\begin{eqnarray}
K = \Phi^\dag\Phi, ~W = \alpha + \lambda \Phi + \frac{m}{2}\Phi^2 + \frac{g}{3}\Phi^3.
\end{eqnarray}

Now, let us consider a gauge transformation 
\begin{eqnarray}
\Phi \rightarrow \Phi_i'= e^{iq_i\Lambda} \Phi_i,
\end{eqnarray}
where $\Lambda$ is chiral. However, the K\"{a}hler potential is not invariant under the gauge transformation because $\Phi_i'^\dag\Phi_i'=e^{iq_i(\Lambda-\Lambda^\dag)}\Phi_i^\dag\Phi_i
\neq \Phi_i^\dag\Phi_i$. 
On the other hand, the following construction is gauge invariant under $V \rightarrow V +i(\Lambda-\Lambda^\dag)$ since
\begin{eqnarray}
 \Phi_i^{\dag} e^{2q_i V} \Phi_i \rightarrow {\Phi'}_i^{\dag} e^{2q_i V'} {\Phi'}_i
=\Phi_i^{\dag} e^{i q_i(\Lambda-\Lambda^{\dag})} e^{2q_i V} e^{-i q_i (\Lambda-\Lambda^{\dag})} \Phi_i = \Phi_i^{\dag} e^{2q_i V} \Phi_i. 
\end{eqnarray}
For the superpotential $W[\Phi_i]$, we have to restrict some terms in the superpotential only if they are not gauge invariant. Then, its action is written by
\begin{eqnarray}
S &=& \int d^4x \int d^4\theta \Big\{    \textrm{Tr}[\mathcal{W}^A[V]\mathcal{W}_A[V]]\delta^2(\bar{\theta})  + \textrm{Tr}[\bar{\mathcal{W}}_{\dot{A}}[V]\bar{\mathcal{W}}^{\dot{A}}[V]]\delta^2(\theta)\nonumber\\
&& + \textrm{Tr}[\Phi^\dag_i e^{2q_iV} \Phi_i] 
+W[\Phi_i]\delta^2(\bar{\theta}) + \bar{W}[\Phi_i^\dag]\delta^2(\theta)\Big\}
\end{eqnarray}
In the component level, for abelian gauge theory, the supersymmetric Lagrangian can be represented by
\begin{eqnarray}
\mathcal{L} &=& [ \Phi^\dag e^{2qV} \Phi ]_D + [W[\Phi]+h.c]_F + [\frac14 \mathcal{W}^\alpha\mathcal{W}_\alpha+h.c.]_F  + [\xi V]_D,\\
(\Phi^\dag e^{2qV} \Phi)_D &=& F^*F + |\partial_\mu \varphi|^2 + i\bar{\psi}\bar{\sigma}^\mu\partial_\mu \psi 
+ q V_\mu (\bar{\psi}\bar{\sigma}_\mu \psi + i\varphi^*\partial_\mu \varphi -i\varphi\partial_\mu \varphi^*) \nonumber\\
&& +\sqrt{2}q(\varphi\bar{\lambda}\bar{\psi}+\varphi^*\lambda\psi)+q(D+q V_\mu V^\mu)|\varphi|^2,\nonumber\\
 (\frac14 \mathcal{W}^\alpha\mathcal{W}_\alpha+h.c.)_F &=& \frac12 D^2 -\frac14 F_{\mu\nu}F^{\mu\nu} -i\lambda \sigma^\mu \partial_\mu \bar{\lambda},\nonumber\\
 (W[\Phi]+h.c)_F &=& F^*F + (\frac{\partial W}{\partial \varphi} F +h.c.) -\frac12 (\frac{\partial^2 W}{\partial \varphi^2}\psi\psi +h.c.)\nonumber\\&&
\end{eqnarray}

%% file: chapters/5.tex
This chapter is mainly based on Refs.~\cite{Superconformal_Freedman,Yamada,Brandt,SUPER_CONFORMAL_TENSORCAL}.

\section{Why Conformal Supergravity?}

In the four-dimensional $\mathcal{N}=1$ {\it Poincar\'{e} supergravity}, the actions of supermultiplets are invariant under transformations of {\bf super-Poincar\'{e} symmetry}, which consists of spacetime diffeomorphisms (i.e. general coordinate transformations), local Lorentz symmetry, and local supersymmetry. This is established as a ``physical'' theory. On the other hand, for the so-called {\it conformal supergravity or superconformal gravity}, its actions are invariant under transformations of {\bf superconformal symmetry}, which is a set of super-Poincar\'{e} symmetries and four additional symmetries (dilatation, chiral $U(1)$ symmetry, special supersymmetry (i.e. $S$-SUSY), and special conformal symmetry). In superconformal theory, supermultiplets are promoted into representations of the superconformal group, so that they become ``{\bf superconformal multiplets}.'' In particular, the four additional symmetries can be broken by gauge-fixing conditions to get a physical theory, meaning that one can obtain the Poincar\'{e} supergravity theory from the broken superconformal theory. At first glance, one may think this conformal approach seems more laborious, and ask: why do we consider conformal supergravity? In fact, there are two main reasons for this.

The first benefit from conformal supergravity is that the superconformal formalism can yield the (physical) Poincar\'{e} supergravity in a convenient manner. For example, in the other approach like the superspace formalism, it is required to make graviton and gravitino to be canonical by taking complicated rescalings. On the contrary, in the superconformal formalism, it is possible to simply obtain the canonical kinetic terms of graviton and gravitino by imposing gauge-fixing conditions on the so-called ``compensator\footnote{The fields removed after gauge fixing are called compensators.}'' multiplet introduced in the superconformal formalism. 

The second benefit is that the conformal supergravity can give us a unified description of Poincar\'{e} supergravity with different types of gravity multiplet. In the superspace formalism, it is difficult to figure out the relationships between {\it different sets of the auxiliary field of gravity multiplet}. However, those relationships can be comprehensively understood in the conformal supergravity in that different sets of auxiliary field in Poincar\'{e} supergravity can correspond to {\it different compensator multiplets} in the conformal supergraivty; that is, {\it Old/New/Non}-Minimal formulation of Poincar\'{e} Supergravity can be obtained by the conformal supergravity with {\it Chiral/Real-Linear/Complex-Linear} compensator multiplet. 

\section{Tensor Calculus of General Gauge Field Theory}

This subsection is particularly based on Refs.~\cite{Superconformal_Freedman,Brandt}.

``{\bf Tensor Calculus}'' centers around the notion of ``{\bf Gauge Covariance}.'' For example, let us consider ``a tensor in general relativity (GR).'' In GR, the general coordinate transformations (GCT) or spacetime diffeomorphism is a gauge symmetry of {\bf local translation} whose gauge parameter and generators are given by $\xi^{\mu}$ and momentum operator $P_{\mu}=\partial_{\mu}$, which generates 
\begin{eqnarray}
x'^{\mu}=x^{\mu}+\xi^{\mu}, \qquad \delta_{GCT}(\xi)x^{\mu}=\xi^{\nu}P_{\nu}x^{\mu}=\xi^{\nu}\partial_{\nu}x^{\mu} = \xi^{\mu}.
\end{eqnarray}
A vector $V^{\mu}$ transfroms under the GCT as
\begin{eqnarray}
V'^{\mu} = \frac{\partial x'^{\mu}}{\partial x^{\nu}} V^{\nu} \approx \Big( \delta_{\nu}^{\mu} + \frac{\partial \xi^{\mu}}{\partial x^{\nu}} \Big) V^{\nu} = V^{\mu} + \lambda_{\nu}^{\mu}(x) V^{\nu} \implies \delta_{GCT}(\lambda) V^{\mu} = \lambda_{\nu}^{\mu}(x) V^{\nu},
\end{eqnarray}
where we define $\lambda_{\nu}^{\mu}(x) \equiv \frac{\partial \xi^{\mu}}{\partial x^{\nu}}$ and take it as a gauge parameter. So, in this notation, we have $\xi^{\mu}=\lambda_{\nu}^{\mu}x^{\nu}$. However, the ordinary derivative of the vector $\partial_{\nu}V^{\mu}$ cannot be covariant. This is because 
\begin{eqnarray}
\partial'_{\nu}V'^{\mu} = \Big(\frac{\partial x^{\rho}}{\partial x'^{\nu}}\partial_{\rho}\Big)\Big(\frac{\partial x'^{\mu}}{\partial x^{\sigma}}V^{\sigma}\Big) = \frac{\partial x^{\rho}}{\partial x'^{\nu}}\frac{\partial x'^{\mu}}{\partial x^{\sigma}}\partial_{\rho}V^{\sigma} + \frac{\partial x^{\rho}}{\partial x'^{\nu}} \bigg[  \frac{\partial}{\partial x^{\rho}}\Big(\frac{\partial x'^{\mu}}{\partial x^{\sigma}}\Big)\bigg]V^{\sigma} \neq \frac{\partial x^{\rho}}{\partial x'^{\nu}}\frac{\partial x'^{\mu}}{\partial x^{\sigma}}\partial_{\rho}V^{\sigma}.
\end{eqnarray}
The point here is that the problematic discrepancy from this gauge transformation of the vector under the GCT comes from the fact that the transformation contains the derivative of the gauge parameter, i.e. $ \frac{\partial}{\partial x^{\rho}}\Big(\frac{\partial x'^{\mu}}{\partial x^{\sigma}}\Big)=\partial_{\rho}\lambda_{\sigma}^{\mu}$. The ``correct'' covariant derivative is found to be $\nabla_{\nu}V^{\mu} \equiv \partial_{\nu}V^{\mu}+\Gamma^{\mu}_{~~\rho\nu}V^{\rho}$ where $\Gamma^{\mu}_{~~\rho\nu}$ is {\it Christoffel symbol} as ``connection'' for the GCT, which as expected can transform covariantly under the GCT. 

Regarding the lesson learned from the example of GCT in GR, we can deduce that {\it any tensor defined with respect to the GCT must properly vary under the GCT whose gauge transformation does not contain any derivatives of the gauge parameter} $\xi^{\mu}$! This statement is a ``punchline'' in this example. Hence, we can state that {\it if an object can have its own gauge transformation without derivatives of gauge parameter for a certain gauge symmetry, then the object is ``{\bf covariant or tensorial}'' with respect to the gauge symmetry and it can thus be defined as a tensor with respect to the symmetry}. We call this property of being {\it covariant} under a (gauge) symmetry as ``{\bf (Gauge) Covariance}\footnote{The precise structure of ``Covariance'' or ``Tensoriality'' can also be understood in the mathematical language through the theory of Principal and Associated Fiber Bundles (See Nakahara's textbook for a review of this viewpoint in Ref. \cite{Nakahara}). In particular, gauge symmetry can be considered as a {\it redundancy of mathematical description for a theory}, which is called {\bf Gauge redundancy}. This aspect is understood in the constrained Hamiltonian systems (see Ref.~\cite{Henneaux} for a comprehensive review on the topic).  References \cite{seahra,Banados} give introductions to symmetries and dynamics of constrained systems. Also, a student-friendly example about the topic is well-described in Ref.~\cite{Brown}. Geometrical analysis of the aspect is explained in Ref. \cite{Jon_Allen}. The general constrained dynamics that considers both bosonic and fermionic variables can be found in Ref.~\cite{Xavier}.}.'' Particularly, I would like to distinguish between a normal tensor defined with respect to only the GCT in GR and another tensor also defined with respect to extra gauge symmetries together with the GCT. I will call the latter as ``{\bf gauge-covariant tensor}'' as a stressed terminology of tensor.

\subsection{Consistency requirements of tensor calculus}

Let us consider a set of {\it Lie symmetries} 
\begin{eqnarray}
\delta(\epsilon) \equiv \epsilon^AT_A,
\end{eqnarray}
where $A$ is an index of generators of the symmetries; $\epsilon^A$ is a transformation parameter, and $T_A$ is a generator indexed by $A$ as the {\it active}  operator acting on the field space. It is very important to be careful about the order of symmetry transformations since these are related to the Lie algebras. In tensor calculus, the product operation is defined by the {\it active operation} 
\begin{eqnarray}
\delta(\epsilon_1)\delta(\epsilon_2)\Phi = \delta(\epsilon_1)\Big(\delta(\epsilon_2)\Phi\Big)=\delta(\epsilon_1)\Big( \epsilon_2^BT_B\Phi \Big) = \epsilon_2^B \delta(\epsilon_1)\Big( T_B\Phi \Big) = \epsilon_2^B\epsilon_1^A T_AT_B\Phi.
\end{eqnarray}

Then, we suppose that the symmetries satisfy the following three consistency conditions \cite{Brandt}
\begin{itemize}
    \item 1. {\bf (Off-shell) Closed algebras of the symmetries}: For any two generators $T_A,T_b$, the following ``(anti)commutator'' holds.
    \begin{eqnarray}
    [T_A,T_B\}=f_{AB}^{~~~C}T_C \Longleftrightarrow
    [\delta(\epsilon_1),\delta(\epsilon_2)\}=(\epsilon^B_2\epsilon_1^Af_{AB}^{~~~C})T_C=\delta(\epsilon_3\equiv \epsilon^B_2\epsilon_1^Af_{AB}^{~~~C}) \label{1st_consi}
    \end{eqnarray} where $[\bullet,\bullet\}$ is defined as the ``graded commutator,'' and the third parameter may be given by a function of the other two parameters and some possible structure constants or functions $f$'s of the algebras. By requiring this closure of algebras, we can find the structure constants or functions.
    \item 2. {\bf Commutativity of Lie and Exterior derivatives} (by geometrical axiom): This condition fixes the gauge transformations of the gauge connections or gauge fields $A_{\mu}^A$'s. 
    \begin{eqnarray}
    [\delta(\epsilon),d]=0~~\Longleftrightarrow ~~[\delta(\epsilon),\partial_{\mu}]=0,\label{2nd_consi_1}
    \end{eqnarray}
    where $d=dx^{\mu}\partial_{\mu} =d\lambda \mathcal{L}_{\xi}$ where $\mathcal{L}_{\xi}$ is a Lie derivative along the vector field $\xi=\xi^{\mu}\partial_{\mu}$. In particular, we can also represent the exterior derivative in terms of {\it covariant generators} $\hat{T}_A$ ($\neq T_A$ from the set of normal generators and local translation) and {\it gauge-connection one-forms} $A^A \equiv A^A_{\mu}dx^{\mu}=A^A_{\mu}\xi^{\mu}d\lambda$ (where $dx^{\mu}=\xi^{\mu}d\lambda$ along the curve of the vector field $\xi$ parameterized by $\lambda$) corresponding to the relevant gauge fields $A^A_{\mu}$ of gauge symmetries $\delta(\epsilon)=\epsilon^A \hat{T}_A$ made by the {\it covariant generators} $\hat{T}_A$ in the following manner \cite{Brandt}
\begin{eqnarray}
d \overset{!}{=} A^A\hat{T}_{A} (= d\lambda \mathcal{L}_{\xi} ).\label{2nd_consi_2}
\end{eqnarray}
This will be used for covariantizing the GCT (i.e. local translation) and equivalently defining a {\bf reformed covariant derivative} ``$\mathcal{D}_a$'' which is comparable with the covariant GCT. Later, we will follow the index convention that {\it small Latin alphabet} ``$a,b,c$'' means {\bf ``tangent'' (local (Lorentz) frame) indices} and {\it small Greek alphabet} ``$\mu,\nu,\rho$'' means {\bf ``world'' (curved frame) indices}. Especially, the reformed covariant derivatives $\mathcal{D}_a$ will be shown throughout all the supergravity sections in this dissertation.
    \item 3. {\bf de Rham Cohomology} (by geometrical theorem): This condition fixes the curvatures of the gauge fields and covariant derivatives.
    \begin{eqnarray}
    d^2 = 0 \quad \Longleftrightarrow \quad [\partial_{\mu},\partial_{\nu}] =0 \quad  \textrm{  on scalar}, \quad [\nabla_{\mu},\nabla_{\nu}]=-R_{\mu\nu}^{~~\rho}\nabla_{\rho}.\label{3rd_consi}
    \end{eqnarray}
\end{itemize}

Here is a remark of the algebras. For any two generators $T_A,T_B$ of the symmetry algebras, we can consider their commutation relation in the form of $\{T_A,T_B]=\sum_{i}f_{AB}^{~~~C_i}T_{C_i}$ where $f_{AB}^{~~~C_{n}}$ is its structure constant (or function if it is field-dependent and forms a soft algebra). In particular, if $C_i$ has $n$ indices $\{a^i_1\cdots a^i_n\}$, then we have to consider ``{\bf symmetry-factor rule of (anti)symmetric-index contraction}'' which is {\it to multiply the symmetry factor ``$\dfrac{1}{n!}$'' in front of the contracting part of (anti)symmetric $n$-indices} to avoid ``double-counting''; for example, for $A^M \in \{A^{ab}\}$ and $B_M \in \{B_{ab}\}$ each of which is the defining elements of the set (e.g. for an antisymmetric rank-2 tensor, $A_{21}=-A_{12}$ where $A_{12}$ is the defining element), the contraction of the two sets can be defined by $A^MB_M \equiv \frac{1}{2!}A^{ab}B_{ab}$. Thus, we have to have the following
\begin{eqnarray}
\{T_A,T_B]=\sum_{i}f_{AB}^{~~~C_i}T_{C_i} \longrightarrow \{T_A,T_B]=\sum_{i} \frac{1}{n_i!} f_{AB}^{~~~a^i_1\cdots a^i_n}T_{a^i_1\cdots a^i_n}
\end{eqnarray}
This commutator algebra should be closed off-shell in general. Normally, given commutation relations of the algebras, it will be needed to read off the structure constants (or functions) from the commutation relations in order to compute relevant quantities of our interest like curvature using them.

\subsection{``Covariant'' quantity, gauge field, and curvature}

Under the transformations $\delta(\epsilon)=\epsilon^AT_A$ of any generators $T_A$ (including the covariant ones $\hat{T}_A$), a {\bf covariant quantity} ``$\Phi$'' is defined by a quantity transforming without derivatives of gauge parameters in the way
\begin{eqnarray}
\delta(\epsilon)\Phi = \epsilon^AT_A\Phi \equiv \epsilon^AK_A(\Phi),
\end{eqnarray}
where we define $K_A(\Phi) \equiv T_A\Phi$ which may be either linearly or non-linearly realized in the quantity $\Phi$. As expected, the ordinary derivative of the quantity which transforms under the relevant symmetries cannot transform covariantly. That is, we face the same issue as we saw from the GCT example in GR:
\begin{eqnarray}
\delta(\epsilon)\partial_{\mu}\Phi = \partial_{\mu}(\epsilon^AK_A)=\epsilon^A(\partial_{\mu}K_A) + (\partial_{\mu}\epsilon^A) K_A
\end{eqnarray}
To obtain the {\bf covariant derivatives} ``$D_{\mu}$,'' it is mandatory to introduce {\bf gauge fields} ``$\mathcal{B}_{\mu}^A$,'' whose transformations are defined by
\begin{eqnarray}
\delta(\epsilon)\mathcal{B}_{\mu}^A \equiv \partial_{\mu} \epsilon^A + \epsilon^C\mathcal{B}_{\mu}^Bf_{BC}^{~~~A}.
\end{eqnarray}
Then, the corresponding gauge-covariant derivatives can be defined by subtracting the symmetry transformations $\delta(\mathcal{B}_{\mu}^A)$ (whose gauge parameter is replaced by the gauge field) from the ordinary derivative $\partial_{\mu}$; that is,
\begin{eqnarray}
D_{\mu} \Phi \equiv (\partial_{\mu}-\mathcal{B}_{\mu}^AT_A)\Phi = (\partial_{\mu}-\delta(\mathcal{B}_{\mu}^A))\Phi.
\end{eqnarray}
Certainly, we can check that 
\begin{eqnarray}
\delta(\epsilon)D_{\mu}\Phi &=& \epsilon^A\partial_{\mu}K_A + (\partial_{\mu})K_A -(\delta(\epsilon)\mathcal{B}_{\mu}^A)K_A -\mathcal{B}_{\mu}^A\delta(\epsilon)K_A \nonumber\\
&=& \epsilon^A\partial_{\mu}K_A - \epsilon^C\mathcal{B}_{\mu}^B f_{BC}^{~~~A}K_A -\mathcal{B}_{\mu}^A\delta(\epsilon)K_A \nonumber\\
&=& \epsilon^A \partial_{\mu}K_A -\epsilon^A \mathcal{B}_{\mu}^B [T_B,T_A]\Phi -\mathcal{B}_{\mu}^B\epsilon^A T_AT_B\Phi = \epsilon^A(\partial_{\mu}K_A-\mathcal{B}_{\mu}^BT_BK_A)=\epsilon^AD_{\mu}K_A. \nonumber\\{}
\end{eqnarray}
Moreover, there is another covariant quantity called {\bf curvature} ``$R_{\mu\nu}^A$'' of the gauge fields $\mathcal{B}_{\mu}^A$, which is defined in the way
\begin{eqnarray}
R_{\mu\nu}^A \equiv \partial_{\mu}\mathcal{B}_{\nu}^A - \partial_{\nu}\mathcal{B}_{\mu}^A +\mathcal{B}_{\nu}^C\mathcal{B}_{\mu}^B f_{BC}^{~~~A} = 2\partial_{[\mu}\mathcal{B}_{\nu]}^A +\mathcal{B}_{\nu}^C\mathcal{B}_{\mu}^B f_{BC}^{~~~A}.
\end{eqnarray}
The {\it gauge transformations of these curvatures} are given by
\begin{eqnarray}
\delta(\epsilon)R_{\mu\nu}^A &=& 2\partial_{[\mu}\partial_{\nu]}\epsilon^A + 2\partial_{[\mu}(\epsilon^C\mathcal{B}_{\mu]}^B)f_{BC}^{~~~A} 
+ \Big( (\partial_{\nu}\epsilon^C+\epsilon^D\mathcal{B}_{\nu}^Ef_{ED}^{~~~C})\mathcal{B}_{\mu}^Bf_{BC}^{~~~A} -(\mu\leftrightarrow\nu)  \Big) \nonumber\\
&=& \epsilon^C (2\partial_{[\mu}\mathcal{B}_{\nu]}^Bf_{BC}^{~~~A})+\epsilon^D\mathcal{B}_{\nu}^E\mathcal{B}_{\mu}^Bf_{ED}^{~~~C}f_{BC}^{~~~A}+\epsilon^D\mathcal{B}_{\nu}^C\mathcal{B}_{\mu}^E f_{ED}^{~~~B}f_{BC}^{~~~A} \nonumber\\
&=& \epsilon^C (2\partial_{[\mu}\mathcal{B}_{\nu]}^Bf_{BC}^{~~~A}) -\epsilon^C\mathcal{B}_{\nu}^E\mathcal{B}_{\mu}^D f_{ED}^{~~~B}f_{BC}^{~~~A}  = \epsilon^C \bigg[ 2\partial_{[\mu}\mathcal{B}_{\nu]}^B -\mathcal{B}_{\nu}^E\mathcal{B}_{\mu}^D f_{ED}^{~~~B}\bigg]f_{BC}^{~~~A}
\nonumber\\
&=& \epsilon^C R_{\mu\nu}^B f_{BC}^{~~~A},\nonumber\\
&& \qquad \implies \quad  \therefore \quad \delta(\epsilon)R_{\mu\nu}^A=\epsilon^C R_{\mu\nu}^B f_{BC}^{~~~A}.\label{R_transform}
\end{eqnarray}
where we take advantage of the Jacobi identity in the third equality:
\begin{eqnarray}
\sum_B \Big( f_{CE}^{~~~B}f_{BD}^{~~~A}+(\textrm{cyclic in C,D,E.})\Big)=\sum_B (f_{CE}^{~~~B}f_{BD}^{~~~A}+f_{DC}^{~~~B}f_{BE}^{~~~A}+f_{ED}^{~~~B}f_{BC}^{~~~A}) = 0.
\end{eqnarray}
Note that the curvatures tranform covariantly under the gauge transformations. Plus, the curvatures satisfy the Bianchi identities 
\begin{eqnarray}
D_{[\rho}R_{\mu\nu]}^A=0.
\end{eqnarray}

\subsection{Covariantization of the local translation: covariant general coordinate transformation (CGCT), and constraints for its consistency}

In the previous sections, we have talked about symmetry algebras, covariant derivatives, gauge fields, and curvatures. Importantly, these symmetries are assumed to be ``internal'' and ``general.'' This means that the gauge fields $\mathcal{B}_{\mu}^A$ that we have so far considered may involve the {\bf vielbein} ``$e^a_{\mu}$'' (or equivalently called ``{\it graviton}'') as the gauge field of the local translation symmetry $P_{a}$ in a theory of gravity. In fact, this fact brings us a ``bad'' problem about covariant derivative of local translation. To see how this arise, let us assume that we have only a local translation, so that the corresponding gauge covariant derivative is given by $D_{\mu} = \partial_{\mu} - e^a_{\mu}P_a$ where $P_a = \partial_a$ is the momentum generator. However, from the properties of vielbein $e^{a}_{\mu}e^{\mu}_b = \delta_b^a$ (invertibility) and $A^a=e^{a}_{\mu}A^{\mu}, ~A_{\mu}=e^a_{\mu}A_a$ (change of reference frame), we reach the fact that 
\begin{eqnarray}
D_{\mu} = \partial_{\mu} - e^a_{\mu}P_a = \partial_{\mu} - e^a_{\mu}\partial_a = \partial_{\mu}-\partial_{\mu}=0.
\end{eqnarray}
Here, we see that {\bf this $D_{\mu}$ is physically ill-defined}, and thus we have to covariantize the local translation $P_a$ in an alternative manner in order to obtain a ``well-defined covariant derivative.'' In fact, there is a clever way of doing this. Let us recall the alternative form of the exterior derivative $d$ in Eg.~\eqref{2nd_consi_2}, which was given by
\begin{eqnarray}
d = d\lambda \mathcal{L}_{\xi} = \mathcal{B}_{\mu} dx^{\mu} =d\lambda \xi^{\mu} \mathcal{B}_{\mu}^A  \hat{T}_A,\label{exterior_trick}
\end{eqnarray}
where we replaced the notation of gauge field $A_{\mu}^A$ by $\mathcal{B}_{\mu}^A$. It is worth noticing that the right-hand side of the relation in Eq.~\eqref{exterior_trick} was defined by a linear combination of all the possible {\it covariant} generators $\hat{T}_A$. Hence, instead of the original local translation $P$, there must be a covariant local translation $\hat{P}$ in the combination. Keeping this in mind, let us assume that the generators $T_A$ and corresponding gauge fields $\mathcal{B}_{\mu}^A$ of all the gauge symmetries including ``covariant local translation'' $\hat{P}_a$ (no normal local translation $P_a$ in the sum of $\mathcal{B}_{\mu}^A\hat{T}_A$ in Eq.~\eqref{exterior_trick}) can be decomposed into 
\begin{eqnarray}
\hat{T}_A = \{ \hat{P}_a, \hat{T}_{A\neq \hat{P}} = T_{A\neq P} (\equiv T_{A\neq \hat{P},P}) \}, \qquad \mathcal{B}_{\mu}^A = \{e^a_{\mu}, \mathcal{B}_{\mu}^{A\neq \hat{P},P}\},
\end{eqnarray}
where $\hat{P}_a \equiv \mathcal{D}_a$ is now defined as a {\bf covariant local translation or covariant GCT} generator that is assigned to the ``vielbein'' gauge field $e^a_{\mu}$, which can also be realized as a ``covariant differential operator'' denoted by $\mathcal{D}_a$ analogous to the conventional gauge covariant derivatives. Then, from Eq.~\eqref{exterior_trick} , we get the following relation
\begin{eqnarray}
 d\lambda \mathcal{L}_{\xi}  =d\lambda \Big( \xi^{\mu} e^a_{\mu} \hat{P}_a + \xi^{\mu}\mathcal{B}_{\mu}^{A\neq \hat{P},P}T_{A\neq \hat{P},P} \Big)  \implies \xi^{\mu} e^a_{\mu} \hat{P}_a =\mathcal{L}_{\xi} - \xi^{\mu}\mathcal{B}_{\mu}^{A\neq \hat{P},P}T_{A\neq \hat{P},P} \textrm{~~~on any tensor~~}
\end{eqnarray}
and thus we are able to define the so-called ``{\bf covariant general coordinate transformation (CGCT)}'' in the way
\begin{eqnarray}
 \delta_{CGCT}(\xi) = \delta_{GCT}(\xi) - \delta_{SG}(\xi^{\mu}\mathcal{B}_{\mu}^{A\neq \hat{P},P})\textrm{~~~on any tensor},\label{CGCT}
\end{eqnarray}
where
\begin{eqnarray}
&& \delta_{CGCT}(\xi) \equiv \xi^{\mu} e^a_{\mu} \hat{P}_a = \xi^a\hat{P}_a,\\
&& \delta_{GCT}(\xi) = \mathcal{L}_{\xi} \qquad \textrm{by definition of the original GCT},\\
&& \delta(\xi^{\mu}\mathcal{B}_{\mu}^{A\neq \hat{P},P}) \equiv \xi^{\mu}\mathcal{B}_{\mu}^{A\neq \hat{P},P}T_{A\neq \hat{P},P}.
\end{eqnarray}
Now let us define the covariant derivative $\mathcal{D}_a$ that can be comparable with the CGCT. Consider the CGCT on a {\it world scalar} $\phi$. Then, we get its CGCT as
\begin{eqnarray}
\delta_{CGCT}(\xi)\phi = \xi^{\mu}\partial_{\mu}\phi -\delta(\xi^{\mu}\mathcal{B}_{\mu}^{A\neq \hat{P},P})\phi = \xi^{\mu} \Big( \partial_{\mu} -\delta(\mathcal{B}_{\mu}^{A\neq \hat{P},P}) \Big)\phi \overset{!}{=} \xi^{\mu} e^a_{\mu}\hat{P}_a \quad  \textrm{on ``world scalar,''}
\end{eqnarray}
where $\delta_{GCT}(\xi)\phi=\xi^{\mu}\partial_{\mu}\phi$. Then, defining a realization of the covariant local translation operator ``$\hat{P}_a$'' as the {\bf well-defined covariant derivative} ``$\mathcal{D}_a$'' that we desire to have, one can obtain
\begin{eqnarray}
e^a_{\mu} \mathcal{D}_a= \Big(\partial_{\mu} - \mathcal{B}_{\mu}^{A\neq \hat{P},P}T_{A\neq \hat{P},P} \Big) \qquad  \textrm{on ``world scalar.''}
\end{eqnarray}
Then, by using the {\it invertibility of the vielbein} $e^a_{\mu}$, we can reach a reformation of the covariant derivative with respect to the covariant local translation in desirable form:
\begin{eqnarray}
 \mathcal{D}_a = e^{\mu}_a (\partial_{\mu} - \mathcal{B}_{\mu}^{A\neq \hat{P},P}T_{A\neq \hat{P},P})=\partial_{a} - \mathcal{B}_{a}^{A\neq \hat{P},P}T_{A\neq \hat{P},P}\qquad  \textrm{on ``world scalar.''}\label{reformed_cov_derivative}
\end{eqnarray}
We emphasize that this new covariant derivative acts on any world scalar living on any tangent space defined over a (curved) spacetime, but the derivative itself rotates under the local Lorentz transformation because it is a local-Lorentz vector. In the other words, {\it the covariant derivative $\mathcal{D}_a$ can be applied to any local-frame tensors as world scalars!} For example, a local-Lorentz vector field $B_b$ has its covariant derivative given by $\mathcal{D}_aB_b$. In the presence of covariant derivatives, the equations of motion for some fields are obtained by simply replacing the ordinary partial derivative with the suitable covariant derivatives \cite{Lewis_cov_deriv}. Then, we call such equations as {\bf (gauge) covariant equations of motion}. The properties of covariant derivative \cite{Lewis_cov_deriv} are as follows: \begin{itemize}
    \item 1. (Operation rule) The gauge covariant derivative acts on objects that respond to a coordinate transformation as well as a gauge transformation.
    \item 2. (Covariant quantity) The gauge covariant derivative of a field transforms as the field itself does whose gauge transformation has no spacetime derivatives of gauge parameter.
    \item 3. (Necessity of gauge field or connection) A gauge field $A_{\mu}$ consisting of gauge connection $A=A_{\mu}dx^{\mu}$ must be embedded in the covariant derivative in order to define covariant quantity under the relevant gauge transformation.
    \item 4. (Product law) The covariant derivative obeys {\it Leibniz rule} with appropriate parity as the product law, i.e. $\mathcal{D}_a(\phi_1\phi_2) = (\mathcal{D}_a\phi_1)\phi_2+(-1)^{|\mathcal{D}_a||\phi_1|}\phi_1(\mathcal{D}_a\phi_2)$ where $|\cdot|=0$ for even or bosonic and $|\cdot|=1$ for odd or fermionic.
    \item 5. (Restoration to ordinary derivative) For a gauge invariant scalar, the gauge covariant derivative is identical to the ordinary partial derivative, i.e. $\mathcal{D}_a(\textrm{inv.scalar})=\partial_a(\textrm{inv.scalar})$.
\end{itemize}

Now, remember that the original local translation ``$P_a$'' was not a spacetime symmetry but one of internal gauge symmetries. However, the covariant local translation ``$\hat{P}_a$'' has been identified with a spacetime symmetry not as an internal one because it becomes the covariant general coordinate transformation. Hence, it is reasonable to check whether or not the reformation of the local translation affects the closure of symmetry algebras. If so, there may be some constraints for its consistency. 

Then, let us evaluate the general coordinate transformation (GCT) of a ``world-indexed vector'' gauge field $\mathcal{B}_{\mu}^A$, which can be considered as a local Lorentz scalar since it has no local frame indices. We point out that the Einstein summation in the GCT includes standard gauge symmetries and the normal local translation $P_a$ ({\it without the covariant local translation $\hat{P}_a$ in the GCT since we have no channel to introduce such a covariant one now.}). That is, consider the origical situation with closed symmetry algebras of the normal local translation and standard gauge symmetries. Then, the GCT of the gauge field with the world vector index $\mu$ is given by its non-trivial Lie derivative 
\begin{eqnarray}
\delta_{GCT}(\xi) \mathcal{B}_{\mu}^A &=& \mathcal{L}_{\xi} \mathcal{B}_{\mu}^A = \xi^{\nu}\partial_{\nu} \mathcal{B}_{\mu}^A+ \mathcal{B}_{\nu}^A\partial_{\mu}\xi^{\nu} 
\nonumber\\
&=& \xi^{\nu}\partial_{\nu} \mathcal{B}_{\mu}^A+ \mathcal{B}_{\nu}^A\partial_{\mu}\xi^{\nu} + (\xi^{\nu}\mathcal{B}_{\nu}^C\mathcal{B}_{\mu}^Bf_{BC}^{~~~A}-\xi^{\nu}\mathcal{B}_{\nu}^C\mathcal{B}_{\mu}^Bf_{BC}^{~~~A}) + (-\xi^{\nu}\partial_{\mu}\mathcal{B}_{\nu}^A+\xi^{\nu}\partial_{\mu}\mathcal{B}_{\nu}^A) \nonumber\\
&=& -\xi^{\nu} (\partial_{\mu}\mathcal{B}_{\nu}^A-\partial_{\nu}\mathcal{B}_{\mu}^A+\mathcal{B}_{\nu}^C\mathcal{B}_{\mu}^Bf_{BC}^{~~~A}) \nonumber\\
&& + \underbrace{\partial_{\mu}(\xi^{\nu}\mathcal{B}_{\nu}^A)
+(\xi^{\nu}\mathcal{B}_{\nu}^C)\mathcal{B}_{\mu}^Bf_{BC}^{~~~A} + \xi^{\nu}\mathcal{B}_{\nu}^B\mathcal{M}_{B\mu}^{~~~A}
}_{\equiv \delta(\xi^{\nu}\mathcal{B}_{\nu}^{M\neq \hat{P}})\mathcal{B}_{\mu}^A}-\xi^{\nu}\mathcal{B}_{\nu}^B\mathcal{M}_{B\mu}^{~~~A}
\nonumber\\
&=& -\xi^{\nu}R_{\mu\nu}^A  + \delta(\xi^{\nu}\mathcal{B}_{\nu}^{M\neq \hat{P}})\mathcal{B}_{\mu}^A -\xi^{\nu}\mathcal{B}_{\nu}^B\mathcal{M}_{\mu B}^{~~~A}
\nonumber\\
&=& \xi^{\nu} R_{\nu\mu}^A-\xi^{\nu}\mathcal{B}_{\nu}^B\mathcal{M}_{\mu B}^{~~~A}  + \delta(\xi^{\nu}\mathcal{B}_{\nu}^{M\neq \hat{P},P})\mathcal{B}_{\mu}^A + \delta(\xi^{\nu}e_{\nu}^{a})\mathcal{B}_{\mu}^A \nonumber\\
&=& \xi^{\nu} \Big(\underbrace{ R_{\nu\mu}^A-2\mathcal{B}_{[\nu}^B\mathcal{M}_{\mu] B}^{~~~A}}_{\equiv \hat{R}_{\nu\mu}^A} \Big) + \delta(\xi^{\nu}\mathcal{B}_{\nu}^{M\neq \hat{P},P})\mathcal{B}_{\mu}^A + \xi^{\nu}e^a_{\nu}P_a\mathcal{B}_{\mu}^A,\label{GCT_of_Gaugefield}
\end{eqnarray}
where we define a {\bf covariant curvature} ``$\hat{R}_{\nu\mu}^A$''
\begin{eqnarray}
\hat{R}_{\nu\mu}^A \equiv  R_{\nu\mu}^A-2\mathcal{B}_{[\nu}^B\mathcal{M}_{\mu] B}^{~~~A},
\end{eqnarray}
and {\bf generalized gauge transformation of gauge field} including possible ``non-gauge field effects'' $\mathcal{M}_{\mu B}^{~~~A}$
\begin{eqnarray}
 \delta(\epsilon)\mathcal{B}_{\mu}^A \equiv \partial_{\mu}\epsilon^A
+\epsilon^C\mathcal{B}_{\mu}^Bf_{BC}^{~~~A} + \epsilon^B\mathcal{M}_{B\mu}^{~~~A}.
\end{eqnarray}
Using the definition of the CGCT, we can rewrite Eq.~\eqref{GCT_of_Gaugefield} as 
\begin{eqnarray}
\delta_{CGCT}(\xi)\mathcal{B}_{\mu}^A\equiv \delta_{GCT}(\xi)\mathcal{B}_{\mu}^A -\delta(\xi^{\nu}\mathcal{B}_{\nu}^{M\neq \hat{P},P})\mathcal{B}_{\mu}^A  =\xi^{\nu}\hat{R}_{\nu\mu}^A + \xi^{\nu}e^a_{\nu}P_a\mathcal{B}_{\mu}^A \overset{!}{=} \xi^{a}\hat{P}_a\mathcal{B}_{\mu}^A.
\end{eqnarray}
Notice that this CGCT of a gauge field is completely different from the gauge transformation of a curvature in Eq. \eqref{R_transform}.

 Hence, for one to be able to take the replacement $P_a \longrightarrow \hat{P}_a$ in the symmetry algebras, we need to identify $\hat{P}_a$ with $P_a$ on the local Lorentz scalar, i.e. $\xi^aP_a\mathcal{B}_{\mu}^A=\xi^a\hat{P}_a\mathcal{B}_{\mu}^A$. As a result, we have to impose the following constraints $C = 0$ 
\begin{eqnarray}
\forall A: C \equiv  \xi^{\nu}\hat{R}_{\mu\nu}^A(T_A) \overset{!}{=} 0, \label{Curvature_constraints}
\end{eqnarray}
where $T_A$ involves the original (i.e. non-covariant) local translation $P_a$ and the other standard gauge symmetries.

From the above constraints \eqref{Curvature_constraints}, one can see that they cannot be preserved under their gauge transformations (i.e. $\delta(\epsilon)R_{\mu\nu}^A=\epsilon^C R_{\mu\nu}^B f_{BC}^{~~~A}\neq 0$ in general) given by Eq.~\eqref{R_transform} as long as additional shifts are not introduced. To understand how this works, let us consider some general situation. Let $C(\omega,\phi)=0$ be a constraint consisting of an independent field $\phi$ and a dependent field $\omega=\omega(\phi)$ which can be solved by the constraint. Then, let us assume that the dependent field $\omega$ has a new transformation, which is modified by an {\it additional shift} ``$\delta_{add}$'' for compensating the remaining changes from the old transformation $\delta_{old} \omega$; that is to say, we have
\begin{eqnarray}
\delta_{new}\omega = \delta_{old} \omega + \delta_{add} \omega.
\end{eqnarray}
Then, consider the transformation of the constraint 
\begin{eqnarray}
\delta_{new}C =  \delta_{old} C + \delta_{add}C =  \delta_{old} C +\frac{\partial C}{\partial \omega} \delta_{add}\omega =0.
\end{eqnarray}
Thus, given $\delta_{old}\omega$ and $\delta_{old}C$, the invertibility of $\partial C/\partial \omega$ allows one to acquire the additional shifts $\delta_{add}\omega$ in this way
\begin{eqnarray}
\delta_{add}\omega(\phi) = -\left(\frac{\partial C}{\partial \omega}\right)^{-1} \delta_{old} C\Big|_{\omega=\omega(\phi)},\label{add_shift}
\end{eqnarray}
where we insert the solution for the dependent field $\omega$, i.e. $\omega = \omega(\phi)$. This trick shows how the constraint can be preserved under the {\it new} transformations modified by the additional shifts. In the following section, we will use this later.

\section{St\"{u}ckelberg Trick and Gauge Equivalence Program}

Superconformal tensor calculus uses the procedure of ``{\bf Gauge Equivalence Program}'' \cite{SUPER_CONFORMAL_TENSORCAL}. This is a program where one utilizes ``more symmetry'' as an extra in constructing a symmetric theory than the number of symmetries that he wants to impose on the final model, and then remove unwanted symmetries by gauge fixing in the end. The point is that this extra symmetry is only used as not physical symmetry but {\it a tool} for building models! In superconformal tensor calculus, we will consider {\it superconformal symmetry group}, but this will not be imposed on the final physical actions of our theory. 

In fact, the gauge equivalence program works based on the {\it St\"{u}ckelberg mechanism} and its underlying philosophy.  

The {\bf St\"{u}ckelberg mechanism (or trick)} is a way of introducing {\it new fields (which are called ``St\"{u}ckelberg or Compensator fields'')} for compensating the remaining shifts of existing fields of a theory that is not gauge-invariant under a (extra) symmetry of interest in order to make this symmetry to become gauge invariance of the theory. Hence, in principle, we can make any theory to be gauge invariant after applying the St\"{u}ckelberg trick to the theory (See Ref. \cite{Stueckelberg} for a detailed review on the St\"{u}ckelberg mechanism). It is worth noting that {\it St\"{u}ckelberg field can be either physical or unphysical in our favor}.

The {\bf Gauge Equivalence Program} inspired by the St\"{u}ckelberg mechanism consists of the following five steps:
\begin{itemize}
    \item 1. ({\bf Full Symmetry with Extras}): Define ``full symmetries'' by introducing ``extra symmetries'' of our interest into the existing ones in our hands (i.e. specify (1) symmetry generators, (2) transformation parameters, (3) gauge fields, (4) their off-shell closed algebras of the symmetry generators with the corresponding structure constants/functions, and (5) their transformation rules). In this step, gauge multiplets containing the extra gauge fields must be defined. Again, transformation rules of all the involved fields under the ``full'' symmetries must be defined.
     \item 2. ({\bf Compensator}): Define St\"{u}ckelberg fields that will not be physical or auxiliary in the end, so that sometimes they may be defined as a ghost with the kinetic terms of a wrong sign (for example, this appears when producing the Einstein curvature term after fixing the dilatation in conformal gravity). Usually, ``unphysical'' St\"{u}ckelberg field is called just {\it compensator}. This compensator must transform under the symmetry in the way to compensate the remaining shifts of the other existing fields. Later on, compensators will be removed by gauge fixing conditions. 
     \item 3. ({\bf Invariant Action}): Construct the actions which are invariant under the full symmetry groups including the extra symmetries. 
     \item 4. ({\bf Gauge Fixing}): Choose gauge fixing conditions for unwanted symmetries. These gauge constraints must be non-invariant under the unwanted symmetries. Different gauge choices lead to the equivalent theories that are merely related by proper ``{\it field redefinitions}.''
     \item 5. ({\bf Rewriting}): Rewrite the actions and transformation rules of the relevant independent quantities by inserting the solutions of dependent quantities which were obtained from the gauge fixing conditions into them. Then, the resulting transformation rules can be given by a linear combination of the existing and extra invariances (into which the solutions of the dependent fields are input) are called ``{\it decomposition laws},'' which leave the gauge fixing conditions to be invariant.
\end{itemize}

\section{Gauge Equivalence Program for Superconformal Symmetry of $\mathcal{N}=1$ Supergravity: Extra symmetry and Compensator}

We are now in a position to apply the gauge equivalence program to the $\mathcal{N}=1$ Supergravity. As a first step of the program, we assume that our {\it full symmetry} group is defined by a {\it superconformal symmetry} which contains not only the super-Poincar\'{e} group of the Poincar\'{e} supergravity but also {\it dilatation, special conformal symmetry, conformal supersymmetry (a.k.a. $S$ supersymmetry), and chiral $U(1)$ symmetry} as extra symmetries. For the second step, we introduce a conformal compensator $S_0$ with the Weyl/chiral weights $(1,1)$ for the old-minimal formulation of supergravity. In fact, these two steps are explained in detail in the appendix \ref{STC} ({\it superconformal tensor calculus of $\mathcal{N}=1$ supergravity}) of this thesis because they correspond to ``technical'' side.

\section{Gauge Equivalence Program for Superconformal Symmetry of $\mathcal{N}=1$ Supergravity: Invariant Action, Gauge Fixing, and Decomposition Laws}

In this section, we perform the remaining three steps of the program. To do this, we will construct a most general superconformal action of the supergravity which is invariant under the superconformal symmetry. Then, by taking superconformal gauge fixing conditions, we find the physical action of the Poincar\'{e} supergravity, and the corresponding decomposition laws.

\subsection{Brief review on K\"{a}hler transformations}

Before going into the main story of supergravity action, we review the necessary concept called ``K\"{a}hler geometry.'' The K\"{a}hler manifold as a target space has the following {\bf K\"{a}hler metric } ``$g_{\alpha\bar{\beta}}$'' characterizing the line element $ds^2$ in the complex coordinates $z^{\alpha} \in \mathbb{C}$:
\begin{eqnarray}
ds^2 = 2g_{\alpha\bar{\beta}}dz^{\alpha}d\bar{z}^{\bar{\beta}},
\end{eqnarray}
where the metric must be identified as
\begin{eqnarray}
g_{\alpha\bar{\beta}} \equiv \partial_{\alpha}\partial_{\bar{\beta}}K(z,\bar{z}),
\end{eqnarray}
where the function $K$ determining the K\"{a}hler metric is called ``{\bf K\"{a}hler potential}.'' In particular, it is possible to compute the corresponding Christoffel symbol, covariant derivatives, and curvatures. Hence, we may have
\begin{eqnarray}
\Gamma^{\alpha}_{\beta\gamma} = g^{\alpha\bar{\delta}}\partial_{\beta}g_{\gamma\bar{\delta}}, 
\end{eqnarray}
Especially, from the definition of the K\"{a}hler metric, we see that there is an isometry given by 
\begin{eqnarray}
K(z,\bar{z}) \longrightarrow K(z,\bar{z}) +f(z) +\bar{f}(\bar{z})
\end{eqnarray}
for some holomorphic function $f(z)$. Then, for a general function $V(z,\bar{z})$, we can generally define the so-called {\bf K\"{a}hler transformation} 
\begin{eqnarray}
V(z,\bar{z}) \longrightarrow V(z,\bar{z}) \exp\Big[-a^{-1}(w_+f(z)+w_-\bar{f}(\bar{z}))\Big],\label{KT}
\end{eqnarray}
where we define ``{\bf K\"{a}hler weights}'' $(w_+,w_-)$. Moreoevr, in fact, this can be related into dilatation and chiral U(1) symmetries. For a scalar function $\mathcal{V}(X,\bar{X})$ on the embedding space with coordinates $(X,\bar{X})$ larger than the target space $(z,\bar{z})$, 
\begin{eqnarray}
\delta \mathcal{V}(X,\bar{X}) = w_+(\lambda_D + i\lambda_T) \mathcal{V}(X,\bar{X})   + w_-(\lambda_D - i\lambda_T)\mathcal{V}(X,\bar{X}) ,
\end{eqnarray}
where $\lambda_{D,T}$ are the gauge parameters of dilatation and chiral U(1) symmetires. Then, in general, it can be represented in terms of the target space variables in this way
\begin{eqnarray}
\mathcal{V}(X,\bar{X}) \equiv S_0^{w_+}\bar{S}_0^{w_-} V(z,\bar{z}),
\end{eqnarray}
where we introduce conformal compensators with the conformal weights, i.e. $S_0$ and $\bar{S}_0$ has $w_+$ and $w_-$, respectively. Notice that {\it the target space function $V(z,\bar{z})$ is inert under the conformal scaling, but must transform under K\"{a}hler transformations}. The {\bf K\"{a}hler covariant derivatives} ``$\nabla_{\alpha}$'' can be defined as
\begin{eqnarray}
&& \nabla_{\alpha} V(z,\bar{z}) \equiv \partial_{\alpha}V(z,\bar{z}) + w_+a^{-1}(\partial_{\alpha}K(z,\bar{z}))V(z,\bar{z}),\\
&& \bar{\nabla}_{\bar{\alpha}} V(z,\bar{z}) \equiv \partial_{\bar{\alpha}}V(z,\bar{z}) + w_-a^{-1}(\partial_{\bar{\alpha}}K(z,\bar{z}))V(z,\bar{z}),
\end{eqnarray}
which produces that
\begin{eqnarray}
[\nabla_{\alpha},\bar{\nabla}_{\bar{\beta}} ]V(z,\bar{z}) =a^{-1}(w_--w_+)g_{\alpha\bar{\beta}}V(z,\bar{z}) ,
\end{eqnarray}
and the covariant derivaives in spacetime
\begin{eqnarray}
\nabla_{\mu} = \partial_{\mu}z^{\alpha}\nabla_{\alpha} + \partial_{\mu}\bar{z}^{\bar{\alpha}}\bar{\nabla}_{\bar{\alpha}}.
\end{eqnarray}
As a concluding remark, it is also certain that $\nabla_{\alpha}V$ is a covariant quantity that can transform as Eq.~\eqref{KT}.

\subsection{Superconformal invariant action of $\mathcal{N}=1$ supergravity}

Here we derive a physical action of the four dimensional $\mathcal{N}=1$ {\bf Poincar\'{e} supergravity} from the action of its {\bf conformal supergravity}. First, let us consider the following {\it general} superconformal action
\begin{eqnarray}
S = [N(Z^I,\bar{Z}^{\bar{I}})]_D + [\mathcal{W}(Z^I)]_F +[f_{AB}(Z^I)\bar{\lambda}^AP_L\lambda^B]_F + c.c.
\end{eqnarray}
where $Z^I\equiv\{X^I,\Omega^I,F^I\}$ are chiral superconformal multiplets with any Weyl/chiral weights $(w,c)$; $N(Z^I,\bar{Z}^{\bar{I}})$ is a ``composite'' real superconformal multiplet with the Weyl/chiral weights $(2,0)$ whose arguments are given by the chiral multiplets $Z^I$'s; $\mathcal{W}(Z^I)$ is a ``composite'' chiral superconformal multiplet with the Weyl/chiral weights $(3,3)$ as a holomorphic function of the chiral multiplets; $\bar{\Lambda}^AP_L\lambda^B$ is a composite chiral superconformal multiplet with the Weyl/chiral weights $(3,3)$ which is a fermionic bilinear of the gauginos of a vector multiplet $V \equiv \{A_{\mu}^A,\lambda^A,D^A\}$ with the Weyl/chiral weights $(0,0)$, and $f_{AB}(Z^I)$ is a composite chiral superconformal multiplet with the Weyl/chiral weights $(0,0)$ as a holomorphic function of the chiral multiplets and called {\bf Gauge kinetic function}.
In particular, the specific forms of $N$ and $\mathcal{W}$ must be given by
\begin{eqnarray}
N(Z^I,\bar{Z}^{\bar{I}}) &\equiv& S_0\bar{S}_0\Phi(Z^{\alpha},\bar{Z}^{\bar{\alpha}}) \equiv -aS_0\bar{S}_0e^{-K(Z^{\alpha},\bar{Z}^{\bar{\alpha}})/a}, \qquad 
\mathcal{W}(Z^I) \equiv S^3_0W(Z^{\alpha}),
\end{eqnarray}
where $a$ is a real constant; the index $I$ runs over $0$ and $\alpha$'s; $S_0 \equiv \{s^0,\chi^0,F^0\}$ is a {\it conformal compensator} chiral multiplet with the Weyl/chiral weights $(1,1)$; $Z^{\alpha}\equiv \{z^{\alpha},\chi^{\alpha},F^{\alpha}\} $'s are {\it matter (physical)} chiral multiplets with the Weyl/chiral weights $(0,0)$; $\Phi(Z^{\alpha},\bar{Z}^{\bar{\alpha}})\equiv -a e^{-K(Z^{\alpha},\bar{Z}^{\bar{\alpha}})/a}$ is a composite real superconformal multiplet with the Weyl/chiral weights $(0,0)$ of the chiral multiplets $Z^I$'s (where $K(Z^{\alpha},\bar{Z}^{\bar{\alpha}})$ is called {\bf K\"{a}hler potential} and defined with the weights $(0,0)$ as a real function of the matter chiral multiplets because it defines the corresponding K\"{a}hler metric.); $W(Z^{\alpha})$ is called {\bf Superpotential} and defined as a holomorphic function of matter chiral multiplets, and which is due to the property of {\it homogeneity} of the functions with respect to their Weyl scaling dimension; that is, $N(\Lambda S_0,\bar{\Lambda }\bar{S}_0) = \Lambda \bar{\Lambda }N(S_0,\bar{S}_0)$ and $\mathcal{W}(\Lambda S_0) = \Lambda^3 \mathcal{W}(S_0)$ (or equivalently $Z^I\mathcal{W}_I=3\mathcal{W}(Z)$) for some chiral multiplet $\Lambda$ with the weights $(0,0)$.

Here are some remarks about gauge symmetry $\delta$. Given Killing vector fields of the symmetry, the physical matter chiral scalars transform as
\begin{eqnarray}
\delta Z^{\alpha} =\theta^A k_A^{\alpha}(Z).
\end{eqnarray}
On the contrary, the conformal compensator must transform as
\begin{eqnarray}
\delta S_0 = a^{-1}\theta^A S_0 r_A(Z).
\end{eqnarray}
Plus, the K\"{a}hler potential can transform as
\begin{eqnarray}
\delta K(Z,\bar{Z}) = f(Z) +\bar{f}(\bar{Z}) = \theta^A(r_A(Z)+\bar{r}_A(Z)) \overset{!}{=} \delta_{K\ddot{a}hler}K,
\end{eqnarray}
which can be identified with the K\"{a}hler transformation. The corresponding moment map $\mathcal{P}_A$ can be given by
\begin{eqnarray}
\mathcal{P}_A = i(k_A^{\alpha}\partial_{\alpha} K - r_A) = c.c.
\end{eqnarray}

Then, using the superconformal tensor calculus, after elimination of all auxiliary fields and lengthy simplifications, the component actions of the conformal supergravity can be found to be
\begin{eqnarray}
e^{-1} \mathcal{L} &=& \frac{1}{6}N\Big[
-R(e,b) + \bar{\psi}_{\mu} R^{\mu} + e^{-1}\partial_{\mu}(e\bar{\psi}\cdot \gamma \psi^{\mu})
\Big]  - V \nonumber\\
&&+ \mathcal{L}_0 + \mathcal{L}_{1/2} + \mathcal{L}_1 + \mathcal{L}_{mass} + \mathcal{L}_{mix} + \mathcal{L}_d + \mathcal{L}_{4f},\label{SUGRA_action_before_gauge_fixing}
\end{eqnarray}
where  $R^{\mu} \equiv \gamma^{\mu\rho\sigma}\Big(
\partial_{\rho} +\frac{1}{2}b_{\rho} +\frac{1}{4}\omega_{\rho}^{~~ab}(e,b) \gamma_{ab} -\frac{3}{2}i\mathcal{A}_{\rho} \gamma_*
\Big)\psi_{\sigma}$ and $\mathcal{A}_{\mu} = i\frac{1}{2N}\Big[
N_{\Bar{I}}\partial_{\mu} \Bar{X}^{\Bar{I}} -N_I\partial_{\mu} X^I 
\Big] + \frac{1}{N}A^A_{\mu}\mathcal{P}_A$
\begin{eqnarray} 
\mathcal{L}_0 &=& -G_{I\bar{J}}D^{\mu}X^ID_{\mu}\bar{X}^{\bar{J}},\\
\mathcal{L}_{1/2} &=& -\frac{1}{2}G_{I\bar{J}}(\bar{\Omega}^{I}\cancel{\hat{D}}^{(0)}\Omega^{\bar{J}}+\bar{\Omega}^{\bar{J}}\cancel{\hat{D}}^{(0)}\Omega^I),\\
\mathcal{L}_{1} &=& (\textrm{Re}f_{AB})\bigg[-\frac{1}{4}F_{\mu\nu}^A F^{\mu\nu B}-\frac{1}{2}\bar{\lambda}^A \cancel{D}^{(0)}\lambda^B\bigg] + \frac{1}{4}i\bigg[(\textrm{Im}f_{AB})F_{\mu\nu}^A \Tilde{F}^{\mu\nu B}+(D_{\mu}\textrm{Im}f_{AB})\bar{\lambda}^A \gamma_* \gamma^{\mu}\lambda^B\bigg]\nonumber\\{}\\
V &=& V_F + V_F = \underbrace{ G^{I\bar{J}}\mathcal{W}_I\bar{\mathcal{W}}_{\bar{J}} }_{\equiv V_F}+ \underbrace{ \frac{1}{2}{(\textrm{Re}f)^{-1}}^{AB}\mathcal{P}_A\mathcal{P}_B}_{\equiv V_D},\\
\mathcal{L}_{mass} &=& \frac{1}{2}\mathcal{W}\bar{\psi}_{\mu}P_R\gamma^{\mu\nu}\psi_{\nu} -\frac{1}{2}\nabla_I\mathcal{W}_J\bar{\Omega}^I\Omega^J +\frac{1}{4}G^{I\bar{J}} \bar{\mathcal{W}}_{\bar{J}}f_{ABI}\bar{\lambda}^AP_L\lambda^B 
\nonumber\\
&&+ \sqrt{2}i\Big(-\partial_I \mathcal{P}_A + \frac{1}{4}f_{ABI}(\textrm{Re}f)^{-1~BC}\mathcal{P}_C\Big)\bar{\lambda}^A\Omega^{I}+h.c.,\\
\mathcal{L}_{mix} &=& \bar{\psi}\cdot \gamma \underbrace{ P_L \Big(\frac{1}{2}i\mathcal{P}_A\lambda^A +\frac{1}{\sqrt{2}}\mathcal{W}_I\Omega^I\Big)}_{=-P_Lv \textrm{ (=Goldstino)}} +h.c.,
\end{eqnarray}
where $\mathcal{W}_I \equiv \partial \mathcal{W}/\partial X^I \equiv \partial_I\mathcal{W}$, $G_{I\bar{J}}\equiv \partial_I\partial_{\bar{J}}N$ and
\begin{eqnarray}
\mathcal{L}_d &=& \frac{1}{8}(\textrm{Re}f_{AB})\bar{\psi}_{\mu}\gamma^{ab}(F^A_{ab}+\hat{F}^A_{ab})\gamma^{\mu}\lambda^B + \frac{1}{\sqrt{2}}\Big\{ G_{I\bar{J}}\bar{\psi}_{\mu} \cancel{D}X^{\bar{J}} \gamma^{\mu}\Omega^I -\frac{1}{4}f_{ABI}\bar{\Omega}^I \gamma^{ab} \hat{F}^A_{ab}\lambda^B\nonumber\\
&&-\frac{2}{3}N_I\bar{\Omega}^I\gamma^{\mu\nu}D_{\mu}\psi_{\nu}+h.c.\Big\},\\
\mathcal{L}_{4f} &=& -\frac{1}{6}N\mathcal{L}_{SG,torsion} \nonumber\\
&& + \bigg\{
-\frac{1}{4\sqrt{2}}f_{ABI}\bar{\psi}\cdot \gamma \Omega^I \bar{\lambda}^A P_L \lambda^B + \frac{1}{8}\nabla_If_{ABJ}\bar{\Omega}^I \Omega^J \bar{\lambda}^A P_L\lambda^B +h.c.
\bigg\}\nonumber\\
&& + \frac{1}{16}ie^{-1}\varepsilon^{\mu\nu\rho\sigma} \bar{\psi}_{\mu} \gamma_{\nu} \psi_{\rho} \Big(
\bar{\Omega}^{\bar{J}}\gamma_{\sigma}\Omega^I +\frac{1}{2}\textrm{Re}f_{AB}\bar{\lambda}^A\gamma_*\gamma_{\sigma}\lambda^B
\Big) -\frac{1}{2}G_{I\bar{J}}\bar{\psi}_{\mu}\Omega^{\bar{J}}\bar{\psi}^{\mu} \Omega^I \nonumber\\
&& +\frac{1}{4}R_{I\bar{K}J\bar{L}}\bar{\Omega}^I\Omega^J \bar{\Omega}^{\bar{K}}\Omega^{\bar{L}} -\frac{1}{16}G^{I\bar{J}}f_{ABI}\bar{\lambda}^{A}P_L\lambda^B \bar{f}_{CD\bar{J}}\bar{\lambda}^C P_R\lambda^D \nonumber\\
&& + \frac{1}{16}(\textrm{Re}f)^{-1 AB}\Big(
f_{AC I}\bar{\Omega}^I - \Bar{f}_{AC\Bar{I}}\Bar{\Omega}^{\Bar{I}}
\Big)\lambda^C\Big(
f_{BD J}\Bar{\Omega}^J-\Bar{f}_{BD\Bar{J}}\Bar{\Omega}^{\Bar{J}}
\Big)\lambda^D+ NA^F_{\mu} {A^F}^{\mu},\nonumber\\
 \mathcal{L}_{SG,torsion} &\equiv& -\frac{1}{16}\Big[
 (\bar{\psi}^{\rho}\gamma^{\mu}\psi^{\nu})(\bar{\psi}_{\rho}\gamma_{\mu}\psi_{\nu}+2\bar{\psi}_{\rho}\gamma_{\nu}\psi_{\mu}) -4(\bar{\psi}_{\mu}\gamma\cdot \psi)(\bar{\psi}^{\mu}\gamma\cdot \psi)
 \Big],
\end{eqnarray}
where $A_{\mu}^F \equiv i\frac{1}{4N}\bigg[
 \sqrt{2}\Bar{\psi}_{\mu} \Big(N_I\Omega^I-N_{\Bar{I}}\Omega^{\Bar{I}}\Big)+N_{I\Bar{J}}\Bar{\Omega}^I\gamma_{\mu} \Omega^{\Bar{J}}+\frac{3}{2}(\textrm{Re}f_{AB})\Bar{\lambda}^A\gamma_{\mu} \gamma_* \lambda^B
 \bigg]$. The relevant covariant derivatives are given by
\begin{eqnarray}
D_{\mu}X^I &=& \partial_{\mu}X^I -b_{\mu}X^I -A^A_{\mu}k_A^I -i\mathcal{A}_{\mu}X^I,\\
D^{(0)}_{\mu}\Omega^I &=& \Big(
\partial_{\mu} -\frac{3}{2}b_{\mu} +\frac{1}{4}\omega_{\mu}^{~~ab}(e,b) \gamma_{ab} +\frac{1}{2}i\mathcal{A}_{\mu} 
\Big)\Omega^I -A^A_{\mu}\partial_Jk_A^I\Omega^J,\\
D^{(0)}_{\mu}\lambda^A &=&  \Big(
\partial_{\mu} -\frac{3}{2}b_{\mu} +\frac{1}{4}\omega_{\mu}^{~~ab}(e,b) \gamma_{ab} -\frac{3}{2}i\mathcal{A}_{\mu} \gamma_*
\Big)\lambda^A -A_{\mu}^C \lambda^B f_{BC}^{~~~A},\\
D_{\mu}\psi_{\nu} &=&  \Big(
\partial_{\mu} +\frac{1}{2}b_{\mu} +\frac{1}{4}\omega_{\mu}^{~~ab}(e,b) \gamma_{ab} -\frac{3}{2}i\mathcal{A}_{\mu} \gamma_*
\Big)\psi_{\nu}.
\end{eqnarray}

\subsection{Superconformal gauge fixing: $K,D,S,A(or T)$-gauge fixings}

Now we are ready to reduce the superconformal action to the physical one by imposing superconformal gauges. First of all, it is possible to fix the special conformal symmetry in the way 
\begin{eqnarray}
\textrm{K-gauge}: \quad b_{\mu} =0 \label{K_gauge}
\end{eqnarray}
since its special conformal transformation is given by $\delta_K(\lambda_K)b_{\mu} = 2{\lambda_K}_{\mu}$. Second, looking at the curvature term $-\frac{N}{6}R(e,b)$ in the superconformal action \eqref{SUGRA_action_before_gauge_fixing}, we notice that if we can fix $N=-\frac{3}{\kappa^2}$ where $\kappa$ is defined as a {\bf dimensionful gravitational coupling constant} such that $\kappa^2 \equiv 8\pi G = M_{pl}^{-2}$ for the {\bf reduced Planck mass} $M_{pl}\equiv 2.4\times 10^{18}\textrm{GeV}$, then we can obtain the {\bf Einstein curvature term} proportional to $\mathcal{L}_{Einstein}\sim \frac{1}{2}R(e)$!
\begin{eqnarray}
\textrm{D-gauge}: \quad N \equiv -3 \kappa^{-2} \implies 3 \kappa^{-2}  \overset{!}{=} a \quad \textrm{and}\quad S_0\Bar{S}_0e^{-K/a} = 1.\label{D_gauge}
\end{eqnarray}
It turns out that this D-gauge choice can eliminate the kinetic mixing terms of ``graviton'' and other ``scalars.'' In addition, we can similarly remove the kinetic mixing terms of ``gravitino'' and other ``fermions'' by fixing the conformal supersymmetry (i.e. $S$-SUSY) through
\begin{eqnarray}
\textrm{S-gauge}: \quad N_I\Omega^I \equiv  0 \Longleftrightarrow N_{\Bar{I}}\Omega^{\Bar{I}} \equiv 0 \implies P_L\chi^0 = \frac{1}{a}\kappa^2 P_L\chi^{\alpha} \label{S_gauge}
\end{eqnarray}
Moreover, we can fix the chiral U(1) symmetry by imposing
\begin{eqnarray}
\textrm{A (or T)-gauge}: \quad S_0 = \Bar{S}_0 \implies s_0 = \Bar{s}_0 = e^{K/2a}.\label{A_gauge}
\end{eqnarray}

\subsection{Physical action of $\mathcal{N}=1$ supergravity}

The physical action of the general $\mathcal{N}=1$ supergraivity theory can now be obtained by the spacetime integral of a {\it gauge-fixed} Lagrangian from the the action in Eq.~\eqref{SUGRA_action_before_gauge_fixing} and constraints in Eqs.~\eqref{K_gauge}, \eqref{D_gauge}, \eqref{S_gauge}, and \eqref{A_gauge}.

Before looking into the physical action, we summarize some points. 
\begin{itemize}
    \item {\bf Chiral multiplets, and K\"{a}hler Potential:} Consider chiral supermultiplets $\{z^{\alpha},P_L\chi^{\alpha}\}$ indexed by $\alpha$, whose kinetic Lagrangians are determined by their K\"{a}hler potential $K(z,\bar{z})$. Then, the K\"{a}hler metric of the scalar target space is invariant udner the K\"{a}hler transformation
    \begin{eqnarray}
    g_{\alpha\bar{\beta}} \equiv \partial_{\alpha}\partial_{\bar{\beta}}K(z,\bar{z}), \qquad K(z,\bar{z}) \longrightarrow K(z,\bar{z}) + f(z) + \bar{f}(\bar{z}),
    \end{eqnarray}
    where $f(z)$ is given as some holomorphic function of the matter scalars. 
    \item {\bf Gauge multiplets:} Consider gauge multiplets $\{A_{\mu}^A,\lambda^A\}$ indexed by $A$, whose kinetic Lagrangians are determined by the symmetric gauge kinetic functions $f_{AB}(z)=f_{BA}(z)$. 
    \item {\bf Superpotential:} Choose a superpotential $W(z)$ of matter scalars. Superpotential is a world scalar, but it transforms under the K\"{a}hler transformation as 
    \begin{eqnarray}
    W(Z) \longrightarrow e^{-\kappa^2f(z)}W(z).
    \end{eqnarray}
    \item {\bf Gauge symmetry (R-symmetry):} Consider a Lie group $G$ with structure constants $f_{AB}^{~~~C}$. Then, the theory can have gauge symmetry provided that $K,W,f_{AB}$ obey some proper conditions by gauge invariance. Let $\theta^A$ be a gauge parameter. Then, the gauge transformation of matter scalar is given by
    \begin{eqnarray}
    \delta_{gauge}z^{\alpha} \equiv \theta^A k_A^{\alpha}(z),
    \end{eqnarray}
    where the gauge Killing vector $k_A^I$ is determined by {\bf Real moment map} ``$\mathcal{P}_A(z,\bar{z})$'' such that
    \begin{eqnarray}
    k_A^{\alpha}(z) = -ig^{\alpha\bar{\beta}}\partial_{\bar{\beta}}\mathcal{P}_A(z,\bar{z}), \quad \nabla_{\alpha}\partial_{\beta}\mathcal{P}_A(z,\bar{z})=0,
    \end{eqnarray}
    in which $\nabla_{\alpha}$ contains the Levi-Chivita connection of the K\"{a}hler metric. In fact, the K\"{a}hler potential $K$ does not have to be invariant under the gauge symmetry because this shift can be considered as a K\"{a}hler transformation. That is,
    \begin{eqnarray}
    && \delta_{gauge}K = \theta^A [r_A(z)+\bar{r}_A(\bar{z})]\overset{!}{=}f(z)+\bar{f}(z)=\delta_{K\ddot{a}hler}K, \\
    && r_A \equiv k_A^{\alpha}\partial_{\alpha}K + i\mathcal{P}_A, \quad f(z)\equiv \theta^Ar_A .
    \end{eqnarray}
    Here is a technical remark. From the definition of $r_A$, we can just read the moment map as 
    \begin{eqnarray}
    \mathcal{P}_A \equiv i(k_A^{\alpha}\partial_{\alpha}K-r_A).
    \end{eqnarray}
    In fact, the fermions can vary under the K\"{a}hler transformation
    \begin{eqnarray}
    && P_L\chi^{\alpha} \longrightarrow e^{i\kappa^2(\textrm{Im}f(z))/2}P_L\chi^{\alpha},\\
    && P_L\psi_{\mu} \longrightarrow e^{-i\kappa^2(\textrm{Im}f(z))/2}P_L\psi_{\mu}, \\
    && P_L\lambda^A \longrightarrow e^{-i\kappa^2(\textrm{Im}f(z))/2}P_L\lambda^A
    \end{eqnarray}
    Plus, the gauge kinetic function must be gauge invariant, i.e. $\delta_{gauge}f_{AB}(z)=0$. Lastly, the superpotential $W(z)$ must satisfy
    \begin{eqnarray}
    &&\delta_{gauge}\mathcal{W}=\mathcal{W}_Ik_A^I=0 \implies k_A^{\alpha}\nabla_{\alpha}W + i\kappa^2 \mathcal{P}_AW = W_{\alpha}k_A^{\alpha}+\kappa^2 r_A W =0, \nonumber\\
    && \therefore ~~ \delta_{gauge}W(z) = W_{\alpha}k_A^{\alpha} = -\kappa^2 r_AW(z) \overset{!}{=} \delta_{K\ddot{a}hler}W(z).
    \end{eqnarray}
    Notice that the gauge transformation of the superpotential is also equivalent to its K\"{a}hler transformation. In particular, when $r_A \equiv i\xi_A$ for some constant $\xi_A$, $r_A$ is called as ``{\bf standard Fayet-Iliopoulos term}.'' Moreover, it is worth noticing that if $r_A =0$, then the symmetry $\delta_{gauge}$ must be realized as {\bf Gauged non-R symmetry}. If $r_A \neq 0$, then the symmetry $\delta_{gauge}$ is realized as a {\bf Gauged R-symmetry}. Hence, for usual gauged non-R symmetries, both K\"{a}hler potential and superpotential must be invariant under them.
\end{itemize}

After lengthy calculations, we reach the following physical action:
\begin{eqnarray}
e^{-1}\mathcal{L} &=& \frac{1}{2\kappa^2} [R(e)-\bar{\psi}_{\mu}R^{\mu}] - g_{\alpha\bar{\beta}}\Big[
\hat{\partial}_{\mu}z^{\alpha}\hat{\partial}^{\mu}\bar{z}^{\bar{\beta}}
+\frac{1}{2}\bar{\chi}^{\alpha}\cancel{D}^{(0)}\chi^{\bar{\beta}} + \frac{1}{2}\bar{\chi}^{\bar{\beta}}\cancel{D}^{(0)}\chi^{\alpha}
\Big] - V \nonumber\\
&&+(\textrm{Re}f_{AB}) \Big[
-\frac{1}{4}F_{\mu\nu}^A F^{\mu\nu B} -\frac{1}{2}\bar{\lambda}^A\cancel{D}^{(0)}\lambda^B
\Big] \nonumber\\
&&+\frac{1}{4}i\bigg[
(\textrm{Im}f_{AB})F_{\mu\nu}^A \Tilde{F}^{\mu\nu B} +(\hat{\partial}_{\mu}\textrm{Im}f_{AB})\bar{\lambda}^A\gamma_*\gamma^{\mu}\lambda^B
\bigg] \nonumber\\
&& +\frac{1}{8}(\textrm{Re}f_{AB}) \bar{\psi}_{\mu}\gamma^{ab}(F_{ab}^A+\hat{F}_{ab}^A)\gamma^{\mu}\lambda^B + \bigg[
\frac{1}{\sqrt{2}}g_{\alpha\bar{\beta}}\bar{\psi}_{\mu}\cancel{\hat{\partial}}\bar{z}^{\bar{\beta}} \gamma^{\mu}\chi^{\alpha} + h.c.
\bigg] \nonumber\\
&& \bigg[
\frac{1}{4\sqrt{2}}f_{AB \alpha}\bar{\lambda}^A\gamma^{ab}\hat{F}_{ab}^B\chi^{\alpha}+h.c.
\bigg] + \mathcal{L}_{mass} + \mathcal{L}_{mix} + \mathcal{L}_{4f},\label{physical_action}
\end{eqnarray}
where the K\"{a}hler metric $g_{\alpha\bar{\beta}}$ is defined by the K\"{a}hler potential $K(z,\bar{z})$ in the way that $g_{\alpha\bar{\beta}} \equiv \partial_{\alpha}\partial_{\bar{\beta}}K(z,\bar{z})$. The first term corresponds to the kinetic terms of graviton and gravitino with $R^{\mu}$
\begin{eqnarray}
R^{\mu} \equiv \gamma^{\mu\rho\sigma} \Big(
\partial_{\rho} +\frac{1}{4}\omega_{\rho}^{~~~ab}(e)\gamma_{ab}-\frac{3}{2}i\mathcal{A}_{\rho}\gamma_*
\Big)\psi_{\sigma}.
\end{eqnarray}
The second term contains the kinetic terms of matter complex scalars and fermions with the following covariant derivatives 
\begin{eqnarray}
\hat{\partial}_{\mu}z^{\alpha} &\equiv& \partial_{\mu}z^{\alpha} -A_{\mu}^A k_A^{\alpha}(z),\\
D_{\mu}^{(0)}\chi^{\alpha} 
&\equiv&
\Big(
\partial_{\rho} +\frac{1}{4}\omega_{\rho}^{~~~ab}(e)\gamma_{ab}+\frac{3}{2}i\mathcal{A}_{\rho}\gamma_*
\Big)\chi^{\alpha} -A_{\mu}^A \frac{\partial k_A^{\alpha}(z)}{\partial z^{\beta}}\chi^{\beta} + \Gamma^{\alpha}_{\beta\gamma} \chi^{\gamma}\hat{\partial}_{\mu}z^{\beta},
\end{eqnarray}
where 
\begin{eqnarray}
\Gamma^{\alpha}_{\beta\gamma} &\equiv& g^{\alpha\bar{\delta}}\partial_{\beta}g_{\gamma\bar{\delta}},\\
\mathcal{A}_{\mu} &\equiv& \frac{1}{6}\kappa^2 \bigg[
\hat{\partial}_{\mu} z^{\alpha} \partial_{\alpha}K - \hat{\partial}_{\mu} \partial_{\bar{\alpha}}K +A_{\mu}^A(r_A-\bar{r}_A).
\bigg]
\end{eqnarray}
Note that the covariant derivatives of scalars only contain the usual Yang-Mills (YM) gauge connections $A_{\mu}^A$, while the covariant derivatives of spinor have K"{a}hler connection $\mathcal{A}_{\mu}$, Lorentz spin connection, YM gauge connection, and Christoffel symbol $\Gamma^{\alpha}_{\beta\bar{\delta}}$.  The K"{a}hler connection $\mathcal{A}_{\mu}$ transforms as
\begin{eqnarray}
\delta\mathcal{A}_{\mu} = -\frac{1}{3}\kappa^{2}\partial_{\mu}(\theta^A \textrm{Im}r_A + \textrm{Im}f).
\end{eqnarray}
The scalar potential $V$ is given by
\begin{eqnarray}
V &=& V_- + V_+ = V_F + V_D, \nonumber\\
V_- &\equiv& -3\kappa^2 e^{\kappa^2 K} W\bar{W}, \quad V_+ \equiv  e^{\kappa^2 K}\nabla_{\alpha}W g^{\alpha\bar{\beta}}\bar{\nabla}_{\bar{\beta}}\bar{W} + \frac{1}{2}(\textrm{Re}f)^{-1 AB}\mathcal{P}_A\mathcal{P}_B,\noindent\\
V_F &\equiv&  e^{\kappa^2 K}\nabla_{\alpha}W g^{\alpha\bar{\beta}}\bar{\nabla}_{\bar{\beta}}\bar{W}-3\kappa^2 e^{\kappa^2 K} W\bar{W} \overset{!}{=} \kappa^{-4} e^{G}(G_{\alpha}G^{\alpha\bar{\beta}}G_{\bar{\beta}}-3), \nonumber\\
V_D &\equiv& \frac{1}{2}(\textrm{Re}f)^{-1 AB}\mathcal{P}_A\mathcal{P}_B,
\end{eqnarray}
where $G \equiv \kappa^2 K + \ln (\kappa^3 W) + \ln(\kappa^3\bar{W})$ is defined as a ``K\"{a}hler-invariant'' {\it (dimensionless) supergravity G-function} (and thus $G_{\alpha\bar{\beta}}=\kappa^2 g_{\alpha\bar{\beta}}$), and supersymmetry breaking scale $M_S$ can be identified by the vacuum expectation value of the positive potential, i.e. $M_S^2 \equiv \left<V_+\right>$. The K\"{a}hler covariant derivative of the superpotential $W$ is given by
\begin{eqnarray}
\nabla_{\alpha}W \equiv \partial_{\mu} + \kappa^2 (\partial_{\alpha}K) W.
\end{eqnarray}
Then, the next two terms in the physical action are the gravitation representation of the gauge multiplet's kinetic terms. The corresponding covariant derivative is given by 
\begin{eqnarray}
D_{\mu}^{(0)}\lambda^A \equiv \Big( 
\partial_{\mu} + \frac{1}{4}\omega_{\mu}^{~~ab}
(e)\gamma_{ab}-\frac{3}{2}i\mathcal{A}_{\mu}\gamma_*\Big)\lambda^A -A_{\mu}^C \lambda^B f_{BC}^{~~~A}.
\end{eqnarray}
The fifth term is the supercurrent coupling of gauge multiplet of form $\bar{\psi}_{\mu}\mathcal{J}^{\mu}$, where the supercovariant gauge curvature is given by
\begin{eqnarray}
\hat{F}_{ab}^A \equiv e^{\mu}_ae^{\nu}_b \Big(
2\partial_{[\mu}A_{\nu]}^A + f_{BC}^{~~~A}A_{\mu}^BA_{\mu}^C +\bar{\psi}_{[\mu}\gamma_{\nu]}\lambda^A 
\Big).
\end{eqnarray}
Here we see that there is a fermionic bilinear contribution in the supercovariant gauge curvature. This is in fact to make it transform covariantly under the local supersymmetry. It is worth noticing that the supergravity equations of motion are supercovariant in the end. Next, let us look at the fermion bilinear mass term 
\begin{eqnarray}
\mathcal{L}_{mass} = \frac{1}{2}m_{3/2}\bar{\psi}_{\mu}P_R\gamma^{\mu\nu}\psi_{\nu} -\frac{1}{2}m_{\alpha\beta}\bar{\chi}^{\alpha}\chi^{\beta} -m_{\alpha A}\bar{\chi}^{\alpha}\lambda^A -\frac{1}{2}m_{AB}\bar{\lambda}^AP_L\lambda^B + h.c.,
\end{eqnarray}
where the gravitino mass parameter $m_{3/2}$ and fermion mass matrix components are given by
\begin{eqnarray}
m_{3/2} &\equiv& \kappa^2 e^{\kappa^2 K/2} W, \\
m_{\alpha\beta} &\equiv &  e^{\kappa^2 K/2}\nabla_{\alpha}\nabla_{\beta}W = e^{\kappa^2 K/2}(\partial_{\alpha} + \kappa^2\partial_{\alpha}K)(\nabla_{\beta}W)
-e^{\kappa^2 K/2}\Gamma_{\alpha\beta}^{\gamma}\nabla_{\gamma}W \nonumber\\
&=&e^{\kappa^2 K/2}(\partial_{\alpha} + \kappa^2\partial_{\alpha}K)(\nabla_{\beta}W) -e^{\kappa^2 K/2} g^{\gamma\bar{\delta}}\partial_{\alpha}g_{\beta\bar{\delta}}\nabla_{\gamma}W ,\\
m_{\alpha A} &\equiv& i\sqrt{2} \Big[ 
\partial_{\alpha}\mathcal{P}_A -\frac{1}{4}f_{AB \alpha} (\textrm{Re}f)^{-1 BC}\mathcal{P}_{C}
\Big] = m_{A \alpha},\\
m_{AB} &\equiv& -\frac{1}{2}e^{\kappa^2 K/2} f_{AB \alpha}g^{\alpha\bar{\beta}}\bar{\nabla}_{\bar{\beta}}\bar{W}.
\end{eqnarray}
Importantly, the location of the minimum of the potential is determined by
\begin{eqnarray}
\partial_{\alpha}V \equiv \frac{\partial V}{\partial z^{\alpha}} =  e^{\kappa^2 K} \Big( -2\bar{m}_{3/2}\nabla_{\alpha}W + m_{\alpha\beta}g^{\beta\bar{\beta}}\bar{\nabla}_{\bar{\beta}}\bar{W}\Big)-\frac{i}{\sqrt{2}}\mathcal{P}_A (\textrm{Re}f)^{-1 AB} m_{\alpha B}.
\end{eqnarray}

The mixing term of gravitino and matter fermions is given by
\begin{eqnarray}
\mathcal{L}_{mix} &\equiv& \bar{\psi}\cdot \gamma \bigg[
\frac{1}{2}iP_L\lambda^A \mathcal{P}_A + \frac{1}{\sqrt{2}}\chi^{\alpha} e^{\kappa^2K/2} \nabla_{\alpha}W
\bigg]  + h.c.
\end{eqnarray}
The final compoent of the action is given by the four-fermion terms 
\begin{eqnarray}
\mathcal{L}_{4f} &\equiv& \frac{1}{2}\kappa^{-2} \mathcal{L}_{SG,torsion}\nonumber\\
&&+\bigg\{
-\frac{1}{4\sqrt{2}}f_{AB\alpha}\bar{\psi}\cdot \gamma \chi^{\alpha} \bar{\lambda}^{A}P_L\lambda^B +\frac{1}{8} (\nabla_{\alpha}f_{AB \beta})\bar{\chi}^{\alpha}\chi^{\beta}\bar{\lambda}^AP_L\lambda^B +h.c.
\bigg\} \nonumber\\
&&+\frac{1}{16}ie^{-1}\varepsilon^{\mu\nu\rho\sigma}\bar{\psi}_{\mu} \gamma_{\nu} \psi_{\rho} \Big( \frac{1}{2}\textrm{Re}f_{AB} \bar{\lambda}^A\gamma_*\gamma_{\sigma}\lambda^B +g_{\alpha\bar{\beta}}\bar{\chi}^{\bar{\beta}}\gamma_{\sigma}\chi^{\alpha} \Big)
-\frac{1}{2}g_{\alpha\bar{\beta}}\bar{\psi}_{\mu} \chi^{\bar{\beta}}\bar{\psi}^{\mu}\chi^{\alpha} \nonumber\\
&& +\frac{1}{4}\Big(
R_{\alpha\bar{\gamma}\beta\bar{\delta}} - \frac{1}{2}\kappa^2 g_{\alpha\bar{\gamma}}g_{\beta\bar{\delta}}
\Big) \bar{\chi}^{\alpha}\chi^{\beta} \bar{\chi}^{\bar{\gamma}}\chi^{\bar{\delta}} \nonumber\\
&& +\frac{3}{64}\kappa^2 \Big[
(\textrm{Re}f_{AB})\bar{\lambda}^A \gamma_{\mu}\gamma_*\lambda^B 
\Big]^2 -\frac{1}{16}f_{AB \alpha}\bar{\lambda}^A P_L\lambda^B g^{\alpha \bar
{\beta}} \bar{f}_{CD\bar{\beta}}\bar{\lambda}^CP_R\lambda^D  \nonumber\\
&& +\frac{1}{16}(\textrm{Re}f)^{-1 AB} \Big(
f_{AC\alpha}\bar{\chi}^{\alpha}-\bar{f}_{AC\bar{\alpha}}\bar{\chi}^{\bar{\alpha}}
\Big)\lambda^C\Big(
f_{BD\alpha}\bar{\chi}^{\beta} -\bar{f}_{BD\bar{\beta}}\bar{\chi}^{\bar{\beta}}
\Big)\lambda^D \nonumber\\
&& -\frac{1}{4}\kappa^2 g_{\alpha\bar{\beta}}(\textrm{Re}f)\bar{\chi}^{\alpha}\lambda^A\bar{\chi}^{\bar{\beta}}\lambda^B
\end{eqnarray}

\subsection{Decomposition laws}

We have seen that after imposing the superconformal gauge, we obtain the physical action of $\mathcal{N}=1$ supergravity. In fact, according to gauge equivalence program, we should get the decomposition laws. Hence, the physical transformations of the fields are then identified as
\begin{eqnarray}
\delta e^{a}_{\mu} &=& \frac{1}{2}\bar{\epsilon}\gamma^a \psi_{\mu},\\
\delta P_L\psi_{\mu} &=& \Big( \partial_{\mu} +\frac{1}{4}\omega_{\mu}^{~~ab}(e)\gamma_{ab} -\frac{3}{2}i\mathcal{A}_{\mu} \Big)P_L\epsilon + \underbrace{\Big(\frac{1}{2}\kappa^2 \gamma_{\mu}e^{\kappa^2 K/2}W\Big)}_{\equiv~ \delta_s P_L\psi_{\mu}}P_R\epsilon+\frac{1}{4}\kappa^2 P_L\psi_{\mu}\theta^A (\bar{r}_A -r_A) \nonumber\\
&&+\textrm{(quadratic in fermions)}\epsilon,\\
\delta z^{\alpha} &=& \frac{1}{\sqrt{2}}\bar{\epsilon}P_L\chi^{\alpha},\\
\delta P_L\chi^{\alpha} &=& \frac{1}{\sqrt{2}} P_L\cancel{\hat{\partial}}z^{\alpha} +  \underbrace{\Big(
 -\frac{1}{\sqrt{2}}e^{\kappa^2K/2}g^{\alpha\bar{\beta}}\bar{\nabla}_{\bar{\beta}}\bar{W}
\Big)}_{\equiv ~ \delta_s P_L\chi^{\alpha}}P_L\epsilon + \theta^A\Big[
\frac{\partial k_A^{\alpha}(z)}{\partial z^{\beta}}\chi^{\beta}+ \frac{1}{4}\kappa^2(\bar{r}_A-r_A)\chi^{\alpha}
\Big]\nonumber\\
&&+ \textrm{(quadratic in fermions)}\epsilon ,\\
\delta A_{\mu}^A &=& -\frac{1}{2}\bar{\epsilon} \gamma_{\mu}\lambda^A + \partial_{\mu}\theta^A +\theta^CA_{\mu}^B f_{BC}^{~~~A},\\
\delta P_L\lambda^A &=& \bigg[
\frac{1}{4}\gamma^{\mu\nu} F_{\mu\nu}^A +\underbrace{\frac{1}{2}i\gamma_*(\textrm{Re}f)^{-1 AB}\mathcal{P}_B}_{\equiv ~ \delta_s P_L\lambda^A}
\bigg]\epsilon + \theta^B \bigg[
\lambda^C f_{CB}^{~~~A} + \frac{1}{4}\kappa^2 \gamma_*(\bar{r}_B-r_B)\lambda^A
\bigg] \nonumber\\
&&+ \textrm{(quadratic in fermions)}\epsilon,
\end{eqnarray}
where we define special parts called ``{\it Fermion shift}'' as the only scalar part of the supersymmetry transformation:
\begin{eqnarray}
\delta_s P_L\psi_{\mu} \equiv \frac{1}{2}\kappa^2 \gamma_{\mu}e^{\kappa^2 K/2}W, \quad \delta_s P_L\chi^{\alpha} \equiv 
 -\frac{1}{\sqrt{2}}e^{\kappa^2K/2}g^{\alpha\bar{\beta}}\bar{\nabla}_{\bar{\beta}}\bar{W}, \quad \delta_s P_L\lambda^A \equiv \frac{1}{2}i\gamma_*(\textrm{Re}f)^{-1 AB}\mathcal{P}_B.
\end{eqnarray}
Then, the scalar potential can be rewritten in terms of these fermion shifts:
\begin{eqnarray}
V = -\frac{1}{2}(D-2)(D-1)\kappa^2 e^{\kappa^2K}|W|^2 + 2(\delta_s P_L\chi^{\alpha})g_{\alpha\bar{\beta}}(\delta_s P_R\chi^{\bar{\beta}}) + 2(\delta_s P_L\lambda^A)\textrm{Re}f_{AB}(\delta_s P_R\lambda^B),\label{fermion_shift_V}
\end{eqnarray}
where we define a general spacetime dimension $D$. Regarding this potential, we observe that supersymmetric solutions lead to 
\begin{eqnarray}
V|_{\textrm{supersymmetric}}=-\frac{1}{2}(D-2)(D-1)\kappa^2 e^{\kappa^2K}|W|^2 \leq 0,
\end{eqnarray}
which means that vacua must be either {\it Minkowski} (if $W=0$) or {\it Anti-de-Sitter (AdS)} (if $W\neq 0$) on the supersymmetric solutions in supergravity.

\subsection{Mass dimension}
Here is a remark about mass dimension. In fact, the ``physical'' scalar $z'^{\alpha}$ must have the canonical mass dimension $[z'^{\alpha}]=1$, so that its superpartner fermion has $[\chi'^{\alpha}]=3/2$, and auxiliary field has $[F'^{\alpha}]=2$. Meanwhile, since scalar fields can normally appear non-linearly in the physical action, it is more convenient to take them as ``dimensionless,'' i.e. $[z^{\alpha}]=0$ through the re-scaling $z'^{\alpha} ~\equiv \kappa^{-1}z$ , so that the other superpartners of $z^{\alpha}$ must have modified mass dimensions, i.e. $[\chi^{\alpha}]=1/2$ and $[F^{\alpha}]=1$. In this unit, then, it turns out that the K\"{a}hler metric $g_{\alpha\bar{\beta}}$ and also K\"{a}hler potential $K(z,\bar{z})$ must be dimension ``2'' from the kinetic term of the scalars $[g_{\alpha\bar{\beta}}\partial_{\mu}z^{\alpha}\partial^{\mu}\bar{z}^{\bar{\beta}}]=4$, i.e. $[g_{\alpha\bar{\beta}}]=[K]=2$. In fact, regardless of the re-scaling, the canonical mass dimensions of the various quantities are given by
\begin{eqnarray}
[\kappa]=-1, \quad [f_{AB}(z)]=0, \quad [\mathcal{A}_{\mu}]=1, \quad 
[K]= [g_{\alpha\bar{\beta}}]= [\mathcal{P}_A]= [r_A] = 2, \quad [W(z)]=3.
\end{eqnarray}

\subsection{Global SUSY limit}

The normalization of the independent functions $K,W,f_{AB},\mathcal{P}_A,r_A$ is defined in such a way that we can obtain the global supersymmetry limit by taking $\kappa = M_{pl}^{-1} =0$ or equivalently $M_{pl}\longrightarrow \infty$ after ``re-scaling of gravitino'' ($\psi \longrightarrow \kappa\psi$) to remove the gravitino dependence from the physical action of global supersymmetry.

\section{Supersymmetry Breaking and Super-Brout-Englert-Higgs (BEH) Effect}

\subsection{Goldstino and the super-BEH effect}

In this section, we are going to discuss a Goldstino mode which plays a critical role in supersymmetry breaking. Previously, in the superconformal approach, we have obtained the Goldstino spinor $P_Lv$ defined by
\begin{eqnarray}
P_Lv \equiv -\frac{1}{\sqrt{2}}\mathcal{W}_I P_L\Omega^I -\frac{1}{2}i\mathcal{P}_A P_L\lambda^A,
\end{eqnarray}
where $\mathcal{W} \equiv s_0^3 W(z)$. After the superconformal gauge fixing, we observe that the mixing term of gravitino and matter fermions is found as
\begin{eqnarray}
\mathcal{L}_{mix} = -\bar{\psi}\cdot \gamma P_Lv + h.c.,
\end{eqnarray}
where the Goldstino becomes 
\begin{eqnarray}
P_Lv = -\frac{1}{2}iP_L\lambda^A \mathcal{P}_A - \frac{1}{\sqrt{2}}P_L\chi^{\alpha} e^{\kappa^2K/2} \nabla_{\alpha}W.
\end{eqnarray}
We can also rewrite this in terms of the fermion shifts
\begin{eqnarray}
P_Lv = P_L\chi^{\alpha}\delta_sP_R\chi_{\alpha} + P_L\lambda^A \delta_sP_R\lambda_A,\label{goldstino_fermion_shift_expression}
\end{eqnarray}
where 
\begin{eqnarray}
\delta_s P_R\chi_{\alpha} \equiv g_{\alpha\bar{\beta}}\delta_sP_R\chi^{\bar{\beta}} = - \frac{1}{\sqrt{2}}e^{\kappa^2K/2} \nabla_{\alpha}W, \qquad \delta_sP_R\lambda_A = (\textrm{Re}f)_{AB}\delta_s P_R \lambda^B = -\frac{1}{2}i\mathcal{P}_A.
\end{eqnarray}
The important point here is that the supersymmetry transforation of the Goldstino can be given by
\begin{eqnarray}
\delta P_Lv = \frac{1}{2}V_+P_L\epsilon + \textrm{vectors, derivatives of scalars, (quadratic terms in fermion)$\epsilon$}.\label{goldstino_shift}
\end{eqnarray}
One can derive this using the decomposition laws and the definition of the positive potential $V_+$. From this shift of the Goldstino, we can conclude the following. If $V_+ >0$ in a supersymmetry breaking vacuum, then the Goldstino goes through the shift $\delta P_Lv$ in Eq.~\eqref{goldstino_shift}, which means that we can set any value of the Goldstino in the direction of the SUSY transformation. Hence, the most convenient gauge can be chosen by
\begin{eqnarray}
v = 0 \qquad \textrm{(supersymmetry gauge)},
\end{eqnarray}
which removes the Goldstino degree of freedom in the physical action by washing out the mixing term $\mathcal{L}_{mix}$ of gravitino and matter fermions. The remaining gravitino mass parameter term $m_{3/2}$ remains the same as that of the massive Rarita-Schwinger action (i.e. massive gravitino action). Thus, in Minkowski space (when $V=0$), we can consider $m_{3/2}$ as the ``physical'' gravitino mass that can be produced by the elimination of Goldstino! 

Of course, the super-BEH effect affects the other fermion masses. To see how this works, let us consider the gravitino terms in the physical action. When assuming that scalars are fixed as constants and neglecting the other irrelevant terms, we can obtain the following gravitino action
\begin{eqnarray}
e^{-1}\mathcal{L} = \frac{1}{2\kappa^2}\bar{\psi}_{\mu} \Big[
-\gamma^{\mu\nu\rho}\partial_{\nu} + (m_{3/2}P_R + \bar{m}_{3/2}P_L)\gamma^{\mu\rho}
\Big]\psi_{\rho} -\bar{\psi}_{\mu}\gamma^{\mu}v.
\end{eqnarray}
Then, let us introduce a massive gravitino state 
\begin{eqnarray}
P_L\Psi_{\mu} \equiv P_L\psi_{\mu} -\frac{2\kappa^2}{3|m_{3/2}|^2}\partial_{\mu}P_Lv -\frac{\kappa^2}{3\bar{m}_{3/2}}\gamma_{\mu}P_Rv.
\end{eqnarray}
Then, by inserting this massive gravitino state in the above action and diagonalizing the action, we obtain
\begin{eqnarray}
e^{-1}\mathcal{L} &=& \frac{1}{2\kappa^2} \bar{\Psi}_{\mu}\Big[
-\gamma^{\mu\nu\rho}\partial_{\nu} + (m_{3/2}P_R+\bar{m}_{3/2}P_L)\gamma^{\mu\rho}
\Big] \Psi_{\rho} \nonumber\\
&& + \frac{\kappa^2}{3|m_{3/2}|^2} \Big[\bar{v}\cancel{\partial}v + 2\bar{v}(m_{3/2}P_R+\bar{m}_{3/2}P_L)v\Big]. \label{Diagonalized_term}
\end{eqnarray}
Then, by inserting Eq.~\eqref{goldstino_fermion_shift_expression} into Eq.~\eqref{Diagonalized_term} and reading the corresponding contributions to the fermion mass matrix, we find that the additional fermion mass by the elimination of the Goldstino are found to be
\begin{eqnarray}
m_{\alpha\beta}^{(v)} &=& -\frac{4\kappa^2}{3m_{3/2}}(\delta_sP_R\chi_{\alpha})(\delta_sP_R\chi_{\beta}),\\
m_{\alpha A}^{(v)} &=& -\frac{4\kappa^2}{3m_{3/2}}(\delta_sP_R\chi_{\alpha})(\delta_sP_R\lambda_A),\\
m_{AB}^{(v)} &=& -\frac{4\kappa^2}{3m_{3/2}} (\delta_sP_R\lambda_A)(\delta_sP_R\lambda_B).
\end{eqnarray}
Therefore, the full fermion mass matrix must be re-written as
\begin{eqnarray}
m_{\alpha\beta}^g \equiv m_{\alpha\beta} + m_{\alpha\beta}^{(v)},\\
m_{\alpha A}^g \equiv m_{\alpha A}+m_{\alpha A}^{(v)},\\
m_{AB}^g\equiv m_{AB}+m_{AB}^{(v)}.
\end{eqnarray}
This consequence of {\it generation of gravitino and matter fermions by elimination of the Goldstino} is called ``{\bf Super-BEH Effect}.''

There is additional remark about the Goldstino. Using the previous results, we can re-express the first field derivative of the scalar potential and gauge invariance condition of the superpotential as follows:
\begin{eqnarray}
&& \partial_{\alpha}V = -\sqrt{2}\bigg[
m_{\alpha\beta}\delta_sP_L\chi^{\beta} + m_{\alpha B}\delta_sP_R\lambda^B -2\bar{m}_{3/2}\delta_sP_R\chi_{\alpha} 
\bigg],\\
&& m_{\beta A}\delta_S P_L\chi^{\beta} -2\bar{m}_{3/2}\delta_s P_R \lambda_A =0,\label{zero_mode_scan_eq}
\end{eqnarray}
where the second line is obtained by the gauge invariance condition
\begin{eqnarray}
\delta_{gauge}\mathcal{W}=\mathcal{W}_Ik_A^I=0 \implies k_A^{\alpha}\nabla_{\alpha}W + i\kappa^2 \mathcal{P}_AW = W_{\alpha}k_A^{\alpha}+\kappa^2 r_A W =0.
\end{eqnarray}
Then, combining Eqs.~\eqref{fermion_shift_V} and the above fermion masses, Eq.~\eqref{zero_mode_scan_eq} can be re-written by
\begin{eqnarray}
&& m_{\alpha\beta}^g \delta_s P_L\chi^{\beta} + m_{\alpha A}^g \delta_s P_L\lambda^B + \frac{2\kappa^2}{3m_{3/2}}V\delta_sP_R\chi_{\alpha} = -\frac{1}{\sqrt{2}}\partial_{\alpha}V,\\
&& m_{A\beta}^g \delta_s P_L\chi^{\beta} + m_{AB}^g \delta_s P_L\lambda^B + \frac{2\kappa^2}{3m_{3/2}}V\delta_sP_R\lambda_A =0.
\end{eqnarray}
Hence, with respect to these results, we find that in the vacuum ($\partial_{\alpha}V=0$) with zero cosmological constant $V=0$, the full fermion mass matrix must have a zero eigenvector with the vanishing masses as expected.

%% file: chapters/5-1.tex
In this chapter, we review recent developments of inflationary models in supergravity. Supergravity based inflation models and their general difficulties are well-reviewed in Ref.~\cite{Yamaguchi}. In recent decades, several supergravity models for covering both inflation and MSSM phenomenology have been investigated in the literature \cite{DS17,DS17_2,InflationMSSM_Chakravarty,InflationMSSM_Pallis,InflationNMSSM_FKLMV,InflationMSSM_KP,InflationMSSM_EMN}. Most recently, Domcke and Schmitz presented a supergravity model unifying high-scale supersymmetry breaking, viable D-term hybrid inflation, spontaneous B-L breaking at the scale of grand unification, baryogenesis via leptogenesis, and standard model neutrino masses \cite{DS17}. Chakravarty, Gupta, Lambiase, and Mohanty in Ref. \cite{InflationMSSM_Chakravarty} found a Higgs inflation model in the supergravity embedding of MSSM by taking into account all higher order non-renormalizable terms to the MSSM superpotential, in which the inflaton is from the $SU(2)$ Higgs doublet. However, the superpotential in this model has no clear connection to its string-theoretical origin. In Ref. \cite{InflationMSSM_Pallis}, Pallis constructed a model that links Starobinsky-type inflation in no-scale supergravity to MSSM by introducing an arbitrarily-chosen superpotential for inflation. Due to this arbitrariness of superpotential, it seems hard to explore its string-theoretical origin as well. 

In Ref. \cite{InflationNMSSM_FKLMV}, Ferrara, Kallosh, Linde, Marrani and Van Proeyen proposed a construction of embedding Next-to MSSM model and realizing inflation in their supergravity setup. However, the authors found a strong tachyonic instability problem during inflation in their model, and mentioned that its remedy is required to support the Higgs-type inflation \cite{InflationNMSSM_FKLMV}. In Ref. \cite{InflationMSSM_KP}, Kaminska and Pachole built a supergravity model of inflation and preheating by considering non-minimal K\"{a}hler potential as well as MSSM and its nonrenormalizable superpotentials to generate MSSM inflation \cite{MSSMinflation,MSSMinflation_2}. In Ref. \cite{InflationMSSM_EMN}, Enqvist, Mether, and Nurmi investigated a supergravity origin of the MSSM inflationary scenarios \cite{MSSMinflation,MSSMinflation_2} with a string-motivated K\"{a}hler potential but with MSSM and its higher order nonrenormalizable corrections to superpotential not related to string theory, where inflaton is identified with a gauge invariant combination of squark or slepton fields. We see that it is not manifest how the superpotentials used in Refs. \cite{InflationMSSM_KP,InflationMSSM_EMN} can be obtained from superstring theory. 

Some ``string-inspired'' supergravity models of inflation based on Type IIB string compactifications have been studied by T. Li, Z. Li, and Nanopoulos \cite{lln1,lln2,lln3}. The authors used KKLT superpotential \cite{KKLT} and first introduced anomalous U(1) gauge symmetry for generating a contribution to D-terms \cite{lln1,lln2,lln3}. Their models focus on realizing viable inflation and moduli stabilizations, while MSSM phenomenology has not been embedded into the models. Therefore, it turns out that a string-inspired supergravity model of inflation compatible with MSSM phenomenology has not yet been sufficiently investigated, causing the strong motivation of our work. 

In terms of supersymmetry (SUSY), there is an issue about rich field degrees of freedom. Supersymmetric theory predicts many field degrees of freedom exceeding the number of fields that standard model (SM) currently has due to their superpartners. This means the SM fermions must pair with superpartner scalars. On the contrary, there are a few scalar bosons including Higgs field in the SM phenomenology and even superparticles have never been discovered yet, signaling that supersymmetry was spontaneously broken at a certain mass scale, say $M_S$, less than $M_{pl}$. 

In addition, there is another problem that many scalar bosons may cause in the early universe cosmology. In fact, single-field slow-roll inflation is strongly supported by recent cosmological observations from Planck 2018 data \cite{Planck2018_review, Planck2018_cosmo_para,Planck2018_infl}. This implies that inflaton, which is a hypothetical scalar boson responsible for inflation, must be the unique scalar boson with mass less than the Hubble scale during inflation among possible scalar multi-fields if they exist in the inflation model of interest. Thus, it turns out that one should be able to integrate out {\it unnecessary} \underline{scalar} degrees of freedom and simultaneously make the relevant mass spectra to be phenomenologically acceptable in a supergravity model of interest.

In fact, Vennin, Koyama, and Wands \cite{LightScalar} found that an effective single-field slow-roll inflation may take place even under the introduction of extra scalars. This argument can be valid only if certain reheating scenario conditions on decay rates and masses of inflaton and extra scalars are satisfied \cite{LightScalar}, implying that numerous extra scalars may lead to a great deal of constraints our model must obey. This may be viewed as another obstruction in constructing phenomenological models. Due to this, we pursue to minimize the number of possible extra light scalars living in our model. That is, remaining a few necessary light scalars, one should be able to make the other superfluous scalars much heavy to obtain minimal reheating scenario constraints to follow by integrating out such scalar modes. Accordingly, in the following chapters, I will utilize the argument of Ref. \cite{LightScalar} when we in our models encounter a light Higgs mass and consider this as such a few {\it extra light scalar}.

Another possibility for enlarging the space of scalar potentials in supergravity can be new Fayet-Iliopoulos (FI) terms without gauging R-symmetry. Recently, such new FI terms have been proposed and used for constructing reliable supergravity models of inflation by many authors \cite{acik,CFTV,oldACIK,Kuzenko,ar,AKK}. Our observation here is that the absence of a gauged R symmetry in the new FI term signals that such new FI terms can be compatible with the KKLT string background. In this sense, in Ref. \cite{jp2}, we first proposed such a supergravity model of inflation and dS vacua by adding a ``constant'' new FI term\footnote{Such new FI term has been long regarded to be banned in supergravity. This is because the only possible FI terms were thought to arise from a gauged R-symmetry \cite{R_symm,R_symm_2,R_symm_3}, require a R-symmetric superpotential \cite{barb} and be subject to quantization conditions when a compact R-symmetry group is gauged \cite{Nathan}.} (i.e. called ACIK-FI term \cite{acik}) to supergravity in the KKLT string background, while appropriate realization of MSSM phenomenology has not been made in the model since there was no truncation of many light scalar modes. So, an improvement on our previous work \cite{jp2} has to be made if we wish to embed MSSM into such supergravity.

%% file: chapters/6.tex
\framebox[1.05\width]{{\large This chapter is based on the author's original work in Ref.~\cite{jp1,jp3}.}} \par
\vspace{1cm}
In this chapter, we compute the component action of liberated $\mathcal{N}=1$ supergravity in superconformal tensor calculus\footnote{Superconformal tensor calculus is reviewed in the appendix \ref{STC} of this dissertation.} \cite{cfgvnv,cfgvnv_2,Linear}.  
We find that the superconformal Lagrangian of liberated $\mathcal{N}=1$ supergravity \cite{fkr} can be written by a D-term
\begin{eqnarray}
\mathcal{L}_{NEW} \equiv \bigg[ \mathcal{Y}^2 \frac{w^2\bar{w}^2}{T(\bar{w}^2)\bar{T}(w^2)} \mathcal{U}(Z,\Bar{Z})\bigg]_D,\label{Superconformal_LSG}
\end{eqnarray}
in which we define $\mathcal{Y} \equiv S_0\bar{S}_0 e^{-K/3}$ where $S_0$ is a conformal compensator with Weyl/chiral weights $(1,1)$ and $K$ is a K\"{a}hler potential with the weights $(2,0)$. The notation for the other fields is as follows: $w^2 \equiv \frac{\mathcal{W}^2(K)}{\mathcal{Y}^2}, \quad \bar{w}^2 \equiv \frac{\bar{\mathcal{W}}^2(K)}{\mathcal{Y}^2}$, $\mathcal{W}^2(K)\equiv \mathcal{W}_{\alpha}(K)\mathcal{W}^{\alpha}(K)$ where $\mathcal{W}_{\alpha}(K)$ is a field strength multiplet with respect to the K\"{a}hler potential; $\mathcal{U}(Z,\Bar{Z})$ is a general function of matter multiplets $Z$, and $T(\Bar{w}^2),\Bar{T}(w^2)$ are chiral projection of $\Bar{w}^2$ and its conjugate respectively. The expression in Eq.~\eqref{Superconformal_LSG} will be a key equation in this chapter.

\section{Embedding Super-Weyl-K\"{a}hler Transformations as an Abelian Gauge Symmetry into the Superconformal Formalism}

In liberated $\mathcal{N}=1$ supergravity \cite{fkr}, a key idea is that the Super-Weyl-K\"{a}hler transformation can be promoted to an Abelian gauge symmetry. Liberated supergravity was constructed in~\cite{fkr} using the superspace formalism, where a K\"{a}hler transformation is introduced to compensate the variation of the action under a super-Weyl rescaling. In this work instead, we want to construct the equivalent liberated supergravity using the superconformal tensor calculus to analyze the fermionic interactions in a systematical and economical way. To do so, we introduce a conformal compensator multiplet, called $S_0$, removing the variation while maintaining the K\"{a}hler potential invariant under superconformal symmetry. Therefore, unlike the superspace formalism, it is essential to define such a gauge transformation independently of superconformal symmetry.

To find the Super-Weyl-K\"{a}hler transformations that are compatible with the superconformal formalism, we recall first 
the Super-Weyl-K\"{a}hler transformations that are used in the superspace formalism. A K\"{a}hler function $K(z,\bar{z})$, whose arguments have the vanishing Weyl/chiral weights, is defined up to a chiral gauge parameter $\Sigma$. The 
redefinition by $\Sigma$ acts on the components of the K\"{a}haler multiplet as
\begin{eqnarray}
&& K \rightarrow K + 6\Sigma+6\bar{\Sigma},\\
&& W \rightarrow We^{-6\Sigma}, \qquad \bar{W} \rightarrow \bar{W} e^{-6\bar{\Sigma}}\\
&& T \rightarrow e^{-4\Sigma+2\bar{\Sigma}}T,\qquad \bar{T} \rightarrow e^{2\Sigma-4\bar{\Sigma}}\bar{T},\\
&& \mathcal{D}_{\alpha}K \rightarrow e^{\Sigma-2\bar{\Sigma}}\mathcal{D}_{\alpha}K, \qquad \mathcal{W}_{\alpha} \rightarrow e^{-3\Sigma}\mathcal{W}_{\alpha}, \qquad \mathcal{W}^2 \rightarrow e^{-6\Sigma}\mathcal{W}^2,\\
&& T(\bar{\mathcal{W}}^2) \rightarrow T(\bar{\mathcal{W}}^2) e^{-4\Sigma -4\bar{\Sigma}},\\
&& \mathcal{D}^{\alpha}\mathcal{W}_{\alpha} \rightarrow e^{-2\Sigma-2\bar{\Sigma}}\mathcal{D}^{\alpha}\mathcal{W}_{\alpha},\\
&& E \rightarrow Ee^{2\Sigma + 2\bar{\Sigma}}, \qquad \mathcal{E} \rightarrow \mathcal{E} e^{6\Sigma}.
\end{eqnarray}
where $E$ and $\mathcal{E}$ are the D- and F-term densities, respectively.

Next, it may be useful to recall the relation between the superspace and superconformal formalisms. The invariant actions from the superconformal formalism are identified with those from the superspace calculus as follows \cite{kyy}:
\begin{eqnarray}
&& [\mathcal{V}]_D = 2 \int d^4\theta E \mathcal{V}, \\
&& [\mathcal{S}]_F = \int d^2\theta \mathcal{E}\mathcal{S} + \int d^2 \bar{\theta} \bar{\mathcal{E}} \bar{\mathcal{S}},
\end{eqnarray}
where $\mathcal{V}$ is a superconformal real multiplet with the Weyl/chirial weights (2,0) and $\mathcal{S}$ is a superconformal chiral multiplet with Weyl/chiral weights (3,3). To make the action invariant under the super-Weyl-K\"{a}hler transformations we should impose that the corresponding superconformal multiplets transform as 
\begin{eqnarray}
\mathcal{V} \rightarrow \mathcal{V} e^{-2\Sigma-2\bar{\Sigma}},\qquad  \mathcal{S} \rightarrow \mathcal{S} e^{-6\Sigma}.
\end{eqnarray}

Instead of considering the K\"{a}hler transformation of the K\"{a}hler potential a superconformal compensator is introduced 
in the superconformal formalism to eliminate the variation of the action transformed by a super-Weyl rescaling (also called Howe-Tucker transformation \cite{SUGRAprimer}). Thus, the compensator must transform as   
\begin{eqnarray}
S_0 \rightarrow S_0 e^{-2\Sigma}, \qquad \bar{S}_0 \rightarrow \bar{S}_0 e^{-2\bar{\Sigma}},
\end{eqnarray}
resulting in
\begin{eqnarray}
S_0\bar{S}_0 e^{-K/3} \rightarrow S_0\bar{S}_0 e^{-K/3} e^{-2\Sigma-2\bar{\Sigma}},
\end{eqnarray}
where $K$ is invariant under the superconformal symmetry (i.e. super-Weyl rescaling), so that the action can be invariant as desired. 

At this point, differently from the usual story of the superconformal symmetry, we require a ``K\"{a}hler transformation'' of the K\"{a}hler potential in order to construct a ``liberated'' supergravity that is invariant under the same Super-Weyl-K\"{a}hler transformations as an abelian gauge symmetry used in \cite{fkr}. Therefore, we assume that the superconformal compensators are inert under the super-Weyl-K\"{a}hler transformations 
\begin{eqnarray}
S_0 \rightarrow S_0, \qquad \bar{S}_0 \rightarrow \bar{S}_0,
\end{eqnarray}
while the K\"{a}hler potential {\it does} transform under the same transformations as above, namely as $K \rightarrow K + 6\Sigma + 6\Bar{\Sigma}$, so that
\begin{eqnarray}
&& \mathcal{Y} \rightarrow \mathcal{Y} e^{-2(\Sigma + \bar{\Sigma})},\\
&& w^2 \equiv \frac{\mathcal{W}^2(K)}{(e^{-K/3} S_0\bar{S}_0 )^2} \rightarrow w^2 e^{-2\Sigma+4\bar{\Sigma}},\\
&& \bar{T} \left(w^2\right) \rightarrow \bar{T}\left(w^2e^{-2\Sigma+4\bar{\Sigma}}\right)  e^{2\Sigma-4\bar{\Sigma}}=\bar{T}\left(w^2\right).
\end{eqnarray}

\section{List of Superconformal Multiplets}

In this section, we present all the superconformal multiplets of the liberated $\mathcal{N}=1$ supergravity following the notations and multiplication laws used in \cite{Linear,Superconformal_Freedman}.

\subsection{K\"{a}hler potential multiplet}

Let us consider $n$ physical chiral multiplets of matter $z^I \equiv \{ z^I , P_L\chi^I, F^I\}$ where $I = 1,2,3,\cdots,n$ and  their anti-chiral multiplets $\bar{z}^{\bar{I}} \equiv \{\bar{z}^{\bar{I}}, P_R\chi^{\bar{I}},\bar{F}^{\bar{I}}\}$.\footnote{The complex conjugates are $\bar{z}^{\bar{I}} \equiv (z^I)^{*}$, $\chi^{\bar{I}} \equiv (\chi^I)^C $, $\bar{\chi}^{\bar{I}} \equiv (\bar{\chi}^I)^C$, and $\bar{F}^{\bar{I}} \equiv (F^I)^{*}$ (The barred index is the complex conjugate index, so that the handedness of fermion becomes opposite, i.e. $(P_{L/R}\chi)^C=(P_{L/R})^C(\chi)^C=P_{R/L}(\chi)^C$.). The chiralities of fermion are specified as $\chi^I \equiv P_L \chi^I$ and  $\chi^{\bar{I}} \equiv P_R \chi^{\bar{I}}$. The Majorana conjugates are $\overline{(P_{L/R}\chi)} =\bar{\chi} P_{L/R}$ (The handedness is preserved.).} Then, according to the superconformal tensor calculus, the K\"{a}hler potential multiplet can be written as follows:

\begin{eqnarray}
K(z,\bar{z}) = \{C_K, \mathcal{Z}_K , \mathcal{H}_K, \mathcal{K}_K, \mathcal{B}_{\mu}^K, \Lambda_K, \mathcal{D}_K \}
\end{eqnarray}
where
\begin{eqnarray}
C_K &=& K(z,\bar{z}),\\
\mathcal{Z}_K  &=& i\sqrt{2}(-K_I \chi^I + K_{\bar{I}}\chi^{\bar{I}}),\\
\mathcal{H}_K &=& -2K_{I}F^I + K_{IJ}\bar{\chi}^I\chi^J,\\
\mathcal{K}_K &=& -2K_{\bar{I}}\bar{F}^{\bar{I}} + K_{\bar{I}\bar{J}}\bar{\chi}^{\bar{I}}\chi^{\bar{J}} = \mathcal{H}_K^{*},\\
\mathcal{B}_{\mu}^K &=& 
iK_I\mathcal{D}_{\mu} z^I -iK_{\bar{I}}\mathcal{D}_{\mu} \bar{z}^{\bar{I}} + i K_{I\bar{J}}\bar{\chi}^I \gamma_{\mu} \chi^{\bar{J}},\\
\Lambda_K &=& P_L\Lambda_K + P_R\Lambda_K 
\\
P_L \Lambda_K &=& -\sqrt{2}iK_{\bar{I}J}[(\cancel{\mathcal{D}}z^{J})\chi^{\bar{I}}-\bar{F}^{\bar{I}}\chi^J] -\frac{i}{\sqrt{2}}K_{\bar{I}\bar{J}K}\chi^{K}\bar{\chi}^{\bar{I}}\chi^J,\\
P_R \Lambda_K &=& \sqrt{2}iK_{I\bar{J}}[(\cancel{\mathcal{D}}\bar{z}^{\bar{J}})\chi^{I}-F^{I}\chi^{\bar{J}}] +\frac{i}{\sqrt{2}}K_{IJ\bar{K}}\chi^{\bar{K}}\bar{\chi}^{I}\chi^{\bar{J}},\\
\mathcal{D}_K &=& 2K_{I\bar{J}} 
\bigg( -\mathcal{D}_{\mu} z^I \mathcal{D}^{\mu} \bar{z}^{\bar{J}} -\frac{1}{2} \bar{\chi}^I P_L \cancel{\mathcal{D}}\chi^{\bar{J}} -\frac{1}{2} \bar{\chi}^{\bar{J}} P_R \cancel{\mathcal{D}} \chi^I + F^I\bar{F}^{\bar{J}}\bigg) \nonumber\\
&&+ K_{IJ\bar{K}} \bigg( -\bar{\chi}^I \chi^J \bar{F}^{\bar{K}} + \bar{\chi}^I (\cancel{\mathcal{D}}z^J)\chi^{\bar{K}}         \bigg) + K_{\bar{I}\bar{J}K} \bigg( -\bar{\chi}^{\bar{I}} \chi^{\bar{J}} F^{K} + \bar{\chi}^{\bar{I}} (\cancel{\mathcal{D}}\bar{z}^{\bar{J}})\chi^{K}\bigg) \nonumber\\
&& + \frac{1}{2} K_{IJ\bar{K}\bar{L}} (\bar{\chi}^IP_L\chi^J )(\bar{\chi}^{\bar{K}} P_R \chi^{\bar{L}}) .
\end{eqnarray}
Here the covariant derivatives\footnote{From Eq. (16.34) in Ref. \cite{Superconformal_Freedman} we find that for a general superconformal chiral multiplet $(z^I,P_L\chi^I,F^I)$ with Weyl/chiral weights $(w,c=w)$ and gauge symmetries with Killing vector fields $k_A^I$, the full superconformal covariant derivatives $\mathcal{D}_{a}$ are given by
\begin{eqnarray}
 \mathcal{D}_{a} z^I &=& e^{\mu}_a\Big[(\partial_{\mu} -wb_{\mu} -wiA_{\mu})z^I -\frac{1}{\sqrt{2}}\bar{\psi}_{\mu} \chi^I -A^A_{\mu} k_A^I\Big],\nonumber\\
 \mathcal{D}_{a} P_L\chi^I &=& e^{\mu}_aP_L\bigg[ \left(\partial_{\mu} +\frac{1}{4}\omega_{\mu}^{ab}\gamma_{ab}-(w+1/2)b_{\mu} + (w-3/2)iA_{\mu}\right)\chi^I -\frac{1}{\sqrt{2}}(\cancel{\mathcal{D}}z^I + F^I)\psi_{\mu} \nonumber\\
&&-\sqrt{2}wz^I  \phi_{\mu} -A^A_{\mu} \chi^J\partial_J k_A^I \bigg]. \nonumber
\end{eqnarray}} 
of chiral multiplets of matter with the weights $(0,0)$ are given by
\begin{eqnarray}
 \mathcal{D}_{a} z^I &=& e^{\mu}_a\Big[\partial_{\mu}z^I -\frac{1}{\sqrt{2}}\bar{\psi}_{\mu} \chi^I\Big],\\
 \mathcal{D}_{a} P_L\chi^I &=& e^{\mu}_aP_L\bigg[ \left(\partial_{\mu} +\frac{1}{4}\omega_{\mu}^{ab}\gamma_{ab}-\frac{1}{2}b_{\mu} -\frac{3}{2}iA_{\mu}\right)\chi^I -\frac{1}{\sqrt{2}}(\cancel{\mathcal{D}}z^I + F^I)\psi_{\mu} \bigg].\nonumber\\{}
\end{eqnarray}

Note that the only bosonic contribution to  $\mathcal{D}_K$ is given by
\begin{eqnarray}
\mathcal{D}_K|_{\textrm{boson}} = 2K_{I\bar{J}} \left( -\partial_{\mu} z^I \partial^{\mu} \bar{z}^{\bar{J}} +F^I\bar{F}^{\bar{J}}\right) \equiv \tilde{\mathcal{F}}
\end{eqnarray}
and we especially denote this by $\tilde{\mathcal{F}}$. We also note that the $\tilde{\mathcal{F}}$ is positive definite up to terms containing spatial gradients; so,  for small spatial gradients,
\begin{eqnarray}
\mathcal{D}_K|_{\textrm{boson}}\equiv \tilde{\mathcal{F}} \sim 2K_{I\bar{J}} \left( \dot{z}^I \dot{\bar{z}}^{\bar{J}} +F^I\bar{F}^{\bar{J}}\right) >0.
\end{eqnarray}

\subsection{Compensator multiplet}

Chiral compensators $S_0,\bar{S}_0$ with the Weyl/chiral weights (1,1) are defined as follows. The chiral supermultiplets $S_0 = \{s_0, P_L\chi^0, F_0\}$ and $\bar{S}_0 = \{s_0^{*}, P_R\chi^0, F_0^{*}\}$ can be embedded into the superconformal formalism as
    \begin{eqnarray}
    && S_0 \equiv \{ s_0, -i\sqrt{2}P_L\chi^0, -2F_0 , 0 , i\mathcal{D}_{\mu} s_0, 0,0 \},\\
    && \bar{S}_0 \equiv \{ s_0^{*}, i\sqrt{2}P_R\chi^0 , 0 ,-2F_0^{*}, -i\mathcal{D}_{\mu} s_0^{*}, 0,0 \}.
    \end{eqnarray}
   Then, the composite real compensator $S_0\bar{S}_0$ is
    \begin{eqnarray}
    S_0\bar{S}_0 = \{ \mathcal{C}_0, \mathcal{Z}_0, \mathcal{H}_0, \mathcal{K}_0, \mathcal{B}_{\mu}^0, \Lambda_0, \mathcal{D}_0 \},
    \end{eqnarray}
    where \footnote{ $P_{L,R}\gamma_{\textrm{odd indices}} =\gamma_{\textrm{odd indices}}P_{R,L}$}
    \begin{eqnarray}
    \mathcal{C}_0 &=& s_0s_0^{*}|_{0f},\\
    \mathcal{Z}_0 &=& i\sqrt{2}(-s_0^{*} P_L\chi^0 + s_0 P_R \chi^0) |_{ 1f},\\
    \mathcal{H}_0 &=& -2 s_0^{*} F^0 |_{0f},\\
    \mathcal{K}_0 &=& -2s_0 F^{*}_0 |_{0f},\\
    \mathcal{B}_{\mu}^0 &=& is_0^{*} \mathcal{D}_{\mu} s_0 - is_0 \mathcal{D}_{\mu} s_0^{*} + i \bar{\chi}_0\gamma_{\mu} P_R \chi^0 =  is_0^{*}\partial_{\mu} s_0 - is_0 \partial_{\mu} s_0^{*}|_{0f}+\cdots,\\
    P_L \Lambda_0 &=& -\sqrt{2}i [(\cancel{\mathcal{D}}s_0^{*})P_R\chi^0 - F_0^{*} P_L\chi^0]=-\sqrt{2}i [(\cancel{\partial}s_0^{*})P_R\chi^0 - F_0^{*} P_L\chi^0]|_{1f}+\cdots,\\
    P_R \Lambda_0 &=& \sqrt{2}i[(\cancel{\mathcal{D}}s_0)P_L\chi^0 - F_0 P_R\chi^0]=\sqrt{2}i[(\cancel{\partial}s_0)P_L\chi^0 - F_0 P_R\chi^0]|_{1f}+\cdots,\\
    \mathcal{D}_0 &=& 2\left( -\mathcal{D}_{\mu}s_0 \mathcal{D}^{\mu}s_0^{*} - \frac{1}{2}\bar{\chi}_0 P_L \cancel{\mathcal{D}}\chi^0-\frac{1}{2}\bar{\chi}_0 P_R \cancel{\mathcal{D}}\chi^0  + F_0F_0^{*} \right) = 2( -\partial_{\mu}s_0 \partial^{\mu}s_0^{*} + F_0F_0^{*} )|_{0f}+\cdots,\nonumber\\{}
    \end{eqnarray}
where the covariant derivatives of the conformal compensator $(s_0,P_L\chi^0,F^0)$ with Weyl/chiral weights $(1,1)$ are given by
\begin{eqnarray}
  \mathcal{D}_{a} s_0 &=&  e^{\mu}_a (\partial_{\mu}s_0 -\frac{1}{\sqrt{2}}\bar{\psi}_{\mu} \chi^0),\\
 \mathcal{D}_{a} P_L\chi^0 &=& e^{\mu}_aP_L\bigg[ \left(\partial_{\mu} +\frac{1}{4}\omega_{\mu}^{ab}\gamma_{ab}-\frac{3}{2}b_{\mu} -\frac{1}{2}iA_{\mu}\right)\chi^0 -\frac{1}{\sqrt{2}}(\cancel{\mathcal{D}}s_0 + F^0)\psi_{\mu} -\sqrt{2}s_0  \phi_{\mu}  \bigg].~\qquad
\end{eqnarray}

\subsection{The composite real compensator $\Upsilon \equiv S_0\bar{S}_0 e^{-K/3}$}
\begin{eqnarray}
\Upsilon = \{ \mathcal{C}_{\Upsilon}, \mathcal{Z}_{\Upsilon}, \mathcal{H}_{\Upsilon}, \mathcal{K}_{\Upsilon}, \mathcal{B}_{\mu}^{\Upsilon}, \Lambda_{\Upsilon}, \mathcal{D}_{\Upsilon} \}
\end{eqnarray}
has components 
\begin{eqnarray}
 \mathcal{C}_{\Upsilon} &=& \Upsilon = s_0\bar{s}_0 e^{-K/3}|_{0f},\\
 \mathcal{Z}_{\Upsilon} &=& e^{-K/3} \mathcal{Z}_0  +\frac{ i\sqrt{2}}{3}\Upsilon(K_I \chi^I - K_{\bar{I}}\chi^{\bar{I}})\nonumber\\
 &=&  i\sqrt{2}\Upsilon(-\frac{1}{s_0} P_L\chi^0 + \frac{1}{s_0^{*}} P_R \chi^0 + \frac{1}{3}K_I \chi^I - \frac{1}{3}K_{\bar{I}}\chi^{\bar{I}})|_{1f},\\
 \mathcal{H}_{\Upsilon} &=& e^{-K/3} \mathcal{H}_0+ \Upsilon \Big(\frac{2}{3}K_{I}F^I + (\frac{1}{9}K_{I}K_{J}-\frac{1}{3}K_{IJ})\bar{\chi}^I\chi^J\Big) -\frac{1}{2}[\bar{\mathcal{Z}_0}P_L\mathcal{Z}_{\Upsilon}+\bar{\mathcal{Z}}_{\Upsilon}P_L\mathcal{Z}_0]\nonumber\\
 &=&-2\Upsilon ( \frac{F^0}{s_0} -\frac{1}{3} K_IF^I) |_{0f}+\cdots,\quad \\
 \mathcal{K}_{\Upsilon} &=& e^{-K/3} \mathcal{H}_0^{*}+ \Upsilon \Big(\frac{2}{3}K_{\bar{I}}\bar{F}^{\bar{I}} + (\frac{1}{9}K_{\bar{I}}K_{\bar{J}}-\frac{1}{3}K_{\bar{I}\bar{J}})\bar{\chi}^{\bar{I}}\chi^{\bar{J}}\Big) -\frac{1}{2}[\bar{\mathcal{Z}_0}P_R\mathcal{Z}_{\Upsilon}+\bar{\mathcal{Z}}_{\Upsilon}P_R\mathcal{Z}_0],\nonumber\\
 &=&-2\Upsilon ( \frac{F^{0*}}{s_0^{*}} -\frac{1}{3} K_{\bar{I}}\bar{F}^{\bar{I}}) |_{0f}+\cdots, \\
 \mathcal{B}_{\mu}^{\Upsilon} &=& e^{-K/3} \mathcal{B}_{\mu}^0 -i\Upsilon \Big(\frac{1}{3}K_I\mathcal{D}_{\mu} z^I -\frac{1}{3}K_{\bar{I}}\mathcal{D}_{\mu} \bar{z}^{\bar{I}} -  (\frac{1}{9}K_{I}K_{\bar{J}}-\frac{1}{3}K_{I\bar{J}})\bar{\chi}^I \gamma_{\mu} \chi^{\bar{J}}\Big)\nonumber\\&&
 + \frac{1}{4}i [\bar{\mathcal{Z}_0}\gamma_*\gamma_{\mu}\mathcal{Z}_{\Upsilon}+\bar{\mathcal{Z}}_{\Upsilon}\gamma_*\gamma_{\mu}\mathcal{Z}_0],\nonumber\\
 &=& i\Upsilon(\frac{1}{s_0}\partial_{\mu} s_0 - \frac{1}{s_0^{*}} \partial_{\mu} s_0^{*} - \frac{1}{3}K_{I}\partial_{\mu}z^I+\frac{1}{3}K_{\bar{I}}\partial_{\mu}\bar{z}^{\bar{I}})|_{0f} + \cdots.
 \end{eqnarray}
 
 \begin{eqnarray}
 \Lambda_{\Upsilon} &=& e^{-K/3}\Lambda_0-i\sqrt{2}\Upsilon\bigg( (\frac{1}{9}K_{\bar{I}}K_{J}-\frac{1}{3}K_{\bar{I}J})[(\cancel{\mathcal{D}}z^{J})\chi^{\bar{I}}-\bar{F}^{\bar{I}}\chi^J] \nonumber\\
&&+\frac{1}{2} (-\frac{1}{27}K_{\bar{I}}K_{\bar{J}}K_K+\frac{1}{9}(K_{\bar{I}\bar{J}}K_K+K_{\bar{I}K}K_{\bar{J}}+K_{\bar{I}}K_{\bar{J}K})-\frac{1}{3}K_{\bar{I}\bar{J}K})\chi^{K}\bar{\chi}^{\bar{I}}\chi^J
\nonumber\\
&&- (\frac{1}{9}K_{I}K_{\bar{J}}-\frac{1}{3}K_{I\bar{J}})[(\cancel{\mathcal{D}}\bar{z}^{\bar{J}})\chi^{I}-F^{I}\chi^{\bar{J}}] \nonumber\\
&&-\frac{1}{2} (-\frac{1}{27}K_{I}K_{J}K_{\bar{K}}+\frac{1}{9}(K_{IJ}K_{\bar{K}}+K_{I\bar{K}}K_{J}+K_{I}K_{J\bar{K}})-\frac{1}{3}K_{IJ\bar{K}})\chi^{\bar{K}}\bar{\chi}^{I}\chi^{\bar{J}} \bigg) \nonumber\\
&&+ \frac{1}{2} \Big( [i\gamma_* \cancel{\mathcal{B}_0}+\textrm{Re}\mathcal{H}_0-i\gamma_* \textrm{Im}\mathcal{H}_0-\cancel{\mathcal{D}}(s_0s_0^{*})]\mathcal{Z}_{\Upsilon}\nonumber\\
 &&+[i\gamma_* \cancel{\mathcal{B}_{\Upsilon}}+\textrm{Re}\mathcal{H}_{\Upsilon}-i\gamma_* \textrm{Im}\mathcal{H}_{\Upsilon}-\cancel{\mathcal{D}}(\Upsilon)]\mathcal{Z}_{0}\Big),\nonumber\\
 &=& -\sqrt{2}ie^{-K/3} [(\cancel{\partial}s_0^{*})P_R\chi^0 - F_0^{*} P_L\chi^0]+\sqrt{2}ie^{-K/3}[(\cancel{\partial}s_0)P_L\chi^0 - F_0 P_R\chi^0]\nonumber\\
 &&-i\sqrt{2}\Upsilon\bigg( (\frac{1}{9}K_{\bar{I}}K_{J}-\frac{1}{3}K_{\bar{I}J})[(\cancel{\partial}z^{J})\chi^{\bar{I}}-\bar{F}^{\bar{I}}\chi^J] - (\frac{1}{9}K_{I}K_{\bar{J}}-\frac{1}{3}K_{I\bar{J}})[(\cancel{\partial}\bar{z}^{\bar{J}})\chi^{I}-F^{I}\chi^{\bar{J}}] \bigg) \nonumber\\
&& + \frac{1}{2}\bigg[ \Big\{ i\gamma_*(is_0^{*}\cancel{\partial}s_0-is_0\cancel{\partial}s_0^{*}) + \textrm{Re}[-2s_0^{*}F^0] -i\gamma_* \textrm{Im}[-2s_0^{*}F^0] -\cancel{\partial}(s_0s_0^{*}) \Big\}\nonumber\\
&&\qquad \times\Big\{ i\sqrt{2}\Upsilon(-s_0^{-1}P_L\chi^0 + s_0^{-1*}P_R\chi^0 + \frac{1}{3}K_{I}\chi^{I}-\frac{1}{3}K_{\bar{I}}\chi^{\bar{I}}) \Big\}  \nonumber\\
&&\qquad  + \Big\{ i\gamma_*\big(i\Upsilon(s_0^{-1}\cancel{\partial}s_0 - s_0^{-1*} \cancel{\partial}s_0^{*} -\frac{1}{3}K_{I}\cancel{\partial}z^{I} + \frac{1}{3}K_{\bar{I}}\cancel{\partial} \bar{z}^{\bar{I}} ) \big)+\textrm{Re}[-2\Upsilon(F^0s_0^{-1}-\frac{1}{3}K_{I}F^{I})] \nonumber\\
&&\qquad\qquad  -i\gamma_* \big(-2\Upsilon (F^0 s_0^{-1}) -\frac{1}{3} K_{I}F^{I}\big) - \cancel{\partial}\Upsilon \Big\} \Big\{ i\sqrt{2}(-s_0^{*} P_L\chi^0 + s_0 P_R \chi^0) \Big\} \bigg] \bigg|_{1f} +\cdots.\nonumber\\{}
 \end{eqnarray}
 
 \begin{eqnarray}
 \mathcal{D}_{\Upsilon} &=& e^{-K/3} \mathcal{D}_0+\Upsilon \bigg[2(\frac{1}{9}K_{I}K_{\bar{J}}-\frac{1}{3}K_{I\bar{J}}) 
\bigg( -\mathcal{D}_{\mu} z^I \mathcal{D}^{\mu} \bar{z}^{\bar{J}} -\frac{1}{2} \bar{\chi}^I P_L \cancel{\mathcal{D}}\chi^{\bar{J}} \nonumber\\&&
-\frac{1}{2} \bar{\chi}^{\bar{J}} P_R \cancel{\mathcal{D}} \chi^I + F^I\bar{F}^{\bar{J}}\bigg) +  (-\frac{1}{27}K_{I}K_{J}K_{\bar{K}}+\frac{1}{9}(K_{IJ}K_{\bar{K}}+K_{I\bar{K}}K_{J}+K_{I}K_{J\bar{K}})\nonumber\\&&-\frac{1}{3}K_{IJ\bar{K}}) \bigg( -\bar{\chi}^I \chi^J \bar{F}^{\bar{K}} + \bar{\chi}^I (\cancel{\mathcal{D}}z^J)\chi^{\bar{K}}         \bigg)
\nonumber\\
&&+  (-\frac{1}{27}K_{\bar{I}}K_{\bar{J}}K_K+\frac{1}{9}(K_{\bar{I}\bar{J}}K_K+K_{\bar{I}K}K_{\bar{J}}+K_{\bar{I}}K_{\bar{J}K})-\frac{1}{3}K_{\bar{I}\bar{J}K}) \nonumber\\&&\times \bigg( -\bar{\chi}^{\bar{I}} \chi^{\bar{J}} F^{K} + \bar{\chi}^{\bar{I}} (\cancel{\mathcal{D}}\bar{z}^{\bar{J}})\chi^{K}\bigg)  + \frac{1}{2}  \bigg\{-\frac{1}{3}K_{IJ\bar{K}\bar{L}}+\frac{1}{81}K_{I}K_{J}K_{\bar{K}}K_{\bar{L}}
\nonumber\\
&&-\frac{1}{27}[K_{IJ}K_{\bar{K}}K_{\bar{L}}+K_{I\bar{K}}K_{J}K_{\bar{L}}+K_{I}K_{J\bar{K}}K_{\bar{L}}+K_{I\bar{L}}K_{J}K_{\bar{K}}+K_{I}K_{J\bar{L}}K_{\bar{K}}+K_{I}K_{J}K_{\bar{K}\bar{L}}] 
\nonumber\\
&&+\frac{1}{9}[K_{IJ\bar{K}}+K_{IJ\bar{L}}K_{\bar{K}}+K_{IJ}K_{\bar{K}\bar{L}}+K_{I\bar{K}\bar{L}}K_{J}+K_{I\bar{K}}K_{J\bar{L}}+K_{I\bar{L}}K_{J\bar{K}}+K_{I}K_{J\bar{K}\bar{L}}] \bigg\} \nonumber\\
&&\times (\bar{\chi}^IP_L\chi^J )(\bar{\chi}^{\bar{K}} P_R \chi^{\bar{L}})\bigg] \nonumber\\
&&+ \frac{1}{2}(\mathcal{H}_0\mathcal{H}^{*}_{\Upsilon}+\mathcal{H}_0^{*}\mathcal{H}_{\Upsilon}-2\mathcal{B}_{\mu}^0\mathcal{B}^{\mu}_{\Upsilon} - 2\mathcal{D}(s_0s_0^{*}) \cdot \mathcal{D} e^{-K/3} - 2\bar{\Lambda}_0\mathcal{Z}_{\Upsilon} -2\bar{\Lambda}_{\Upsilon}\mathcal{Z}_{0}- \bar{\mathcal{Z}}_0 \cancel{\mathcal{D}} \mathcal{Z}_{\Upsilon}-\bar{\mathcal{Z}}_{\Upsilon} \cancel{\mathcal{D}} \mathcal{Z}_{0}).\nonumber\\
&=& \bigg[ e^{-K/3}2( -\partial_{\mu}s_0 \partial^{\mu}s_0^{*} + F_0F_0^{*} ) + 2\Upsilon (\frac{1}{9}K_{I}K_{\bar{J}}-\frac{1}{3}K_{I\bar{J}})(-\partial_{\mu}z^I\partial^{\mu}\bar{z}^{\bar{J}}+F^I\bar{F}^{\bar{J}}) \nonumber\\
&& + \frac{1}{2}(-2s_0^{*}F^0)\Big(-2\Upsilon ( \frac{F^{0*}}{s_0^{*}} -\frac{1}{3} K_{\bar{I}}\bar{F}^{\bar{I}})\Big)+ \frac{1}{2}(-2s_0F^{0*})\Big(-2\Upsilon ( \frac{F^0}{s_0} -\frac{1}{3} K_{I}F^{I})\Big)\nonumber\\
&&-(is_0^{*}\partial_{\mu} s_0 - is_0 \partial_{\mu} s_0^{*})\Big(i\Upsilon(\frac{1}{s_0}\partial^{\mu} s_0 - \frac{1}{s_0^{*}} \partial^{\mu} s_0^{*} - \frac{1}{3}K_{I}\partial^{\mu}z^I+\frac{1}{3}K_{\bar{I}}\partial^{\mu}\bar{z}^{\bar{I}})\Big) \nonumber\\
&&- (s_0^{*}\partial s_0+s_0\partial s_0^{*})\Big( -\frac{1}{3}e^{-K/3}(K_I\partial z^I + K_{\bar{I}}\partial\bar{z}^{\bar{I}})\Big)  \bigg]_{0f}+\cdots.
\end{eqnarray}

The conventional superconformal gauge is defined by the choice
\begin{eqnarray}
&&\mathcal{C}_{\Upsilon} = \Upsilon = 1,\\ &&\mathcal{Z}_{\Upsilon} = 0 \implies P_L\chi^0 - \frac{1}{3}e^{K/6} K_I P_L\chi^I =0,\\
&& b_{\mu} = 0.
\end{eqnarray}

\subsection{\texorpdfstring{$\mathcal{W}^2(K) \equiv \mathcal{W}_{\alpha}(K)\mathcal{W}^{\alpha}(K)$}{} composite chiral multiplet: (Weyl/Chiral) weights \texorpdfstring{$=(3,3)$}{}}

Let us define 
\begin{eqnarray}
&& \mathcal{W}^2(K) \equiv \{\mathcal{C}_{W},\mathcal{Z}_{W},\mathcal{H}_{W},\mathcal{K}_{W},\mathcal{B}^{W}_{\mu},\Lambda_{W},\mathcal{D}_{W} \},\\
&& \bar{\mathcal{W}}^{2}(K) \equiv \{\mathcal{C}_{W}^{*},\mathcal{Z}_{W}^C,\mathcal{K}_{W}^{*},\mathcal{H}_{W}^{*},(\mathcal{B}^{W}_{\mu})^{*},\Lambda_{W}^C,\mathcal{D}_{W}^{*} \}
\end{eqnarray}
where
\begin{eqnarray}
\mathcal{C}_{W} &=& \Bar{\Lambda}_K P_L \Lambda_K \nonumber\\
&=&
-2\Big[ K_{\bar{I}'J'}\Bar{\chi}^{\bar{I}'}\cancel{\partial}z^{J'} K_{\bar{I}J}\cancel{\partial}z^J\chi^{\bar{I}} - K_{\bar{I}'J'}\Bar{\chi}^{\bar{I}'}\cancel{\partial}z^{J'} K_{\bar{I}J}\bar{F}^{\bar{I}}\chi^J\nonumber\\
&&-K_{\bar{I}'J'}\bar{F}^{\bar{I}'}\Bar{\chi}^{J'}K_{\bar{I}J}\cancel{\partial}z^J\chi^{\bar{I}}+ K_{\bar{I}'J'}\bar{F}^{\bar{I}'}\Bar{\chi}^{J'}K_{\bar{I}J}\bar{F}^{\bar{I}}\chi^J \Big]\Big|_{2f}
 + \cdots,\\
\mathcal{Z}_{W} &=&  -i2 P_L (-\frac{1}{2}\gamma\cdot \hat{F}_K + i\mathcal{D}_K)\Lambda_K = -2\sqrt{2}i\tilde{\mathcal{F}}K_{\bar{I}J}P_L[\cancel{\partial}z^J\chi^{\bar{I}}-\bar{F}^{\bar{I}}\chi^J]\Big|_{1f}+\cdots,\\
\mathcal{H}_{W} &=& -2(2\Bar{\Lambda}_KP_L \cancel{\mathcal{D}}\Lambda_K + \hat{F}_K^-\cdot \hat{F}_K^- - \mathcal{D}^2_K) = 2\tilde{\mathcal{F}}^2|_{0f} +\cdots \equiv -2F^W,\\
\mathcal{K}_{W} &=& 0,\\
\mathcal{B}^{W}_{\mu} &=& i\mathcal{D}_{\mu}(\Bar{\Lambda}_K P_L \Lambda_K) = i\partial_{\mu}(\Bar{\Lambda}_K P_L \Lambda_K)|_{2f}+\cdots,\\
\Lambda_{W} &=& 0,\\
\mathcal{D}_{W} &=& 0. \label{Multiplet_W1}
\end{eqnarray}
Note that $\mathcal{C}_{W} = \mathcal{C}_{W}|_{2f}+\cdots, \mathcal{Z}_{W} = \mathcal{Z}_{W}|_{1f} +\cdots, \mathcal{H}_{W} = \mathcal{H}_{W}|_{0f} + \cdots, \mathcal{K}_{W} = 0, \mathcal{B}^{W}_{\mu} = \mathcal{B}^{W}_{\mu}|_{2f} + \cdots, \Lambda_{W}=\mathcal{D}_{W} =0$.

and
\begin{eqnarray}
\mathcal{C}_{\bar{W}}&=& \mathcal{C}_{W}^{*} = \Bar{\Lambda}_K P_R \Lambda_K,\\
\mathcal{Z}_{\bar{W}}&=&\mathcal{Z}_{W}^C = i2 P_R (-\frac{1}{2}\gamma\cdot \hat{F}_K -i\mathcal{D}_K)\Lambda_K,\\
\mathcal{H}_{\bar{W}}&=&\mathcal{K}_{W}^{*} = 0,\\
\mathcal{K}_{\bar{W}}&=&\mathcal{H}_{W}^{*} = -2(2\Bar{\Lambda}_KP_R \cancel{\mathcal{D}}\Lambda_K + \hat{F}_K^+\cdot \hat{F}_K^+ - \mathcal{D}^2_K) \equiv -2\bar{F}^{\bar{W}},\\
\mathcal{B}^{\bar{Q}}_{\mu}&=&(\mathcal{B}^{W}_{\mu})^{*} = -i\mathcal{D}_{\mu}(\Bar{\Lambda}_K P_R \Lambda_K),\\
\Lambda_{\bar{W}}&=&\Lambda_{W}^C = 0,\\
\mathcal{D}_{\bar{W}}&=&\mathcal{D}_{W}^{*} = 0.\label{Multiplet_W2}
\end{eqnarray}
We also defined:
\begin{eqnarray}
&& \hat{F}_{ab}^K \equiv e_a^{\mu}e_b^{\nu} (2\partial_{[\mu }\mathcal{B}^K_{\nu]} + \bar{\psi}_{[\mu}\gamma_{\nu]}\Lambda_K),\\
&& \Tilde{\hat{F}}_{ab} \equiv -i\frac{1}{2} \varepsilon_{abcd}\hat{F}^{cd}, \qquad \hat{F}_{ab}^{\pm} \equiv \frac{1}{2} (\hat{F}_{ab} \pm \tilde{\hat{F}}_{ab}),\qquad   (\hat{F}_{ab}^{\pm})^{*}= \hat{F}_{ab}^{\mp}\\
&& \cancel{\mathcal{D}}\Lambda_K \equiv \gamma \cdot \mathcal{D} \Lambda_K,\\
&& \mathcal{D}_{\mu} \Lambda_K \equiv \Big(\partial_{\mu} -\frac{3}{2}b_{\mu} + \frac{1}{4}\omega_{\mu}^{ab} \gamma_{ab} -\frac{3}{2}i\gamma_* \mathcal{A}_{\mu} \Big) \Lambda_K - \Big(\frac{1}{4}\gamma^{ab} \hat{F}_{ab}^K + \frac{1}{2}i\gamma_* \mathcal{D}_K \Big)\psi_{\mu}.
\end{eqnarray}

Next, we shall consider a gauge fixing that is equivalent to the gauge condition given by $\eta =0$ in Ref. \cite{fkr}. In fact, this can be obtained by imposing $\Lambda_K =0$. We will call this gauge the ``Liberated SUGRA gauge''. Then, the only non-vanishing superconformal components of the multiplets $\mathcal{W}^2(K),\bar{\mathcal{W}}^2(K)$ in this gauge are given by  $\mathcal{H}_{W} = -2(\hat{F}_K^-\cdot\hat{F}_K^- -\mathcal{D}_K^2)$ and $\mathcal{K}_{\bar{W}} = -2(\hat{F}_K^+\cdot\hat{F}_K^+ -\mathcal{D}_K^2) = \mathcal{H}_{W}^{*}$.

Another representation of the chiral multiplet is
\begin{eqnarray}
\mathcal{W}^2(K) &=& \left(\mathcal{C}_{W}, \frac{i}{\sqrt{2}}\mathcal{Z}_{W},-\frac{1}{2}\mathcal{H}_{W}\right) \nonumber\\
&=& \left( \Bar{\Lambda}_K P_L \Lambda_K,~ \sqrt{2} P_L (-\gamma\cdot \hat{F}_K + 2i\mathcal{D}_K)\Lambda_K,~ 2\Bar{\Lambda}_KP_L \cancel{\mathcal{D}}\Lambda_K + \hat{F}_K^-\cdot \hat{F}_K^- - \mathcal{D}^2_K\right) \nonumber\\
&\equiv& \left( X^W, P_L\chi^W, F^W \right),\\
\bar{\mathcal{W}}^2(K) &=& \left(\mathcal{C}_{W}^{*}, -\frac{i}{\sqrt{2}}\mathcal{Z}_{W}^C,-\frac{1}{2}\mathcal{H}_{W}^{*}\right) \nonumber\\
&=& \left( \Bar{\Lambda}_K P_R \Lambda_K,~ \sqrt{2} P_R (-\gamma\cdot \hat{F}_K - 2i\mathcal{D}_K^{*})\Lambda_K^C,~ 2\Bar{\Lambda}_KP_R \cancel{\mathcal{D}}\Lambda_K + \hat{F}_K^+ \cdot \hat{F}_K^+ - (\mathcal{D}^{*}_K)^2\right) \nonumber\\
&\equiv& \left( \bar{X}^{\bar{W}}, P_R\chi^{\bar{W}}, \bar{F}^{\bar{W}} \right)
\end{eqnarray}

We will also need the following definitions:
\begin{eqnarray}
X^W = \bar{\Lambda}_K P_L \Lambda_K \equiv W &=&  -2K_{\bar{I}J}[\bar{\chi}^{J}(\overline{\cancel{\mathcal{D}}z^{\bar{I}}})-\bar{F}^{\bar{I}}\bar{\chi}^{J}]K_{{\bar{I}}'J'}[(\cancel{\mathcal{D}}z^{{\bar{I}}'})\chi^{J'}-F^{J'}\chi^{{\bar{I}}'}] \nonumber\\
&& - K_{{\bar{I}}J}[\bar{\chi}^{J}(\overline{\cancel{\mathcal{D}}z^{\bar{I}}})-\bar{F}^{\bar{I}}\bar{\chi}^{J}] K_{\bar{I}'\bar{J}'K'}[\chi^{K'}\bar{\chi}^{\bar{I}'}\chi^{\bar{J}'}] \nonumber\\
&& - K_{\bar{I}\bar{J}K}[\bar{\chi}^{\bar{J}}\chi^{\bar{I}}\bar{\chi}^{K}]K_{{\bar{I}}'J'}[(\cancel{\mathcal{D}}z^{{\bar{I}}'})\chi^{J'}-F^{J'}\chi^{{\bar{I}}'}] \nonumber\\
&& - \frac{1}{2}K_{\bar{I}\bar{J}K}[\bar{\chi}^{\bar{J}}\chi^{\bar{I}}\bar{\chi}^{K}]K_{\bar{I}'\bar{J}'K'}[\chi^{K'}\bar{\chi}^{\bar{I}'}\chi^{\bar{J}'}],\\
\bar{X}^{\bar{W}} =  \bar{\Lambda}_K P_R \Lambda_K \equiv \bar{W} &=& (\bar{\Lambda}_K P_L \Lambda_K)^C,
\end{eqnarray}
and
\begin{eqnarray}
\mathcal{D}_KP_L\Lambda_K &=& \bigg[2K_{I\bar{J}} 
\bigg( -\mathcal{D}_{\mu} z^I \mathcal{D}^{\mu} \bar{z}^{\bar{J}} -\frac{1}{2} \bar{\chi}^I P_L \cancel{\mathcal{D}}\chi^{\bar{J}} -\frac{1}{2} \bar{\chi}^{\bar{J}} P_R \cancel{\mathcal{D}} \chi^I + F^I\bar{F}^{\bar{J}}\bigg) \nonumber\\
&&+ K_{IJ\bar{K}} \bigg( -\bar{\chi}^I \chi^J \bar{F}^{\bar{K}} + \bar{\chi}^I (\cancel{\mathcal{D}}z^J)\chi^{\bar{K}}         \bigg) + K_{\bar{I}\bar{J}K} \bigg( -\bar{\chi}^{\bar{I}} \chi^{\bar{J}} F^{K} + \bar{\chi}^{\bar{I}} (\cancel{\mathcal{D}}\bar{z}^{\bar{J}})\chi^{K}\bigg) \nonumber\\
&& + \frac{1}{2} K_{IJ\bar{K}\bar{L}} (\bar{\chi}^IP_L\chi^J )(\bar{\chi}^{\bar{K}} P_R \chi^{\bar{L}})\bigg]
\nonumber\\&&\times
\bigg[-\sqrt{2}iK_{\bar{I}'J'}[(\cancel{\mathcal{D}}z^{J'})\chi^{\bar{I}'}-\bar{F}^{\bar{I}'}\chi^{J'}] -\frac{i}{\sqrt{2}}K_{\bar{I}'\bar{J}'K'}\chi^{K'}\bar{\chi}^{\bar{I}'}\chi^{J'}\bigg].
\end{eqnarray}
\begin{eqnarray}
\mathcal{D}_K^2 &=& \bigg[2K_{I\bar{J}} 
\bigg( -\mathcal{D}_{\mu} z^I \mathcal{D}^{\mu} \bar{z}^{\bar{J}} -\frac{1}{2} \bar{\chi}^I P_L \cancel{\mathcal{D}}\chi^{\bar{J}} -\frac{1}{2} \bar{\chi}^{\bar{J}} P_R \cancel{\mathcal{D}} \chi^I + F^I\bar{F}^{\bar{J}}\bigg) \nonumber\\
&&+ K_{IJ\bar{K}} \bigg( -\bar{\chi}^I \chi^J \bar{F}^{\bar{K}} + \bar{\chi}^I (\cancel{\mathcal{D}}z^J)\chi^{\bar{K}}         \bigg) + K_{\bar{I}\bar{J}K} \bigg( -\bar{\chi}^{\bar{I}} \chi^{\bar{J}} F^{K} + \bar{\chi}^{\bar{I}} (\cancel{\mathcal{D}}\bar{z}^{\bar{J}})\chi^{K}\bigg) \nonumber\\
&& + \frac{1}{2} K_{IJ\bar{K}\bar{L}} (\bar{\chi}^IP_L\chi^J )(\bar{\chi}^{\bar{K}} P_R \chi^{\bar{L}})\bigg]^2.
\end{eqnarray}

\paragraph{$w^2,\bar{w}^2$ Composite Complex Multiplets: (Weyl/Chiral) weights $=(-1,\pm 3)$}

\begin{eqnarray}
&&  w^2 \equiv \dfrac{\mathcal{W}^2(K)}{\Upsilon^2} 
= \{\mathcal{C}_w,\mathcal{Z}_w,\mathcal{H}_w,\mathcal{K}_w,\mathcal{B}^w_{\mu},\Lambda_w,\mathcal{D}_w\},\\
&&   \bar{w}^2 \equiv \dfrac{\bar{\mathcal{W}}^2(K)}{\Upsilon^2} 
= \{\mathcal{C}_{\bar{w}},\mathcal{Z}_{\bar{w}},\mathcal{H}_{\bar{w}},\mathcal{K}_{\bar{w}},\mathcal{B}^{\bar{w}}_{\mu},\Lambda_{\bar{w}},\mathcal{D}_{\bar{w}}\}
\end{eqnarray}
where
\begin{eqnarray}
\mathcal{C}_{w} &=& \frac{\mathcal{C}_{W}}{\Upsilon^2},\\
\mathcal{Z}_{w} &=& \frac{1}{\Upsilon^2}\mathcal{Z}_{W} -2\frac{\mathcal{C}_{W}}{\Upsilon^3}\mathcal{Z}_{\Upsilon},\\
\mathcal{H}_{w} &=&  \frac{1}{\Upsilon^2}\mathcal{H}_{W} -2\frac{\mathcal{C}_{W}}{\Upsilon^3}\mathcal{H}_{\Upsilon} -\frac{1}{2}\bigg[ -4\frac{1}{\Upsilon^3} (\bar{\mathcal{Z}}_{W}P_L\mathcal{Z}_{\Upsilon}+\bar{\mathcal{Z}}_{\Upsilon}P_L\mathcal{Z}_{W})+6\frac{\mathcal{C}_{W}}{\Upsilon^4}\bar{\mathcal{Z}}_{\Upsilon}P_L\mathcal{Z}_{\Upsilon}\bigg],\\
\mathcal{K}_{w} &=&  \frac{1}{\Upsilon^2}\mathcal{K}_{W} -2\frac{\mathcal{C}_{W}}{\Upsilon^3}\mathcal{K}_{\Upsilon} -\frac{1}{2}\bigg[ -4\frac{1}{\Upsilon^3} (\bar{\mathcal{Z}}_{W}P_R\mathcal{Z}_{\Upsilon}+\bar{\mathcal{Z}}_{\Upsilon}P_R\mathcal{Z}_{W})+6\frac{\mathcal{C}_{W}}{\Upsilon^4}\bar{\mathcal{Z}}_{\Upsilon}P_R\mathcal{Z}_{\Upsilon}\bigg],\\
\mathcal{B}^{w}_{\mu} &=& \frac{1}{\Upsilon^2}\mathcal{B}^{W}_{\mu} -2\frac{\mathcal{C}_{W}}{\Upsilon^3}\mathcal{B}^{\Upsilon}_{\mu} + \frac{1}{2}i  \bigg[ -4\frac{1}{\Upsilon^3} (\bar{\mathcal{Z}}_{W}P_L\gamma_{\mu}\mathcal{Z}_{\Upsilon}+\bar{\mathcal{Z}}_{\Upsilon}P_L\gamma_{\mu}\mathcal{Z}_{W})+6\frac{\mathcal{C}_{W}}{\Upsilon^4}\bar{\mathcal{Z}}_{\Upsilon}P_L\gamma_{\mu}\mathcal{Z}_{\Upsilon}\bigg],\qquad\quad\\
\Lambda_{w} &=& \frac{1}{\Upsilon^2}\Lambda_{W} -2\frac{\mathcal{C}_{W}}{\Upsilon^3}\Lambda_{\Upsilon} + \frac{1}{2} 
\bigg[ -\frac{2}{\Upsilon^3} (i\gamma_*\cancel{\mathcal{B}}_{W}+P_L\mathcal{K}_{W}+P_R\mathcal{H}_{W}-\cancel{\mathcal{D}}\mathcal{C}_{W})\mathcal{Z}_{\Upsilon} \nonumber\\
&&-\frac{2}{\Upsilon^3}(i\gamma_*\cancel{\mathcal{B}}_{\Upsilon}+P_L\mathcal{K}_{\Upsilon}+P_R\mathcal{H}_{\Upsilon}-\cancel{\mathcal{D}}\mathcal{C}_{\Upsilon})\mathcal{Z}_{W} 
+ 6\frac{\mathcal{C}_{W}}{\Upsilon^4}(i\gamma_*\cancel{\mathcal{B}}_{\Upsilon}+P_L\mathcal{K}_{\Upsilon}+P_R\mathcal{H}_{\Upsilon}-\cancel{\mathcal{D}}\mathcal{C}_{\Upsilon})\mathcal{Z}_{\Upsilon}   \bigg],\nonumber\\
&& -\frac{1}{4}\bigg[ \frac{6}{\Upsilon^4}\frac{3!}{2!} \mathcal{Z}_{(W}\bar{\mathcal{Z}}_{\Upsilon}\mathcal{Z}_{\Upsilon)} -24 \frac{\mathcal{C}_{W}}{\Upsilon^5}\mathcal{Z}_{\Upsilon}\bar{\mathcal{Z}}_{\Upsilon}\mathcal{Z}_{\Upsilon}  \bigg],\\ 
\mathcal{D}_{w} &=& \frac{1}{\Upsilon^2}\mathcal{D}_{W} -2\frac{\mathcal{C}_{W}}{\Upsilon^3}\mathcal{D}_{\Upsilon} + \frac{1}{2}  \bigg[ -\frac{2}{\Upsilon^3} 2!(\mathcal{K}_{(W}\mathcal{H}_{\Upsilon)}-\mathcal{B}_{(W}\cdot \mathcal{B}_{\Upsilon)} -\mathcal{D}\mathcal{C}_{(W}\cdot \mathcal{D}\mathcal{C}_{\Upsilon)} - 2\bar{\Lambda}_{(W} \mathcal{Z}_{\Upsilon)} - \bar{\mathcal{Z}}_{(W} \cancel{\mathcal{D}}\mathcal{Z}_{\Upsilon)})\nonumber\\
&&+6\frac{\mathcal{C}_{W}}{\Upsilon^4}(\mathcal{K}_{\Upsilon}\mathcal{H}_{\Upsilon}-\mathcal{B}_{\Upsilon}\cdot \mathcal{B}_{\Upsilon} -\mathcal{D}\mathcal{C}_{\Upsilon}\cdot \mathcal{D}\mathcal{C}_{\Upsilon} - 2\bar{\Lambda}_{\Upsilon} \mathcal{Z}_{\Upsilon} - \bar{\mathcal{Z}}_{\Upsilon} \cancel{\mathcal{D}}\mathcal{Z}_{\Upsilon}) \bigg]\nonumber\\
&& -\frac{1}{4}\bigg[ \frac{6}{\Upsilon^4}\frac{3!}{2!} \bar{\mathcal{Z}}_{(W}(i\gamma_*\cancel{\mathcal{B}}_{\Upsilon}+P_L\mathcal{K}_{\Upsilon}+P_R\mathcal{H}_{\Upsilon})\mathcal{Z}_{\Upsilon)} -24\frac{\mathcal{C}_{W}}{\Upsilon^5}\bar{\mathcal{Z}}_{\Upsilon}(i\gamma_*\cancel{\mathcal{B}}_{\Upsilon}+P_L\mathcal{K}_{\Upsilon}+P_R\mathcal{H}_{\Upsilon})\mathcal{Z}_{\Upsilon}  \bigg],\nonumber\\
&& +\frac{1}{8} \bigg[ -\frac{24}{\Upsilon^5} \frac{4!}{3!} \bar{\mathcal{Z}}_{(W} P_L \mathcal{Z}_{\Upsilon} \bar{\mathcal{Z}}_{\Upsilon} P_R \mathcal{Z}_{\Upsilon)} +120\frac{\mathcal{C}_{W}}{\Upsilon^6}
\bar{\mathcal{Z}}_{\Upsilon} P_L \mathcal{Z}_{\Upsilon} \bar{\mathcal{Z}}_{\Upsilon} P_R \mathcal{Z}_{\Upsilon} 
\bigg] .
\end{eqnarray}
Note that we have to insert $\mathcal{K}_{W}=\Lambda_{W}=\mathcal{D}_{W}=0$ and $\mathcal{H}_{\bar{W}}=\Lambda_{\bar{W}}=\mathcal{D}_{\bar{W}}=0$ as given in Eqs. \eqref{Multiplet_W1} and \eqref{Multiplet_W2}. In addition, the complex conjugate multiplet $\bar{w}^2$ can be obtained by taking a replacement $Q \rightarrow \bar{Q}$ in the above expressions.

In the Liberated SUGRA gauge ($\Lambda_K=0$), the non-vanishing components are given by
\begin{eqnarray}
\mathcal{H}_{w}|_{\Lambda_K=0} &=& \frac{\mathcal{H}_{W}}{\Upsilon^2},\quad \mathcal{K}_{\bar{w}}|_{\Lambda_K=0} = \frac{\mathcal{K}_{\bar{W}}}{\Upsilon^2}\\
\Lambda_{w}|_{\Lambda_K=0} &=& -\frac{1}{\Upsilon^3} P_R \mathcal{H}_{W} \mathcal{Z}_{\Upsilon},\quad \Lambda_{\bar{w}}|_{\Lambda_K=0} = -\frac{1}{\Upsilon^3} P_L \mathcal{K}_{\bar{W}} \mathcal{Z}_{\Upsilon}\\
\mathcal{D}_{w}|_{\Lambda_K=0} &=& -\frac{1}{\Upsilon^3}\mathcal{K}_{\Upsilon}\mathcal{H}_{W} -\frac{3}{2\Upsilon^4} \bar{\mathcal{Z}}_{} P_R\mathcal{H}_{W} \mathcal{Z}_{\Upsilon},\quad \mathcal{D}_{\bar{w}}|_{\Lambda_K=0} = -\frac{1}{\Upsilon^3}\mathcal{K}_{\bar{W}}\mathcal{H}_{\Upsilon} -\frac{3}{2\Upsilon^4} \bar{\mathcal{Z}}_{} P_L\mathcal{K}_{\bar{W}} \mathcal{Z}_{\Upsilon}. \nonumber\\{} 
\end{eqnarray}

Furthermore, with the superconformal gauge choice $\mathcal{Z}_{\Upsilon}=0$, the non-vanishing components are
\begin{eqnarray}
\mathcal{H}_{w}|_{\Lambda_K=0,\mathcal{Z}_{\Upsilon}=0} &=& \frac{\mathcal{H}_{W}}{\Upsilon^2},\quad \mathcal{K}_{\bar{w}}|_{\Lambda_K=0,\mathcal{Z}_{\Upsilon}=0} = \frac{\mathcal{K}_{\bar{W}}}{\Upsilon^2}\\
\mathcal{D}_{w}|_{\Lambda_K=0,\mathcal{Z}_{\Upsilon}=0} &=& -\frac{1}{\Upsilon^3}\mathcal{K}_{\Upsilon}\mathcal{H}_{W},\quad \mathcal{D}_{\bar{w}}|_{\Lambda_K=0,\mathcal{Z}_{\Upsilon}=0} = -\frac{1}{\Upsilon^3}\mathcal{K}_{\bar{W}}\mathcal{H}_{\Upsilon}. \nonumber\\{} 
\end{eqnarray}

\subsection{\texorpdfstring{$T(\bar{w}^2),\bar{T}(w^2)$}{} chiral projection multiplets: (Weyl/Chiral) weights \texorpdfstring{$=(0,0)$}{}}

\begin{eqnarray}
T(\bar{w}^2) &=& \left( -\frac{1}{2}\mathcal{K}_{\bar{w}}, -\frac{1}{2} \sqrt{2} iP_L (\cancel{\mathcal{D}}\mathcal{Z}_{\bar{w}}+\Lambda_{\bar{w}}), \frac{1}{2}(\mathcal{D}_{\bar{w}}+\square^C \mathcal{C}_{\bar{w}} + i\mathcal{D}_a \mathcal{B}^a_{\bar{w}}) \right),\\
\bar{T}(w^2) &=& \left( -\frac{1}{2}\mathcal{K}_{\bar{w}}^{*}, \frac{1}{2} \sqrt{2} iP_R (\cancel{\mathcal{D}}\mathcal{Z}_{\bar{w}}^C+\Lambda_{\bar{w}}^C), \frac{1}{2}(\mathcal{D}_{\bar{w}}^{*}+\square^C \mathcal{C}_{\bar{w}}^{*} - i\mathcal{D}_a (\mathcal{B}^a_{\bar{w}})^{*}) \right) .
\end{eqnarray}
The superconformal setting for this and its complex conjugate are then given by
\begin{eqnarray}
T \equiv T(\bar{w}^2) &=& \{\mathcal{C}_T,\mathcal{Z}_T,\mathcal{H}_T,\mathcal{K}_T,\mathcal{B}_{\mu}^T,\Lambda_T,\mathcal{D}_T\} ,
\nonumber\\
\bar{T} \equiv \bar{T}(w^2) &=& \{\mathcal{C}_{\bar{T}},\mathcal{Z}_{\bar{T}},\mathcal{H}_{\bar{T}},\mathcal{K}_{\bar{T}},\mathcal{B}_{\mu}^{\bar{T}},\Lambda_{\bar{T}},\mathcal{D}_{\bar{T}}\},
\end{eqnarray}
where
\begin{eqnarray}
\mathcal{C}_T &=&  -\frac{1}{2} \mathcal{K}_{\bar{w}} = -\frac{1}{2\Upsilon^2}\mathcal{K}_{\bar{W}} +\frac{\mathcal{C}_{\bar{W}}}{\Upsilon^3}\mathcal{K}_{\Upsilon} +\frac{1}{4}\bigg[ -4\frac{1}{\Upsilon^3} (\bar{\mathcal{Z}}_{\bar{W}}P_R\mathcal{Z}_{\Upsilon}+\bar{\mathcal{Z}}_{\Upsilon}P_R\mathcal{Z}_{\bar{W}})+6\frac{\mathcal{C}_{\bar{W}}}{\Upsilon^4}\bar{\mathcal{Z}}_{\Upsilon}P_R\mathcal{Z}_{\Upsilon}\bigg],\nonumber\\{} \\
\mathcal{Z}_T &=& -P_L(\cancel{\mathcal{D}}\mathcal{Z}_{\bar{w}}+\Lambda_{\bar{w}}),\\
\mathcal{H}_T &=& -(\mathcal{D}_{\bar{w}}+\square^C \mathcal{C}_{\bar{w}} + i\mathcal{D}_a \mathcal{B}^a_{\bar{w}}),\\
\mathcal{K}_T &=& 0,\\
\mathcal{B}^T_{\mu} &=& -\frac{1}{2} i \mathcal{D}_{\mu} \mathcal{K}_{\bar{w}},\\
\Lambda_T &=& 0 ,\\
\mathcal{D}_T &=& 0.
\end{eqnarray}
and
\begin{eqnarray}
\mathcal{C}_{\bar{T}} &=&  -\frac{1}{2} \mathcal{K}_{\bar{w}}^{*},\\
\mathcal{Z}_{\bar{T}} &=& -P_R(\cancel{\mathcal{D}}\mathcal{Z}_{\bar{w}}^C+\Lambda_{\bar{w}}^C),\\
\mathcal{H}_{\bar{T}} &=& 0,\\
\mathcal{K}_{\bar{T}} &=& -(\mathcal{D}_{\bar{w}}^{*}+\square^C \mathcal{C}_{\bar{w}}^{*} - i\mathcal{D}_a (\mathcal{B}^a_{\bar{w}})^{*}),\\
\mathcal{B}^{\bar{T}}_{\mu} &=& \frac{1}{2} i \mathcal{D}_{\mu} \mathcal{K}_{\bar{w}}^{*},\\
\Lambda_{\bar{T}} &=& 0 ,\\
\mathcal{D}_{\bar{T}} &=& 0.
\end{eqnarray}

In the Liberated SUGRA gauge, the non-vanishing components are given by
\begin{eqnarray}
\mathcal{C}_T|_{\Lambda_K=0} &=&  -\frac{1}{2} \mathcal{K}_{\bar{w}}, \\
\mathcal{Z}_T|_{\Lambda_K=0} &=& -P_L\Lambda_{\bar{w}},\\
\mathcal{H}_T|_{\Lambda_K=0} &=& -\mathcal{D}_{\bar{w}},\\
\mathcal{B}^T_{\mu}|_{\Lambda_K=0} &=& -\frac{1}{2} i \mathcal{D}_{\mu} \mathcal{K}_{\bar{w}}.
\end{eqnarray}
With the superconformal gauge choice $\mathcal{Z}_{\Upsilon} =0$, we have 
\begin{eqnarray}
\mathcal{C}_T|_{\Lambda_K=0,\mathcal{Z}_{\Upsilon} =0} &=&  -\frac{1}{2} \mathcal{K}_{\bar{w}},\qquad \mathcal{C}_{\bar{T}}|_{\Lambda_K=0,\mathcal{Z}_{\Upsilon} =0} =  -\frac{1}{2} \mathcal{H}_{w} = -\frac{1}{2} \mathcal{K}_{\bar{w}}^{*} \\
\mathcal{H}_T|_{\Lambda_K=0,\mathcal{Z}_{\Upsilon} =0} &=& -\mathcal{D}_{\bar{w}},\qquad \mathcal{K}_{\bar{T}}|_{\Lambda_K=0,\mathcal{Z}_{\Upsilon} =0} = -\mathcal{D}_{w}=-\mathcal{D}_{\bar{w}}^{*}\\
\mathcal{B}^T_{\mu}|_{\Lambda_K=0,\mathcal{Z}_{\Upsilon} =0} &=& -\frac{1}{2} i \mathcal{D}_{\mu} \mathcal{K}_{\bar{w}},\qquad \mathcal{B}^{\bar{T}}_{\mu}|_{\Lambda_K=0,\mathcal{Z}_{\Upsilon} =0} = \frac{1}{2} i \mathcal{D}_{\mu} \mathcal{H}_{w}= \frac{1}{2} i \mathcal{D}_{\mu} \mathcal{K}_{\bar{w}}^{*}.
\end{eqnarray}

\paragraph{Final form of the composite multiplet of liberated $\mathcal{N}=1$ supergravity}

To construct the final composite multiplet which will give the new term in the action of liberated supergravity, we first collect the 
following chiral multiplets:
\begin{eqnarray}
\Big( X^i \equiv  \{S_0,z^I,T(\bar{w}^2),\mathcal{W}^2(K)\},~\chi^i \equiv  \{\chi^0,\chi^I,\chi^T,\chi^W\},~F^i \equiv \{F^{0},F^{I},F^{T},F^{W}\} \Big),\nonumber\\{}
\end{eqnarray}
where
\begin{eqnarray}
S_0 &=& \left(s_0, P_L\chi^0, F^0 \right),\qquad  
\bar{S}_0 = \left(s_0^{*}, P_R\chi^0, F^{0*} \right),\\
z^I &=& \left( z^I, P_L\chi^I, F^I \right),\qquad  
 \bar{z}^{\bar{I}} = \left( \bar{z}^{\bar{I}}, P_R\chi^{\bar{I}}, \bar{F}^{\bar{I}} \right),
\end{eqnarray}

\begin{eqnarray}
\mathcal{W}^2(K) &=& \left( \Bar{\Lambda}_K P_L \Lambda_K,~ \sqrt{2} P_L (-\gamma\cdot \hat{F}_K + 2i\mathcal{D}_K)\Lambda_K,~ 2\Bar{\Lambda}_KP_L \cancel{\mathcal{D}}\Lambda_K + \hat{F}_K^-\cdot \hat{F}_K^- - \mathcal{D}^2_K\right) \nonumber\\
&\equiv& \left( X^W, P_L\chi^W, F^W \right),\\
\bar{\mathcal{W}}^2(K) &=& \left( \Bar{\Lambda}_K P_R \Lambda_K,~ \sqrt{2} P_R (-\gamma\cdot \hat{F}_K - 2i\mathcal{D}_K^{*})\Lambda_K^C,~ 2\Bar{\Lambda}_KP_R \cancel{\mathcal{D}}\Lambda_K + \hat{F}_K^+\cdot \hat{F}_K^+ - (\mathcal{D}^{*}_K)^2\right) \nonumber\\
&\equiv& \left( \bar{X}^{\bar{W}}, P_R\chi^{\bar{W}}, \bar{F}^{\bar{W}} \right),
\end{eqnarray}

\begin{eqnarray}
T(\bar{w}^2) &=& \left( -\frac{1}{2}\mathcal{K}_{\bar{w}}, -\frac{1}{2} \sqrt{2} iP_L (\cancel{\mathcal{D}}\mathcal{Z}_{\bar{w}}+\Lambda_{\bar{w}}), \frac{1}{2}(\mathcal{D}_{\bar{w}}+\square^C \mathcal{C}_{\bar{w}} + i\mathcal{D}_a \mathcal{B}^a_{\bar{w}}) \right) \equiv \left( X^T, P_L\chi^T, F^T \right),\nonumber\\{}\\
\bar{T}(w^2) &=& \left( -\frac{1}{2}\mathcal{K}_{\bar{w}}^{*}, \frac{1}{2} \sqrt{2} iP_R (\cancel{\mathcal{D}}\mathcal{Z}_{\bar{w}}^C+\Lambda_{\bar{w}}^C), \frac{1}{2}(\mathcal{D}_{\bar{w}}^{*}+\square^C \mathcal{C}_{\bar{w}}^{*} - i\mathcal{D}_a (\mathcal{B}^a_{\bar{w}})^{*}) \right)
\equiv \left( \bar{X}^{\bar{T}}, P_R\chi^{\bar{T}}, \bar{F}^{\bar{T}} \right). \nonumber\\{}
\end{eqnarray}

\section{Liberated Term}
Then, the lowest component of the final composite real multiplet is
\begin{eqnarray}
N &\equiv& \Upsilon^2 \frac{w^2\bar{w}^2}{T(\bar{w}^2)\bar{T}(w^2)}\mathcal{U} = \Upsilon^{-2}\dfrac{\mathcal{W}^2(K)\bar{\mathcal{W}}^2(K)}{T(w^2)\bar{T}(\bar{w}^2)} \mathcal{U} \nonumber\\
&=&  (X^0\bar{X}^0e^{-K(z^I,\bar{z}^{\bar{I}})/3})^{-2} X^W\bar{X}^{\bar{W}} (X^T)^{-1} (\bar{X}^{\bar{T}})^{-1}\mathcal{U}(z^I,\bar{z}^{\bar{I}}).
\end{eqnarray}
Then, since $N$ is now expressed as a function of superconformal chiral multiplets, we can represent the full off-shell expression of the superconformal Lagrangian of the liberated $\mathcal{N}=1$ supergravity via Eq. (17.19) of Freedman and Van Proeyen~\cite{fvp} as follows:
\begin{eqnarray}
  \mathcal{L}_{NEW} &\equiv& [\mathbb{N}]_De^{-1} \nonumber\\
  &=&
 \frac{1}{2}\mathcal{D}_N - \frac{1}{4}\bar{\psi}\cdot \gamma i\gamma_* \Lambda_N -\frac{1}{6} \mathcal{C}_N R(\omega) + \frac{1}{12} \left(\mathcal{C}_N\bar{\psi}_{\mu} \gamma^{\mu\rho\sigma} - i \bar{\mathcal{Z}}_N  \gamma^{\rho\sigma} \gamma_*\right) R_{\rho\sigma}'(Q)\nonumber\\
&&+\frac{1}{8} \varepsilon^{abcd} \bar{\psi}_{a}\gamma_b \psi_c (\mathcal{B}_{d}^N -\frac{1}{2}\bar{\psi}_d \mathcal{Z}_N) 
,\\
 &=& N_{i\bar{j}} \bigg( -\mathcal{D}_{\mu}X^i\mathcal{D}^{\mu}\bar{X}^{\bar{j}} - \frac{1}{2} \bar{\chi}^i \cancel{\mathcal{D}} \chi^{\bar{j}} - \frac{1}{2} \bar{\chi}^{\bar{j}}\cancel{\mathcal{D}}\chi^i + F^i\bar{F}^{\bar{j}}\bigg)
\nonumber\\
&& +\frac{1}{2}\bigg[ N_{ij\bar{k}} \Big( -\bar{\chi}^i\chi^j \bar{F}^{\bar{k}} + \bar{\chi}^i (\cancel{\mathcal{D}}X^j)\chi^{\bar{k}} \Big) +h.c. \bigg]
+ \frac{1}{4}N_{ij\bar{k}\bar{l}} \bar{\chi}^i\chi^j \bar{\chi}^{\bar{k}}\chi^{\bar{l}}
\nonumber\\
&& +\bigg[\frac{1}{2\sqrt{2}}\bar{\psi}\cdot \gamma 
\bigg( N_{i\bar{j}} F^i \chi^{\bar{j}} - N_{i\bar{j}} \cancel{\mathcal{D}}\bar{X}^{\bar{j}}\chi^i -\frac{1}{2}N_{ij\bar{k}} \chi^{\bar{k}}\bar{\chi}^i\chi^j  \bigg)  \nonumber\\
&& +\frac{1}{8} i\varepsilon^{\mu\nu\rho\sigma} \bar{\psi}_{\mu} \gamma_{\nu} \psi_{\rho} \bigg( N_i \mathcal{D}_{\sigma}X^i + \frac{1}{2} N_{i\bar{j}} \bar{\chi}^i \gamma_{\sigma} \chi^{\bar{j}} + \frac{1}{\sqrt{2}} N_i \bar{\psi}_{\sigma} \chi^i  \bigg)+h.c. \bigg] 
\nonumber\\
&&+\frac{1}{6} N \left( -R(\omega) +\frac{1}{2} \bar{\psi}_{\mu} \gamma^{\mu \nu\rho} R'_{\nu\rho}(Q) \right)
-\frac{1}{6\sqrt{2}} \left( N_i\bar{\chi}^i + N_{\bar{i}}\bar{\chi}^{\bar{i}} \right)\gamma^{\mu\nu} R'_{\mu\nu}(Q),\nonumber\\{}\label{NEW_Lagrangian}
\end{eqnarray}
where $i,j = 0,I,W,T$ and $\bar{j},\bar{k},\bar{l} = \Bar{0},\Bar{I},\Bar{W},\Bar{T}$ and
\begin{eqnarray}
&& R_{\mu\nu ab}(\omega) \equiv  \partial_{\mu} \omega_{\nu ab} - \partial_{\nu} \omega_{\mu ab} + \omega_{\mu ac}\omega_{\nu ~ b}^{~c} - \omega_{\nu ac}\omega_{\mu ~ b}^{~c} ,\\
&& R_{\mu\nu}'(Q) \equiv 2 \left(\partial_{[\mu} + \frac{1}{2} b_{[\mu} -\frac{3}{2} A_{[\mu} \gamma_* + \frac{1}{4} \omega_{[\mu}^{ab}(e,b,\psi)\gamma_{ab} \right)\psi_{\nu]},\\
&& \omega_{\mu}^{ab}(e,b,\psi) = \omega_{\mu}^{ab}(e,b) + \frac{1}{2} \psi_{\mu} \gamma^{[a} \psi^{b]} + \frac{1}{4} \bar{\psi}^a \gamma_{\mu} \psi^b.    
\end{eqnarray}

\subsection{Bosonic lagrangians}

The bosonic contribution to the scalar potential can be found from the term $\mathcal{D}_N$. Using the above results and Eq. \eqref{NEW_Lagrangian}, we find the equivalent bosonic contribution to the scalar potential as follows: 
\begin{eqnarray}
\mathcal{L}_B \supset N_{i\bar{j}}F^i\bar{F}^{\bar{j}} \sim N_{W\bar{W}}F^W\bar{F}^{\bar{W}} = \Upsilon^{-2}\frac{1}{\mathcal{C}_{T}\mathcal{C}_{\bar{T}}}\mathcal{U} F^W\bar{F}^{\bar{W}}.
\end{eqnarray}
Since $\mathcal{C}_{T} =- \frac{1}{2}\mathcal{K}_{\bar{w}} \sim -\frac{1}{2} \frac{\mathcal{K}_{\bar{W}}}{\Upsilon^2} =-\frac{1}{2} \frac{(-2\bar{F}^{\bar{W}})}{\Upsilon^2} = \frac{\bar{F}^{\bar{W}}}{\Upsilon^2}$ and $\mathcal{C}_{\bar{T}} = \frac{F^W}{\Upsilon^2}$, we have
\begin{eqnarray}
\mathcal{L}_B \supset \Upsilon^{-2}\frac{\Upsilon^2\Upsilon^2}{\bar{F}^{\bar{W}}F^W}\mathcal{U} F^W\bar{F}^{\bar{W}} = \Upsilon^2 \mathcal{U}.
\end{eqnarray}

In the superconformal gauge $\Upsilon=1$, we get
\begin{eqnarray}
V_{NEW} =  \mathcal{U}.
\end{eqnarray}

\subsection{Fermionic lagrangians}

In this section, we investigate fermionic terms in the liberated $\mathcal{N}=1$ supergravity Lagrangian. We focus in particular 
on the matter-coupling and on the most divergent fermionic terms, in order to explore interesting interactions and check the limits
of validity of the liberated $\mathcal{N}=1$ supergravity as an effective theory. 

First of all, let us recall that the final composite multiplet $N$ in terms of the superconformal chiral multiplets $\{S_0, z^I, W \equiv \mathcal{W}^2(K), T \equiv T(\bar{w}^2)\}$ are:
\begin{eqnarray}
N = \left(s_0s_0^{*}e^{-K(z^I,\bar{z}^{\bar{I}})/3}\right)^{-2}\dfrac{W\bar{W}}{T\bar{T}}\mathcal{U}(z^I,\bar{z}^{\bar{I}}).
\end{eqnarray}
We we also denote their lowest components as $W \equiv \bar{\Lambda}_KP_L\Lambda_K$, $\bar{W} \equiv \bar{\Lambda}_KP_R\Lambda_K$, $T\equiv \mathcal{C}_{T}$, and $\bar{T} \equiv \mathcal{C}_{\bar{T}}$. Note that the final composite multiplet consists of the four superconformal chiral multiplets only. Remember that $\mathcal{C}_{T} \sim \dfrac{\bar{F}^{\bar{W}}}{\Upsilon^2}$ and $F^W \propto \mathcal{D}_K^2|_{\textrm{boson}}\equiv \tilde{\mathcal{F}}^2$. 

Generically, the matter couplings are found from the following contributions:
\begin{eqnarray}
\mathcal{L}_{F}^{\textrm{matter}}|_{\psi=0} &=& N_{i\bar{j}} \bigg( -\mathcal{D}_{\mu}X^i\mathcal{D}^{\mu}\bar{X}^{\bar{j}} - \frac{1}{2} \bar{\chi}^i \cancel{\mathcal{D}} \chi^{\bar{j}} - \frac{1}{2} \bar{\chi}^{\bar{j}}\cancel{\mathcal{D}}\chi^i + F^i\bar{F}^{\bar{j}}\bigg) - \frac{N}{6}R(\omega)|_{\psi=0}
\nonumber\\
&& +\frac{1}{2}\bigg[ N_{ij\bar{k}} \Big( -\bar{\chi}^i\chi^j \bar{F}^{\bar{k}} + \bar{\chi}^i (\cancel{\mathcal{D}}X^j)\chi^{\bar{k}} \Big) +h.c. \bigg]
+ \frac{1}{4}N_{ij\bar{k}\bar{l}} \bar{\chi}^i\chi^j \bar{\chi}^{\bar{k}}\chi^{\bar{l}}\bigg|_{\psi=0},\nonumber\\
&=& \mathcal{L}_{F1}+\mathcal{L}_{F2}+\bar{\mathcal{L}}_{F2}+\mathcal{L}_{F3} + (\mathcal{L}_{F4}+\mathcal{L}_{F5}+h.c.)+\mathcal{L}_{F6} + \mathcal{L}_{F7}
\end{eqnarray}
where
\begin{eqnarray}
 \mathcal{D}_{\mu} X^i|_{\psi=0} &=& (\partial_{\mu} -w_ib_{\mu} -w_iA_{\mu})X^i,\nonumber\\
 \mathcal{D}_{\mu} P_L\chi^i|_{\psi=0} &=& \left(\partial_{\mu} +\frac{1}{4}\omega_{\mu}^{ab}\gamma_{ab}-(w_i+1/2)b_{\mu} + (w_i-3/2)iA_{\mu}\right) P_L\chi^i -\sqrt{2}w_iX^i   P_L\phi_{\mu}. \nonumber
\end{eqnarray}

Note that the matter couplings of fermions can be classified into seven types:
\begin{eqnarray}
\mathcal{L}_{F1} &\equiv&  -N_{i\bar{j}}\mathcal{D}_{\mu}X^i\mathcal{D}^{\mu}\bar{X}^{\bar{j}}\Big|_{\psi=0},\\
\mathcal{L}_{F2} &\equiv&  -\frac{1}{2} N_{i\bar{j}} \bar{\chi}^i \cancel{\mathcal{D}} \chi^{\bar{j}}\Big|_{\psi=0}  ,\\
\mathcal{L}_{F3} &\equiv&   -N_{i\bar{j}} F^i\bar{F}^{\bar{j}} \Big|_{\psi=0},\\
\mathcal{L}_{F4} &\equiv& -\frac{1}{2} N_{ij\bar{k}} \bar{\chi}^i\chi^j \bar{F}^{\bar{k}}\Big|_{\psi=0},\\
\mathcal{L}_{F5} &\equiv& \frac{1}{2}  N_{ij\bar{k}} \bar{\chi}^i (\cancel{\mathcal{D}}X^j)\chi^{\bar{k}}\Big|_{\psi=0} ,\\
\mathcal{L}_{F6} &\equiv&  \frac{1}{4}N_{ij\bar{k}\bar{l}} \bar{\chi}^i\chi^j \bar{\chi}^{\bar{k}}\chi^{\bar{l}} \Big|_{\psi=0} ,\\
\mathcal{L}_{F7} &\equiv& - \frac{N}{6}R(\omega)|_{\psi=0}.
\end{eqnarray}

The derivatives of the $N$ are given in general by
\begin{eqnarray}
N^{(r=q+p+m+k)}_{q,p,m,k} &=& (\partial_{0}^{q}\partial_{W}^{p}\partial_{T}^m \partial_{I}^{k} N ) \nonumber\\
&=& \bigg[(\partial_0^q\partial_I^{(k-n)}\Upsilon^{-2})_{(N)}(\Upsilon^{2})_{(\mathcal{C}_T)}(\Upsilon^{2})_{(\mathcal{C}_{\bar{T}})}
\nonumber\\
&&
\times (\Upsilon^{2m_1})_{(\partial_T^{m_1}\mathcal{C}_T)}(\Upsilon^{2m_2})_{(\partial_{\bar{T}}^{m_2}\mathcal{C}_{\bar{T}})} 
(W)^{1-p_1}(\bar{W})^{1-p_2}\mathcal{U}^{(n)} \bigg]\nonumber\\
&&/\bigg[ (\tilde{\mathcal{F}}^{2})_{(\mathcal{C}_{T})}(\tilde{\mathcal{F}}^{2})_{(\mathcal{C}_{\bar{T}})}(\tilde{\mathcal{F}}^{2m_1})_{(\partial_T^{m_1}\mathcal{C}_T)}(\tilde{\mathcal{F}}^{2m_2})_{(\partial_{\bar{T}}^{m_2}\mathcal{C}_{\bar{T}})} \bigg]. \nonumber\\
&=& (\partial_0^q\partial_I^{(k-n)}\Upsilon^{-2})\Upsilon^{4+2m_1+2m_2}\frac{\mathcal{U}^{(n)}}{\tilde{\mathcal{F}}^{2+2+2m_1+2m_2}}W^{1-p_1}\bar{W}^{1-p_2} \nonumber\\
&=& (\partial_0^q\partial_I^{(k-n)}\Upsilon^{-2})\Upsilon^{4+2m}\frac{\mathcal{U}^{(n)}}{\tilde{\mathcal{F}}^{4+2m}}W^{1-p_1}\bar{W}^{1-p_2},
\end{eqnarray}
where $\tilde{\mathcal{F}} \equiv  2K_{I\bar{\jmath}} \left( -\partial_{\mu} z^I \partial^{\mu} \bar{z}^{\bar{\jmath}} +F^I\bar{F}^{\bar{\jmath}}\right)$; 
$\mathcal{U}^{(n)}$ ($0 \leq n \leq 4$) is the $n$-th derivative of the 
function $\mathcal{U}(z^I,\bar{z}^{\bar{\imath}})$ with respect to $z^I,\bar{z}^I$, 
which are the lowest component of the matter chiral  multiplets; $q=q_1+q_2$ where $q_1$ 
($q_2$) is the order of the derivative with respect to the compensator scalar $s_0$ ($s_0^{*}$); $p=p_1+p_2$ where $p_1$ ($p_2$) is the order of the derivative with respect to the field strength multiplet scalar $W$ ($\bar{W}$); $m=m_1+m_2$ where $m_1$ ($m_2$) is the 
order of the derivative with respect to the chiral projection multiplet scalar $T(\bar{w}^2)$ ($\bar{T}(w^2)$); $k$ is the order of the derivative with respect to the matter multiplet scalar $z^I$; 
$n$ is the order of the derivative acting on the new term $\mathcal{U}$ with respect to the matter multiplet; $q$ is the total order of 
derivative with respect to the compensator scalars $s_0$ and $s_0^{*}$. An explicit form of the derivatives of $N$ is given in
Appendix A.

The mass dimension\footnote{the mass dimensions of the multiplets' lowest components are $[s_0]=1,[z^I]=0,[W]=3,[T]=0$, which gives $[\tilde{\mathcal{F}}]=2$.} of the derivatives of the $N$ is $[N^{(r=q+p+m+k)}_{q,p,m,k}]=-3p-4m-2$. This implies that the mass dimension of the operator coupled to $N^{(r=q+p+m+k)}_{q,p,m,k}$ must be equal to $3p+4m+2$.

Now, let us focus on the case such that $q=0$ and $k=n$ which gives the most singular fermionic terms in the limit
that the D-term vanishes. The most singular terms are those that contain the highest power of the auxiliary field $D$ in the 
denominator and therefore are the nonrenormalizable operators associated with the smallest UV cutoff mass scale. That is, we consider that the matter scalar derivatives act only on the new term $\mathcal{U}$ and there are no the derivatives with respect to the compensator scalar. Then, we have 
\begin{eqnarray}
N^{(r=p+m+n)}_{p,m,n} = \Upsilon^{2+2m}\frac{\mathcal{U}^{(n)}}{\tilde{\mathcal{F}}^{4+2m}}W^{1-p_1}\bar{W}^{1-p_2}.
\end{eqnarray}
In particular, since $r = p+m+n$ ($0 \leq r \leq 4$), it reduces to
\begin{eqnarray}
N_{i \ldots l}^{(r)} = N^{(r)}_{p,m,n} = \Upsilon^{2(1+r-n-p)}\frac{\mathcal{U}^{(n)}}{\tilde{\mathcal{F}}^{2(2+r-n-p)}}W^{1-p_1}\bar{W}^{1-p_2}.
\end{eqnarray}
Remember that we called $r$ the total order of the derivatives acting on the $N$. Here $n$ is the number of 
derivatives acting on the 
matter scalars in the new term $\mathcal{U}$; they produce $\mathcal{U}^{(n)}$. Finally, $p=p_1+p_2$ is the sum of the number of derivatives w.r.t the multiplets $(W,\bar{W})$ acting on $\mathcal{U}$.

\paragraph{Structure of the fermionic components}

Next, let us explore the detailed structure of the chiral fermions of the superconformal multiplets. First of all, the compensator and matter fermions are given by
\begin{eqnarray}
 \chi^0 &=& P_L\chi^0,\qquad \chi^I = P_L\chi^I.
 \end{eqnarray}
Note that $\chi^0$ and $\chi^I$ are fundamental fermions, and later in the $S$-gauge, the compensator chiral fermions will be replaced by the matter ones according to: $P_L\chi^0 = \frac{1}{3}s_0 K_I P_L\chi^I$ and $P_R\chi^0 = \frac{1}{3}s_0^{*} K_{\bar{I}} P_R\chi^{\bar{I}}$.
 
On the contrary, the other fermions $\chi^W$ and $\chi^T$ are composite. Hence, we need to find their specific structure.
 
First, the $\mathcal{W}(K)^2$-multiplet fermion, say $\chi^W$, is found to be 
\begin{eqnarray}
 \chi^W &=& P_L \chi^W = \sqrt{2}  P_L(-\gamma\cdot \hat{F}_K + 2i\mathcal{D}_K)\Lambda_K
 \nonumber\\
 &=&  -2\sqrt{2} P_L\gamma^{\mu\nu} \partial_{[\mu}\mathcal{B}_{\nu]}^K\Lambda_K-\sqrt{2}P_L\gamma^{\mu\nu} \Bar{\psi}_{[\mu} \gamma_{\nu]}\Lambda_K\Lambda_K + 2\sqrt{2}iP_L\mathcal{D}_K\Lambda_K
 \nonumber\\
 &=& -2\sqrt{2} P_L\gamma^{\mu\nu} \partial_{[\mu}(iK_I\mathcal{D}_{\nu]} z^I -iK_{\bar{I}}\mathcal{D}_{\nu]} \bar{z}^{\bar{I}} + i K_{I\bar{J}}\bar{\chi}^I \gamma_{\nu]} \chi^{\bar{J}})\Lambda_K+ 2\sqrt{2}iP_L\mathcal{D}_K\Lambda_K
 \nonumber\\ 
 &=& 2 i \gamma^{\mu\nu} \partial_{[\mu}(K_I \Bar{\psi}_{\nu]}\chi^I -K_{\bar{I}}\Bar{\chi}^{\bar{I}}\psi_{\nu]} - \sqrt{2} K_{I\bar{J}}\bar{\chi}^I \gamma_{\nu]} \chi^{\bar{J}})P_L\Lambda_K
 \nonumber\\&& 
 + 2\sqrt{2}i\bigg[2K_{I\bar{J}} 
\bigg( -\mathcal{D}_{\mu} z^I \mathcal{D}^{\mu} \bar{z}^{\bar{J}} -\frac{1}{2} \bar{\chi}^I P_L \cancel{\mathcal{D}}\chi^{\bar{J}} -\frac{1}{2} \bar{\chi}^{\bar{J}} P_R \cancel{\mathcal{D}} \chi^I + F^I\bar{F}^{\bar{J}}\bigg) \nonumber\\
&&+ K_{IJ\bar{K}} \bigg( -\bar{\chi}^I \chi^J \bar{F}^{\bar{K}} + \bar{\chi}^I (\cancel{\mathcal{D}}z^J)\chi^{\bar{K}}         \bigg) + K_{\bar{I}\bar{J}K} \bigg( -\bar{\chi}^{\bar{I}} \chi^{\bar{J}} F^{K} + \bar{\chi}^{\bar{I}} (\cancel{\mathcal{D}}\bar{z}^{\bar{J}})\chi^{K}\bigg) \nonumber\\
&& + \frac{1}{2} K_{IJ\bar{K}\bar{L}} (\bar{\chi}^IP_L\chi^J )(\bar{\chi}^{\bar{K}} P_R \chi^{\bar{L}})\bigg]P_L\Lambda_K
 \nonumber\\ 
 &=& 2\sqrt{2}i\tilde{\mathcal{F}} (P_L\Lambda_K)_{1f} + \cdots + 7~\textrm{fermions}\nonumber\\
 &=& 2\sqrt{2}i \tilde{\mathcal{F}} (-\sqrt{2}iK_{\bar{I}J}[(\cancel{\mathcal{D}}z^{J})_{0f}\chi^{\bar{I}}-\bar{F}^{\bar{I}}\chi^J]) + \cdots + 7~\textrm{fermions} \nonumber\\
 &=& 4 \tilde{\mathcal{F}} K_{\bar{I}J}[(\cancel{\partial}z^{J})\chi^{\bar{I}}-\bar{F}^{\bar{I}}\chi^J] + \cdots + 7~\textrm{fermions},
\end{eqnarray}
where $P_L \Lambda_K = -\sqrt{2}iK_{\bar{I}J}[(\cancel{\mathcal{D}}z^{J})\chi^{\bar{I}}-\bar{F}^{\bar{I}}\chi^J] -\frac{i}{\sqrt{2}}K_{\bar{I}\bar{J}K}\chi^{K}\bar{\chi}^{\bar{I}}\chi^J$. Note that the composite fermion $\chi^W$ contains powers of the
matter fermions ranging from one to seven, and it is nonvanishing on-shell. 

Second, the chiral projection multiplet $T(\bar{w}^2)$ fermions are found to be as follows:
\begin{eqnarray}
 \chi^T &=& P_L \chi^T = -\frac{i}{\sqrt{2}} P_L(\cancel{\mathcal{D}}\mathcal{Z}_{\bar{w}}+\Lambda_{\bar{w}}) = -\frac{i}{\sqrt{2}}\bigg[ \frac{(\cancel{\mathcal{D}}\mathcal{Z}_{\bar{W}})}{\Upsilon^2} - \frac{(\cancel{\mathcal{D}}\Upsilon)\mathcal{Z}_{\bar{W}}}{\Upsilon^3} -2\frac{\mathcal{C}_{\bar{W}}}{\Upsilon^3}P_L\Lambda_{\Upsilon} -\frac{i\cancel{\mathcal{B}}_{\Upsilon}\mathcal{Z}_{\bar{W}}}{\Upsilon^3} \bigg]_{\mathcal{Z}_{\Upsilon}=0}\nonumber\\
  &=& \frac{1}{\Upsilon^2}\bigg[~ \cancel{\mathcal{D}}\chi^{\bar{W}} - \frac{(\cancel{\mathcal{D}}\Upsilon)\chi^{\bar{W}}}{\Upsilon} -2\frac{\Bar{\Lambda}_KP_R\Lambda_K}{\Upsilon}P_L\Lambda_{\Upsilon} -\frac{i\cancel{\mathcal{B}}_{\Upsilon}\chi^{\bar{W}}}{\Upsilon}~ \bigg]_{\mathcal{Z}_{\Upsilon}=0}
  \nonumber\\
  &=& \frac{1}{\Upsilon^2}\bigg[~ (\cancel{\mathcal{D}}\chi^{\bar{W}})_{1f} - \frac{[(\cancel{\mathcal{D}}\Upsilon)_{0f}+i(\cancel{\mathcal{B}}_{\Upsilon})_{0f}]}{\Upsilon}(\chi^{\bar{W}})_{1f}~ \bigg] + \cdots + 9~\textrm{fermions} ,\\
  &=& \frac{1}{\Upsilon^2}\bigg[~  4 (\cancel{\partial}\tilde{\mathcal{F}}) K_{\bar{I}J}[(\cancel{\partial}z^{J})\chi^{\bar{I}}-\bar{F}^{\bar{I}}\chi^J] \nonumber\\
  &&- \Big(\frac{2}{s_0^{*}}\cancel{\partial}s_0^{*} -\frac{2}{3}K_{\bar{K}}\cancel{\partial}\bar{z}^{\bar{K}} - 2\gamma^{\mu}(b_{\mu}+iA_{\mu})\Big)4 \tilde{\mathcal{F}} K_{\bar{I}J}[(\cancel{\partial}z^{J})\chi^{\bar{I}}-\bar{F}^{\bar{I}}\chi^J] ~ \bigg]_{1f}+ \cdots + 9~\textrm{fermions} ,\nonumber\\
  &=& \frac{4}{\Upsilon^2}  (\cancel{\partial}\tilde{\mathcal{F}}) K_{\bar{I}J}[(\cancel{\partial}z^{J})\chi^{\bar{I}}-\bar{F}^{\bar{I}}\chi^J] + (\chi^T)_{1f}|_{\tilde{\mathcal{F}}^1} + \cdots + 9~\textrm{fermions},
\end{eqnarray}
where $\tilde{\mathcal{F}} \equiv  2K_{I\bar{J}} \left( -\partial_{\mu} z^I \partial^{\mu} \bar{z}^{\bar{J}} +F^I\bar{F}^{\bar{J}}\right)$; ``$|_{\tilde{\mathcal{F}}^1}$'' denotes the terms proportional to $\tilde{\mathcal{F}}^1$, and
\begin{eqnarray}
\cancel{\partial} \tilde{\mathcal{F}}&=& 2(\cancel{\partial}K_{I\bar{J}}) \left( -\partial_{\mu} z^I \partial^{\mu} \bar{z}^{\bar{J}} +F^I\bar{F}^{\bar{J}}\right)
\nonumber\\
&&+2K_{I\bar{J}} \left( -(\cancel{\partial}\partial_{\mu} z^I) \partial^{\mu} \bar{z}^{\bar{J}}-\partial_{\mu} z^I (\cancel{\partial}\partial^{\mu} \bar{z}^{\bar{J}}) +(\cancel{\partial}F^I)\bar{F}^{\bar{J}}+F^I(\cancel{\partial}\bar{F}^{\bar{J}})\right).
\end{eqnarray}
Note that $\chi^T$ is also nonvanishing on-shell. In the above calculation we have used the formula:
\begin{eqnarray}
\mathcal{D}_{\mu} P_L\chi^W &=&  \left(\partial_{\mu} +\frac{1}{4}\omega_{\mu}^{ab}\gamma_{ab}-\frac{7}{2}b_{\mu} + \frac{3}{2}iA_{\mu}\right)P_L\chi^W -\frac{1}{\sqrt{2}}P_L(\cancel{\mathcal{D}}W + F^W)\psi_{\mu} \nonumber\\
&&-3\sqrt{2}W  P_L\phi_{\mu}.
\end{eqnarray}
to find
\begin{eqnarray}
(\cancel{\mathcal{D}}\chi^W)_{1f} &=& (\cancel{\mathcal{D}}P_L\chi^W)_{1f} \nonumber\\
&=& \gamma^{\mu}\left(\partial_{\mu} +\frac{1}{4}\omega_{\mu}^{ab}\gamma_{ab}-\frac{7}{2}b_{\mu} + \frac{3}{2}iA_{\mu}\right)(P_L\chi^W)_{1f} + \frac{1}{\sqrt{2}}P_L (F^W)_{0f} \gamma^{\mu} \psi_{\mu} \nonumber\\
&=& \cancel{\partial}(P_L\chi^W)_{1f}+\gamma^{\mu}\left(\frac{1}{4}\omega_{\mu}^{ab}\gamma_{ab}-\frac{7}{2}b_{\mu} + \frac{3}{2}iA_{\mu}\right)(P_L\chi^W)_{1f} + \frac{1}{\sqrt{2}}P_L \tilde{\mathcal{F}}^2 \gamma^{\mu} \psi_{\mu}\nonumber\\
&=& 4 (\cancel{\partial}\tilde{\mathcal{F}}) K_{\bar{I}J}[(\cancel{\partial}z^{J})\chi^{\bar{I}}-\bar{F}^{\bar{I}}\chi^J] + 4 \tilde{\mathcal{F}} (\cancel{\partial}K_{\bar{I}J})[(\cancel{\partial}z^{J})\chi^{\bar{I}}-\bar{F}^{\bar{I}}\chi^J]\nonumber\\
&&+4 \tilde{\mathcal{F}} K_{\bar{I}J}[(\square z^{J})\chi^{\bar{I}}+\cancel{\partial}z^{J}\cancel{\partial}\chi^{\bar{I}}-(\cancel{\partial}\bar{F}^{\bar{I}})\chi^J-\bar{F}^{\bar{I}}(\cancel{\partial}\chi^J)] \nonumber\\
&&+\gamma^{\mu}\left(\frac{1}{4}\omega_{\mu}^{ab}\gamma_{ab}-\frac{7}{2}b_{\mu} + \frac{3}{2}iA_{\mu}\right)(4 \tilde{\mathcal{F}} K_{\bar{I}J}[(\cancel{\partial}z^{J})\chi^{\bar{I}}-\bar{F}^{\bar{I}}\chi^J] ) + \frac{1}{\sqrt{2}}P_L \tilde{\mathcal{F}}^2 \gamma^{\mu} \psi_{\mu} \nonumber\\
&=& 4 (\cancel{\partial}\tilde{\mathcal{F}}) K_{\bar{I}J}[(\cancel{\partial}z^{J})\chi^{\bar{I}}-\bar{F}^{\bar{I}}\chi^J] + (\cancel{\mathcal{D}}\chi^W)_{1f}|_{\tilde{\mathcal{F}}^1} + (\cancel{\mathcal{D}}\chi^W)_{1f}|_{\tilde{\mathcal{F}}^2}.
\end{eqnarray}
where $F^W = 2\Bar{\Lambda}_KP_L \cancel{\mathcal{D}}\Lambda_K + \hat{F}_K^-\cdot \hat{F}_K^- - \mathcal{D}^2_K$, $(\hat{F}_K^-)_{0f}=0$, and $(F^W)_{0f} = -(\mathcal{D}_K^2)_{0f} = -\tilde{\mathcal{F}}^2$, and
\begin{eqnarray}
(\cancel{\partial} \chi^W)_{1f} &=& (\cancel{\partial} P_L\chi^W)_{1f} = 4 (\cancel{\partial}\tilde{\mathcal{F}}) K_{\bar{I}J}[(\cancel{\partial}z^{J})\chi^{\bar{I}}-\bar{F}^{\bar{I}}\chi^J] + 4 \tilde{\mathcal{F}} (\cancel{\partial}K_{\bar{I}J})[(\cancel{\partial}z^{J})\chi^{\bar{I}}-\bar{F}^{\bar{I}}\chi^J]\nonumber\\
&&+4 \tilde{\mathcal{F}} K_{\bar{I}J}[(\cancel{\partial}\cancel{\partial}z^{J})\chi^{\bar{I}}+\cancel{\partial}z^{J}\cancel{\partial}\chi^{\bar{I}}-(\cancel{\partial}\bar{F}^{\bar{I}})\chi^J-\bar{F}^{\bar{I}}(\cancel{\partial}\chi^J)]\nonumber\\
&=& 4 (\cancel{\partial}\tilde{\mathcal{F}}) K_{\bar{I}J}[(\cancel{\partial}z^{J})\chi^{\bar{I}}-\bar{F}^{\bar{I}}\chi^J] + 4 \tilde{\mathcal{F}} (\cancel{\partial}K_{\bar{I}J})[(\cancel{\partial}z^{J})\chi^{\bar{I}}-\bar{F}^{\bar{I}}\chi^J]\nonumber\\
&&+4 \tilde{\mathcal{F}} K_{\bar{I}J}[(\square z^{J})\chi^{\bar{I}}+\cancel{\partial}z^{J}\cancel{\partial}\chi^{\bar{I}}-(\cancel{\partial}\bar{F}^{\bar{I}})\chi^J-\bar{F}^{\bar{I}}(\cancel{\partial}\chi^J)]
\nonumber\\
&\approx& 4 (\cancel{\partial}\tilde{\mathcal{F}}) K_{\bar{I}J}[(\cancel{\partial}z^{J})\chi^{\bar{I}}-\bar{F}^{\bar{I}}\chi^J] + 4 \tilde{\mathcal{F}} (\cancel{\partial}K_{\bar{I}J})[(\cancel{\partial}z^{J})\chi^{\bar{I}}-\bar{F}^{\bar{I}}\chi^J]\nonumber\\
&&-4 \tilde{\mathcal{F}} K_{\bar{I}J}(\cancel{\partial}\bar{F}^{\bar{I}})\chi^J .
\end{eqnarray}
Here $\approx$ means equality up to terms proportional to the equations of motion of free massless matter fields. Such 
terms produce only terms that contain additional factors of the matter field masses in the numerator and therefore give rise to either renormalizable operators or nonrenormalizable operators weighted by a mass scale higher than that associated to those
terms that do not vanish on shell. Some details of the above calculation are as follows:
\begin{eqnarray}
\mathcal{D}_{\mu} z^I &=& \partial_{\mu}z^I -\frac{1}{\sqrt{2}}\bar{\psi}_{\mu} \chi^I,\\
\mathcal{C}_{\bar{W}}  &\equiv& \bar{W} = \Bar{\Lambda}_KP_R\Lambda_K =  2~\textrm{fermions} + 4~\textrm{fermions} + 6~\textrm{fermions} ,\\
P_L \Lambda_K &=& -\sqrt{2}iK_{\bar{I}J}[(\cancel{\mathcal{D}}z^{J})\chi^{\bar{I}}-\bar{F}^{\bar{I}}\chi^J] -\frac{i}{\sqrt{2}}K_{\bar{I}\bar{J}K}\chi^{K}\bar{\chi}^{\bar{I}}\chi^J,\\
(\cancel{\mathcal{D}}\chi^{\bar{W}})_{1f} &=& 2\sqrt{2}i(\cancel{\mathcal{D}}\tilde{\mathcal{F}})_{0f}(P_L\Lambda_K)_{1f} + 2\sqrt{2}i \tilde{\mathcal{F}}  (P_L\cancel{\mathcal{D}}\Lambda_K)_{1f},\\
 \cancel{\mathcal{D}}\Lambda_K &=& \gamma \cdot \mathcal{D} \Lambda_K,\\
 \mathcal{D}_{\mu} \Lambda_K &=& \Big(\partial_{\mu} -\frac{3}{2}b_{\mu} + \frac{1}{4}\omega_{\mu}^{ab} \gamma_{ab} -\frac{3}{2}i\gamma_* \mathcal{A}_{\mu} \Big) \Lambda_K - \Big(\frac{1}{4}\gamma \cdot  \hat{F}^K + \frac{1}{2}i\gamma_* \mathcal{D}_K \Big)\psi_{\mu},\\
 (P_L\cancel{\mathcal{D}}\Lambda_K)_{1f} &=& \gamma^{\mu}\mathcal{D}_{\mu}|_{\psi=0} (P_R\Lambda_K)_{1f} + \frac{i}{2} \tilde{\mathcal{F}} \gamma^{\mu}\psi_{\mu},\\
 (\gamma \cdot \hat{F}^K)_{0f} &=& 0,\qquad \mathcal{D}_K|_{0f} = \tilde{\mathcal{F}}\\
 (\hat{F}^K_{\mu\nu})_{0f} &=& (2\partial_{[\mu} \mathcal{B}_{\nu]}^K)_{0f} = 2i\partial_{[\mu}(K_I\partial_{\nu]} z^I -K_{\bar{I}}\partial_{\nu]} \bar{z}^{\bar{I}}) = 2i(\partial_{[\mu}K_I\partial_{\nu]} z^I -\partial_{[\mu}K_{\bar{I}}\partial_{\nu]} \bar{z}^{\bar{I}}) \nonumber\\
 &=& 2i(K_{IJ}\partial_{[\mu}z^{(J}\partial_{\nu]} z^{I)} -K_{\bar{I}\bar{J}}\partial_{[\mu}z^{(\bar{J}}\partial_{\nu]} \bar{z}^{\bar{I})}) = 0,\\
 (\cancel{\mathcal{D}}\Upsilon)_{0f} &=&  (\cancel{\partial} -2\gamma^{\mu}b_{\mu} -2i\gamma^{\mu}A_{\mu})\Upsilon \nonumber\\
 &=& (\cancel{\partial}s_0)s_0^{*}e^{-K/3} + s_0(\cancel{\partial}s_0^{*})e^{-K/3} + s_0s_0^{*}e^{-K/3}(-\frac{1}{3}\cancel{\partial}K) -2\gamma^{\mu}(b_{\mu}+iA_{\mu})\Upsilon \nonumber\\
 &=&  \Upsilon(\frac{1}{s_0}\cancel{\partial} s_0 + \frac{1}{s_0^{*}} \cancel{\partial} s_0^{*} - \frac{1}{3}K_{I}\cancel{\partial}z^I-\frac{1}{3}K_{\bar{I}}\cancel{\partial}\bar{z}^{\bar{I}}-2\gamma^{\mu}(b_{\mu}+iA_{\mu})) 
 ,\\
 i(\cancel{\mathcal{B}}_{\Upsilon})_{0f} &=&  \Upsilon(-\frac{1}{s_0}\cancel{\partial} s_0 + \frac{1}{s_0^{*}} \cancel{\partial} s_0^{*} + \frac{1}{3}K_{I}\cancel{\partial}z^I-\frac{1}{3}K_{\bar{I}}\cancel{\partial}\bar{z}^{\bar{I}}),\nonumber\\
\frac{(\cancel{\mathcal{D}}\Upsilon)_{0f}+i(\cancel{\mathcal{B}}_{\Upsilon})_{0f} }{\Upsilon}  &=& \frac{2}{s_0^{*}}\cancel{\partial}s_0^{*} -\frac{2}{3}K_{\bar{I}}\cancel{\partial}\bar{z}^{\bar{I}} - 2\gamma^{\mu}(b_{\mu}+iA_{\mu}).
\end{eqnarray}

Finally, we present here the chiral fermions of the superconformal multiplets up to multiple fermion terms.
\begin{eqnarray}
\chi^i =
\begin{cases}
 \chi^0,\\
 \chi^I,\\
 \chi^W = 4 \tilde{\mathcal{F}} K_{\bar{I}J}[(\cancel{\partial}z^{J})\chi^{\bar{I}}-\bar{F}^{\bar{I}}\chi^J] + \cdots 7~\textrm{fermions},\\
 \chi^T = \frac{1}{\Upsilon^2}\bigg[~  4 (\cancel{\partial}\tilde{\mathcal{F}}) K_{\bar{I}J}[(\cancel{\partial}z^{J})\chi^{\bar{I}}-\bar{F}^{\bar{I}}\chi^J]- \Big(\frac{2}{s_0^{*}}\cancel{\partial}s_0^{*} -\frac{2}{3}K_{\bar{K}}\cancel{\partial}\bar{z}^{\bar{K}} - 2\gamma^{\mu}(b_{\mu}+iA_{\mu})\Big)
 \\ \qquad\qquad\quad  \times 4 \tilde{\mathcal{F}} K_{\bar{I}J}[(\cancel{\partial}z^{J})\chi^{\bar{I}}-\bar{F}^{\bar{I}}\chi^J] ~ \bigg]_{1f}+ \cdots + 9~\textrm{fermions} ,
\end{cases}
 \nonumber
\end{eqnarray}
where 
\begin{eqnarray}
\tilde{\mathcal{F}} &=&  2K_{I\bar{J}} \left( -\partial_{\mu} z^I \partial^{\mu} \bar{z}^{\bar{J}} +F^I\bar{F}^{\bar{J}}\right)
\nonumber\\
\cancel{\partial} \tilde{\mathcal{F}}&=& 2(\cancel{\partial}K_{I\bar{J}}) \left( -\partial_{\mu} z^I \partial^{\mu} \bar{z}^{\bar{J}} +F^I\bar{F}^{\bar{J}}\right)
\nonumber\\
&&+2K_{I\bar{J}} \left( -(\cancel{\partial}\partial_{\mu} z^I) \partial^{\mu} \bar{z}^{\bar{J}}-\partial_{\mu} z^I (\cancel{\partial}\partial^{\mu} \bar{z}^{\bar{J}}) +(\cancel{\partial}F^I)\bar{F}^{\bar{J}}+F^I(\cancel{\partial}\bar{F}^{\bar{J}})\right).\nonumber
\end{eqnarray}
Notice that none of the chiral fermions $\chi^i$ vanish on-shell, and that only $\chi^T$ dependens on $\tilde{\mathcal{F}}$ and includes the factor of $\Upsilon^{-2}$. All of these properties affect the mass dimension of the expansion coefficients of the nonrenormalizable Lagrangians.

We finally expand $W$ and $\Bar{W}$ as follows:
\begin{eqnarray}
 W &=&  -2K_{\bar{\imath}J}[\bar{\chi}^{J}(\overline{\cancel{\mathcal{D}}\bar{z}^{\bar{\imath}}})-\bar{F}^{\bar{\imath}}\bar{\chi}^{J}]K_{{\bar{\imath}}'J'}[(\cancel{\mathcal{D}}\bar{z}^{{\bar{\imath}}'})\chi^{J'}-F^{J'}\chi^{{\bar{\imath}}'}] 
 \nonumber\\
 && - K_{{\bar{\imath}}J}[\bar{\chi}^{J}(\overline{\cancel{\mathcal{D}}\bar{z}^{\bar{\imath}}})-\bar{F}^{\bar{\imath}}\bar{\chi}^{J}] K_{\bar{\imath}'\bar{\jmath}'K'}[\chi^{K'}\bar{\chi}^{\bar{\imath}'}\chi^{\bar{\jmath}'}] \nonumber\\
&& - K_{\bar{\imath}\bar{\jmath}K}[\bar{\chi}^{\bar{\jmath}}\chi^{\bar{\imath}}\bar{\chi}^{K}]K_{{\bar{\imath}}'J'}[(\cancel{\mathcal{D}}\bar{z}^{{\bar{\imath}}'})\chi^{J'}-F^{J'}\chi^{{\bar{\imath}}'}]  \nonumber\\
&&- \frac{1}{2}K_{\bar{\imath}\bar{\jmath}K}[\bar{\chi}^{\bar{\jmath}}\chi^{\bar{\imath}}\bar{\chi}^{K}]K_{\bar{\imath}'\bar{\jmath}'K'}[\chi^{K'}\bar{\chi}^{\bar{\imath}'}\chi^{\bar{\jmath}'}],\\
\bar{W} &=& (W)^{*}.
\end{eqnarray}
Notice that $W$ and $\Bar{W}$ can be represented by products of two, four, and six fundamental fermions. 

\subsection{Additional gauge fixing for physical theory in the liberated supergravity}

First of all, we consider the conventional superconformal gauge which is chosen by
\begin{eqnarray}
&&\mathcal{C}_{\Upsilon} = \Upsilon = 1 \Longleftrightarrow s_0\bar{s}_0e^{-K/3}=1,\\ &&\mathcal{Z}_{\Upsilon} = 0 \implies P_L\chi^0 - \frac{1}{3}s_0 K_I P_L\chi^I =0 \quad \& \quad P_R\chi^0 - \frac{1}{3}s_0^{*} K_{\bar{I}} P_R\chi^{\bar{I}} =0, \label{conv1}\\
&& s_0 = \bar{s}_0 \implies s_0 = \bar{s}_0 = e^{K/6}, \label{conv2} \\
&& b_{\mu} = 0. \label{conv3}
\end{eqnarray}
Note that the first condition is the $D$-gauge fixing which gets us to the Einstein frame; the second one is the improved $S$-gauge fixing; the third one is the $A$-gauge fixing; the last one is the $K$-gauge fixing.

To compare our results with the formulation of liberated supergravity in~\cite{fkr}, we choose a gauge given by
\begin{eqnarray}
\Lambda_K =0 \Longleftrightarrow \chi^{W}=\chi^{T}=0.\label{gauge-fixing-lib}
\end{eqnarray}

In both the conventional superconformal gauge~(\ref{conv1},\ref{conv2},\ref{conv3})  and in~\eqref{gauge-fixing-lib}, the relevant multiplets are
\begin{eqnarray}
S_0 &=& \left(e^{K/6}, \frac{1}{3}e^{K/6} K_I P_L\chi^I, F^0 \right),\qquad  
\bar{S}_0 = \left(e^{K/6}, \frac{1}{3}e^{K/6} K_{\bar{I}} P_R\chi^{\bar{I}}, F^{0*} \right),\\
z^I &=& \left( z^I, P_L\chi^I, F^I \right),\qquad  
 \bar{z}^{\bar{I}} = \left( \bar{z}^{\bar{I}}, P_R\chi^{\bar{I}}, \bar{F}^{\bar{I}} \right),\\
\mathcal{W}^2(K) &=&  \left( W, P_L\chi^W, F^W \right)=\left( 0, 0, \hat{F}_K^-\cdot \hat{F}_K^- - \mathcal{D}^2_K\right),\\
\bar{\mathcal{W}}^2(K) &=&  \left( \bar{W}, P_R\chi^{\bar{W}}, \bar{F}^{\bar{W}} \right)= \left( 0,0, \hat{F}_K^+ \cdot \hat{F}_K^+ - (\mathcal{D}^{*}_K)^2\right),\\
T(\bar{w}^2) &=& \left( T, P_L\chi^T, F^T \right) = \left( -\frac{1}{2}\mathcal{K}_{\bar{w}}, 0, \frac{1}{2}\mathcal{D}_{\bar{w}} \right),\\
\bar{T}(w^2) &=& \left( \bar{X}^{\bar{T}}, P_R\chi^{\bar{T}}, \bar{F}^{\bar{T}} \right) = \left( -\frac{1}{2}\mathcal{K}_{\bar{w}}^{*}, 0, \frac{1}{2}\mathcal{D}_{\bar{w}}^{*} \right). 
\end{eqnarray}
Let us further observe that all the terms coupled to $N$ and its derivatives with respect to $0,I,T$ vanish because $N$ contains the product of $W\bar{W}$, which is zero in the liberated gauge, i.e. $W=\bar{W}=0$. We see that all the fermionic terms coupled to the derivatives of $N$ with respect to $W$ or $T$ vanish as well, because these terms always couple to the fermions $\chi^W$ and $\chi^T$ which vanish in the gauge. The only non-vanishing contribution is given by only the bosonic term, $N_{W\bar{W}}F^W\bar{F}^{\bar{W}}$ which gives us the new term $\mathcal{U}$ to the scalar potential. Therefore, in both gauges, the liberated supergravity Lagrangian is specified by
\begin{eqnarray}
\mathcal{L}_{Lib} = \mathcal{L}_{SUGRA} + \mathcal{L}_{NEW},
\end{eqnarray}
where $\mathcal{L}_{SUGRA}$ is the standard supergravity Lagrangian which contains the auxiliary fields $F^0,F^I$ and $\mathcal{L}_{NEW} = \mathcal{U}(z^I,\bar{z}^{\bar{I}})$. Then, with this Lagrangian, after solving the equations of motion for the auxiliary fields, we can obtain the usual supergravity action in addition to the new term.

\section{Spectroscophy for Non-renormalizable Interactions in the Liberated $\mathcal{N}=1$ Supergravity}

In this section, we investigate the suppression of the nonrenormalizable fermionic terms in the liberated $\mathcal{N}=1$ supergravity in the superconformal formalism. To do this, we need to recall the EFT expansion reviewed in Sec. \ref{EFT_expansion}, and consider the structure of EFT expansion in Eq.~\eqref{EFT_alternative} and constraint given in Eq.~\eqref{EFT_constraint}. Hence, we have
\begin{eqnarray}
\mathcal{L}_{EFT} \supset \sum_{\delta \geq 0}^{\textrm{Finite N}} \frac{C^{\delta}}{M^{\delta-d}} \mathcal{O}^{(\delta)} \sim \sum_{\delta \geq 0}^{\textrm{Finite N}} \frac{1}{\Lambda_{cut}^{\delta-d}} \mathcal{O}^{(\delta)}  \implies \frac{C^{\delta}}{M^{\delta-d}} \lesssim \frac{1}{\Lambda_{cut}^{\delta-d}} \nonumber
\end{eqnarray}
where $\Lambda_{cut}$ is a cutoff scale; $M$ is a characteristic mass scale of a theory; $C^{\delta}$ is a dimensionless Wilson coefficient, and $\mathcal{O}^{(\delta)}$ is an effective field operator with the mass dimension $\delta$. In our model, the dimensionless Wilson coefficient will be dependent on the liberated term $\mathcal{U}$, and the mass scale $M$ will depend on $\mathcal{F}$. Hence, the key result we will find in this section is given by
\begin{eqnarray}
 \mathcal{U}^{(n)} \lesssim \begin{cases}
  \mathcal{F}^{2(4-n)} \left(\dfrac{M_{pl}}{\Lambda_{cut}}\right)^{2(4-n)}  \quad \textrm{where}\quad0\leq n \leq 2\quad\textrm{for}~N_{mat} =1,\\
\mathcal{F}^{2(6-n)}\left(\dfrac{M_{pl}}{\Lambda_{cut}}\right)^{2(6-n)}  \quad \textrm{where}\quad0\leq n \leq 4\quad\textrm{for}~N_{mat} \geq 2,
  \end{cases}
\end{eqnarray}
where $n$ is the total number of derivatives with respect to the chiral matter multiplets $z^I$; $\mathcal{F} \equiv \left<K_{I\Bar{J}}F^I\Bar{F}^{\Bar{J}}\right>$ is the dimensionless vacuum expectation value of $\tilde{\mathcal{F}}$; $\mathcal{U}(z,\Bar{z})$ is called ``liberated term'' defined as a dimensionless, gauge-invariant, general real functon of the matter scalars $z^I$'s, and $\mathcal{U}^{(n)}$ is the $n$-th derivative of the liberated term.

\subsection{Fermionic terms in liberated $\mathcal{N}=1$ supergravity}

The component Lagrangians of the liberated supergravity is given by
\begin{eqnarray}
 && \mathcal{L}_{NEW} \equiv [\mathbb{N}]_De^{-1} \nonumber\\
 &&= N_{i\bar{\jmath}} \bigg( -\mathcal{D}_{\mu}z^i\mathcal{D}^{\mu}\bar{z}^{\bar{\jmath}} - \frac{1}{2} \bar{\chi}^i \cancel{\mathcal{D}} \chi^{\bar{\jmath}} - \frac{1}{2} \bar{\chi}^{\bar{\jmath}}\cancel{\mathcal{D}}\chi^i + F^i\bar{F}^{\bar{\jmath}}\bigg)
+\frac{1}{2}\bigg[ N_{ij\bar{k}} \Big( -\bar{\chi}^i\chi^j \bar{F}^{\bar{k}} + \bar{\chi}^i (\cancel{\mathcal{D}}z^j)\chi^{\bar{k}} \Big) +h.c. \bigg]
\nonumber\\&&+ \frac{1}{4}N_{ij\bar{k}\bar{l}} \bar{\chi}^i\chi^j \bar{\chi}^{\bar{k}}\chi^{\bar{l}}
\nonumber\\
&& +\bigg[\frac{1}{2\sqrt{2}}\bar{\psi}\cdot \gamma 
\bigg( N_{i\bar{\jmath}} F^i \chi^{\bar{\jmath}} - N_{i\bar{\jmath}} \cancel{\mathcal{D}}\bar{z}^{\bar{\jmath}}\chi^i -\frac{1}{2}N_{ij\bar{k}} \chi^{\bar{k}}\bar{\chi}^i\chi^j\bigg)   +\frac{1}{8} i\varepsilon^{\mu\nu\rho\sigma} \bar{\psi}_{\mu} \gamma_{\nu} \psi_{\rho} \bigg( N_i \mathcal{D}_{\sigma}z^i + \frac{1}{2} N_{i\bar{\jmath}} \bar{\chi}^i \gamma_{\sigma} \chi^{\bar{\jmath}} \nonumber\\
&&+ \frac{1}{\sqrt{2}} N_i \bar{\psi}_{\sigma} \chi^i  \bigg)+h.c. \bigg] 
+\frac{1}{6} N \left( -R(\omega) +\frac{1}{2} \bar{\psi}_{\mu} \gamma^{\mu \nu\rho} R'_{\nu\rho}(Q) \right)
-\frac{1}{6\sqrt{2}} \left( N_i\bar{\chi}^i + N_{\bar{\imath}}\bar{\chi}^{\bar{\imath}} \right)\gamma^{\mu\nu} R'_{\mu\nu}(Q),\label{componentLagrangian}
\end{eqnarray}
First of all, focusing only on matter couplings, i.e. looking at terms independent of $\psi$, we read the following terms from Eq.~\eqref{componentLagrangian}
\begin{eqnarray}
\mathcal{L}_{F1} &\equiv&  -N_{i\bar{\jmath}}\mathcal{D}_{\mu}z^i\mathcal{D}^{\mu}\bar{z}^{\bar{\jmath}}\Big|_{\psi=0},\quad 
\mathcal{L}_{F2} \equiv  -\frac{1}{2} N_{i\bar{\jmath}} \bar{\chi}^i \cancel{\mathcal{D}} \chi^{\bar{\jmath}}\Big|_{\psi=0}  ,\nonumber\\
\mathcal{L}_{F3} &\equiv&   -N_{i\bar{\jmath}} F^i\bar{F}^{\bar{\jmath}} \Big|_{\psi=0},\quad 
\mathcal{L}_{F4} \equiv -\frac{1}{2} N_{ij\bar{k}} \bar{\chi}^i\chi^j \bar{F}^{\bar{k}}\Big|_{\psi=0},\nonumber\\
\mathcal{L}_{F5} &\equiv& \frac{1}{2}  N_{ij\bar{k}} \bar{\chi}^i (\cancel{\mathcal{D}}z^j)\chi^{\bar{k}}\Big|_{\psi=0} ,\nonumber\\
 \mathcal{L}_{F6} &\equiv&  \frac{1}{4}N_{ij\bar{k}\bar{l}} \bar{\chi}^i\chi^j \bar{\chi}^{\bar{k}}\chi^{\bar{l}} \Big|_{\psi=0} ,\quad 
\mathcal{L}_{F7} \equiv - \frac{N}{6}R(\omega)|_{\psi=0}.\label{fermi_Lagrangian}
\end{eqnarray}
Here, we observe that the fermionic terms in the  effective Lagrangian contain couplings to the function $\mathcal{U}$ 
and its derivatives since $N \propto \mathcal{U}$.  

The general structure of the fermionic terms can be found as a power series in derivatives of the composite multiplet $N$ (i.e. $N_i,N_{i\bar{\jmath}},N_{ij\bar{k}}$ and $N_{ij\bar{k}\bar{l}}$). The $r$-th derivative of $N$, denoted with 
$N_{i\ldots\bar{l}}^{(r)}$ has the following generic form
\begin{eqnarray}
N_{i\ldots\bar{l}}^{(r)} &=& N^{(r=q+p+m+k)}_{q,p,m,k} 
= (\partial_0^q\partial_I^{(k-n)}\Upsilon^{-2})\Upsilon^{4+2m}\frac{\mathcal{U}^{(n)}}{\tilde{\mathcal{F}}^{4+2m}}W^{1-p_1}\bar{W}^{1-p_2}
\nonumber\\
&=&\Big(
(-1)^{q_1+q_2} (q_1+1)!(q_2+1)! s_0^{-(2+q_1)} {s_0^*}^{-(2+q_2)}(\partial_I^{(k-n)}e^{2K/3})\Big) \Upsilon^{4+2m}\frac{\mathcal{U}^{(n)}}{\tilde{\mathcal{F}}^{4+2m}}W^{1-p_1}\bar{W}^{1-p_2}.
\nonumber\\{} \label{Gen_str_N_deriv}
\end{eqnarray}
where 
\begin{eqnarray}
 W &=&  -2K_{\bar{\imath}J}[\bar{\chi}^{J}(\overline{\cancel{\mathcal{D}}\bar{z}^{\bar{\imath}}})-\bar{F}^{\bar{\imath}}\bar{\chi}^{J}]K_{{\bar{\imath}}'J'}[(\cancel{\mathcal{D}}\bar{z}^{{\bar{\imath}}'})\chi^{J'}-F^{J'}\chi^{{\bar{\imath}}'}]  - K_{{\bar{\imath}}J}[\bar{\chi}^{J}(\overline{\cancel{\mathcal{D}}\bar{z}^{\bar{\imath}}})-\bar{F}^{\bar{\imath}}\bar{\chi}^{J}] K_{\bar{\imath}'\bar{\jmath}'K'}[\chi^{K'}\bar{\chi}^{\bar{\imath}'}\chi^{\bar{\jmath}'}] \nonumber\\
&& - K_{\bar{\imath}\bar{\jmath}K}[\bar{\chi}^{\bar{\jmath}}\chi^{\bar{\imath}}\bar{\chi}^{K}]K_{{\bar{\imath}}'J'}[(\cancel{\mathcal{D}}\bar{z}^{{\bar{\imath}}'})\chi^{J'}-F^{J'}\chi^{{\bar{\imath}}'}]  - \frac{1}{2}K_{\bar{\imath}\bar{\jmath}K}[\bar{\chi}^{\bar{\jmath}}\chi^{\bar{\imath}}\bar{\chi}^{K}]K_{\bar{\imath}'\bar{\jmath}'K'}[\chi^{K'}\bar{\chi}^{\bar{\imath}'}\chi^{\bar{\jmath}'}],\\
\bar{W} &=& (W)^*,
\end{eqnarray}
$\tilde{\mathcal{F}} \equiv  2K_{I\bar{\jmath}} \left( -\partial_{\mu} z^I \partial^{\mu} \bar{z}^{\bar{\jmath}} +F^I\bar{F}^{\bar{\jmath}}\right)$; 
$\mathcal{U}^{(n)}$ ($0 \leq n \leq 4$) is the $n$-th derivative of the 
function $\mathcal{U}(z^I,\bar{z}^{\bar{\imath}})$ with respect to $z^I,\bar{z}^I$, 
which are the lowest component of the matter chiral  multiplets; $q=q_1+q_2$ where $q_1$ 
($q_2$) is the order of the derivative w.r.t. the compensator scalar $s_0$ ($s_0^*$); $p=p_1+p_2$ where $p_1$ ($p_2$) is the order of the derivative w.r.t. the field strength multiplet scalar $W$ ($\bar{W}$); $m=m_1+m_2$ where $m_1$ ($m_2$) is the 
order of the derivative w.r.t. the chiral projection multiplet scalar $T(\bar{w}^2)$ ($\bar{T}(w^2)$); $k$ is the order of the derivative w.r.t. the matter multiplet scalar $z^I$; 
$n$ is the order of the derivative acting on the new term $\mathcal{U}$ w.r.t. the matter multiplet; $q$ is the total order of 
derivative w.r.t. the compensator scalars $s_0$ and $s_0^*$. 

To find restrictions on $V_{NEW}$ coming from fermionic terms, we have to identify the most singular terms in the 
Lagrangian. These terms can be found using the fact that powers of $\tilde{\mathcal{F}}$ in the denominator may lead to a 
singularity which gets stronger when $m$ increases by taking more derivatives with respect to the lowest component of the
 multiplet $T(\bar{w}^2)$ as seen from Eq.~\eqref{Gen_str_N_deriv}. Hence, we will investigate the fermionic terms containing only derivatives with respect to the chiral projection
 and matter scalar indices, i.e. $T$ and $I$, in order to find the terms coupled to $\mathcal{U}^{(n)}$ that contain the 
 maximal inverse powers of $\tilde{\mathcal{F}}$. They are those with $q=p=0$ and $k=n$. We note in particular that 
 if our theory has a single chiral matter multiplet then the most singular terms are 
 found to be the couplings to the derivatives proportional to $N_{T\bar{T}}$, $N_{WT\bar{T}}$, $N_{W\bar{W}T\bar{T}}$
 while for two or more chiral matter multiplets they are $N_{TT\bar{T}\bar{T}}$. The latter terms vanish identically for a 
 single multiplet because of Fermi statistics.
 
First of all, let us examine the single matter chiral multiplet case. Due to Fermi statistics, the possible fermionic terms are 
proportional only to three tems, $\mathcal{U}^{(0)}$, $\mathcal{U}^{(1)}$, and $\mathcal{U}^{(2)}$, so that the maximal
 order of the derivative with respect to the chiral projection that can appear in the Lagrangians scalar is two and
 appears in the terms 
 proportional to $N_{T\bar{T}}$, $N_{WT\bar{T}}$, and $N_{W\bar{W}T\bar{T}}$. To show that such terms do not 
 vanish consider 
\begin{eqnarray}
\mathcal{L}_{F2}|_{q=p=0,k=n} &\supset&   \Upsilon^{2+2m}\frac{\mathcal{U}^{(n)}}{\tilde{\mathcal{F}}^{4+2m}}W\bar{W}  \bigg[-\frac{1}{2} \Big((\bar{\chi}^I)^{n_1} (\Upsilon^{-2} 4 (\cancel{\partial}\tilde{\mathcal{F}}) K_{\bar{\imath}J}(\cancel{\partial}z^{J})\bar{\chi}^{\bar{\imath}}P_R)^{m_1}\Big)
\nonumber\\
&&\times
\Big((\cancel{\mathcal{D}}\chi^I)^{n_2} (\cancel{\mathcal{D}}(\Upsilon^{-2} 4 (\cancel{\partial}\tilde{\mathcal{F}}) K_{\bar{\imath}J}(\cancel{\partial}z^{J})P_R\chi^{\bar{\imath}}))^{m_2}\Big)\bigg]_{\psi=0},
\end{eqnarray}
where $m = m_1 + m_2$, $n = n_1 + n_2$, and $2=m+n$. Restoring the mass dimensions by fixing the super-Weyl gauge\footnote{Here, we use the convention of the superconformal formalism that all physical bosonic and fermionic matter
 fields have dimensions 0 and 1/2 respectively and  $\mathcal{F}$ has dimension 2 while the compensator 
 $s_0$ has dimension 1~\cite{Superconformal_Freedman}. Through dimensional analysis, we find $[\mathcal{D}_{\mu}]=1,[z^i]\equiv 0+[i],[\chi^i]\equiv \frac{1}{2}+[i],[F^i]\equiv 1+[i]$ where $i=0,I,W,T$ and
 $[0]=1,[I]=0,[W]=3,[T]=0$.} (i.e. $\Upsilon = M_{pl}^2$, $s_0=s_0^*=M_{pl} e^{K/6}$, $P_L\chi^0 = \frac{1}{3}s_0 K_I P_L\chi^I= \frac{1}{3}M_{pl} e^{K/6} K_I P_L\chi^I$, and $b_{\mu}=0$), we obtain
\begin{eqnarray}
\mathcal{L}_{F2}|_{q=p=0,k=n} &\supset&   M_{pl}^{2(2+2m)}\frac{\mathcal{U}^{(n)}}{\tilde{\mathcal{F}}^{4+2m}} W\bar{W}  \bigg[-\frac{1}{2} \Big((\bar{\chi}^I)^{n_1} (M_{pl}^{-4} 4 (\cancel{\partial}\tilde{\mathcal{F}}) K_{\bar{\imath}J}(\cancel{\partial}z^{J})\bar{\chi}^{\bar{\imath}}P_R)^{m_1}\Big)
\nonumber\\
&&\times
\Big((\cancel{\mathcal{D}}\chi^I)^{n_2} (M_{pl}^{-4}( 4\cancel{\mathcal{D}} (\cancel{\partial}\tilde{\mathcal{F}}) K_{\bar{\imath}J}(\cancel{\partial}z^{J})P_R\chi^{\bar{\imath}}))^{m_2}\Big)\bigg]_{\psi=0}
\approx M_{pl}^{4} \frac{\mathcal{U}^{(n)}}{\tilde{\mathcal{F}}^{4+2m}} \mathcal{O}_F^{(\delta)},\quad 
\end{eqnarray}
where we require $m_a+n_a = 1$ for $a=1,2$ since we are studying the second derivative term $N_{i\bar{\jmath}}$ coupled to $\bar{\chi}^i$ and $\cancel{\mathcal{D}}\chi^{\bar{\jmath}}$. 
Redefining $\tilde{\mathcal{F}}$ to be dimensionless by $\tilde{\mathcal{F}} \rightarrow M_{pl}^2 \tilde{\mathcal{F}}$, we obtain
\begin{eqnarray}
 \mathcal{L}_{F2}|_{q=p=0,k=n} \supset M_{pl}^{-4-2m} \frac{\mathcal{U}^{(n)}}{\tilde{\mathcal{F}}^{4+2m}} 
 \mathcal{O}_F'^{(\delta)}. 
\end{eqnarray}
We then find $\delta = 8+4m$ by trivial dimensional analysis because the Lagrangian has mass dimension 4.
Then, since $2 = m+ n$, we find that the most singular term is  
\begin{eqnarray}
 \mathcal{L}_{F2}|_{q=p=0,k=n} \supset M_{pl}^{2(n-4)} \frac{\mathcal{U}^{(n)}}{\tilde{\mathcal{F}}^{2(4-n)}} \mathcal{O}_F'^{(2(6-n))}.\label{SingleChiralMatterCase} 
\end{eqnarray}

Next we consider the general case with several multiplets. We shall focus on the fourth derivative term denoted by $N_{ij\bar{k}\bar{l}}$, which gives a four-fermion term. Also, we have to consider the four-fermion product made only of 
the 
chiral fermions with $i=0,I,T$ because they do not contribute one 
power of the F-term $\tilde{\mathcal{F}}$ in the numerator, which would reduce the number of  inverse power of the F-term
 $\tilde{\mathcal{F}}$. This is because the overall factor of $\chi^W$ contains such linear dependence. 
 The effective fermionic Lagrangian \eqref{fermi_Lagrangian} reads then as follows:
\begin{eqnarray}
 \mathcal{L}_{F6}|_{q=p=0,k=n}
&\supset& \Upsilon^{2+2m}\frac{\mathcal{U}^{(n)}}{\tilde{\mathcal{F}}^{4+2m}} W\bar{W}
 \Big(\frac{1}{4}(\chi^I)^n(4 \Upsilon^{-2} (\cancel{\partial}\tilde{\mathcal{F}}) K_{\bar{\imath}J}(\cancel{\partial}z^{J})P_R\chi^{\bar{\imath}})^m\Big).
\end{eqnarray}
After the super-Weyl gauge fixing  we obtain 
\begin{eqnarray}
 \mathcal{L}_{F6}|_{q=p=0,k=n}
&\supset& M_{pl}^{2(2+2m)}\frac{\mathcal{U}^{(n)}}{\tilde{\mathcal{F}}^{4+2m}} W\bar{W}
 \Big(\frac{1}{4}(\chi^I)^n(4 M_{pl}^{-4} (\cancel{\partial}\tilde{\mathcal{F}}) K_{\bar{\imath}J}(\cancel{\partial}z^{J})P_R\chi^{\bar{\imath}})^m\Big) \nonumber\\
&\approx& c M_{pl}^{4}\frac{\mathcal{U}^{(n)}}{\tilde{\mathcal{F}}^{4+2m}} \mathcal{O}_F^{(\delta)}. \nonumber\\{}
\end{eqnarray}
where $ \mathcal{O}(1)\lesssim c \lesssim \mathcal{O}(10^3)$. After doing the same dimensional analysis as in the 
single-multiplet case, we obtain $\delta = 8+4m$ and
\begin{eqnarray}
\mathcal{L}_{F6}|_{q=p=0,k=n} \supset c M_{pl}^{-4-2m}\frac{\mathcal{U}^{(n)}}{\tilde{\mathcal{F}}^{4+2m}} \mathcal{O}_F'^{(8+2m)}.
\end{eqnarray}
Then, since $4 = m+ n$, we can write the most singular terms as 
\begin{eqnarray}
\mathcal{L}_{F6}|_{q=p=0,k=n} \supset cM_{pl}^{2(n-6)}\frac{\mathcal{U}^{(n)}}{\tilde{\mathcal{F}}^{2(6-n)}} \mathcal{O}_F'^{(2(8-n))},\label{GeneralChiralMatterCases}
\end{eqnarray}
 
 Since we are going to use liberated supergravity to describe time-dependent backgrounds such as slow-roll inflation
  we need to look more closely at the structure of $\tilde{\mathcal{F}}$. From its definition 
 $\tilde{\mathcal{F}} \equiv  2K_{I\bar{\jmath}} \left( -\partial_{\mu} z^I \partial^{\mu} \bar{z}^{\bar{\jmath}} +F^I\bar{F}^{\bar{\jmath}}\right)$, 
 we find $\tilde{\mathcal{F}} \equiv  2K_{I\bar{\jmath}} \left( \dot{z}^I\dot{\bar{z}}^{\bar{\jmath}} +F^I\bar{F}^{\bar{\jmath}}\right) >0$ 
 whenever spatial gradients can be neglected. We see that the most singular behaviors of the fermionic terms arises 
 when $\dot{z}^I=0$. By expanding $\tilde{F}$ around a static vacuum and 
 reserving the notation $\mathcal{F}$ for the expectation value $\left<K_{I\bar{\jmath}}F^I\bar{F}^{\bar{\jmath}}\right>$, the 
 effective Lagrangian can finally be rewritten as
 \begin{itemize}
    \item For the  single chiral matter multiplet case, 
    \begin{eqnarray}
     \mathcal{L}_{F2}|_{q=p=0,k=n} \supset  M_{pl}^{2(n-4)} \frac{\mathcal{U}^{(n)}}{\mathcal{F}^{2(4-n)}} \mathcal{O}_F'^{(2(6-n))}.\label{SingleChiralMatterCase_Vac}
    \end{eqnarray}
    \item For two or more chiral matter multiplets, 
    \begin{eqnarray}
     \mathcal{L}_{F6}|_{q=p=0,k=n} \supset c'M_{pl}^{2(n-6)}\frac{\mathcal{U}^{(n)}}{\mathcal{F}^{2(6-n)}} \mathcal{O}_F'^{(2(8-n))}.\label{GeneralChiralMatterCases_Vac}
    \end{eqnarray}
    where $ \mathcal{O}(10^{-2}) \lesssim c' \lesssim  \mathcal{O}(1)$.
\end{itemize}

The effective operators we obtained are generically nonzero even after considering possible cancellations due to Fermi 
statistics or nonlinear field redefinitions. As an example we can take terms containing $\chi^i$. They are made of two 
composite
 chiral multiplets $\chi^W$ and $\chi^T$ and these produce terms that do not vanish on shell (i.e. imposing $\cancel{\partial}P_L\chi^I\approx 0$ for matter fermions). 
For instance, in a theory with only one matter chiral multiplet $(z,P_L\chi,F)$, we have 
$W = -2K_{\bar{z}z}[\{(\cancel{\partial}z)^2- F^*\cancel{\partial}z\}(\bar{\chi}P_R\chi)+\{{F^*}^2-F^*\cancel{\partial}z\} 
(\bar{\chi}P_L\chi) ]+2K_{\bar{z}z}K_{\bar{z}\bar{z}z}(\cancel{\partial}z-F^*)(\bar{\chi}P_L\chi) (\bar{\chi}P_R\chi)$, and 
$\bar{W} = (W)^*$, so that $W\bar{W} = 4K_{\bar{z}z}^4(|\cancel{\partial}z|^2+|F|^2)||\cancel{\partial}z-F^*|^2(\bar{\chi}
P_L\chi)(\bar{\chi}P_R\chi)$. Hence, looking at the possible fermionic terms from $\mathcal{L}_{F1}$, when $i=z,\bar{\jmath}
=\bar{z}$ (i.e. $q=p=m=0,k=n=2$), we get 
\begin{eqnarray}
\mathcal{L}_{F1} &\supset&  \Upsilon^2\frac{\mathcal{U}^{(2)}}{\tilde{\mathcal{F}}^4}W\bar{W}(\partial_{\mu} z\partial^{\mu}\bar{z})
= \Upsilon^2\frac{\mathcal{U}^{(2)}}{\tilde{\mathcal{F}}^4}4K_{\bar{z}z}^4(|\cancel{\partial}z|^2+|F|^2)||\cancel{\partial}z-F^*|^2(\partial_{\mu} z\partial^{\mu}\bar{z})(\bar{\chi}P_L\chi)(\bar{\chi}P_R\chi) . \nonumber\\
\end{eqnarray}
 It is easy to see that this operator does not vanish on the mass shell of the matter scalars, $\square z \approx 0$. 
As another example, from $\mathcal{L}_{F2}$ we get terms containing up to three matter fermions when we consider $q=m=p_1=0,k=n=1,p_2=p=1$
\begin{eqnarray}
\mathcal{L}_{F2} &\supset& \Upsilon^2\frac{\mathcal{U}^{(1)}}{\tilde{\mathcal{F}}^4} \frac{1}{2} W \bar{\chi} \cancel{\partial}P_R\chi^{\bar{W}} \approx  \Upsilon^2\frac{\mathcal{U}^{(1)}}{\tilde{\mathcal{F}}^4} 4K_{\bar{z}z}^2(\cancel{\partial}\tilde{\mathcal{F}})(\cancel{\partial}z)(\bar{\chi}P_R\chi) P_L(\cancel{\partial}z-F^*)^2\chi|_{\mbox{\small 3-fermion terms}}
+\cdots.\nonumber\\{}
\end{eqnarray}

Back to the results in Eqs. \eqref{SingleChiralMatterCase_Vac} and \eqref{GeneralChiralMatterCases_Vac}, the general effective Lagrangians can be cast in the form 
\begin{eqnarray}
\mathcal{L}_F = \Lambda_{cut}^{4-\delta}\mathcal{O}_F'^{(\delta)}= 
\begin{cases}
M_{pl}^{2(n-4)} \dfrac{\mathcal{U}^{(n)}}{\mathcal{F}^{2(4-n)}} \mathcal{O}_F'^{(\delta  = 2(6-n))}~\qquad \textrm{for}~N_{mat} =1,\\
 c'M_{pl}^{2(n-6)}\dfrac{\mathcal{U}^{(n)}}{\mathcal{F}^{2(6-n)}} \mathcal{O}_F'^{(\delta = 2(8-n))} ~\qquad\textrm{for}~N_{mat} \geq 2.
\end{cases}
\end{eqnarray}
where $ \mathcal{O}(10^{-2}) \lesssim c' \lesssim \mathcal{O}(1)$; $N_{mat}$ and $\Lambda_{cut}$ are defined to be the number of chiral multiplets of matter, and the cutoff scale of our effective theory, respectively.

If we demand that our effective theory describe physics up to the energy scale  $\Lambda_{cut}$ we obtain the following
 inequalities:
\begin{eqnarray}
 \mathcal{U}^{(n)} \lesssim \begin{cases}
  \mathcal{F}^{2(4-n)} \left(\dfrac{M_{pl}}{\Lambda_{cut}}\right)^{2(4-n)}  \quad \textrm{where}\quad0\leq n \leq 2\quad\textrm{for}~N_{mat} =1,\\
\mathcal{F}^{2(6-n)}\left(\dfrac{M_{pl}}{\Lambda_{cut}}\right)^{2(6-n)}  \quad \textrm{where}\quad0\leq n \leq 4\quad\textrm{for}~N_{mat} \geq 2.
  \end{cases}\label{pre_inequality}
\end{eqnarray}

A conventional definition of the supersymmetry breaking scale $M_S$ is in terms of F-term expectation value so we define 
$M_S^4=M_{pl}^4\mathcal{F}$, so the constraints on $\mathcal{U}^{(n)}$ become 
\begin{eqnarray}
 \mathcal{U}^{(n)} \lesssim \begin{cases}
  \left(\dfrac{M_S}{M_{pl}}\right)^{8(4-n)}\left(\dfrac{M_{pl}}{\Lambda_{cut}}\right)^{2(4-n)} \quad \textrm{where}\quad0\leq n \leq 2\quad\textrm{for}~N_{mat} =1,\\
\left(\dfrac{M_S}{M_{pl}}\right)^{8(6-n)}\left(\dfrac{M_{pl}}{\Lambda_{cut}}\right)^{2(6-n)} \quad\textrm{where}\quad0\leq n \leq 4\quad\textrm{for}~N_{mat} \geq 2.
  \end{cases}\label{generic_constraints_here}
\end{eqnarray}
Equation~\eqref{generic_constraints_here} is the crucial one in our paper, as it constrains precisely the new 
function $\mathcal{U}$ 
introduced by liberated $\mathcal{N}=1$ supergravity. The constraint depends on the reduced Planck scale $M_{pl}$, the
supersymmetry breaking scale $M_S$, $\Lambda_{cut}$ and the number of chiral multiplets of matter in the theory. 
Of course, when we push both the cutoff and supersymmetry breaking scales to the reduced Planck scale, i.e. $\Lambda_{cut} \sim M_S \sim M_{pl}$, we obtain a model-independent universal constraint
\begin{eqnarray}
 \forall n:~ \mathcal{U}^{(n)} \lesssim 1. 
\label{Universal_Constraint1}
\end{eqnarray}

A model where supersymmetry is broken at the Planck scale is hardly the most interesting. In the more interesting case that 
$M_S\ll M_{pl}$ we need the constraints \eqref{generic_constraints_here} again, so we need to first determine how many matter 
chiral multiplets we have in our theory. The constraints will then depend only on our choice  of 
$\Lambda_{cut}$ and $M_S$. 

In the rest of this section we will examine the constraints in two cases. The first is the true, post-inflationary vacuum of the
 theory. To make a supergravity theory meaningful we want it to be valid at least 
 up to energies $\Lambda_{cut}\gtrsim M_S$. 
The second is slow-roll inflation. In this case we must have $\Lambda_{cut}\gtrsim H$, with $H$ the Hubble constant 
during inflation.

For the post-inflationary vacuum the interesting regime is when $M_S$ is relatively small, say 
$M_S\sim 10\,\mathrm{TeV} \approx 10^{-15} M_{pl}$ and the effective theory is valid up to an energy scale not smaller 
than $M_S$, i.e. $\Lambda_{cut}\gtrsim M_S$. If $\Lambda_{cut} < M_S$ liberated supergravity would be a useless 
complication, since in its domain of validity supersymmetry would be always nonlinearly realized. 
In the post-inflationary vacuum, for the single matter chiral multiplet case, the constraints~\eqref{generic_constraints_here} thus give for
\bea
&& \mathcal{U}^{(0)} \lesssim \left(\frac{M_S}{M_{pl}}\right)^{32}\left( \frac{M_{pl}}{M_S} \right)^{8} \implies \mathcal{U}^{(0)} \sim 10^{-360},\\
&& \mathcal{U}^{(1)} \lesssim \left(\frac{M_S}{M_{pl}}\right)^{24}\left( \frac{M_{pl}}{M_S} \right)^{6}\implies \mathcal{U}^{(1)} \sim 10^{-270},\\
&& \mathcal{U}^{(2)} \lesssim \left(\frac{M_S}{M_{pl}}\right)^{16}\left( \frac{M_{pl}}{M_S} \right)^{4}\implies \mathcal{U}^{(2)} \sim 10^{-180}.
\eea{Option1_Const}
Notice that $\mathcal{U}^{(3)},\mathcal{U}^{(4)}$ are not restricted. For two or more matter chiral multiplets, the
constraints  are given by
\bea
&& \mathcal{U}^{(0)} \lesssim \left(\frac{M_S}{M_{pl}}\right)^{48}\left( \frac{M_{pl}}{M_S} \right)^{12} \implies \mathcal{U}^{(0)} \sim 10^{-540},\\
&& \mathcal{U}^{(1)} \lesssim \left(\frac{M_S}{M_{pl}}\right)^{40}\left( \frac{M_{pl}}{M_S} \right)^{10}\implies \mathcal{U}^{(1)} \sim 10^{-450},\\
&& \mathcal{U}^{(2)} \lesssim \left(\frac{M_S}{M_{pl}}\right)^{32}\left( \frac{M_{pl}}{M_S} \right)^{8}\implies \mathcal{U}^{(2)} \sim 10^{-360}.
\eea{Option2_Const}
From the constraints on 
$\mathcal{U}^{(0)}$ and $\mathcal{U}^{(2)}$, we find that the liberated scalar potential contributes only a negligibly small 
cosmological constant and negligibly small corrections to the mass terms 
of the chiral multiplet scalars. For the single chiral multiplet case, restoring dimensions we get a vacuum energy density 
\beq
\mathcal{U}^{(0)} \lesssim 10^{-360} M_{pl}^4
\eeq{vac-en}
and scalar masses
\beq
M_z \lesssim M_{pl} \sqrt{|\mathcal{U}^{(2)}|} = 10^{-90}M_{pl}.
\eeq{masses}
These constraints become even tighter if the theory contains more than one chiral multiplet, but the ones we obtained are
already so stringent as to rule out any observable contribution to the cosmological constant and scalar masses from the
new terms made possible by liberated supergravity. We can say that  Eqs.~(\ref{vac-en}) and (\ref{masses}) already send 
liberated supergravity back to prison after the end of inflation.

The constraints during inflation instead can be easily satisfied if during inflation the supersymmetry breaking scale is 
very high, say $M_S = M_{pl}$. In that case, $\mathcal{U}^{(0)} \lesssim \mathcal{O}(1)$. 
After inflation the ``worst case scenario'' constraints coming from Eq.~\eqref{generic_constraints_here} with $N_{mat}\geq 2$ and
$M_S = 10^{-15}M_{pl}$ are 
\begin{eqnarray}
 \forall n:~\mathcal{U}^{(n)} \lesssim 10^{-120(6-n)}.\label{Option2_Const_after_inf}
\end{eqnarray}

A simple way to satisfy all these constraints is to choose a no scale structure for the supersymmetric part of the scalar potential. This ensures the vanishing of the F-term contribution to the potential independently of the magnitude of the 
F-terms.
\begin{eqnarray}
V_F =e^G(G_IG^{I\bar{\jmath}}G_{\bar{\jmath}}-3) =0,\label{No_scale_structure}
\end{eqnarray}
The total scalar potential is then given by $V = V_D+V_{NEW}$. Thanks to the no-scale structure, we can have 
both $M_S\sim M_{pl}$ and $\mathcal{U}^{(0)} \sim H^2 \sim 10^{-10}$ during inflation. 

Our scenario has $M_S = M_{pl}$ during inflation and $M_S = 10^{-15}M_{pl}$ at the true vacuum in the post-inflation phase, so we see that to satisfy all constraints a transition between the two different epochs must occur, for which the scale of the composite F-term $\mathcal{F}$ changes from $\mathcal{O}(M_{pl})$ to $\mathcal{O}(10^{-15}M_{pl})$.

%% file: chapters/7.tex
\framebox[1.05\width]{{\large This chapter is based on the author's original work in Ref.~\cite{jp3}.}} \par
\vspace{1cm}

\section{Component Action of New FI Term in Superconformal Tensor Calculus}

In this chapter, we review the component action of a new, K\"{a}hler invariant Fayet-Iliopoulos term proposed by Antoniadis, Chatrabhuti, Isono, and Knoops \cite{acik}, using again the superconformal tensor calculus. The full Lagrangian with the new FI terms is given by 
\begin{eqnarray}
\mathcal{L} &=& -3[S_0\bar{S}_0e^{-K(z,\bar{z})}]_D + ([S_0^3W(z)]_F -\frac{1}{4} [\bar{\lambda}P_L\lambda]_F+h.c.) \nonumber\\
&&- \xi \left[(S_0\bar{S}_0e^{-K(z,\bar{z})})^{-3}\frac{(\bar{\lambda}P_L\lambda)(\bar{\lambda}P_R\lambda)}{T(\bar{w}'^2)\bar{T}(w'^2)}(V)_D\right]_D,
\end{eqnarray}
where the last term coupled to the parameter $\xi$ corresponds to the new FI terms.

We consider matter chiral multiplets $Z^i$, the chiral compensator $S_0$, a real multiplet $V$, and another real multiplet 
$(V)_D$, whose lowest component is the auxiliary D term of the real multiplet $V$. Their superconformal multiplets are given as follows:
\begin{eqnarray}
&& V = \{0,0,0,0,A_{\mu},\lambda,D\} ~\textrm{in the Wess-Zumino gauge,~i.e.}~v=\zeta=\mathcal{H}=0, \\
&& Z^i = (z^i,-i\sqrt{2}P_L\chi^i,-2F^i,0,+i\mathcal{D}_{\mu}z^i,0,0) = \{ z^i, P_L\chi^i,F^i\},\\
&& \bar{Z}^{\bar{i}} = (\bar{z}^{\bar{i}},+i\sqrt{2}P_R\chi^{\bar{i}},0,-2\bar{F}^{\bar{i}},-i\mathcal{D}_{\mu}\bar{z}^{\bar{i}},0,0) = \{ \bar{z}^{\bar{i}}, P_R\chi^{\bar{i}},\bar{F}^{\bar{i}}\},\\
&& S_0 = (s_0,-i\sqrt{2}P_L\chi^0,-2F_0,0,+i\mathcal{D}_{\mu}s_0,0,0) = \{ s_0, P_L\chi^0,F_0\},\\
&& \bar{S}_0 = (\bar{s}_0,+i\sqrt{2}P_R\chi^0,0,-2\bar{F}_0,-i\mathcal{D}_{\mu}\bar{s}_0,0,0) = \{\bar{s}_0, P_R\chi^0,\bar{F}_0\},\\
&& \bar{\lambda}P_L\lambda = (\bar{\lambda}P_L\lambda,-i\sqrt{2}P_L\Lambda,2D_-^2,0,+i\mathcal{D}_{\mu}(\bar{\lambda}P_L\lambda),0,0) = \{\bar{\lambda}P_L\lambda, P_L\Lambda,-D_-^2\},\\
&& \bar{\lambda}P_R\lambda = (\bar{\lambda}P_R\lambda,+i\sqrt{2}P_R\Lambda,0,2D_+^2,-i\mathcal{D}_{\mu}(\bar{\lambda}P_R\lambda),0,0) = \{\bar{\lambda}P_R\lambda, P_R\Lambda,-D_+^2\},\\
&& (V)_D = (D,\cancel{\mathcal{D}}\lambda,0,0,\mathcal{D}^{b}\hat{F}_{ab},-\cancel{\mathcal{D}}\cancel{\mathcal{D}}\lambda,-\square^CD),
\end{eqnarray}
where
\begin{eqnarray}
&& P_L\Lambda \equiv \sqrt{2}P_L(-\frac{1}{2}\gamma\cdot \hat{F} + iD)\lambda,\qquad P_R\Lambda \equiv \sqrt{2}P_R(-\frac{1}{2}\gamma\cdot \hat{F} - iD)\lambda,\\
&& D_-^2 \equiv D^2 - \hat{F}^-\cdot\hat{F}^- - 2  \bar{\lambda}P_L\cancel{\mathcal{D}}\lambda,\qquad D_+^2 \equiv D^2 - \hat{F}^+\cdot\hat{F}^+ - 2  \bar{\lambda}P_R\cancel{\mathcal{D}}\lambda,\\
&& \mathcal{D}_{\mu}\lambda \equiv \bigg(\partial_{\mu}-\frac{3}{2}b_{\mu}+\frac{1}{4}w_{\mu}^{ab}\gamma_{ab}-\frac{3}{2}i\gamma_*\mathcal{A}_{\mu}\bigg)\lambda - \bigg(\frac{1}{4}\gamma^{ab}\hat{F}_{ab}+\frac{1}{2}i\gamma_* D\bigg)\psi_{\mu}
\\
 && \hat{F}_{ab} \equiv F_{ab} + e_a^{~\mu}e_b^{~\nu} \bar{\psi}_{[\mu}\gamma_{\nu]}\lambda,\qquad F_{ab} \equiv e_a^{~\mu}e_b^{~\nu} (2\partial_{[\mu}A_{\nu]}),\\
 && \hat{F}^{\pm}_{\mu\nu} \equiv \frac{1}{2}(\hat{F}_{\mu\nu}\pm \tilde{\hat{F}}_{\mu\nu}), \qquad \tilde{\hat{F}}_{\mu\nu} \equiv -\frac{1}{2} i\epsilon_{\mu\nu\rho\sigma}\hat{F}^{\rho\sigma} .
\end{eqnarray}

\subsection{\texorpdfstring{$w'^2,\bar{w}'^2$}{} Composite complex multiplets: (Weyl/Chiral) weights \texorpdfstring{$= (-1,3)$}{} and \texorpdfstring{$(-1,-3)$}{}}

We show here the components of the first superconformal {\it composite} complex multiplets $w'^2$ and $\Bar{w}'^2$ with Weyl/chiral weights $(-1,3)$ and $(-1,-3)$ respectively. These composite multiplets are defined to be
\begin{eqnarray}
&& w'^2 \equiv \frac{\bar{\lambda}P_L\lambda}{(S_0\bar{S}_0e^{-K/3})^2} = \{\mathcal{C}_w,\mathcal{Z}_w,\mathcal{H}_w,\mathcal{K}_w,\mathcal{B}^w_{\mu},\Lambda_w,\mathcal{D}_w\} \\
&& \bar{w}'^2 \equiv \frac{\bar{\lambda}P_R\lambda}{(S_0\bar{S}_0e^{-K/3})^2}
= \{\mathcal{C}_{\bar{w}},\mathcal{Z}_{\bar{w}},\mathcal{H}_{\bar{w}},\mathcal{K}_{\bar{w}},\mathcal{B}^{\bar{w}}_{\mu},\Lambda_{\bar{w}},\mathcal{D}_{\bar{w}}\}.
\end{eqnarray}
where
\begin{eqnarray}
\mathcal{C}_w &=& h \equiv \frac{\bar{\lambda}P_L\lambda}{(s_0\bar{s}_0e^{-K(z,\bar{z})/3})^2},\\
\mathcal{Z}_w &=& i\sqrt{2}(-h_a\Omega^a + h_{\bar{a}}\Omega^{\bar{a}}),\\
\mathcal{H}_w &=& -2h_aF^a + h_{ab}\bar{\Omega}^a\Omega^b,\\ 
\mathcal{K}_w &=& -2h_{\bar{a}}F^{\bar{a}} + h_{\bar{a}\bar{b}}\bar{\Omega}^{\bar{a}}\Omega^{\bar{b}},\\ 
\mathcal{B}^w_{\mu} &=& ih_a\mathcal{D}_{\mu}X^a-ih_{\bar{a}}\mathcal{D}_{\mu}\bar{X}^{\bar{a}}+ih_{a\bar{b}}\bar{\Omega}^{a}\gamma_{\mu}\Omega^{\bar{b}},\\ 
P_L\Lambda_w &=& -\sqrt{2}ih_{\bar{a}b}[(\cancel{\mathcal{D}}X^b)\Omega^{\bar{a}}-F^{\bar{a}}\Omega^b]-\frac{i}{\sqrt{2}}h_{\bar{a}\bar{b}c}\Omega^c\bar{\Omega}^{\bar{a}}\Omega^{\bar{b}},\\
P_R\Lambda_w &=& \sqrt{2}ih_{a\bar{b}}[(\cancel{\mathcal{D}}\bar{X}^{\bar{b}})\Omega^{a}-F^{a}\Omega^{\bar{b}}]+\frac{i}{\sqrt{2}}h_{ab\bar{c}}\Omega^{\bar{c}}\bar{\Omega}^{a}\Omega^{b},\\
\mathcal{D}_w &=& 2h_{a\bar{b}}\Big(-\mathcal{D}_{\mu}X^a\mathcal{D}^{\mu}\bar{X}^{\bar{b}}-\frac{1}{2}\bar{\Omega}^aP_L\cancel{\mathcal{D}}\Omega^{\bar{b}}-\frac{1}{2}\bar{\Omega}^{\bar{b}}P_R\cancel{\mathcal{D}}\Omega^a+F^aF^{\bar{b}}\Big) \nonumber\\
&&+h_{ab\bar{c}}(-\bar{\Omega}^a\Omega^bF^{\bar{c}}+\bar{\Omega}^a(\cancel{\mathcal{D}}X^b)\Omega^{\bar{c}})+ h_{\bar{a}\bar{b}c}(-\bar{\Omega}^{\bar{a}}\Omega^{\bar{b}}F^{c}+\bar{\Omega}^{\bar{a}}(\cancel{\mathcal{D}}\bar{X}^{\bar{b}})\Omega^{c}) \nonumber\\
&&+ \frac{1}{2}h_{ab\bar{c}\bar{d}}(\bar{\Omega}^aP_L\Omega^b)(\bar{\Omega}^{\bar{c}}P_R\Omega^{\bar{d}}).
\end{eqnarray}
Notice that when finding the multiplet $\bar{w}'^2$, we can just replace $h$ by its complex conjugate $h^{*}$.

\subsection{\texorpdfstring{$T(\bar{w}'^2),\bar{T}(w'^2)$}{} chiral projection multiplets: (Weyl/Chiral) weights \texorpdfstring{$=(0,0)$}{}}

The second superconformal multiplets that we need are the {\it composite} chiral projection multiplets $T(\Bar{w}'^2)$ and $\Bar{T}(w'^2)$ with Weyl/chiral weights $(0,0)$. From their component supermultiplets defined by 
\begin{eqnarray}
T(\bar{w}'^2) &=& \left( -\frac{1}{2}\mathcal{K}_{\bar{w}}, -\frac{1}{2} \sqrt{2} iP_L (\cancel{\mathcal{D}}\mathcal{Z}_{\bar{w}}+\Lambda_{\bar{w}}), \frac{1}{2}(\mathcal{D}_{\bar{w}}+\square^C \mathcal{C}_{\bar{w}} + i\mathcal{D}_a \mathcal{B}^a_{\bar{w}}) \right),\\
\bar{T}(w'^2) &=& \left( -\frac{1}{2}\mathcal{K}_{\bar{w}}^{*}, \frac{1}{2} \sqrt{2} iP_R (\cancel{\mathcal{D}}\mathcal{Z}_{\bar{w}}^C+\Lambda_{\bar{w}}^C), \frac{1}{2}(\mathcal{D}_{\bar{w}}^{*}+\square^C \mathcal{C}_{\bar{w}}^{*} - i\mathcal{D}_a (\mathcal{B}^a_{\bar{w}})^{*}) \right)
\end{eqnarray}
we find the corresponding superconformal multiplets and their complex conjugates as follows:
\begin{eqnarray}
T \equiv T(\bar{w}'^2) &=& \{\mathcal{C}_T,\mathcal{Z}_T,\mathcal{H}_T,\mathcal{K}_T,\mathcal{B}_{\mu}^T,\Lambda_T,\mathcal{D}_T\} \nonumber\\
\bar{T} \equiv \bar{T}(w'^2) &=& \{\mathcal{C}_{\bar{T}},\mathcal{Z}_{\bar{T}},\mathcal{H}_{\bar{T}},\mathcal{K}_{\bar{T}},\mathcal{B}_{\mu}^{\bar{T}},\Lambda_{\bar{T}},\mathcal{D}_{\bar{T}}\},
\end{eqnarray}
whose superconformal components are given by
\begin{eqnarray}
\mathcal{C}_T &=&  -\frac{1}{2} \mathcal{K}_{\bar{w}} = h^{*}_{\bar{a}}F^{\bar{a}} -\frac{1}{2} h^{*}_{\bar{a}\bar{b}}\bar{\Omega}^{\bar{a}}\Omega^{\bar{b}} \equiv C_T\\
\mathcal{Z}_T &=& -\sqrt{2}iP_L\bigg[\cancel{\mathcal{D}}(-h^{*}_a\Omega^a + h^{*}_{\bar{a}}\Omega^{\bar{a}})-h^{*}_{\bar{a}b}[(\cancel{\mathcal{D}}X^b)\Omega^{\bar{a}}-F^{\bar{a}}\Omega^b]-\frac{1}{2}h^{*}_{\bar{a}\bar{b}c}\Omega^c\bar{\Omega}^{\bar{a}}\Omega^{\bar{b}}\nonumber\\
&&\qquad\qquad+h^{*}_{a\bar{b}}[(\cancel{\mathcal{D}}\bar{X}^{\bar{b}})\Omega^{a}-F^{a}\Omega^{\bar{b}}]+\frac{1}{2}h^{*}_{ab\bar{c}}\Omega^{\bar{c}}\bar{\Omega}^{a}\Omega^{b}\bigg] \equiv -\sqrt{2}iP_L\Omega_T ,\\
\mathcal{H}_T &=& -2\bigg[h^{*}_{a\bar{b}}\Big(-\mathcal{D}_{\mu}X^a\mathcal{D}^{\mu}\bar{X}^{\bar{b}}-\frac{1}{2}\bar{\Omega}^aP_L\cancel{\mathcal{D}}\Omega^{\bar{b}}-\frac{1}{2}\bar{\Omega}^{\bar{b}}P_R\cancel{\mathcal{D}}\Omega^a+F^aF^{\bar{b}}\Big) \nonumber\\
&&+\frac{1}{2}h^{*}_{ab\bar{c}}(-\bar{\Omega}^a\Omega^bF^{\bar{c}}+\bar{\Omega}^a(\cancel{\mathcal{D}}X^b)\Omega^{\bar{c}})+\frac{1}{2} h^{*}_{\bar{a}\bar{b}c}(-\bar{\Omega}^{\bar{a}}\Omega^{\bar{b}}F^{c}+\bar{\Omega}^{\bar{a}}(\cancel{\mathcal{D}}\bar{X}^{\bar{b}})\Omega^{c}) \nonumber\\
&&+ \frac{1}{4}h^{*}_{ab\bar{c}\bar{d}}(\bar{\Omega}^aP_L\Omega^b)(\bar{\Omega}^{\bar{c}}P_R\Omega^{\bar{d}})+\frac{1}{2}\square^C h^{*} + \frac{1}{2}i\mathcal{D}^{\mu} (ih^{*}_a\mathcal{D}_{\mu}X^a-ih^{*}_{\bar{a}}\mathcal{D}_{\mu}\bar{X}^{\bar{a}}+ih^{*}_{a\bar{b}}\bar{\Omega}^{a}\gamma_{\mu}\Omega^{\bar{b}})\bigg] \nonumber\\
&\equiv& -2F_T,\\
\mathcal{K}_T &=& 0,\\
\mathcal{B}^T_{\mu} &=& -i\mathcal{D}_{\mu}\mathcal{C}_T,\\
\Lambda_T &=& 0 ,\\
\mathcal{D}_T &=& 0,
\end{eqnarray}
where we used $a,b,c,d = 0,i(\equiv z^i),W (\equiv \bar{\lambda}P_L\lambda)$. This gives the superfield components of the chiral projection multiplet $T$:
\begin{eqnarray}
T(\bar{w}'^2) = ( C_T, P_L\Omega_T, F_T )
\end{eqnarray}
where 
\begin{eqnarray}
C_T &=&  h^{*}_{\bar{a}}F^{\bar{a}} -\frac{1}{2} h^{*}_{\bar{a}\bar{b}}\bar{\Omega}^{\bar{a}}\Omega^{\bar{b}},\\
P_L\Omega_T &=& \cancel{\mathcal{D}}(-h^{*}_a\Omega^a + h^{*}_{\bar{a}}\Omega^{\bar{a}})-h^{*}_{\bar{a}b}[(\cancel{\mathcal{D}}X^b)\Omega^{\bar{a}}-F^{\bar{a}}\Omega^b]-\frac{1}{2}h^{*}_{\bar{a}\bar{b}c}\Omega^c\bar{\Omega}^{\bar{a}}\Omega^{\bar{b}}\nonumber\\
&&+h^{*}_{a\bar{b}}[(\cancel{\mathcal{D}}\bar{X}^{\bar{b}})\Omega^{a}-F^{a}\Omega^{\bar{b}}]+\frac{1}{2}h^{*}_{ab\bar{c}}\Omega^{\bar{c}}\bar{\Omega}^{a}\Omega^{b},\\
F_T &=&  h^{*}_{a\bar{b}}\Big(-\mathcal{D}_{\mu}X^a\mathcal{D}^{\mu}\bar{X}^{\bar{b}}-\frac{1}{2}\bar{\Omega}^aP_L\cancel{\mathcal{D}}\Omega^{\bar{b}}-\frac{1}{2}\bar{\Omega}^{\bar{b}}P_R\cancel{\mathcal{D}}\Omega^a+F^aF^{\bar{b}}\Big) \nonumber\\
&&+\frac{1}{2}h^{*}_{ab\bar{c}}(-\bar{\Omega}^a\Omega^bF^{\bar{c}}+\bar{\Omega}^a(\cancel{\mathcal{D}}X^b)\Omega^{\bar{c}})+\frac{1}{2} h^{*}_{\bar{a}\bar{b}c}(-\bar{\Omega}^{\bar{a}}\Omega^{\bar{b}}F^{c}+\bar{\Omega}^{\bar{a}}(\cancel{\mathcal{D}}\bar{X}^{\bar{b}})\Omega^{c}) \nonumber\\
&&+ \frac{1}{4}h^{*}_{ab\bar{c}\bar{d}}(\bar{\Omega}^aP_L\Omega^b)(\bar{\Omega}^{\bar{c}}P_R\Omega^{\bar{d}})+\frac{1}{2}\square^C h^{*} - \frac{1}{2}\mathcal{D}^{\mu} (h^{*}_a\mathcal{D}_{\mu}X^a-h^{*}_{\bar{a}}\mathcal{D}_{\mu}\bar{X}^{\bar{a}}+h^{*}_{a\bar{b}}\bar{\Omega}^{a}\gamma_{\mu}\Omega^{\bar{b}}). \nonumber\\{}
\end{eqnarray}

Morever,
\begin{eqnarray}
\bar{T}(w'^2) = \{ C_T^{*}, P_R\Omega_T, F_T^{*}\}
\end{eqnarray}
where
\begin{eqnarray}
C_T^{*} &=&  h_{a}F^{a} -\frac{1}{2} h_{ab}\bar{\Omega}^{a}\Omega^{b},\\
P_R\Omega_T &=& \cancel{\mathcal{D}}(-h_a\Omega^a + h_{\bar{a}}\Omega^{\bar{a}})-h_{\bar{a}b}[(\cancel{\mathcal{D}}X^b)\Omega^{\bar{a}}-F^{\bar{a}}\Omega^b]-\frac{1}{2}h_{\bar{a}\bar{b}c}\Omega^c\bar{\Omega}^{\bar{a}}\Omega^{\bar{b}}\nonumber\\
&&+h_{a\bar{b}}[(\cancel{\mathcal{D}}\bar{X}^{\bar{b}})\Omega^{a}-F^{a}\Omega^{\bar{b}}]+\frac{1}{2}h_{ab\bar{c}}\Omega^{\bar{c}}\bar{\Omega}^{a}\Omega^{b},\\
F_T^{*} &=&  h_{a\bar{b}}\Big(-\mathcal{D}_{\mu}X^a\mathcal{D}^{\mu}\bar{X}^{\bar{b}}-\frac{1}{2}\bar{\Omega}^aP_L\cancel{\mathcal{D}}\Omega^{\bar{b}}-\frac{1}{2}\bar{\Omega}^{\bar{b}}P_R\cancel{\mathcal{D}}\Omega^a+F^aF^{\bar{b}}\Big) \nonumber\\
&&+\frac{1}{2}h_{ab\bar{c}}(-\bar{\Omega}^a\Omega^bF^{\bar{c}}+\bar{\Omega}^a(\cancel{\mathcal{D}}X^b)\Omega^{\bar{c}})+\frac{1}{2} h_{\bar{a}\bar{b}c}(-\bar{\Omega}^{\bar{a}}\Omega^{\bar{b}}F^{c}+\bar{\Omega}^{\bar{a}}(\cancel{\mathcal{D}}\bar{X}^{\bar{b}})\Omega^{c}) \nonumber\\
&&+ \frac{1}{4}h_{ab\bar{c}\bar{d}}(\bar{\Omega}^aP_L\Omega^b)(\bar{\Omega}^{\bar{c}}P_R\Omega^{\bar{d}})+\frac{1}{2}\square^C h - \frac{1}{2}\mathcal{D}^{\mu} (h_a\mathcal{D}_{\mu}X^a-h_{\bar{a}}\mathcal{D}_{\mu}\bar{X}^{\bar{a}}+h_{a\bar{b}}\bar{\Omega}^{\bar{b}}\gamma_{\mu}\Omega^{a}). \nonumber\\{}
\end{eqnarray}

\subsection{Composite real multiplet \texorpdfstring{$\mathcal{R}$}{}: (Weyl/Chiral) weights \texorpdfstring{$=(0,0)$}{}}

We present here a superconformal composite real multiplet $\mathcal{R}$ with Weyl/chiral weights $(0,0)$. Defining some chiral multiplets $\mathcal{X}^A \equiv \{X^A,P_L\Omega^A,F^A\}$ where $A=\{ S_0,Z^i,\bar{\lambda}P_L\lambda,T(\bar{w}'^2)$\} and their conjugates, we represent the composite one $\mathcal{R}$ as
\begin{eqnarray}
\mathcal{R} \equiv (S_0\bar{S}_0e^{-K/3})^{-3} \frac{(\bar{\lambda}P_L\lambda)(\bar{\lambda}P_R\lambda)}{T(\bar{w}'^2) \bar{T}(w'^2)} 
\end{eqnarray}
whose lowest component is 
\begin{eqnarray}
\mathcal{C}_{\mathcal{R}}  \equiv (s_0\bar{s}_0e^{-K/3})^{-3}\frac{(\bar{\lambda}P_L\lambda)(\bar{\lambda}P_R\lambda)}{C_TC_{\bar{T}}}\equiv f(X^A,\bar{X}^{\bar{A}}) \label{def-f}
\end{eqnarray}
where $C_T = -D_+^2 \Delta^{-2}$; $C_{\bar{T}} = -D_-^2\Delta^{-2}$, and $\Delta \equiv s_0\bar{s}_0e^{-K/3}$. Then, the superconformal multiplet of the new Fayet-Iliopoulos term can be written by using
\begin{eqnarray}
\mathcal{R}\cdot (V)_D = \{\tilde{\mathcal{C}},\tilde{\mathcal{Z}},\tilde{\mathcal{H}},\tilde{\mathcal{K}},\tilde{\mathcal{B}}_{\mu},\tilde{\Lambda},\tilde{\mathcal{D}}\},
\end{eqnarray}
whose superconformal multiplet components are as follows:
\begin{eqnarray}
\tilde{\mathcal{C}} &=& Df,\\
\tilde{\mathcal{Z}} &=& f\cancel{\mathcal{D}}\lambda+Di\sqrt{2}(-f_{A}\Omega^A+f_{\bar{A}}\Omega^{\bar{A}}),\\
\tilde{\mathcal{H}} &=& D(-2f_AF^A + f_{AB}\bar{\Omega}^A\Omega^B)-i\sqrt{2}(-f_{A}\bar{\Omega}^A+f_{\bar{A}}\bar{\Omega}^{\bar{A}})P_L\cancel{\mathcal{D}}\lambda,\\
\tilde{\mathcal{K}} &=& D(-2f_{\bar{A}}F^{\bar{A}} + f_{\bar{A}\bar{B}}\bar{\Omega}^{\bar{A}}\Omega^{\bar{B}})-i\sqrt{2}(-f_{A}\bar{\Omega}^A+f_{\bar{A}}\bar{\Omega}^{\bar{A}})P_R\cancel{\mathcal{D}}\lambda,\\
\tilde{\mathcal{B}} &=& (\mathcal{D}^{\nu}\hat{F}_{\mu\nu})f+D(if_A\mathcal{D}_{\mu}X^A-if_{\bar{A}}\mathcal{D}_{\mu}\bar{X}^{\bar{A}}+if_{A\bar{B}}\bar{\Omega}^A\gamma_{\mu}\Omega^{\bar{B}}),\\
\tilde{\Lambda} &=& -f\cancel{\mathcal{D}}\cancel{\mathcal{D}}\lambda + D(P_L\Lambda^f+P_R\Lambda^f)+\frac{1}{2}\Big(\gamma_*(-f_A\cancel{\mathcal{D}}X^A+f_{\bar{A}}\cancel{\mathcal{D}}\bar{X}^{\bar{A}}-f_{A\bar{B}}\bar{\Omega}^A\cancel{\gamma}\Omega^{\bar{B}})\nonumber\\
&&+P_L(-2f_{\bar{A}}F^{\bar{A}} + f_{\bar{A}\bar{B}}\bar{\Omega}^{\bar{A}}\Omega^{\bar{B}})+P_R(-2f_AF^A + f_{AB}\bar{\Omega}^A\Omega^B) -\cancel{\mathcal{D}}f\Big)\cancel{\mathcal{D}}\lambda\nonumber\\
&& +\frac{1}{2}\Big( i\gamma_*\gamma^{\mu}\mathcal{D}^{\nu}\hat{F}_{\mu\nu}   -\cancel{\mathcal{D}}D\Big)i\sqrt{2}(-f_{A}\Omega^A+f_{\bar{A}}\Omega^{\bar{A}}) ,\\
\tilde{\mathcal{D}} &=&-f\square^C D + D 
\bigg\{ 2f_{A\bar{B}}(-\mathcal{D}_{\mu}X^A\mathcal{D}^{\mu}\bar{X}^{\bar{B}}-\frac{1}{2}\bar{\Omega}^AP_L\cancel{\mathcal{D}}\Omega^{\bar{B}}-\frac{1}{2}\cancel{\Omega}^{\bar{B}}P_R\cancel{\mathcal{D}}\Omega^A+F^AF^{\bar{B}}) \nonumber\\
&&+f_{AB\bar{C}}(-\bar{\Omega}^A\Omega^B F^{\bar{C}} + \bar{\Omega}^A(\cancel{\mathcal{D}}X^B)\Omega^{\bar{C}}) 
+f_{\bar{A}\bar{B}C}(-\bar{\Omega}^{\bar{A}}\Omega^{\bar{B}} F^C + \bar{\Omega}^{\bar{A}}(\cancel{\mathcal{D}}\bar{X}^{\bar{B}})\Omega^C) \nonumber\\
&&+\frac{1}{2}f_{AB\bar{C}\bar{D}} (\bar{\Omega}^AP_L\Omega^B)(\bar{\Omega}^{\bar{C}}P_R\Omega^{\bar{D}}) \bigg\}\nonumber\\
&& -(\mathcal{D}_{\nu}\hat{F}^{\mu\nu})(if_A\mathcal{D}_{\mu}X^A-if_{\bar{A}}\mathcal{D}_{\mu}\bar{X}^{\bar{A}}+if_{A\bar{B}}\bar{\Omega}^A\gamma_{\mu}\Omega^{\bar{B}}) \nonumber\\
&& + \bigg( \sqrt{2}if_{\bar{A}B}[(\cancel{\mathcal{D}}X^B)\Omega^{\bar{A}}-F^{\bar{A}}\Omega^B]  +\frac{i}{\sqrt{2}}f_{\bar{A}\bar{B}C} \Omega^C\bar{\Omega}^{\bar{A}}\Omega^{\bar{B}} \bigg)\cancel{\mathcal{D}}\lambda\nonumber\\
&& - \bigg( \sqrt{2}if_{A\bar{B}}[(\cancel{\mathcal{D}}\bar{X}^{\bar{B}})\Omega^{A}-F^{A}\Omega^{\bar{B}}]  +\frac{i}{\sqrt{2}}f_{AB\bar{C}} \Omega^{\bar{C}}\bar{\Omega}^{A}\Omega^{B} \bigg)\cancel{\mathcal{D}}\lambda\nonumber\\
&&-(\mathcal{D}_{\mu}f)(\mathcal{D}^{\mu}D)-\frac{1}{2}\cancel{\mathcal{D}}[i\sqrt{2}(-f_{A}\Omega^A+f_{\bar{A}}\Omega^{\bar{A}})](\cancel{\mathcal{D}}\lambda)+\frac{1}{2}i\sqrt{2}(-f_{A}\Omega^A+f_{\bar{A}}\Omega^{\bar{A}})(\cancel{\mathcal{D}}\cancel{\mathcal{D}}\lambda),\nonumber\\{}
\end{eqnarray}
where the indices $A,B,C,D$ run over $0,i,W,T$. The component action of the new FI term is then given by the D-term density formula 
\begin{eqnarray}
\mathcal{L}_{NEW} \equiv -[\xi \mathcal{R}\cdot (V)_D]_D &=& -\frac{\xi}{4}\int d^4x e \bigg[  \tilde{\mathcal{D}} -\frac{1}{2}\bar{\psi}\cdot \gamma i\gamma_* \tilde{\Lambda} -\frac{1}{3}\tilde{\mathcal{C}}R(\omega)\nonumber\\
&&+\frac{1}{6}\Big(\tilde{\mathcal{C}}\bar{\psi}_{\mu}\gamma^{\mu\rho\sigma}-i\bar{\tilde{\mathcal{Z}}}\gamma^{\rho\sigma}\gamma_*\Big)R'_{\rho\sigma}(Q)\nonumber\\
&&+\frac{1}{4}\varepsilon^{abcd}\bar{\psi}_{a}\gamma_b\psi_c\Big(\tilde{\mathcal{B}}_{d}-\frac{1}{2}\bar{\psi}_d\tilde{\mathcal{Z}}\Big)\bigg]+\textrm{h.c.}.
\end{eqnarray}

\subsection{Bosonic term of the new FI term}

The new FI term is obtained from the term $D^2f_{W\bar{W}}F^WF^{\bar{W}}$ inside the D-term $\tilde{\mathcal{D}}$. Thus, we get
\begin{eqnarray}
 \mathcal{L}_{NEW} &\supset& -\xi D f_{W\bar{W}}F^WF^{\bar{W}} = -\xi D \frac{ (s_0\bar{s}_0e^{-K/3})^{-3}}{C_TC_T^{*}}F^WF^{\bar{W}}
 \supset
-\xi D\frac{ (s_0\bar{s}_0e^{-K/3})^{-3}}{(h_WF^W)(h^{*}_{\bar{W}}F^{\bar{W}})}F^WF^{\bar{W}}
\nonumber\\&=& - \xi D\frac{ (s_0\bar{s}_0e^{-K/3})^{-3}}{(s_0\bar{s}_0e^{-K/3})^{-4}F^{W}F^{\bar{W}}}F^WF^{\bar{W}} = -\xi D(s_0\bar{s}_0e^{-K/3}).
\end{eqnarray}
Hence, in the superconformal gauge ($s_0\bar{s}_0e^{-K/3}=M_{pl}^2=1$), we obtain 
\begin{eqnarray}
 \mathcal{L}_{\textrm{new FI}}/e = -\xi D,
\end{eqnarray}
or
\begin{eqnarray}
 \mathcal{L}_{\textrm{new FI}}/e = -M_{pl}^2\xi D.
\end{eqnarray}

\section{Spectroscophy for Non-renormalizable Interactions in the New FI Terms}\label{Spectroscophy}

In this section, we carefully analyze suppression of the nonrenormalizable fermionic interactions from the new FI term from the perspective of effective field theory. To do this, as we did in the previous section of the liberated supergravity, we also need to recall the EFT expansion reviewed in Sec. \ref{EFT_expansion}, and consider the structure of EFT expansion in Eq.~\eqref{EFT_alternative} and constraint given in Eq.~\eqref{EFT_constraint}. Hence, we have
\begin{eqnarray}
\mathcal{L}_{EFT} \supset \sum_{\delta \geq 0}^{\textrm{Finite N}} \frac{C^{\delta}}{M^{\delta-d}} \mathcal{O}^{(\delta)} \sim \sum_{\delta \geq 0}^{\textrm{Finite N}} \frac{1}{\Lambda_{cut}^{\delta-d}} \mathcal{O}^{(\delta)}  \implies \frac{C^{\delta}}{M^{\delta-d}} \lesssim \frac{1}{\Lambda_{cut}^{\delta-d}}, 
\end{eqnarray}
where $\Lambda_{cut}$ is a cutoff scale; $M$ is a characteristic mass scale of a theory; $C^{\delta}$ is a dimensionless Wilson coefficient, and $\mathcal{O}^{(\delta)}$ is an effective field operator with the mass dimension $\delta$. In the case of the new FI terms, we will see that the constraint depends on the nw FI terms $\xi$ and thus the auxiliary field $D$ of the vector multiplet. The key equation we will see in this section is given by
\begin{eqnarray}
 \Lambda_{cut} \lesssim H^{\alpha},\nonumber
\end{eqnarray}
where $H$ is the Hubble scale, and $\alpha$ is a parameter that depends on the mass dimension of the effective field operator. Since $\Lambda_{cut},~ H\sim 10^{-5}M_{pl} < M_{pl}=1$, we explore whether $\alpha <1$ holds. After all, we will find that the maximum of $\alpha$ is 2/3.

In particular, for the later use, we generalize the new FI constant $\xi$ into a real gauge-invariant function $\mathcal{U}(z,\bar{z})$ of matter fields. Thus, for ACIK FI term, we can just put $\mathcal{U}(z,\bar{z}) \equiv \xi$. First, after solving the equations of motion for the auxiliary fields and fixing the superconformal gauge, the component multiplets ``on-shell'' are given by 
\begin{eqnarray}
&& Z^i  = \{ z^i, P_L\Omega^i,F^i\},\\
&& S_0  = \{ s_0, \quad \frac{1}{3}e^{K/6}K_iP_L\Omega^i,\quad F_0\},\\
&& W \equiv \bar{\lambda}P_L\lambda = \Big\{\bar{\lambda}P_L\lambda,\quad \sqrt{2}P_L(-\frac{1}{2}\gamma\cdot \hat{F} + iD)\lambda,\quad -D^2 + \hat{F}^-\cdot\hat{F}^- \Big\},
\end{eqnarray}
where
\begin{eqnarray}
 && \hat{F}_{ab} \equiv F_{ab} + e_a^{~\mu}e_b^{~\nu} \bar{\psi}_{[\mu}\gamma_{\nu]}\lambda,\qquad F_{ab} \equiv e_a^{~\mu}e_b^{~\nu} (2\partial_{[\mu}A_{\nu]}),\\
 && \hat{F}^{\pm}_{\mu\nu} \equiv \frac{1}{2}(\hat{F}_{\mu\nu}\pm \tilde{\hat{F}}_{\mu\nu}), \qquad \tilde{\hat{F}}_{\mu\nu} \equiv -\frac{1}{2} i\epsilon_{\mu\nu\rho\sigma}\hat{F}^{\rho\sigma},
\end{eqnarray}
and the solutions for the auxiliary fields are identified as
\begin{eqnarray}
D&=& \mathcal{U} + \frac{1}{\mathcal{U}^2}
\bigg[
\bigg(
-F^0\mathcal{U}e^{-K/6}  + \mathcal{U}_IF^I +\frac{1}{3}\mathcal{U}K_IF^I
\bigg)(\bar{\lambda}P_L\lambda)  -i 
\frac{\mathcal{U}_I}{\sqrt{2}}(\bar{\Omega}^IP_L\lambda) + h.c.
\bigg]\nonumber\\
&&+\textrm{higher order terms},\\
F^0 &=& e^{2K/3}\bar{W} - \frac{1}{3}e^{K/6}(\bar{\lambda}P_R\lambda),\\
F^{\bar{J}} &=&- 3e^{K/2}G^{I\bar{J}}\nabla_IW -G^{I\bar{J}}\Big(9\frac{\mathcal{U}_I }{\mathcal{U}}+ 3K_I\Big)(\bar{\lambda}P_L\lambda)
\end{eqnarray}
Plus, the relevant chiral projection multiplet $T(\bar{w}'^2)$ is given by
\begin{eqnarray}
T(\bar{w}'^2) &=& ( C_T, P_L\Omega_T, F_T ),\\
 C_T &=&  h^{*}_{\bar{a}}F^{\bar{a}} -\frac{1}{2} h^{*}_{\bar{a}\bar{b}}\bar{\Omega}^{\bar{a}}\Omega^{\bar{b}},\\
P_L\Omega_T &=& \cancel{\mathcal{D}}(-h^{*}_a\Omega^a + h^{*}_{\bar{a}}\Omega^{\bar{a}})-h^{*}_{\bar{a}b}[(\cancel{\mathcal{D}}X^b)\Omega^{\bar{a}}-F^{\bar{a}}\Omega^b]-\frac{1}{2}h^{*}_{\bar{a}\bar{b}c}\Omega^c\bar{\Omega}^{\bar{a}}\Omega^{\bar{b}}\nonumber\\
&&+h^{*}_{a\bar{b}}[(\cancel{\mathcal{D}}\bar{X}^{\bar{b}})\Omega^{a}-F^{a}\Omega^{\bar{b}}]+\frac{1}{2}h^{*}_{ab\bar{c}}\Omega^{\bar{c}}\bar{\Omega}^{a}\Omega^{b},\\
F_T &=&  h^{*}_{a\bar{b}}\Big(-\mathcal{D}_{\mu}X^a\mathcal{D}^{\mu}\bar{X}^{\bar{b}}-\frac{1}{2}\bar{\Omega}^aP_L\cancel{\mathcal{D}}\Omega^{\bar{b}}-\frac{1}{2}\bar{\Omega}^{\bar{b}}P_R\cancel{\mathcal{D}}\Omega^a+F^aF^{\bar{b}}\Big) \nonumber\\
&&+\frac{1}{2}h^{*}_{ab\bar{c}}(-\bar{\Omega}^a\Omega^bF^{\bar{c}}+\bar{\Omega}^a(\cancel{\mathcal{D}}X^b)\Omega^{\bar{c}})+\frac{1}{2} h^{*}_{\bar{a}\bar{b}c}(-\bar{\Omega}^{\bar{a}}\Omega^{\bar{b}}F^{c}+\bar{\Omega}^{\bar{a}}(\cancel{\mathcal{D}}\bar{X}^{\bar{b}})\Omega^{c}) \nonumber\\
&&+ \frac{1}{4}h^{*}_{ab\bar{c}\bar{d}}(\bar{\Omega}^aP_L\Omega^b)(\bar{\Omega}^{\bar{c}}P_R\Omega^{\bar{d}})+\frac{1}{2}\square^C h^{*} - \frac{1}{2}\mathcal{D}^{\mu} (h^{*}_a\mathcal{D}_{\mu}X^a-h^{*}_{\bar{a}}\mathcal{D}_{\mu}\bar{X}^{\bar{a}}+h^{*}_{a\bar{b}}\bar{\Omega}^{a}\gamma_{\mu}\Omega^{\bar{b}}),\nonumber\\{}
\end{eqnarray}
where  $C_T,C_{\bar{T}} = -D^2 \Delta^{-2} + \textrm{fermions including } \lambda$; $\Delta \equiv s_0\bar{s}_0e^{-K/3}$; $a = 0,i,W(=\bar{\lambda}P_L\lambda)$, and
\begin{eqnarray}
 h^* \equiv \frac{\bar{\lambda}P_R\lambda}{(s_0\bar{s}_0e^{-K(z,\bar{z})/3})^2}.
\end{eqnarray}
We note that in this chiral projection multiplet, the terms ``$h^*_{\bar{W}}\bar{F}^{\bar{W}}$ of $C_T$,'' ``$h^*_{\bar{W}k}\bar{F}^{\bar{W}}\Omega^{k}$ of $P_L\Omega_T$,'' and ``$F^k\bar{F}^{\bar{W}}$ and $h^*_{jk\bar{W}}\bar{F}^{\bar{W}}\bar{\Omega}^j\Omega^k$ of $F_T$'' (where $j,k=0,i$) have no $\lambda$, while the other terms must have at least one $\lambda$.

In summary, the components of the multiplets can be found to be
\begin{eqnarray}
 &&S_0:\quad s_0 \sim e^{K/6},\quad  \cancel{\mathcal{D}}s_0 \sim
 K_i\mathcal{O}_i^{(2)}+\mathcal{O}^{(1)}+K_i\mathcal{O}^{(4)}_{i\bar{\psi}}
 ,  \nonumber\\
 && \qquad P_L\Omega^0 \sim K_ie^{K/6}\mathcal{O}^{(3/2)}_i, \quad F^0 \sim We^{2K/3},\\
 &&Z^i:\quad z^i \sim z^i ,\quad \cancel{\mathcal{D}}z^i\sim \mathcal{O}^{(2)}+\mathcal{O}^{(4)}_{\bar{\psi}}, \quad P_L\Omega^i \sim \mathcal{O}^{(3/2)}_i , \quad F^i \sim F^i,\\
 &&W:\quad (\Bar{\lambda}P_L\lambda) \sim \mathcal{O}^{(3)}_{\lambda\lambda
 }, \quad \cancel{\mathcal{D}}(\Bar{\lambda}P_L\lambda) \sim \mathcal{O}^{(4)}_{\Bar{\lambda}\lambda} + \mathcal{O}^{(5)}_{\bar{\psi}\lambda}+D\mathcal{O}^{(3)}_{\Bar{\psi}\lambda},
 \nonumber\\
 &&\qquad \Omega^W \sim \mathcal{O}^{(7/2)}_{\lambda} + D \mathcal{O}^{3/2}_{\lambda}, \quad F^W \sim D^2 + \mathcal{O}^{(4)}\nonumber\\{}\\
&&T:\quad  C_T \supset  h^*_{\bar{W}}\bar{F}^{\bar{W}}|_{\textrm{lowest}} \sim D^2 + DK_i \mathcal{O}^{(3)}_{i\lambda} + K_i \mathcal{O}^{(5)}_{i\lambda},\\
&&{}\qquad\cancel{\mathcal{D}}C_T \sim \cancel{\partial}C_T \sim DD_i \mathcal{O}^{(2)}_i + D_i K_j\mathcal{O}^{(5)}_{ij\lambda} + DK_{ij}\mathcal{O}^{(5)}_{ij\lambda} +K_{ij}\mathcal{O}^{(7)}_{ij\lambda} \\
&&{}\qquad P_L \Omega_T \supset  P_L\cancel{\mathcal{D}}\mathcal{Z}_{\Bar{w}}+ h^*_{\bar{W}k}\bar{F}^{\bar{W}}\Omega^{k}|_{\textrm{lowest}}\sim D^2K_i\mathcal{O}^{(3/2)}_i + D\mathcal{O}_{\lambda}^{(5/2)}+K_i\mathcal{O}^{(11/2)}_{i\lambda},\\
&&{}\qquad F_T \supset F^k\bar{F}^{\bar{W}}+h^*_{jk\bar{W}}\bar{F}^{\bar{W}}\bar{\Omega}^j\Omega^k |_{\textrm{lowest}} \sim
D^2 F^k + D^2(K'^2+K'')\mathcal{O}^{(3)}
\end{eqnarray}
where we used the following superconformal covariant derivatives of the lowest component scalar $\mathcal{C}$ with the weights $(w,c)$ and the first fermion $\mathcal{Z}$ in a superconformal multiplet:
\begin{eqnarray}
 \mathcal{D}_{\mu}\mathcal{C} &=& (\partial_{\mu}-wb_{\mu}-icA_{\mu})\mathcal{C}-\frac{i}{2}\bar{\psi}_{\mu}\gamma_*\mathcal{Z},\\
P_L\mathcal{D}_{\mu}\mathcal{Z} &=&
\Big(\partial_{\mu}-(w+1/2)b_{\mu}-i(c-3/2)A_{\mu}+\frac{1}{4}\omega_{\mu}^{ab}\gamma_{ab}\Big)\mathcal{Z} \nonumber\\
&&-\frac{1}{2}P_L(i\mathcal{H}-\gamma^a\mathcal{B}_a-i\gamma^a\mathcal{D}_a\mathcal{C})\psi_{\mu} -i(w+c)P_L\phi_{\mu}\mathcal{C}
\end{eqnarray}

The full action of the new FI term ``on-shell'' (i.e. $\mathcal{D}^{\nu}\hat{F}_{\mu\nu}=0,~ \cancel{\mathcal{D}}\lambda=0$) is given by
\begin{eqnarray}
\mathcal{L}_{\textrm{new FI}} \equiv -[ \mathcal{R}\cdot (V)_D]_D &=& -\frac{1}{4}\int d^4x e \bigg[  \tilde{\mathcal{D}} -\frac{1}{2}\bar{\psi}\cdot \gamma i\gamma_* \tilde{\Lambda} -\frac{1}{3}\tilde{\mathcal{C}}R(\omega)\nonumber\\
&&+\frac{1}{6}\Big(\tilde{\mathcal{C}}\bar{\psi}_{\mu}\gamma^{\mu\rho\sigma}-i\bar{\tilde{\mathcal{Z}}}\gamma^{\rho\sigma}\gamma_*\Big)R'_{\rho\sigma}(Q)\nonumber\\
&&+\frac{1}{4}\varepsilon^{abcd}\bar{\psi}_{a}\gamma_b\psi_c\Big(\tilde{\mathcal{B}}_{d}-\frac{1}{2}\bar{\psi}_d\tilde{\mathcal{Z}}\Big)\bigg]+\textrm{h.c.},\label{action}
\end{eqnarray}
where the relevant components are given, for $f \equiv  (s_0\bar{s}_0e^{-K(z^i,\bar{z}^i)/3})^{-3}\frac{(\bar{\lambda}P_L\lambda)(\bar{\lambda}P_R\lambda)}{C_TC_{\bar{T}}}\mathcal{U}(z^i,\bar{z}^i)$ (where $C_T,C_{\bar{T}} = -D^2 \Delta^{-2} + \textrm{fermions including } \lambda$; $\Delta \equiv s_0\bar{s}_0e^{-K/3}$), as follows
\begin{eqnarray}
\tilde{\mathcal{C}} &=& Df,\\
\tilde{\mathcal{Z}} &=& Di\sqrt{2}(-f_{A}\Omega^A+f_{\bar{A}}\Omega^{\bar{A}}),\\
\tilde{\mathcal{H}} &=& D(-2f_AF^A + f_{AB}\bar{\Omega}^A\Omega^B),\\
\tilde{\mathcal{K}} &=& D(-2f_{\bar{A}}F^{\bar{A}} + f_{\bar{A}\bar{B}}\bar{\Omega}^{\bar{A}}\Omega^{\bar{B}}),\\
\tilde{\mathcal{B}} &=& D(if_A\mathcal{D}_{\mu}X^A-if_{\bar{A}}\mathcal{D}_{\mu}\bar{X}^{\bar{A}}+if_{A\bar{B}}\bar{\Omega}^A\gamma_{\mu}\Omega^{\bar{B}}),\\
\tilde{\Lambda} &=&   D\bigg(  -\sqrt{2}if_{\bar{A}B}[(\cancel{\mathcal{D}}X^B)\Omega^{\bar{A}}-F^{\bar{A}}\Omega^B]-\frac{i}{\sqrt{2}}f_{\bar{A}\bar{B}C}\Omega^C\bar{\Omega}^{\bar{A}}\Omega^{\bar{B}} \nonumber\\
  &&+ \sqrt{2}if_{A\bar{B}}[(\cancel{\mathcal{D}}\bar{X}^{\bar{B}})\Omega^{A}-F^{A}\Omega^{\bar{B}}]+\frac{i}{\sqrt{2}}f_{AB\bar{C}}\Omega^{\bar{C}}\bar{\Omega}^{A}\Omega^{B}\bigg) \nonumber\\
  &&+\frac{1}{2}\Big( -\cancel{\mathcal{D}}D\Big)i\sqrt{2}(-f_{A}\Omega^A+f_{\bar{A}}\Omega^{\bar{A}}) ,\\
\tilde{\mathcal{D}} &=&-f\square^C D + D 
\bigg\{ 2f_{A\bar{B}}(-\mathcal{D}_{\mu}X^A\mathcal{D}^{\mu}\bar{X}^{\bar{B}}-\frac{1}{2}\bar{\Omega}^AP_L\cancel{\mathcal{D}}\Omega^{\bar{B}}-\frac{1}{2}\cancel{\Omega}^{\bar{B}}P_R\cancel{\mathcal{D}}\Omega^A+F^AF^{\bar{B}}) \nonumber\\
&&+f_{AB\bar{C}}(-\bar{\Omega}^A\Omega^B F^{\bar{C}} + \bar{\Omega}^A(\cancel{\mathcal{D}}X^B)\Omega^{\bar{C}}) 
+f_{\bar{A}\bar{B}C}(-\bar{\Omega}^{\bar{A}}\Omega^{\bar{B}} F^C + \bar{\Omega}^{\bar{A}}(\cancel{\mathcal{D}}\bar{X}^{\bar{B}})\Omega^C) \nonumber\\
&&+\frac{1}{2}f_{AB\bar{C}\bar{D}} (\bar{\Omega}^AP_L\Omega^B)(\bar{\Omega}^{\bar{C}}P_R\Omega^{\bar{D}}) \bigg\}-(\mathcal{D}_{\mu}f)(\mathcal{D}^{\mu}D),
\end{eqnarray}
where the indices $A,B,C,D$ run over $0,i,W,T$. Plus, we have 
\begin{eqnarray}
 && \square^CD= e^{a\mu}[\partial_{\mu}\mathcal{D}_aD -3b_{\mu}\mathcal{D}_aD + \omega_{\mu ab} \mathcal{D}^b D + 4 f_{\mu a}D], \nonumber\\
 && \mathcal{D}_aD|_{on-shell}=e_a^{\mu}(\partial_{\mu}-2b_{\mu})D, \quad \delta D = 2\lambda_D D,\nonumber\\
 && \mathcal{D}_aD |_{b=0}=  \mathcal{U}_i \partial_{\mu}z^i \sim \mathcal{U}_i \mathcal{O}^{(2)} \textrm{  and  } D e^{a\mu} f_{\mu a} = -D\frac{R}{12} \sim D\mathcal{O}^{(2)}.
\end{eqnarray}
Since we consider $D|_{bos}=\mathcal{U}$ as a quadratic function of matter fields, we can assume that $\mathcal{U}_i \sim 0$ at the minimum of the potential. Then, the components reduce to
\begin{eqnarray}
\tilde{\mathcal{C}} &=& Df,\\
\tilde{\mathcal{Z}} &=& Di\sqrt{2}(-f_{A}\Omega^A+f_{\bar{A}}\Omega^{\bar{A}}),\\
\tilde{\mathcal{H}} &=& D(-2f_AF^A + f_{AB}\bar{\Omega}^A\Omega^B),\\
\tilde{\mathcal{K}} &=& D(-2f_{\bar{A}}F^{\bar{A}} + f_{\bar{A}\bar{B}}\bar{\Omega}^{\bar{A}}\Omega^{\bar{B}}),\\
\tilde{\mathcal{B}} &=& D(if_A\mathcal{D}_{\mu}X^A-if_{\bar{A}}\mathcal{D}_{\mu}\bar{X}^{\bar{A}}+if_{A\bar{B}}\bar{\Omega}^A\gamma_{\mu}\Omega^{\bar{B}}),\\
\tilde{\Lambda} &=&   D\bigg(  -\sqrt{2}if_{\bar{A}B}[(\cancel{\mathcal{D}}X^B)\Omega^{\bar{A}}-F^{\bar{A}}\Omega^B]-\frac{i}{\sqrt{2}}f_{\bar{A}\bar{B}C}\Omega^C\bar{\Omega}^{\bar{A}}\Omega^{\bar{B}} \nonumber\\
  &&+ \sqrt{2}if_{A\bar{B}}[(\cancel{\mathcal{D}}\bar{X}^{\bar{B}})\Omega^{A}-F^{A}\Omega^{\bar{B}}]+\frac{i}{\sqrt{2}}f_{AB\bar{C}}\Omega^{\bar{C}}\bar{\Omega}^{A}\Omega^{B}\bigg),\\
\tilde{\mathcal{D}} &=&\frac{DRf}{3}+ D 
\bigg\{ 2f_{A\bar{B}}(-\mathcal{D}_{\mu}X^A\mathcal{D}^{\mu}\bar{X}^{\bar{B}}-\frac{1}{2}\bar{\Omega}^AP_L\cancel{\mathcal{D}}\Omega^{\bar{B}}-\frac{1}{2}\cancel{\Omega}^{\bar{B}}P_R\cancel{\mathcal{D}}\Omega^A+F^AF^{\bar{B}}) \nonumber\\
&&+f_{AB\bar{C}}(-\bar{\Omega}^A\Omega^B F^{\bar{C}} + \bar{\Omega}^A(\cancel{\mathcal{D}}X^B)\Omega^{\bar{C}}) 
+f_{\bar{A}\bar{B}C}(-\bar{\Omega}^{\bar{A}}\Omega^{\bar{B}} F^C + \bar{\Omega}^{\bar{A}}(\cancel{\mathcal{D}}\bar{X}^{\bar{B}})\Omega^C) \nonumber\\
&&+\frac{1}{2}f_{AB\bar{C}\bar{D}} (\bar{\Omega}^AP_L\Omega^B)(\bar{\Omega}^{\bar{C}}P_R\Omega^{\bar{D}}) \bigg\},
\end{eqnarray}

Now we represent the possible interaction terms using the following expression
\begin{eqnarray}
\mathcal{L}_{\textrm{new FI}} &\supset& 
\partial_{i}^m \partial_0^c \partial_W^b \partial_T^p \bigg[\frac{W\Bar{W}}{T\Bar{T}}Y\bigg]\cdot D \cdot \hat{\mathcal{O}}^{(\mathcal{D}_0;\Lambda_{\Bar{\psi}3/2};\mathcal{C}_{R(\omega)2};\mathcal{C}_{\Bar{\psi}R'(Q)4}; \mathcal{Z}_{R'(Q)5/2};\mathcal{B}_{\Bar{\psi}\psi 3};\mathcal{Z}_{\Bar{\psi}\psi\Bar{\psi}9/2})} \nonumber\\
&&\times  \{ (\cancel{D}S_0)^{c_1}(\Omega^0)^{c_2}(F^0)^{c_3}\}\{ (\cancel{D}z^i)^{m_1}(\Omega^i)^{m_2}(F^i)^{m_3}\} \nonumber\\
&&\times \{ (\cancel{D}W)^{b_1}(\Omega^W)^{b_2}(F^W)^{b_3}\}
\{ (\cancel{D}T)^{p_1}(\Omega^T)^{p_2}(F^T)^{p_3}\},
\end{eqnarray}
where $\Delta \equiv s_0\bar{s}_0e^{-K/3}$; $Y\equiv \Delta^{-3}\mathcal{U}(z,\Bar{z})$, and several parameters are defined as the number of derivatives with respect to the corresponding variable, such as $c=c_1+c_2+c_3$, $m=m_1+m_2+m_3$, $b=b_1+b_2+b_3 \leq 2$ (Here, the number $b$ must be constrained by ``2'' because each of the bilinear terms $W$ and $\Bar{W}$ appears only once in the numerator.), and $p=p_1+p_2+p_3$. Also, $\Gamma \equiv m+c+b+p \leq 4$ since we can take the four field-derivatives at most. The notation $\Lambda_{\Bar{\psi}3/2}$ means that the operator $\hat{\mathcal{O}}$ is given by the dimension-(3/2) operator $\Bar{\psi}$ which couples to the $\tilde{\Lambda}$ of the new FI term.

Plus, the operator $\hat{\mathcal{O}}$ denotes the additional contribution to the effective operator from the couplings to gravitino $\psi_{\mu}$ or curvatures $R(\omega)$ and $R'(Q)$ in the final action of new FI term in Eq.~\eqref{action}.

Then, taking the derivatives and inserting $T \sim D^2$, we obtain
\begin{eqnarray}
 \mathcal{L}_{\textrm{new FI}}|_{\textrm{with no numerical factors}} &\supset& 
 D^{-3-2p}(\partial^{m+c}Y)\mathcal{O}^{(2\delta_{\Gamma,0})}_R\mathcal{O}^{(6_{4\lambda}-3_{2\lambda}b)}\cdot  \hat{\mathcal{O}} \nonumber\\
&&\times  \{ (\cancel{D}S_0)^{c_1}(\Omega^0)^{c_2}(F^0)^{c_3}\}\{ (\cancel{D}z^i)^{m_1}(\Omega^i)^{m_2}(F^i)^{m_3}\} \nonumber\\
&&\times \{ (\cancel{D}W)^{b_1}(\Omega^W)^{b_2}(F^W)^{b_3}\}
\{ (\cancel{D}T)^{p_1}(\Omega^T)^{p_2}(F^T)^{p_3}\},
\end{eqnarray}
where we define $\delta_{\Gamma,0}=1$ if $\Gamma =0$, and $W^{1-b_{2\lambda}}\Bar{W}^{1-\bar{b}_{2\lambda}} \equiv \mathcal{O}^{(3_{2\lambda}(1-b_{2\lambda})+3_{2\lambda}(1-\bar{b}_{2\lambda}))} \equiv \mathcal{O}^{(6_{4\lambda}-3_{2\lambda}b)}$; that is, $6_{4\lambda}$ means that the dimension ``6'' comes from the four $\lambda$-fermions. Next, we introduce particular parameters that constitute the powers of the terms like $c_1,c_2,c_3$ in detail. Since the component actions depend only on $\cancel{\mathcal{D}}z^i$, $P_L\Omega^i$, and $F^i$, we define $d_i,f_i,a_i$ as the powers of ``each term'' (like $\mathcal{O}^{(7/2)}_{\lambda}$ and $D \mathcal{O}^{3/2}_{\lambda}$ of $\Omega^W$) within the expression of $\cancel{\mathcal{D}}z^i$ (``d''erivative), $P_L\Omega^i$ (``f''ermion), and $F^i$ (``a''uxiliary field). Equivalently, we define the correponding parameters as $d_s,f_s,a_s$, $d_W,f_W,a_W$, and $d_T,f_T,a_T$. Then, we find 
\begin{eqnarray}
 \{ (\cancel{D}S_0)^{c_1}(\Omega^0)^{c_2}(F^0)^{c_3}\} \sim K'^{d_{s1}+d_{s3}^{\psi}+f_{s1}}{F^0}^{a_{s1}}\mathcal{O}^{(2d_{s1}+d_{s2}+4d_{s3}^{\psi}+\frac{3}{2}f_{s1})},
\end{eqnarray}
where $d_{s1}+d_{s2}+d_{s3}^{\psi}+f_{s1}+a_{s1} =c_1+c_2+c_3 =c \leq 4$. For example, if $d_{s3}^{\psi}=2$ in the effective operator $\mathcal{O}$, then it means that there are two gravitinos in the operator. The next one is 
\begin{eqnarray}
 \{ (\cancel{D}z^i)^{m_1}(\Omega^i)^{m_2}(F^i)^{m_3}\} \sim {F^i}^{a_{z1}}\mathcal{O}^{(2d_{z1}+4d_{z2}^{\psi}+\frac{3}{2}f_{z1})},
\end{eqnarray}
where $d_{z1}+d_{z2}^{\psi}+f_{z1}+a_{z1}=m_1+m_2+m_3=m \leq 4$. The $W$ case is 
\begin{eqnarray}
  \{ (\cancel{D}W)^{b_1}(\Omega^W)^{b_2}(F^W)^{b_3}\} \sim 
  D^{d_{W3}^{\psi\lambda}+f_{W2}^{\lambda}+2a_{W1}} \mathcal{O}^{(4d_{W1}^{\lambda\lambda}+5d_{W2}^{\psi\lambda}+3d_{W3}^{\psi\lambda}+\frac{7}{2}f_{W1}^{\lambda}+\frac{3}{2}f_{W2}^{\lambda}+4a_{W2})},
\end{eqnarray}
where $d_{W1}^{\lambda\lambda}+d_{W2}^{\psi\lambda}+d_{W3}^{\psi\lambda}+f_{W1}^{\lambda}+f_{W2}^{\lambda}+a_{W1}+a_{W2} =b_1+b_2 +b_3= b\leq2$. As an example, also, if $d_{W1}^{\lambda\lambda}=2$ in $\mathcal{O}$, then it means that there are four $\lambda$ fermions in the effective operator. Next, the chiral projection multiplet $T$ case is given by
\begin{eqnarray}
 \{ (\cancel{D}T)^{p_1}(\Omega^T)^{p_2}(F^T)^{p_3}\} &\sim& 
 D^{d_{T1}+d_{T3}^{\lambda}+2f_{T1}+f_{T2}^{\lambda}+2a_{T1}+2a_{T2}}
 D_i^{d_{T1}+d_{T2}^{\lambda}} \nonumber\\
 && \times K'^{d_{T2}^{\lambda}+f_{T1}+f_{T3}^{\lambda}+2a_{T2}}
 K''^{d_{T3}^{\lambda}+d_{T4}^{\lambda}+a_{T2}} {F^i}^{a_{T1}} \nonumber\\
 &&\times \mathcal{O}^{(2d_{T1}+5d_{T2}^{\lambda}+5d_{T3}^{\lambda}+7d_{T4}^{\lambda}+\frac{3}{2}f_{T1}+\frac{5}{2}f_{T2}^{\lambda}+\frac{11}{2}f_{T3}^{\lambda}+3 a_{T2})},
\end{eqnarray}
where $d_{T1}+d_{T2}^{\lambda}+d_{T3}^{\lambda}+d_{T4}^{\lambda}+f_{T1}+f_{T2}^{\lambda}+f_{T3}^{\lambda}+a_{T1}+a_{T2}=p_1+p_2+p_3=p\leq 4$. In addition, we can also get 
\begin{eqnarray}
 \partial^{m+c}Y &=&  \partial^{m+c} \Big(\Delta^{-3} \mathcal{U}(z,\Bar{z}) \Big) = \partial^s \Big( (s_0\Bar{s})^{-3} \Big) \partial^m (e^K\mathcal{U}) \nonumber\\
 && \sim (K')^{k_1}(K'')^{k_2}(K''')^{k_3}(K'''')^{k_1} (\mathcal{U})^{u_0}(\mathcal{U}^{(1)})^{u_1}(\mathcal{U}^{(2)})^{u_2}(\mathcal{U}^{(3)})^{u_3}(\mathcal{U}^{(4)})^{u_4},
\end{eqnarray}
where $k_1+k_2+k_3+k_4 + u_1 + u_2 + u_3 + u_4 = m$ (here, we define $k_i,u_i$ to be present only when $i\leq  m$ and $0\leq k_i\leq m$, $0\leq u_i \leq 1$) and $u_0 =\delta_{m,k_{tot}}$ (which is defined as a Kronecker delta) for $k_{tot}\equiv k_1+k_2+k_3+k_4 $.  In particular, we can consider $\mathcal{U} \sim D$. Combining these results, we can rewrite the terms as
\begin{eqnarray}
 \mathcal{L}_{\textrm{new FI}} &\supset& (K')^{k_1}(K'')^{k_2}(K''')^{k_3}(K'''')^{k_1} (\mathcal{U})^{u_0}(\mathcal{U}^{(1)})^{u_1}(\mathcal{U}^{(2)})^{u_2}(\mathcal{U}^{(3)})^{u_3}(\mathcal{U}^{(4)})^{u_4}\nonumber\\
 && \times 
 D^{-3-2p}(\partial^{m+c}Y)\mathcal{O}^{(2n_R)}_R\mathcal{O}^{(6_{4\lambda}-3_{2\lambda}b)}\cdot  \hat{\mathcal{O}} \nonumber\\
&&\times K'^{d_{s1}+d_{s3}^{\psi}+f_{s1}}{F^0}^{a_{s1}}\mathcal{O}^{(2d_{s1}+d_{s2}+4d_{s3}^{\psi}+\frac{3}{2}f_{s1})} \nonumber\\
&&\times {F^i}^{a_{z1}}\mathcal{O}^{(2d_{z1}+4d_{z2}^{\psi}+\frac{3}{2}f_{z1})} \nonumber\\
&&\times  D^{d_{W3}^{\psi\lambda}+f_{W2}^{\lambda}+2a_{W1}} \mathcal{O}^{(4d_{W1}^{\lambda\lambda}+5d_{W2}^{\psi\lambda}+3d_{W3}^{\psi\lambda}+\frac{7}{2}f_{W1}^{\lambda}+\frac{3}{2}f_{W2}^{\lambda}+4a_{W2})}\nonumber\\
&&\times D^{d_{T1}+d_{T3}^{\lambda}+2f_{T1}+f_{T2}^{\lambda}+2a_{T1}+2a_{T2}}
 D_i^{d_{T1}+d_{T2}^{\lambda}} \nonumber\\
 && \times K'^{d_{T2}^{\lambda}+f_{T1}+f_{T3}^{\lambda}+2a_{T2}}
 K''^{d_{T3}^{\lambda}+d_{T4}^{\lambda}+a_{T2}} {F^i}^{a_{T1}} \nonumber\\
 &&\times \mathcal{O}^{(2d_{T1}+5d_{T2}^{\lambda}+5d_{T3}^{\lambda}+7d_{T4}^{\lambda}+\frac{3}{2}f_{T1}+\frac{5}{2}f_{T2}^{\lambda}+\frac{11}{2}f_{T3}^{\lambda}+3 a_{T2})}.
\end{eqnarray}

The simplified expression can be given by
\begin{eqnarray}
 \mathcal{L}_{\textrm{new FI}} &\supset&
 (K''')^{k_3}(K'''')^{k_1} (\mathcal{U}^{(1)})^{\alpha}(\mathcal{U}^{(2)})^{u_2}(\mathcal{U}^{(3)})^{u_3}(\mathcal{U}^{(4)})^{u_4} \nonumber\\
 &&\times (D)^{-n}(F)^{r}(K')^{\beta}(K'')^{\gamma} \mathcal{O}^{(\delta_{pre})}\hat{\mathcal{O}}^{(\delta_{add})},
\end{eqnarray}
where 
\begin{eqnarray}
 n &=&-u_0+ 3+2p-(d_{W3}^{\psi\lambda}+f_{W2}^{\lambda}+2a_{W1}+d_{T1}+d_{T3}^{\lambda}+2f_{T1}+f_{T2}^{\lambda}+2a_{T1}+2a_{T2}) \nonumber\\
 &=& -u_0+3 + 2(d_{T1}+d_{T2}^{\lambda}+d_{T3}^{\lambda}+d_{T4}^{\lambda}+f_{T1}+f_{T2}^{\lambda}+f_{T3}^{\lambda}+a_{T1}+a_{T2}) \nonumber\\ &&-(d_{W3}^{\psi\lambda}+f_{W2}^{\lambda}+2a_{W1}+d_{T1}+d_{T3}^{\lambda}+2f_{T1}+f_{T2}^{\lambda}+2a_{T1}+2a_{T2}) \nonumber\\
 &=&-u_0+ 3 + d_{T1} + 2d_{T2}^{\lambda} + d_{T3}^{\lambda} + 2d_{T4}^{\lambda} - d_{W3}^{\psi\lambda} + f_{T2}^{\lambda} + 2f_{T3}^{\lambda} -f_{W2}^{\lambda}-2a_{W1},\\
 r &=& a_{s1} +a_{z1} + a_{T1}.
 \end{eqnarray}
\begin{eqnarray}
 \alpha &=& u_1+d_{T1} + d_{T2}^{\lambda},\\
 \beta &=& k_1+ d_{s1}+d_{s3}^{\psi}+f_{s1}+d_{T2}^{\lambda}+f_{T1}+f_{T3}^{\lambda}+2a_{T2},\\
  \gamma &=&k_2 + d_{T3}^{\lambda}+d_{T4}^{\lambda}+a_{T2}
\end{eqnarray}
\begin{eqnarray}
 \delta_{pre} &=& 2\delta_{\Gamma,0}+6_{4\lambda}-3_{2\lambda}b +2d_{s1}+d_{s2}+4d_{s3}^{\psi}+\frac{3}{2}f_{s1}+2d_{z1}+4d_{z2}^{\psi}+\frac{3}{2}f_{z1} \nonumber\\
 && + 4d_{W1}^{\lambda\lambda}+5d_{W2}^{\psi\lambda}+3d_{W3}^{\psi\lambda}+\frac{7}{2}f_{W1}^{\lambda}+\frac{3}{2}f_{W2}^{\lambda}+4a_{W2} \nonumber\\
 && +2d_{T1}+5d_{T2}^{\lambda}+5d_{T3}^{\lambda}+7d_{T4}^{\lambda}+\frac{3}{2}f_{T1}+\frac{5}{2}f_{T2}^{\lambda}+\frac{11}{2}f_{T3}^{\lambda}+3 a_{T2}
\end{eqnarray}
\begin{eqnarray}
 \delta_{add} &=& \textrm{one of  } \mathcal{D}_0;\Lambda_{\Bar{\psi}3/2};\mathcal{C}_{R(\omega)2};\mathcal{C}_{\Bar{\psi}R'(Q)4}; \mathcal{Z}_{R'(Q)5/2};\mathcal{B}_{\Bar{\psi}\psi 3};\mathcal{Z}_{\Bar{\psi}\psi\Bar{\psi}9/2}.
\end{eqnarray}
Plus, we define a parameter $\Gamma$ as the number of derivatives among $\partial_i^m$, $\partial_0^c$, $\partial_W^b$, and $\partial_T^p$ as follows:
\begin{eqnarray}
\Gamma \equiv c+m+b+p&=& d_{s1}+d_{s2}+d_{s3}^{\psi}+f_{s1}+a_{s1}+d_{z1}+d_{z2}^{\psi}+f_{z1}+a_{z1} \nonumber\\
&& +d_{W1}^{\lambda\lambda}+d_{W2}^{\psi\lambda}+d_{W3}^{\psi\lambda}+f_{W1}^{\lambda}+f_{W2}^{\lambda}+a_{W1}+a_{W2} \nonumber\\
&& +d_{T1}+d_{T2}^{\lambda}+d_{T3}^{\lambda}+d_{T4}^{\lambda}+f_{T1}+f_{T2}^{\lambda}+f_{T3}^{\lambda}+a_{T1}+a_{T2} \leq 4.
\end{eqnarray}
Furthermore, we define selection rules for the parameter set $\{(d,f,a)\}$ such that $\Gamma=\sum_{all} (d+f+a)$:
\begin{eqnarray}
\textrm{For  } (d,f,a),
&& \Tilde{\mathcal{C}}:\quad (0,0,0) \implies \Gamma(\Tilde{\mathcal{C}})=0 \qquad \delta_{add} = 2,4,\nonumber\\
&& \Tilde{\mathcal{Z}}:\quad {\color{red}(0, 1,0)}\implies \Gamma(\Tilde{\mathcal{Z}})=1\qquad \delta_{add} =5/2,~9/2,\nonumber\\
&& \Tilde{\mathcal{B}_a}:\quad {\color{orange} (1,0,0)},{\color{blue}(0,2,0)}\implies \Gamma(\Tilde{\mathcal{B}_a})=1,2\qquad \delta_{add} =3,\nonumber\\
&& \Tilde{\Lambda}:\quad {\color{red}(1,1,0)},{\color{red}(1,1,1)},{\color{blue}(0,3,0)}\implies \Gamma(\Tilde{\Lambda})=2,3\qquad \delta_{add} =3/2,\nonumber\\
&& \Tilde{\mathcal{D}}:\quad (0,0,0),{\color{orange}(2,0,0)},{\color{blue}(0,2,0)},{\color{orange}(0,0,2)},\nonumber\\
&& \qquad \qquad {\color{blue}(0,2,1)},{\color{blue}(1,2,0)},{\color{blue}(0,4,0)}\implies \Gamma(\Tilde{\mathcal{D}})=0,2,3,4\qquad \delta_{add} =0. \nonumber\label{selection_rule}
\end{eqnarray}
For example, when we consider the case $\Gamma=1$ (which means that we look at the terms coupling to $f_A$ in the action \eqref{action}.), then we can only consider the following selections $(1,0,0)$ (i.e. only one $d$ and others are zero.) from $\tilde{\mathcal{B}}_a$ or $(0,1,0)$ (i.e. only one $f$ and others are zero.) from $\tilde{\mathcal{Z}}$, and the couplings to $R'_{\rho\sigma}(Q)$ and $\Bar{\psi}\psi$ in the action \eqref{action}.

Since we are interested in $\mathcal{U}$ as a quadratic function of matter fields, we have $u_3=u_4=0$, so that 
\begin{eqnarray}
 \mathcal{L}_{\textrm{new FI}} &\supset&
 (K''')^{k_3}(K'''')^{k_1} (\mathcal{U}^{(1)})^{\alpha}(\mathcal{U}^{(2)})^{u_2}\nonumber\\
 &&\times (D)^{-n}(F)^{r}(K')^{\beta}(K'')^{\gamma} \mathcal{O}^{(\delta_{pre})}\hat{\mathcal{O}}^{(\delta_{add})}.
\end{eqnarray}
Since we consider the case of $D \sim F$, it reduces to 
\begin{eqnarray}
 \mathcal{L}_{\textrm{new FI}} &\supset&
 (K''')^{k_3}(K'''')^{k_1} (\mathcal{U}^{(1)})^{\alpha}(\mathcal{U}^{(2)})^{u_2}\nonumber\\
 &&\times (D)^{-n+r}(K')^{\beta}(K'')^{\gamma} \mathcal{O}^{(\delta_{pre})}\hat{\mathcal{O}}^{(\delta_{add})}.
\end{eqnarray}
In particular, we can make these terms to be small due to $\mathcal{U}^{(1)} \sim z^i \sim 0$ along the inflationary trajectory, we can set $\alpha = u_1+d_{T1} + d_{T2}^{\lambda} =0$, which means $u_1=d_{T1}= d_{T2}^{\lambda}=0$. Moreover, the derivatives of the K\"{a}hler potential with respect to the matter fields can be small as well if it includes the quadratic function of some matter fields. Of course, the second derivative cannot be neglected like the case of $\mathcal{U}^{(2)}$. However, the moduli fields will not be small due to the logarithmic dependence in the K\"{a}hler potential. Nevertheless, it is not a problem to ignore the derivatives here because they will be suppressed for large inflaton field, while they will be of order of $\mathcal{O}(1)$ in Planck unit after inflation. In any case, we can drop the dependence of the derivatives of the K\"{a}hler potential but the second derivatives. Furthermore, the second derivatives of the K\"{a}hler potential with respect to the moduli fields can be of order of $\mathcal{O}(1)$ after inflation, so that ignoring the factor of $K''$, we get 
\begin{eqnarray}
 \mathcal{L}_{\textrm{new FI}} &\supset&
(\mathcal{U}^{(2)})^{u_2} (D)^{-N} \mathcal{O}^{(\delta_{pre})}\hat{\mathcal{O}}^{(\delta_{add})} \equiv (\mathcal{U}^{(2)})^{u_2} (D)^{-N} \mathcal{O}^{(\delta_{tot})},
\end{eqnarray}
where $N\equiv n-r$ and
\begin{eqnarray}
 N &=& 3  + d_{T3}^{\lambda} + 2d_{T4}^{\lambda} + f_{T2}^{\lambda} + 2f_{T3}^{\lambda} \nonumber\\
 &&- d_{W3}^{\psi\lambda} -f_{W2}^{\lambda}-2a_{W1} -a_{s1} -a_{z1} - a_{T1}-u_0,\\
 \delta_{pre} &=& 2\delta_{\Gamma,0}+6_{4\lambda}-3_{2\lambda}b +2d_{s1}+d_{s2}+4d_{s3}^{\psi}+\frac{3}{2}f_{s1}+2d_{z1}+4d_{z2}^{\psi}+\frac{3}{2}f_{z1} \nonumber\\
 && + 4d_{W1}^{\lambda\lambda}+5d_{W2}^{\psi\lambda}+3d_{W3}^{\psi\lambda}+\frac{7}{2}f_{W1}^{\lambda}+\frac{3}{2}f_{W2}^{\lambda}+4a_{W2} \nonumber\\
 &&+5d_{T3}^{\lambda}+7d_{T4}^{\lambda}+\frac{3}{2}f_{T1}+\frac{5}{2}f_{T2}^{\lambda}+\frac{11}{2}f_{T3}^{\lambda}+3 a_{T2},\\
 \delta_{add} &=& \textrm{one of  } \mathcal{D}_0;\Lambda_{\Bar{\psi}3/2};\mathcal{C}_{R(\omega)2};\mathcal{C}_{\Bar{\psi}R'(Q)4}; \mathcal{Z}_{R'(Q)5/2};\mathcal{B}_{\Bar{\psi}\psi 3};\mathcal{Z}_{\Bar{\psi}\psi\Bar{\psi}9/2},\\
 \delta_{tot} &\equiv& \delta_{pre} + \delta_{add}.
\end{eqnarray}
In the meantime, we had $k_1+k_2+k_3+k_4  + u_1 + u_2 + u_3 + u_4 = m$ and $d_{z1}+d_{z2}^{\psi}+f_{z1}+a_{z1}=m_1+m_2+m_3=m \leq 4$. Since we now have $u_1=u_3=u_4=0$, we get 
\begin{eqnarray}
 u_2 =d_{z1}+d_{z2}^{\psi}+f_{z1}+a_{z1}-k_{tot},\label{u2_value}
\end{eqnarray}
where we define $k_{tot}\equiv k_1+k_2+k_3+k_4$. Here, we point out that if the mass-squared from $\mathcal{U}^{(2)}$ is of order of the electroweak scale, then the non-renormalizable terms can be suppressed strongly. If not, we have to keep it.

Looking at the above terms, we observe that the non-renormalizable terms will have the following form 
\begin{eqnarray}
\mathcal{L}_{\textrm{new FI}}  \supset  \frac{(\mathcal{U}^{(2)})^{u_2}}{D^N}  \mathcal{O}^{(\delta_{tot})} \implies  \frac{(\mathcal{U}^{(2)})^{u_2}}{D^N}  \lesssim \frac{1}{\Lambda_{cut}^{\delta_{tot}-4}} \implies \Lambda_{cut} \lesssim \left(\frac{D^N}{(\mathcal{U}^{(2)})^{u_2}}\right)^{1/(\delta_{tot}-4)}
\end{eqnarray}
where we set $M_{pl}=1$ for simplicity. Since we are interested in $D\sim \mathcal{U}^{(2)}\sim H$ where $H \sim 10^{-5}$ is the Hubble scale, the constraint can be given by
\begin{eqnarray}
 \Lambda_{cut} \lesssim H^{(N-u_2)/(\delta_{tot}-4)} = H^{\alpha}.
\end{eqnarray}
We note that in order for the non-renormalizable terms to well-suppressed for $H<1$, we have to require that 
\begin{eqnarray}
\alpha \equiv \frac{N-u_2}{\delta_{tot}-4} \leq 1,
\end{eqnarray}
where $\delta_{tot}-4>0$ for nonrenormalizable operators. In particular, the restrictions on the cutoff will come from the first condition only since the upperbound of the second condition exceeds $\mathcal{O}(1) \sim M_{pl} =1$. In the meantime, since we consider the non-renormalizable operators, we always have $\delta_{tot}>4$. That is, if there exists a combination of the parameters satisfying the above condition with maximal value 1, then the cutoff must be constrained as $\Lambda_{cut} \lesssim H$, which will be out of our interest.

Inserting $u_0=\delta_{m,k_{tot}}$ and the value of $u_2$ in Eq.~\eqref{u2_value}, the relevant parameters are as follow:
\begin{eqnarray}
 N -u_2&=& 3  + d_{T3}^{\lambda} + 2d_{T4}^{\lambda} + f_{T2}^{\lambda} + 2f_{T3}^{\lambda} \nonumber\\
 &&- d_{W3}^{\psi\lambda} -f_{W2}^{\lambda}-2a_{W1} -a_{s1} -a_{z1} - a_{T1}-u_0-u_2,\\
 \delta_{tot}-4 &=&2\delta_{\Gamma,0}+ 6_{4\lambda}-3_{2\lambda}b +2d_{s1}+d_{s2}+4d_{s3}^{\psi}+\frac{3}{2}f_{s1}+2d_{z1}+4d_{z2}^{\psi}+\frac{3}{2}f_{z1} \nonumber\\
 && + 4d_{W1}^{\lambda\lambda}+5d_{W2}^{\psi\lambda}+3d_{W3}^{\psi\lambda}+\frac{7}{2}f_{W1}^{\lambda}+\frac{3}{2}f_{W2}^{\lambda}+4a_{W2} \nonumber\\
 &&+5d_{T3}^{\lambda}+7d_{T4}^{\lambda}+\frac{3}{2}f_{T1}+\frac{5}{2}f_{T2}^{\lambda}+\frac{11}{2}f_{T3}^{\lambda}+3 a_{T2} \nonumber\\
 &&+\delta_{add}-4 \label{del-4}.
\end{eqnarray}
Here is a remark. Since the derivatives of $\mathcal{U}$ but $\mathcal{U}^{(2)}$ can give us the small number from the vacuum expectation values of matter fields, we consider the case of either $m=k_{tot}=d_{z1}+f_{z1}+a_{z1}+d_{z2}^{\psi}$ (which implies that $u_i=0$ for $i=1,2,3,4$ and $u_0=\delta_{m,k_{tot}}=1$) or $u_2=1$. Thus, it reduces to
\begin{eqnarray}
 N -u_2&=& 2 + d_{T3}^{\lambda} + 2d_{T4}^{\lambda} + f_{T2}^{\lambda} + 2f_{T3}^{\lambda} - d_{W3}^{\psi\lambda} -f_{W2}^{\lambda}-2a_{W1} -a_{s1} -a_{z1} - a_{T1}\label{N-u2}.
\end{eqnarray}
Specifically, we obtain the following different cases.
\begin{itemize}
    \item Case of $b=0$ (with no derivatives):
    \begin{eqnarray}
     (N -u_2)_{b=0}&=& 2-a_{s1} -a_{z1} - a_{T1} ,\\
    (\delta_{tot}-4)_{b=0} &=&2+ 6_{4\lambda} +\delta_{add}-4
    \end{eqnarray}
     \item Case of $b=0$ (with derivatives):
    \begin{eqnarray}
     (N -u_2)_{b=0}&=& 2-a_{s1} -a_{z1} - a_{T1} ,\\
    (\delta_{tot}-4)_{b=0} &=&6_{4\lambda} +2d_{s1}+d_{s2}+4d_{s3}^{\psi}+\frac{3}{2}f_{s1}+2d_{z1}
    \nonumber\\
    && +4d_{z2}^{\psi}+\frac{3}{2}f_{z1} +\frac{3}{2}f_{T1}+3 a_{T2}+\delta_{add}-4
    \end{eqnarray}
    \item  Case of $b=1$:
    \begin{eqnarray}
     (N -u_2)_{b=1=f_{W1}^{\lambda}}&=& 2 + d_{T3}^{\lambda} + 2d_{T4}^{\lambda} + f_{T2}^{\lambda} + 2f_{T3}^{\lambda}-a_{s1} -a_{z1} - a_{T1} ,\\
     (\delta_{tot}-4)_{b=1=f_{W1}^{\lambda}} &=&3_{2\lambda} +2d_{s1}+d_{s2}+4d_{s3}^{\psi}+\frac{3}{2}f_{s1}+2d_{z1}+4d_{z2}^{\psi}+\frac{3}{2}f_{z1} \nonumber\\
 &&+5d_{T3}^{\lambda}+7d_{T4}^{\lambda}+\frac{3}{2}f_{T1}+\frac{5}{2}f_{T2}^{\lambda}+\frac{11}{2}f_{T3}^{\lambda}+3 a_{T2} +\delta_{add}-\frac{1}{2},
    \end{eqnarray}
    \begin{eqnarray}
  (N -u_2)_{b=1=f_{W2}^{\lambda}}&=& 1  + d_{T3}^{\lambda} + 2d_{T4}^{\lambda} + f_{T2}^{\lambda} + 2f_{T3}^{\lambda} -a_{s1} -a_{z1} - a_{T1},\\
  (\delta_{tot}-4)_{b=1=f_{W2}^{\lambda}} &=& 3_{2\lambda} +2d_{s1}+d_{s2}+4d_{s3}^{\psi}+\frac{3}{2}f_{s1}+2d_{z1}+4d_{z2}^{\psi}+\frac{3}{2}f_{z1} \nonumber\\
 &&+5d_{T3}^{\lambda}+7d_{T4}^{\lambda}+\frac{3}{2}f_{T1}+\frac{5}{2}f_{T2}^{\lambda}+\frac{11}{2}f_{T3}^{\lambda}+3 a_{T2} +\delta_{add}-\frac{5}{2},
    \end{eqnarray}
    \item Case of $b=2$:
    \begin{eqnarray}
    (N -u_2)_{b=2=f_{W1}^{\lambda}} &=& 2  + d_{T3}^{\lambda} + 2d_{T4}^{\lambda} + f_{T2}^{\lambda} + 2f_{T3}^{\lambda}-a_{s1} -a_{z1} - a_{T1}  ,\\
    (\delta_{tot}-4)_{b=2=f_{W1}^{\lambda}} &=& 
    2d_{s1}+d_{s2}+4d_{s3}^{\psi}+\frac{3}{2}f_{s1}+2d_{z1}+4d_{z2}^{\psi}+\frac{3}{2}f_{z1}  \nonumber\\
 &&+5d_{T3}^{\lambda}+7d_{T4}^{\lambda}+\frac{3}{2}f_{T1}+\frac{5}{2}f_{T2}^{\lambda}+\frac{11}{2}f_{T3}^{\lambda}+3 a_{T2} +\delta_{add}+3,
    \end{eqnarray}
    \begin{eqnarray}
    (N -u_2)_{b=2=f_{W2}^{\lambda}} &=&  d_{T3}^{\lambda} + 2d_{T4}^{\lambda} + f_{T2}^{\lambda} + 2f_{T3}^{\lambda} -a_{s1} -a_{z1} - a_{T1},\\
    (\delta_{tot}-4)_{b=2=f_{W2}^{\lambda}} &=& 
    2d_{s1}+d_{s2}+4d_{s3}^{\psi}+\frac{3}{2}f_{s1}+2d_{z1}+4d_{z2}^{\psi}+\frac{3}{2}f_{z1} \nonumber\\
 &&+5d_{T3}^{\lambda}+7d_{T4}^{\lambda}+\frac{3}{2}f_{T1}+\frac{5}{2}f_{T2}^{\lambda}+\frac{11}{2}f_{T3}^{\lambda}+3 a_{T2} +\delta_{add}-1,
    \end{eqnarray}
    \begin{eqnarray}
    (N -u_2)_{b=2,f_{W1}^{\lambda}=f_{W2}^{\lambda}=1} &=&1  + d_{T3}^{\lambda} + 2d_{T4}^{\lambda} + f_{T2}^{\lambda} + 2f_{T3}^{\lambda} -a_{s1} -a_{z1} - a_{T1},\\
    (\delta_{tot}-4)_{b=2,f_{W1}^{\lambda}=f_{W2}^{\lambda}=1} &=& 
   2d_{s1}+d_{s2}+4d_{s3}^{\psi}+\frac{3}{2}f_{s1}+2d_{z1}+4d_{z2}^{\psi}+\frac{3}{2}f_{z1} \nonumber\\
 &&+5d_{T3}^{\lambda}+7d_{T4}^{\lambda}+\frac{3}{2}f_{T1}+\frac{5}{2}f_{T2}^{\lambda}+\frac{11}{2}f_{T3}^{\lambda}+3 a_{T2} +\delta_{add}+1. \nonumber\\{}
    \end{eqnarray}
\end{itemize}

The following parameter condition must be satisfied \begin{eqnarray}
\Gamma \equiv c+m+b+p&=& d_{s1}+d_{s2}+d_{s3}^{\psi}+f_{s1}+a_{s1}+\underbrace{d_{z1}+d_{z2}^{\psi}+f_{z1}+a_{z1}}_{=k_{tot} + u_0 + u_2} \nonumber\\
&& +\underbrace{d_{W1}^{\lambda\lambda}+d_{W2}^{\psi\lambda}+d_{W3}^{\psi\lambda}+f_{W1}^{\lambda}+f_{W2}^{\lambda}+a_{W1}+a_{W2}}_{=~ b\leq 2} \nonumber\\
&& +d_{T3}^{\lambda}+d_{T4}^{\lambda}+f_{T1}+f_{T2}^{\lambda}+f_{T3}^{\lambda}+a_{T1}+a_{T2} \quad \leq 4.
\end{eqnarray}
Again, $\Gamma$ is defined as the number of derivatives among $\partial_i^m$, $\partial_0^c$, $\partial_W^b$, and $\partial_T^p$. In the appendix \ref{alpha_cal}, the largest value of $\alpha$ is found by $\alpha =2/3$, so that
\begin{eqnarray}
\Lambda_{cut} \lesssim H^{2/3} \sim 10^{-5(2/3)} = 10^{-2.66} < M_S \sim 10^{-2.5}.
\end{eqnarray}
Hence, the theory of new FI terms is basically an effective field theory with broken supersymmetry.

%% file: chapters/8.tex
\framebox[1.05\width]{{\large This chapter is based on the author's original work in Ref.~\cite{jp4}.}} \par
\vspace{1cm}

In this chapter, the key result we will see is given by
\begin{eqnarray}
V = V_D + V_F - \mathcal{U}, \qquad \frac{1}{\mathcal{U}}
\lesssim \frac{1}{\Lambda_{cut}^4},
\end{eqnarray}
where $V_D,V_F$ are the supergravity D- and F-term scalar potentials and $\mathcal{U}$ is defined as a gauge-invariant generl real function of matter scalars $z$'s. In particular, we will discuss how to relax the scalar potential $V$ that can satisfy the constraint above. 

\section{Introduction}

The application of supergravity to inflationary cosmology has recently been of great interest and studied by many authors as discussed in Ch.~\ref{ch5-1}. However, it remains still challenging to build viable models of a certain phenomenology in the context of supergravity. For instance, it is not straightforward to realize both inflationary dynamics and minimal supersymmetric standard model (MSSM) at the same time in a unified setup. The first reason for the difficulty is due to the large hierarchy between Hubble scale $H\sim 10^{-5}M_{pl}$ for inflation and electroweak (TeV) scale of order $10^{-15}M_{pl}$ for the observable-sector dynamics of standard model (SM). The second is because {\it standard} supergravity predicts the complicated structure of the F-term scalar potential, i.e. $V_F=e^G(G_IG^{I\bar{J}}G_{\bar{J}}-3)$ where $G$ is the supergravity G-function defined by $G \equiv K + \ln W + \ln \bar{W}$ which consists of K\"{a}hler potential $K$ and superpotential $W$. This implies that one must always explore a proper choice of the supergravity G-function, which is unfortunately nontrivial in general. For these reasons, it is very demanding to construct phenomenologically-desirable scalar potentials within standard supergravity. 

The $\eta$ problem \cite{eta_problem} is an example of such a difficulty\footnote{This issue gives rise to a hardship for obtaining a very small slow-roll parameter $\eta$ such that $\eta \ll 1$ due to the exponentially growing behavior of the F-term potential.}. No-scale supergravity \cite{Ellis_no_scale} may be a solution to the $\eta$ problem because the corresponding F-term potential can exactly vanish, i.e. $V_F =0$. This is established by a clever choice of the supergravity G-function. In fact, the gravitino mass term ``$-3e^G$'' plays a critical role in the no-scale cancellation. Interestingly, one can easily have such no-scale structure through a logarithmic K\"{a}hler potential of the volume modulus fields and constant superpotential in string theory \cite{Ellis_no_scale,KKLT,KKLMMT}. However, certain choice of superpotential may spoil the ``exact'' cancellation of the F-term potential yielding a remnant as shown in Eq. (14) of Ref. \cite{FP}. Hence, no-scale supergravity is very sensitive to the given form of both K\"{a}hler potential and superpotential. Moreover, no-scale supergravity may cause a vast number of moduli, which correspond to degenerate vacua being along flat directions in scalar potential. This turns out that moduli stabilization, which is necessary to obtain a unique vacuum, is still required in no-scale supergravity. Thus, current no-scale supergravity is not a complete strategy for model building.

Recent developments in modification of the supergraivty scalar potential have been made, e.g. liberated supergravity recently proposed by Farakos, Kehagias, and Riotto \cite{fkr} and various types of new Fayet-Iliopoulos (FI) terms proposed by many authors \cite{acik,CFTV,oldACIK,Kuzenko,ar,AKK}. In particular, liberated supergravity was the first attempt to allows us to have a general scalar potential. In fact, inspired by liberated supergravity, we investigate such a general scalar potential in the other fashion in this work. However, it has recently been found that liberated supergravity is not liberated literally due to strong constraints on the general function \cite{jp1,jp3}. On the contrary, new FI terms can modify only D-term potentials, which still have non-trivial field dependence and can give us only the non-negative-definite contribution to the scalar potential. Consequently, the recent studies do not have full generality of scalar potential.

Obviously, it has very long been thought of that a negative-definite term in scalar potential can be given only by the gravitino-mass term ``$-3e^G$'' in the standard $\mathcal{N}=1$ supergravity, and is not present in global supersymmetry (SUSY) \cite{Superconformal_Freedman}. It is thus inevitable to acquire another type of cancellation in scalar potential through a new negative term so that we have a general scalar potential being beyond no-scale supergravity and the recent works. In that sense, it remains very intriguing to answer the following open questions: How can we obtain a new negative-definite potential term in supergravity? To what extent can we reform the supergravity scalar potential in a general fashion? We affirmatively answer these questions throughout this letter.

Our work is organized as follows. In Sec. \ref{Nogo}, we revisit the higher order corrections in the minimal supergravity models of inflation constructed by Ferrara, Kallosh, Linde, and Porrati (FKLP) \cite{HOC}. We firstly identify a no-go theorem for the higher order corrections. The no-go theorem is supported by the fact that canonical kinetic term of the gauge field in the vector multiplet must be present in the supergravity lagrangian. In Sec. \ref{Novel_class}, we propose how to relax the strongly-constrained standard form of the scalar potential by adding a special choice of higher order correction to the standard supergravity. Here we discover a new negative-definite scalar potential as a general function. In addition, we find essential constraints on the new negative term by inspecting the suppression of nonrenormalizable lagrangians to ensure that our theory is self-consistent as an effective field theory. This leads to a cutoff which is identified with the high-scale SUSY breaking mass $M_S$ \cite{High_SUSY}. Next, using the new negative term, we present a relaxing procedure for generating a general scalar potential\footnote{In this sense, our proposal reserves the name ``Relaxed supergravity.''}. Plus, we compare our theory with the liberated supergravity by Farakos, Kehagias, and Riotto. Next, we briefly discuss the global SUSY limit of relaxed supergravity. Finally, in Sec. \ref{con_o}, we summarize our findings and give outlook on this work.

\section{No-go Theorem for Higher Order Corrections in FKLP Model}\label{Nogo}

In this section, we revisit {\it higher order corrections} in the minimal supergravity models of inflation proposed by  Ferrara, Kallosh, Linde, and Porrati (FKLP) \cite{HOC}. First, we start with considering a vector multiplet $V$, its field strength multiplet $\bar{\lambda}P_L\lambda$, and real linear multiplet $(V)_D$ whose lowest component is given by the auxiliary field $D$ of the vector multiplet $V$ as follows:
\begin{eqnarray}
&& V = \{0,0,0,0,A_{\mu},\lambda,D\} ~\textrm{in the Wess-Zumino gauge,~i.e.}~v=\zeta=\mathcal{H}=0,\\
&& \bar{\lambda}P_L\lambda = (\bar{\lambda}P_L\lambda,-i\sqrt{2}P_L\Lambda,2D_-^2,0,+i\mathcal{D}_{\mu}(\bar{\lambda}P_L\lambda),0,0) = \{\bar{\lambda}P_L\lambda, P_L\Lambda,-D_-^2\},\\
&& \bar{\lambda}P_R\lambda = (\bar{\lambda}P_R\lambda,+i\sqrt{2}P_R\Lambda,0,2D_+^2,-i\mathcal{D}_{\mu}(\bar{\lambda}P_R\lambda),0,0) = \{\bar{\lambda}P_R\lambda, P_R\Lambda,-D_+^2\},\\
&& (V)_D = (D,\cancel{\mathcal{D}}\lambda,0,0,\mathcal{D}^{b}\hat{F}_{ab},-\cancel{\mathcal{D}}\cancel{\mathcal{D}}\lambda,-\square^CD),
\end{eqnarray}
where we have used the following notations\footnote{We follow the sign convention $(-,+,\cdots,+)$ for spacetime metric, and notations used in Ref. \cite{Superconformal_Freedman}:
\begin{eqnarray}
&& \mathcal{D}_{\mu}\lambda \equiv \bigg(\partial_{\mu}-\frac{3}{2}b_{\mu}+\frac{1}{4}w_{\mu}^{ab}\gamma_{ab}-\frac{3}{2}i\gamma_*\mathcal{A}_{\mu}\bigg)\lambda - \bigg(\frac{1}{4}\gamma^{ab}\hat{F}_{ab}+\frac{1}{2}i\gamma_* D\bigg)\psi_{\mu}\nonumber
\\
 && \hat{F}_{ab} \equiv F_{ab} + e_a^{~\mu}e_b^{~\nu} \bar{\psi}_{[\mu}\gamma_{\nu]}\lambda,\qquad F_{ab} \equiv e_a^{~\mu}e_b^{~\nu} (2\partial_{[\mu}A_{\nu]}),\nonumber\\
 && \hat{F}^{\pm}_{\mu\nu} \equiv \frac{1}{2}(\hat{F}_{\mu\nu}\pm \tilde{\hat{F}}_{\mu\nu}), \qquad \tilde{\hat{F}}_{\mu\nu} \equiv -\frac{1}{2} i\epsilon_{\mu\nu\rho\sigma}\hat{F}^{\rho\sigma}.\nonumber
\end{eqnarray}}
\begin{eqnarray}
&& P_L\Lambda \equiv \sqrt{2}P_L(-\frac{1}{2}\gamma\cdot \hat{F} + iD)\lambda,\qquad P_R\Lambda \equiv \sqrt{2}P_R(-\frac{1}{2}\gamma\cdot \hat{F} - iD)\lambda,\\
&& D_-^2 \equiv D^2 - \hat{F}^-\cdot\hat{F}^- - 2  \bar{\lambda}P_L\cancel{\mathcal{D}}\lambda,\qquad D_+^2 \equiv D^2 - \hat{F}^+\cdot\hat{F}^+ - 2  \bar{\lambda}P_R\cancel{\mathcal{D}}\lambda.
\end{eqnarray}
Then, after making the corrections to be generic as K\"{a}hler-invariant and field-dependent form, the higher order corrections to the standard supergravity action, which is given by Eq. (3.17) of Ref. \cite{HOC}, can be rewritten as 
\begin{eqnarray}
\mathcal{L}_{n} \supset \frac{(\bar{\lambda}P_L\lambda)^2(\bar{\lambda}P_R\lambda)^2}{(S_0\bar{S}_0e^{-K/3})^2}
T^k\Big( \frac{(\bar{\lambda}P_R\lambda)^2}{(S_0\bar{S}_0e^{-K/3})^2}\Big)\bar{T}^l\Big( \frac{(\bar{\lambda}P_L\lambda)^2}{(S_0\bar{S}_0e^{-K/3})^2}\Big)
\Big( \frac{(V)_D}{(S_0\bar{S}_0e^{-K/3})} \Big)^p\Psi_n(Z,\bar{Z})|_D, 
\end{eqnarray}
where $K(Z,\bar{Z})$ is a K\"{a}hler potential; $T$ is the chiral projection; $n=4+2k+2l+p$ with $n\geq 4$, and $\Psi_n(Z,\bar{Z})$ is a general real function of matter fields $Z$'s:
\begin{eqnarray}
&& Z^i = (z^i,-i\sqrt{2}P_L\chi^i,-2F^i,0,+i\mathcal{D}_{\mu}z^i,0,0) = \{ z^i, P_L\chi^i,F^i\},\\
&& \bar{Z}^{\bar{i}} = (\bar{z}^{\bar{i}},+i\sqrt{2}P_R\chi^{\bar{i}},0,-2\bar{F}^{\bar{i}},-i\mathcal{D}_{\mu}\bar{z}^{\bar{i}},0,0) = \{ \bar{z}^{\bar{i}}, P_R\chi^{\bar{i}},\bar{F}^{\bar{i}}\}.
\end{eqnarray}
We also use the superconformal compensator multiplet $S_0$: 
\begin{eqnarray}
&& S_0 = (s_0,-i\sqrt{2}P_L\chi^0,-2F_0,0,+i\mathcal{D}_{\mu}s_0,0,0) = \{ s_0, P_L\chi^0,F_0\},\\
&& \bar{S}_0 = (\bar{s}_0,+i\sqrt{2}P_R\chi^0,0,-2\bar{F}_0,-i\mathcal{D}_{\mu}\bar{s}_0,0,0) = \{\bar{s}_0, P_R\chi^0,\bar{F}_0\}.
\end{eqnarray}

Using the superconformal tensor calculus \cite{SUPER_CONFORMAL_TENSORCAL,Linear} in the appendix \ref{STC}, we find the corresponding bosonic lagrangian as
\begin{eqnarray}
\mathcal{L}_n|_B \supset (F^{+2}-D^2 )^{1+k}(F^{-2}-D^2)^{1+l}D^p \Psi_n(z,\bar{z})|_D= \Big(\frac{F^2}{2}-D^2\Big)^{(n-p)/2}D^p\Psi_n(z,\bar{z})|_D,\label{highorder}
\end{eqnarray}
where $F^2 \equiv F_{\mu\nu}F^{\mu\nu}$ is the square of Maxwell tensor; $F^{\pm}_{\mu\nu} \equiv \frac{1}{2}(F_{\mu\nu} \pm \Tilde{F}_{\mu\nu})$, and $\Tilde{F}_{\mu\nu} \equiv -\frac{i}{2}\varepsilon_{\mu\nu\rho\sigma}F^{\rho\sigma}$ is the dual tensor. In the last line of Eq. \eqref{highorder}, we have used $F^{\pm2} = \frac{1}{2}F^2$ and $(n-p)/2 = 2+k+l$. We note that the lagrangian of many higher order corrections can be given by a polynomial of the terms with various powers of $n,p$. 

Now, let us consider the case when $p=0$. Then, defining $\hat{D}\equiv \frac{F^2}{2}-D^2$ and $\Psi_n\equiv \Psi_n(z,\bar{z})|_D$, we rewrite the bosonic lagrangian as 
\begin{eqnarray}
\mathcal{L}_{\textrm{higher order}}^{(p=0)}|_B = \sum_{n=4}^N \hat{D}^{n/2}\Psi_n = \hat{D}^2 \Psi_4 + \hat{D}^{5/2} \Psi_5+ \hat{D}^3 \Psi_6 + \cdots.
\end{eqnarray}
The standard supergravity is specified by the following superconformal action
\begin{eqnarray}
 \mathcal{L}_{standard} &=& -3[S_0\Bar{S}_0e^{-K(Z^A,\bar{Z}^{\bar{A}})/3}]_D + [S_0^3W(Z^A)]_F -\beta[ \bar{\lambda}P_L\lambda]_F + h.c.,
\end{eqnarray}
where we have used $\beta$ as a general normalization of the kinetic term of the vector field. The corresponding D-term lagrangian is then found to be
 \begin{eqnarray}
 \mathcal{L}_{standard}|_B = 2\beta D^2 - \beta (F^{+2}+F^{-2}) = -2\beta \Big(\frac{F^2}{2}-D^2\Big) \equiv -2\beta \hat{D}.
 \end{eqnarray}
 Taking both the standard and higher order terms, we find the general D-term lagrangian as
\begin{eqnarray}
\mathcal{L}_{tot}|_{B} = -2\beta \hat{D} + \hat{D}^2 \Psi_4 + \hat{D}^{5/2} \Psi_5+ \hat{D}^3 \Psi_6 + \cdots \equiv P(\hat{D}).
\end{eqnarray}
Notice that this lagrangian is a polynomial of $\hat{D}$. When solving the equation of motion for $D$, we gain 
\begin{eqnarray}
\frac{\partial \mathcal{L}_{tot}|_{B}}{\partial D} = \frac{\partial \hat{D}}{\partial D} \frac{\partial \mathcal{L}_{tot}|_{B}}{\partial \hat{D}} = 0 \quad \implies \quad D=0  \quad \textrm{ or } \quad \frac{\partial \mathcal{L}_{tot}|_{B}}{\partial \hat{D}}= \frac{\partial P(\hat{D})}{\partial \hat{D}}= 0.
\end{eqnarray}
If the trivial solution $D=0$ is unstable or supersymmetry is broken, we have to consider the non-vanishing solution for $D$. The non-trivial solution for $\hat{D}$ can be found by
\begin{eqnarray}
\hat{D} = \hat{D}(\Psi_4,\Psi_5,\Psi_6,\cdots).
\end{eqnarray}
Notice that there is no any dependence on Maxwell tensor term in the solution for $\hat{D}$! After integrating out $D$, we face the {\it unphysical} situation that the kinetic term of the vector field is always absent in the lagrangian for any $N$. We point out that this case must be physically excluded. So, we propose a no-go theorem for the higher order corrections in FKLP supergravity model of inflation as follows:
\begin{theorem}[No-go theorem for higher order corrections in FKLP model]~\\
An arbitrary combination of the standard term and higher order corrections without any power of the real linear multiplet $(V)_D$, i.e. $p$, cannot produce the gauge kinetic term, and thus must be excluded in a physical theory.\label{no_go_theorem}
\end{theorem}
Based on this no-go theorem for the higher order corrections, we speculate that one has to include some non-vanishing powers of $(V)_D$, i.e. $p$, in the higher order corrections in order to generate the correct kinetic term.

\section{Novel Class of $\mathcal{N}=1$ Supergravity: ``Relaxed Supergravity''}\label{Novel_class}

In this section, we propose a novel class of $\mathcal{N}=1$ supergravity, called ``Relaxed Supergravity,'' that enlarges the space of scalar potentials by considering the higher order correction in FKLP minimal supergravity models of inflation \cite{HOC}.

\subsection{Discovery of a new negative-definite term of scalar potential in supergravity}

For the vector multiplet $V$, as a setup, we suppose three ``NO'' things when constructing a superconformal action of supergravity containing some higher order corrections as follows:
\begin{itemize}
    \item {\bf No Fayet-Iliopoulos term.} There is no term linear in the auxiliary field $D$, i.e. no any Fayet-Iliopoulos $D$ term.
    \item {\bf No gauging.} The vector multiplet $V$ is {\it not} gauged.
    \item {\bf No-go theorem.} There must be some powers of the real linear multiplet $(V)_D$ in the higher order corrections in order to satisfy the no-go theorem \eqref{no_go_theorem}.
\end{itemize} These three assumptions will play a role in finding a new contribution to the scalar potential. Notice that the three conditions are {\it not applicable} for the other vector multiplets associated with conventional gauge groups.

Now, we are ready to consider a superconformal action of a certain higher order correction. We define 
\begin{eqnarray}
\mathcal{L}_{RS} \equiv \bigg[ -\frac{1}{4}(S_0\bar{S}_0 e^{-K/3})^{-4}  (\bar{\lambda}P_L\lambda)(\bar{\lambda}P_R\lambda) ((V)_D)^2 \frac{1}{\mathcal{U}(Z,\Bar{Z})^2}\bigg]_D,\label{RS}
\end{eqnarray}
where $S_0$ is the conformal compensator; $K(Z^I,\bar{Z}^{\bar{\imath}})$ is the supergravity K\"{a}hler potential of the matter chiral multiplets $Z^I$'s; $\bar{\lambda}P_L\lambda$ is the field strength multiplet corresponding to a vector multiplet $V$ whose fermionic superpartner is given by $\lambda$; $(V)_D$ is a real multiplet whose lowest component is given by the auxiliary field $D$ of the vector multiplet $V$, and $\mathcal{U}$ is defined as a general {\it gauge-invariant} real function of the matter multiplets. Therefore, including the standard supergravity terms, we reach the total superconformal action of our supergravity as
\begin{eqnarray}
 \mathcal{L} &=& -3[S_0\Bar{S}_0e^{-K(Z^A,\bar{Z}^{\bar{A}})/3}]_D + \bigg[ -\frac{1}{4}(S_0\bar{S}_0 e^{-K/3})^{-4}  (\bar{\lambda}P_L\lambda)(\bar{\lambda}P_R\lambda) ((V)_D)^2 \frac{1}{\mathcal{U}(Z,\Bar{Z})^2}\bigg]_D \nonumber\\
 && + [S_0^3W(Z^A)]_F -\frac{3}{4}[ \bar{\lambda}P_L\lambda]_F + h.c.
\end{eqnarray}
Notice that the numerical factor of the kinetic term for the vector multiplet $V$ is not $1/4$ but $3/4$. This different factor is set to yield the canonically normalized kinetic term of the vector field. The bosonic lagrangian of the auxiliary field $D$ is then found by
\begin{eqnarray}
 \mathcal{L}_D &=& -\frac{3}{4} \Big(-D^2 + \frac{F^2}{2}-D^2 + \frac{F^2}{2}\Big) - \frac{1}{2\mathcal{U}^2}\Big(-D^2 + \frac{F^2}{2}\Big)\Big(-D^2 + \frac{F^2}{2}\Big)D^2 \nonumber\\
 &=& \frac{3}{2}D^2 -\frac{3}{4}F^2 +  \left( -\frac{D^6}{2\mathcal{U}^2} + \frac{D^4F^2}{2\mathcal{U}^2} -\frac{D^2F^4}{8\mathcal{U}^2}\right),
\end{eqnarray}
where $F^2 \equiv F_{\mu\nu}F^{\mu\nu}$. This gives the potential in $D$
\begin{eqnarray}
 V(D) = \frac{3}{4}F^2-\frac{3}{2}D^2 + \left( \frac{D^2F^4}{8\mathcal{U}^2} - \frac{D^4F^2}{2\mathcal{U}^2}+\frac{D^6}{2\mathcal{U}^2}\right).\label{Dpotential}
\end{eqnarray}

Here is a crucial remark. We should be careful about presence of the kinetic term for the auxiliary field $D$. Let us look at the composite superconformal multiplet $\mathcal{V}\equiv ((V)_D)^2$. Its highest component $\mathcal{D}_{\mathcal{V}}$ contains the second derivative term of the field $D$ with respect to the spacetime coordinates, i.e. $\mathcal{D}_{\mathcal{V}} \supset -2D\square^CD \sim (\partial D)^2 + \textrm{total derivative}$. We find that since $\mathcal{L}_{RS} \equiv -[\mathcal{R}\cdot \mathcal{V}]$ where $\mathcal{R}$ is defined to be the remaining parts except for $\mathcal{V}\equiv ((V)_D)^2$ and a minus sign, the relaxed supergravity action gives us a kinetic term of the field $D$ with the canonical sign, i.e. $\mathcal{L}_{RS} \supset -(\partial D)^2$ in the spacetime-metric convention $(-,+,\cdots,+)$. That is, $D$ is not a ghost but a physical field. From Eq.~\eqref{Dpotential}, we see that the canonically normalized field ``$\tilde{D}$'' such that $\Tilde{D} \equiv D/M_{pl}$ has a mass of Planck scale as follows: 
\begin{eqnarray}
\frac{\partial^2 V(D)}{\partial D^2}\bigg|_{D\sim \sqrt{\mathcal{U}}} = -3 + 15 \frac{D^4}{\mathcal{U}^2}\bigg|_{D\sim \sqrt{\mathcal{U}}}= -3 + 15 \equiv \frac{m_{\Tilde{D}}^2}{M_{pl}^2}.
\end{eqnarray}
 Accordingly, we are able to integrate out the $D$ degree of freedom with the Planck mass in the first place. 

After solving the equation of motion for $D$, we obtain the following solutions 
\begin{eqnarray}
 D= 0 , \qquad D^2 = \mathcal{U} \sqrt{1 + \frac{F^4}{36\mathcal{U}^2}} + \frac{F^2}{3},\label{Dsolution}
\end{eqnarray}
where the first corresponds to the supersymmetric case, while the latter corresponds to the non-supersymmetric case. Looking at the potential for $D$ in Eq.~\eqref{Dpotential}, we observe that the point at $D=0$ is unstable, and the vacua is located at $D \neq 0$. Therefore, in our model, supersymmetry is spontanesously broken like the Higgs mechanism.  

We point out that since $D^2>0$, the general function must be non-negative-definite, i.e. $\mathcal{U}>0$. Next, integrating out the field $D$, we obtain the bosonic lagrangian as 
\begin{eqnarray}
 \mathcal{L}_D = -\frac{1}{4}F^2 +  \mathcal{U} \sqrt{1 + \frac{F^4}{36\mathcal{U}^2}} + \frac{F^4}{36\mathcal{U}}\sqrt{1 + \frac{F^4}{36\mathcal{U}^2}} + \frac{F^6}{24\mathcal{U}^2}.\label{RS_Lagrangian}
\end{eqnarray}
Then, expanding Eq.~\eqref{RS_Lagrangian}, we have the following 
\begin{eqnarray}
 \mathcal{L}_D = -\frac{1}{4}F^2 + \mathcal{U} + \frac{F^4}{24\mathcal{U}} + \frac{F^6}{24\mathcal{U}^2} + \frac{F^8}{2592\mathcal{U}^3} + \textrm{higher order terms in}~ F^2.\label{RS_Lagrangian_expanded}
\end{eqnarray}
Notice that the lagrangian produces the correct kinetic term for the vector $V$, and a new negative contribution to the scalar potential
\begin{eqnarray}
 V_{RS} \equiv -\mathcal{U},
\end{eqnarray}
where $\mathcal{U}>0$. Hence, the total scalar potential can be written in general by
\begin{eqnarray}
 V_{tot} = V_D + V_F - \mathcal{U},\label{Total_General_Scalar_Potential}
\end{eqnarray}
where $V_D$ and $V_F$ are the standard D- and F-term potentials. Again, $\mathcal{U}$ is a positive generic function, so that $V_{RS}$ is a purely negative-definite.

\subsection{Constraints on the new negative term}

In this section, by inspecting the most singular nonrenormalizable lagrangians, we find constraints on $\mathcal{U}$. We use the same analysis already done in our previous study \cite{jp1,jp3}. We identify the most singular terms by checking four fermions (i.e. $\sim (\bar{\chi}P_L\chi)(\bar{\chi}P_R\chi)$) and derivative terms:
\begin{eqnarray}
 \mathcal{L}_F^{\textrm{on U}} &\supset&  \bigg\{ \frac{(\mathcal{U}^{(1)})^4}{\mathcal{U}^5}, ~ 
 \frac{(\mathcal{U}^{(1)})^2\mathcal{U}^{(2)}}{\mathcal{U}^5} ,~\frac{(\mathcal{U}^{(2)})^2}{\mathcal{U}^3}, 
 ~ \frac{\mathcal{U}^{(1)}\mathcal{U}^{(3)}}{\mathcal{U}^3}, ~
 \frac{\mathcal{U}^{(4)}}{\mathcal{U}^2}
 \bigg\} \mathcal{O}_F^{(d=12)}, \nonumber\\
 \mathcal{L}_F^{\textrm{on U}} &\supset&  \bigg\{ \frac{(\mathcal{U}^{(1)})^4}{\mathcal{U}^6}, ~ 
 \frac{(\mathcal{U}^{(1)})^2\mathcal{U}^{(2)}}{\mathcal{U}^6} ,~\frac{(\mathcal{U}^{(2)})^2}{\mathcal{U}^4}, 
 ~ \frac{\mathcal{U}^{(1)}\mathcal{U}^{(3)}}{\mathcal{U}^4}, ~
 \frac{\mathcal{U}^{(4)}}{\mathcal{U}^3}
 \bigg\} [\mathcal{O}_F^{(d=12)}F^2]^{(d=16)}, \nonumber\\
  \mathcal{L}_F^{\textrm{on K}} &\supset&  \bigg\{ \frac{(K^{(1)})^4}{\mathcal{U}M_{pl}^4}, ~ 
 \frac{(K^{(1)})^2K^{(2)}}{\mathcal{U}M_{pl}^2} ,~\frac{(K^{(2)})^2}{\mathcal{U}}, 
 ~ \frac{K^{(1)}K^{(3)}}{\mathcal{U}}, ~ 
 \frac{K^{(4)}}{\mathcal{U}}M_{pl}^2
 \bigg\} \mathcal{O}_F^{(d=12)}, \nonumber\\
 \mathcal{L}_F^{\textrm{on K}} &\supset&  \bigg\{ \frac{(K^{(1)})^4}{\mathcal{U}^2M_{pl}^8}, ~ 
 \frac{(K^{(1)})^2K^{(2)}}{\mathcal{U}^2M_{pl}^6} ,~\frac{(K^{(2)})^2}{\mathcal{U}^2M_{pl}^4}, 
 ~ \frac{K^{(1)}K^{(3)}}{\mathcal{U}^2M_{pl}^4}, ~ 
 \frac{K^{(4)}}{\mathcal{U}^2M_{pl}^2}
 \bigg\} [\mathcal{O}_F^{(d=12)}F^2]^{(d=16)}, \nonumber\\
 \mathcal{L}_F^{\textrm{on S}} &\supset& \frac{1}{\mathcal{U}}M_{pl}^{-4} \mathcal{O}_F^{(d=12)},\nonumber\\
 \mathcal{L}_F^{\textrm{on S}} &\supset& \frac{1}{\mathcal{U}^2}M_{pl}^{-4} [\mathcal{O}_F^{(d=12)}F^2]^{(d=16)},\nonumber\\
  \mathcal{L}_D &=& -\frac{1}{4}F^2 + \mathcal{U} + \frac{F^4}{24\mathcal{U}} + \frac{F^6}{24\mathcal{U}^2} + \frac{F^8}{2592\mathcal{U}^3} + \textrm{higher order terms in}~ F^2 \nonumber.
\end{eqnarray}
where $K$ is the K\"{a}hler potential; $\mathcal{U}$ is the general function; $\mathcal{O}_F^{(d=12)}$ only includes fermions, and $F^2 \equiv F_{\mu\nu}F^{\mu\nu}$. We denote $\mathcal{L}_F^{\textrm{on U/K}}$ by the lagrangians of the derivatives of $\mathcal{U}$ and $K$ with respect to the matter fields, while $\mathcal{L}_F^{\textrm{on S}}$ by those of the derivatives of $s_0\bar{s}_0$ with respect to the conformal compensator field. We observe that the strongest constraint comes from the D-term lagrangian 
\begin{eqnarray}
\mathcal{L}_D \supset \frac{F^4}{24\mathcal{U}} \implies \frac{1}{\mathcal{U}} \lesssim \frac{1}{\Lambda_{cut}^{4}} \implies \Lambda_{cut} \sim \mathcal{U}^{1/4} = M_S \lesssim M_{pl}, \label{main_constraint}
\end{eqnarray}
where the last inequality is given by the fact that $\Lambda_{cut} \lesssim M_{pl}$. This means that relaxed supergravity has a cutoff exactly at the SUSY breaking scale, and supersymmetry may be broken at high scale \cite{High_SUSY} according to the cutoff. Therefore, our model is basically an effective field theory with broken SUSY valid up to the low energies below the SUSY breaking scale $M_S$.

\subsection{Relaxation of scalar potential beyond no-scale supergravity}

In the previous section 3.2, we have seen that the total scalar potential in our theory is given by Eq. \eqref{Total_General_Scalar_Potential}
\begin{eqnarray}
V_{tot}(z^I,\bar{z}^{\bar{I}}) = V_D(z^I,\bar{z}^{\bar{I}}) + V_F(z^I,\bar{z}^{\bar{I}}) - \mathcal{U}(z^I,\bar{z}^{\bar{I}}),\nonumber
\end{eqnarray}
where the potentials are functions of matter fields $z^I$'s, and the new term is moderately constrained by Eq.~\eqref{main_constraint}. To analyze the new potential term, let us begin with a decomposition of matter multiplets as follows:
\begin{eqnarray}
Z^I \equiv (Z^s, Z^i),
\end{eqnarray}
where $Z^s$ is supposed to control the SUSY breaking scale $M_S$ in a hidden sector, while $Z^i$ are the normal matter ones that may belong to an observable sector. Next, we define
\begin{eqnarray}
\mathcal{U}(z^I,\bar{z}^{\bar{I}}) \equiv V_{\mathcal{U}}^{\cancel{S}}(z^I,\bar{z}^{\bar{I}}) - \sum_{a \neq \cancel{S}}V_{\mathcal{U}}^a(z^I,\bar{z}^{\bar{I}}) >0,\\
V_D(z^I,\bar{z}^{\bar{I}}) \equiv V_D^{\cancel{S}}(z^I,\bar{z}^{\bar{I}}) + \sum_{A\neq \cancel{S}} V_D^A(z^I,\bar{z}^{\bar{I}}) >0, 
\end{eqnarray}
where each potential $V_{\mathcal{U}}^a$ can be either negative or positive definite, and has a different energy scale such that $|V_{\mathcal{U}}^{\cancel{S}}| > |\sum_{a=1}V_{\mathcal{U}}^a|$. On the other hand, each D-term potential is positive semi-definite. Then, the total scalar potential is rewritten as
\begin{eqnarray}
V_{tot} =  V_D^{\cancel{S}}+ \Big( \sum_{A\neq \cancel{S}} V_D^A + V_F \Big) + \sum_{a}V_{\mathcal{U}}^a  - V_{\mathcal{U}}^{\cancel{S}}.
\end{eqnarray}
In order to have a maximally relaxed scalar potential, we may take the following choice
\begin{eqnarray}
V_{\mathcal{U}}^{\cancel{S}} \overset{!}{=}  V_D^{\cancel{S}} + \Big( \sum_{A\neq \cancel{S}} V_D^A + V_F \Big),\label{SUSY_scale_eq1}
\end{eqnarray}
which provides us the most general function form of the scalar potential
\begin{eqnarray}
V_{tot} = \sum_{a \neq \cancel{S}} V_{\mathcal{U}}^a(z^I,\bar{z}^I) < |V_{\mathcal{U}}^{\cancel{S}}|,
\end{eqnarray}
and the SUSY breaking scale $M_S$ such that
\begin{eqnarray}
M_S^4 = \mathcal{U} \equiv V_D^{\cancel{S}} + \Big( \sum_{A\neq \cancel{S}} V_D^A + V_F \Big) - V_{tot}.
\end{eqnarray}

Now, we have to explore under which conditions the general scalar potential can be well established. We may consider the following four suppositions:
\begin{itemize}
\item {\bf Partitioned gauge symmetries.} All $Z^i$'s must be neutral under any gauge group in which $Z^s$ is charged, and {\it vice versa}:
\begin{eqnarray}
&& G^{\cancel{S}}:~Z^s \rightarrow Z^s e^{q_s\Sigma} , \quad Z^i \rightarrow Z^i,\nonumber\\
&& G^i:~Z^s \rightarrow Z^s , \quad Z^i \rightarrow Z^i e^{q_i\Omega}
\end{eqnarray}
where $\Sigma$ and $\Omega$ are chiral multiplets as gauge parameters of the gauge groups $G^{\cancel{S}}$ and $G^i$, respectively. 
\item {\bf SUSY-breaking-scale cutoff dominance.} The scale of $V_D^{\cancel{S}}$ far exceeds the magnitude of any combination of the other potentials $V_D^A,V_F,V_{tot}$, so that the combination cannot cancel out $V_D^{\cancel{S}}$ and this solely controls the SUSY breaking scale $M_S$, i.e.
\begin{eqnarray}
|V_D^{\cancel{S}}| \gg \Big|\Big( \sum_{A\neq \cancel{S}} V_D^A + V_F \Big)-V_{tot}\Big| \implies \Lambda_{cut} = M_S = \mathcal{U}^{1/4} \approx |V_D^{\cancel{S}}|^{1/4} \neq0.\label{Dominance}
\end{eqnarray}
\item {\bf Broken supersymmetry.} We must have proper values of $z^s$ and $z^i$ such that $V_D^{\cancel{S}}\neq 0$ to protect broken SUSY all the times. 
\item {\bf Decomposition of scalar potential for moduli stabilization.} The total scalar potential must be decomposed into $z^s$-dependent and $z^s$-independent sectors in order to perform moduli stabilization for the fields $z^s$ in the simplest way, i.e.
\begin{eqnarray}
V_{tot} = V_{\mathcal{U}}^{s-depen}(z^s,\bar{z}^s) + V_{\mathcal{U}}^{s-indepen}(z^i,z^i).
\end{eqnarray}
If $V_{\mathcal{U}}^{s-depen}(z^s,\bar{z}^s)=0$, then we can choose any value of $z^s$ such that $V_D^{\cancel{S}}\neq0$, and $z^s$ becomes massless.
\end{itemize}
As long as the above conditions are satisfied, we are able to have the maximally relaxed scalar potential
\begin{eqnarray}
V_{tot} = \sum_{a \neq \cancel{S}}V_{\mathcal{U}}^a(z^I,\bar{z}^I)= V_{\mathcal{U}}^{s-depen}(z^s,\bar{z}^s) + \sum_{a\neq \cancel{S},s-depen}V_{\mathcal{U}}^a(z^i,\bar{z}^i) < |\underbrace{V_D^{\cancel{S}} + \Big( \sum_{A\neq \cancel{S}} V_D^A + V_F \Big)}_{=V_{\mathcal{U}}^{\cancel{S}}}|,
\end{eqnarray}
where the inequality comes from the condition $\mathcal{U}>0$.
 In the meantime, the corresponding constraint is given by
\begin{eqnarray}
\Lambda_{cut}^4=M_S^4= \underbrace{V_D^{\cancel{S}} + \Big( \sum_{A\neq \cancel{S}} V_D^A + V_F \Big)}_{=V_{\mathcal{U}}^{\cancel{S}}} - V_{tot}  \lesssim M_{pl}^4 \implies M_S^4 \sim V_D^{\cancel{S}}\lesssim M_{pl}^4,
\end{eqnarray}
where the last inequality is due to the dominance condition in Eq.~\eqref{Dominance}. Notice that $V_{D}^{\cancel{S}}$ is parametrically free up to the Planck scale $M_{pl}$, while the total scalar potential is parametrically free up to the SUSY breaking scale $M_S$.

Of course, one may wish to utilize the normal structures of the D- and F-term potentials in supergravity for some reasons. In this case, one can recover them by respecting the above assumptions in the following way: 
\begin{eqnarray}
&& V_{\mathcal{U}}^a(z^i) \supset V_F' \equiv A\cdot V_F(z^s=0)  = A\cdot e^G(G_IG^{I\bar{J}}G_{\bar{J}}-3) |_{z^s=0},\\
&& V_{\mathcal{U}}^a(z^i) \supset V_D' \equiv B\cdot V_D(z^s=0),
\end{eqnarray}
in which we have put $z^s=0$ in the usual D- and F-term potentials, and multiplied them by some arbitrary constants $A,B$ for generality. Thus, we have extra freedom in adjusting the scales of the D- and F-term potentials.

For example, the simplest toy model of relaxed supergravity can be given by the following. Let us consider an abelian $U_s(1)$ gauge symmetry. Assume that only a single matter field $z^s$ is charged under the $U_s(1)$, say $z^s \rightarrow e^{iq_s\theta}z^s$. Then, for a K\"{a}hler potential $K = -3\ln[T+\bar{T}-\frac{|z^s|^2+\delta_{i\bar{j}}z^i\bar{z}^{\bar{j}}}{3}]$ and a superpotential $W(T,z^i)$, a corresponding D-term potential is given by
\begin{eqnarray}
V_D^{\cancel{S}} = \frac{1}{2}g^2 q_s^2\frac{(z^s\bar{z}^s)^2}{(T+\bar{T})^2},\label{toy}
\end{eqnarray}
and the total scalar potential is given by
\begin{eqnarray}
V_{tot} = \underbrace{\sigma (|z^s|^2-\rho^2)^2}_{\textrm{s-dependent part}} + \underbrace{\sum_{a} V_{\mathcal{U}}^a(T,z^i)}_{\textrm{s-independent part}}  < V_D^{\cancel{S}}(z^s,T),\label{toy2}
\end{eqnarray}
where $\sigma,\rho$ are some constants, and we have used a potential $V_{\mathcal{U}}^a \supset \sigma (|z^s|^2-\rho^2)^2$ for producing a mass of $z^s$ in general. For this potential, we observe that $\left<z^s\right>=\rho \neq0$. Of course, it is straightforward for Eq.~\eqref{toy} to obey the dominance condition in Eq.~\eqref{Dominance} by choosing a large value of $\rho$. We see that SUSY breaking scale is determined by Eq.~\eqref{toy} while we have generic potentials in Eq.~\eqref{toy2}.

In this section, using a special choice in Eq.~\eqref{SUSY_scale_eq1}, we have treated a particular mechanism to derive a general scalar potential. However, there can be other mechanisms. These possibilities deserve further investigation in the future.

\subsection{``Relaxed'' supergravity versus ``Liberated'' supergravity}
Here we compare our relaxed supergraivity (RS) with the liberated supergravity (LS). First, let us recall the main result of the constraints on the liberated supergravity \cite{jp1}. The liberated term $\mathcal{U}$ that is added to the supergravity scalar potential as a general function of the matter fields is severely constrained by
\begin{eqnarray}
 \mathcal{U}^{(n)} \lesssim \begin{cases}
  \left(\dfrac{M_S}{M_{pl}}\right)^{8(4-n)}\left(\dfrac{M_{pl}}{\Lambda_{cut}}\right)^{2(4-n)} \quad \textrm{where}\quad0\leq n \leq 2\quad\textrm{for}~N_{mat} =1,\\
\left(\dfrac{M_S}{M_{pl}}\right)^{8(6-n)}\left(\dfrac{M_{pl}}{\Lambda_{cut}}\right)^{2(6-n)} \quad\textrm{where}\quad0\leq n \leq 4\quad\textrm{for}~N_{mat} \geq 2,
  \end{cases}\label{generic_constraints}
\end{eqnarray}
where $n$ is the order of the derivative with respect to the matter field, and $N_{mat}$ is the number of matter multiplets involved in a liberated supergravity theory of interest. The constraints correspond to the case when the matter fields are at their vacua. The scalar potential in the liberated supergravity must obey
\begin{eqnarray}
V_{LS} \equiv \mathcal{U} \lesssim \begin{cases}
  \left(\dfrac{M_S}{M_{pl}}\right)^{32}\left(\dfrac{M_{pl}}{\Lambda_{cut}}\right)^{8} \quad\textrm{for}~N_{mat} =1,\\
\left(\dfrac{M_S}{M_{pl}}\right)^{48}\left(\dfrac{M_{pl}}{\Lambda_{cut}}\right)^{12} \quad\textrm{for}~N_{mat} \geq 2,
  \end{cases}\label{generic_constraints}
\end{eqnarray}
On the other hand, in relaxed supergravity, we found that 
\begin{eqnarray}
V_{RS} = \sum_{a\neq \cancel{S}} V_{\mathcal{U}}^a < V_{\mathcal{U}}^{\cancel{S}} \sim V_D^{\cancel{S}} \sim M_S^4 \lesssim M_{pl}^4.
\end{eqnarray}
For instance, when we consider $\Lambda_{cut} = 10^{-2} M_{pl}$, we obtain
\begin{eqnarray}
 V_{RS} < 10^{-8}M_{pl}^4, \quad V_{LS} \lesssim 10^{-64}M_{pl}^4 \quad  \textrm{for}\quad N_{mat}=1, \quad  V_{LS} \lesssim 10^{-96}M_{pl}^4 \quad  \textrm{for}\quad N_{mat}\geq 2.
\end{eqnarray} 
Note that $V_{RS}$ can describe the inflation scale $\mathcal{O}(H^2M_{pl}^2)$ since it is bounded by parametrically free $V_{\mathcal{U}}^{\cancel{S}}$ up to the Planck scale $M_{pl}$, while any of $V_{LS}$'s cannot. This shows that relaxed supergravity excels the liberated one in defining a scalar potential at a desired energy level.  

\subsection{The first negative term of scalar potential in global supersymmetry}

We briefly discuss an intriguing physical implication on our findings in relaxed supergravity. It is well known that when supergravity is turned off (i.e. $M_{pl} \rightarrow \infty$), the scalar potential of the standard supergravity reduces to that of global SUSY. This is because in the limit we have $M_{pl}^4e^{G} \rightarrow 0$ and $M_{pl}^4e^G(G_IG^{I\bar{J}}G_{\bar{J}}) \rightarrow W_IK^{I\bar{J}}W_{\bar{J}}$ in the F-term potential $V_F$ where $G \equiv \frac{K}{M_{pl}^2} + \ln \frac{W}{M_{pl}^3}+ \ln \frac{\bar{W}}{M_{pl}^3}$ after recovering the Planck mass dimension $M_{pl}$. In particular, the relaxing term $\mathcal{U}$ can be alive in global SUSY since the bosonic lagrangians in Eqs.~\eqref{RS_Lagrangian} and \eqref{RS_Lagrangian_expanded} are independent of Planck mass $M_{pl}$. Of course, the general function $\mathcal{U}$ changes the total scalar potential in the same way as follows:
\begin{eqnarray}
V_{tot} = \underbrace{\frac{1}{2}|D^a|^2 + |W_I|^2}_{\textrm{standard global SUSY}} -~ \mathcal{U},\label{New_Global_SUSY}
\end{eqnarray}
where $D^a$ is the D-term solution with respect to some gauge killing vector fields $k^a(z)$, and $W_I \equiv \partial W/\partial z^I$ is the field derivative of superpotential. The result in Eq.~\eqref{New_Global_SUSY} is surprising in that it gives us the first {\it negative} contribution to the scalar potential in global supersymmery, allowing us to have any of Minkowski and (Anti) de Sitter spacetimes. Surely, it has long been regarded that there is only positive potentials in global SUSY, and thus either Minkowski or de Sitter spacetime is possible to exist. We expect that this new aspect may alter some known arguments led by the fact that the global SUSY scalar potential is always semi-positive, i.e. $V_{tot} = \frac{1}{2}|D^a|^2 + |F^I|^2 \geq 0$. We do not explore this here since it is beyond the scope of our purpose in this letter.

\section{Conclusion and Outlook}\label{con_o}

We have presented a relaxing procedure of the scalar potential by requiring four conditions. The first is that $z^s$ ($z^i$) is charged but $z^i$ ($z^s$) is neutral under a gauge group $G^{\cancel{S}}$ ($G^i$). The second is that the scale of $V_{D}^{\cancel{S}}$ governing the SUSY-breaking scale $M_S$ rather exceeds those of the other potentials $V_D^A,V_F,V_{tot}$ satisfying Eq.~\eqref{Dominance}. The third is that values of $z^s$ and $z^i$ must hold non-vanishing $V_{D}^{\cancel{S}}$. The last is that the total scalar potential is decomposed into $z^s$-dependent and independent sectors to do moduli stabilization for the fields $z^s$ in the simplest way.  

Lastly, we discuss outlook on relaxed supergravity. First, one may wish to explain some phenomenologies from particle physics to cosmology in the context of either locally or globally supersymmetric theory. We suggest that our proposal can be utilized for constructing both supergravity and globally supersymmetric models of particle and cosmological phenomenologies. This is based on the fact that our supergravity predicts a general scalar potential up to the Planck energy $M_{pl}$, and the relaxing term can emerge in both theories in a consistent fashion. Second, we remark that the string realization of the superconformal action of the relaxing term deserves future investigation like the work of Ref. \cite{StringRealization}. Third, it would be worth studying to explore if other relaxing mechanisms can possibly exist in different setups beyond this work. Fourth, since our model has a cutoff $\Lambda_{cut}$ equal to the SUSY breaking scale $M_S$, one may study improved versions of relaxed supergravity which has a sufficiently large hierarchy between cutoff and SUSY breaking scale in order to recover the naturalness in the future. The last is that one may explore physical implications which are deduced by the first negative scalar potential in global SUSY.

%% file: chapters/9.tex
\framebox[1.05\width]{{\large This chapter is based on the author's original work in Ref.~\cite{jp1}.}} \par
\vspace{1cm}

In Ch.~\ref{ch7}, we have argued that liberated $\mathcal{N}=1$ supergravity can be an effective field theory for 
describing the inflationary dynamics while at the same time satisfying all the constraints if a transition that changes the 
supersymmetry breaking scale at the end of inflation is allowed. 
Note that due to the no-scale structure, the scalar potential is given only by an eventual D-term supersymmetric potential 
$V_D$ and the ``liberated'' term $\mathcal{U}$. In this chapter, we present an explicit minimal model of single-field, 
slow-roll inflation in 
 liberated $\mathcal{N}=1$ supergravity which obeys the inequality $H \ll \Lambda_{cut} = M_{pl} =1$. 

To begin with, let us consider a chiral multiplet $T$ with K\"{a}hler potential  $K(T,\bar{T}) = -3\ln[T+\bar{T}]$ 
and a constant superpotential $W_0$. Then, the supergravity G-function~\cite{cfgvnv} is given by 
\begin{eqnarray}
G \equiv K + \ln |W|^2 = -3\ln[T+\bar{T}] + \ln W_0 + \ln \bar{W}_0.\label{sugra}
\end{eqnarray}
It automatically produces a no-scale structure in which the F-term potential vanishes identically: $V_F =0$. 

Next, let us find the canonically normalized degrees of freedom of the theory. From the kinetic term corresponding to the G
 function~\eqref{sugra}, we read
\bea
\mathcal{L}_K &=& \frac{3}{(T+\bar{T})^2}\partial T\partial \bar{T} \nonumber\\
&=& \frac{3}{4(\textrm{Re}T)^2}(\partial \textrm{Re}T)^2 + \frac{3}{4(\textrm{Re}T)^2}(\partial \textrm{Im}T)^2 \nonumber\\
&=& \frac{1}{2} (\partial \chi)^2 + \frac{1}{2} e^{-2\sqrt{2/3}\chi}(\partial \phi)^2,
\eea{can-nor}
where we have used the following field redefinition 
\begin{eqnarray}
T = \textrm{Re}T + i \textrm{Im}T = \frac{1}{2}e^{\sqrt{2/3}\chi} + i\frac{\phi}{\sqrt{6}}.\label{field_redefinition}
\end{eqnarray}
Note that the $\mathbb{Z}_2$ symmetry $\chi \rightarrow -\chi$ is already explicitly broken by the kinetic Lagrangian, while 
the symmetry $\phi\rightarrow -\phi$ is unbroken. However, even the latter symmetry will be broken by the inflationary 
potential. The field $\chi$ is always canonically normalized while $\phi$ has a canonical kinetic term only at  $\chi =0$.

The composite 
F-term is given by $\mathcal{F} = e^GG_TG^{T\bar{T}}G_{\bar{T}}$ after solving the equation of motion for the auxiliary 
fields $F^I$. For our G function we obtain an exponentially decreasing function 
$\mathcal{F} = 3|W_0|^2/(T+\bar{T})^3= 3|W_0|^2e^{-3\sqrt{2/3}\chi}$. This is what we want to get a viable 
supersymmetry breaking mechanism. The reason
is that we look for a supersymmetry breaking scale during inflation  $M_S^i\sim M_{pl}=1$, while the final scale should be 
parametrically lower than the Planck scale --for instance $M_S^f = 10^{-15}M_{pl}$. To achieve this large difference of
scales, the vacuum expectation value of the field $\chi$ should change during the phase transition. 
On the other hand, the cutoff scale of our model can remain $\mathcal{O}(M_{pl})$ both before and after the 
phase transition. 

We will achieve this with a potential that changes from ($\phi \neq 0, \chi=0$) during inflation to 
$\chi \neq 0, \phi =0$ after inflation. We will also choose $\phi$ as the inflaton field and $\chi$ as the field that controls the 
supersymmetry breaking scale. 

A function $\mathcal{U}$ producing a correct inflationary dynamics is
\begin{eqnarray}
 \mathcal{U} \equiv \alpha (1-e^{-\sqrt{2/3}\phi})^2(1+\frac{1}{2}\sigma\chi^2),\label{infl-pot}
\end{eqnarray}
where $\alpha,\beta,\gamma,\sigma$ are arbitrary positive constants. This is a key result in this section.

Next, we assume that the mass $m_{\chi}$ of $\chi$ is greater than the Hubble scale $H$ during inflation; 
this is necessary to 
describe a single-field slow-roll inflation governed only by the inflaton field $\phi$. Hence, we impose that during inflation
$m_{\chi}^2 = \alpha \sigma \gg H^2$. Since $\alpha \sim H^2 \sim 10^{-10}$, the condition reduces to 
$\sigma \gg 1$. 

We must also analyze the vacuum structure of the potential. First of all, we explore the minima with respect to $\chi$. 
By computing $\frac{\partial \mathcal{U}}{\partial \chi} = 0$ and defining
 $ V_{\textrm{inf}}  \equiv \alpha (1-e^{-\sqrt{2/3}\phi})^2$ we find that during inflation 
 (where $\phi \neq 0$) the equation of motion for $\chi$ is given by 
 $\sigma \chi V_{\textrm{inf}}  = 0$ 
 so it gives a unique minimum at $\chi =0$. On the other hand after inflation we have $\phi = 0$ and  $V_{\textrm{inf}}=0$, 
 so the equation of motion gives a flat potential in $\chi$. The final position of the field $\chi$ is then determined either
 by the initial conditions on $\chi$ or by small corrections to the either the liberated supergravity potential $\mathcal{U}$ 
 or to $V_F$. Here we content
 ourselves with pointing out that the simple potential~\eqref{infl-pot} already achieves the goal of making the final 
 supersymmetry breaking scale different from $M_S^i$. 
   
Before studying supersymmetry breaking we notice that a deformation of the scalar potential such as 
$\mathcal{U}$ was obtained using an off-shell \underline{linear} realization of supersymmetry in~\cite{Linear}.  
Therefore, for the new term 
to be consistent, supersymmetry must be broken as usual by some nonvanishing auxiliary field belonging to
the standard chiral 
multiplets and moreover the K\"{a}hler metric of the scalar manifold must be positive~\cite{Linear}. 
So, in spite of the presence of the
new term $\mathcal{U}$, the analysis of supersymmetry breaking is completely standard.  
Since the supersymmetry breaking scale $M_S$ comes from the positive potential part $V_+$, as shown in the 
Goldstino SUSY transformation $\delta_{\epsilon} P_Lv = \frac{1}{2}V_+ P_L \epsilon$ that is constructed with the fermion 
shifts with respect to the auxiliary scalar contributions \cite{Superconformal_Freedman}, we have
\begin{eqnarray}
V_+ = e^{G}G_TG^{T\bar{T}}G_{\bar{T}}=\frac{3|W_0|^2}{(T+\bar{T})^3}  = 3|W_0|^2e^{-3\sqrt{2/3}\chi}.\nonumber\\{}
\end{eqnarray}

During inflation we demand that the initial supersymmetry breaking scale is $M_{pl}$, so we identify $W_0 \equiv \dfrac{(M_S^i)^2}{\sqrt{3}}$ and therefore $V_+ = (M_S^i)^4 e^{-3\sqrt{2/3}\chi}$. 
Because $\chi=0$ during
inflation we indeed have $V_+|_{\chi=0,\phi\neq0} = (M_S^i)^4=1  \gg H^2 = \mathcal{O}(10^{-10}M_{pl}^2)$. 

On the other hand, we want to get a much smaller SUSY breaking scale $M_S^f \approx 10^{-15}M_{pl}$ 
around the true vacuum at the 
end of inflation. Thus, at the true vacuum (i.e. $\chi = C$ and $\phi=0$) where $\mathcal{U} = 0$, we get $ V_+|_{\chi=C,\phi=0} \approx (M_S^i)^4 e^{-3\sqrt{2/3}C} \equiv (M_S^f)^4 $. From this, we find where the location of the true vacuum in the $\chi$ direction should be (recall that $\chi$ is a flat direction after inflation) 
\begin{eqnarray}
 C = \sqrt{\frac{8}{3}} \ln \frac{M_S^i}{M_S^f},
\end{eqnarray}
where $M_S^f$ is a free parameter, which we set to be approximately $10^{-15}$ in Planck units. 

The proposed potential $\mathcal{U}$ vanishes after inflation hence it already trivially satisfies the constraints (\ref{generic_constraints}).  So all we need to do is to check that it also satisfies~\eqref{Universal_Constraint1}.
Using $\mathcal{F} = e^{-3\sqrt{2/3}\chi}$, which gives $M_S^i = M_{pl}=1$ during inflation ($\chi =0$), we first have $\mathcal{U}^{(n)}|_{\chi=0} \ll e^{-3m\sqrt{2/3}\chi}\mathcal{O}(1)|_{\chi=0} = \mathcal{O}(1)$. Using Eq. (\ref{field_redefinition}), we find $\partial_T = \sqrt{6}(-i\partial_{\phi} + e^{-\sqrt{2/3}\chi}\partial_{\chi})$ and 
$\partial_{\bar{T}} = \sqrt{6}(i\partial_{\phi} + e^{-\sqrt{2/3}\chi}\partial_{\chi})$. Note that 
$\mathcal{U}^{(n)}|_{\chi=0} \equiv \partial_T^k \partial_{\bar{T}}^l \mathcal{U}(T,\bar{T})|_{\chi=0}$ 
where $n = k+l$. In particular, since the functional dependence
 on $\chi$ does not produce any singularity at $\chi =0$, it is sufficient to check that 
 $\partial_{\phi}^n \mathcal{U} \ll \mathcal{O}(1)$. Thus, because the dependence on $\phi$ is solely given by the 
 Starobinsky inflationary potential, i.e. $V \sim \alpha (1-e^{-\sqrt{2/3}\phi})$, we will get that its derivatives are always less 
 than the coefficient $\alpha$, thanks to the decreasing exponential factor $e^{-\sqrt{2/3}\phi}$.  This implies that the 
 constraint is automatically satisfied since $\alpha \sim 10^{-10} < \mathcal{O}(1)$. So, all consistency conditions can be 
 satisfied by a liberated supergravity potential.

%% file: chapters/10.tex
\framebox[1.05\width]{{\large This chapter is based on the author's original work in Ref.~\cite{jp2}.}} \par
\vspace{1cm}

\section{Introduction}
It is rather challenging to describe inflation, supersymmetry (SUSY) breaking, and de Sitter (dS) vacua in simple supergravity models and even more so in string theory. In string theory, the 
Kachru-Kallosh-Linde-Trivedi (KKLT) model \cite{KKLT} is a prototype
that can give de Sitter (dS) vacua, under certain assumptions about moduli stabilization. The effective field theory description of the
KKLT model is a supergravity with a no-scale K\"ahler potential for its volume modulus 
and with a superpotential that differs from its constant no-scale form because of two non-perturbative 
corrections\footnote{The corrections come from either Euclidean D3 branes in type IIB compactifications or from 
gaugino condensation due to D7 branes.}.  
The superpotential produces a supersymmetric Anti-de-Sitter (AdS) vacuum. In ref.~\cite{KKLT}, a mechanism was
 proposed for generating dS
 vacua through the addition of anti-D3 brane contributions to the superpotential, that uplifts the AdS vacuum to dS. While the 
additional correction by anti-D3 branes creates dS vacua, it also deforms the shape of the scalar potential creating a ``bump'' which 
gives rise to a moduli stabilization problem~\cite{KKLT}. As an attempt to improve on KKLT, Kachru, Kallosh, Linde, 
Maldacena, McAllister
 and Trivedi (KKLMMT) proposed a model that modifies KKLT by introducing a contribution arising from the anti-D3 tension in a highly warped compactifications~\cite{KKLMMT}.

Both models,  KKLT and the KKLMMT, contain anti-D3 branes, whose known effective field theory description uses nonlinear 
realizations of supersymmetry. The presence of  nonlinearly realized
 supersymmetry means that if supersymmetry is restored at energies below the string scale, $M_{string}$, then 
 the known description of 
 KKLT cannot accurately describe the whole energy range $E\lesssim M_{string}$\footnote{We assume $M_{string}< M_{pl}$, with 
 $M_{pl}$ the Planck scale.}. On the other hand, nothing in principle forbids the existence of {\em some} 
 effective field theory description even in that
 energy range, but such description must employ a linear realization of supersymmetry, which would necessarily employ only whole multiplets. 
A natural question to ask from an effective field theory point of view is whether such a description is possible. Said differently: does
a supergravity with the same K\"ahler potential and superpotential as KKLT exists, that breaks supersymmetry, gives rise to an 
inflationary potential and a dS post-inflationary vacuum, and is valid even at energy scales where supersymmetry is
restored?
We answer affirmatively to this question by adding to the KKLT effective theory a new Fayet-Iliopoulos (FI) term, in the form proposed 
by Antoniadis, Chatrabhuti, Isono and Knoops~\cite{acik}. 
We will show that this FI term also generates irrelevant operators that introduce a cutoff scale for the effective theory. We will
also show that differently from nonlinear realizations, this cutoff can be made larger than the supersymmetry breaking scale --and in
fact even larger than the Planck scale.     

Our construction begins with the observation that, in the absence of anti-D3 branes, the supergravity scalar potential of 
the KKLT model has a supersymmetric AdS vacuum and it becomes flat for large values of the volume modulus field. The flat direction could be used 
for constructing a viable model of inflation without eta problem, if the scalar potential minimum $V_0$ could be simply
 translated upward by a constant, 
$V_0\rightarrow V_0+ \mathrm{constant}$. This could happen if a constant positive FI term existed. This term was long thought to be 
forbidden in supergravity, since the only possible FI terms were thought to arise from gauging the R-symmetry~\cite{R_symm,R_symm_2,R_symm_3},
require an R-invariant superpotential~\cite{barb} and be subject to quantization conditions when the gauged R-symmetry is 
compact~\cite{Nathan}.
On the other hand, recently FI terms not associated with R-symmetry were proposed, starting 
with ref.~\cite{CFTV}.
We use here the K\"{a}hler-invariant FI term proposed in~\cite{acik} and we 
call it ``ACIK-FI'' to distinguish it from many other new FI terms
 suggested in the literature (for instance in~\cite{CFTV,oldACIK,AKK,Kuzenko}). 
To find an approximately flat potential for inflation and a dS post-inflationary vacuum, we add an ACKI-FI term  to
the  $\mathcal{N}=1$ supergravity describing the KKLT model without anti-D3 branes. 
 We must remark that a field-dependent generalization of the new K\"{a}hler-invariant FI term has 
 been introduced recently in ref.~\cite{ar}, which also studies the cosmological consequences of such a term.

In our model supersymmetry is spontaneously broken in a hidden sector at a very high but still sub-Planckian scale 
$M_{pl} \gg M_S \gg 10^{-15}M_{pl}$. 
We employ gravity mediation (see e.g. the review~\cite{GM}) to communicate the SUSY breaking to the observable sector, where supersymmetry breaking manifests itself through the existence of explicit soft 
SUSY breaking terms, characterized by an energy scale $M_{observable} \ll M_S$.
    The reason for a high $M_S$
  is that $M_S$ controls the magnitude of non-renormalizable fermionic terms that determine the cutoff of the effective theory. This is a
  feature that the ACIK-FI term shares with liberated supergravity (see e.g.~\cite{fkr,jp1}).
   
 The purpose of our work is to find an effective field theory of inflation, de-Sitter moduli stabilization, and supersymmetry breaking as a
  cosmological application to the effective theory of KKLT of the ACIK-FI term proposed in~\cite{acik}. 
  The string theory origin of one particular type of the new FI 
 terms  has recently been investigated through a supersymmetric Born-Infeld action~\cite{CFT}, so it would be of clear 
 interest to study a possible string-theoretical origin of the ACIK-FI term. 

This work is organized as follows. In Sec. \ref{2} we show how to add an ACIK-FI term to the $\mathcal{N}=1$ supergravity 
effective theory of the KKLT model. Next we add matter, which we divide into a hidden sector and an observable sector. Supersymmetry 
is broken in the hidden sector and the SUSY breaking is communicated to the observable sector via gravity mediation.
  In Secs. \ref{3} and \ref{4} we probe the hidden-sector dynamics of our model. In Sec. \ref{3}, we construct a minimal supergravity model of 
 plateau-potential inflation --sometimes called in the literature ``Starobinsky'' or ``Higgs'' inflation-- with high scale SUSY breaking and dS vacua, using the results from Sec. \ref{2}. In Sec. \ref{4}, we explore the gravitino mass, which is very high, being
 well above the EeV-scale. We also study possible constraints on the ACIK-FI term by investigating the nonrenormalizable fermionic terms in the Lagrangian based on Ch. \ref{Spectroscophy}. In Sec. \ref{5} we 
  study the observable-sector dynamics of our model by computing its soft SUSY breaking terms. A few final observations are collected
  in Sec. \ref{6}.

\section{Adding a K\"{a}hler-Invariant Fayet-Iliopoulous Term to KKLT-type $\mathcal{N}=1$ Supergravity}\label{2}

In this section, we propose an $\mathcal{N}=1$ supergravity model that can describe the low energy effective field theory of inflation and moduli stabilization in KKLT-type backgrounds \cite{KKLT}. To do so, we first add an ACIK-FI term to an $\mathcal{N}=1$ supergravity that is compatible with the KKLT model.

In general, an ACIK-FI term can be introduced  into an $\mathcal{N}=1$ supergravity without requiring a gauged R-symmetry \cite{acik,ar}. In our proposal, we will introduce instead only 
an  ordinary U(1) symmetry (under which the superpotential is invariant) which will be gauged by a vector multiplet $V$. Inflation will come from the same potential as in the KKLT 
scenario. KKLT~\cite{KKLT} argues that in string theory some moduli can develop a non-perturbative superpotential of the form
\begin{eqnarray}
W = W_0 + A e^{-aT}, \label{kklt}
\end{eqnarray}
where $T$ is a ``volume'' modulus field, which is a chiral superfield, and $W_0,A$ are constants. 
For our construction it is sufficient to compute the component action of $\mathcal{N}=1$ supergravity characterized by the 
superpotential~\eqref{kklt} and by an ACIK-FI term. Notice that
 Antoniadis and Rondeau have recently studied  cosmological applications of generalized ACIK-FI terms by considering no-scale models
  with a constant superpotential $W=W_0$ \cite{ar}. Differently from that model, ours uses the KKLT-type superpotential~\eqref{kklt}. 

The key assumption that we will use is that both the volume modulus $T$ and the other matter fields that may exist in the 
superpotential are gauge-invariant under the U(1) that is used to introduce the ACIK-FI term. In this paper, we use superconformal 
tensor calculus \cite{cfgvnv} to calculate the action. 

The goal of this work is to find a modestly 
realistic minimal supergravity model of inflation with realistic moduli stabilization and supersymmetry
breaking pattern. The study of irrelevant operators generated by the ACIK-FI term will show that a low energy supersymmetry breaking
is incompatible with demanding that the cutoff for the effective field theory is higher than the Hubble constant during inflation. So, we 
take an alternative approach and break supersymmetry at a high scale in the hidden sector (as in {\it e.g.}~\cite{HighSUSYGUT} )
while keeping some of the scales of supersymmetry breaking interactions in the observable sector low~\cite{softSUSYbreaking}.

To do so, we first decompose matter into a hidden sector and an observable sector.  We will discuss them separately in Sections 3, 4
and 5. So we separate the field coordinates $y^A$ into 
\begin{eqnarray}
y^A \equiv (T,z^{\hat{I}}) \equiv (\{T,z^I\}^h,\{z^i\}^o),
\end{eqnarray}
where $\hat{I} \equiv (I,i)$ and $\{T,z^I\}^h$ are hidden-sector fields, while $\{z^i\}^o$ are the observable-sector ones. In addition to this, we write a generic superpotential $W$ as a sum of a hidden-sector term $W^h$ and observable-sector term $W^o$:
\begin{eqnarray}
W(y^A) \equiv W^h(T,z^I) + W^o(z^i).
\end{eqnarray}

We further assume that the hidden-sector superpotential carries a high energy scale compared to the observable-sector one. This 
implies that we decompose the F-term scalar potential into two different parts: a hidden sector F-term potential characterized by a
 high energy scale and observable-sector F-term potential containing only  low scale SUSY-breaking soft terms.

Next, to introduce an ACIK-FI term into our theory we suppose that the volume modulus multiplet $T$ and all observable-sector chiral matter multiplets $Z^i$ are neutral under an ordinary (non-R) 
U(1) gauge symmetry, while the hidden-sector chiral matter multiplets $z^I$ are charged, {\it i.e.} they transform as
\begin{eqnarray}
Z^i \rightarrow Z^i, \quad T \rightarrow T, \quad Z^I \rightarrow e^{-q_I\Omega}Z^I.
\end{eqnarray}
Here $q_I$ denote the U(1) gauge charges of the hidden-sector chiral multiplets $Z^I$ and $\Omega$ is the chiral multiplet containing 
in its lowest component the ordinary gauge parameter. We make these choices because we 
will introduce both a new FI term generated by a gauge vector multiplet and  a KKLT superpotential, which depends on  the volume modulus $T$ and must be gauge invariant under all gauge symmetries.

The superconformal action of the ACIK-FI term \cite{acik,ar} is defined by
\begin{eqnarray}
\mathcal{L}_{\textrm{NEW FI}} \equiv - \xi \left[(S_0\bar{S}_0e^{-K(Ze^{qV},\bar{Z})})^{-3}\frac{(\mathcal{W}_{\alpha}(V)\mathcal{W}^{\alpha}(V))(\bar{\mathcal{W}}_{\dot{\alpha}}(V)\bar{\mathcal{W}}^{\dot{\alpha}}(V))}{T(\bar{w}^2)\bar{T}(w^2)}(V)_D\right]_D,\label{newFIterm}
\end{eqnarray}
and the corresponding superconformal action of $\mathcal{N}=1$ supergravity with superpotential~\eqref{kklt}  and the 
new the FI term is 
\begin{eqnarray}
\mathcal{L} &=& -3[S_0\Bar{S}_0e^{-K(Ze^{qV},\bar{Z})/3}]_D + [S_0^3W(Z,Z')]_F + \frac{1}{2g^2}[
\mathcal{W}_{\alpha}(V)\mathcal{W}^{\alpha}(V)]_F + c.c. \nonumber\\
&&- \xi \left[(S_0\bar{S}_0e^{-K(Ze^{qV},\bar{Z})})^{-3}\frac{(\mathcal{W}_{\alpha}(V)\mathcal{W}^{\alpha}(V))(\bar{\mathcal{W}}_{\dot{\alpha}}(V)\bar{\mathcal{W}}^{\dot{\alpha}}(V))}{T(\bar{w}^2)\bar{T}(w^2)}(V)_D\right]_D. \label{newFI2}
\end{eqnarray}
In Eqs.~(\ref{newFIterm},\ref{newFI2}) $S_0$ is the conformal compensator with Weyl/chiral weights (1,1); $Z^A = (T,Z^I;Z^i)$ and $V$ are chiral matter and vector multiplets with weights $(0,0)$; $K(Ze^{qV},\bar{Z})$ is a K\"{a}hler potential gauged by a vector multiplet $V$; $W(Z,Z')$ is a 
superpotential;
$\mathcal{W}_{\alpha}(V)$ is the field strength of the vector multiplet $V$; $\xi$ is the constant coefficient of ACIK-FI term; 
$w^2 \equiv \frac{\mathcal{W}_{\alpha}(V)\mathcal{W}^{\alpha}(V)}{(S_0\bar{S}_0e^{-K(Z,\bar{Z})})^{2}}$ and $\bar{w}^2 \equiv \frac{\bar{\mathcal{W}}_{\dot{\alpha}}(V)\bar{\mathcal{W}}^{\dot{\alpha}}(V)}{(S_0\bar{S}_0e^{-K(Z,\bar{Z})})^{2}}$ are composite 
multiplets, $T(X), \bar{T}(X)$ are chiral projectors, and $(V)_D$ is a real multiplet, whose lowest component is the auxiliary 
field $D$ of the vector multiplet $V$.  

Next, we write the following K\"ahler potential, invariant under the same U(1) that generates the ACIK-FI term
\begin{eqnarray}
K(Z^Ae^{qV},\bar{Z}^{\bar{A}})\equiv -3\ln[T+\Bar{T} - \Phi(Z^Ie^{qV},\bar{Z}^{\bar{I}}; Z^i,\bar{Z}^{\bar{i}})/3] ,
\end{eqnarray}
where $\Phi$ is a real function of the matter multiplets $Z^i,Z^I$ and the two terms in the superpotential 
$W \equiv W^h + W^o$ are the hidden-sector term
\begin{eqnarray}
W^h(T) \equiv W_0 + Ae^{-aT}
\end{eqnarray}
and the observable-sector superpotential 
\begin{eqnarray}
W^o(Z^i) \equiv B_0 + S_iZ^i + M_{ij}Z^iZ^j + Y_{ijk}Z^iZ^jZ^k + \cdots,
\end{eqnarray}
where $B_0,S_i,M_{ij},Y_{ijk}$ are constant coefficients.
We will choose $\Phi$ to be sum of a term containing only hidden-sector fields and one containing only those of the
observable sector 
\begin{equation}
\Phi=\Phi^h(Z^Ie^{qV},\bar{Z}^{\bar{I}})+ \Phi^o(Z^i,\bar{Z}^{\bar{i}}) . \label{factor}
\end{equation}

The supergravity G-function corresponding to our model is then 
\begin{eqnarray}
G(y^A,\bar{y}^{\bar{A}}) \equiv K(y^A,\bar{y}^{\bar{A}}) + \ln|W(y^A)|^2 = -3\ln[T+\bar{T}-\frac{\Phi(z^{\hat{I}},\bar{z}^{\bar{\hat{I}}})}{3}] + \ln|W^h(T,z^I)+W^o(z^i)|^2.
\end{eqnarray}
The F-term supergravity scalar potential is given by the formula $V_F \equiv e^G(G_AG^{A\bar{B}}G_{\bar{B}}-3)$, which in our case 
reads
\begin{eqnarray}
V_F &=& 
-\frac{1}{X^{2}}[(W^h+W^o)\bar{W}^h_{\bar{T}}+(\bar{W}^h+\bar{W}^o)W^h_{T}]
\nonumber\\&&
+\frac{1}{3}\frac{|W^h_T|^2}{X^{2}}+\frac{1}{9}\frac{|W^h_T|^2}{X^{2}}[\Phi_{I}\Phi^{I\bar{J}}\Phi_{\bar{J}}+\Phi_{i}\Phi^{i\bar{j}}\Phi_{\bar{j}}] \nonumber\\&&
+ \frac{1}{3}\frac{1}{X^{2}}
[W^h_T(\Phi_{I}\Phi^{I\bar{J}}\bar{W}^h_{\bar{J}}+\Phi_{i}\Phi^{i\bar{j}}\bar{W}^o_{\bar{j}})
+\bar{W}^h_{\bar{T}}(W^h_{I}\Phi^{I\bar{J}}\Phi_{\bar{J}}+W^o_{i}\Phi^{i\bar{j}}\Phi_{\bar{j}})] \nonumber\\&&
+\frac{1}{X^{2}}[W^h_I\Phi^{I\bar{J}}\bar{W}^h_{\bar{J}}+W^o_i\Phi^{i\bar{j}}\bar{W}^o_{\bar{j}}]. \label{scal-pot}
\end{eqnarray}

When matter scalars are charged under a gauge group there exists also a D-term contribution to the scalar potential, $V_D$.
 In our model, we find it to be
\beq
V_D = \frac{1}{2}g^2  \Big(\xi+\sum_I(q_Iz^IG_I+q_I\bar{z}^{\bar{I}}G_{\bar{I}})\Big)^2 = 
\frac{1}{2}g^2   \Big(\xi+\frac{q_Iz^I\Phi_I+q_I\bar{z}^{\bar{I}}\Phi_{\bar{I}}}{X}\Big)^2,
\eeq{D-term}
where $X \equiv T+\bar{T}-\Phi/3$, $g$ is the gauge coupling constant and $\xi$ is the ACIK-FI constant. Remember that only hidden-sector chiral matter multiplets are charged under the U(1). 
The scalar potential is the sum of two terms. One, $V_h$  contains the D-term contribution and the F-term potential of the hidden 
sector, depends on the high mass scale $M_S$ and is $O(H^2M_{pl}^2)$ during inflation; the other, $V_{soft}$ contains the observable sector scalars and depends only on low mass scales: 
\begin{eqnarray}
V = V_h + V_{soft},
\end{eqnarray}
where
\begin{eqnarray}
V_h &\equiv& V_D - \frac{W^h_T\bar{W}^h+\bar{W}^h_{\bar{T}}W^h}{X^{2}} + \frac{|W^h_T|^2}{3X^{2}}\Big(X+\frac{1}{3}\Phi_I\Phi^{I\bar{J}}\Phi_{\bar{J}}\Big) \nonumber\\
&& + \frac{1}{3}\frac{1}{X^{2}}
[W^h_T\Phi_{I}\Phi^{I\bar{J}}\bar{W}^h_{\bar{J}}
+\bar{W}^h_{\bar{T}}W^h_{I}\Phi^{I\bar{J}}\Phi_{\bar{J}}]+\frac{1}{X^{2}}W^h_I\Phi^{I\bar{J}}\bar{W}^h_{\bar{J}} ,\\
V_{soft} &\equiv&  
-\frac{1}{X^{2}}[W^o\bar{W}^h_{\bar{T}}+\bar{W}^oW^h_{T}]
+\frac{1}{9}\frac{|W^h_T|^2}{X^{2}}\Phi_{i}\Phi^{i\bar{j}}\Phi_{\bar{j}} \nonumber\\&&
+ \frac{1}{3}\frac{1}{X^{2}}
[W^h_T\Phi_{i}\Phi^{i\bar{j}}\bar{W}^o_{\bar{j}}
+\bar{W}^h_{\bar{T}}W^o_{i}\Phi^{i\bar{j}}\Phi_{\bar{j}}] 
+\frac{1}{X^{2}}W^o_i\Phi^{i\bar{j}}\bar{W}^o_{\bar{j}}.
\end{eqnarray}

\section{Hidden Sector Dynamics 1: A Minimal Supergravity Model of Inflation, High-Scale Supersymmetry Breaking, and de Sitter Vacua}\label{3}

In this section, we explore a minimal supergravity model of high-scale supersymmetry breaking and plateau-potential inflation through gravity mediation and no-scale K\"ahler potential. We investigate first the hidden sector dynamics. We have assumed that the hidden-sector potential depends on a high energy scale and dominates over the observable-sector one. Hence, it is reasonable to minimize the hidden-sector potential first. 
Let us compute now the F-term potential in the hidden sector. Recalling that the KKLT superpotential is
\begin{eqnarray}
W^h(T) \equiv W_0 + Ae^{-aT},
\end{eqnarray}
and redefining $W_0 \equiv -cA$, we rewrite it as
\begin{eqnarray}
W^h(T) = A(e^{-aT}-c),
\end{eqnarray}
where $a,c,A$ are positive constants. Note that $W^h_I = \partial W^h/\partial z^I = 0$.

Since we defined $X \equiv T+\bar{T}-\Phi/3$, the KKLT superpotential gives 
\begin{eqnarray}
&& W^h_T = -aAe^{-aT},\quad |W^h_T|^2 = a^2A^2 e^{-a(T+\bar{T})}= a^2A^2e^{-a(X+\Phi/3)}, 
\end{eqnarray}
\begin{eqnarray}
W^h_T\bar{W}^h+\bar{W}^h_{\bar{T}}W^h &=& -aA^2e^{-aT}(e^{-a\bar{T}}-c) -aA^2e^{-a\bar{T}}(e^{-aT}-c) \nonumber\\
&=& -2aA^2e^{-a(T+\bar{T})}+aA^2c(e^{-aT}+e^{-a\bar{T}}) \nonumber\\
&=& -2aA^2e^{-a(X+\Phi/3)}+aA^2c(e^{-a(\textrm{Re}T+i\textrm{Im}T)}+e^{-a(\textrm{Re}T-i\textrm{Im}T)}) \nonumber\\
&=& -2aA^2e^{-a(X+\Phi/3)}+2aA^2c e^{-a\textrm{Re}T} \cos(a\textrm{Im}T)
\nonumber\\
&=& -2aA^2e^{-a(X+\Phi/3)}+2acA^2 e^{-a(X+\Phi/3)/2} \cos(a\textrm{Im}T)
\end{eqnarray}
\begin{eqnarray}
|W^h|^2 &=& A^2|e^{-aT}-c|^2 = A^2 (e^{-aT}-c)(e^{-a\bar{T}}-c)= A^2(e^{-a(T+\bar{T})}-c(e^{-aT}+e^{-a\bar{T}})+c^2) \nonumber\\
&=& A^2 (e^{-a(X+\Phi/3)}-2ce^{-a(X+\Phi/3)/2} \cos(a\textrm{Im}T)+c^2).
\end{eqnarray}
Here we have used the following transformation from the complex coordinate $T$ to two real coordinates $X,\textrm{Im}T$:
\begin{eqnarray}
T = \textrm{Re}T+i\textrm{Im}T = \frac{1}{2}\Big(X+\frac{\Phi}{3}\Big) + i\textrm{Im}T, 
\end{eqnarray}
which gives $e^{-aT} = e^{-\frac{a}{2}(X+\frac{\Phi}{3})}e^{-a i\textrm{Im}T}$. Remember that $X \equiv T+\bar{T}-\Phi/3$.

Then, since $W^h_I=0$, the corresponding hidden-sector F-term scalar potential is given by
\begin{eqnarray}
V^h_F &=& - \frac{W^h_T\bar{W}^h+\bar{W}^h_{\bar{T}}W^h}{X^{2}} + \frac{|W^h_T|^2}{3X^{2}}\Big(X+\frac{1}{3}\Phi_I\Phi^{I\bar{J}}\Phi_{\bar{J}}\Big) \nonumber\\
&=& - \frac{1}{X^2}\bigg(-2aA^2e^{-a(X+\Phi/3)}+2acA^2 e^{-a(X+\Phi/3)/2} \cos(a\textrm{Im}T)\bigg)
\nonumber\\&&+ \frac{1}{3X^{2}}\Big(X+\frac{1}{3}\Phi_I\Phi^{I\bar{J}}\Phi_{\bar{J}}\Big) a^2A^2 e^{-a(X+\Phi/3)}.
\end{eqnarray}

Since we assume that the D-term potential belongs to the hidden sector, the hidden-sector total scalar potential can be written as
\begin{eqnarray}
V_h &=& V_D+V_F^h  \nonumber\\
&=& \frac{1}{2}g^2   \Big(\xi+\frac{q_Iz^I\Phi_I+q_I\bar{z}^{\bar{I}}\Phi_{\bar{I}}}{X}\Big)^2 - \frac{1}{X^2}\bigg(-2aA^2e^{-a(X+\Phi/3)}+2acA^2 e^{-a(X+\Phi/3)/2} \cos(a\textrm{Im}T)\bigg)
\nonumber\\&&+ \frac{1}{3X^{2}}\Big(X+\frac{1}{3}\Phi_I\Phi^{I\bar{J}}\Phi_{\bar{J}}\Big) a^2A^2 e^{-a(X+\Phi/3)}.
\end{eqnarray}
We define the SUSY breaking scale $M_S$ in terms of the scalar potential $V_h$ and the gravitino mass $m_{3/2}$ as
\begin{eqnarray}
V_+ \equiv M_S^4 = V_h+3m_{3/2}^2 = V_h + \frac{3}{X^3}A^2 (e^{-a(X+\Phi/3)}-2ce^{-a(X+\Phi/3)/2} \cos(a\textrm{Im}T)+c^2).
\end{eqnarray}

To investigate the moduli stabilization, we identify the canonically normalized fields by inspection of the kinetic terms, which are given by
\begin{eqnarray}
\mathcal{L}_K &=& \frac{\Phi_{\hat{I}\bar{\hat{J}}}}{X}g_{\mu\nu} D_{\mu}z^{\hat{I}}D^{\nu}\bar{z}^{\bar{\hat{J}}} + \frac{3}{4X^2}g_{\mu\nu}\partial_{\mu}X\partial_{\nu}X
\nonumber\\
&&+\frac{3}{X^2}g_{\mu\nu}[\partial_{\mu}\textrm{Im}T-(\textrm{Im}D_{\mu}z^{\hat{I}}\Phi_{\hat{I}}/3)][\partial_{\nu}\textrm{Im}T-(\textrm{Im}D_{\nu}z^{\hat{I}}\Phi_{\hat{I}}/3)] ,
\end{eqnarray}
where $D_{\mu} \equiv \partial_{\mu} - iq_{\hat{I}}A_{\mu}$ is the U(1) gauge covariant derivative for the matter multiplets $z^{\hat{I}} = (z^I,z^i)$ with gauge charge $q_{\hat{I}}=(q_I\neq0,q_i=0)$, and $A_{\mu}$ is the corresponding gauge field.

After performing another field redefinition $X \equiv e^{\sqrt{2/3}\phi}$, we find 
\begin{eqnarray}
\mathcal{L}_K &=& \Phi_{I\bar{J}}e^{-\sqrt{2/3}\phi}g_{\mu\nu} D_{\mu}z^ID^{\nu}\bar{z}^{\bar{J}} + \frac{1}{2}g_{\mu\nu}\partial_{\mu}\phi\partial_{\nu}\phi
\nonumber\\
&&+3e^{-2\sqrt{2/3}\phi}g_{\mu\nu}[\partial_{\mu}\textrm{Im}T-(\textrm{Im}D_{\mu}z^I\Phi_I/3)][\partial_{\nu}\textrm{Im}T-(\textrm{Im}D_{\nu}z^I\Phi_I/3)].\label{kinetic_term}
\end{eqnarray}
Notice that $\phi$ is canonically normalized, while the other fields $z^I, \textrm{Im}T$ are so only when $\phi$ is small.

Now, let us investigate the scalar potential vacuum. First of all, we find the minimum with respect to the matter scalars $z^{\hat{I}}$, 
\begin{eqnarray}
\frac{\partial V}{\partial z^{\hat{I}}} = 0 \implies \Phi_{{\hat{I}}} = 0. 
\end{eqnarray}
If we choose a real function such that $\Phi_{\hat{I}}=0$ implies $\Phi =0$ together with $z^{\hat{I}}=0$ then at this vacuum the scalar potential becomes~\footnote{The observable-sector superpotential $W^o$ can shift the VEVs
of the scalars in the observable sector $z^i$, but since those VEVs must be in any case small compared to $H$ and $M_{pl}$ we can
approximately set $z^i=0$. Moreover, in our toy example in Section 5 we will choose a superpotential that indeedgives a 
minimum at $z^i=0$.}
\begin{eqnarray}
V_h &=& \frac{1}{2}g^2\xi^2 - \frac{1}{X^2}\bigg(-2aA^2e^{-aX}+2acA^2 e^{-aX/2} \cos(a\textrm{Im}T)\bigg)+ \frac{1}{3X} a^2A^2 e^{-aX}.
\end{eqnarray}
Next, we consider the vacuum with respect to the $\textrm{Im}T$ field. We find the vacuum at $a\textrm{Im}T=n\pi$, where $n$ is an 
even integer, leading to $\cos(a\textrm{Im}T)=1$\footnote{When $\cos(a\textrm{Im}T)=1$, the second derivative of the potential can be positive, which means that the stationary point is a minimum.} and
\begin{eqnarray}
V_h &=& \frac{1}{2}g^2\xi^2 +\frac{2aA^2}{X^2}e^{-aX}- \frac{2acA^2}{X^2} e^{-aX/2}+ \frac{a^2A^2}{3X}  e^{-aX}.
\end{eqnarray}

Next, let us find the vacuum with respect to the $\phi$ field. Recalling that  $X=e^{\sqrt{2/3}\phi}$, calling 
$\langle \phi \rangle$ the vacuum expectation value of $\phi$ and setting
$\phi=\left<\phi\right> = \sqrt{\frac{3}{2}}\ln \left<X\right> = \sqrt{\frac{3}{2}}\ln x$, where  $X = \left<X\right> \equiv x$, we have
\begin{eqnarray}
\frac{\partial V_h}{\partial \phi}\bigg|_{\phi=\left<\phi\right>} = \frac{\partial V_h}{\partial X}\bigg|_{X=x}\frac{\partial X}{\partial \phi} \bigg|_{\phi=\left<\phi\right>}= 0 \implies \frac{\partial V_h}{\partial X}\bigg|_{X=x} = 0,
\end{eqnarray}
which gives 
\begin{eqnarray}
\frac{\partial V_h}{\partial X}\bigg|_{X=x}=-\frac{4aA^2}{x^3} e^{-ax} - \frac{2a^2A^2}{x^2} e^{-ax} + \frac{4acA^2}{x^3}e^{-ax/2} + \frac{a^2cA^2}{x^2}e^{-ax/2} - \frac{a^2A^2}{3x^2}e^{-ax} - \frac{a^3A^2}{3x}e^{-ax} = 0. \nonumber\\
{}
\end{eqnarray}
At first glance, this equation seems a little complicated, but after a short calculation, we can obtain the following simple relation
\begin{eqnarray}
\frac{\partial V_h}{\partial X}\bigg|_{X=x} = 0 \implies c = \Big(1+\frac{ax}{3}\Big)e^{-ax/2}.
\end{eqnarray}

Inserting the value of $c$ into $V_h$, we obtain the following equation
\begin{eqnarray}
V_h &=& \frac{1}{2}g^2\xi^2 +\frac{2aA^2}{X^2}e^{-aX}- \frac{2aA^2}{X^2}\Big(1+\frac{ax}{3}\Big)e^{-ax/2} e^{-aX/2}+ \frac{a^2A^2}{3X}  e^{-aX},
\end{eqnarray}
where $X = e^{\sqrt{2/3}\phi}$.

Then, the vacuum energy at $X=x$ is given by
\begin{eqnarray}
V_h|_{X=x} = \frac{1}{2}g^2\xi^2 - \frac{a^2A^2e^{-ax}}{3x} \equiv \Lambda,
\end{eqnarray}
where $\Lambda$ is defined to be the post-inflationary cosmological constant, and the SUSY breaking scale is given by
\begin{eqnarray}
V_+|_{X=x} &=& V_h|_{X=x} + \frac{3}{X^3}A^2 (e^{-aX}-2ce^{-aX/2}+c^2)\bigg|_{X=x} \nonumber\\
&=& \Lambda + \frac{3}{x^3}A^2 (e^{-ax/2}-c)^2 = \Lambda + \frac{3A^2}{x^3} \frac{a^2x^2e^{-ax}}{9}\nonumber\\
&=& \Lambda + \frac{a^2A^2e^{-ax}}{3x} = \frac{1}{2}g^2\xi^2 \equiv M_S^4 ,
\end{eqnarray}
where $M_S$ is by definition the SUSY breaking scale.  
We can set $\Lambda$ to any value we wish, in particular we can choose it to be 
$\Lambda\sim 10^{-120}$. 

Here, we point out that the term $ \frac{1}{2}g^2\xi^2$ governs the magnitude of the total scalar potential, and simultaneously controls the scale of spontaneously supersymmetry breaking. Hence, if we want that the scalar potential describes inflation, we need to impose 
\begin{eqnarray}
M_S^4 =  \frac{1}{2}g^2\xi^2 \overset{!}{=} H^2M_{pl}^2 \equiv M_I^4,
\end{eqnarray}
where $H$ is the Hubble parameter, and $M_I$ is defined to be the mass scale of inflation. 

We then identify 
\begin{eqnarray}
A = \sqrt{\frac{3x(M_I^4-\Lambda)e^{ax}}{a^2}}, \qquad W_0 = -cA = -\Big(1+\frac{ax}{3}\Big)\sqrt{\frac{3x(M_I^4-\Lambda)}{a^2}}.
\end{eqnarray}

Substituting the above parameters $A,W_0,M_I$ into the hidden-sector potential, we can obtain a plateau inflation potential 
\begin{eqnarray}
V_h = M_I^4 - (M_I^4-\Lambda)x \bigg[ \frac{6e^{-a(X-x)/2}}{aX^2}\left(1-e^{-a(X-x)/2}+\frac{ax}{3}\right)-\frac{e^{-a(X-x)}}{X}  \bigg],
\end{eqnarray}
where $X = e^{\sqrt{2/3}\phi}$ and $\phi$ is defined to be the inflaton. Notice that the inflaton mass after inflation is of order of the 
Hubble scale, {\it i.e.} $m_{\phi}^2 \sim H^2 = 10^{-10}M_{pl}^2$.

We note that the hidden-sector scalar potential $V_h$ has a plateau, so it is of HI type (in the notations of ref.~\cite{LightScalar}).
Furthermore, it depends only on four parameters, which are: the vacuum expectation value of $X$ ({\it i.e.} $x \equiv \left<X\right>$); the KKLT parameter $a$ in the superpotential, which will be determined according to the type of the nonperturbative correction we choose\footnote{ For example, if we consider a nonperturbative correction due to gaugino condensation, then we find $a = \frac{2\pi}{N_c}$ for a non-abelian gauge group $SU(N_c)$ where $N_c$ is interpreted as the number of coincident D7 branes being stacked \cite{KKLT}.}; the inflation scale $M_I$, and the post-inflationary cosmological constant $\Lambda$. 
At $X=x$, the potential indeed reduces to the post-inflationary cosmological constant. As an additional remark, we observe that for fixed $x,M_I,\Lambda$ inflation ends earlier when $a$ is smaller. When the  
nonperturbative corrections to the KKLT superportential come from gaugino condensation~\cite{KKLT}, a smaller parameter $a$
corresponds to more D7 branes being stacked.

\section{Hidden Sector Dynamics 2: Super-EeV Gravitino Mass}\label{4}

In this section, we investigate some physical implications that can be obtained from our model. The SUSY breaking scale is identified with $M_S^4 \sim H^2M_{pl}^2$. We find the gravitino mass after inflation, which is
 generated by the high-scale SUSY breaking in the hidden sector. It is given by
\begin{eqnarray}
m_{3/2}^2 &=& e^G = \frac{|W^h|^2}{X^3} = \frac{A^2}{X^3}(e^{-a(X+\Phi/3)}-2ce^{-a(X+\Phi/3)/2} \cos(a\textrm{Im}T)+c^2)\bigg|_{z^I=0,a\textrm{Im}T=0,X=x} \nonumber\\
&=& \frac{3x(M_I^4-\Lambda)e^{ax}}{a^2x^3}(e^{-ax}-2ce^{-ax/2}+c^2)
=\frac{3x(M_I^4-\Lambda)e^{ax}}{a^2x^3}(c-e^{-ax/2})^2 \nonumber\\
&=& \frac{(M_I^4-\Lambda)}{3} \implies m_{3/2} \approx \frac{H}{\sqrt{3}}  = 10^{-6}M_{pl} \sim 10^{12}~\textrm{GeV} = 10^3 ~\textrm{EeV},
\end{eqnarray}
which is compatible with the case of EeV-scale gravitino cold dark matter candidates. This is not surprising because we are considering
 the same high-scale supersymmetry breaking scale as in \cite{EeVGravitino,Inf_HighSUSY_EeVGrav,HeavyGravitino,GDM}, where
 that scenario was proposed. The possibility of direct detection for such heavy dark matter candidates has recently been studied in ref. \cite{HDM}. Notice that in our model the gravitino mass is always $\mathcal{O}(H)$, irrespective of the ultraviolet cutoff.

\section{Observable Sector Dynamics: Low Scale Soft Supersymmetry Breaking Interactions}\label{5}

In this section we investigate the mass scales of the soft supersymmetry-breaking interactions in the observable sector. We
need to find under which conditions our model could be phenomenologically realistic. A full investigation of the detailed structure of the 
soft interactions in the observable sector requires a study that goes beyond the scope of this work, so here we will limit ourselves to
general remarks and a coarse-grained analysis of necessary conditions for the viability of our model. We focus our
analysis on the soft masses. 

Restoring the mass dimension (so that the $T,z^i$ have  canonical mass dimension 1), the soft-term potential becomes
\begin{eqnarray}
V_{soft} &\equiv&  
-\frac{1}{M_{pl}X^{2}}[W^o\bar{W}^h_{\bar{T}}+\bar{W}^oW^h_{T}]
+\frac{1}{9}\frac{|W^h_T|^2}{M_{pl}^2X^{2}}\Phi_{i}\Phi^{i\bar{j}}\Phi_{\bar{j}} \nonumber\\&&
+ \frac{1}{3}\frac{1}{M_{pl}X^{2}}
[W^h_T\Phi_{i}\Phi^{i\bar{j}}\bar{W}^o_{\bar{j}}
+\bar{W}^h_{\bar{T}}W^o_{i}\Phi^{i\bar{j}}\Phi_{\bar{j}}] 
+\frac{1}{X^{2}}W^o_i\Phi^{i\bar{j}}\bar{W}^o_{\bar{j}}.
\end{eqnarray}
This formula is obtained by taking the following low-energy limit:  $F^h, M_{pl} \rightarrow \infty$ (where $F^h$ are the hidden-sector auxiliary F-term fields) while $m_{3/2}$=constant~\cite{GM}. Elegant examples of gravity mediation and soft SUSY breaking are simply explained in {\it e.g.}~\cite{Superconformal_Freedman}. 

In addition, the hidden-sector superpotential can be written as 
\begin{eqnarray}
W^h &=& A(e^{-aT/M_{pl}}-c)=M_{pl}\sqrt{\frac{3x(M_I^4-\Lambda)e^{ax}}{a^2}}(e^{-aT/M_{pl}}-(1+ax/3)e^{-a/2}) \nonumber\\
&=& M_{pl}\sqrt{\frac{3x(M_I^4-\Lambda)e^{ax}}{a^2}}(e^{-a(X+\Phi/3M_{pl}^2)/2}e^{-ia\textrm{Im}T/M_{pl}}-(1+ax/3)e^{-ax/2}),
\end{eqnarray}
 where we have used $e^{-aT} = e^{-\frac{a}{2}(X+\Phi/3M_{pl}^2)}e^{-a i\textrm{Im}T/M_{pl}}$.

Then, using
\begin{eqnarray}
W^h_T = -\frac{1}{M_{pl}}aAe^{-a(X+\Phi/3M_{pl}^2)/2}e^{-ia\textrm{Im}T/M_{pl}},\quad |W^h_T|^2 = \frac{1}{M_{pl}^2}a^2A^2e^{-a(X+\Phi/3M_{pl}^2)},
\end{eqnarray}
we obtain
\begin{eqnarray}
V_{soft} &\equiv&  
\frac{aAe^{-a(X+\Phi/3M_{pl}^2)/2}}{M_{pl}^2X^{2}}[W^oe^{ia\textrm{Im}T/M_{pl}}+\bar{W}^oe^{-ia\textrm{Im}T/M_{pl}}]
+\frac{1}{9}\frac{a^2A^2e^{-a(X+\Phi/3M_{pl}^2)}}{M_{pl}^4X^{2}}\Phi_{i}\Phi^{i\bar{j}}\Phi_{\bar{j}} \nonumber \\ &&
- \frac{1}{3}\frac{aAe^{-a(X+\Phi/3M_{pl}^2)/2}}{M_{pl}^2X^{2}}
[e^{-ia\textrm{Im}T/M_{pl}}\Phi_{i}\Phi^{i\bar{j}}\bar{W}^o_{\bar{j}}
+e^{ia\textrm{Im}T/M_{pl}}W^o_{i}\Phi^{i\bar{j}}\Phi_{\bar{j}}] 
+\frac{1}{X^{2}}W^o_i\Phi^{i\bar{j}}\bar{W}^o_{\bar{j}}. \nonumber  \\ &&
\end{eqnarray}

At the true vacuum we have $a\textrm{Im}T/M_{pl}=n\pi, ~ X=x,~z^I=0$ where $n$ is an even integer, so the soft terms become
\begin{eqnarray}
V_{soft} &\equiv&  
\frac{aAe^{-ax/2}}{M_{pl}^2x^{2}}[W^o+\bar{W}^o]
+\frac{1}{9}\frac{a^2A^2e^{-ax}}{M_{pl}^4x^{2}}\Phi_{i}\Phi^{i\bar{j}}\Phi_{\bar{j}} \nonumber\\&&
- \frac{1}{3}\frac{aAe^{-ax/2}}{M_{pl}^2x^{2}}
[\Phi_{i}\Phi^{i\bar{j}}\bar{W}^o_{\bar{j}}
+W^o_{i}\Phi^{i\bar{j}}\Phi_{\bar{j}}] 
+\frac{1}{x^{2}}W^o_i\Phi^{i\bar{j}}\bar{W}^o_{\bar{j}}.
\end{eqnarray}
From $A = \sqrt{\frac{3x(M_I^4-\Lambda)e^{ax}}{a^2}}M_{pl} \approx \frac{\sqrt{3}}{a}x^{1/2}e^{ax/2}M_I^2M_{pl}$, we find $aAe^{-ax/2} = \sqrt{3}x^{1/2}M_I^2M_{pl}$. Inserting this expression into the soft-terms potential, we get
\begin{eqnarray}
V_{soft} &\equiv&  
\frac{\sqrt{3}x^{1/2}M_I^2M_{pl}}{M_{pl}^2x^{2}}[W^o+\bar{W}^o]
+\frac{1}{9}\frac{(\sqrt{3}x^{1/2}M_I^2M_{pl})^2}{M_{pl}^4x^{2}}\Phi_{i}\Phi^{i\bar{j}}\Phi_{\bar{j}} \nonumber\\&&
- \frac{1}{3}\frac{\sqrt{3}x^{1/2}M_I^2M_{pl}}{M_{pl}^2x^{2}}
[\Phi_{i}\Phi^{i\bar{j}}\bar{W}^o_{\bar{j}}
+W^o_{i}\Phi^{i\bar{j}}\Phi_{\bar{j}}] 
+\frac{1}{x^{2}}W^o_i\Phi^{i\bar{j}}\bar{W}^o_{\bar{j}}.
\end{eqnarray}
The soft-terms potential thus reduces to
\begin{eqnarray}
V_{soft} &\equiv&  
\frac{\sqrt{3}x^{-3/2}M_I^2}{M_{pl}}[W^o+\bar{W}^o]
+\frac{1}{3}\frac{M_I^4}{M_{pl}^2x}\Phi_{i}\Phi^{i\bar{j}}\Phi_{\bar{j}} \nonumber\\&& 
- \frac{1}{\sqrt{3}}\frac{x^{-3/2}M_I^2}{M_{pl}}  
[\Phi_{i}\Phi^{i\bar{j}}\bar{W}^o_{\bar{j}}  
+W^o_{i}\Phi^{i\bar{j}}\Phi_{\bar{j}}] 
+\frac{1}{x^{2}}W^o_i\Phi^{i\bar{j}}\bar{W}^o_{\bar{j}}. 
\end{eqnarray}

Next, let us consider a general expansion of the observable-sector superpotential $W^o$
\begin{eqnarray}
W^o(z^i) =\sum_{n=0} \frac{W^o_{i\cdots k}}{n!}  z^i\cdots z^k  = B_0 + S_iz^i + M_{ij}z^iz^j + Y_{ijk}z^iz^jz^k + \cdots, \label{ossp}
\end{eqnarray}
where $W^o_{i\cdots k} \equiv \partial^n W^o(z^i)/\partial z^i \cdots \partial z^k$ and $B_0,S_i,M_{ij},Y_{ijk}$ are constant parameters
determining masses and interactions.

We won't perform a full analysis of all possible ranges of values
for $B_0$, $S_i$, $M_{ij}$ and $Y_{ijk}$; instead, we will simplify out analysis by setting $S_i=0$, so that 
the vacuum of the observable sector is at $z^i=0$, assume that for all $i,j,k$ all $M_{ij}$ and 
$Y_{ijk}$ are of the same order, and set $B_0=0$.
The soft terms in the scalar potential are then generated only by following terms in the expansion of $W^{o}$~\cite{softSUSYbreaking}.
\begin{eqnarray}
W^o(z^i) = M_{ij}z^iz^j + Y_{ijk}z^iz^jz^k,
\end{eqnarray}
where the $M_{ij}$ have mass dimension one and the $Y_{ijk}$ are dimensionless.

This choice also implies that such superpotential does not significantly change the cosmological constant because all 
the minima of the $z^i$ are located at zero. We will choose  the U(1) gauge-invariant K\"ahler function of matter fields as follows
\begin{eqnarray}
\Phi =  \delta_{I\bar{J}}z^{I}\bar{z}^{\bar{J}}+ \delta_{i\bar{j}}z^{i}\bar{z}^{\bar{j}},
\end{eqnarray}
where the first (second) term corresponds to hidden (observable) sector. 

With our simplifying assumptions we obtain
\begin{eqnarray}
V_{soft} &=& \frac{2\sqrt{3}}{3} \frac{x^{-3/2}M_{I}^2}{M_{pl}}\Big[ (M_{ij}z^iz^j+Y_{ijk}z^iz^jz^k) +c.c. \Big] + \frac{M_I^4x^{-1}}{3M_{pl}^2}\delta_{i\bar{j}}z^i\bar{z}^{\bar{j}} \nonumber\\
&&+x^{-2}[ M_{ij}z^j+Y_{ijk}z^jz^k]\delta^{i\bar{j}}[\bar{M}_{\bar{i}\bar{j}}\bar{z}^{\bar{i}}+\bar{Y}_{\bar{i}\bar{j}\bar{k}}\bar{z}^{\bar{i}}\bar{z}^{\bar{k}}]. \label{V_soft}
\end{eqnarray}

We also find the magnitude of the corresponding soft parameters from Eq.~\eqref{V_soft} as 
\begin{eqnarray}
&& \frac{2\sqrt{3}}{3}\frac{M_I^2}{M_{pl}}x^{-3/2}|M_{ij} | \equiv m_{s1}^2, \quad \frac{2\sqrt{3}}{3}\frac{M_I^2}{M_{pl}}x^{-3/2}|Y_{ijk}| \equiv m_{s2}, \quad \frac{1}{3}\frac{M_I^4}{M_{pl}^2}x^{-1} \equiv m_{s3}^2, \nonumber\\
&& |M_{ij}|^2x^{-2} \equiv m_{s4}^2, \quad |Y_{ijk}|^2x^{-2} \equiv \frac{m_{s5}}{M_{pl}}, \quad |M_{ij}||Y_{ijk}|x^{-2} \equiv m_{s6}.
\end{eqnarray}
We observe that during inflation (for large $X$ or $\phi$) all the soft mass parameters are very small. Also, the above result give us the following relations
\begin{eqnarray}
&& x = \frac{M_I^4}{3m_{s3}^2M_{pl}^2} = \frac{H^2}{3m_{s3}^2},\quad| M_{ij}| =\frac{1}{6} \frac{M_I^4}{M_{pl}^2}\frac{m_{s1}^2}{m_{s3}^3}= \frac{1}{6}H^2 \frac{m_{s1}^2}{m_{s3}^3} = \frac{\sqrt{3}}{2}\frac{m_{s1}^2}{H}x^{3/2}, \quad |Y_{ijk}| =\frac{1}{6} \frac{M_I^4}{M_{pl}^2}\frac{m_{s2}}{m_{s3}^3}, \nonumber\\
&& m_{s4} = \frac{1}{4} \frac{m_{s1}^2}{m_{s3}},\quad m_{s5} =\frac{1}{4}  \left(\frac{m_{s2}}{m_{s3}}\right)^2M_{pl}, \quad m_{s6} = \frac{1}{4} \left(\frac{m_{s1}}{m_{s3}}\right)^2m_{s2}. \label{softmasses}
\end{eqnarray}
We note that only $m_{s1},m_{s2},m_{s3}$ are free parameters. However, when we examine the kinetic term in Eq.~\eqref{kinetic_term}
 we observe that at $x=1$ the kinetic terms of the matter multiplets are canonically normalized. The condition $x=1$ then gives 
 $m_{s3} = \frac{H}{\sqrt{3}} = 10^{-6}M_{pl} \sim m_{3/2}$. So in this case, the free parameters reduce to $m_1$ and $m_2$ only. 
Notice that in the regime $m_{s3} \sim m_{3/2}$, the parameter $m_{s1}$ determines the magnitude of $|M_{ij}|$ and $m_{s4}$, 
while $m_{s2}$ determines that of $|Y_{ijk}|$, $m_{s5}$, and $m_{s6}$.


Finally, let us investigate further 
the physical masses of matter scalars in the observable sector. Here, we are going to look only at the matter scalar masses and leave a detailed study of fermion masses and interactions to a future work,
since the purpose of this section is to demonstrate the existence of light scalars in the observable sector, 
whose masses can be smaller than that of the gravitino. Because of the soft mass parameters we found, we expect that some scalars will be as heavy as the gravitino, while other scalars could be much  lighter.

To compute the scalar masses we must remember to include contributions coming from the expansion of the hidden-sector 
potential to second order in the observable-sector scalars $z^i$: $V_h(z^I,z^i)=V_h(0.0) + V_{h\; i{}\bar{j}} z^i {\bar{z}}^{{}\bar{j}}$. We thus consider the general expression for the total scalar potential, which is written with the canonical mass dimensions by 
\begin{eqnarray}
V &=& V_D+V_F^h + V_{soft} 
\nonumber\\
&=& \frac{1}{2}g^2  M_{pl}^4 \Big(\xi+\frac{q_Iz^I\Phi_I+q_I\bar{z}^{\bar{I}}\Phi_{\bar{I}}}{XM_{pl}^2 }\Big)^2 \nonumber\\
&&- \frac{1}{X^2M_{pl}^2}\bigg(-2aA^2e^{-a(X+\Phi/3M_{pl}^2)}+2acA^2 e^{-a(X+\Phi/3M_{pl}^2)/2} \cos(a\textrm{Im}T/M_{pl})\bigg)
\nonumber\\
&&+ \frac{1}{3X^{2}M_{pl}^2}\Big(X+\frac{1}{3M_{pl}^2}\Phi_I\Phi^{I\bar{J}}\Phi_{\bar{J}}\Big) a^2A^2 e^{-a(X+\Phi/3M_{pl}^2)}
\nonumber\\
&&+\frac{aAe^{-a(X+\Phi/3M_{pl}^2)/2}}{M_{pl}^2X^{2}}[W^oe^{ia\textrm{Im}T/M_{pl}}+\bar{W}^oe^{-ia\textrm{Im}T/M_{pl}}]  +\frac{1}{9}\frac{a^2A^2e^{-a(X+\Phi/3M_{pl}^2)}}{M_{pl}^4X^{2}}\Phi_{i}\Phi^{i\bar{j}}\Phi_{\bar{j}} \nonumber \\ &&
- \frac{1}{3}\frac{aAe^{-a(X+\Phi/3M_{pl}^2)/2}}{M_{pl}^2X^{2}}
[e^{-ia\textrm{Im}T/M_{pl}}\Phi_{i}\Phi^{i\bar{j}}\bar{W}^o_{\bar{j}}
+e^{ia\textrm{Im}T/M_{pl}}W^o_{i}\Phi^{i\bar{j}}\Phi_{\bar{j}}] 
+\frac{1}{X^{2}}W^o_i\Phi^{i\bar{j}}\bar{W}^o_{\bar{j}}.
\end{eqnarray}

First, we find that masses of the hidden-sector matter scalars $z^I$ and $\textrm{Im}T$ can be independently defined by tuning the magnitude of the U(1) gauge charge $\forall I:q_I \equiv q$ and the parameter $a$ respectively such that they are positive definite. This implies that the hidden-sector fields can be heavy as much as we wish. Thus, to get an effective single-field slow-roll inflation we should make the hidden-sector matter scalars much heavier than the Hubble scale during slow roll. Their masses can be lighter than the Hubble scale before the onset of the slow-roll period, that is for
very large values of $X$. Second, it is obvious that the inflaton mass is of the same order as the Hubble scale, i.e. $m_{\phi} \sim H$, since the scalar potential is of ``HI'' or ``Starobinsky''form and has a de Sitter vacuum, as we have
seen in the previous sections.

Next, we investigate masses of the observable-sector fields. We can simplify further our analysis to make our point 
clearer by assuming that the quadratic term in the superpotential is diagonal $M_{ij}=\delta_{ij}M$. From the total scalar potential, we find the observable-sector squared mass matrix $M_{obs}^2$ at the vacuum specified by the conditions that $a\textrm{Im}T=n\pi$, $z^I=0$, and $z^i=0$
\begin{eqnarray}
M^2_{obs} \equiv \begin{pmatrix}
V_{i\bar{j}} & V_{ij} \\ 
V_{\bar{i}\bar{j}} & V_{\bar{i}j} 
\end{pmatrix} 
\end{eqnarray}
where
\begin{eqnarray}
V_{i\bar{j}} &=& -\frac{2a^2A^2}{3X^2}\delta_{i\bar{j}}e^{-aX} 
+ \frac{ca^2A^2}{3X^2}\delta_{i\bar{j}}e^{-aX/2}  - \frac{a^3A^2}{9X}e^{-aX} \delta_{i\bar{j}} + \frac{a^2A^2}{9X^2}e^{-aX}\delta_{i\bar{j}} 
+\frac{1}{X^2} W_{il}^o\Phi^{l\bar{n}}\bar{W}_{\bar{n}\bar{j}}^o \nonumber\\
&=& \frac{1}{X^2} W_{il}^o\Phi^{l\bar{n}}\bar{W}_{\bar{n}\bar{j}}^o -\frac{2a^2A^2}{9X^2}e^{-aX}\delta_{i\bar{j}} = \Big(\frac{M^2}{X^2}-\frac{2a^2A^2}{9X^2}e^{-aX}\Big)\delta_{i\bar{j}},
\nonumber\\
V_{ij} &=& \frac{aA}{3X^2}e^{-aX/2}W_{ij}^o  = \frac{aA}{3X^2}e^{-aX/2} M \delta_{ij}. 
\end{eqnarray}

Restoring the mass dimension, the mass eigenvalues are 
\begin{eqnarray}
    m_{\pm}^2 &\equiv& \Big(\frac{M^2}{X^2}-\frac{2a^2A^2}{9X^2M_{pl}^4}e^{-aX}\Big)  \pm \frac{aA}{3X^2M_{pl}^2}e^{-aX/2} M  \nonumber\\
    &=& \frac{1}{X^2}\left(M\pm \frac{aA}{6M_{pl}^2}e^{-aX/2}\right)^2 -\frac{a^2A^2}{4X^2M_{pl}^4}e^{-aX}. \label{GeneralScalarMass}
\end{eqnarray}
We observe that if $M \sim aA/M_{pl}^2$ (which is equivalent to the condition that $m_{s1} \sim H$), then both masses $m_{\pm}$ are positive definite for all $X=e^{\sqrt{2/3}\phi}>0$ (or all $\phi$), which means that during inflation the matter scalar masses are well defined (and become very light for large values of $X$ or $\phi$). This will be confirmed in the following.

Let us check the values of the scalar masses on the post-inflationary vacuum. Using the relations $A=M_{pl}\sqrt{\frac{3x(M_I^4-\Lambda)e^{ax}}{a^2}}$, $c=(1+\frac{ax}{3})e^{-ax/2}$, and $M = \frac{\sqrt{3}}{2}\frac{m_{s1}^2}{H}x^{3/2}$ in Eq. \eqref{softmasses} and setting $\Lambda \approx 0$ at $X=x=1$, where the kinetic terms of the matter scalars are canonically normalized, we obtain 
\begin{eqnarray}
      m_{\pm}^2 = \frac{3}{4}\frac{M_{pl}^2}{M_{I}^4}m^4_{s1} -\frac{2}{3}\frac{M_I^4}{M_{pl}^2} \pm \frac{1}{2}m_{s1}^2 = \Big(\frac{3}{4}k^2 \pm k - \frac{2}{3}\Big)H^2,\label{MassEigenvalues}
\end{eqnarray}
in which we define $m_{s1}^2 \equiv k H^2$ (where $k>0$ and $M_I^2/M_{pl} = H$). We notice that physical masses of scalars are determined only by the ``free'' parameter $m_{s1}$ (or $k$) and the Hubble mass $H$. Positivity of the physical masses ``$m_\pm$'' imposes the inequality
\begin{eqnarray}
m_+:~-\frac{2}{3}+\frac{2}{3}\sqrt{3} <  k, \quad     m_-:~\frac{2}{3}+\frac{2}{3}\sqrt{3} <  k \implies \frac{2}{3}+\frac{2}{3}\sqrt{3} <  k \label{LightScalarCondition}
\end{eqnarray}
Then, with this inequality, we can choose an arbitrary value of $k$ such that
\begin{eqnarray}
 \frac{2}{3}+\frac{2}{3}\sqrt{3} <    k = \frac{2}{3} + \frac{2}{3}\sqrt{3(1+m_-^2/H^2)} = -\frac{2}{3} + \frac{2}{3}\sqrt{3(1+m_+^2/H^2)}
\end{eqnarray}
allowing one physical mass $m_-$ to be parametrically lighter than the other physical mass $m_+$ as
\begin{eqnarray}
m_+^2 = \frac{4H^2\left(1+\sqrt{3(1+m_-^2/H^2)}\right)+3m_-^2}{3}.
\end{eqnarray}
We note that $m_{s1} = \sqrt{k}H \sim H$ when $m_- \ll H$, implying that $m_{\pm}^2 >0$ is indeed satisfied. In this limit, we find that one physical mass $m_-$ can be much smaller than the Hubble scale, while the other physical mass are of the order of the Hubble scale: 
\begin{eqnarray}
  m_- \ll H, \quad m_+ \gtrsim  H.
\end{eqnarray}

From this we note that in the observable sector after inflation (that is at $x=1$) one physical mass $m_-$ can be lighter than that of the gravitino, while the other physical mass $m_+$ becomes of the same order of the gravitino mass. 
We also note that the matter scalar with 
masses of order of the super-EeV gravitino mass ($\sim 10^{-6}M_{pl}$) may be a candidate for heavy dark matter candidate, because
it is in the mass range $10^{-8}M_{pl} \leq m_{\chi} \leq M_{pl}$, which is outside the excluded region shown in Figs. 2, 3, and 4 of 
ref.~\cite{HDM}.  
To summarize, we found the following constraints on soft masses. First, $m_{s1}$ must satisfy Eq.~\eqref{LightScalarCondition} to 
allow for some light scalars while $m_{s3}$ is of the same order as the gravitino mass $m_{3/2}$ and $m_{s4}$ is determined by the 
chosen value of $m_{s1}$. Notice that all these mass parameters are subject to strict constraints such as Eq.~\eqref{LightScalarCondition}. Furthermore, $m_{s2}$, $m_{s5}$, and $m_{s6}$ can be arbitrarily small, $m_{s5}$ and $m_{s6}$ are proportional to $m_{s2}^2$ and $m_{s2}$ respectively, and $m_{s2}$ is a free parameter.

It is worth noticing that the observable sector masses $m_-$ are compatible with the ``Case 1'' reheating-scenario condition of ref.~\cite{LightScalar}, for which single-field plateau-potential inflation is robust under the introduction of light scalars. 
The parameters characterizing the reheating scenario are
\begin{eqnarray}
    \Gamma_{\phi} < \Gamma_{z^i} < m_{z^i} \sim m_{-} < H , \quad \frac{\left<z^i\right>}{M_{pl}}  \ll 1,
\end{eqnarray}
where $\Gamma_{\phi},\Gamma_{z^i}$ are the decay rates of $\phi$ and $z^i$ during the reheating phase and $\left<z^i\right>$ are the expectation values of matter scalars $z^i$ after inflation. We note that $\frac{\left<z^i\right>}{M_{pl}}  \ll 1$ implies that the slow-roll inflation should begin around the minima of matter scalars, so that at the end of inflation the corresponding vacuum expectation values will be much smaller than the Planck scale $M_{pl}$. Hence, as long as the above ``Case 1'' reheating-scenario condition is satisfied, the slow-roll inflation in our model will effectively be driven by a single inflaton field $\phi$ along the minima of the matter scalars.

%% file: chapters/11.tex
\framebox[1.05\width]{{\large This chapter is based on the author's original work in Ref.~\cite{jp5}.}} \par
\vspace{1cm}

The main goal of this work is to propose a string-inspired supergravity model that can describe both an effective single-field slow-roll inflation with dS vacua and the low energy MSSM phenomenology at once in a single model in the context of 4D $\mathcal{N}=1$ supergravity. To do this, we consider generalized (i.e. field-dependent) K\"{a}hler invariant Fayet-Iliopoulos terms\footnote{This is also constructed without gauging R-symmetry.} proposed by Aldabergenova, Ketov, and Knoops (AKK) \cite{AKK} (say AKK-FI term) and KKLT string superpotential \cite{KKLT}. We point out that the new FI term will play a role in making the extra scalars sufficiently heavier than the Hubble scale in order to integrate them out in the end. This is a key to make MSSM induced by supergravity models be phenomenologically reliable. We then take advantage of gravity-mediated supersymmetry breaking \cite{GM} to reproduce the soft lagrangians \cite{softSUSYbreaking} and masses of MSSM. In addition, putting the naturalness issue away, we consider supersymmetry breaking at high scale \cite{High_SUSY,Inf_HighSUSY_EeVGrav} via D-term, which is the order of $M_S \sim \sqrt{HM_{pl}}$ where $H$ is the Hubble scale at $\mathcal{O}(10^{-5})M_{pl}$ (instead of the low scale $M_S \sim \mathcal{O}(10^{-15})M_{pl}$) in order that the scale of inflation in our model is ensured. 

Of course, one may ask the string theoretical origin of the new FI terms. Regarding this, Cribiori, Farakos, and Tournoy have presented a possible string-theoretical construction of new FI terms by treating supersymmetric Born-Infeld actions that have a second non-linear supersymmetry \cite{CFT}. We moderately mention that the AKK-FI term we used in this work may be conjectured to be originated from a similar construction as shown in Ref. \cite{CFT}, but a precise investigation on it is still required in the future. What we would like to emphasize here is that if a string realization of the new FI terms is discovered, then our proposal can be a low energy effective supergravity description of realistic superstring thoery as a promising bridge between infrared physics and its UV completion. This motivates us to explore a fully string theoretical construction of our model.

\section{$4D~\mathcal{N}=1$ Supergravity with Generalized K\"{a}hler-Invariant Fayet-Iliopoulos Terms and KKLT String Background}

In this section, we construct a superconformal action of $\mathcal{N}=1$ supergravity equipped with generalized K\"{a}hler invariant Fayet-Iliopoulos (FI) terms \cite{AKK} to explore possible dynamics that the action may possess in superconformal tensor calculus \cite{Superconformal_Freedman,Linear}. To do this, we start with a combination of standard $\mathcal{N}=1$ supergravity action and certain new FI terms studied in Ref. \cite{jp2}. We write the superconformal action as 
\begin{eqnarray}
\mathcal{L} &=& -3[S_0\Bar{S}_0e^{-K(Z^A,\bar{Z}^{\bar{A}})/3}]_D + [S_0^3W(Z^A)]_F + \frac{1}{2}[
\mathcal{W}_{\alpha}(V)\mathcal{W}^{\alpha}(V)]_F + c.c. + \mathcal{L}_{new~FI} \label{newFI2}
\end{eqnarray}
where $S_0$ is the conformal compensator with Weyl/chiral weights (1,1); $Z^A$ and $V$ are chiral matter and vector multiplets with weights $(0,0)$; $K(Z,\bar{Z})$ is a K\"{a}hler potential gauged by a vector multiplet $V$, $W(Z)$ is a superpotential, and
$\mathcal{W}_{\alpha}(V)$ is the field strength of the vector multiplet $V$. 

Then, we decompose matter multiplets $Z^A$'s into hidden and observable sectors, denoting $T$ as the volume modulus multiplet, and $Z^i$ as observable sector matter multiplets. We then consider the corresponding decomposition of superpotential as
\begin{eqnarray}
 W(T) \equiv W^h(T)+W^o(Z^i), 
\end{eqnarray}
where
\begin{eqnarray}
 W^h(T) &\equiv& W_0 + Ae^{-aT}, \\
 W^o(Z^i) &\equiv& W_{\textrm{MSSM}} + W_{BSM'} ,\\
 W_{MSSM} &\equiv&  -Y_u \hat{U}_R \hat{H}_u \cdot \hat{Q} + Y_d \hat{D}_R \hat{H}_d \cdot \hat{Q} + Y_e \hat{E}_R \hat{H}_u \cdot \hat{L}  + \mu \hat{H}_u \cdot \hat{H}_d \nonumber\\
 &=&  -Y_u \tilde{u}_R (H^+_u \tilde{d}_L -H^0_u\tilde{u}_L) + Y_d \tilde{d}_R (H^0_d \tilde{d}_L -H^-_d\tilde{u}_L)\nonumber\\
 && + Y_e \tilde{e}_R (H^+_u \tilde{\nu}_L -H^0_u\tilde{e}^-_L)  + \mu (H^+_uH^-_d-H^0_uH^0_d),
\end{eqnarray} 
where $W_0,A,a,Y_u,Y_d,Y_e,\mu$ are constants; $\hat{A} \cdot \hat{B} \equiv \epsilon_{ab}A^aB^b $ is the product between $SU(2)_L$ doublets in which $\epsilon_{12}=1=-\epsilon_{21}$ and $a,b$ are $SU(2)_L$ indices; $\tilde{u},\tilde{d},\tilde{e},\tilde{\nu}$ are the component fields of the superfield $SU(2)_L$ doublets $\hat{Q},\hat{L}$ as superpartners correponding to the SM quarks and leptons. In particular, $W_{BSM'}$ may be added in the observable sector, but in this work, we will not treat it.

In our setup, we define the hidden sector superpotential $W^h$ using the nonperturbative corrections which are obtained by either Euclidean D3 branes in type IIB compactifications or gaugino condensation due to D7 branes \cite{KKLT}. We then define the observable sector superpotential by the one for the minimal supersymmetric standard model (MSSM)\footnote{We follow the conventions used in Ref. \cite{RPP}.} consisting of supermultiplets of quarks, leptons, and Higgs fields. Meanwhile, we consider a K\"{a}hler potential of the volume modulus $T$, the same as in KKLT string background \cite{KKLT}, given by
\begin{eqnarray}
 K = -3\ln[T+\bar{T}-\Phi(Z^i,\bar{Z}^{\bar{i}})/3],
\end{eqnarray}
where $\Phi(Z^i,\bar{Z}^{\bar{i}})$ is a real function of the observable-sector matter multiplets $Z^i$'s. In terms of the real function $\Phi$, we suppose that  
\begin{eqnarray}
 \Phi(Z^i,\bar{Z}^{\bar{i}}) = \delta_{i\bar{j}}Z^{i}\bar{Z}^{\bar{j}},
\end{eqnarray}
in which $\Phi_i\Phi^{i\bar{j}}\Phi_{\bar{j}}=\Phi$ where $\Phi_i \equiv \partial \Phi / \partial z^i$ and $\Phi_{i\bar{j}}\Phi^{l\bar{j}}=\delta_{i}^{l}$.

The next step we need to do is going to determine which type of new FI terms we are going to use. In this work, we employ ``field-dependent'' K\"{a}hler-invariant Fayet-Iliopoulos (FI) terms proposed by Aldabergenova, Ketov, and Knoops (AKK) \cite{AKK}. We refer to this FI term as AKK-FI term to distinguish it from many other FI terms. The AKK-FI terms are written as
\begin{eqnarray}
\mathcal{L}_{1} \equiv -  \left[(S_0\bar{S}_0e^{-K/3})^{-3}\frac{(\mathcal{W}_{\alpha}(V)\mathcal{W}^{\alpha}(V))(\bar{\mathcal{W}}_{\dot{\alpha}}(V)\bar{\mathcal{W}}^{\dot{\alpha}}(V))}{T(\bar{w}^2)\bar{T}(w^2)}(V)_D (\xi_1+U_1(\Phi, \bar \Phi, H, \bar H, V))\right]_D,\label{newFIterm}
\end{eqnarray}
and
\begin{eqnarray} 
\mathcal{L}_{2} \equiv  -  \frac{1}{4} \left[ S_0 e^{-\hat{K}/3} \bar S_0  \frac{(\mathcal{W}_{\alpha}(V)\mathcal{W}^{\alpha}(V))(\bar{\mathcal{W}}_{\dot{\alpha}}(V)\bar{\mathcal{W}}^{\dot{\alpha}}(V))}{ \left( (V)_D \right)^3 } 
 \left( \xi_2 + U_2(\Phi, \bar \Phi, H, \bar H, V) \right) \right]_D~
\end{eqnarray}
where 
$w^2 \equiv \frac{\mathcal{W}_{\alpha}(V)\mathcal{W}^{\alpha}(V)}{(S_0\bar{S}_0e^{-K(Z,\bar{Z})})^{2}}$ and $\bar{w}^2 \equiv \frac{\bar{\mathcal{W}}_{\dot{\alpha}}(V)\bar{\mathcal{W}}^{\dot{\alpha}}(V)}{(S_0\bar{S}_0e^{-K(Z,\bar{Z})})^{2}}$ are composite 
multiplets, $T(X), \bar{T}(X)$ are chiral projectors, and $(V)_D$ is a real multiplet, whose lowest component is the auxiliary 
field $D$ of the vector multiplet $V$; $\xi_1,\xi_2$ are non-vanishing constants such that $\xi_2/(\xi_1+\xi_2) < 1/4$ which ensures the positivity of kinetic term of the vector multipelt $V$, and $U_1,U_2$ are chosen such that $\left<U_1\right>=0$ and  $\left<U_2\right>=0$ only  if $\xi_1,\xi_2$ are true vacuum expectation values out of $U_1,U_2$. We note that we may rewrite $\xi_i+U_i$ as a simple decomposition $\xi_i+U_i$ such that $\xi_i+U_i >0$ everywhere and $\left<\xi_i+U_i\right> = \xi_i' \neq \xi_i$ since $\left<U_i\right> \neq 0$. Hence, the condition can also be rewritten as
\begin{eqnarray}
\xi'_2/(\xi'_1+\xi'_2) < 1/4.
\end{eqnarray}

Keeping this in mind, with both terms, we find the solution for the auxiliary field $D$ for the vector multiplet as 
\begin{eqnarray}
 D/g^2  = \xi_1+\xi_2 + U_1 + U_2 \equiv \xi + U, 
\end{eqnarray}
 where $\xi \equiv \xi_1 + \xi_2$ and $U \equiv U_1 + U_2$. Of course, if we identify the vector $V_R$ with a gauge multiplet for some R or non-R $U(1)$ gauge symmetry, we may include the corresponding contribution to the D-term, so that
 \begin{eqnarray}
   D/g^2  = \xi + \xi' + U, 
 \end{eqnarray}
 where $\xi' \equiv k_A^I(z)G_I+c.c.$, and $g,k^I_A(z)$ are the corresponding gauge coupling and killing vector field for the field $z^I$, and $G_I \equiv \partial G/\partial z^I$ for some scalar $z^I$. Since we are interested in the situation without gauged R-symmetry, we can only consider the conventional non-R gauge symmetry in this work. 
 
Then, the $D$-term scalar potential is given by
\begin{eqnarray}
 V_D = \frac{1}{2}g^2D^2 = \frac{1}{2}g^2(\xi + U)^2 = \frac{1}{2}g^2\xi^2 + g^2\xi U + \frac{1}{2}g^2U^2.
\end{eqnarray}

Besides, the corresponding Lagrangian of the new FI term can be rewritten as
\begin{eqnarray}
\mathcal{L}_{\textrm{newFI}} \equiv -  \left[(S_0\bar{S}_0e^{-K/3})^{-3}\frac{(\mathcal{W}_{\alpha}(V_R)\mathcal{W}^{\alpha}(V_R))(\bar{\mathcal{W}}_{\dot{\alpha}}(V_R)\bar{\mathcal{W}}^{\dot{\alpha}}(V_R))}{T(\bar{w}^2_R)\bar{T}(w^2_R)}(V_R)_D \mathcal{U} \right]_D,
\end{eqnarray}
where we defined $\mathcal{U} \equiv \xi + U$ and rescaled the vector multiplet $V$ into $V_R \equiv \gamma V$ in order to introduce additional parameter $\gamma$ for generality, and $w^2_R \equiv \frac{ \mathcal{W}_{\alpha}(V_R)\mathcal{W}^{\alpha}(V_R)}{(S_0\bar{S}_0 e^{-K/3})^2}$.

Moreover, we may extend the form of $D$ by taking into account two different contributions to the generic function $U$ as 
\begin{eqnarray}
 U \equiv U^h + U^o \implies D/g^2 = \mathcal{U} = \xi+U=\xi + U^h + U^o,
\end{eqnarray}
where $U^h,U^o$ are the generic functions that are involved in the hidden and observable sectors respectively.

Basically, we are able to consider both $\mathcal{L}_1$ and $\mathcal{L}_2$. However, we are going to take $\mathcal{L}_1$ for simplicity of our model. Hence, we automatically satisfy the condition on the vacuum expectation values $\xi'_1,\xi'_2$, i.e. $\xi'_2/(\xi'_1+\xi'_2) < 1/4$ found in Ref. \cite{AKK}, because we have $\xi'_2/(\xi'_1+\xi'_2) =0 < 1/4$ with the assumption $\xi_2 = U_2=0$ (i.e. $\xi'_2=0$). 

In the meantime, we may take into account additional D-term potentials from independent vector multiplets corresponding to some gauge symmetries of interest. That is, for different vector multiplets $V_A$, we may have
\begin{eqnarray}
 V_D' \equiv \sum_A V_D^A = \sum_A \frac{g_A^2}{2}(k^I_A(z) G_I+c.c.)^2, 
\end{eqnarray} 
In the end, we will consider the gauge groups $U(1)_Y,SU(2)_L,SU(3)_C$ for $V_D'$ in the observable sector.

Since the total supergravity scalar potential is given by the sum of F- and D-term potentials ($V_D$ and $V_F = e^G(G_AG^{A\bar{B}}G_{\bar{B}}-3)$), we find 
\begin{eqnarray}
 V =V_D + V_F = \Big( \frac{g^2}{2}(\xi+U^h+U^o)^2 + V_D' \Big) + V_F \equiv  V_h + V_{soft}.
\end{eqnarray}
where we decompose the scalar potential into hidden-sector and soft contributions according to the separation of the superpotential $W = W^h + W^o$, and thus the most dominant terms with $\sim |W^h|^2$ are classified into the hidden sector potential $V^h$:
\begin{eqnarray}
V_h &\equiv&\frac{g^2}{2}(\xi+U^h)^2 - \frac{W^h_T\bar{W}^h+\bar{W}^h_{\bar{T}}W^h}{X^{2}} + \frac{|W^h_T|^2}{3X^{2}}\Big(X+\frac{1}{3}\Phi_i\Phi^{i\bar{j}}\Phi_{\bar{j}}\Big) \nonumber\\
&& + \frac{1}{3}\frac{1}{X^{2}}
[W^h_T\Phi_{i}\Phi^{i\bar{j}}\bar{W}^h_{\bar{j}}
+\bar{W}^h_{\bar{T}}W^h_{i}\Phi^{i\bar{j}}\Phi_{\bar{j}}]+\frac{1}{X^{2}}W^h_i\Phi^{i\bar{j}}\bar{W}^h_{\bar{j}} ,\\
V_{soft} &\equiv& g^2(\xi+U^h)(U^o) + \frac{g^2}{2}(U^o)^2 + V_D'-\frac{1}{X^{2}}[W^o\bar{W}^h_{\bar{T}}+\bar{W}^oW^h_{T}] \nonumber\\&&
+ \frac{1}{3}\frac{1}{X^{2}}
[W^h_T\Phi_{i}\Phi^{i\bar{j}}\bar{W}^o_{\bar{j}}
+\bar{W}^h_{\bar{T}}W^o_{i}\Phi^{i\bar{j}}\Phi_{\bar{j}}]
+\frac{1}{X^{2}}W^o_i\Phi^{i\bar{j}}\bar{W}^o_{\bar{j}},
\end{eqnarray}
where $X \equiv T+\bar{T}-\Phi(z,\bar{z})/3$ and $W_I \equiv \partial W/\partial z^I$ for $I=T,i$. Inserting the hidden sector superpotentials, i.e. $W = W^h(T)+W^o(z)$ where $W^h(T) = W_0 + Ae^{-aT}$ and $W^o(z)$, into the above, similarly to Ref. \cite{jp2}, we obtain
\begin{eqnarray}
V_h &=& \Big( \frac{1}{2}g^2\xi^2 + g^2\xi U^h  + \frac{1}{2}g^2{U^h}^2 \Big)  - \frac{1}{X^2}\bigg(-2aA^2e^{-a(X+\Phi/3)}+2acA^2 e^{-a(X+\Phi/3)/2} \cos(a\textrm{Im}T)\bigg)
\nonumber\\&&+ \frac{1}{3X^{2}}\Big(X+\frac{1}{3}\Phi_i\Phi^{i\bar{j}}\Phi_{\bar{j}}\Big) a^2A^2 e^{-a(X+\Phi/3)},\\
V_{soft} &\equiv& g^2(\xi+U^h)(U^o) + \frac{g^2}{2}(U^o)^2 +  \sum_A \frac{g_A^2}{2}(k^I_A(V_A) G_I+c.c.)^2 \nonumber\\
&&+
\frac{aAe^{-a(X+\Phi/3)/2}}{X^{2}}[W^oe^{ia\textrm{Im}T}+\bar{W}^oe^{-ia\textrm{Im}T}]
+\frac{1}{9}\frac{a^2A^2e^{-a(X+\Phi/3)}}{X^{2}}\Phi_{i}\Phi^{i\bar{j}}\Phi_{\bar{j}} \nonumber \\ &&
- \frac{1}{3}\frac{aAe^{-a(X+\Phi/3)/2}}{X^{2}}
[e^{-ia\textrm{Im}T}\Phi_{i}\Phi^{i\bar{j}}\bar{W}^o_{\bar{j}}
+e^{ia\textrm{Im}T}W^o_{i}\Phi^{i\bar{j}}\Phi_{\bar{j}}] 
+\frac{1}{X^{2}}W^o_i\Phi^{i\bar{j}}\bar{W}^o_{\bar{j}},
\end{eqnarray}
in which we use a redefinition of the constant, $W_0 \equiv -cA$ where $c$ is a constant.

\section{Hidden-Sector Dynamics: Starobinsky-type Inflation and de Sitter Vacua}

In this section, we derive an inflationary potential and explore its hidden-sector dynamics. Let us begin with the general potential $V=V_h+V_{soft}$, which is given by
\begin{eqnarray}
 V &=& \Big( \frac{1}{2}g^2\xi^2 + g^2\xi U^h  + \frac{1}{2}g^2{U^h}^2 \Big)  - \frac{1}{X^2}\bigg(-2aA^2e^{-a(X+\Phi/3)}+2acA^2 e^{-a(X+\Phi/3)/2} \cos(a\textrm{Im}T)\bigg)
\nonumber\\&&+ \frac{1}{3X^{2}}\Big(X+\frac{1}{3}\Phi_i\Phi^{i\bar{j}}\Phi_{\bar{j}}\Big) a^2A^2 e^{-a(X+\Phi/3)} + V_{soft}.
\end{eqnarray}
Now we assume that the hidden-sector part of the real function, say $U^h$, is defined by
\begin{eqnarray}
 U^h \equiv C_iz^i\bar{z}^i,
\end{eqnarray}
where $z^i$'s are the matter scalars (except for Higgs fields) involved in our supergravity model and $C_i$'s are coupling constants. It is easy to see that the minima of the total scalar potential $V=V_h + V_{soft}$ with respect to the matter scalars $z^i$'s without Higgs ones are placed at $z^i=0$. 

To explore the inflationary trajectory in the direction of inflaton field $\phi$ (or $X \equiv e^{\sqrt{2/3}\phi}$), we focus on the path along the minima at $z^i=0$ where $i \neq \textrm{Higgs}$; $\textrm{Im}T=0$, and $H_u^+=H_d^-=0, H_u^0=v_u/\sqrt{2},H_d^0=v_d/\sqrt{2}$ where $v_u,v_d$ are non-zero constants. Then, along the path, the total scalar potential can be written as
\begin{eqnarray}
  V|_{minima} &=& \frac{1}{2}g^2\xi^2  - \frac{1}{X^2}\bigg(-2aA^2e^{-a(X+v^2/6)}+2acA^2 e^{-a(X+v^2/6)/2} \bigg)
  \nonumber\\
  &&+ \frac{1}{3X^{2}}\Big(X+\frac{v^2}{6}\Big) a^2A^2 e^{-a(X+v^2/6)}+ V_{soft}|_{minima},
\end{eqnarray}
where we defined $v^2 \equiv v_u^2 + v_d^2$. 
Basically, we can further simplify the form of this potential using the fact that $v = 246 \textrm{GeV} \sim 10^{-16} M_{pl} \ll X \sim \mathcal{O}(M_{pl})$ all the time during and after inflation. That is, we can take some limits $X \gg v^2/6$ and $V^h \gg V_{soft}$ during and after inflation, which produces 
\begin{eqnarray}
   V|_{minima} \approx \frac{1}{2}g^2\xi^2  - \frac{1}{X^2}\bigg(-2aA^2e^{-aX}+2acA^2 e^{-aX/2} \bigg)+ \frac{1}{3X} a^2A^2 e^{-aX}.
\end{eqnarray}
The vacuum with respect to the direction $X$ can be found at $X=x$ such that $c=(1+ax/3)e^{-ax/2}$ (see Ref. \cite{jp2} for the derivation of $c$). In fact, the scale of $g^2\xi^2$ must be of order of the inflation energy since we want to describe inflation using that potential in the end. That is, we must require that
\begin{eqnarray}
 \frac{1}{2}g^2\xi^2 \overset{!}{=} M_I^4 \equiv H^2M_{pl}^2,
\end{eqnarray}
where $M_I$ and $H$ are denoted by the inflation and Hubble scale respectively. Using $X=e^{\sqrt{2/3}\phi}$, we rewrite the potential as
\begin{eqnarray}
   V|_{minima} \approx M_I^4  - e^{-2\sqrt{2/3}\phi}\bigg(-2aA^2e^{-ae^{\sqrt{2/3}\phi}}+2acA^2 e^{-ae^{\sqrt{2/3}\phi}/2} \bigg) + \frac{1}{3}a^2A^2e^{-\sqrt{2/3}\phi}  e^{-ae^{\sqrt{2/3}\phi}}.
\end{eqnarray}
It is worth noticing that this result exactly coincides with that of Ref. \cite{jp2}. Also notice that this potential resembles the Starobinsky potential in feature. 

What we need to do next is to find the de Sitter vacua in our theory because we wish to obtain the observed cosmological constant $\Lambda \sim 10^{-120}M_{pl}$. Now considering an exact value of the soft potential at the {\it vacua} when $X=x$, $i \neq \textrm{Higgs}$; $\textrm{Im}T=0$, and $H_u^+=H_d^-=0, H_u^0=v_u/\sqrt{2},H_d^0=v_d/\sqrt{2}$, we can determine what the constant $g^2\xi^2/2$ must be. At the vacua, if we define the vacuum with respect to $X$ (or $\phi$) at $x=1$ (or $\phi=0$), the potential is given by
\begin{eqnarray}
 V|_{vacua} = \frac{1}{2}g^2\xi^2-\frac{a^2A^2e^{-a}}{3}+ \Lambda_{soft} \equiv \Lambda,
\end{eqnarray}
where we define $\Lambda_{soft} = \left<V_{soft}\right>$ and impose that the VEV of the potential is equal to the cosmological constant $\Lambda$. Hence, we determine $g^2\xi^2/2$ as
\begin{eqnarray}
 \frac{1}{2}g^2\xi^2 = \frac{a^2A^2e^{-a}}{3}+\Lambda- \Lambda_{soft}.
\end{eqnarray}

Now let us investigate supersymmetry (SUSY) breaking of our model. The SUSY breaking scale, say $M_S$, can be found by the positive contributions to both D and F terms
\begin{eqnarray}
 V_+|_{vacua} = (V + 3e^G)|_{vacua} = \Lambda + \frac{a^2A^2e^{-a}}{3} \equiv M_S^4,
\end{eqnarray}
which gives
\begin{eqnarray}
 \frac{a^2A^2e^{-a}}{3} = M_S^4 - \Lambda \implies \frac{1}{2}g^2\xi^2 = M_S^4 - \Lambda_{soft} = M_I^4 \implies M_S^4 = M_I^4 + \Lambda_{soft}.
\end{eqnarray}
This means that putting the naturalness issue away, we have to require a high-scale supersymmetry breaking \cite{High_SUSY} because the SUSY breaking mass $M_S$ is at high scale as given by
\begin{eqnarray}
 M_S = (H^2M_{pl}^2+\Lambda_{soft})^{1/4} \sim \mathcal{O}(\sqrt{HM_{pl}}) = 10^{-2.5}M_{pl},
\end{eqnarray}
where we note that $H^2M_{pl}^2 \gg \Lambda_{soft}$. Consequently, we have seen that the final form we obtained is identified as an inflationary potential of the Hubble scale order.

\section{Observable-Sector Dynamics: Minimal Supersymmetric Standard Model (MSSM)} 

In this section, we embed minimal supersymmetric standard model (MSSM) into the observable sector of our supergravity, which describes both inflation and de Sitter vacua in the hidden sector as shown in the previous section. To do so, we explore a supersymmetric Higgs potential to find the observed Higgs mass of electroweak scale in MSSM.

\subsection{Supersymmetric Higgs potential using new Fayet-Iliopoulos terms}

Here we focus on finding a supersymmetric Higgs potential to realize MSSM phenomenology in our supergravity model of inflation. To generate both Higgs and matter masses which are phenomenologically favored, we assume that the generic function $U=U^h+U^o$ is defined by
\begin{eqnarray}
 U^h &=& C_i|z^i|^2\quad\textrm{for non-Higgs matters}, \\
 U^o &=& b[(|H_u^+|^2+|H_u^0|^2) - (|H_d^0|^2+|H_d^-|^2)] \quad\textrm{for Higgs},
\end{eqnarray}
where $b$ is a free parameter. Notice that these are gauge invariant under the SM gauge groups. We can then identify the supersymmetric Higgs potential from the soft potential, which is specified with
\begin{eqnarray}
 V_{soft} &=& \xi g^2U^o + \frac{g^2}{2}{U^o}^2 + V_{U(1)_Y}+V_{SU(2)_L} + V_{SU(3)_c} \nonumber\\
 &&+ \frac{2}{3}\frac{aAe^{-a(X+\Phi/3)}/2}{X^2}(W^o+\bar{W}^o) + \frac{1}{9}\frac{a^2A^2e^{-a(X+\Phi/3)}}{X^2}\Phi + \frac{W_i^o\delta^{i\bar{j}}\bar{W}^o_{\bar{j}}}{X^2}.
\end{eqnarray}

Since the Higgs doublets do not transform under $U(1)'$, $U(1)_Y$, and $SU(2)_L$, the parts of the Higgs potential are given by
\begin{eqnarray}
 g^2\xi U^o &=&  g^2\xi b(|H_u^+|^2+|H_u^0|^2 - |H_d^0|^2-|H_d^-|^2)  ,\\
 \frac{g^2}{2}{U^o}^2 &=& \frac{g^2b^2}{2}(|H_u^+|^2+|H_u^0|^2 - |H_d^0|^2-|H_d^-|^2)^2  ,\\
 V_{U(1)_Y} &\supset& \frac{g_1^2}{8X^2} (|H^+_u|^2+|H^0_u|^2-|H_d^0|^2-|H_d^-|^2)^2,\\
 V_{SU(2)_L} &\supset& \frac{g_2^2}{2X^2}|\bar{H}^0_uH^+_u+\bar{H}_d^-H_d^0|^2 + \frac{g_2^2}{8X^2} (|H^+_u|^2+|H^0_u|^2-|H_d^0|^2-|H_d^-|^2)^2.
\end{eqnarray}
In addition, we can find the other part of the Higgs potential from the F-term part of the soft potential $ V_{soft}|_F $, which provides
\begin{eqnarray}
V_{soft}|_F &\supset& \frac{2}{3}\frac{aAe^{-a(X+\Phi/3)}/2}{X^2}(W^o+\bar{W}^o) + \frac{1}{9}\frac{a^2A^2e^{-a(X+\Phi/3)}}{X^2}\Phi + \frac{W_i^o\delta^{i\bar{j}}\bar{W}^o_{\bar{j}}}{X^2}\nonumber\\
&=& \frac{2}{3}\frac{aAe^{-a(X+\Phi/3)}/2}{X^2}\mu (H_u^+H^-_d -H^0_u H^0_d + h.c.)\nonumber\\
&&+\bigg(\frac{1}{9}\frac{a^2A^2e^{-a(X+\Phi/3)}}{X^2}+\frac{|\mu|^2}{X^2}\bigg)(|H^+_u|^2+|H^0_u|^2+|H_d^0|^2+|H_d^-|^2)
\end{eqnarray}

Therefore, the final form of the Higgs potential at the non-Higgs matter minima $z^i=0$ (where $i\neq$ Higgs) is given by
\begin{eqnarray}
 V_H &=&\frac{g_2^2}{2X^2}|\bar{H}^0_uH^+_u+\bar{H}_d^-H_d^0|^2 + \Big(\frac{g_1^2+g_2^2}{8X^2} + \frac{g^2b^2}{2}\Big) (|H^+_u|^2+|H^0_u|^2-|H_d^0|^2-|H_d^-|^2)^2\nonumber\\
 &&+\frac{4}{3}\frac{aAe^{-a(X+(|H^+_u|^2+|H^0_u|^2+|H_d^0|^2+|H_d^-|^2)/3)}/2}{X^2}\mu \textrm{Re}(H_u^+H^-_d -H^0_u H^0_d)\nonumber\\
&&+\bigg(\frac{1}{9}\frac{a^2A^2e^{-a(X+(|H^+_u|^2+|H^0_u|^2+|H_d^0|^2+|H_d^-|^2)/3)}}{X^2}+\frac{|\mu|^2}{X^2} + g^2\xi b \bigg)(|H^+_u|^2+|H^0_u|^2)\nonumber\\
&& +\bigg(\frac{1}{9}\frac{a^2A^2e^{-a(X+(|H^+_u|^2+|H^0_u|^2+|H_d^0|^2+|H_d^-|^2)/3)}}{X^2}+\frac{|\mu|^2}{X^2}-g^2\xi b \bigg)(|H_d^0|^2+|H_d^-|^2)
\end{eqnarray}
If we take an assumption that $X \gg H^{\pm,0}_{u,d}$, then the potential can be approximated into
\begin{eqnarray}
  V_H &=&\frac{g_2^2}{2X^2}|\bar{H}^0_uH^+_u+\bar{H}_d^-H_d^0|^2 + \Big(\frac{g_1^2+g_2^2}{8X^2} + \frac{g^2b^2}{2}\Big) (|H^+_u|^2+|H^0_u|^2-|H_d^0|^2-|H_d^-|^2)^2\nonumber\\
 &&+\frac{4}{3}\frac{aAe^{-aX/2}}{X^2}\mu \textrm{Re}(H_u^+H^-_d -H^0_u H^0_d)\nonumber\\
&&+\bigg(\frac{1}{9}\frac{a^2A^2e^{-aX}}{X^2}+\frac{|\mu|^2}{X^2} +g^2\xi b \bigg)(|H^+_u|^2+|H^0_u|^2)\nonumber\\
&& +\bigg(\frac{1}{9}\frac{a^2A^2e^{-aX}}{X^2}+\frac{|\mu|^2}{X^2}-g^2\xi b \bigg)(|H_d^0|^2+|H_d^-|^2)
\end{eqnarray}
We then find the minima at $H_u^+=H_d^-=0$, which gives
\begin{eqnarray}
  V_H &=& \Big(\frac{g_1^2+g_2^2}{8X^2} + \frac{g^2b^2}{2}\Big) (|H^0_u|^2-|H_d^0|^2)^2-\frac{4}{3}\frac{aAe^{-aX/2}}{X^2}\mu \textrm{Re}(H^0_u H^0_d)\nonumber\\
&&+\bigg(\frac{1}{9}\frac{a^2A^2e^{-aX}}{X^2}+\frac{|\mu|^2}{X^2} +g^2\xi b \bigg)|H^0_u|^2 +\bigg(\frac{1}{9}\frac{a^2A^2e^{-aX}}{X^2}+\frac{|\mu|^2}{X^2}-g^2\xi b \bigg)|H_d^0|^2.
\end{eqnarray}
In terms of the approximated potential, the vacuum solutions can be found as those of the MSSM. That is, 
\begin{eqnarray}
 \left<H^{0}_{u}\right> = \frac{v_{u}}{\sqrt{2}}(= v_2),\quad \left<H^{0}_{d}\right> = \frac{v_{d}}{\sqrt{2}}(= v_1), \quad \left<H^{+}_{u}\right>= \left<H^{-}_{d}\right> = 0 \implies H^0_i \approx \left<H^{0}_{i}\right> + \varphi_i,
\end{eqnarray}
where $\varphi_i$ are fluctuations of the Higgs fields $H_i^0$ around the vacuum ($i=u,d$). We take here the same definitions used in the MSSM
\begin{eqnarray}
 v^2 \equiv v_u^2 + v_d^2 = (246~\textrm{GeV})^2, \quad \tan\beta = v_2/v_1= v_u/v_d,
\end{eqnarray}
where $\beta$ is a free parameter such that $0\leq \beta \leq \pi/2$. Hence, we can merely recall the MSSM results when we compute scalar masses. 

Recalling some results of Sec. 28.5 in Ref. \cite{WeinbergSUSY}, we can identify the following correspondences 
\begin{eqnarray}
 && \frac{g^2+g'^2}{8} \rightarrow \frac{g_1^2+g_2^2}{8X^2} + \frac{g^2b^2}{2}, \quad  m_1^2 \rightarrow \frac{1}{9}\frac{a^2A^2e^{-aX}}{X^2} -g^2\xi b, \quad  m_2^2 \rightarrow \frac{1}{9}\frac{a^2A^2e^{-aX}}{X^2} +g^2\xi b,\nonumber\\{}\\
 && |\mu|^2  \rightarrow \frac{|\mu|^2}{X^2}, \quad  B\mu \rightarrow \frac{4}{3}\frac{aAe^{-aX/2}}{X^2}\mu , \\
 && m_Z^2 = \frac{1}{2}(g^2+g'^2)(v_1^2+v_2^2) \rightarrow m_Z'^2 =\left(\frac{g_1^2+g_2^2}{2X^2} + 2g^2b^2\right)(v_1^2+v_2^2)= \left(X^{-2}+\frac{4g^2b^2}{g_1^2+g_2^2}\right)m_Z^2 ,\nonumber\\{}\\
 && m_A^2 = 2|\mu|^2 + m_1^2 + m_2^2 \rightarrow m_A'^2 =  \frac{2}{9}\frac{a^2A^2e^{-aX}}{X^2}+\frac{2|\mu|^2}{X^2},
\end{eqnarray}
and the vacuum solutions produces the following relations
\begin{eqnarray}
 B\mu = {m'}_A^2 \sin 2\beta, \quad m_1^2 - m_2^2 = - ({m'}_A^2+{m'}_Z^2)\cos 2\beta = -2g^2\xi b, \quad \tan \beta = v_2/v_1. 
\end{eqnarray}

\subsection{Soft supersymmetry breaking masses of scalars comparable with single-field inflation}

The scalar masses are determined as follows. The normal matter masses are found to be 
\begin{eqnarray}
 m_z^2|_{vac}  &=& V_{z\bar{z}}|_{z=0} =  \Big(g^2\xi U_{z\bar{z}} + g^2(U_zU_{\bar{z}}+UU_{z\bar{z}}) + (V_F + V_D')_{z\bar{z}}\Big)\Big|_{z=0}
 \nonumber \\
 &=& (g^2\xi U_{z\bar{z}}+ (V_F + V_D')_{z\bar{z}})|_{z=0} \gg H^2 \nonumber\\
 &\implies& g^2\xi U_{z\bar{z}}|_{z=0} \gg H^2
\end{eqnarray}
along the vacua when $z=0,a\textrm{Im}T=0$. Now we may suppose the form of the general function $U$ as 
\begin{eqnarray}
 U \supset  C_{i} z^i\bar{z}^i,
\end{eqnarray}
where $z^i$'s are the matter fields without the Higgs fields. This leads to in the end
\begin{eqnarray}
  g^2\xi U_{z\bar{z}} = g^2\xi C_i \gg H^2 \implies C_i \gg \frac{H^2}{g^2\xi}>0.
\end{eqnarray}
Notice that $U$ is positive definite, so that $D = \xi + U >0$ is nowhere vanishing. Here the point is that the matter scalars can be as mush heavy as we want during and after inflation, enabling us to integrate out them easily. 
 
Next, let us identify the Higgs masses. Let us recall the W and Z gauge boson masses
\begin{eqnarray}
  m_W^2 = \frac{g_2^2v^2}{4} , \quad m_Z^2 = \frac{g_1^2+g_2^2}{4} v^2
\end{eqnarray}
The eigenvalues of the Higgs fields in the previous two models are 
\begin{eqnarray}
 m_H^2 &=& \frac{1}{2}({m'}_A^2+{m'}_Z^2 + \sqrt{({m'}_A^2+{m'}_Z^2)^2 - 4{m'}_A^2{m'}_Z^2 \cos^22\beta}),\\
  m_h^2 &=& \frac{1}{2}({m'}_A^2+{m'}_Z^2 - \sqrt{({m'}_A^2+{m'}_Z^2)^2 - 4{m'}_A^2{m'}_Z^2 \cos^22\beta}) \approx \frac{{m'}_A^2{m'}_Z^2\cos^2 2\beta}{{m'}_A^2+{m'}_Z^2} \nonumber\\
  &\approx& \frac{(m_1^2-m_2^2)^2}{{m'}_A^2{m'}_Z^2} \quad \textrm{if}\quad {m'}_A \gg {m'}_Z \implies {m'}_A^2 \approx  \frac{(m_1^2-m_2^2)^2}{m_h^2{m'}_Z^2} 
\end{eqnarray}
where $\mu, B,m_1^2,m_2^2$ are the MSSM soft parameters. We note that now the MSSM soft parameters are functions of the inflaton field $\phi$ via $X \equiv e^{\sqrt{2/3}\phi}$ whose vacuum is at $\phi=0$ (or $X=1$). 

First, let us check the Higgs masses after inflation at $X=1$. Then, the above relation implies that
\begin{eqnarray}
 {m'}_A^2 = \frac{2a^2A^2e^{-a}}{9} + 2|\mu|^2 \approx \frac{(m_1^2-m_2^2)^2}{m_h^2{m'}_Z^2}, 
\end{eqnarray}
so that we can determine the parameter $\mu$ as
\begin{eqnarray}
 |\mu|^2  \approx \frac{(m_1^2-m_2^2)^2}{2m_h^2{m'}_Z^2} - \frac{1}{3}(M_I^4-\Lambda) 
\end{eqnarray}
since $3(M_I^4-\Lambda) = a^2A^2e^{-a} = 3m_{3/2}^2$. Since we have 
\begin{eqnarray}
 && m_1^2 - m_2^2 = -2g^2\xi b \approx -2g^2 b \frac{\sqrt{2}M_I^2}{g} \sim 2\sqrt{2}bgM_I^2,
\end{eqnarray}
we obtain
\begin{eqnarray}
 && |\mu|^2  \sim  \bigg( \frac{4 b^2g^2}{m_h^2m_Z^2 \left(1+\frac{4g^2b^2}{g_1^2+g_2^2}\right)} -\frac{1}{3} \bigg)  M_I^4 >0.
\end{eqnarray}
We remark that it is necessary to consider the new FI term in this model since it helps us to acquire different values of $m_1^2$ and $m_2^2$, which determines non-vanishing of the light Higgs scalar mass $m_h$. Furthermore, we observe that we can integrate out the degree of freedom for the heavy Higgs scalar with $m_H^2$ because this is of order of Hubble scale, while the light Higgs scalar can be set up as the observed Higgs degree of freedom using the cancellation between the first and second terms in the mass formula. Next, let us inspect the Higgs masses during inflation for $X \gg 1$. In this phase, we have
\begin{eqnarray}
 {m'}_A^2 \rightarrow 0, \quad {m'}_Z^2 \rightarrow \frac{4g^2b^2}{g_1^2+g_2^2}m_Z^2 \implies m_H^2 \rightarrow  \frac{4g^2b^2}{g_1^2+g_2^2}m_Z^2, \quad m_h^2 \rightarrow 0.
\end{eqnarray}
We thus need to impose 
\begin{eqnarray}
 \frac{4g^2b^2}{g_1^2+g_2^2}m_Z^2 =  \frac{g^2b^2}{v^2} \gg H^2 \implies g \gg \frac{vH}{b}.
\end{eqnarray}
Since we have $g \sim M_S^{-2}$ and $M_S^2 \sim H \sim 10^{-5}$, it reduces to
\begin{eqnarray}
b \gg  v H M_S^2 \sim 10^{-26},
\end{eqnarray}
We observe that the parameter $b$ indeed corresponds to the scale of the low energy observable sector if the parameter $b$ is in $\xi \sim M_S^4 = M_I^4 = H^2 \sim 10^{-10} \gg b \gg v H M_S^2 \sim 10^{-26}$. In the limit, the $\mu$ term becomes 
\begin{eqnarray}
  |\mu|^2  \sim  \bigg( \frac{4v^2}{m_h^2} -\frac{1}{3} \bigg)  M_I^4  >0 \implies \mu \sim \mathcal{O}(H).
\end{eqnarray}
Hence, we need to obey a constraint 
\begin{eqnarray}
 v > \frac{m_h}{2\sqrt{3}},
\end{eqnarray}
which can be satisfied since we already have $v>m_h$ with the observed values, $v = 246~\textrm{GeV}$ and $m_h = 125~\textrm{GeV}$. Meanwhile, regarding Higgs mass, the impact of quantum loop corrections to the Higgs mass must be evaluated. However, we leave this issue for future investigation since this is out of scope of this work. 

We now summarize spectra of the scalar masses. We find that only the light Higgs scalar mass $m_h$ varies from almost zero during inflation to the observed Higgs mass $m_h \sim 125~\textrm{GeV}$ at the true vacua after inflation. On the other hand, the other scalar masses in this model can be much heavier than the Hubble scale during and after inflation, so that they do not occur extra slow-roll inflation along the directions of those scalar fields. 

As for the light Higgs mass during inflation, at first glance, this seems to be against the effective single field slow roll inflation. However, according to Ref. \cite{LightScalar}, it is possible to have a robust slow-roll inflation under the introduction of extra scalars when some reheating scenario conditions are satisfied. We find that our model may be allowed to carry either ``Case-5'' or ``Case-8'' reheating scenario, which are strongly favoured according to Ref. \cite{LightScalar}. The corresponding conditions are found as follows:
\begin{eqnarray}
&& \textrm{Case-5}:~ \Gamma_{h} < \Gamma_{\phi} < m_h < H, \qquad \left(\frac{\Gamma_h}{\Gamma_{\phi}}\right)^{1/4}\ll \frac{\left<h\right>}{M_{pl}}\sim \frac{v}{M_{pl}} \ll 1,\\
&& \textrm{Case-8}:~ \Gamma_{h} < m_h < \Gamma_{\phi} < H, \qquad \left(\frac{\Gamma_h}{m_h}\right)^{1/4}\ll \frac{\left<h\right>}{M_{pl}}\sim \frac{v}{M_{pl}} \ll 1,
\end{eqnarray}
where $\Gamma_{\phi},\Gamma_h$ are the decay rates of inflaton $\phi$ and light Higgs $h$ during the reheating phase, and $v$ is the VEV of the Higgs after inflation. Note that the decay rate of Higgs has to be the smallest.

We also note that unlike our previous model in Ref. \cite{jp2}, we can specify the reheating scenario conditions using the observed values\footnote{i.e. $v=246~\textrm{GeV} \sim 10^{-16}M_{pl}$ and $m_h = 125~\textrm{GeV} \sim 10^{-16} M_{pl}$ while $H \sim 10^{-5}M_{pl}$} and make them to be a problem of determination of the decay rates. Therefore, it is worthwhile to investigate how big the decay rates are and which scenario will win. So, we leave this as another further study.

\subsection{Fermion mass matrix comparable with light SM fermions}

In this section, we compute fermionic masses in our supergravity model. First, we recall the superpotential in our model 
\begin{eqnarray}
 W(T) \equiv W^h(T)+W^o(Z^i), 
\end{eqnarray}
where
\begin{eqnarray}
 W^h(T) &\equiv& W_0 + Ae^{-aT}, \\
 W^o(Z^i) &\equiv& W_{MSSM} =  -Y_u \hat{U}_R \hat{H}_u \cdot \hat{Q} + Y_d \hat{D}_R \hat{H}_d \cdot \hat{Q} + Y_e \hat{E}_R \hat{H}_u \cdot \hat{L}  + \mu \hat{H}_u \cdot \hat{H}_d \nonumber\\
 &=&  -Y_u \tilde{u}_R (H^+_u \tilde{d}_L -H^0_u\tilde{u}_L) + Y_d \tilde{d}_R (H^0_d \tilde{d}_L -H^-_d\tilde{u}_L)\nonumber\\
 && + Y_e \tilde{e}_R (H^+_u \tilde{\nu}_L -H^0_u\tilde{e}^-_L)  + \mu (H^+_uH^-_d-H^0_uH^0_d).
\end{eqnarray}
The most general fermion masses $m^{(g)}$ are given by all the contributions from the standard supergravity, new FI terms, and the super-Higgs effects to the fermion mass, which are written as \eqref{General_Fermion_Masses}. Here, we point out that if the gauge kinetic function is purely a constant, then the gaugino masses almost vanish at the vacuum. In particular, when a gauged R-symmetry is imposed, gauginos can get massive enough thanks to the $U_R(1)$ anomaly cancellation between one-loop quantum correction to the Lagrangian and the shift of Green-Schwarz term by the presence of a linear term in some charged moduli in the gauge kinetic function. However, in our model, we consider a model without gauging a R-symmetry. Thus, we can just take advantage of adding a linear term in the gauge kinetic function as follows:
\begin{eqnarray}
 f_{AB}(T) = \delta_{AB}\Big( \frac{1}{\sqrt{g_Ag_B}} + \sqrt{\beta_A\beta_B} T\Big), 
\end{eqnarray}
where $T$ is the modulus field and $\delta_{AB}$ is the Kronecker delta. We also assume that the coefficient $\beta_A$ can be sufficiently small such that
\begin{eqnarray}
 g^{-2}_A \gg \beta_A T \implies g^{-2}_A \gg \beta_A  ~~\textrm{at the vacuum where } T \sim 1,
\end{eqnarray}
so that the gauge kinetic Lagrangians can be considered as canonically normalized. We note that in any case we can consider the scale of $\beta g^2$ \begin{eqnarray}
g \equiv 10^{-n}, \beta \equiv 10^m \implies \beta g^2=10^{m-2n} \ll 1 \implies m < 2n.
\end{eqnarray}
which will be used for estimating the gaugino masses. Especially, for example, when the gauge coupling can already be sufficiently small, i.e. $g=10^{-n} \ll 1$, we may consider the order of the constant $\beta$ like $\beta \sim \mathcal{O}(10^m)$ where $0<m<2n$. This will contribute to the fermion masses as a big number in our model. The weaker $g$ gets, the larger $\beta$ can get.

Then, the correspondimg fermion mass expressions reduce to the following
\begin{eqnarray}
 m_{3/2} &=& We^{K/2},\\
  m_{IJ}^{(g)} &=&  e^{K/2}(W_{IJ} + K_{IJ}W+K_JW_I+K_IW_J + K_IK_JW)
\nonumber\\
&& -e^{K/2} G^{K\bar{L}}\partial_I G_{J\bar{L}}(W_K + K_KW) -\frac{2}{3}  (W_I+K_IW)(W_J+K_JW),\nonumber\\
m_{IA}^{(g)} &=& 
 i\sqrt{2}[\partial_I\mathcal{P}_A-\frac{1}{4}\delta_{A}^C\sqrt{\beta_A\beta_C}\delta_{IT} \Big( \frac{1}{\sqrt{g_Ag_C}} + \sqrt{\beta_A\beta_C} \textrm{Re}T\Big)^{-1}\mathcal{P}_C]-i\frac{2}{3\sqrt{2}W}(W_I+K_IW) \mathcal{P}_A
\nonumber\\{}\\
m_{AB}^{(g)} &=&  -\frac{1}{2}e^{K/2}\delta_{AB}\sqrt{\beta_A\beta_B} G^{T\bar{J}}(\bar{W}_{\bar{J}}+K_{\bar{J}}\bar{W}) + \frac{1}{3e^{K/2}W} \mathcal{P}_A\mathcal{P}_B\\
m_{I\lambda}^{(g)} &=&  -\frac{i}{\sqrt{2}}\frac{\mathcal{U}_I}{\mathcal{U}}-\frac{i\sqrt{2}}{3W}(W_I+K_IW) \mathcal{U} = m_{\lambda I}^{(g)},\\
m_{\lambda\lambda}^{(g)} &=&  -e^{K/2} \left( \bar{W} + 4G^{I\bar{J}}\left(\frac{\mathcal{U}_I}{\mathcal{U}}+\frac{K_I}{3}\right)(\bar{W}_{\bar{J}}+K_{\bar{J}}\bar{W}) \right) +\frac{\mathcal{U}^2}{3e^{K/2}W},
\end{eqnarray}
where $\lambda^A$ is the gaugino corresponding to the gauge multiplet $V_A$ ($A=SU(3)_c,SU(2)_L,U(1)_Y$), and $\lambda$ is the superpartner of the new FI term vector multiplet $V$. Remember that $\mathcal{U}$ is nowhere vanishing by definition; that is, $\mathcal{U} = \xi + U>0$ with $U\geq 0$ and $\xi \neq 0$. The detailed derivation of the masses is present in the appendix \ref{deri}. We note that gravitino in this model has the Hubble mass $H$, i.e. super-EeV-scale gravition, which may be heavy dark matter candidates explored in Refs. \cite{EeVGravitino,HeavyGravitino,GDM}.

We have checked that only neutral components $H_u^0,H_d^0$ of the Higgs fields have non-vanishing vacuum expectation values (VEV), while the other matters have vanishing VEVs. In addition, we have supposed that $\mathcal{U}|_{vac}=(\xi+U)|_{vac}>0$. Then, denoting $H_u^0,H_d^0$ by an index $a$ and $z^i$ including $H_u^+,H^-_d$ (where $i\neq a$) by $i'$, we have at the vacuum ($\left<H_a^0\right> = v_a/\sqrt{2}$ and $\left<z^{i'}\right>=0$): \begin{eqnarray}
 && \Phi_{a}|_{vac} = \frac{v^2_a}{2}, \quad \Phi_{i'}|_{vac}  = 0, \quad \mathcal{U}|_{vac} \approx \xi \sim M_S^4 \sim H^2, \quad X|_{vac} = 1, \quad  W^o_{i'} |_{vac} =0, \quad W^o_{i'b}|_{vac}=0 ,\nonumber\\
 && m_{3/2} = e^{G/2}|_{vac} = \sqrt{|W|^2}|_{vac} \sim H,\quad \mu \sim \mathcal{O}(H).
\end{eqnarray}

The moment maps with respect to the gauge groups of SM are given by
\begin{eqnarray}
 \mathcal{P}_{U(1)_Y} &=& \frac{g_1}{X} \bigg[ \sum_{i=gen} \Big( \frac{1}{6}\tilde{Q}^{\dag}_i\tilde{Q}_i-\frac{1}{2}\tilde{L}^{\dag}_i\tilde{L}_i -\frac{2}{3} \tilde{u}^{\dag}_{R_i}\tilde{u}_{R_i} + \frac{1}{3} \tilde{d}^{\dag}_{R_i}\tilde{d}_{R_i} + \tilde{l}^{\dag}_{R_i}\tilde{l}_{R_i}
 \Big) + \frac{1}{2}H^{\dag}_u H_u -\frac{1}{2}H^{\dag}_d H_d\bigg],\nonumber\\{}\\
  \mathcal{P}_{SU(2)_L} &=&  \frac{g_2}{X} \bigg[\sum_{i=gen} \Big(\tilde{Q}^{\dag}_i\frac{\vec{\sigma}}{2}\tilde{Q}_i  +\tilde{L}^{\dag}_i\frac{\vec{\sigma}}{2}\tilde{L}_i \Big) + H^{\dag}_u \frac{\vec{\sigma}}{2} H_u+H^{\dag}_d \frac{\vec{\sigma}}{2} H_d\bigg],\\
   \mathcal{P}_{SU(3)_c} &=& \frac{g_3}{X} \bigg[ \sum_{i=gen} \Big( \tilde{Q}^{\dag}_i\frac{\vec{\lambda}}{2}\tilde{Q}_i  - \tilde{u}^{\dag}_{R_i}\frac{\vec{\lambda}}{2}\tilde{u}_{R_i}-
   \tilde{d}^{\dag}_{R_i}\frac{\vec{\lambda}}{2}\tilde{d}_{R_i}\Big)\bigg],
\end{eqnarray}
where tilded fields are superpartner scalars to the SM fermions; $\vec{\sigma}$ and $\vec{\lambda}$ are Pauli and Gell-Mann matrices; $g_1,g_2,g_3$ are gauge couplings, and the index $i$ runs over the three generations of particle in the SM. Their vacuum expectation values are found by
\begin{eqnarray}
 \left<\mathcal{P}_{U(1)_Y}\right> = \frac{g_1}{4}(v_u^2 - v_d^2),\quad 
 \left<\mathcal{P}_{SU(2)_L}\right> = -\frac{g_2}{4}(v_u^2 - v_d^2),\quad 
  \left<\mathcal{P}_{SU(3)_c}\right> =  0, 
\end{eqnarray}
and
\begin{eqnarray}
&& \left<\partial_{H_u^0}\mathcal{P}_{U(1)_Y}\right> = \frac{g_1v_u}{2\sqrt{2}} + \frac{g_1v_u}{12\sqrt{2}}(v_u^2 - v_d^2), \quad
 \left<\partial_{H_d^0}\mathcal{P}_{U(1)_Y}\right> = -\frac{g_1v_d}{2\sqrt{2}}+ \frac{g_1v_d}{12\sqrt{2}}(v_u^2 - v_d^2),\\
&& \left<\partial_{H_u^0}\mathcal{P}_{SU(2)_L}\right>  = - \frac{g_2v_u}{2\sqrt{2}} - \frac{g_2v_u}{12\sqrt{2}}(v_u^2 - v_d^2) , \quad 
  \left<\partial_{H_d^0}\mathcal{P}_{SU(2)_L}\right>  =  \frac{g_2v_d}{2\sqrt{2}} -\frac{g_2v_d}{12\sqrt{2}}(v_u^2 - v_d^2),
\end{eqnarray}
where $\left<\partial_{I}\mathcal{P}_{A}\right>=0$ for others.

Now we are ready to estimate the scales of the fermionic masses. First, we are estimating masses of the matter fermions. Given the supergravity G-function $G = -3\ln[T+\bar{T}-\Phi/3] + \ln W + \ln\bar{W}$ with the superpotential $W = W^h(T) +W^o(z^i)$, the components of the fermion mass matrix are as follows: 
\begin{eqnarray}
m_{ij}^{(g)} &=& \sqrt{\frac{1}{X^3}}\Big[W_{ij}^o
+\frac{2}{3X}(W_i^o\Phi_j+\Phi_iW_j^o)+\frac{2}{3X^2}\Phi_i\Phi_jW
\nonumber\\
&&+ \frac{2\Phi_i\Phi_j}{9X^2}(\Phi-\Phi_{m}\Phi^{m\bar{l}}\Phi_{\bar{l}})\left(\frac{W_T^h}{3}-\frac{W}{X}\right) 
\Big] - \frac{2}{3} \Big(W_i^o+\frac{\Phi_iW}{X}\Big)\Big(W_j^o+\frac{\Phi_jW}{X}\Big),\\
 m_{iT}^{(g)} &=& \sqrt{\frac{1}{X^3}}\Big[
 -\frac{W_i^o}{X} + \frac{2\Phi_i}{X}\left(\frac{W_T^h}{3}-\frac{W}{X}\right) - \left(\frac{W_T^h}{3}-\frac{W}{X}\right)(\Phi-\Phi_{m}\Phi^{m\bar{l}}\Phi_{\bar{l}})\frac{2\Phi_i}{3X^2}
 \Big] \nonumber\\
 &&-\frac{2}{3}\Big(W_i^o+\frac{\Phi_i}{X}W\Big)\Big(W_T^h-\frac{3}{X}W\Big),\\
 m_{TT}^{(g)} &=& \frac{6}{X}\left(\frac{W}{X}-\frac{W_T^h}{3}\right)\Big(
 1+\frac{1}{3X}(\Phi-\Phi_{m}\Phi^{m\bar{l}}\Phi_{\bar{l}})
 \Big),
\end{eqnarray}

If $\Phi= \delta_{i\bar{j}}z^i\bar{z}^{\bar{j}}$, then $\Phi=\Phi_{m}\Phi^{m\bar{l}}\Phi_{\bar{l}}$. Thus, the components reduce to 
\begin{eqnarray}
m_{ij}^{(g)} &=& \sqrt{\frac{1}{X^3}}\Big[W_{ij}^o
+\frac{2}{3X}(W_i^o\Phi_j+\Phi_iW_j^o)+\frac{2}{3X^2}\Phi_i\Phi_jW\Big]
 - \frac{2}{3} \Big(W_i^o+\frac{\Phi_iW}{X}\Big)\Big(W_j^o+\frac{\Phi_jW}{X}\Big),\nonumber\\{}\\
 m_{iT}^{(g)} &=& \sqrt{\frac{1}{X^3}}\Big[ 
 -\frac{W_i^o}{X} + \frac{2\Phi_i}{X}\left(\frac{W_T^h}{3}-\frac{W}{X}\right) 
 \Big] -\frac{2}{3}\Big(W_i+\frac{\Phi_i}{X}W\Big)\Big(W_T^h-\frac{3}{X}W\Big),\\
 m_{TT}^{(g)} &=& \frac{6}{X}\left(\frac{W}{X}-\frac{W_T^h}{3}\right).
\end{eqnarray}
The non-trivial components at the vacuum are then given by
\begin{eqnarray}
 m_{i'j'}^{(g)} &=& W_{i'j'}^o \approx \frac{v}{\sqrt{2}}Y_{i'j'},\\
  m_{uu}^{(g)} &=&-\frac{2}{3}\mu v_uv_d+\frac{1}{6}v_u^2W
 - \frac{2}{3} \Big(-\mu \frac{v_d}{\sqrt{2}}+\frac{v_u^2W}{2}\Big)^2\approx \mu v^2 \sim \mathcal{O}(H v^2) \sim m_{dd}^{(g)},\\
  m_{ud}^{(g)} &=& \Big[W_{ud}^o
+\frac{1}{3}(W_u^ov_d+v_uW_d^o)+\frac{1}{6}v_uv_dW\Big]
 - \frac{2}{3} \Big(W_u^o+\frac{v_u^2W}{2}\Big)\Big(W_d^o+\frac{v_d^2W}{2}\Big) \nonumber\\
 &\approx& -\mu \sim -H \sim -m_{+-}^{(g)},\\
  m_{uT}^{(g)} &=& \Big[ 
 -W_u^o + v_u^2\left(\frac{W_T^h}{3}-W\right)\Big] -\frac{2}{3}\Big(W_u+\frac{v_u^2}{2}W\Big)\Big(W_T^h-3W\Big) \approx \mu \frac{v_d}{\sqrt{2}}\sim Hv,\\
   m_{dT}^{(g)} &=& \Big[ 
 -W_d^o + v_d^2\left(\frac{W_T^h}{3}-W\right) 
 \Big] -\frac{2}{3}\Big(W_d+\frac{v_d^2}{2}W\Big)\Big(W_T^h-3W\Big)\approx \mu \frac{v_u}{\sqrt{2}} \sim Hv,\\
 m_{TT}^{(g)} &=& 6W-2W_T^h \approx m_{3/2} \sim H,
\end{eqnarray}
where $i'$'s are denoted by non-higgs matters, and the indices $\pm$ mean $H^+_u$ and $H^-_d$ respectively. 

Next, let us estimate the other mass parameters. We find
\begin{eqnarray}
 m_{uB}^{(g)} \sim m_{dB}^{(g)} &\approx& g_B(iv - i \frac{(-\mu v + v H)}{H} v^2 ) \sim ig_Bv,\\
m_{AB}^{(g)} &\approx& g_Ag_B\frac{v^2}{H} - H \delta_{AB}\sqrt{\beta_A\beta_B} \sim -\mathcal{O}(\beta H), \\
m_{AT}^{(g)} &\approx& g_A\mathcal{O}(v^2)- \beta_A g_A^2 \left<\mathcal{P}^A\right>,\\
m_{A\lambda}^{(g)} &\approx& g_A\mathcal{O}(v^2),\\
m_{u\lambda}^{(g)} &\approx& -i\frac{bv_u}{\xi} -i \frac{-\mu v_d+v_uH}{H} \xi \sim  -i\frac{bv}{H} \sim m_{d\lambda}^{(g)},\\
m_{T\lambda}^{(g)} &\approx&  \xi \sim H,\\
m_{\lambda\lambda}^{(g)} &\approx& m_{3/2} + \frac{\xi^2}{m_{3/2}} \sim H + \frac{H^2}{H} \sim H ,
\end{eqnarray}
where $A,B = 1,2$ for $U(1)_Y$ and $SU(2)_L$ respectively. In terms of $m_{u\lambda},m_{d\lambda}$, since $10^{-26} \ll b \ll 10^{-10}$, we have 
\begin{eqnarray}
 10^{-32} \ll m_{u\lambda}, m_{d\lambda} \ll 10^{-16} \ll H.
\end{eqnarray}

In summary, the fermion mass matrix is represented by
\begin{eqnarray}
 M_f &\equiv& 
 \begin{pmatrix}
 m_{i'j'}^{(g)} &  m_{i'u}^{(g)} &  m_{i'd}^{(g)} &  m_{i'+}^{(g)} &  m_{i'-}^{(g)} & m_{i'T}^{(g)} & m_{i'B}^{(g)}  & m_{i'\lambda}^{(g)} \\
 m_{uj'}^{(g)} &  m_{uu}^{(g)} &  m_{ud}^{(g)} &  m_{u+}^{(g)} &  m_{u-}^{(g)} &  m_{uT}^{(g)} & m_{uB}^{(g)}  & m_{u\lambda}^{(g)} \\
 m_{dj'}^{(g)} &  m_{du}^{(g)} &  m_{dd}^{(g)} &  m_{d+}^{(g)} &  m_{d-}^{(g)} &  m_{dT}^{(g)} & m_{dB}^{(g)}  & m_{d\lambda}^{(g)} \\
 m_{+j'}^{(g)} &  m_{+u}^{(g)} &  m_{+d}^{(g)} &  m_{++}^{(g)} &  m_{+-}^{(g)} &  m_{+T}^{(g)} & m_{+B}^{(g)}  & m_{+\lambda}^{(g)} \\
 m_{-j'}^{(g)} &  m_{-u}^{(g)} &  m_{-d}^{(g)} &  m_{-+}^{(g)} &  m_{--}^{(g)} &  m_{-T}^{(g)} & m_{-B}^{(g)}  & m_{-\lambda}^{(g)} \\
 m_{Tj'}^{(g)} &  m_{Tu}^{(g)} &  m_{Td}^{(g)} &  m_{T+}^{(g)} &  m_{T-}^{(g)}  &  m_{TT}^{(g)} & m_{TB}^{(g)}  & m_{T\lambda}^{(g)} \\
 m_{Aj'}^{(g)} &  m_{Au}^{(g)} &  m_{Ad}^{(g)} &  m_{A+}^{(g)} &  m_{A-}^{(g)} &  m_{AT}^{(g)} & m_{AB}^{(g)}  & m_{A\lambda}^{(g)} \\
 m_{\lambda j'}^{(g)} &  m_{\lambda u}^{(g)} &  m_{\lambda d}^{(g)} &  m_{\lambda +}^{(g)} &  m_{\lambda -}^{(g)}  &  m_{\lambda T}^{(g)} & m_{\lambda B}^{(g)}  & m_{\lambda\lambda}^{(g)} \\
 \end{pmatrix}
 \nonumber\\
 &\approx&  
 \begin{pmatrix}
\frac{v}{\sqrt{2}}Y_{ij} & 0 &  0 &  0 &  0 & 0 & 0  & 0 \\
0 &  \mathcal{O}(Hv^2) &   -H &  0 &  0 & Hv & ivg_B  & -i\frac{bv}{H}\\
0 &  -H & \mathcal{O}(Hv^2)  & 0 &  0 &  Hv & ivg_B  & -i\frac{bv}{H} \\
0 & 0 & 0 & 0 & H & 0 & 0 & 0 \\
0 & 0 & 0 & H & 0 & 0 & 0 & 0 \\
0 & Hv &  Hv &  0 &  0 & H & \mathcal{O}(v^2)g_B & H \\
 0 & ivg_A &  ivg_A &  0 &  0 & \mathcal{O}(v^2)g_A & -\mathcal{O}(\beta H) & 0 \\
0& -i\frac{bv}{H} &  -i\frac{bv}{H} &  0 &  0 & H & 0  & H \\
 \end{pmatrix} \nonumber\\{}
\end{eqnarray}
where $m_{A\lambda}^{(g)}=m_{\lambda B}^{(g)} =0$ since there are no couplings between the relevant vector multiplets. Keeping the Yukawa masses of the matter fermions and dropping the other terms much less than the Hubble scale, the fermion mass matrix can be approximated into
\begin{eqnarray}
 M_f \approx 
  \begin{pmatrix}
\frac{v}{\sqrt{2}}Y_{ij} & 0 &  0 & 0 &  0 & 0 & 0  & 0 \\
0 &  0 &   -H & 0 & 0 &  0 &0  & 0\\
0 &  -H & 0  &  0 & 0 &  0 &0  & 0 \\
0 & 0 & 0 & 0 & H & 0 & 0 & 0 \\
0 & 0 & 0 & H & 0 & 0 & 0 & 0 \\
0 & 0 & 0 & 0 &  0 & H & 0 & H \\
0 & 0 & 0 & 0 &  0 & 0 & -\beta H  & 0 \\
0 & 0 & 0 & 0 &  0 & H & 0  & H \\
 \end{pmatrix}
\end{eqnarray}
We observe that masses of the SM matter fermions can be matched with the observed values by adjusting Yukawa couplings which is free parameters. Diagonalizing the fermion mass matrices may produce negative mass eigenvalues, but the masses can be made to be positive by absorbing the negative sign into the mixing matrices that get imaginary \cite{Fermion_mass_matrix}. We note that the chargino, neutralino, and gaugino masses at the true vacua after inflation are of the order of Hubble scale $\mathcal{O}(10^{-5})M_{pl} \sim \mathcal{O}(10^{13})~\textrm{GeV}$, implying that they may be candidate of the so-called supermassive dark matter ``WIMPZILLA'' \cite{WIMPZILLA1,WIMPZILLA2,WIMPZILLA3,WIMPZILLA4} (or superheavy dark matter in Ref. \cite{SuperheavyDM}). We are not going to further investigate details of phenomenology related to these fermions since it is beyond the scope of our purpose in this work. Nevertheless, it would be worthwhile to study possible phenomenological implications about those fermions in the future. 

Lastly, we summarize all the parameters in our supergravity model of inflation compatible with MSSM as follows:
\begin{itemize}
    \item Hubble Scale $H$ ($\sim \mu, g^{-1}, M_S^2,A$) for inflation,
    \item Yukawa couplings $Y_{i'j'}$ for fermion masses,
    \item Neutral Higgs VEVs $v_u,v_d$ such that $v \equiv \sqrt{v_u^2 + v_d^2}$ for Higgs mechanism,
    \item Angle between $v_u$ and $v_d$, i.e. $\tan \beta = v_u/v_d$,
    \item Gauge couplings $g_1,g_2,g_3$ for strong, weak, and hypercharge interactions in the SM,
    \item New-FI-term hidden-sector parameters $C_i$'s for producing scalar bosons heavier than Hubble,
    \item New-FI-term observable-sector parameter $b$ for generating supersymmetric Higgs potential,
    \item Hidden-sector superpotential parameter $a$ for KKLT superpotential,
\end{itemize}
which determine all the other parameters in our model. In particular, fully-free parameters among them are $\beta,C_i,b,a$.

%% file: chapters/12.tex
In this thesis, we have investigated how to construct and examine locally supersymmetric effective field theories of inflation in various aspects in the superconformal formalism.

In Ch. \ref{ch6}, we have confirmed that the liberated $\mathcal{N}=1$ supergravity can be defined in the superconformal formalism as well, which were originally proposed in the superspace formalism by Farakos, Kehagias, and Riotto. The key fact of the liberated supergravity is that a new general scalar potential can be introduced by promoting K\"{a}hler transformations to abelian gauge symmetry and contriving a new D-term action with the K\"{a}hler-potential real multiplet. As a similar effort to this, relaxed supergraivty is proposed in this research. We also proved that liberated supergravity is not literally liberated due to some strong constraints on the liberated terms, which is done by inspecting effective-field-theoretical suppression of the nonrenormalizable fermionic Lagrangians. There, we relate the ``liberated'' scalar potential $\mathcal{U}$ to the UV cutoff. Nevertheless, in Ch.~\ref{ch9}, we show that it is shown that a possible toy model of inflation can consistently be built if taking into account a special phase transition of supersymmetry breaking scale from Planck during inflation to electroweak scale after inflation. 

In Ch.~\ref{ch7}, we have revisited the component action of a K\"{a}hler-invariant new FI term (called ``ACIK-FI term'') in the superconformal formalism. We used the superconformal tensor calculus to constrain the size of new supergravity terms that are present in the ACIK-FI term, but absent in standard supergravity. Specifically, we derive the constraints on the ACIK-FI term used in~\cite{jp2}. We have seen that ACIK-FI term has a cutoff such that $\Lambda_{cut} \sim H^{2/3} = 10^{-2.66}M_{pl}$. What makes our constraints powerful is that differently from standard supergravity, both liberated supergravity and the ACIK-FI terms introduce nonrenormalizable interactions proportional to inverse powers of the supersymmetry breaking scale $M_S$. This makes it impossible to send $M_S$ to zero while keeping the UV cutoff of the theory finite. The most singular nonrenormalizable interactions in the limit $M_S\rightarrow 0$ are cumbersome, multi-fermion operators, but they can be found and studied using the superconformal tensor calculus in a systematic and economical way.

In Ch.~\ref{ch8}, we have found a no-go theorem for the higher order corrections in FKLP minimal supergravity models of inflation. The no-go theorem tells us that the higher order corrections in the field strength of a vector multiplet $V$ must include some powers of the real linear multiplet $(V)_D$ whose lowest component is given by the auxiliary field $D$ of the vector multiplet $V$. We discovered a new negative-definite generic potential term. We also found moderate constraints on the new negative term by evaluating the suppression of nonrenormalizable interactions with respect to a cutoff scale $\Lambda_{cut}$. These constraints identify the cutoff $\Lambda_{cut}$ with the SUSY breaking scale $M_S$. It turns out that we have to take into account high-scale SUSY breaking putting the naturalness away. Then, we showed a comparison between relaxed and liberated supergravities. As an example, we observed that we can generate the inflation energy of order $10^{-10}M_{pl}^4$ through relaxed supergravity since the total scalar potential is bounded above by $V_{D}^{\cancel{S}}$ less than the Planck scale $M_{pl}$, while we cannot do through the liberated supergravity because only the scales below $10^{-64}M_{pl}^4$ or $10^{-96}M_{pl}^4$ can be allowed. In this sense, relaxed supergravity is truly liberated than the original liberated supergravity.

In Ch.~\ref{ch10}, we have seen that our model can naturally produce plateau-potential inflation at the Hubble scale with a high scale spontaneously supersymmetry breaking in the hidden sector and low scale soft supersymmetry breaking interactions with various soft 
masses in the observable sector. We also obtain naturally a super-EeV gravitino, which is compatible with  
constraint for heavy gravitino cold dark matter ({\it i.e.} $0.1 ~\textrm{EeV} \lesssim m_{3/2} \lesssim 1000~\textrm{EeV}$) 
\cite{Inf_HighSUSY_EeVGrav}. In this work, we have not investigated the specific structure and dynamics of observable-sector 
interactions or the detailed construction of a realistic low energy effective theory of the observable sector. It would be of obvious interest
to see how far this scenario could be pursued and how to incorporate in it a supersymmetric extension of the Standard Model or a
Grand Unified Theory. Models with a dynamically generated FI term, realistic observable sector, 
D-term inflation, and high-scale supersymmetry breaking have been studied in~\cite{DS17,DS17_2}; other uses of FI terms for inflation were 
presented in~\cite{lln1,lln2,lln3}. It would be interesting to reproduce the phenomenologically desirable features of~\cite{DS17,DS17_2,lln1,lln2,lln3}  and other 
models proposed in the literature in our scenario.
On a different note, it would be extremely interesting to see if the new FI term in general and in our KKLT-type
scenario in particular 
can be obtained in string theory. In other words, it would be interesting to see if our model belongs to the string landscape or
the string swampland \cite{Swampland} (See refs. \cite{Recent_Review_Swampland1,Recent_Review_Swampland2} for recent reviews of  swampland conjectures).

In Ch.~\ref{ch11}, we have explored a possible effective field theory of single-field Starobinsky-type inflation and minimal supersymmetric standard model in four-dimensional $\mathcal{N}=1$ supergraivty with the KKLT string background and new K\"{a}hler-invariant Fayet-Iliopoulos (FI) terms without gauging R-symmetry. We takes advantage of gravity mediation for supersymmetry breaking, which is broken at the inflation scale $M_S \sim 10^{-2.5}M_{pl}$. That is, this model is effective field theory of inflation with broken supersymmetry. In the hidden sector, we observe that slow-roll inflation occurs with the Hubble scale $H \sim 10^{-5}M_{pl}$ through the combination of the new FI term and F-term scalar potentials in de Sitter space, whose vacuum energy is given by the cosmological constant $\Lambda \sim 10^{-120}M_{pl}^4$. In particular, we have examined suppression of the nonrenormalizable fermionic Lagrangians arising from the new FI term, and then imposed a constraint on the new FI term. We have seen that our model can be valid as effective field theory if we require a proper choice of the gauge kinetic function for the vector multiplet responsible for the new FI term. We showed a specific example about this, and proved that the case is indeed comparable with a weak coupling $g$ of the new FI term, and does not ruin the kinetic term of the new-FI-term vector field.   

Remarkably, every scalar except for inflaton and lightest Higgs in the observable sector can be supermassive compared to the Hubble mass $H \sim 10^{-5}M_{pl}$. Because of the introduction of the new FI terms, we have no issue of the tachyonic mass, and no excessive number of unnecessary extra light scalars in our model. Indeed, this implies that the unique inflationary plateau can be naturally created. Moreover, we have seen that stabilization of Higgs fields can easily be done since the new FI terms allows us to modify the scalar potential in our favor. Regarding the standard-model fermions, they can be light up to whatever we wish through the free parameters of Yukawa coupling. Particularly, the other supersymmetric fermions including gravitino can be massive up to the Hubble due to the F-term of high scale. Furthermore, we can make gauginos to be heavy by considering the linear term of the moduli $T$ in the gauge kinetic functions corresponding to the standard model gauge groups. 

At last, I put some research directions for future investigation. First, it would be attractive to explore how quantum one-loop effective potentials in supergravity under broken supersymmetry will affect the theories discussed in this thesis. Specifically, it would be interesting to see what conditions determine their validity. In addition, it is also interesting to devise a general ``spectroscopy'' scheme for one to be able to easily scrutinize suppression of non-renormalizable terms using the superconformal tensor calculus in order to read off the exact cutoff scale of the theory in a systematic and efficient manner. Besides, it would be fascinating to search for possible phenomenological consequences from supergravity in order to explain the remaining open questions in cosmology, such as large non-gaussianity, production of primordial black holes as possible candidate of dark matter relic, or normally dark-sector particle physics. Lastly, I would like to mention that the conventional slow-roll inflation could not be compatible with string theory if de Sitter swampland conjecture (dSC) \cite{de_Sitter_Conjecture} is accepted. Certainly, this is the main limitation of this dissertation. In particular, I would like to mention that minimal warm inflation (MWI) \cite{MWI} may be a good inflationary model according to some recent works, which say that MWI can be compatible with some swampland conditions like de Sitter swampland conjecture \cite{MWI_dSC} and Trans-planckian censorship conjecture (TCC) \cite{MWI_TCC}. Therefore, it would be worth studying how we can obtain other possible inflationary models which can satisfy such swampland conjectures, and how such models including minimal warm inflation can be constructed in supergravity.

%% file: chapters/spinor_algebra.tex
Here, I follow the same convention used in Ref. \cite{Superconformal_Freedman}, which this chapter is mainly based on.  

\section{Clifford Algebra in General Diemnsions}

\subsection{The generating Gamma matrices of Clifford algebra}

\begin{itemize}
\item Pauli matrices $\in SL(2,\mathbb{C})$: They are Hermitian/Involutory\footnote{Involutory matrix is the matrix that is its own inverse.}/Traceless(HIT).
\begin{eqnarray}
\sigma_1 = 
\begin{pmatrix}
0 & 1 \\
1 & 0 \\
\end{pmatrix},\quad 
\sigma_2 = 
\begin{pmatrix}
0 & -i \\
i & 0 \\
\end{pmatrix},\quad
\sigma_3 = 
\begin{pmatrix}
1 & 0 \\
0 & -1 \\
\end{pmatrix}
\end{eqnarray}
\item Pauli-matrix 4-vector $\in SO(1,3)$:
\begin{eqnarray}
\sigma^{\mu} \equiv (\mathbf{1},\sigma_i) = \Bar{\sigma}_{\mu}, \quad \sigma_{\mu} \equiv (-\mathbf{1},\sigma_i) = \Bar{\sigma}^{\mu}
\end{eqnarray}
\item Homomorphisms between $SL(2,\mathbb{b})$ and $SO(1,3)$:
\begin{eqnarray}
&& \vec{x} = \Bar{\sigma}_{\mu} x^{\mu} \Longleftrightarrow x^{\mu} = \frac{1}{2}\textrm{Tr}[\sigma^{\mu}\vec{x}],\\
&& \vec{x}' = A\vec{x}A^{\dag} \Longleftrightarrow x'^{\mu} = \frac{1}{2}\textrm{Tr}[\sigma^{\mu}A\Bar{\sigma}_{\nu}A^{\dag}] x^{\nu}
\end{eqnarray}
\\
\item Euclidean(or Spatial) Gamma Matrices: They are Hermitian/Involutory/Anti-commuting. \\
The Euclidean Gamma matrices in $D=2m >0$ (or $D=2m+1>1$) dimensional spacetime can be constructed by taking a tensor-product of `$m$' Pauli \& identity matrices, so that they become a $2^m$-by-$2^m$ matrices.

\begin{eqnarray}
&& \gamma^1 \equiv \underbrace{\sigma_1 \otimes \mathbf{1} \otimes \mathbf{1} \otimes \mathbf{1} \otimes \cdots}_{`m'},\nonumber\\
&& \gamma^2 \equiv \underbrace{\sigma_2 \otimes \mathbf{1} \otimes \mathbf{1} \otimes \mathbf{1} \otimes \cdots}_{`m'},\nonumber\\
&& \gamma^3 \equiv \underbrace{\sigma_3 \otimes \sigma_1 \otimes \mathbf{1} \otimes \mathbf{1} \otimes \cdots}_{`m'},\nonumber\\
&& \gamma^4 \equiv \underbrace{\sigma_3 \otimes \sigma_2 \otimes \mathbf{1} \otimes \mathbf{1} \otimes \cdots}_{`m'},\nonumber\\
&& \vdots\\
&& \gamma^{\mu} \equiv \underbrace{\sigma_3 \otimes \sigma_3 \otimes \cdots \otimes \sigma_3 \otimes \sigma_{i}}_{`m'},\nonumber\\
\end{eqnarray}
While the last gamma matrix will end with $\sigma_2$ in even dimension $D=2m$, it will end with $\sigma_3$ in odd dimension $D=2m+1$.

\item Lorentzian Gamma Matrices: time-direction gamma matrix is anti-Hermitian/anti-involutory. \\
To obtain the Lorentzian Gamma matrix, \\
(1) Decompose $A$ Euclidean Gamma matrices into $A = (a,\alpha)$, \\
(2) Pick any single Euclidean Gamma matrix $\gamma^a$, \\
(3) Multiply it by $i$ and re-label this as $\gamma^a \rightarrow \gamma^0 \equiv i\gamma^a$ \\(We call this as ``timelike-direction'' Lorentzian Gamma matrix). \\
(4) Re-label the remaining Euclidean matrices as $\gamma^i \equiv \gamma^{\alpha}$ \\(We call this as ``spacelike-direction'' Lorentzian Gamma matrix).\\
(5) The final Lorentzian Gamma Matrices $\gamma^{\mu} = (\gamma^0, \gamma^i)$, where $0 \leq \mu \leq D-1$, satisfy the Hermitian representation of the Gamma matrix $(\gamma^{\mu})^{\dag} = \gamma^0 \gamma^{\mu} \gamma^0$.

\item Conjugancy of Gamma matrix: for some unitary matrix $S$, two representations that differ by the unitary matrix are equivalent, i.e. $\gamma'^{\mu} = S \gamma^{\mu} S^{-1}$. Up to this conjugancy, there is a unique irreducible representation of the Clifford algebra by the Gamma matrices in `even' dimension. In odd dimension, there are two inequivalent irreducible representations, which are determined by the sign of the final gamma matrix $\gamma^{D=2m+1}$, i.e. $\pm\gamma^{D=2m+1}$.

\item Orthogonal Basis of Clifford Algebra in Even Dimensional Spacetime:
\begin{eqnarray}
\{\Gamma^A \equiv \mathbf{1},~\gamma^{\mu},~\gamma^{\mu_1\mu_2},~\cdots,~\gamma^{\mu_1 \cdots \mu_D} \} ,\qquad \{\Gamma_A \equiv \mathbf{1},~\gamma_{\mu},~\gamma_{\mu_2\mu_1},~\cdots,~\gamma_{\mu_D \cdots \mu_1} \} 
\end{eqnarray}
where $\gamma^{\mu_r\cdots \mu_1} = (-1)^{r(r-1)/2}\gamma^{\mu_1\cdots \mu_r}$ ($r= 2,3~\textrm{mod}~4$) for some rank-$r$ gamma matrix $\gamma^{\mu_1\cdots \mu_r}$, and the index values follow the conditions $\mu_1 < \mu_2 < \cdots < \mu_r$. At each rank $r$ of Gamma matrix (i.e. $\gamma^{\mu_1\cdots \mu_r}$), there are ${}_DC_r$ different choices of index values like $\mu=3$ for some indices $\mu_1\cdots \mu_r$ of a rank-$r$ Gamma matrix $\gamma^{\mu_1\cdots \mu_r}$, so that the total number of different ``unindexed'' matrices, like $\gamma^{123}$, in the set of $\{\Gamma^A\}$ is $2^D$ (of course, in the set $\{\Gamma^A\}$, there are $D+1$ ``indexed'' matrices like $\gamma^{\mu\nu}$). The upper/lower indices can be interchanged with each other by contracting it to the spacetime metric $g^{\mu\nu}$. For any matrix $M$ with the same dimension of Gamma matrices, it satisfies the following properties
\begin{eqnarray}
&& \textrm{Trace orthogonality:~~~}  \textrm{Tr}[\Gamma^A \Gamma_B] = 2^m \delta^A_B,\label{trace_ortho}\\
&& \textrm{Gamma-matrix expansion:~~~}  M = \sum_A m_A \Gamma^A = \sum_{k=0}^{[D]}\frac{1}{k!}m_{\mu_k\ldots \mu_1}\gamma^{\mu_1\ldots \mu_k},\label{gam_expansion} \\ 
&& \textrm{Expansion coefficient:~~~}  m_A = \frac{1}{2^m} \textrm{Tr}[M\Gamma_A] \implies m_{\mu_k\ldots \mu_1}=\frac{1}{2^m} \textrm{Tr}[M\gamma_{\mu_k\ldots \mu_1}] ,\label{gam_expansion_coeff}
\end{eqnarray}
where $[D]=D$ for even $D$, while $[D]=(D-1)/2$ for odd $D$.  

\item The highest rank Clifford algebra component $\gamma_*$: This is Hermitian/Involutory/Traceless, and commutes with all even-rank Clifford elements (e.g. $\gamma^{\mu\nu},\gamma^{\mu\nu\rho\sigma}$, etc.) but anti-commutes with all odd-rank Clifford elements (e.g. $\gamma^{\mu},\gamma^{\mu\nu\rho}$, etc.). Plus, this provides the link between even and odd dimensions, and is closely realted to the ``chirality'' of fermions. 
\begin{eqnarray}
\gamma_* \equiv (-i)^{m+1} \prod_{\mu=0}^{D-1}\gamma_{\mu} = (-i)^{m+1}\gamma_0\gamma_1\cdots\gamma_{D-1}.
\end{eqnarray}
For any order of $\mu_i$, one can write 
\begin{eqnarray}
\gamma_{\mu_1\cdots \mu_D} = i^{m+1}\varepsilon_{\mu_1\cdots\mu_D} \gamma_*.
\end{eqnarray}

\item Chiral-Projection operators $P_L,P_R$: they are idempotent (e.g. $P_{L/R}^2=P_{L/R}$), and orthogonal to each other ($P_LP_R=P_RP_L=0$). 
\begin{eqnarray}
P_L \equiv \frac{1+\gamma_*}{2},\qquad P_R \equiv \frac{1-\gamma_*}{2}.
\end{eqnarray}
For a Dirac (or Majorana) spinor $\Psi \equiv \begin{pmatrix}
\psi \\ \Bar{\chi}
\end{pmatrix}$ consisting of two Weyl fields $(\psi,\Bar{\chi})$, these can be found as chiral-projected fields from the Dirac (or Majorana) spinor as
\begin{eqnarray}
P_L\Psi = \begin{pmatrix}
\psi \\ 0
\end{pmatrix}, \qquad P_R\Psi = \begin{pmatrix}
0 \\ \Bar{\chi}
\end{pmatrix}
\end{eqnarray}

\item Symmetries of Clifford elements \\
- Charge-conjugation unitary matrix $C$ and Transpose of Gamma matrix:

In the Clifford algebra of the $2^m$-by-$2^m$ matrices, for both even and odd dimensions $D$, we can distinguish the `symmetric' and `anti-symmetric' ones in the following way. 

There exists a unitary matrix $C$ called ``Charge Conjugation'' such that each matrix $C\Gamma^{(r)}$ for some rank-$r$ Clifford element is either symmetric or anti-symmetric:
\begin{eqnarray}
(C\Gamma^{(r)})^T = -t_r C\Gamma^{(r)} \quad \textrm{where} \quad t_r = \pm1 \quad\textrm{such that}\quad t_r^2 =1.
\end{eqnarray}
For rank-0 and 1, we obtain
\begin{eqnarray}
C^T = -t_0 C \implies C^{\dag} = -t_0 C^* = C^{-1}, \qquad {\gamma^{\mu}}^T = t_0t_1 C\gamma^{\mu}C^{-1}.
\end{eqnarray}
These determine the symmetries of all the other $C\gamma^{\mu_1\cdots \mu_r}$ Clifford elements via the iteration of $(t_0,t_1,t_2=-t_0,t_3=-t_1)$ over the $t_r$'s, e.g. $t_4=t_0,t_5=t_1,t_6=-t_0,t_7=-t_1$. The values of $t_0,t_1$ are determined by the spacetime dimension modulo 8 ($D~\textrm{mod}~8$) and the rank modulo 4 ($r~\textrm{mod}~4$). For example, $D=4 \implies (t_0=1,t_1=-1)$, $D=10 \implies (t_0=t_1=-1)$, $D=11 \implies (t_0=1,t_1=-1)$. We can also represent this as follows: 
\begin{eqnarray}
t_r = 
\begin{cases}
{}~~t_0 \quad \textrm{when}\quad r = 0~\textrm{mod} ~4\\
{}~~t_1 \quad \textrm{when}\quad r = 1~\textrm{mod} ~4\\
-t_0 \quad \textrm{when}\quad r = 2~\textrm{mod} ~4\\
-t_1 \quad \textrm{when}\quad r = 3~\textrm{mod} ~4\\
\end{cases}.
\end{eqnarray}

For even dimension, there are two choices according to the product $t_0t_1$.
\begin{eqnarray}
&& C(t_0t_1=-1)\equiv C_- = (\sigma_2 \otimes \sigma_1) \otimes (\sigma_2 \otimes \sigma_1) \otimes \cdots \equiv C.\\
&& C(t_0t_1=+1)\equiv C_+ = (\sigma_1 \otimes \sigma_2) \otimes (\sigma_1 \otimes \sigma_2) \otimes \cdots. = C\gamma_*.
\end{eqnarray}
For odd dimension, there is a unique $C$.

- Spinor-Charge-Conjugation unitary matrix $B$ and Complex-conjugates of Gamma matrix:

The complex conjugate of gamma matrix is determined by the Hermitian and transpose properties of gamma matrix using a unitary matrix $B$:
\begin{eqnarray}
{\gamma^{\mu}}^* = {\gamma^0}^T {\gamma^{\mu}}^T {\gamma^0}^T = -t_0t_1 B\gamma^{\mu} B^{-1} \qquad \textrm{where} \qquad B \equiv it_0C\gamma^0  \quad \textrm{satisfying}\quad  B^*B = -t_1\mathbf{1}.
\end{eqnarray}
proof)
\begin{eqnarray}
{\gamma^{\mu}}^{\dag} = \gamma^0 \gamma^{\mu} \gamma^0 \implies {\gamma^{\mu}}^* 
&=& {\gamma^0}^T {\gamma^{\mu}}^T {\gamma^0}^T \nonumber\\
&=& (t_0t_1)^3 (C\gamma^0C^{-1}) (C \gamma^{\mu} C^{-1}) (C\gamma^0C^{-1}) \nonumber\\
&=& t_0t_1 C \gamma^0 \gamma^{\mu} \gamma^0 C^{-1} = t_0t_1 C \gamma^0 \gamma^{\mu} (-(\gamma^0)^{-1}) C^{-1}\nonumber\\
&=& - t_0t_1 C \gamma^0 \gamma^{\mu} (\gamma^0)^{-1} C^{-1}=  - t_0t_1 (it_0C \gamma^0) \gamma^{\mu} (it_0C\gamma^0)^{-1} \nonumber\\
&=& -t_0t_1 B \gamma^{\mu} B^{-1}.  \qquad\blacksquare
\end{eqnarray}
\begin{eqnarray}
B^*B &=& (-it_0C^*{\gamma^0}^*)(it_0C\gamma^0) = C^* (-{\gamma^0}^T) C\gamma^0 \nonumber\\
&=& (-t_0 C^{-1})(-t_0t_1C\gamma^0C^{-1}) C \gamma^0 = -t_1 (\gamma^0)^2 = -t_1\mathbf{1}.  \qquad\blacksquare
\end{eqnarray}

Note that the transpose and complex conjugate properties of the gamma matrix hold for the Hermitian representation ${\gamma^{\mu}}^{\dag} = \gamma^0\gamma^{\mu}\gamma^0$ via the unitary matrices $C$ and $B$. In another representation given by $\gamma'^{\mu} = S\gamma^{\mu} S^{-1}$, the corresponding new $C'$ and $B'$ are given by $C' = {S^{-1}}^T C S^{-1}$ and $B' = {S^{-1}}^T B S^{-1}$.
\end{itemize}
\section{Spinors in General Dimensions}

\subsection{Spinor and spinor-bilinears}
\begin{itemize}
\item Majorana Conjugate: for any spinor $\lambda$, we define ``Majorana conjugate'' as $\Bar{\lambda} \equiv \lambda^T C$.

\item Majorana Flip condition of Spinor-Bilinear Scalar: \\
Using the Majorana conjugate definition and $(C\Gamma^{(r)})^T = -t_r C\Gamma^{(r)}$, we find the following relation:
\begin{eqnarray}
\Bar{\lambda}\gamma_{\mu_1\cdots\mu_r}\chi = t_r \Bar{\chi}\gamma_{\mu_1\cdots\mu_r}\lambda.
\end{eqnarray}
proof)
\begin{eqnarray}
\Bar{\lambda}\gamma_{\mu_1\cdots\mu_r}\chi &=& \lambda^TC \gamma_{\mu_1\cdots\mu_r}\chi = -(\chi^T[C\gamma_{\mu_1\cdots\mu_r}]^T  \lambda)^T \quad \because \lambda,\chi~\textrm{are anti-commuting!} \nonumber\\
&=& t_r (\chi^T C\gamma_{\mu_1\cdots\mu_r} \lambda)^T = t_r (\Bar{\chi}\gamma_{\mu_1\cdots\mu_r} \lambda)^T = t_r\Bar{\chi}\gamma_{\mu_1\cdots\mu_r} \lambda  \qquad\blacksquare
\end{eqnarray}
\begin{itemize}
\item Corallaries:
\begin{eqnarray}
\Bar{\lambda} (\Gamma^{(r_1)}\cdots \Gamma^{(r_p)}) \chi = (t_0^{p-1}) (t_{r_1}\cdots t_{r_p}) \Bar{\chi} (\Gamma^{(r_p)}\cdots \Gamma^{(r_1)}) \lambda.
\end{eqnarray}
\item For Majorana dimensions $D=2,3,4 ~\textrm{mod}~ 8$, 
\begin{eqnarray}
\Bar{\lambda} \gamma^{\mu_1} \cdots \gamma^{\mu_p} \chi = (-1)^p \Bar{\chi}   \gamma^{\mu_p}\cdots  \gamma^{\mu_1} \lambda.
\end{eqnarray}
\begin{eqnarray}
&& \chi_{\mu_1\cdots\mu_r} \equiv \gamma_{\mu_1\cdots\mu_r}\lambda \implies \Bar{\chi}_{\mu_1\cdots\mu_r} = t_0t_r \Bar{\lambda}\gamma_{\mu_1\cdots\mu_r},\\
&& \chi \equiv \Gamma^{(r_1)} \cdots \Gamma^{(r_p)} \lambda \implies \Bar{\chi} =  t_0^p t_{r_1}\cdots t_{r_p} \Bar{\lambda} \Gamma^{(r_p)} \cdots \Gamma^{(r_1)}
\end{eqnarray}
\end{itemize}

\item Chirality of the conjugate spinor in ``even'' dimensions: \\
In ``even'' dimensional spacetime, we can define the chirality (left-handed/right-handed parts) of spinors using the chiral projection operators $P_L$ and $P_R$. Since $\chi_* = \gamma_* \lambda \implies \Bar{\chi}_* = t_0t_D \Bar{\lambda} \gamma_*$, we obtain
\begin{eqnarray}
\chi = P_L\lambda \implies \Bar{\chi}= \overline{P_L\lambda} = \begin{cases}
\Bar{\lambda}P_L \quad \textrm{for} \quad D= 0~\textrm{mod}~4~(: t_0t_D=+1),   \\
\Bar{\lambda}P_R \quad \textrm{for} \quad D= 2~\textrm{mod}~4~(: t_0t_D=-1).   \\
\end{cases}
\end{eqnarray}
\end{itemize}
\subsection{Spinor indices}
\begin{itemize}
\item The fundamental spinor $\lambda$ and its transpose $\lambda^T$:\\ $(\lambda)_{\alpha} \equiv \lambda_{\alpha}$ and $(\lambda^T)_{\alpha} = \lambda_{\alpha}$.\\
\item The Majonara-conjugate (barred) spinor $\Bar{\lambda}$ and its transpose $\Bar{\lambda}^T$: \\
$(\bar{\lambda})^{\alpha} \equiv \bar{\lambda}^{\alpha}\equiv  \lambda^{\alpha}$ and $(\bar{\lambda}^T)^{\alpha} = \bar{\lambda}^{\alpha}\equiv  \lambda^{\alpha}$ \footnote{We can omit the barred notation in the indexed form.}\\
Note that `transpose' changes the alignment of the entry components of a spinor between horizontal and vertical alignments, so that the position of the index remains same.\\
\item Majorana-spinor-metric matrix $\mathcal{C}$ (i.e. Rasing/Lowering matrices) generated by Charge conjugation matrix $C$: \\

(1) From the definition of the Majorana conjugate $\bar{\lambda} \equiv \lambda^TC$ and using the transpose operation \footnote{Refer to the deefinition of Transpose of a matrix $A$, $[A^T]_{ij} \equiv ([A]_{ij})^T \equiv [A]_{ji}$. This tells us that the transpose operation can directly act on the indices, resulting in the exchange of the indices for the original matrix after the transpose operation. The symmetry of a matrix $A$ is defined to be the condition that $A^T = A \implies [A^T]_{ij} \overset{!}{=} [A]_{ij} \implies A_{ji}=A_{ij}$, while the anti-symmetry of a matrix $A$ is defined to be the condition that $A^T = -A \implies [A^T]_{ij} \overset{!}{=} - [A]_{ij} \implies A_{ji}=-A_{ij}$.}, 
\begin{eqnarray}
&& (\bar{\lambda})^{\alpha} = (\lambda^TC)^{\alpha} = \lambda_{\beta} (C)^{\beta\alpha} = (C^T)^{\alpha\beta} \lambda_{\beta}  \equiv \mathcal{C}^{\alpha\beta}\lambda_{\beta} = \lambda^{\alpha} \nonumber\\
&& \implies (C^T)^{\alpha\beta} \equiv \mathcal{C}^{\alpha\beta} \qquad \textrm{such that} \qquad \lambda^{\alpha} \equiv \mathcal{C}^{\alpha\beta}\lambda_{\beta}.
\end{eqnarray}

(2) Of course, from $C^T = -t_0 C \implies C = -t_0 C^T$, we find
\begin{eqnarray}
 (C)^{\alpha\beta} = -t_0 (C^T)^{\alpha\beta} \implies (C)^{\alpha\beta} = -t_0\mathcal{C}^{\alpha\beta}.
\end{eqnarray}

(3) Also, from $\bar{\lambda} \equiv \lambda^TC$, we have $\bar{\lambda}C^{-1} \equiv \lambda^T$, which gives
\begin{eqnarray}
&& (\lambda^T)_{\alpha} = (\bar{\lambda}C^{-1})_{\alpha} = \lambda^{\beta} (C^{-1})_{\beta\alpha} \equiv \lambda^{\beta} \mathcal{C}_{\beta\alpha} = \lambda_{\alpha}\nonumber\\
&& \implies (C^{-1})_{\alpha\beta} \equiv \mathcal{C}_{\alpha\beta} \qquad \textrm{such that} \qquad \lambda_{\alpha} \equiv \lambda^{\beta}\mathcal{C}_{\beta\alpha}
\end{eqnarray}
(4) Likewise, from $C^{-1} = -t_0 C^* \implies C^* = -t_0 C^{-1}$, we find
\begin{eqnarray}
  (C^*)_{\alpha\beta} = -t_0 (C^{-1})_{\alpha\beta} \implies (C^*)_{\alpha\beta} = -t_0\mathcal{C}_{\alpha\beta}.
\end{eqnarray}

Note that both $C$ and $C^T$ have upper indices, while both $C^{-1}$ and $C^*$ have lower indices. Plus, all the {\it positive} contractions occur according to the NW-SE convention (I will also call this contraction as ``Back-Slash contraction''). 

(5) In order for the two equations to be consistent with each other, we need to impose
\begin{eqnarray}
\lambda^{\alpha} = \mathcal{C}^{\alpha\beta}\lambda_{\beta} = \mathcal{C}^{\alpha\beta}\lambda^{\gamma}\mathcal{C}_{\gamma\beta} \overset{!}{=} \lambda^{\gamma}\delta_{\gamma}^{\alpha} \implies
\mathcal{C}^{\alpha\beta}\mathcal{C}_{\gamma\beta} = \delta_{\gamma}^{\alpha}, \qquad \mathcal{C}_{\beta\alpha}\mathcal{C}^{\beta\gamma} = \delta^{\gamma}_{\alpha}
\end{eqnarray}
Notice that the identity matrix has mixed indices.

(6) Moreover, since taking the complex conjugate of $C^{-1} = C^{\dag}$ (due to the unitarity of $C$ (i.e. $CC^{\dag}=C^TC^*=1$)) which has lower indices produces the $C^T$ which has upper indices, we obtain the fact that 
\begin{eqnarray}
 \mathcal{C}^{\alpha\beta}=(\mathcal{C}_{\alpha\beta})^* ~\&~ \mathcal{C}_{\alpha\beta}=(\mathcal{C}^{\alpha\beta})^*.
\end{eqnarray}
Notice that the complex conjugate operation on the charge conjugation matrix changes the position of indices between upper and lower ones.

(7) The (anti-)symmetry property of the metric is given as 
\begin{eqnarray}
 (C)^{\alpha\beta} = [(C^T)^T]^{\alpha\beta} = (C^T)^{\beta\alpha} = \mathcal{C}^{\beta\alpha} \overset{!}{=} -t_0 \mathcal{C}^{\alpha\beta} \implies \therefore ~\mathcal{C}^{\alpha\beta} = -t_0\mathcal{C}^{\beta\alpha} ~~\&~~\mathcal{C}_{\alpha\beta} = -t_0\mathcal{C}_{\beta\alpha}
\end{eqnarray}

\item Gamma matrix indices:

(1) Gamma matrix has mixed indices: the spinor-bilinear scalar is represented in index by
\begin{eqnarray}
 \bar{\chi}\gamma_{\mu}\lambda = \chi^{\alpha} (\gamma_{\mu})_{\alpha}^{~\beta}\lambda_{\beta}.
\end{eqnarray}
(2) But, one can also raise or lower the indices of gamma matrix using the metric as follows:
\begin{eqnarray}
 (\gamma_{\mu})_{\alpha\beta} = (\gamma_{\mu})_{\alpha}^{~\gamma}\mathcal{C}_{\gamma\beta}.
\end{eqnarray}
 
(3) These Gamma matrices with indices at the ``same level'' (i.e. fully-upper or fully-lower indices) have a definite (anti-)symmetry property: 
\begin{eqnarray}
 (\Gamma^{(r)})^{\alpha\beta} = -t_r  (\Gamma^{(r)})^{\beta\alpha} \quad \& \quad  (\Gamma^{(r)})_{\alpha\beta} = -t_r  (\Gamma^{(r)})_{\beta\alpha}.
\end{eqnarray}
proof)
\begin{eqnarray}
 && (C\Gamma^{(r)})^T = -t_r C\Gamma^{(r)} \implies  [(C\Gamma^{(r)})^T]^{\alpha\beta} = -t_r [C\Gamma^{(r)}]^{\alpha\beta}  \implies  [C\Gamma^{(r)}]^{\beta\alpha} =  -t_r [C\Gamma^{(r)}]^{\alpha\beta} \nonumber\\
 && \implies (C)^{\beta\gamma}(\Gamma^{(r)})_{\gamma}^{~\alpha} = -t_r (C)^{\alpha\gamma}(\Gamma^{(r)})_{\gamma}^{~\beta} \implies (-t_0\mathcal{C}^{\beta\gamma}) (\Gamma^{(r)})_{\gamma}^{~\alpha}  = -t_r (-t_0\mathcal{C}^{\alpha\gamma}) (\Gamma^{(r)})_{\gamma}^{~\beta} \nonumber\\
 && \implies \therefore (\Gamma^{(r)})^{\alpha\beta} = -t_r  (\Gamma^{(r)})^{\beta\alpha} \quad \& \quad  (\Gamma^{(r)})_{\alpha\beta} = -t_r  (\Gamma^{(r)})_{\beta\alpha} \qquad \blacksquare
\end{eqnarray}
For example,
\begin{eqnarray}
 (\gamma_{\mu_1\cdots \mu_r})_{\alpha\beta} = -t_r (\gamma_{\mu_1\cdots \mu_r})_{\beta\alpha}.
\end{eqnarray}

\item Rasing or lowering a contracted index produces ``$-t_0$'' factor:
\begin{eqnarray}
 \lambda^{\alpha}\chi_{\alpha} = -t_0\lambda_{\alpha}\chi^{\alpha}.
\end{eqnarray}
proof)
\begin{eqnarray}
  \lambda^{\alpha}\chi_{\alpha} = (\mathcal{C}^{\alpha\gamma}\lambda_{\gamma})( \chi^{\sigma}\mathcal{C}_{\sigma\alpha}) = \lambda_{\gamma}\chi^{\sigma} \mathcal{C}^{\alpha\gamma}(\mathcal{C}_{\sigma\alpha})=\lambda_{\gamma}\chi^{\sigma} \mathcal{C}^{\alpha\gamma}(-t_0\mathcal{C}_{\alpha\sigma})=-t_0 \lambda_{\gamma}\chi^{\sigma} \delta^{\gamma}_{~\sigma} = -t_0\lambda_{\alpha}\chi^{\alpha} \quad \blacksquare~
\end{eqnarray}
Notice that $(\textrm{Back-Slash contraction}) = -t_0 (\textrm{Slash contraction})$.
\end{itemize}

\subsection{Fierz rearrangement}

In this section, we consider even dimensions $D=2m$. The following arguments can also be applied to odd dimensions $D=2m+1$ if the sum over $A$ is restricted to rank $r_A \leq m$ of the Gamma matrix basis $\Gamma^A$. ``Fierz rearrangement'' is a procedure of changing the pairing of spinors in products of spinor bilinears. To figure out this, let us recall the results of Gamma-matrix expansion from Eqs.~\eqref{trace_ortho},\eqref{gam_expansion}, and \eqref{gam_expansion_coeff}:
\begin{eqnarray}
&& \textrm{Trace orthogonality:~~~}  \textrm{Tr}[\Gamma^A \Gamma_B] = 2^m \delta^A_B,\nonumber\\
&& \textrm{Gamma-matrix expansion:~~~}  M = \sum_A m_A \Gamma^A = \sum_{k=0}^{[D]}\frac{1}{k!}m_{\mu_k\ldots \mu_1}\gamma^{\mu_1\ldots \mu_k}, \nonumber\\ 
&& \textrm{Expansion coefficient:~~~}  m_A = \frac{1}{2^m} \textrm{Tr}[M\Gamma_A] \implies m_{\mu_k\ldots \mu_1}=\frac{1}{2^m} \textrm{Tr}[M\gamma_{\mu_k\ldots \mu_1}] ,\nonumber
\end{eqnarray}
where $[D]=D$ for even $D$, while $[D]=(D-1)/2$ for odd $D$. {\it The key idea of Fierz rearrangement is that instead of $M,m_A$ with no explicit indices on them, one may consider any matrix $(M)_{\alpha}^{~~\delta}$ and its corresponding expansion coefficient $(m_A)_{\alpha}^{~~\delta}$ with spectator indices $\alpha,\delta$}. Then, the indexed expansion is written by
\begin{eqnarray}
M_{\alpha}^{~~\delta}= \sum_A (m_A)_{\alpha}^{~~\delta} \Gamma^A \implies (M_{\alpha}^{~~\delta})^{~~\beta}_{\gamma} = \sum_A (m_A)_{\alpha}^{~~\delta} (\Gamma^A)_{\gamma}^{~~\beta},
\end{eqnarray} 
and the indexed expansion coefficient is given by
\begin{eqnarray}
(m_A)_{\alpha}^{~~\delta} =\frac{1}{2^m} (\textrm{Tr}[(M_{\alpha}^{~~\delta})(\Gamma_A)])=\frac{1}{2^m}((M_{\alpha}^{~~\delta})^{~~\beta}_{\gamma}(\Gamma_A)_{\beta}^{~~\gamma}).
\end{eqnarray}
We are now in a position to find some useful Fierz identities. As the most simple example, let us find the basic Fierz identity. First, we specify $(M_{\alpha}^{~~\delta})^{~~\beta}_{\gamma}$ in the following way
\begin{eqnarray}
(M_{\alpha}^{~~\delta})^{~~\beta}_{\gamma} \equiv \delta_{\alpha}^{\beta}\delta_{\gamma}^{\delta}.
\end{eqnarray}
Then, by inserting this into the expansion coefficients formula, we get
\begin{eqnarray}
(m_A)_{\alpha}^{~~\delta}= \frac{1}{2^m} \delta_{\alpha}^{\beta}\delta_{\gamma}^{\delta} (\Gamma_A)_{\beta}^{~~\gamma} =  \frac{1}{2^m} (\Gamma_A)_{\alpha}^{~~\delta}. 
\end{eqnarray}
Lastly, by putting this back to the expansion formula, we obtain the basic Fierz identity 
\begin{eqnarray}
\delta_{\alpha}^{\beta}\delta_{\gamma}^{\delta} = \frac{1}{2^m} \sum_A (\Gamma_A)_{\alpha}^{~~\delta}(\Gamma^A)_{\gamma}^{~~\beta}.
\end{eqnarray}
For example, by applying $\lambda_3^{\alpha}\lambda_{4\beta}\lambda_1^{\gamma}\lambda_{2\delta}= -\lambda_3^{\alpha}\lambda_{2\delta}\lambda_1^{\gamma}\lambda_{4\beta}$ to the above basic Fierz identity, we can immediately find
\begin{eqnarray}
(\bar{\lambda}_1\lambda_2)(\bar{\lambda}_3\lambda_4) = - \frac{1}{2^m} \sum_A (\bar{\lambda}_1 \Gamma^A \lambda_4 )(\bar{\lambda}_3\Gamma_A\lambda_2)
\end{eqnarray}
for any set of four anticommuting spinors. Hence, following this logic, one may obtain various Fierz rearrangements. Other Fierz identity can be given by
\begin{eqnarray}
(\gamma^{\mu})_{\alpha}^{~~\beta}(\gamma_{\mu})_{\gamma}^{~~\delta} = \frac{1}{2^m}\sum_A (-1)^{r_A}(D-2r_A) (\Gamma_A)_{\alpha}^{~~\delta}(\Gamma^A)_{\gamma}^{~~\beta},\label{special_Fierz}
\end{eqnarray}
where $r_A$ is a tensor rank of the Gamma matrix basis $\Gamma_A$. Especially, let us consider the case of $(\gamma^{\mu})_{\alpha\beta}(\gamma_{\mu})_{\gamma\delta}$ and completely symmetric parts in $(\beta\gamma\delta)$. Then, the left-hand side of Eq.~\eqref{special_Fierz} can be non-vanishing only for dimensions with $t_1=-1$. However, its right-hand side can be given only by rank-1 gamma matrices when $D=3,4$. When $D=3,4$, Eq.~\eqref{special_Fierz} implies a cyclic identity 
\begin{eqnarray}
(\gamma_{\mu})_{\alpha(\beta}(\gamma^{\mu})_{\gamma\delta)}=0
\end{eqnarray}
which can hold for $D=2,10$ when contracted with chiral spinors, and thus gives rise to 
\begin{eqnarray}
\gamma_{\mu}\lambda_{[1}\bar{\lambda}_2\gamma^{\mu}\lambda_{3]}=0.
\end{eqnarray}
Notice that this can be obtained by multiplying $\lambda_1^{\beta},\lambda_2^{\gamma},\lambda_3^{\delta}$ to the cyclic identity. Another Fierz identities can be found as follows:
\begin{eqnarray}
(\bar{\lambda}_1 M \lambda_2 )(\bar{\lambda}_3N\lambda_4) = -\frac{1}{2^m}\sum_A (\bar{\lambda}_1 \Gamma_A N \lambda_4)(\bar{\lambda}_3\Gamma^A M\lambda_2),
\end{eqnarray}
and for $D=4$ we have
\begin{eqnarray}
&& P_L\chi \bar{\lambda}P_L = -\frac{1}{2}P_L(\bar{\lambda}P_L\chi) + \frac{1}{8}P_L\gamma^{\mu\nu}(\bar{\lambda}\gamma_{\mu\nu}P_L\chi),\\
&& P_L\chi \bar{\lambda}P_R = -\frac{1}{2}P_L\gamma^{\mu} (\bar{\lambda}\gamma_{\mu}P_L\chi),
\end{eqnarray}
and also for $D=5$, we have
\begin{eqnarray}
\chi\bar{\lambda}-\lambda\bar{\chi} = \gamma_{\mu\nu} (\bar{\lambda}\gamma^{\mu\nu}\chi).
\end{eqnarray}

\subsection{Reality}
\begin{itemize}
\item Charge and Complex conjugations of spinors and matrices:

(1) Complex conjugation of any ``Scalar'' is equal to the charge conjugation of that scalar; i.e. $(S)^C = (S)^*$ for some scalar $S$. 

(2) Complex conjugation of any ``Grassmann number'': for any two Grassmann numbers $z$ and $w$, the complex conjugate of $(zw)$ is determined by their Grassmann parities $|z|$ and $|w|$ as follows:
\begin{eqnarray}
   (zw)^* = w^*z^* = (-1)^{|z||w|} z^*w^*,
\end{eqnarray}
where $|z|=0$ if $z$ is commuting (bosonic), and $|z|=1$ if $z$ is anti-commuting (fermionic).

(3) For any spinor $\lambda$, its charge conjugate is defined by
\begin{eqnarray}
  \lambda^C \equiv B^{-1}\lambda^*
\end{eqnarray}
and using $B=it_0C\gamma^0$, $\bar{\lambda}\equiv \lambda^TC$,  $B^*B=-t_1\mathbf{1}$ and $(\gamma^0)^2=-1$, the barred charge conjugate spinor is then determined as
\begin{eqnarray}
    \overline{\lambda^C} = (-t_0t_1) \underbrace{i\lambda^{\dag} \gamma^0}_{\textrm{Dirac conj.}} = (\bar{\lambda})^C
\end{eqnarray}
in which we find the relation between Dirac and Majorana conjugations as
\begin{eqnarray}
(\textrm{Dirac conj. of}~\lambda) \equiv i\lambda^{\dag} \gamma^0 = (-t_0t_1) (\textrm{Majorana conj. of}~\lambda^C ) = (-t_0t_1) \overline{\lambda^C}.
\end{eqnarray}
proof)
First, let us find the left-hand side.
\begin{eqnarray}
  \overline{\lambda^C} &=& (\lambda^C)^TC =  (B^{-1}\lambda^*)^TC = \lambda^{\dag}(B^{-1})^TC = \lambda^{\dag}B^*C = -t_1 \lambda^{\dag}B^{-1}C \nonumber\\
  &=& -t_1\lambda^{\dag} (it_0(\gamma^0)^{-1}C^{-1})C = (-t_0t_1)i\lambda^{\dag} \gamma^0   \qquad \blacksquare
\end{eqnarray}
Next, let us prove the right-hand side.
\begin{eqnarray}
  (\bar{\lambda}\chi)^C &=& (\bar{\lambda}\chi)^* = -\bar{\lambda}^*\chi^* = -(\lambda^TC)^* B\chi^C = -\lambda^{\dag} C^* B \chi^C = -\lambda^{\dag} (-t_0C^{-1}) (it_0C\gamma^0)\chi^C \nonumber\\
  &=& i\lambda^{\dag}\gamma^0 \chi^C \overset{!}{=} (-t_0t_1) (\bar{\lambda})^C(\chi)^C \implies (\bar{\lambda})^C = (-t_0t_1)i\lambda^{\dag}\gamma^0 \overset{!}{=} \overline{\lambda^C} \quad \blacksquare
\end{eqnarray}

{\bf Note that $(-t_0t_1) = +1$ in 2, 3, 4, 10, or 11 Dimensions.}

(4) For any $2^m$-by-$2^m$ matrices $M$ and $N$, its charge conjugate is 
\begin{eqnarray}
  M^C \equiv B^{-1}M^* B , \qquad (MN)^C = M^CN^C.
\end{eqnarray}
- In fact, the matrices $M$ are usually given by the generating Gamma matrices. 
\begin{eqnarray}
  (\gamma_{\mu})^C = B^{-1} \gamma_{\mu}^* B = (-t_0t_1) \gamma_{\mu}
\end{eqnarray}
proof)
\begin{eqnarray}
   (\gamma_{\mu})^C = B^{-1} \gamma_{\mu}^* B = B^{-1} (-t_0t_1 B\gamma_{\mu} B^{-1}) B =  (-t_0t_1)\gamma_{\mu} \qquad \blacksquare
\end{eqnarray}
- For the highest gamma matrix $\gamma_* \equiv (-i)^{m+1} \gamma_0 \gamma_1 \cdots \gamma_{D-1}$, 
\begin{eqnarray}
   (\gamma_*)^C = (-1)^{D/2+1}\gamma_*
\end{eqnarray}
proof)
\begin{eqnarray}
  (\gamma_*)^C &=& [ (-i)^{m+1} \gamma_0 \gamma_1 \cdots \gamma_{D-1}]^C = [(-i)^{m+1}]^C \gamma_0^C \gamma_1^C \cdots \gamma_{D-1}^C \nonumber\\
  &=& (i)^{m+1}(-t_0t_1)^D \gamma_0 \gamma_1 \cdots \gamma_{D-1} = \frac{(-i)^{m+1}}{(-i)^{m+1}}(i)^{m+1}(-t_0t_1)^D \gamma_0 \gamma_1 \cdots \gamma_{D-1} \nonumber\\
  &=&  \frac{(i)^{m+1}}{(-i)^{m+1}}(-t_0t_1)^D \gamma_* = (-1)^{m+1}(-t_0t_1)^D \gamma_* = (-1)^{D/2+1}(-t_0t_1)^{2m} \gamma_* \nonumber\\
  &=& (-1)^{D/2+1} \gamma_* \qquad \blacksquare
\end{eqnarray}

(5) Complex(or Charge) conjugation of spinor-bilinear scalar:
\begin{eqnarray}
  (\bar{\chi}M\lambda)^* \equiv (\bar{\chi}M\lambda)^C = (-t_0t_1) \overline{\chi^C}M^C \lambda^C.
\end{eqnarray}
proof) using $\bar{\chi} = \chi^TC$, $C^*=-t_0C^{-1}$, $\lambda^C=B^{-1}\lambda^*$, and $M^C = B^{-1}M^*B$, we have
\begin{eqnarray}
   (\bar{\chi}M\lambda)^* &=& (-1)^{|\bar{\chi}||\lambda|}\bar{\chi}^*M^*\lambda^*=-\bar{\chi}^*M^*\lambda^* = -(\chi^TC)^*(BM^CB^{-1})(B\lambda^C)\nonumber\\
   &=&  - \chi^{\dag} (-t_0C^{-1}) (BM^CB^{-1})B\lambda^C = t_0 \chi^{\dag} C^{-1}BM^C\lambda^C
\end{eqnarray}
Since $B = it_0C\gamma^0$, we get
\begin{eqnarray}
   (\bar{\chi}M\lambda)^* &=&  t_0 \chi^{\dag} C^{-1}(it_0C\gamma^0)M^C\lambda^C
   = (i\chi^{\dag}\gamma^0) M^C\lambda^C = (-t_0t_1)\overline{\chi^C} M^C\lambda^C \qquad \blacksquare
\end{eqnarray}

(6) For any spinor $\lambda$, 
\begin{eqnarray}
   (\lambda^C)^C = - t_1 \lambda
\end{eqnarray}
proof)
\begin{eqnarray}
    (\lambda^C)^C = (B^{-1}\lambda^*)^C = B^{-1} (B^{-1}\lambda^*)^* = B^{-1} {B^{-1}}^* \lambda = (B^*B)^{-1}\lambda =( -t_1 \mathbf{1})^{-1}\lambda = -t_1 \lambda \quad \blacksquare \quad 
\end{eqnarray}
\end{itemize}

\section{Majorana Spinors}

In even dimension, Dirac fields have $2^m$ complex components. Weyl fields have $2^{m-1}$ complex components due to the equations of motion. Majorana fields have also $2^{m-1}$ complex components due to a reality condition.

\subsection{Definition and properties of Majorana spinor}
\begin{itemize}
\item Majorana condition: the reality constraint we may impose is given by
\begin{eqnarray}
  \psi=  \psi^C = B^{-1}\psi^*  \implies \psi^* = B\psi.\label{reality}
\end{eqnarray}
Taking a complex conjugate of $\psi^*$ gives
\begin{eqnarray}
 && (\psi^*)^* = B^*\psi^* = B^* B \psi = -t_1 \psi \overset{!}{=} \psi \implies B^*B = \mathbf{1}  \nonumber\\
 && \implies |B|^2 = \mathbf{1} \implies ~\therefore ~t_1 = -1 \quad \textrm{and}\quad  B = \mathbf{1} \quad \textrm{up to a phase factor $e^{i\theta}$}
\end{eqnarray}
This constraint then leads to
\begin{eqnarray}
&& C = t_0 i\gamma^0, \quad C^T = -t_0 C,\quad C^* = -t_0 C^{-1},\nonumber\\
&& (\gamma^{\mu})^{T} = -t_0 C \gamma^{\mu} C^{-1} = -t_0 \gamma^0\gamma^{\mu}\gamma^0,\quad (\gamma^{\mu})^{C} = (\gamma^{\mu})^{*} = t_0 \gamma^{\mu} \nonumber\\
&& M^C = M^*,\quad \lambda^C = \lambda^* = \lambda
\end{eqnarray}

However, we have to still consider ``$t_0=\pm1$'' in general.

(1) {\bf Majorana (M) spinor} ($t_0=+1,t_1=-1$):\\
\begin{itemize}
    \item $t_0=+1$ holds for the spacetime dimensions $D=2,3,4~\textrm{mod}~8$.    
    \item When $t_0=+1,t_1=-1$, Majorana conjugate is equal to Dirac conjugate.
    \begin{eqnarray}
     \bar{\lambda} \overset{!}{=} \overline{\lambda^C} = i\lambda^{\dag}\gamma^0 \overset{!}{\equiv} \bar{\lambda}_D  \qquad \blacksquare
    \end{eqnarray}
    \item Really real representations of the Gamma matrices are allowed for $D=2,3,4 ~\textrm{mod}~8$. In $D=4$, 
    \begin{eqnarray}
    && \gamma^0 = 
    \begin{pmatrix}
    0 & \mathbb{1} \\
    -\mathbb{1} & 0 \\
    \end{pmatrix} = i\sigma_2 \otimes \mathbb{1}, \qquad
     \gamma^1 = 
    \begin{pmatrix}
     \mathbb{1} & 0 \\
     0 & -\mathbb{1}  \\
    \end{pmatrix} = \sigma_3 \otimes \mathbb{1}, \nonumber\\ && \gamma^2 = 
    \begin{pmatrix}
    0 & \sigma_1 \\
   \sigma_1 & 0 \\
    \end{pmatrix} = \sigma_1 \otimes \sigma_1, \qquad \gamma^3 = 
    \begin{pmatrix}
    0 & \sigma_3 \\
    \sigma_3 & 0 \\
    \end{pmatrix} = \sigma_1 \otimes \sigma_3.
    \end{eqnarray}
    In this really real representation, all the gamma matrices are real-valued. Plus, the spatial gamma matrices $\gamma^i$ are symmetric, while the time-direction gamma matrix $\gamma^0$ is anti-symmetric. In particular, in this representation, 
    \begin{eqnarray}
    && B = \mathbf{1}, \quad C = i\gamma^0, \quad \psi^C=\psi^* = \psi, \quad M^C = M^*, \quad (\gamma_{\mu})^C =\gamma_{\mu}^* =\gamma_{\mu} \nonumber\\
    && (\gamma_{\mu})^T = \gamma^0\gamma_{\mu}\gamma^0,\qquad    (\bar{\chi} \gamma_{\mu_1\cdots\mu_r}\psi)^* = (\bar{\chi} \gamma_{\mu_1\cdots\mu_r}\psi)^C = \overline{\chi^C} (\gamma_{\mu_1\cdots\mu_r})^C \psi^C = \bar{\chi} \gamma_{\mu_1\cdots\mu_r} \psi. \nonumber\\
    &&{}
    \end{eqnarray}
\end{itemize}
    
(2) {\bf pseudo-Majorana spinor} ($t_0=-1,t_1=-1$): $t_0=-1$ holds for the spacetime dimensiosn $D=8,9$. In this case, we have $(\gamma^{\mu})^*=-\gamma^{\mu}$. Thus, the reality properties of bilinears by pseudo-Majorana spinor are different from those of bilinears by Majorana spinor. In spite of this distinction, it is common not to distinguish between pseudo-Majorana and Majorana spinors since the core property of reality of Majorana spinor still holds.  
~\\
~~\\
(3) {\bf Majorana-Weyl (MW) spinor}: When $D=2~\textrm{mod}~8$ (i.e. $D=2,10$), we have $(\gamma_*\psi)^C=\gamma_*\psi$, which gives rise to 
\begin{eqnarray}
\textrm{Majorana:~} \psi^C = \psi \overset{\textrm{compatible!}}{\&} \textrm{Weyl:~} P_{L,R}\psi = \psi \implies (P_{L,R}\psi)^C = P_{L,R}\psi.\label{MW}
\end{eqnarray}
Each chiral projection of Majorana spinor (i.e. $P_{L,R}\psi$) satisfying both Majorana and Weyl conditions is called Majorana-Weyl spinor, which has $2^{m-1}$ ``real'' components. On the other hand, when $D=4~\textrm{mod}~4$, we have 
\begin{eqnarray}
(P_{L,R}\psi)^C = P_{R,L}\psi.
\end{eqnarray}
Notice that this property of Majorana-Weyl spinor is completely different from that of Weyl spinors $P_{L,R}\psi_D$ from Dirac spinor $\psi_D$ since $\psi_D$ does not satisfy the reality conditon of Majorana spinor! 
~\\
~~\\
(4) {\bf Symplectic-Majorana (S) spinor}: When $t_1=1$, it is not possible to define a Majorana spinor. However, we can define a so-called Symplectic-Majorana spinor, which follows modified reality condition given by
\begin{eqnarray}
\chi^i = \varepsilon^{ij}(\chi^j)^C = \varepsilon^{ij} B^{-1}(\chi^j)^*,
\end{eqnarray}
where $\varepsilon^{ij}$ is a non-singular antisymmetric matrix such that $\varepsilon^{ij}\varepsilon_{kj}=\delta_k^i$. In particular, in dimensions $D=6~\textrm{mod}~8$, using $(\gamma_*)^C = (-)^{D/2+1}\gamma_*$, it is possible to show that the symplectic Majorana condition is compatible with chirality, and thus we have 
\begin{eqnarray}
P_L\chi^i = \varepsilon^{ij}(P_L\chi^j)^C = \varepsilon^{ij} B^{-1}(P_L\chi^j)^*,
\end{eqnarray}
which is called {\bf symplectic-Majorana-Weyl (SW) spinor}.
\end{itemize}

\section{Majorana Spinors in Physical Theories}

\subsection{Variation of a Majorana lagrangian}

Consider a Majorana spinor field $\Psi$ in dimensions $D=2,3,4~\textrm{mod}~8$. While Majorana and Dirac spinors transform in the same way under Lorentz transformation, since Majorana fermions have half of degrees of freedom of Dirac fermion, the Majorana action is given by
\begin{eqnarray}
S[\Psi] = -\frac{1}{2} \int d^Dx \bar{\Psi}(x)[\gamma^{\mu}\partial_{\mu}-m]\Psi(x).
\end{eqnarray}
Regarding this Majorana action, there is an interesting remark. If the field components of Majorana spinor $\Psi$ are given by conventioanal commuting variables, then the action vanishes because $m\bar{\Psi}(x)\Psi(x)=m\Psi^TC\Psi=0$ where $C$ is antisymmetric, and because $\bar{\Psi}(x)\gamma^{\mu}\partial_{\mu}\Psi(x)=\Psi^T(x)C\gamma^{\mu}\partial_{\mu}\Psi(x) \sim \partial_{\mu} [\Psi^T(x)C\gamma^{\mu}\Psi(x)]$ where $C\gamma^{\mu}$ is symmetric. Hence, when the field components are given by commuting variables, it is not able to have any dynamics from such Majorana action $S[\Psi]$. This means that we have to consider ``anti-commuting Grassmann variables'' as the field components of Majorana spinor! Then, holding this, after Majorana flip and partial integration, we find the equation of motion for Majorana spinor as
\begin{eqnarray}
\delta S [\Psi] = -\int d^D x \delta \bar{\Psi} [\gamma^{\mu}\partial_{\mu}-m]\Psi = 0 \implies [\gamma^{\mu}\partial_{\mu}-m]\Psi = 0,
\end{eqnarray}
which means that Majorana field satisfies the conventional Dirac equation of motion.

\subsection{Relation of Majorana and Weyl spinor theories}

In even dimensions $D=0,2,4~\textrm{mod}~8$, both Majorana and Weyl fields can be present. When $D=4$, the action can be written as
\begin{eqnarray}
S[\Psi] &=& -\frac{1}{2} \int d^4x \bar{\Psi}(x)[\gamma^{\mu}\partial_{\mu}-m]\Psi(x)= -\frac{1}{2} \int d^4x \bar{\Psi}(x)[\gamma^{\mu}\partial_{\mu}-m](P_L+P_R)\Psi(x) \nonumber\\
&=& -\int d^4x \Big(\bar{\Psi}\gamma^{\mu}\partial_{\mu} P_L\Psi -\frac{1}{2}m\bar{\Psi}P_L\Psi -\frac{1}{2}m\bar{\Psi}P_R\Psi\Big),
\end{eqnarray}
which gives the equations of motion
\begin{eqnarray}
\cancel{\partial}P_L\Psi = mP_R\Psi, \quad \cancel{\partial}P_R\Psi = mP_L\Psi \implies \square P_{L,R}\Psi = m^2P_{L,R}\Psi =0. 
\end{eqnarray}
In particular, from the Majorana condition $\Psi = B^{-1}\Psi^* =\gamma^0\gamma^1\gamma^3\Psi^*$, we find that 
\begin{eqnarray}
\Psi  = \begin{pmatrix}
\phi_1 \\
\phi_2 \\
\phi_2^* \\
-\phi_1^* \\
\end{pmatrix}= \begin{pmatrix}
\psi \\
\tilde{\psi} \\
\end{pmatrix} = \begin{pmatrix}
\psi \\
0 \\
\end{pmatrix} + \begin{pmatrix}
0 \\
\tilde{\psi}  \\
\end{pmatrix} \equiv P_L\Psi + P_R \Psi.
\end{eqnarray}
The equations of motion for two-component Weyl fields $\psi,\tilde{\psi}$ are given by
\begin{eqnarray}
\bar{\sigma}^{\mu}\partial_{\mu}\psi = m\tilde{\psi},\quad \sigma^{\mu}\partial_{\mu}\tilde{\psi} =m\psi.
\end{eqnarray}

\subsection{U(1) symmetries of a Majorana field}

It is easy to see that the $U(1)$ symmetry given by $\Psi'=e^{i\theta}\Psi$ cannot be compatible with Majorana condition. On the contrary, the axial transformation or chiral $U(1)$ symmetry given by $\Psi'=e^{i\gamma_*\theta}\Psi$ can be compatible with Majorana condition thanks to $(i\gamma_*)^C=i\gamma_*$. However, the corresponding Majorana action can be invariant only if the Majorana field is massless because the variation of the action under the chiral $U(1)$ symmetry is given by 
\begin{eqnarray}
\delta S[\Psi] = i\theta m \int d^4x \bar{\Psi} \gamma_*\Psi ,
\end{eqnarray}
which can vanish only for a massless Majorana spinor.

%% file: chapters/SCC.tex
The purpose of {\it Tensor calculus} of supergravity is to construct models in the way that the symmetry algebras are ensured to be closed. There are two representative tensor calculus, which are {\it super-Poincar\'{e} tensor calculus} and {\it superconformal tensor calculus}.

In tensor calculus, there are three steps. One is to construct multiplets as representations of symmetry algebras like super-Poincar\'{e} or superconformal ones. The second is to make new (composite) multiplet in terms of the fields of the other multiplet. The last is to use a density formula which gives invariant actions as a function of fields of a multiplet on the composite multiplet built in the second step. 

Then, one may ask: which tensor calculus is better? Of course, the two ways of tensor calculus have their own benefits. Nevertheless, I would like to point out that superconformal tensor calculus has more advantages compared to super-Poincar\'{e} tensor calculus from the user-friendly perspective. Superconformal tensor calculus has {\it simplicity} in that one does not have to specify which set of auxiliary fields at the beginning. To do this, one can start with a larger symmetry group which keep expression simpler, which means that superconformal tensor calculus has {\it generality} in that respect. The tensor calculus has {\it flexibility} since one does not have to do laborious rescalings of fields, which must be done in the final calculations of the super-Poincar\'{e} tensor calculus. 

\section{Superconformal Algebras}

In this section, we specify superconformal algebras, and we find their representations as the covariant quantities of the gauge theory of superconformal symmetry. Then, we construct the invariant action of representations under the superconformal transformations.

Following the Gauge Equivalence Program, we first define the superconformal symmetry transformation $\delta_{sc}$ as ``extra'' symmetries in the way
\begin{eqnarray}
\delta_{sc} \equiv \xi^a P_a + \epsilon^{\alpha}Q_{\alpha} + \frac{1}{2}\lambda^{ab}M_{ab}+\lambda_D D + \theta A + \eta^{\alpha}S_{\alpha} + \lambda_K^a k_a,
\end{eqnarray}
where symmetry generators and their transformation parameters are given as follows: $P_a$ is spacetime translation with $\xi^a$; $Q_{\alpha}$ is supersymmetry with $\epsilon^{\alpha}$ (a.k.a. $Q$-SUSY); $M_{ab}$ is local Lorentz symmetry with $\lambda^{ab}$; $D$ is dilatation with $\lambda_D$; $A$ is chiral U(1) symmetry with $\theta$; $S_{\alpha}$ is conformal supersymmetry with $\eta^{\alpha}$ (a.k.a. $S$-SUSY), and $K_a$ is special conformal symmetry with $\lambda_K^a$. Then, these generators form the ``17'' non-trivial commutation relations as follows:
\begin{eqnarray}
&&1.~[P_a,M_{bc}]=2\eta_{a[b}P_{c]} \implies f_{PM}^{~~~ P} = f_{a[bc]}^{~~~~~~ d} = 2\eta_{a[b}\delta_{c]}^d, \label{PM}\\
&&2.~ [P_a,D] = -P_a \implies f_{PD}^{~~~ P} = f_a^{~ b} = -\delta_a^b,\label{PD}\\
&&3.~ [P_a,S_{\alpha}] = (\gamma_aQ)_{\alpha} \implies f_{PS}^{~~~ Q} = f_{a\alpha}^{~~~ \beta} = (\gamma_a)_{\alpha}^{~~\beta},\label{PS}\\
&&4.~ \{Q_{\alpha},Q_{\beta}\} = -\frac{1}{2}(\gamma^a)_{\alpha}^{~~\gamma}C^{-1}_{\gamma\beta} P_a \implies f_{QQ}^{~~~P} = f_{\alpha\beta}^{~~~ a} = -\frac{1}{2}(\gamma^a)_{\alpha\beta},\label{QQ}\\
&&5.~ [Q_{\alpha},M_{ab}] = \frac{1}{2}(\gamma_{ab}Q)_{\alpha} \implies f_{QM}^{~~~Q} = f_{\alpha [ab]}^{~~~~~\beta} = \frac{1}{2}(\gamma_{ab})_{\alpha}^{~~\beta},\label{QM}\\
&&6.~ [Q_{\alpha},D]=-\frac{1}{2}Q_{\alpha} \implies f_{QD}^{~~~Q} = f_{\alpha \bullet}^{~~\beta} = -\frac{1}{2}\delta_{\alpha}^{\beta},\label{QD}\\
&&7.~ [Q_{\alpha},A]=\frac{3}{2}i(\gamma_*Q)_{\alpha}  \implies f_{QA}^{~~~Q} = f_{\alpha \bullet}^{~~\beta}= \frac{3}{2}i(\gamma_*)_{\alpha}^{~~\beta},\label{QA}\\
&&8.~\{Q_{\alpha},S_{\beta}\} = -\frac{1}{2}C^{-1}_{\alpha\beta}D -\frac{1}{4}(\gamma^{ab})_{\alpha\beta}M_{ab} +\frac{i}{2}(\gamma_*)_{\alpha\beta}A \nonumber\\
&&\implies f_{QS}^{~~~D}=f_{\alpha\beta}^{~~~\bullet} = -\frac{1}{2}C^{-1}_{\alpha\beta},\quad f_{QS}^{~~~M}=f_{\alpha\beta}^{~~[ab]} = -\frac{1}{2}(\gamma^{ab})_{\alpha\beta},\quad f_{QS}^{~~~A}=f_{\alpha\beta}^{~~~\bullet} = \frac{i}{2}(\gamma_*)_{\alpha\beta},\nonumber\\{}\label{QS}\\
&& 9.~[Q_{\alpha},K_a]=-(\gamma_aS)_{\alpha} \implies f_{QK}^{~~~S} = f_{\alpha a}^{~~\beta} = -(\gamma_a)_{\alpha}^{~~\beta},\label{QK}\\
&& 10.~[M_{ab},M_{cd}] = 4\eta_{[a[c}M_{d]b]} \implies f_{MM}^{~~~M} = f_{[ab],[cd]}^{\quad \quad[ef]} = 8\eta_{[c[b}\delta_{a]}^{[e}\delta_{d]}^{f]},\label{MM}\\
&& 11.~[M_{ab},S_{\alpha}]= -\frac{1}{2}(\gamma_{ab}S)_{\alpha} \implies f_{MS}^{~~~S}=f_{[ab],\alpha}^{~~~~~~~\beta}= -\frac{1}{2}(\gamma_{ab})_{\alpha}^{~~\beta},\label{MS}\\
&&12.~[K_a,M_{bc}] = 2\eta_{a[b}K_{c]} \implies f_{KM}^{~~~K} = f_{a,[bc]}^{~~~~~~~d} = 2\eta_{a[b}\delta_{c]}^{d},\label{KM}\\
&&13.~[D,S_{\alpha}] = -\frac{1}{2}S_{\alpha} \implies f_{DS}^{~~~S} = f_{\bullet \alpha}^{~~~\beta} = -\frac{1}{2}\delta_{\alpha}^{\beta},\label{DS}\\
&&14.~[D,K_a] = -K_a \implies f_{DK}^{~~~K} = f_{\bullet a}^{~~b} = -\delta_a^b,\label{DK}\\
&&15.~[A,S_{\alpha}] = \frac{3}{2}i(\gamma_*S)_{\alpha} \implies f_{AS}^{~~~S} = f_{\bullet \alpha}^{~~~\beta} = \frac{3}{2}i(\gamma_*)_{\alpha}^{~~\beta},\label{AS}\\
&&16.~\{S_{\alpha},S_{\beta}\} = -\frac{1}{2}(\gamma^a)_{\alpha}^{~~\gamma}C^{-1}_{\gamma\beta}K_a \implies f_{SS}^{~~~K} = f_{\alpha\beta}^{~~~a} = -\frac{1}{2}(\gamma^a)_{\alpha\beta},\label{SS}\\
&&17.~[P_a,K_b] = 2\eta_{ab}D + 2M_{ab} \implies f_{PK}^{~~~D}=f_{ab}^{~~~\bullet} = 2\eta_{ab}, \quad f_{PK}^{~~~[cd]} = 4\delta_a^{[c}\delta_{b}^{d]},\label{PK}
\end{eqnarray}
where we used the charge conjugation matrix $C_{\alpha\beta}$ such that 
\begin{eqnarray}
&& C^T=-C, \quad C^{\dag}=C^{-1}, \quad C^* = -C^{-1}, \quad (\gamma^{\mu})^T = -C \gamma^{\mu} C^{-1} \\
&& \psi^{\alpha}=(C^T)^{\alpha\beta}\psi_{\beta}, \quad \psi_{\alpha} = \psi^{\beta}(C^{-1})_{\beta\alpha} , \quad \psi^{\alpha}(\gamma^a)_{\alpha}^{~~\beta}\psi_{\beta} = -\psi^{\alpha}(\gamma^a)_{\alpha\beta}\psi^{\beta} \equiv \Bar{\psi}\gamma^a\psi 
\end{eqnarray}
The other commutators of the generators vanish.

\section{Gauge Fields and Curvatures}

We define a set of gauge fields that correspond to the superconformal symmetry. 
\begin{eqnarray}
B_{\mu}^A T_A \equiv e^a_{\mu}P_a + \psi^{\alpha}_{\mu} Q_{\alpha} + \frac{1}{2}\omega_{\mu}^{ab}M_{ab} + b_{\mu}D + A_{\mu}A + \phi^{\alpha}_{\mu} S_{\alpha} + f_{\mu}^a K_a,
\end{eqnarray}
where $e^a_{\mu},\psi^{\alpha}_{\mu},\omega_{\mu}^{ab},b_{\mu},A_{\mu},\phi^{\alpha}_{\mu},f_{\mu}^a$ are gauge fields corresponding to the generators, respectively. Then, it is possible to compute the corresponding curvatures using
\begin{eqnarray}
R_{\mu}^{~~A}(T_A) \equiv \partial_{\mu}B_{\nu}^A - \partial_{\nu}B_{\mu}^A + B_{\nu}^CB_{\mu}^Bf_{BC}^{~~~A}.
\end{eqnarray}
\begin{itemize}
    \item 1. Curvature of {\it Local spacetime translation} ``$R_{\mu\nu}^a(P_a)$'' is calculated by using the structure constants from the commutators \eqref{PM} for PM, \eqref{PD} for PD, \eqref{QQ} for QQ since $P_a$ operator can be only found in the right-hand side of the three commutators, so that
    \begin{eqnarray}
    R_{\mu}^a(P_a) = 2\partial_{[\mu}e^a_{\nu]}+2b_{[\mu}e^a_{\nu]} + 2\omega_{[\mu}^{ab}e_{\nu]b} - \frac{1}{2}\Bar{\psi}_{\mu}\gamma^a\psi_{\nu}.\label{curvature_P}
    \end{eqnarray}
    \item 2. Curvature of {\it Local Q-SUSY} ``$R_{\mu\nu}^{\alpha}(Q_{\alpha})$'' is calculated by using the structure constants \eqref{PS} for PS, \eqref{QM} for QM, \eqref{QD} for QD, and \eqref{QA} for QA since $Q_{\alpha}$ operator can be found only in the right-hand side of the four commutators, so that
    \begin{eqnarray}
    R_{\mu\nu}^{\alpha}(Q_{\alpha}) = 2\partial_{[\mu}\psi^{\alpha}_{\nu]}-2\gamma_{[\mu}\phi^{\alpha}_{\nu]} + \frac{1}{2}\omega_{[\mu}^{ab}(\gamma_{ab}\psi_{\nu]})^{\alpha}+b_{[\mu}\psi_{\nu]}^{\alpha}-3iA_{[\mu}(\gamma_*\psi_{\nu]})^{\alpha}.\label{curvature_Q}
    \end{eqnarray}
    \item 3. Curvature of {\it Local Lorentz transformation} ``$R_{\mu\nu}^{ab}(M)$'' is calculated by using the structure constants from the commutators \eqref{QS} for QS, \eqref{MM} for MM, and \eqref{PK} for PK since $M_{ab}$ operator can be found only in the right-hand side of the three commutators, so that
    \begin{eqnarray}
    R_{\mu\nu}^{ab}(M) = 2\partial_{[\mu}\omega_{\nu]}^{ab} -2\omega_{[\mu c}^a\omega_{\nu]}^{cb}+8f_{[\mu}^{[a}e^{b]}_{\nu]} -\Bar{\psi}_{[\mu}\gamma^{ab}\phi_{\nu]}.\label{curvature_M}
    \end{eqnarray}
    \item 4. Curvature of {\it Local Dilatation} ``$R_{\mu\nu}(D)$'' is calculated by using the structure constants from the commutators \eqref{QS} for QS and \eqref{PK} for PK since $D$ operator can be found only in the right-hand side of the two commutators, so that
    \begin{eqnarray}
    R_{\mu\nu}(D) = 2\partial_{[\mu}b_{\nu]} - \Bar{\psi}_{[\mu}\phi_{\nu]}-4f_{[\mu}^a e_{\nu]a}.\label{curvature_D}
    \end{eqnarray}
    \item 5. Curvature of {\it Local Chiral U(1) symmetry} ``$R_{\mu\nu}(A)$'' is calculated by using the structure constants from the commutator \eqref{QS} for QS since $A$ operator can be found only in the right-hand side of the commutator, so that
    \begin{eqnarray}
    R_{\mu\nu}(A) = 2\partial_{[\mu}A_{\nu]} + i\Bar{\psi}_{[\mu}\gamma_*\phi_{\nu]}.\label{curvature_A}
    \end{eqnarray}
    \item 6. Curvature of {\it Local $S$-SUSY} ``$R_{\mu\nu}^{\alpha}(S_{\alpha})$'' is calculated by using the structure constants from the commutators \eqref{QK} for QK, \eqref{MS} for MS, \eqref{DS} for DS, and \eqref{AS} for AS since $S_{\alpha}$ operator can be found only in the right-hand side of the four commutators, so that
    \begin{eqnarray}
    R_{\mu\nu}^{\alpha}(S_{\alpha}) = 2\partial_{[\mu}\phi_{\nu]}^{\alpha} -b_{[\mu}\phi_{\nu]}^{\alpha} -3iA_{[\mu}(\gamma_*\phi_{\nu]})^{\alpha} + \frac{1}{2}\omega^{ab}_{[\mu}(\gamma_{ab}\phi_{\nu]})^{\alpha} -2f_{[\mu}(\gamma_a\psi_{\nu]})^{\alpha}.
    \label{curvature_S}
    \end{eqnarray}
    \item 7. Curvature of {\it Local Special Conformal symmetry} ``$R_{\mu\nu}^a(K_a)$'' is calculated by using the structure constnats from the commutators \eqref{KM} for KM, \eqref{DK} for DK, and \eqref{SS} for SS since $K_a$ operator can be found only in the right-hand side of the three commutators, so that
    \begin{eqnarray}
    R_{\mu\nu}^a(K_a) = 2\partial_{[\mu}f_{\nu]}^a -2b_{[\mu}f_{\nu]}^a + 2\omega_{[\mu}^{ab}f_{\nu]b} -\frac{1}{2}\Bar{\phi}_{\mu}\gamma^a \phi_{\nu}.\label{curvature_K}
    \end{eqnarray}
\end{itemize} 

\section{Covariant Local Translation under Superconformal Symmetry, and Introduction of Gravitational Curvature}

In this section, we find some constraints on the curvatures since we wish to use the covariant local translation in Eq.~\eqref{reformed_cov_derivative} under the superconformal symmetry. Thus, let us recall the general curvature constraint from Eq.~\eqref{Curvature_constraints}, i.e.
\begin{eqnarray}
\xi^{\nu}\hat{R}_{\nu\mu}^A(T_A) = 0,
\end{eqnarray}
where $T_A$'s are given by the non-covariant local translation $P_a$ and the other standard gauge symmetries. Certainly, for any $A$ in $\hat{R}_{\mu\nu}^A$, the constraint must hold in order to take the replacement $P_a \longrightarrow \hat{P}_a$ in the symmetry algebras. 

Getting back to the superconformal algebras in Eqs. \eqref{PM} to \eqref{PK}, we notice that only the right-hand side of the commutator of QQ in Eq.~\eqref{QQ} includes a single $P_a$. Thus, we need to examine this commutator. Let us apply this to a gauge field $e^a_{\mu}$. Then, we have
\begin{eqnarray}
[\delta_Q(\epsilon_1),\delta_Q(\epsilon_2)]e^{a}_{\mu}&=& [\epsilon_1^{\alpha}Q_{\alpha}\epsilon_2^{\beta}Q_{\beta}-   \epsilon_2^{\beta}Q_{\beta}\epsilon_1^{\alpha}Q_{\alpha}]e^{a}_{\mu} = \epsilon_2^{\beta}\epsilon_1^{\alpha}\{Q_{\alpha},Q_{\beta}\}e^a_{\mu} \nonumber\\
&=& -\frac{1}{2} (\gamma^b)_{\alpha}^{~~~\gamma}C^{-1}_{\gamma\beta}\epsilon_2^{\beta}\epsilon_1^{\alpha}P_b e^a_{\mu} = \frac{1}{2}\epsilon_1^{\alpha} (\gamma^b)_{\alpha}^{~~~\gamma}C^{-1}_{\gamma\beta}\epsilon_2^{\beta}P_b e^a_{\mu}\nonumber\\
&=& -\frac{1}{2}\bar{\epsilon}_1\gamma^b\epsilon_2 P_be^{a}_{\mu} = \frac{1}{2}\bar{\epsilon}_2\gamma^b\epsilon_1 P_be^{a}_{\mu} \equiv \xi^{b}P_be^a_{\mu},
\end{eqnarray}
where $\xi^b \equiv \frac{1}{2}\bar{\epsilon}_2\gamma^b\epsilon_1$. Hence, since the non-gauge field effect $\mathcal{M}_{\mu B}^{~~a}=0$ does not exist for the vielbein, we can impose
\begin{eqnarray}
\xi^{\nu}R^a_{\mu\nu}(P_b)=\frac{1}{2}\bar{\epsilon}_2\gamma^{\nu}\epsilon_1R^a_{\mu\nu}(P_b) = 0 \implies R^a_{\mu\nu}(P_b) = 0,
\end{eqnarray}
which is because $R^a_{\mu\nu}(P)$ is purely bosonic. Regarding this constraint, there is a problem, which is that the constraint is not invariant under the $Q$ supersymmetry. That is, if we obey the original transformation rules under the Q-SUSY, we obtain
\begin{eqnarray}
\delta_Q(\epsilon) R^a_{\mu\nu}(P) = \frac{1}{2}\bar{\epsilon}\gamma^a R_{\mu\nu}(Q) \neq 0.
\end{eqnarray}
The solution to this problem may be to modify the supersymmetry transformation of $\omega_{\mu}^{ab}$. This is because $\omega_{\mu}^{ab}$ is no longer an independent field but a dependent one due to the constraint $R^a_{\mu\nu}(P_b) = 0$. This means that the Q-SUSY transformation law of this dependent field $\omega_{\mu}^{ab}$ must be consistent with the constraint. Hence, we assume that there is an additional Q-SUSY transformation of $\omega_{\mu}^{ab}$ denoted by $\delta_Q'$ for compensating the remaining shift of $\frac{1}{2}\bar{\epsilon}\gamma^a R_{\mu\nu}(Q)$. That is, we have 
\begin{eqnarray}
\delta_Q(\epsilon)R^a_{\mu\nu} = \frac{1}{2}\bar{\epsilon}\gamma^a R_{\mu\nu}(Q) + 2\delta_Q'(\epsilon)\omega_{[\mu}^{ab}e_{\nu]b} \equiv 0,
\end{eqnarray}
which gives 
\begin{eqnarray}
\delta_Q'(\epsilon)\omega_{\mu}^{ab} = -\frac{1}{2}\bar{\epsilon}\gamma^{[a}R^{b]}_{\mu}(Q) -\frac{1}{4}\bar{\epsilon}\gamma_{\mu}R^{ab}(Q),
\end{eqnarray}
which is a special example of Eq.~\eqref{add_shift}. In fact, $\omega_{\mu}^{ab},\phi_{\mu}^{\alpha},f_{\mu}^a$ are dependent fields. This is because the other two constraints are given by
\begin{eqnarray}
e^{\nu}_b \hat{R}_{\mu\nu}(M^{ab})=0, \qquad \gamma^{\mu}R_{\mu\nu}(Q) =0,
\end{eqnarray}
where the first one can produce a solution for the gauge field $f_{\mu}^a$ of the special conformal symmetry $K_a$, while the second one can yield the solution for the gauge field $\phi_{\mu}^{\alpha}$ of the conformal supersymmetry $S_{\alpha}$. The covariant curvature of $M^{ab}$ is given by
\begin{eqnarray}
\hat{R}_{\mu\nu}(M^{ab}) &=& \hat{R}_{\mu\nu}^{~~~ab}(M) + 8f_{[\mu}^{[a}e^{b]}_{\nu]} ,\\
\hat{R}_{\mu\nu}^{~~~ab}(M) &=& r_{\mu\nu}^{~~~ab}(M) + \bar{\psi}_{[\mu}\gamma^{[a}R_{\nu]}^{b]}(Q) + \frac{1}{2}\bar{\psi}_{[\mu}\gamma_{\nu]}R^{ab}(Q),\\
r_{\mu\nu}^{~~~ab}(M) &=& 2\partial_{[\mu}\omega_{\nu]}^{~~~ab} + 2\omega_{\mu~~c}^{~a}\omega_{\nu]}^{~~~cd} -\bar{\psi}_{[\mu}\gamma^{ab}\phi_{\nu]}.
\end{eqnarray}
We can also compute the corresponding Ricci tensor and scalar 
\begin{eqnarray}
\hat{R}(M) = e^{\mu}_a \hat{R}_{\mu}^{~~a}(M), \qquad \hat{R}_{\mu}^{~~a}(M) = \hat{R}_{\mu\nu}^{~~~ab}(M)e_{b}^{\nu}.
\end{eqnarray}
The curvature of $Q$ is 
\begin{eqnarray}
R_{\mu\nu}(Q) &=& R'_{\mu\nu}(Q) - 2\gamma_{[\mu}\phi_{\nu]},\\
R'_{\mu\nu}(Q) &=& 2\Big(
\partial_{[\mu} + \frac{1}{2}b_{[\mu} -\frac{3}{2}iA_{[\mu}\gamma_* +\frac{1}{4}\omega_{[\mu}^{~~~ab}\gamma_{ab}
\Big)\psi_{\nu]}.
\end{eqnarray}
Then, we are now able to solve for the gauge fields $f_{\mu}^a$ and $\phi_{\mu}$, which are found to be
\begin{eqnarray}
f_{\mu}^{~a} &=& -\frac{1}{4}\hat{R}_{\mu}^{~a}(M) + \frac{1}{24}e_{\mu}^a\hat{R}(M), \\
\phi_{\mu} &=& -\frac{1}{2}\gamma^{\nu}R'_{\mu\nu}(Q) +\frac{1}{12}\gamma_{\mu}\gamma^{ab}R'_{ab}(Q),
\end{eqnarray}
which gives rise to
\begin{eqnarray}
f_{\mu}^{~\mu} &=& -\frac{1}{12}\hat{R}(M) = -\frac{1}{12}(R(\omega)-\bar{\psi}_a\gamma^{ab}\phi_b),\\
\gamma^{\mu}\phi_{\mu} &=& -\frac{1}{6} \gamma^{\mu\nu}R'_{\mu\nu}(Q),\\
\gamma^{ab}\phi_b &=& -\frac{1}{4}\gamma^{abc}R'_{bc}(Q).
\end{eqnarray}
Here we note that {\bf gravitational curvature} ``$R(\omega)$'' (i.e. related to the kinetic term of spin-2 graviton for Einstein-Hilbert action) is introduced into the superconformal action of supergravity via the dependent gauge field of the special conformal symmetry, i.e. $f_{\mu}^a$.

\section{Superconformal Multiplets}

\subsection{General complex superconformal multiplet}

Here we construct the representations of the superconformal algebras. First of all, let us recall a supermultiplet $\Phi$. In SUSY, it is possible to construct finite number of its supersymmetric descendant fields by applying the supersymmetry operators $Q,\bar{Q}$ to the supermultiplet $\Phi$, such as $Q\Phi$, $\bar{Q}\Phi$, $Q^2\bar{Q}^2\Phi$, etc. That is, {\it all the supersymmetric descendant fields can appear as the component fields of the supermultiplet $\Phi$}. However, it is not guaranteed for a supermultiplet to be a superconformal multiplet. This is largely because the fields of superconformal multiplet transform under the superconformal symmetry, not just a supersymmetry. For example, let us consider the $S$-SUSY transformation of $Q\Phi$. When we apply $S_{\beta}$ to $Q_{\alpha}\Phi$, we obtain  
\begin{eqnarray}
S_{\beta}(Q_{\alpha}\Phi)=\{S_{\beta},Q_{\alpha}\}\Phi -Q_{\alpha}(S_{\beta}\Phi) = \Big( -\frac{1}{2}C^{-1}_{\alpha\beta}D -\frac{1}{4}(\gamma^{ab})_{\alpha\beta}M_{ab} +\frac{i}{2}(\gamma_*)_{\alpha\beta}A\Big)\Phi - Q_{\alpha}(S_{\beta}\Phi),
\end{eqnarray}
where we used $\{Q,S\} \sim M+D+A$ from the commutator \eqref{QS} for QS, i.e. $\{Q_{\alpha},S_{\beta}\} = -\frac{1}{2}C^{-1}_{\alpha\beta}D -\frac{1}{4}(\gamma^{ab})_{\alpha\beta}M_{ab} +\frac{i}{2}(\gamma_*)_{\alpha\beta}A$. From the above result, we notice that the $S$-transformation of $(Q\Phi)$ is dependent on the $Q$-transformation of a new field $S\Phi$! This is problematic because the field $S\Phi$ is beyond the components of $\Phi$; that is, $S\Phi$ is not a superpartner of $\Phi$! Hence, in this sense, a supermultiplet may be ``not supersymmetric'' in the superconformal algebras.

Interestingly, there is a clever way of defining a superconformal multiplet $\mathcal{V}$ as a supermultiplet. This is to impose the two conditions to the lowest component $\mathcal{C}$ of a superconformal multiplet $\mathcal{V}$:
\begin{eqnarray}
S_{\alpha}\mathcal{C}=0, \qquad K_a\mathcal{C}=0,
\end{eqnarray}
which means that the lowest component $\mathcal{C}$ of a superconformal multiplet $\mathcal{V}$ is {\it inert under both S-SUSY and Special conformal symmetry} of the superconformal symmetry.

Keeping this in mind, let us start with a superconformal multiplet $\mathcal{V}$ whose lowest component $\mathcal{C}$ transforms under 
\begin{eqnarray}
\delta_Q\mathcal{C} = \frac{i}{2}\bar{\epsilon}\gamma_*\mathcal{Z}, \quad \delta_M\mathcal{C}=0, \quad \delta_D=w\lambda_D \mathcal{C} , \quad \delta_AC=ic\theta \mathcal{C}, \quad \delta_S \mathcal{C} =0 , \quad \delta_K \mathcal{C} =0,\label{C_trans_laws}
\end{eqnarray}
where $\delta_I \equiv \epsilon^IT_I$ without summation for $I$, and importantly, we introduce {\it a ``supersymmetric'' descendant field $\mathcal{Z}$ as an arbitrary spinor whose transformation law is determined by the ``superconformal'' algebras}. In addition, note that we have introduced the so-called {\bf Weyl weight} ``$w$'' and {\bf Chiral weight} ``$c$,'' which usually make a pair as $(w,c)$ that is called ``{\bf Weyl/Chiral weights}.'' These weights are very crucial in that they completely characterize a superconformal multiplet $\mathcal{V}$.

Next, let us find the $S$-transformation of $\mathcal{Z}$. From the commutator \eqref{QS} for QS, we have 
\begin{eqnarray}
[\delta_S(\eta),\delta_Q(\epsilon)] &=& -\eta^{\beta}\epsilon^{\alpha}\{S_{\beta},Q_{\alpha}\} 
= \frac{1}{2}\eta^{\beta}\epsilon^{\alpha}C^{-1}_{\alpha\beta}D +\frac{1}{4}\eta^{\beta}\epsilon^{\alpha}(\gamma^{ab})_{\alpha\beta}M_{ab} -\frac{i}{2}\eta^{\beta}\epsilon^{\alpha}(\gamma_*)_{\alpha\beta}A \nonumber\\
&=& \frac{1}{2}\eta^{\beta}\epsilon_{\beta} D -\frac{1}{4}\epsilon^{\alpha}(\gamma^{ab})_{\alpha\beta}\eta^{\beta}M_{ab} +\frac{i}{2}\epsilon^{\alpha}(\gamma_*)_{\alpha\beta}\eta^{\beta}A \nonumber\\
&=& \frac{1}{2}\eta^{\beta}\epsilon_{\beta} D +\frac{1}{4}\epsilon^{\alpha}(\gamma^{ab})_{\alpha}^{~~\beta}\eta_{\beta}M_{ab} -\frac{i}{2}\epsilon^{\alpha}(\gamma_*)_{\alpha}^{~~\beta}\eta_{\beta}A \nonumber\\
&=& \frac{1}{2}\bar{\eta}\epsilon D +\frac{1}{4}\bar{\epsilon}\gamma^{ab}\eta M_{ab} -\frac{i}{2}\bar{\epsilon}\gamma_*\eta A  \nonumber\\
&=& \delta_D\Big(\frac{1}{2}\bar{\eta}\epsilon\Big) + \delta_M\Big(\frac{1}{4}\bar{\epsilon}\gamma^{ab}\eta\Big)  + \delta_A\Big(-\frac{i}{2}\bar{\epsilon}\gamma_*\eta\Big), 
\end{eqnarray}
where we used the properties that $\eta,\epsilon,S,Q$ all are anticommuting with each other, and switching the upper and lower positions of two spinor indices in their contraction produces a minus sign, i.e. $A^{\alpha}B_{\alpha}=-A_{\alpha}B^{\alpha}$. Then, we are ready to consider the following computation for the lowest component $\mathcal{C}$, i.e.
\begin{eqnarray}
\delta_S(\eta)\delta_Q(\epsilon)\mathcal{C} = [\delta_S(\eta),\delta_Q(\epsilon)]\mathcal{C} + \delta_Q(\epsilon)\delta_S(\eta)\mathcal{C}. \label{SQ_transform}
\end{eqnarray}
The left-hand side of Eq.~\eqref{SQ_transform} gives
\begin{eqnarray}
\delta_S(\eta)\delta_Q(\epsilon)\mathcal{C} = \delta_S(\eta)\Big(\frac{i}{2}\bar{\epsilon}\gamma_*\mathcal{Z}\Big)=\frac{i}{2}\bar{\epsilon}\gamma_*[\delta_S(\eta)\mathcal{Z}].
\end{eqnarray}
The right-hand side of Eq.~\eqref{SQ_transform} gives
\begin{eqnarray}
 [\delta_S(\eta),\delta_Q(\epsilon)]\mathcal{C} + \delta_Q(\epsilon)\delta_S(\eta)\mathcal{C}
 &=& \delta_D\Big(\frac{1}{2}\bar{\eta}\epsilon\Big)\mathcal{C} + \delta_M\Big(\frac{1}{4}\bar{\epsilon}\gamma^{ab}\eta\Big)\mathcal{C}  + \delta_A\Big(-\frac{i}{2}\bar{\epsilon}\gamma_*\eta\Big)\mathcal{C} \nonumber\\&&+\delta_Q(\epsilon)\delta_S(\eta)\mathcal{C} \nonumber\\
 &=& \frac{1}{2}\bar{\eta}\epsilon w\mathcal{C} + 0 -ic\frac{i}{2}\bar{\epsilon}\gamma_*\eta\mathcal{C} + 0 =  \frac{1}{2}\bar{\epsilon}\gamma_*^2\eta w\mathcal{C}  -ic\frac{i}{2}\bar{\epsilon}\gamma_*\eta\mathcal{C} \nonumber\\
 &=& \frac{i}{2}\bar{\epsilon}\gamma_* \Big( -iw\gamma_*-ic\Big)\eta \mathcal{C},
\end{eqnarray}
where we used $\gamma_*^2=1$, which defines the projection operators $P_L\equiv \frac{1+\gamma_*}{2}$ and $P_R\equiv \frac{1-\gamma_*}{2}$. By equating these two results, we can determine the $S$-transformation of the new field $\mathcal{Z}$ we introduced in Eq. \eqref{C_trans_laws}:
\begin{eqnarray}
 \delta_S(\eta)\mathcal{Z} = -i(w\gamma_*+c)\eta \mathcal{C}.
\end{eqnarray}
The point here is that {\it transformation laws of a descendant field can be uniquely determined by the transformation laws of the ascendant fields}. In particular, we can also take the decomposition of spinor $\psi = P_L\psi + P_R \psi$ using the projection operators $P_L \equiv (1+\gamma_*)/2,P_R\equiv (1-\gamma_*)/2$ such that $P_L^2=P_R^2=1,~ P_LP_R=P_RP_L=0,~ \gamma_*P_L=P_L,~\gamma_*P_R=-P_R$. Thus, we can also get
\begin{eqnarray}
 \delta_S(\eta)P_L\mathcal{Z} = -i(w+c)P_L\eta \mathcal{C}, \qquad \delta_S(\eta)P_R\mathcal{Z} = i(w-c)P_R\eta \mathcal{C}.
\end{eqnarray}

Another important computation is the $Q$-transformation of $\mathcal{Z}$. Let us consider 
\begin{eqnarray}
 [\delta_Q(\epsilon_1),\delta_Q(\epsilon_2)]\mathcal{C} = -\frac{1}{2}\bar{\epsilon}_1\gamma^a\epsilon_2 \mathcal{D}_a \mathcal{C},
\end{eqnarray}
where $\mathcal{D}_a$ is a relevant covariant derivative. To calculate the left-hand side of this commutator, let us consider the most general form of $\delta_Q\mathcal{Z}$. In the basis of $\gamma$-matrices, we can express $\delta_Q\mathcal{Z}$ as
\begin{eqnarray}
&& \delta_Q\mathcal{Z} = (\Phi^0+\Phi^1\gamma^a+\Phi^2_{ab}\gamma^{ab}+\gamma^a\gamma_*\Phi^3_a+\gamma_*\Phi^4)\epsilon,\\
 && \delta_QP_L\mathcal{Z} =P_L (\Phi^0+\Phi^1\gamma^a+\Phi^2_{ab}\gamma^{ab}-\gamma^a\Phi^3_a+\Phi^4)\epsilon,\\
&& \delta_QP_R\mathcal{Z} =P_R (\Phi^0+\Phi^1\gamma^a+\Phi^2_{ab}\gamma^{ab}+\gamma^a\Phi^3_a-\Phi^4)\epsilon,
\end{eqnarray}
where $\Phi^A$ ($A=0,1,2,3,4$) are bosonic fields with(out) the local Lorentz indices, and we used $P_{L/R}\gamma^a=\gamma^aP_{R/L}$ given by $\{\gamma^a,\gamma_*\}=0$. In fact, the expansion coefficients, $\Phi^I$'s, correspond to all the possible supersymmetric descendant fields of $\mathcal{Z}$. Then, the commutator result can be given by
\begin{eqnarray}
 && [\delta_Q(\epsilon_1),\delta_Q(\epsilon_2)]\mathcal{C}= \frac{i}{2}\bar{\epsilon}_2\gamma_*\delta_Q(\epsilon_1)\mathcal{Z} - (\epsilon_1 \leftrightarrow \epsilon_2) 
 \nonumber\\
 &&= \frac{i}{2}(\bar{\epsilon}_2\gamma_*\epsilon_1\Phi^0 + \bar{\epsilon}_2\gamma_*\gamma^a\epsilon_1\Phi_a^1 + \bar{\epsilon}_2\gamma_*\gamma^{ab}\epsilon_1\Phi_{ab}^2 + \bar{\epsilon}_2\gamma_*\gamma^a\gamma_*\epsilon_1\Phi_a^3 + \bar{\epsilon}_2\epsilon_1\Phi^4) - (\epsilon_1 \leftrightarrow \epsilon_2)\qquad \nonumber\\
 &&=i\bar{\epsilon}_2\gamma_*\gamma^{ab}\epsilon_1\Phi_{ab}^2 + i\bar{\epsilon}_1\gamma^a\epsilon_2\Phi_a^3 \overset{!}{=} -\frac{1}{2}\bar{\epsilon}_1\gamma^a\epsilon_2 \mathcal{D}_a \mathcal{C}
 \nonumber\\
 && \implies \therefore ~ \Phi_{ab}^2 =0 \textrm{~~and~~} \Phi_a^3 \equiv \frac{i}{2} \mathcal{D}_a \mathcal{C}
\end{eqnarray}
The undetermined coefficients $\Phi^0,\Phi_a^1,\Phi^4$ will be considered as the supersymmetric descendant fields of $\mathcal{Z}$. Conventionally, we represent them as (following the notation of FKVW \cite{Linear}\footnote{The Kugo notation $H,K$ can be written in terms of our notation as $H \equiv \frac{1}{2}(\mathcal{H}+\mathcal{K})$ and $K \equiv \frac{i}{2}(\mathcal{K}-\mathcal{H})$.}):
\begin{eqnarray}
 \Phi^0 \equiv -\frac{i}{4}(\mathcal{K}-\mathcal{H}), \quad \Phi_a^1 \equiv -\frac{1}{2}\mathcal{B}_a , \quad \Phi^4 \equiv \frac{i}{4}(\mathcal{H}+\mathcal{K}).
\end{eqnarray}
Therefore, the $Q$-transformation of $\mathcal{Z}$ can be written as
\begin{eqnarray}
&& \delta_Q(\epsilon)\mathcal{Z} = \frac{1}{2}(i\mathcal{H}-\gamma^a\mathcal{B}_a -i\gamma_*\gamma^a\mathcal{D}_a\mathcal{C})\epsilon,\\
&&  \delta_Q(\epsilon)P_L\mathcal{Z} = \frac{1}{2}P_L(i\mathcal{H}-\gamma^a\mathcal{B}_a -i\gamma^a\mathcal{D}_a\mathcal{C})\epsilon,\\
&& \delta_Q(\epsilon)P_R\mathcal{Z} = \frac{1}{2}P_R(-i\mathcal{K}-\gamma^a\mathcal{B}_a +i\gamma^a\mathcal{D}_a\mathcal{C})\epsilon.
\end{eqnarray}

Following this procedure, it is possible to identify all the components of the superconformal multiplet $\mathcal{V}$ and its complex conjugate multiplet $\mathcal{V}^*$, which are usually represented by the following collection
\begin{eqnarray}
&& \mathcal{V}_{(w,c)} \equiv \Big(\mathcal{C},\mathcal{Z},\mathcal{H},\mathcal{K},\mathcal{B}_a,\Lambda,\mathcal{D}\Big),\\
&&  \mathcal{V}_{(w,-c)}^* \equiv \Big(\mathcal{C}^*,\mathcal{Z}^C,\mathcal{K}^*,{\mathcal{H}^*},\mathcal{B}_a,\Lambda^C,\mathcal{D}^*\Big), 
\end{eqnarray}
where $C$ denotes the charge conjugation on the spinor; for the Majorana spinor represention, it becomes the complex conjugate, and $\mathcal{V}_{(w,n)}$ means its lowest component has the Weyl/chiral weights ($w,c$). The transformation laws of all the components of the superconformal multiplet $\mathcal{V}_{(w,c)}$ can be found follows:
\begin{itemize}
    \item $\mathcal{C}~(w,c)$: 
    \begin{eqnarray}
     \delta_Q\mathcal{C} = \frac{i}{2}\bar{\epsilon}\gamma_*\mathcal{Z}, \quad \delta_M\mathcal{C}=0,\quad \delta_D\mathcal{C}=w\lambda_D\mathcal{C},\quad \delta_A\mathcal{C}=in\theta \mathcal{C}, \quad \delta_S\mathcal{C}=0, \quad \delta_K\mathcal{C}=0.
    \end{eqnarray}
    \item $P_L\mathcal{Z}~(w+1/2,c-3/2)$, $P_R\mathcal{Z}~(w+1/2,c+ 3/2)$:
    \begin{eqnarray}
    &&  \delta_Q(\epsilon)P_L\mathcal{Z} = \frac{1}{2}P_L(i\mathcal{H}-\gamma^a\mathcal{B}_a -i\gamma^a\mathcal{D}_a\mathcal{C})\epsilon, \quad \delta_MP_L\mathcal{Z} = -\frac{1}{4}\lambda^{ab}\gamma_{ab}P_L\mathcal{Z}, \nonumber\\
     && \delta_DP_L\mathcal{Z}=(w+1/2)\lambda_DP_L\mathcal{Z}, \quad \delta_AP_L\mathcal{Z} = i(c-3/2)\theta P_L\mathcal{Z}, \nonumber\\
     && \delta_S P_L\mathcal{Z}= -i(w+c)P_L\eta \mathcal{C}, \quad \delta_KP_L \mathcal{Z}=0. ,\\
     &&  \delta_Q(\epsilon)P_R\mathcal{Z} = \frac{1}{2}P_R(-i\mathcal{K}-\gamma^a\mathcal{B}_a +i\gamma^a\mathcal{D}_a\mathcal{C})\epsilon, \quad \delta_MP_R\mathcal{Z} = -\frac{1}{4}\lambda^{ab}\gamma_{ab}P_R\mathcal{Z}, \nonumber\\
     && \delta_DP_R\mathcal{Z}=(w+1/2)\lambda_DP_R\mathcal{Z}, \quad \delta_AP_R\mathcal{Z} = i(c+3/2)\theta P_R\mathcal{Z}, \nonumber\\
     && \delta_S P_R\mathcal{Z}= i(w-c)P_R\eta \mathcal{C}, \quad \delta_KP_R \mathcal{Z}=0. \qquad 
    \end{eqnarray}
    \item $\mathcal{H}~(w+1,c-3)$:
    \begin{eqnarray}
    && \delta_Q\mathcal{H} = -i\bar{\epsilon}P_R(\gamma^a\mathcal{D}_a\mathcal{Z}+\Lambda), \quad \delta_M\mathcal{H} = 0,\quad \delta_D\mathcal{H}=(w+1)\lambda_D\mathcal{H},\quad \delta_A\mathcal{H}=i\theta (c-3)\mathcal{H},\nonumber\\
    && \delta_S\mathcal{H} = i\bar{\eta}(w-2+c)P_L\mathcal{Z},\quad \delta_K\mathcal{H}=0.
    \end{eqnarray}
    \item $\mathcal{K}~(w+1,c+3)$:
    \begin{eqnarray}
    && \delta_Q\mathcal{K} = i\bar{\epsilon}P_L(\gamma^a\mathcal{D}_a\mathcal{Z}+\Lambda), \quad \delta_M\mathcal{K} = 0,\quad \delta_D\mathcal{K}=(w+1)\lambda_D\mathcal{K},\quad \delta_A\mathcal{K}=i\theta (c+3)\mathcal{K},\nonumber\\
    && \delta_S\mathcal{K} = i\bar{\eta}(-w+2+c)P_R\mathcal{Z},\quad \delta_K\mathcal{K}=0.
    \end{eqnarray}
    \item $\mathcal{B}_a~(w+1,c)$:
    \begin{eqnarray}
    && \delta_Q\mathcal{B}_a = -\frac{1}{2}\bar{\epsilon}(\gamma_a\Lambda+\mathcal{D}_a\mathcal{Z}), \quad \delta_M\mathcal{B}_a = - \lambda_a^{~~b}\mathcal{B}_b, \quad \delta_D\mathcal{B}_a = (w+1)\lambda_D\mathcal{B}_a, \quad \delta_A\mathcal{B}_a = ic\theta \mathcal{B}_a,\nonumber\\
    && \delta_S\mathcal{B}_a = \frac{1}{2}\bar{\eta}[(w+1)+c\gamma_*]\gamma_a\mathcal{Z},\quad \delta_K\mathcal{B}_a = -2i\lambda_{K,a} c\mathcal{C}
    \end{eqnarray}
    \item $P_L\Lambda~(w+3/2,c-3/2)$, $P_R\Lambda~(w+3/2,c+3/2)$:
    \begin{eqnarray}
    && \delta_QP_L\Lambda = \frac{1}{2}[\gamma^{ab}(\mathcal{D}_a\mathcal{B}_b-i\mathcal{D}_a\mathcal{D}_b\mathcal{C})+i\mathcal{D}]P_L\epsilon, \quad \delta_MP_L\Lambda=-\frac{1}{4} \lambda^{ab}\gamma_{ab}P_L\Lambda, \nonumber\\
    && \delta_DP_L\Lambda = (w+3/2)\lambda_D P_L\Lambda, \quad \delta_AP_L\Lambda=i\theta(c+3/2)P_L\Lambda, \nonumber\\
    && \delta_SP_L\Lambda=
    -\frac{1}{2}P_L(i\mathcal{K}+\gamma^a\mathcal{B}_a+i\gamma^a\mathcal{D}_a\mathcal{C})(w+c\gamma_*)\eta, \quad \delta_KP_L\Lambda = \lambda_K^a(w+c)\gamma_aP_R\mathcal{Z},\\
    && \delta_QP_R\Lambda = \frac{1}{2}[\gamma^{ab}(\mathcal{D}_a\mathcal{B}_b+i\mathcal{D}_a\mathcal{D}_b\mathcal{C})-i\mathcal{D}]P_R\epsilon, \quad \delta_MP_R\Lambda=-\frac{1}{4} \lambda^{ab}\gamma_{ab}P_R\Lambda, \nonumber\\
    && \delta_DP_R\Lambda = (w-3/2)\lambda_D P_R\Lambda, \quad \delta_AP_R\Lambda=i\theta(c-3/2)P_R\Lambda, \nonumber\\
    && \delta_SP_R\Lambda=
    \frac{1}{2}P_R(i\mathcal{H}-\gamma^a\mathcal{B}_a+i\gamma^a\mathcal{D}_a\mathcal{C})(w+c\gamma_*)\eta, \quad \delta_KP_R\Lambda = \lambda_K^a(w-c)\gamma_aP_L\mathcal{Z}
    \end{eqnarray}
    \item $\mathcal{D}~(w+2,c)$:
    \begin{eqnarray}
    && \delta_Q\mathcal{D} = \frac{i}{2}\bar{\epsilon}\gamma_*\gamma^a\mathcal{D}_a\Lambda, \quad \delta_M\mathcal{D} =0, \quad \delta_D\mathcal{D}=(w+2)\lambda_D \mathcal{D}, \quad \delta_A\mathcal{D}=ic\theta\mathcal{D},\nonumber\\
    && \delta_S\mathcal{D}=i\bar{\eta}(w\gamma_*+c)(\frac{1}{2}\gamma^a\mathcal{D}_a\mathcal{Z}+\Lambda), \quad \delta_K\mathcal{D}= 2\lambda_K^a(w\mathcal{D}_a\mathcal{C}+ic\mathcal{B}_a).
    \end{eqnarray}
\end{itemize}
If there are internal symmetries $\delta_G\mathcal{C}^I=\theta^Ak_A^I(\mathcal{C}^I)$ of the lowest components $\mathcal{C}^I$ from several superconformal multiplets $\mathcal{V}^I$ labeled by an index $I$, then we have to add additional contributions to the above transformations, which are given by 
\begin{eqnarray}
\delta_Q\mathcal{B}_a^I &=& \frac{i}{2}\bar{\epsilon}\gamma_a\gamma_*(\Lambda^G)^Ak_A^I(\mathcal{C}),\\
\delta_Q\Lambda^I &=& \bigg[-\frac{1}{2}\mathcal{D}^A_Gk_A^I(\mathcal{C})+\frac{1}{4}\{(\bar{\Lambda}^G)^A\gamma^a\mathcal{Z}^J\}\partial_Jk_A^I\gamma_a+\frac{1}{4}
\{(\bar{\Lambda}^G)^A\gamma_*\gamma_a\mathcal{Z}^J\}\partial_Jk_A^I\gamma_*\gamma_a
\bigg]\epsilon,\\
\delta_Q\mathcal{D}^I &=& \frac{1}{2}\bar{\epsilon}\mathcal{Z}^J\partial_Jk_A^I(\mathcal{C})\mathcal{D}_G^A + \frac{i}{2}\bar{\epsilon}\gamma_*\gamma^a\mathcal{B}_a^I(\Lambda^G)^A\partial_Jk_A^I(\mathcal{C})-\frac{1}{2}\bar{\epsilon}\gamma^a\mathcal{D}_a(k_A^I(\mathcal{C})(\Lambda^G)^A),\label{additional_shifts_gauged_case}
\end{eqnarray}
where $(\Lambda^G)^A,\mathcal{D}_G^A$ are from the components of gauge multiplets $\mathcal{V}_G$.

The {\bf superconformal covariant derivatives} of the fields of a superconformal multiplet $\mathcal{V}$ under the superconformal symmetry are as follows:
\begin{eqnarray}
\mathcal{D}_{\mu}\mathcal{C} &=& (\partial_{\mu}-wb_{\mu}-icA_{\mu})\mathcal{C}-\frac{i}{2}\bar{\psi}_{\mu}\gamma_*\mathcal{Z},\\
P_L\mathcal{D}_{\mu}\mathcal{Z} &=&
\Big(\partial_{\mu}-(w+1/2)b_{\mu}-i(c-3/2)A_{\mu}+\frac{1}{4}\omega_{\mu}^{ab}\gamma_{ab}\Big)\mathcal{Z} \nonumber\\
&&-\frac{1}{2}P_L(i\mathcal{H}-\gamma^a\mathcal{B}_a-i\gamma^a\mathcal{D}_a\mathcal{C})\psi_{\mu} -i(w+c)P_L\phi_{\mu}\mathcal{C},\\
P_R\mathcal{D}_{\mu}\mathcal{Z} &=&
\Big(\partial_{\mu}-(w+1/2)b_{\mu}-i(c+3/2)A_{\mu}+\frac{1}{4}\omega_{\mu}^{ab}\gamma_{ab}\Big)\mathcal{Z} \nonumber\\
&&-\frac{1}{2}P_R(-i\mathcal{K}-\gamma^a\mathcal{B}_a+i\gamma^a\mathcal{D}_a\mathcal{C})\psi_{\mu} -i(w-c)P_R\phi_{\mu}\mathcal{C},\\
\mathcal{D}_a\mathcal{B}_b &=& e^{\mu}_a\Big[
(\partial_{\mu}-(w+1)b_{\mu}-icA_{\mu})\mathcal{B}_b+\omega_{\mu bc}\mathcal{B}^c \nonumber\\
&&+\frac{1}{2}\bar{\psi}_{\mu}(\mathcal{D}_b\mathcal{Z}+\gamma_b\Lambda)-\frac{1}{2}\bar{\phi}_{\mu}(w+1+c\gamma_*)\gamma_b \mathcal{Z}+2ic\mathcal{C}f_{\mu b}
\Big],\\
P_L\mathcal{D}_{\mu}\Lambda &=& P_L\Big(
\partial_{\mu}-(w+3/2)b_{\mu} -i(c+3/2)A_{\mu}+\frac{1}{4}\omega_{\mu}^{ab}\gamma_{ab}
\Big)\Lambda \nonumber\\
&&-\frac{1}{2}[
\gamma^{ab}(\mathcal{D}_a\mathcal{B}_b-i\mathcal{D}_a\mathcal{D}_b\mathcal{C})+i\mathcal{D}]P_L\psi_{\mu} \nonumber\\
&&+\frac{1}{2}P_L(
i\mathcal{K}+\gamma^a\mathcal{B}_a+i\gamma^a\mathcal{D}_a\mathcal{C})(w+c\gamma_*)\phi_{\mu} 
-(w+c)\gamma_a P_R \mathcal{Z}f_{\mu}^a,\\
P_R\mathcal{D}_{\mu}\Lambda &=& P_R\Big(
\partial_{\mu}-(w+3/2)b_{\mu} -i(c-3/2)A_{\mu}+\frac{1}{4}\omega_{\mu}^{ab}\gamma_{ab}
\Big)\Lambda \nonumber\\
&&-\frac{1}{2}[
\gamma^{ab}(\mathcal{D}_a\mathcal{B}_b+i\mathcal{D}_a\mathcal{D}_b\mathcal{C})-i\mathcal{D}]P_R\psi_{\mu} \nonumber\\
&&-\frac{1}{2}P_R(
i\mathcal{H}-\gamma^a\mathcal{B}_a+i\gamma^a\mathcal{D}_a\mathcal{C})(w+c\gamma_*)\phi_{\mu} 
-(w-c)\gamma_a P_L \mathcal{Z}f_{\mu}^a,\\
\mathcal{D}_{[a}\mathcal{D}_{b]}\mathcal{C} &=& -\frac{1}{2}[wR_{ab}(D)+icR_{ab}(A)]\mathcal{C} -\frac{i}{4}\overline{R_{ab}(Q)}\gamma_*\mathcal{Z}.
\end{eqnarray}

\subsection{Chiral superconformal multiplet}

A chiral supermultiplet is given by
\begin{eqnarray}
\mathcal{X} = \{X,P_L\Omega,F\}.
\end{eqnarray}
In the superconformal formalism, this can be represented by imposing chiral condition $P_R\mathcal{Z} \equiv 0$, so that the other consistency conditions are followed as
\begin{eqnarray}
P_R\mathcal{Z} \equiv 0 \implies \mathcal{K}=\Lambda=\mathcal{D}=0, \quad \mathcal{B}_{\mu} = i\mathcal{D}_{\mu}\mathcal{C},
\end{eqnarray}
which gives the following collection as a {\bf chiral superconformal multiplet} 
\begin{eqnarray}
\mathcal{X} = \Big(X,-i\sqrt{2}P_L\Omega,-2F,0,i\mathcal{D}_{\mu}X,0,0\Big).
\end{eqnarray}
The complex conjugate of this superconformal multiplet can be obtained by imposing the chiral condition $P_L\mathcal{Z}=0$, leading to
\begin{eqnarray}
P_L\mathcal{Z} \equiv 0 \implies \mathcal{H}=\Lambda=\mathcal{D}=0, \quad \mathcal{B}_{\mu} = -i\mathcal{D}_{\mu}\mathcal{C}^*,
\end{eqnarray}
and
\begin{eqnarray}
\mathcal{X}^* = \Big(X^*,i\sqrt{2}P_R\Omega,0,-2F^*,-i\mathcal{D}_{\mu}X^*,0,0\Big).
\end{eqnarray}

In fact, the chiral superconformal multiplet $\mathcal{X}$ can only be obtained when Weyl weight is equal to chiral weight, i.e. $w=c$. The reason for this is that when we impose the chiral confition $P_R\mathcal{Z}=0$, its $S$-transformation must vanish as well, resulting in
\begin{eqnarray}
\delta_SP_R\mathcal{Z}=i(w-c)P_R\eta\mathcal{C} \overset{!}{=}0 \implies w=c.
\end{eqnarray}

Meanwhile, there is another way of obtaining a chiral superconformal multiplet from the genral one. Let us take a look at the superconformal transformations of $\mathcal{K}$. If we impose $c=w-2$, the transformation laws of a scalar  $\mathcal{C}'\equiv -\frac{1}{2}\mathcal{K}$ are given by
\begin{eqnarray}
    && \delta_Q\mathcal{C}' = -\frac{1}{2} i\bar{\epsilon}P_L(\gamma^a\mathcal{D}_a\mathcal{Z}+\Lambda), \quad \delta_M\mathcal{C}' = 0,\quad \delta_D\mathcal{C}'=(w+1)\lambda_D\mathcal{C}',\nonumber\\
    &&  \delta_A\mathcal{C}'=i\theta (w+1)\mathcal{C}',\quad\delta_S\mathcal{C}' =0,\quad \delta_K\mathcal{C}'=0.
\end{eqnarray}
We notice that these transformations are those of a chiral superconformal multiplet whose lowest component is given by $\mathcal{C}'=-\frac{1}{2}\mathcal{K}$ and the chiral fermion is identified with $\mathcal{Z}' = -P_L(\gamma^a\mathcal{D}_a\mathcal{Z}+\Lambda)$. Therefore, we can define an operation called {\bf Chiral projection} $T$ such that
\begin{eqnarray}
&&T(\mathcal{C}_{(w,c=w-2)})_{(w+1,w+1)} \equiv \mathcal{X}' = \Big( -\frac{1}{2}\mathcal{K}, -\frac{i}{\sqrt{2}}P_L(\gamma^a\mathcal{D}_a\mathcal{Z}+\Lambda),  \frac{1}{2}(\mathcal{D}+\square^C\mathcal{C}+i\mathcal{D}^a\mathcal{B}_a)  \Big),\\
&&T(\mathcal{C}^*_{(w,c=w-2)})_{(w+1,w+1)} \equiv \mathcal{X}' = \Big( -\frac{1}{2}{\mathcal{H}^*}, \frac{i}{\sqrt{2}}P_R(\gamma^a\mathcal{D}_a\mathcal{Z}^C+\Lambda^C),  \frac{1}{2}(\mathcal{D}^*+\square^C\mathcal{C}^*-i\mathcal{D}^a\mathcal{B}_a^*)  \Big),\nonumber\\{}
\end{eqnarray}
Note that the Weyl/chiral weights changes from $(w,c=w-2)$ to $(w+1,c=w+1)$. Hence, we can say that {\it $T$ carries the Weyl/chiral weights $(1,3)$!}

Plus, there is a special case that the chiral projection operation $T(\cdot)$ can be applied to a chiral multiplet with $w=1$, i.e. $\mathcal{X}_{(1,1)}\equiv\{X,P_L\Omega,F\}$. Then, the corresponding chiral projection is 
\begin{eqnarray}
T(\mathcal{X}_{(1,1)})_{(2,2)} = \Big(F^*,\quad \gamma^a\mathcal{D}_a P_R \Omega, \quad \square^CX^*\Big).
\end{eqnarray}

\subsection{Real superconformal multiplet}

It is also possible to obtain a real superconformal multiplet by imposing a real condition that 
\begin{eqnarray}
&& \mathcal{C}^*=\mathcal{C}\equiv C,\quad (P_L\mathcal{Z})^C = P_R\mathcal{Z}\equiv P_R\zeta, \quad (P_R\mathcal{Z})^C = P_L \mathcal{Z}\equiv P_L\zeta, \quad {\mathcal{H}^*} = \mathcal{K}, \nonumber\\
&& \mathcal{B}_{a}^*=\mathcal{B}_{a}\equiv B_a, \quad (P_L\Lambda)^C = P_R\Lambda \equiv  P_R\lambda, \quad (P_R\Lambda)^C = P_L \Lambda\equiv  P_L\lambda, \quad \mathcal{D}^*=\mathcal{D}\equiv D.
\end{eqnarray}
The real multiplt $V$ is then given by
\begin{eqnarray}
V = \Big(C,\zeta, \mathcal{H}, {\mathcal{H}^*}, B_a,\lambda,D \Big).
\end{eqnarray}

\subsection{Linear superconformal multiplet}

Using the chiral projection operation, we can define a complex linear superconformal multiplet $\mathcal{V}_L$ by imposing two conditions $c=w-2$ and
\begin{eqnarray}
T(\mathcal{V}_L) \equiv 0.
\end{eqnarray}
In particular, by requiring reality to $L$ in addition to the two conditions, we get a real linear superconformal multiplet given by
\begin{eqnarray}
\mathcal{V}_L  = \Big( C^L, \zeta^L, 0,0,B_a^L,-\gamma^a\mathcal{D}_a \zeta^L,-\square^C C^L \Big),
\end{eqnarray}
where $C^L$ is a real scalar, $\zeta^L$ is a Majorana spinor, and $B_a^L$ is a real vector such that $\mathcal{D}^aB_a^L=0$.

\subsection{Gauge superconformal multiplet}

Another crucial multiplet is a {\bf Gauge superconformal multiplet}. To get this, we can start with a real superconformal multiplet with the Weyl/chiral weight (0,0):
\begin{eqnarray}
V^A_G = \Big(C^A,\zeta^A, \mathcal{H}^A, {\mathcal{H}^*}^A, B_a^A,(\lambda^G)^A,(D^G)^A \Big).
\end{eqnarray}
Then, let us look at the following field redefinition
\begin{eqnarray}
\hat{B}_{\mu} \equiv e^{a}_{\mu}B_a -\frac{1}{2}\bar{\psi}_{\mu}\zeta.
\end{eqnarray}
Then, the $Q$-transformation of this vector field is given by
\begin{eqnarray}
\delta_Q \hat{B}_{\mu} = -\frac{1}{2}\bar{\epsilon}\gamma_{\mu}\lambda - \frac{1}{2}\partial_{\mu}(\bar{\epsilon}\zeta).
\end{eqnarray}
Here we see that the last term may be thought as a $U(1)$ gauge transformation of a gauge field $\hat{B}_{\mu} \longrightarrow \hat{B}_{\mu} - \partial_{\mu}\theta$ where $\theta$ is a real scalar gauge parameter. We then establish a superconformal multiplet defined by
\begin{eqnarray}
V^A_G = \Big(0,0, 0, 0, (\hat{B}_a)^A,(\lambda^G)^A,(D^G)^A \Big),
\end{eqnarray}
where $(\hat{B}_a)^A$ is {\it gauge (vector) field}; $(\lambda^G)^A$ is its fermionic superpartner called {\it gaugino}, and $(D^G)^A$ is {\it auxiliary real scalar} of the gauge multiplet. Their transformations are specified as follows:
\begin{itemize}
    \item $(\hat{B}_{\mu})^A$ ($1,0$):
    \begin{eqnarray}
    \delta_Q(\hat{B}_{\mu})^A = -\frac{1}{2}\bar{\epsilon}\gamma_{\mu}(\lambda^G)^A, \quad \delta_M(\hat{B}_{\mu})^A = \delta_D(\hat{B}_{\mu})^A=\delta_A(\hat{B}_{\mu})^A= \delta_S(\hat{B}_{\mu})^A=\delta_K(\hat{B}_{\mu})^A=0
    \end{eqnarray}
    Here notice that gauge field $(\hat{B}_{\mu})^A$ transforms only under the Q-SUSY transformation, while it is inert under the others.
    \item $(\lambda^G)^A$ ($3/2,\pm 3/2$):
    \begin{eqnarray}
    && \delta_Q(\lambda^G)^A = \Big(\frac{i}{2}\gamma_*(D^G)^A+\frac{1}{4}\gamma^{ab}(\hat{F}^G_{ab})^A\Big)\epsilon, \quad \delta_M(\lambda^G)^A = -\frac{1}{4}\lambda^{ab}\gamma_{ab}(\lambda^G)^A ,\nonumber\\
    && \delta_D(\lambda^G)^A = \frac{3}{2}\lambda_D(\lambda^G)^A ,\quad \delta_A(\lambda^G)^A =\frac{3}{2}i\theta \gamma_*(\lambda^G)^A , \quad \delta_S (\lambda^G)^A  = \delta_K(\lambda^G)^A =0,\nonumber\\
    &&\textrm{where~~} (\hat{F}_{ab}^G)^A \equiv e^{\mu}_ae^{\nu}_b\Big(2\partial_{[\mu}(\hat{B}_{\nu]})^A+f_{BC}^{~~~A}(\hat{B}_{\mu})^B(\hat{B}_{\nu})^C+\bar{\psi}_{[\mu}\gamma_{\nu]}(\lambda^G)^A\Big).
    \end{eqnarray}
    \item $(D^G)^A$ ($2,0$):
    \begin{eqnarray}
    && \delta_Q(D^G)^A = \frac{i}{2}\bar{\epsilon}\gamma_*(\gamma^a\mathcal{D}_a\lambda^G)^A, \quad \delta_M(D^G)^A=0, \quad \delta_D(D^G)^A=2\lambda_D(D^G)^A, \nonumber\\
    && \delta_s(D^G)^A=\delta_K(D^G)^A=0.
    \end{eqnarray}
\end{itemize}
It is obvious that the Q-SUSY transformations of a general superconformal multiplet will change due to the additional shifts caused by internal gauge symmetries. When a multiplet is charged under gauge symmetries, a partial derivative must be replaced by a proper covariant derivative which involves the gauge-coupling part given by $\partial_{\mu} \longrightarrow D_{\mu} \supset -(\hat{B}_{\mu})^Ak_A^I(\mathcal{C})$. Since the gauge field transforms under Q-SUSY, a transformation of the covariant derivative of some fields of the gauged multiplet must depend on the gaugino, which is completely new compared to the Q-SUSY transformations of the un-gauged case. The additional corrections to the Q-SUSY transformations are shown in Eq.~\eqref{additional_shifts_gauged_case}. In addition, a gauged chiral superconformal multiplet is given by
\begin{eqnarray}
\mathcal{X}_{gauged} = \Big(X, -i\sqrt{2}P_L\Omega, -2F, -2iF, i\mathcal{D}_aX, -2iP_R(\lambda^G)^Ak_A(X), -ik_A(X)(D^G)^A \Big),
\end{eqnarray}
where $k_A(X)$ is the Killing vector of gauge symmetries.

\section{Multiplication Laws of Superconformal multiplets}

In this section, I introduce the multiplication laws of the superconformal multiplets from Ref. \cite{Linear}. Multiplication laws for a composite multiplet whose arguments are given only by {\bf complex} superconformal multiplets:
\begin{eqnarray}
\Tilde{\mathcal{C}} &=& f(\mathcal{C}^i) ,\\
\Tilde{\mathcal{Z}} &=& f_i\mathcal{Z}^i,\\
\Tilde{\mathcal{H}} &=& f_i\mathcal{H}^i -\frac{1}{2}f_{ij}\bar{\mathcal{Z}}^iP_L\mathcal{Z}^j,\\
\Tilde{\mathcal{K}} &=& f_i\mathcal{K}^i -\frac{1}{2}f_{ij}\bar{\mathcal{Z}}^iP_R\mathcal{Z}^j,\\
\Tilde{\mathcal{B}}_{\mu} &=& f_i\mathcal{B}_{\mu}^i +\frac{i}{4}f_{ij}\bar{\mathcal{Z}}^i\gamma_*\gamma_{\mu}\mathcal{Z}^j = f_i\mathcal{B}_{\mu}^i+\frac{i}{2}f_{ij}\bar{\mathcal{Z}}^i P_L\gamma_{\mu}\mathcal{Z}^j,\\
\Tilde{\Lambda} &=& f_i\lambda^i +\frac{1}{2}f_{ij}\Big[
i\gamma_*\cancel{\mathcal{B}}^i +P_L\mathcal{K}^i + P_R\mathcal{H}^i -\cancel{\mathcal{D}}\mathcal{C}^i
\Big]\mathcal{Z}^j -\frac{1}{4}f_{ijk}\mathcal{Z}^i\bar{\mathcal{Z}}^j\mathcal{Z}^k,\\
\Tilde{\mathcal{D}} &=& f_i\mathcal{D}^i + \frac{1}{2}f_{ij}\Big(
\mathcal{K}^i\mathcal{H}^j -\mathcal{B}^i\cdot \mathcal{B}^j -\mathcal{D}\mathcal{C}^i \cdot \mathcal{D}\mathcal{C}^j - 2\bar{\Lambda}\mathcal{Z}^j -\bar{\mathcal{Z}}^i \cancel{\mathcal{D}}\mathcal{Z}^j
\Big) \nonumber\\
&&-\frac{1}{4}f_{ijk}\bar{\mathcal{Z}}^i\Big(
i\gamma_*\cancel{\mathcal{B}}^i +P_L\mathcal{K}^i + P_R\mathcal{H}^i\Big)\mathcal{Z}^k +\frac{1}{8}f_{ijkl}(\bar{\mathcal{Z}}^i P_L\mathcal{Z}^j)(\bar{\mathcal{Z}}^kP_R\mathcal{Z}^l).
\end{eqnarray}
Multiplication laws fora composite multiplet whose arguments are given only by {\bf chiral} superconformal multiplets:
\begin{eqnarray}
\Tilde{\mathcal{C}} &=& f(X^{\alpha},\bar{X}^{\bar{\alpha}})  ,\\
\Tilde{\mathcal{Z}} &=& i\sqrt{2}(-f_{\alpha}\Omega^{\alpha} + f_{\bar{\alpha}}\Omega^{\bar{\alpha}}),\\
\Tilde{\mathcal{H}} &=& -2f_{\alpha}F^{\alpha} + f_{\alpha\beta}\bar{\Omega}^{\alpha}\Omega^{\beta},\\ 
\Tilde{\mathcal{K}} &=& -2f_{\bar{\alpha}}F^{\bar{\alpha}} + f_{\bar{\alpha}\bar{\beta}}\bar{\Omega}^{\bar{\alpha}}\Omega^{\bar{\beta}},\\ 
\Tilde{\mathcal{B}}_{\mu} &=& if_{\alpha}\mathcal{D}_{\mu}X^{\alpha}-if_{\bar{\alpha}}\mathcal{D}_{\mu}\bar{X}^{\bar{\alpha}}+if_{\alpha\bar{\beta}}\bar{\Omega}^{\alpha}\gamma_{\mu}\Omega^{\bar{\beta}},\\ 
P_L\Tilde{\Lambda} &=& -\sqrt{2}if_{\bar{\alpha}\beta}[(\cancel{\mathcal{D}}X^{\beta})\Omega^{\bar{\alpha}}-F^{\bar{\alpha}}\Omega^{\beta}]-\frac{i}{\sqrt{2}}f_{\bar{\alpha}\bar{\beta}\gamma}\Omega^{\gamma}\bar{\Omega}^{\bar{\alpha}}\Omega^{\bar{\beta}},\\
P_R\Tilde{\Lambda}&=& \sqrt{2}if_{\alpha\bar{\beta}}[(\cancel{\mathcal{D}}\bar{X}^{\bar{\beta}})\Omega^{\alpha}-F^{\alpha}\Omega^{\bar{\beta}}]+\frac{i}{\sqrt{2}}f_{\alpha\beta\bar{\gamma}}\Omega^{\bar{\gamma}}\bar{\Omega}^{\alpha}\Omega^{\beta},\\
\Tilde{\mathcal{D}} &=& 2f_{\alpha\bar{\beta}}\Big(-\mathcal{D}_{\mu}X^{\alpha}\mathcal{D}^{\mu}\bar{X}^{\bar{\beta}}-\frac{1}{2}\bar{\Omega}^{\alpha}P_L\cancel{\mathcal{D}}\Omega^{\bar{\beta}}-\frac{1}{2}\bar{\Omega}^{\bar{\beta}}P_R\cancel{\mathcal{D}}\Omega^{\alpha}+F^{\alpha}F^{\bar{\beta}}\Big) \nonumber\\
&&+f_{\alpha\beta\bar{\gamma}}(-\bar{\Omega}^{\alpha}\Omega^{\beta}F^{\bar{\gamma}}+\bar{\Omega}^{\alpha}(\cancel{\mathcal{D}}X^{\beta})\Omega^{\bar{\gamma}})+ f_{\bar{\alpha}\bar{\beta}\gamma}(-\bar{\Omega}^{\bar{\alpha}}\Omega^{\bar{\beta}}F^{\gamma}+\bar{\Omega}^{\bar{\alpha}}(\cancel{\mathcal{D}}\bar{X}^{\bar{\beta}})\Omega^{\gamma}) \nonumber\\
&&+ \frac{1}{2}f_{\alpha\beta\bar{\gamma}\bar{\delta}}(\bar{\Omega}^{\alpha}P_L\Omega^{\beta})(\bar{\Omega}^{\bar{\gamma}}P_R\Omega^{\bar{\delta}}).
\end{eqnarray}
Multiplication laws for a composite multiplet whose arguments are given only by {\bf real} superconformal multiplets:
\begin{eqnarray}
\Tilde{\mathcal{C}} &=& f(C^i) ,\\
\Tilde{\mathcal{Z}} &=& f_i\zeta^i,\\
\Tilde{\mathcal{H}} &=& f_i\mathcal{H}^i -\frac{1}{2}f_{ij}\bar{\zeta}^iP_L\zeta^j,\\
\Tilde{\mathcal{K}} &=& f_i{\mathcal{H}^*}^i -\frac{1}{2}f_{ij}\bar{\zeta}^iP_R\zeta^j,\\
\Tilde{\mathcal{B}}_{\mu} &=& f_iB_{\mu}^i +\frac{i}{4}f_{ij}\bar{\zeta}^i\gamma_*\gamma_{\mu}\zeta^j,\\
\Tilde{\Lambda} &=& f_i\lambda^i +\frac{1}{2}f_{ij}\Big[
i\gamma_*\cancel{B}^i +\textrm{Re}\mathcal{H}^i-i\gamma_*\textrm{Im}\mathcal{H}^i -\cancel{\mathcal{D}}C^i
\Big]\zeta^j -\frac{1}{4}f_{ijk}\zeta^i\bar{\zeta}^j\zeta^k,\\
\Tilde{\mathcal{D}} &=& f_iD^i + \frac{1}{2}f_{ij}\Big(
\mathcal{H}^i{\mathcal{H}^*}^j -B^i\cdot B^j -\mathcal{D}C^i \cdot \mathcal{D}C^j - 2\bar{\lambda}^i\zeta^j -\bar{\zeta}^i \cancel{\mathcal{D}}\zeta^j
\Big) \nonumber\\
&&-\frac{1}{4}f_{ijk}\bar{\zeta}^i\Big(
i\gamma_*\cancel{B}^i +\textrm{Re}\mathcal{H}^i-i\gamma_*\textrm{Im}\mathcal{H}^i\Big)\zeta^k +\frac{1}{8}f_{ijkl}(\bar{\zeta}^i P_L\zeta^j)(\bar{\zeta}^kP_R\zeta^l).
\end{eqnarray}

\section{Density Formulae}

\begin{itemize}
   \item {\bf F-term Density Formula}: For any chiral superconformal multiplet $\mathcal{X}$ with the Weyl/chiral weights $(3,3)$ whose supermultiplet form is given by $\{X,P_L\Omega,F\}$, the corresponding invariant action ``$[X]_F$'' via F-term sector can be obtained by the following expression:
\begin{eqnarray}
[X]_F &\equiv& \int d^4x e \bigg[F + \frac{1}{\sqrt{2}}\bar{\psi}_{\mu}\gamma^{\mu}P_L\Omega + \frac{1}{2}X\bar{\psi}_{\mu}\gamma^{\mu\nu}P_R\psi_{\nu} \bigg] + h.c.
\end{eqnarray}
where $e \equiv \det(e^a_{\mu})$. This F-term density formula form is the same as that of the Kugo's notation.
    \item {\bf D-term Density Formula}: For any real superconformal multiplet $\mathcal{V}$ with Weyl/chiral weights $(2,0)$ whose lowest component is given by a real scalar $C$, we can always construct a chiral superconformal multiplet with the Weyl/chiral weights $(3,3)$ by applying the chiral projection operation $T$ to the real multiplet $\mathcal{V}$. This means that it is possible to compute the corresponding invariant action using the above F-term density formula. Here, we define the D-term density formula in terms of the chiral projection and F-term density formula as follows:
    \begin{eqnarray}
    [\mathcal{C}]_D \equiv \frac{1}{2}[T(\mathcal{C})]_F.
    \end{eqnarray}
    Also, we have $[T(\mathcal{C})]_F=[T(\mathcal{C}^*)]$.
    After integrating by parts, we reach the following equivalent invariant action ``$[C]_D$'' that can also be calculated via D-term sector:
\begin{eqnarray}
[C]_D &\equiv& {\color{red} \frac{1}{2}} \int d^4x e \bigg[
D + \mathcal{D}^a\mathcal{D}_aC + \frac{1}{2}\Big(i\bar{\psi}\cdot\gamma P_R(\lambda+\gamma^a\mathcal{D}_a\zeta)-\frac{1}{4}\bar{\psi}_{\mu}P_L\gamma^{\mu\nu}\psi_{\nu}\mathcal{H}+h.c.\Big)
\bigg] \nonumber\\
&=& {\color{red} \frac{1}{2}} \int d^4x e \bigg[
D-\frac{1}{2}\bar{\psi}\cdot \gamma i\gamma_*\lambda -\frac{1}{3}CR(\omega)+\frac{1}{6}\Big(C\bar{\psi}_{\mu}\gamma^{\mu\rho\sigma}-i\bar{\zeta}\gamma^{\rho\sigma}\gamma_*\Big)R'_{\rho\sigma}(Q) \nonumber\\
&&\qquad \qquad \quad  +\frac{1}{4}\varepsilon^{abcd}\bar{\psi}_a\gamma_b\psi_c(B_d-\frac{1}{2}\bar{\psi}_d\zeta)
\Bigg],
\end{eqnarray}
where $R'_{\rho\sigma}(Q) \equiv 2(\partial_{[\mu}+\frac{1}{4}\omega_{[\mu}^{ab}\gamma_{ab}+\frac{1}{2}b_{[\mu}-\frac{3}{2}iA_{[\mu}\gamma_*)\psi_{\nu]}$, and $R(\omega)\equiv R_{\mu\nu}^{cov}(M)g^{\mu\nu}$. It is worth noticing that there is a factor of 1/2 in this D-term density formula in our notation (which follows Ferrara's one), while in the Kugo's notation, there is no the factor of 1/2. 
\item {\bf Theorems on the chiral projection}: $T(Z\mathcal{C})=Z T(\mathcal{C})$ is satisfied if $Z$ is a chiral multiplet and $\mathcal{C}$ has the Weyl/chiral weights $(w,w-2)$. Plus, for any two chiral multiplets $\Lambda$ with $(0,0)$ and $Z$ with $(1,1)$, $[(\Lambda+\Lambda^*)ZZ^*]_D=\frac{1}{2}[T\Big((\Lambda+\Lambda^*)ZZ^*\Big)]_F=[\Lambda ZT(Z^*)]_F$ can be satisfied. 
\end{itemize}

%% file: chapters/deri.tex
We consider matter chiral multiplets $Z^i$, the chiral compensator $S_0$, a real multiplet $V$, and another real multiplet 
$(V)_D$, whose lowest component is the auxiliary D term of the real multiplet $V$. Their superconformal multiplets are given as follows:
\begin{eqnarray}
&& V = \{0,0,0,0,A_{\mu},\lambda,D\} ~\textrm{in the Wess-Zumino gauge,~i.e.}~v=\zeta=\mathcal{H}=0, \\
&& Z^i = (z^i,-i\sqrt{2}P_L\chi^i,-2F^i,0,+i\mathcal{D}_{\mu}z^i,0,0) = \{ z^i, P_L\chi^i,F^i\},\\
&& \bar{Z}^{\bar{i}} = (\bar{z}^{\bar{i}},+i\sqrt{2}P_R\chi^{\bar{i}},0,-2\bar{F}^{\bar{i}},-i\mathcal{D}_{\mu}\bar{z}^{\bar{i}},0,0) = \{ \bar{z}^{\bar{i}}, P_R\chi^{\bar{i}},\bar{F}^{\bar{i}}\},\\
&& S_0 = (s_0,-i\sqrt{2}P_L\chi^0,-2F_0,0,+i\mathcal{D}_{\mu}s_0,0,0) = \{ s_0, P_L\chi^0,F_0\},\\
&& \bar{S}_0 = (\bar{s}_0,+i\sqrt{2}P_R\chi^0,0,-2\bar{F}_0,-i\mathcal{D}_{\mu}\bar{s}_0,0,0) = \{\bar{s}_0, P_R\chi^0,\bar{F}_0\},\\
&& \bar{\lambda}P_L\lambda = (\bar{\lambda}P_L\lambda,-i\sqrt{2}P_L\Lambda,2D_-^2,0,+i\mathcal{D}_{\mu}(\bar{\lambda}P_L\lambda),0,0) = \{\bar{\lambda}P_L\lambda, P_L\Lambda,-D_-^2\},\\
&& \bar{\lambda}P_R\lambda = (\bar{\lambda}P_R\lambda,+i\sqrt{2}P_R\Lambda,0,2D_+^2,-i\mathcal{D}_{\mu}(\bar{\lambda}P_R\lambda),0,0) = \{\bar{\lambda}P_R\lambda, P_R\Lambda,-D_+^2\},\\
&& (V)_D = (D,\cancel{\mathcal{D}}\lambda,0,0,\mathcal{D}^{b}\hat{F}_{ab},-\cancel{\mathcal{D}}\cancel{\mathcal{D}}\lambda,-\square^CD),
\end{eqnarray}
where
\begin{eqnarray}
&& P_L\Lambda \equiv \sqrt{2}P_L(-\frac{1}{2}\gamma\cdot \hat{F} + iD)\lambda,\qquad P_R\Lambda \equiv \sqrt{2}P_R(-\frac{1}{2}\gamma\cdot \hat{F} - iD)\lambda,\\
&& D_-^2 \equiv D^2 - \hat{F}^-\cdot\hat{F}^- - 2  \bar{\lambda}P_L\cancel{\mathcal{D}}\lambda,\qquad D_+^2 \equiv D^2 - \hat{F}^+\cdot\hat{F}^+ - 2  \bar{\lambda}P_R\cancel{\mathcal{D}}\lambda,\\
&& \mathcal{D}_{\mu}\lambda \equiv \bigg(\partial_{\mu}-\frac{3}{2}b_{\mu}+\frac{1}{4}w_{\mu}^{ab}\gamma_{ab}-\frac{3}{2}i\gamma_*\mathcal{A}_{\mu}\bigg)\lambda - \bigg(\frac{1}{4}\gamma^{ab}\hat{F}_{ab}+\frac{1}{2}i\gamma_* D\bigg)\psi_{\mu}
\\
 && \hat{F}_{ab} \equiv F_{ab} + e_a^{~\mu}e_b^{~\nu} \bar{\psi}_{[\mu}\gamma_{\nu]}\lambda,\qquad F_{ab} \equiv e_a^{~\mu}e_b^{~\nu} (2\partial_{[\mu}A_{\nu]}),\\
 && \hat{F}^{\pm}_{\mu\nu} \equiv \frac{1}{2}(\hat{F}_{\mu\nu}\pm \tilde{\hat{F}}_{\mu\nu}), \qquad \tilde{\hat{F}}_{\mu\nu} \equiv -\frac{1}{2} i\epsilon_{\mu\nu\rho\sigma}\hat{F}^{\rho\sigma} .
\end{eqnarray}

Next, we show the components of the first superconformal {\it composite} complex multiplets $w'^2$ and $\Bar{w}'^2$ with Weyl/chiral weights $(-1,3)$ and $(-1,-3)$ respectively. These composite multiplets are defined to be
\begin{eqnarray}
&& w'^2 \equiv \frac{\bar{\lambda}P_L\lambda}{(S_0\bar{S}_0e^{-K/3})^2} = \{\mathcal{C}_w,\mathcal{Z}_w,\mathcal{H}_w,\mathcal{K}_w,\mathcal{B}^w_{\mu},\Lambda_w,\mathcal{D}_w\} \\
&& \bar{w}'^2 \equiv \frac{\bar{\lambda}P_R\lambda}{(S_0\bar{S}_0e^{-K/3})^2}
= \{\mathcal{C}_{\bar{w}},\mathcal{Z}_{\bar{w}},\mathcal{H}_{\bar{w}},\mathcal{K}_{\bar{w}},\mathcal{B}^{\bar{w}}_{\mu},\Lambda_{\bar{w}},\mathcal{D}_{\bar{w}}\}.
\end{eqnarray}
where
\begin{eqnarray}
\mathcal{C}_w &=& h \equiv \frac{\bar{\lambda}P_L\lambda}{(s_0\bar{s}_0e^{-K(z,\bar{z})/3})^2},\\
\mathcal{Z}_w &=& i\sqrt{2}(-h_a\Omega^a + h_{\bar{a}}\Omega^{\bar{a}}),\\
\mathcal{H}_w &=& -2h_aF^a + h_{ab}\bar{\Omega}^a\Omega^b,\\ 
\mathcal{K}_w &=& -2h_{\bar{a}}F^{\bar{a}} + h_{\bar{a}\bar{b}}\bar{\Omega}^{\bar{a}}\Omega^{\bar{b}},\\ 
\mathcal{B}^w_{\mu} &=& ih_a\mathcal{D}_{\mu}X^a-ih_{\bar{a}}\mathcal{D}_{\mu}\bar{X}^{\bar{a}}+ih_{a\bar{b}}\bar{\Omega}^{a}\gamma_{\mu}\Omega^{\bar{b}},\\ 
P_L\Lambda_w &=& -\sqrt{2}ih_{\bar{a}b}[(\cancel{\mathcal{D}}X^b)\Omega^{\bar{a}}-F^{\bar{a}}\Omega^b]-\frac{i}{\sqrt{2}}h_{\bar{a}\bar{b}c}\Omega^c\bar{\Omega}^{\bar{a}}\Omega^{\bar{b}},\\
P_R\Lambda_w &=& \sqrt{2}ih_{a\bar{b}}[(\cancel{\mathcal{D}}\bar{X}^{\bar{b}})\Omega^{a}-F^{a}\Omega^{\bar{b}}]+\frac{i}{\sqrt{2}}h_{ab\bar{c}}\Omega^{\bar{c}}\bar{\Omega}^{a}\Omega^{b},\\
\mathcal{D}_w &=& 2h_{a\bar{b}}\Big(-\mathcal{D}_{\mu}X^a\mathcal{D}^{\mu}\bar{X}^{\bar{b}}-\frac{1}{2}\bar{\Omega}^aP_L\cancel{\mathcal{D}}\Omega^{\bar{b}}-\frac{1}{2}\bar{\Omega}^{\bar{b}}P_R\cancel{\mathcal{D}}\Omega^a+F^aF^{\bar{b}}\Big) \nonumber\\
&&+h_{ab\bar{c}}(-\bar{\Omega}^a\Omega^bF^{\bar{c}}+\bar{\Omega}^a(\cancel{\mathcal{D}}X^b)\Omega^{\bar{c}})+ h_{\bar{a}\bar{b}c}(-\bar{\Omega}^{\bar{a}}\Omega^{\bar{b}}F^{c}+\bar{\Omega}^{\bar{a}}(\cancel{\mathcal{D}}\bar{X}^{\bar{b}})\Omega^{c}) \nonumber\\
&&+ \frac{1}{2}h_{ab\bar{c}\bar{d}}(\bar{\Omega}^aP_L\Omega^b)(\bar{\Omega}^{\bar{c}}P_R\Omega^{\bar{d}}).
\end{eqnarray}
Notice that when finding the multiplet $\bar{w}'^2$, we can just replace $h$ by its complex conjugate $h^{*}$.

The second superconformal multiplets that we need are the {\it composite} chiral projection multiplets $T(\Bar{w}'^2)$ and $\Bar{T}(w'^2)$ with Weyl/chiral weights $(0,0)$. From their component supermultiplets defined by 
\begin{eqnarray}
T(\bar{w}'^2) &=& \left( -\frac{1}{2}\mathcal{K}_{\bar{w}}, -\frac{1}{2} \sqrt{2} iP_L (\cancel{\mathcal{D}}\mathcal{Z}_{\bar{w}}+\Lambda_{\bar{w}}), \frac{1}{2}(\mathcal{D}_{\bar{w}}+\square^C \mathcal{C}_{\bar{w}} + i\mathcal{D}_a \mathcal{B}^a_{\bar{w}}) \right),\\
\bar{T}(w'^2) &=& \left( -\frac{1}{2}\mathcal{K}_{\bar{w}}^{*}, \frac{1}{2} \sqrt{2} iP_R (\cancel{\mathcal{D}}\mathcal{Z}_{\bar{w}}^C+\Lambda_{\bar{w}}^C), \frac{1}{2}(\mathcal{D}_{\bar{w}}^{*}+\square^C \mathcal{C}_{\bar{w}}^{*} - i\mathcal{D}_a (\mathcal{B}^a_{\bar{w}})^{*}) \right)
\end{eqnarray}
we find the corresponding superconformal multiplets and their complex conjugates as follows:
\begin{eqnarray}
T \equiv T(\bar{w}'^2) &=& \{\mathcal{C}_T,\mathcal{Z}_T,\mathcal{H}_T,\mathcal{K}_T,\mathcal{B}_{\mu}^T,\Lambda_T,\mathcal{D}_T\} \nonumber\\
\bar{T} \equiv \bar{T}(w'^2) &=& \{\mathcal{C}_{\bar{T}},\mathcal{Z}_{\bar{T}},\mathcal{H}_{\bar{T}},\mathcal{K}_{\bar{T}},\mathcal{B}_{\mu}^{\bar{T}},\Lambda_{\bar{T}},\mathcal{D}_{\bar{T}}\},
\end{eqnarray}
whose superconformal components are given by
\begin{eqnarray}
\mathcal{C}_T &=&  -\frac{1}{2} \mathcal{K}_{\bar{w}} = h^{*}_{\bar{a}}F^{\bar{a}} -\frac{1}{2} h^{*}_{\bar{a}\bar{b}}\bar{\Omega}^{\bar{a}}\Omega^{\bar{b}} \equiv C_T\\
\mathcal{Z}_T &=& -\sqrt{2}iP_L\bigg[\cancel{\mathcal{D}}(-h^{*}_a\Omega^a + h^{*}_{\bar{a}}\Omega^{\bar{a}})-h^{*}_{\bar{a}b}[(\cancel{\mathcal{D}}X^b)\Omega^{\bar{a}}-F^{\bar{a}}\Omega^b]-\frac{1}{2}h^{*}_{\bar{a}\bar{b}c}\Omega^c\bar{\Omega}^{\bar{a}}\Omega^{\bar{b}}\nonumber\\
&&\qquad\qquad+h^{*}_{a\bar{b}}[(\cancel{\mathcal{D}}\bar{X}^{\bar{b}})\Omega^{a}-F^{a}\Omega^{\bar{b}}]+\frac{1}{2}h^{*}_{ab\bar{c}}\Omega^{\bar{c}}\bar{\Omega}^{a}\Omega^{b}\bigg] \equiv -\sqrt{2}iP_L\Omega_T ,\\
\mathcal{H}_T &=& -2\bigg[h^{*}_{a\bar{b}}\Big(-\mathcal{D}_{\mu}X^a\mathcal{D}^{\mu}\bar{X}^{\bar{b}}-\frac{1}{2}\bar{\Omega}^aP_L\cancel{\mathcal{D}}\Omega^{\bar{b}}-\frac{1}{2}\bar{\Omega}^{\bar{b}}P_R\cancel{\mathcal{D}}\Omega^a+F^aF^{\bar{b}}\Big) \nonumber\\
&&+\frac{1}{2}h^{*}_{ab\bar{c}}(-\bar{\Omega}^a\Omega^bF^{\bar{c}}+\bar{\Omega}^a(\cancel{\mathcal{D}}X^b)\Omega^{\bar{c}})+\frac{1}{2} h^{*}_{\bar{a}\bar{b}c}(-\bar{\Omega}^{\bar{a}}\Omega^{\bar{b}}F^{c}+\bar{\Omega}^{\bar{a}}(\cancel{\mathcal{D}}\bar{X}^{\bar{b}})\Omega^{c}) \nonumber\\
&&+ \frac{1}{4}h^{*}_{ab\bar{c}\bar{d}}(\bar{\Omega}^aP_L\Omega^b)(\bar{\Omega}^{\bar{c}}P_R\Omega^{\bar{d}})+\frac{1}{2}\square^C h^{*} + \frac{1}{2}i\mathcal{D}^{\mu} (ih^{*}_a\mathcal{D}_{\mu}X^a-ih^{*}_{\bar{a}}\mathcal{D}_{\mu}\bar{X}^{\bar{a}}+ih^{*}_{a\bar{b}}\bar{\Omega}^{a}\gamma_{\mu}\Omega^{\bar{b}})\bigg] \nonumber\\
&\equiv& -2F_T,\\
\mathcal{K}_T &=& 0,\\
\mathcal{B}^T_{\mu} &=& -i\mathcal{D}_{\mu}\mathcal{C}_T,\\
\Lambda_T &=& 0 ,\\
\mathcal{D}_T &=& 0,
\end{eqnarray}
where we used $a,b,c,d = 0,i(\equiv z^i),W (\equiv \bar{\lambda}P_L\lambda)$. This gives the superfield components of the chiral projection multiplet $T$:
\begin{eqnarray}
T(\bar{w}'^2) = ( C_T, P_L\Omega_T, F_T )
\end{eqnarray}
where 
\begin{eqnarray}
C_T &=&  h^{*}_{\bar{a}}F^{\bar{a}} -\frac{1}{2} h^{*}_{\bar{a}\bar{b}}\bar{\Omega}^{\bar{a}}\Omega^{\bar{b}},\\
P_L\Omega_T &=& \cancel{\mathcal{D}}(-h^{*}_a\Omega^a + h^{*}_{\bar{a}}\Omega^{\bar{a}})-h^{*}_{\bar{a}b}[(\cancel{\mathcal{D}}X^b)\Omega^{\bar{a}}-F^{\bar{a}}\Omega^b]-\frac{1}{2}h^{*}_{\bar{a}\bar{b}c}\Omega^c\bar{\Omega}^{\bar{a}}\Omega^{\bar{b}}\nonumber\\
&&+h^{*}_{a\bar{b}}[(\cancel{\mathcal{D}}\bar{X}^{\bar{b}})\Omega^{a}-F^{a}\Omega^{\bar{b}}]+\frac{1}{2}h^{*}_{ab\bar{c}}\Omega^{\bar{c}}\bar{\Omega}^{a}\Omega^{b},\\
F_T &=&  h^{*}_{a\bar{b}}\Big(-\mathcal{D}_{\mu}X^a\mathcal{D}^{\mu}\bar{X}^{\bar{b}}-\frac{1}{2}\bar{\Omega}^aP_L\cancel{\mathcal{D}}\Omega^{\bar{b}}-\frac{1}{2}\bar{\Omega}^{\bar{b}}P_R\cancel{\mathcal{D}}\Omega^a+F^aF^{\bar{b}}\Big) \nonumber\\
&&+\frac{1}{2}h^{*}_{ab\bar{c}}(-\bar{\Omega}^a\Omega^bF^{\bar{c}}+\bar{\Omega}^a(\cancel{\mathcal{D}}X^b)\Omega^{\bar{c}})+\frac{1}{2} h^{*}_{\bar{a}\bar{b}c}(-\bar{\Omega}^{\bar{a}}\Omega^{\bar{b}}F^{c}+\bar{\Omega}^{\bar{a}}(\cancel{\mathcal{D}}\bar{X}^{\bar{b}})\Omega^{c}) \nonumber\\
&&+ \frac{1}{4}h^{*}_{ab\bar{c}\bar{d}}(\bar{\Omega}^aP_L\Omega^b)(\bar{\Omega}^{\bar{c}}P_R\Omega^{\bar{d}})+\frac{1}{2}\square^C h^{*} - \frac{1}{2}\mathcal{D}^{\mu} (h^{*}_a\mathcal{D}_{\mu}X^a-h^{*}_{\bar{a}}\mathcal{D}_{\mu}\bar{X}^{\bar{a}}+h^{*}_{a\bar{b}}\bar{\Omega}^{a}\gamma_{\mu}\Omega^{\bar{b}}). \nonumber\\{}
\end{eqnarray}

Morever,
\begin{eqnarray}
\bar{T}(w'^2) = \{ C_T^{*}, P_R\Omega_T, F_T^{*}\}
\end{eqnarray}
where
\begin{eqnarray}
C_T^{*} &=&  h_{a}F^{a} -\frac{1}{2} h_{ab}\bar{\Omega}^{a}\Omega^{b},\\
P_R\Omega_T &=& \cancel{\mathcal{D}}(-h_a\Omega^a + h_{\bar{a}}\Omega^{\bar{a}})-h_{\bar{a}b}[(\cancel{\mathcal{D}}X^b)\Omega^{\bar{a}}-F^{\bar{a}}\Omega^b]-\frac{1}{2}h_{\bar{a}\bar{b}c}\Omega^c\bar{\Omega}^{\bar{a}}\Omega^{\bar{b}}\nonumber\\
&&+h_{a\bar{b}}[(\cancel{\mathcal{D}}\bar{X}^{\bar{b}})\Omega^{a}-F^{a}\Omega^{\bar{b}}]+\frac{1}{2}h_{ab\bar{c}}\Omega^{\bar{c}}\bar{\Omega}^{a}\Omega^{b},\\
F_T^{*} &=&  h_{a\bar{b}}\Big(-\mathcal{D}_{\mu}X^a\mathcal{D}^{\mu}\bar{X}^{\bar{b}}-\frac{1}{2}\bar{\Omega}^aP_L\cancel{\mathcal{D}}\Omega^{\bar{b}}-\frac{1}{2}\bar{\Omega}^{\bar{b}}P_R\cancel{\mathcal{D}}\Omega^a+F^aF^{\bar{b}}\Big) \nonumber\\
&&+\frac{1}{2}h_{ab\bar{c}}(-\bar{\Omega}^a\Omega^bF^{\bar{c}}+\bar{\Omega}^a(\cancel{\mathcal{D}}X^b)\Omega^{\bar{c}})+\frac{1}{2} h_{\bar{a}\bar{b}c}(-\bar{\Omega}^{\bar{a}}\Omega^{\bar{b}}F^{c}+\bar{\Omega}^{\bar{a}}(\cancel{\mathcal{D}}\bar{X}^{\bar{b}})\Omega^{c}) \nonumber\\
&&+ \frac{1}{4}h_{ab\bar{c}\bar{d}}(\bar{\Omega}^aP_L\Omega^b)(\bar{\Omega}^{\bar{c}}P_R\Omega^{\bar{d}})+\frac{1}{2}\square^C h - \frac{1}{2}\mathcal{D}^{\mu} (h_a\mathcal{D}_{\mu}X^a-h_{\bar{a}}\mathcal{D}_{\mu}\bar{X}^{\bar{a}}+h_{a\bar{b}}\bar{\Omega}^{\bar{b}}\gamma_{\mu}\Omega^{a}). \nonumber\\{}
\end{eqnarray}

We then present a superconformal composite real multiplet $\mathcal{R}$ with Weyl/chiral weights $(0,0)$. Defining some chiral multiplets $\mathcal{X}^A \equiv \{X^A,P_L\Omega^A,F^A\}$ where $A=\{ S_0,Z^i,\bar{\lambda}P_L\lambda,T(\bar{w}'^2)$\} and their conjugates, we represent the composite one $\mathcal{R}$ as
\begin{eqnarray}
\mathcal{R} \equiv (S_0\bar{S}_0e^{-K/3})^{-3} \frac{(\bar{\lambda}P_L\lambda)(\bar{\lambda}P_R\lambda)}{T(\bar{w}'^2) \bar{T}(w'^2)} \mathcal{U}
\end{eqnarray}
whose lowest component is 
\begin{eqnarray}
\mathcal{C}_{\mathcal{R}}  \equiv (s_0\bar{s}_0e^{-K/3})^{-3}\frac{(\bar{\lambda}P_L\lambda)(\bar{\lambda}P_R\lambda)}{C_TC_{\bar{T}}}\mathcal{U} 
\equiv f(X^A,\bar{X}^{\bar{A}}) \label{def-f}
\end{eqnarray}
where $C_T = -D_+^2 \Delta^{-2}$; $C_{\bar{T}} = -D_-^2\Delta^{-2}$, and $\Delta \equiv s_0\bar{s}_0e^{-K/3}$, and $K,\mathcal{U}$ are functions of the matter multiplets $Z^i$'s.

\begin{eqnarray}
\mathcal{C}_{\mathcal{R}} &=& f \equiv  (s_0\bar{s}_0e^{-K/3})^{-3}\frac{(\bar{\lambda}P_L\lambda)(\bar{\lambda}P_R\lambda)}{C_TC_{\bar{T}}}\mathcal{U} ,\\
\mathcal{Z}_{\mathcal{R}} &=& i\sqrt{2}(-f_A\Omega^A + f_{\bar{A}}\Omega^{\bar{A}}),\\
\mathcal{H}_{\mathcal{R}} &=& -2f_AF^A + f_{AB}\bar{\Omega}^A\Omega^B,\\ 
\mathcal{K}_{\mathcal{R}} &=& -2f_{\bar{A}}F^{\bar{A}} + f_{\bar{A}\bar{B}}\bar{\Omega}^{\bar{A}}\Omega^{\bar{B}},\\ 
\mathcal{B}^{\mathcal{R}}_{\mu} &=& if_A\mathcal{D}_{\mu}X^A-if_{\bar{A}}\mathcal{D}_{\mu}\bar{X}^{\bar{A}}+if_{A\bar{B}}\bar{\Omega}^{A}\gamma_{\mu}\Omega^{\bar{B}},\\ 
P_L\Lambda_{\mathcal{R}} &=& -\sqrt{2}if_{\bar{A}B}[(\cancel{\mathcal{D}}X^B)\Omega^{\bar{A}}-F^{\bar{A}}\Omega^B]-\frac{i}{\sqrt{2}}f_{\bar{A}\bar{B}C}\Omega^C\bar{\Omega}^{\bar{A}}\Omega^{\bar{B}},\\
P_R\Lambda_{\mathcal{R}} &=& \sqrt{2}if_{A\bar{B}}[(\cancel{\mathcal{D}}\bar{X}^{\bar{B}})\Omega^{A}-F^{A}\Omega^{\bar{B}}]+\frac{i}{\sqrt{2}}f_{AB\bar{C}}\Omega^{\bar{C}}\bar{\Omega}^{A}\Omega^{B},\\
\mathcal{D}_{\mathcal{R}} &=& 2f_{A\bar{B}}\Big(-\mathcal{D}_{\mu}X^A\mathcal{D}^{\mu}\bar{X}^{\bar{B}}-\frac{1}{2}\bar{\Omega}^AP_L\cancel{\mathcal{D}}\Omega^{\bar{B}}-\frac{1}{2}\bar{\Omega}^{\bar{B}}P_R\cancel{\mathcal{D}}\Omega^A+F^AF^{\bar{B}}\Big) \nonumber\\
&&+f_{AB\bar{C}}(-\bar{\Omega}^A\Omega^BF^{\bar{C}}+\bar{\Omega}^A(\cancel{\mathcal{D}}X^B)\Omega^{\bar{C}})+ f_{\bar{A}\bar{B}C}(-\bar{\Omega}^{\bar{A}}\Omega^{\bar{B}}F^{C}+\bar{\Omega}^{\bar{A}}(\cancel{\mathcal{D}}\bar{X}^{\bar{B}})\Omega^{C}) \nonumber\\
&&+ \frac{1}{2}f_{AB\bar{C}\bar{D}}(\bar{\Omega}^AP_L\Omega^B)(\bar{\Omega}^{\bar{C}}P_R\Omega^{\bar{D}}).
\end{eqnarray}

Then, the superconformal multiplet of the new Fayet-Iliopoulos term can be written by using
\begin{eqnarray}
\mathcal{R}\cdot (V)_D = \{\tilde{\mathcal{C}},\tilde{\mathcal{Z}},\tilde{\mathcal{H}},\tilde{\mathcal{K}},\tilde{\mathcal{B}}_{\mu},\tilde{\Lambda},\tilde{\mathcal{D}}\},
\end{eqnarray}
whose superconformal multiplet components are as follows:
\begin{eqnarray}
\tilde{\mathcal{C}} &=& Df,\\
\tilde{\mathcal{Z}} &=& f\cancel{\mathcal{D}}\lambda+Di\sqrt{2}(-f_{A}\Omega^A+f_{\bar{A}}\Omega^{\bar{A}}),\\
\tilde{\mathcal{H}} &=& D(-2f_AF^A + f_{AB}\bar{\Omega}^A\Omega^B)-i\sqrt{2}(-f_{A}\bar{\Omega}^A+f_{\bar{A}}\bar{\Omega}^{\bar{A}})P_L\cancel{\mathcal{D}}\lambda,\\
\tilde{\mathcal{K}} &=& D(-2f_{\bar{A}}F^{\bar{A}} + f_{\bar{A}\bar{B}}\bar{\Omega}^{\bar{A}}\Omega^{\bar{B}})-i\sqrt{2}(-f_{A}\bar{\Omega}^A+f_{\bar{A}}\bar{\Omega}^{\bar{A}})P_R\cancel{\mathcal{D}}\lambda,\\
\tilde{\mathcal{B}} &=& (\mathcal{D}^{\nu}\hat{F}_{\mu\nu})f+D(if_A\mathcal{D}_{\mu}X^A-if_{\bar{A}}\mathcal{D}_{\mu}\bar{X}^{\bar{A}}+if_{A\bar{B}}\bar{\Omega}^A\gamma_{\mu}\Omega^{\bar{B}}),\\
\tilde{\Lambda} &=& -f\cancel{\mathcal{D}}\cancel{\mathcal{D}}\lambda + D(P_L\Lambda_{\mathcal{R}}+P_R\Lambda_{\mathcal{R}})+\frac{1}{2}\Big(\gamma_*(-f_A\cancel{\mathcal{D}}X^A+f_{\bar{A}}\cancel{\mathcal{D}}\bar{X}^{\bar{A}}-f_{A\bar{B}}\bar{\Omega}^A\cancel{\gamma}\Omega^{\bar{B}})\nonumber\\
&&+P_L(-2f_{\bar{A}}F^{\bar{A}} + f_{\bar{A}\bar{B}}\bar{\Omega}^{\bar{A}}\Omega^{\bar{B}})+P_R(-2f_AF^A + f_{AB}\bar{\Omega}^A\Omega^B) -\cancel{\mathcal{D}}f\Big)\cancel{\mathcal{D}}\lambda\nonumber\\
&& +\frac{1}{2}\Big( i\gamma_*\gamma^{\mu}\mathcal{D}^{\nu}\hat{F}_{\mu\nu}   -\cancel{\mathcal{D}}D\Big)i\sqrt{2}(-f_{A}\Omega^A+f_{\bar{A}}\Omega^{\bar{A}}) ,\\
\tilde{\mathcal{D}} &=&-f\square^C D + D 
\bigg\{ 2f_{A\bar{B}}(-\mathcal{D}_{\mu}X^A\mathcal{D}^{\mu}\bar{X}^{\bar{B}}-\frac{1}{2}\bar{\Omega}^AP_L\cancel{\mathcal{D}}\Omega^{\bar{B}}-\frac{1}{2}\cancel{\Omega}^{\bar{B}}P_R\cancel{\mathcal{D}}\Omega^A+F^AF^{\bar{B}}) \nonumber\\
&&+f_{AB\bar{C}}(-\bar{\Omega}^A\Omega^B F^{\bar{C}} + \bar{\Omega}^A(\cancel{\mathcal{D}}X^B)\Omega^{\bar{C}}) 
+f_{\bar{A}\bar{B}C}(-\bar{\Omega}^{\bar{A}}\Omega^{\bar{B}} F^C + \bar{\Omega}^{\bar{A}}(\cancel{\mathcal{D}}\bar{X}^{\bar{B}})\Omega^C) \nonumber\\
&&+\frac{1}{2}f_{AB\bar{C}\bar{D}} (\bar{\Omega}^AP_L\Omega^B)(\bar{\Omega}^{\bar{C}}P_R\Omega^{\bar{D}}) \bigg\}\nonumber\\
&& -(\mathcal{D}_{\nu}\hat{F}^{\mu\nu})(if_A\mathcal{D}_{\mu}X^A-if_{\bar{A}}\mathcal{D}_{\mu}\bar{X}^{\bar{A}}+if_{A\bar{B}}\bar{\Omega}^A\gamma_{\mu}\Omega^{\bar{B}}) \nonumber\\
&& + \bigg( \sqrt{2}if_{\bar{A}B}[(\cancel{\mathcal{D}}X^B)\Omega^{\bar{A}}-F^{\bar{A}}\Omega^B]  +\frac{i}{\sqrt{2}}f_{\bar{A}\bar{B}C} \Omega^C\bar{\Omega}^{\bar{A}}\Omega^{\bar{B}} \bigg)\cancel{\mathcal{D}}\lambda\nonumber\\
&& - \bigg( \sqrt{2}if_{A\bar{B}}[(\cancel{\mathcal{D}}\bar{X}^{\bar{B}})\Omega^{A}-F^{A}\Omega^{\bar{B}}]  +\frac{i}{\sqrt{2}}f_{AB\bar{C}} \Omega^{\bar{C}}\bar{\Omega}^{A}\Omega^{B} \bigg)\cancel{\mathcal{D}}\lambda\nonumber\\
&&-(\mathcal{D}_{\mu}f)(\mathcal{D}^{\mu}D)-\frac{1}{2}\cancel{\mathcal{D}}[i\sqrt{2}(-f_{A}\Omega^A+f_{\bar{A}}\Omega^{\bar{A}})](\cancel{\mathcal{D}}\lambda)+\frac{1}{2}i\sqrt{2}(-f_{A}\Omega^A+f_{\bar{A}}\Omega^{\bar{A}})(\cancel{\mathcal{D}}\cancel{\mathcal{D}}\lambda),
\end{eqnarray}
where the indices $A,B,C,D$ run over $0,i,W,T$. The component action of the new FI term is then given by the D-term density formula 
\begin{eqnarray}
\mathcal{L}_{NEW} \equiv -[ \mathcal{R}\cdot (V)_D]_D &=& -\frac{1}{4}\int d^4x e \bigg[  \tilde{\mathcal{D}} -\frac{1}{2}\bar{\psi}\cdot \gamma i\gamma_* \tilde{\Lambda} -\frac{1}{3}\tilde{\mathcal{C}}R(\omega)\nonumber\\
&&+\frac{1}{6}\Big(\tilde{\mathcal{C}}\bar{\psi}_{\mu}\gamma^{\mu\rho\sigma}-i\bar{\tilde{\mathcal{Z}}}\gamma^{\rho\sigma}\gamma_*\Big)R'_{\rho\sigma}(Q)\nonumber\\
&&+\frac{1}{4}\varepsilon^{abcd}\bar{\psi}_{a}\gamma_b\psi_c\Big(\tilde{\mathcal{B}}_{d}-\frac{1}{2}\bar{\psi}_d\tilde{\mathcal{Z}}\Big)\bigg]+\textrm{h.c.}.
\end{eqnarray}

Using $f=\Delta^{-3}\frac{W\bar{W}}{C_TC_{\bar{T}}}\mathcal{U} $, $W\equiv(\bar{\lambda}P_L\lambda)$, $\bar{W}\equiv (\bar{\lambda}P_R\lambda)$, $\Omega^W \sim \sqrt{2}iDP_L\lambda$, $\bar{\Omega}^T \sim 2D^2\Delta^{-2}(\frac{\bar{\Omega}^0}{s_0}-\frac{K_I\bar{\Omega}^I}{3}) $, $C_T \sim -D^2\Delta^{-2}$, $F^W \sim -D^2$, $F^{\bar{T}} \sim 2D^2\Delta^{-2}( \frac{F^{\bar{0}}}{\bar{s}_0} - \frac{1}{3}K_{\bar{J}}F^{\bar{J}})$ where $\Delta \equiv s_0\bar{s}_0e^{-K/3}$,
\begin{eqnarray}
&& \mathcal{L}_{\textrm{newFI}}^{(\textrm{2f})}e^{-1} \nonumber\\
&&= -D f_{0\bar{W}}F^0F^{\bar{W}}  -D f_{I\bar{W}}F^IF^{\bar{W}} -D f_{T\bar{W}}F^TF^{\bar{W}} \nonumber\\
&&\quad + \frac{1}{2}f_{0W\bar{W}}\bar{\Omega}^0\Omega^WF^{\bar{W}}
+ \frac{1}{2}f_{IW\bar{W}}\bar{\Omega}^I\Omega^WF^{\bar{W}} 
+ \frac{1}{2}f_{TW\bar{W}}\bar{\Omega}^T\Omega^WF^{\bar{W}} 
- \frac{D\sqrt{2}}{4} \bar{\psi}_{\mu} \gamma^{\mu} f_{\bar{W}W}F^{\bar{W}}\Omega^W 
+c.c..,\nonumber\\
&&=  -3\Delta\frac{(\bar{\lambda}P_L\lambda)}{D}\mathcal{U}\frac{F^{0}}{s_0} +\Delta\frac{(\bar{\lambda}P_L\lambda)}{D}(K_I\mathcal{U}+\mathcal{U}_I)F^{I}  +2\Delta\frac{(\bar{\lambda}P_L\lambda)}{D}\mathcal{U}( \frac{F^0}{s_0} - \frac{1}{3}K_{I}F^{I})\nonumber\\
&&\quad + \frac{3i}{\sqrt{2}}\frac{\Delta}{Ds_0}\mathcal{U}(\bar{\Omega}^0P_L\lambda) - \frac{i}{\sqrt{2}}\frac{\Delta}{D}(K_I\mathcal{U}+\mathcal{U}_I)(\bar{\Omega}^IP_L\lambda) + \sqrt{2}i \frac{\Delta}{Ds_0}\mathcal{U}(\bar{\Omega}^0P_L\lambda)
-\frac{\sqrt{2}i}{3} \frac{\Delta}{D}\mathcal{U}K_I(\bar{\Omega}^IP_L\lambda)\nonumber\\
&&\quad + \frac{i}{2}\Delta \mathcal{U} (\bar{\psi}_{\mu} \gamma^{\mu}P_L\lambda) + c.c.,\nonumber\\
&&= \frac{\Delta}{D}\bigg(
-\frac{F^0\mathcal{U}}{s_0}  + \mathcal{U}_IF^I +\frac{1}{3}\mathcal{U}K_IF^I
\bigg)(\bar{\lambda}P_L\lambda) 
+\frac{5}{\sqrt{2}}i \frac{\Delta}{Ds_0}\mathcal{U}(\bar{\Omega}^0P_L\lambda) 
\nonumber\\
&&\quad -i \frac{\Delta}{D}  \bigg(
\frac{5}{3\sqrt{2}} K_I \mathcal{U} +\frac{1}{\sqrt{2}}\mathcal{U}_I
\bigg)(\bar{\Omega}^IP_L\lambda) + \frac{i}{2}\Delta \mathcal{U} (\bar{\psi}_{\mu} \gamma^{\mu}P_L\lambda) + h.c.
\end{eqnarray}

At the superconformal gauge (i.e. $P_L\Omega^0 = \frac{1}{3}e^{K/6}K_IP_L\Omega^I$, $s_0=\bar{s}_0=e^{K/6}$, $\Delta = 1$), the lagrangian is rewritten by
\begin{eqnarray}
&&\mathcal{L}_{\textrm{newFI}}^{(\textrm{2f})}e^{-1} \nonumber\\
&&= \frac{1}{D}\bigg(
-F^0\mathcal{U}e^{-K/6}  + \mathcal{U}_IF^I +\frac{1}{3}\mathcal{U}K_IF^I
\bigg)(\bar{\lambda}P_L\lambda) 
 -\frac{i}{D}  
\frac{\mathcal{U}_I}{\sqrt{2}}(\bar{\Omega}^IP_L\lambda) + \frac{i}{2} \mathcal{U} (\bar{\psi}_{\mu} \gamma^{\mu}P_L\lambda) + h.c. \nonumber\\
\end{eqnarray}

The D-term lagrangian is found to be
\begin{eqnarray}
\mathcal{L}_De^{-1} &\supset& \frac{1}{2}D^2 - \mathcal{U}D + \frac{1}{D}\bigg(
-F^0\mathcal{U}e^{-K/6}  + \mathcal{U}_IF^I +\frac{1}{3}\mathcal{U}K_IF^I
\bigg)(\bar{\lambda}P_L\lambda) 
 -\frac{i}{D}  
\frac{\mathcal{U}_I}{\sqrt{2}}(\bar{\Omega}^IP_L\lambda)\nonumber\\
&& + \frac{1}{D}\bigg(
-F^{\bar{0}}\mathcal{U}e^{-K/6}  + \mathcal{U}_{\bar{J}}F^{\bar{J}} +\frac{1}{3}\mathcal{U}K_{\bar{J}}F^{\bar{J}}
\bigg)(\bar{\lambda}P_R\lambda) 
 +\frac{i}{D}  
\frac{\mathcal{U}_{\bar{J}}}{\sqrt{2}}(\bar{\Omega}^{\bar{J}}P_R\lambda)
\end{eqnarray}
The solution for $D$ can be obtained by
\begin{eqnarray}
D= \mathcal{U} + \frac{1}{\mathcal{U}^2}
\bigg[
\bigg(
-F^0\mathcal{U}e^{-K/6}  + \mathcal{U}_IF^I +\frac{1}{3}\mathcal{U}K_IF^I
\bigg)(\bar{\lambda}P_L\lambda)  -i 
\frac{\mathcal{U}_I}{\sqrt{2}}(\bar{\Omega}^IP_L\lambda) + h.c.
\bigg]+\textrm{higher order terms},\nonumber\\{}
\end{eqnarray}
Then, we find
\begin{eqnarray}
\mathcal{L}_{\textrm{newFI}}^{(2f)}e^{-1} = \frac{1}{\mathcal{U}}
\bigg[
\bigg(
-F^0\mathcal{U}e^{-K/6}  + \mathcal{U}_IF^I +\frac{1}{3}\mathcal{U}K_IF^I
\bigg)(\bar{\lambda}P_L\lambda)  -i 
\frac{\mathcal{U}_I}{\sqrt{2}}(\bar{\Omega}^IP_L\lambda) + h.c.
\bigg]
\end{eqnarray}
The total lagrangian containing the auxiliary fields $F^0$ and $F^I$ is given by
\begin{eqnarray}
\mathcal{L}e^{-1} &=& -3e^{-K/3}F^0F^{\bar{0}} + 3e^{K/3}WF^0 + 3e^{K/3}\bar{W}F^{\bar{0}} + \frac{1}{9}G_{I\bar{J}}F^IF^{\bar{J}} \nonumber\\
&& + \frac{1}{3}e^{K/2}\nabla_IWF^I + \frac{1}{3}e^{K/2}\nabla_{\bar{J}}\bar{W}F^{\bar{J}}\nonumber\\
&& + \frac{1}{\mathcal{U}}
\bigg[
\bigg(
-F^0\mathcal{U}e^{-K/6}  + \mathcal{U}_IF^I +\frac{1}{3}\mathcal{U}K_IF^I
\bigg)(\bar{\lambda}P_L\lambda)  -i 
\frac{\mathcal{U}_I}{\sqrt{2}}(\bar{\Omega}^IP_L\lambda) 
\bigg]
\nonumber\\
&& + \frac{1}{\mathcal{U}}
\bigg[
\bigg(
-F^{\bar{0}}\mathcal{U}e^{-K/6}  + \mathcal{U}_{\bar{J}}F^{\bar{J}} +\frac{1}{3}\mathcal{U}K_{\bar{J}}F^{\bar{J}}
\bigg)(\bar{\lambda}P_R\lambda)  +i 
\frac{\mathcal{U}_{\bar{J}}}{\sqrt{2}}(\bar{\Omega}^{\bar{J}}P_R\lambda)
\bigg],
\end{eqnarray}
where $\nabla_IW \equiv W_I + K_I W$. By solving the equations of motion for the auxiliary fields, we find 
\begin{eqnarray}
F^0 &=& e^{2K/3}\bar{W} - \frac{1}{3}e^{K/6}(\bar{\lambda}P_R\lambda),\\
F^{\bar{J}} &=&- 3e^{K/2}G^{I\bar{J}}\nabla_IW -G^{I\bar{J}}\Big(9\frac{\mathcal{U}_I }{\mathcal{U}}+ 3K_I\Big)(\bar{\lambda}P_L\lambda)
\end{eqnarray}
and also read the mass $m_{I\lambda}$
\begin{eqnarray}
m_{I\lambda}^{FI} &=& -\frac{i}{\sqrt{2}}\frac{\mathcal{U}_I}{\mathcal{U}},\\
m_{\lambda\lambda}^{FI} &=& -e^{K/2} \left( \bar{W} + 4G^{I\bar{J}}\left(\frac{\mathcal{U}_I}{\mathcal{U}}+\frac{K_I}{3}\right)(\bar{W}_{\bar{J}}+K_{\bar{J}}\bar{W}) \right)
\end{eqnarray}

The gravitino mixing term is given by
\begin{eqnarray}
\mathcal{L}_{\textrm{mix}} e^{-1} =
\frac{1}{\sqrt{2}}\nabla_IW e^{K/2} \bar{\psi}_{\mu} \gamma^{\mu} P_L\Omega^I + \frac{i}{2} \mathcal{P}_A \bar{\psi}_{\mu} \gamma^{\mu} P_L\lambda^A + \frac{i}{2}\mathcal{U}\bar{\psi}_{\mu} \gamma^{\mu} P_L\lambda + h.c = -\bar{\psi}_{\mu}\gamma^{\mu} P_Lv+h.c.,
\end{eqnarray}
which gives the goldstino 
\begin{eqnarray}
P_Lv = -\frac{1}{\sqrt{2}}\nabla_IW e^{K/2}  P_L\Omega^I - \frac{i}{2} \mathcal{P}_A P_L\lambda^A -\frac{i}{2}\mathcal{U} P_L\lambda ,
\end{eqnarray}
where $\lambda^A$ is the gaugino corresponding to the gauge multiplet $V_A$, and $\lambda$ is the superpartner of the new FI term vector multiplet $V$.

The fermionic masses from standard $\mathcal{N}=1$ supergravity are found by
\begin{eqnarray}
m_{3/2} &=& e^{K/2}W,\\
m_{IJ}^{(0)} &=& e^{K/2}(\partial_I+K_I)(W_J+K_JW)-e^{K/2} G^{K\bar{L}}\partial_IG_{J\bar{L}}(W_K+K_KW),\\
m_{IA}^{(0)} &=& i\sqrt{2}[\partial_I\mathcal{P}_A-\frac{1}{4}f_{ABI}(\textrm{Re}f)^{-1~BC}\mathcal{P}_C],\\
m_{AB}^{(0)} &=& -\frac{1}{2}e^{K/2}f_{ABI} G^{I\bar{J}}(\bar{W}_{\bar{J}}+K_{\bar{J}}\bar{W}),\\
m_{I\lambda}^{(0)} &=& 0,\\
m_{\lambda\lambda}^{(0)} &=& 0.
\end{eqnarray}
The fermionic masses from super-Higgs effect are given by
\begin{eqnarray}
m_{IJ}^{(\nu)} &=& -\frac{2}{3W} e^{K/2} (W_I+K_IW)(W_J+K_JW),\\
m_{IA}^{(\nu)} &=& -i\frac{2}{3\sqrt{2}W}(W_I+K_IW) \mathcal{P}_A ,\\
m_{AB}^{(\nu)} &=& \frac{1}{3e^{K/2}W} \mathcal{P}_A\mathcal{P}_B ,\\
m_{I\lambda}^{(\nu)} &=& -i\frac{2}{3\sqrt{2}W}(W_I+K_IW) \mathcal{U},\\
m_{\lambda\lambda}^{(\nu)} &=& \frac{\mathcal{U}^2}{3e^{K/2}W}
\end{eqnarray}
The fermionic masses from the new FI term are found to be
\begin{eqnarray}
 m_{IJ}^{FI} &=& 0 ,\\
m_{IA}^{FI} &=&  0,\\
m_{AB}^{FI} &=&  0,\\
m_{I\lambda}^{FI} &=& -\frac{i}{\sqrt{2}}\frac{\mathcal{U}_I}{\mathcal{U}},\\
m_{\lambda\lambda}^{FI} &=& -e^{K/2} \left( \bar{W} + 4G^{I\bar{J}}\left(\frac{\mathcal{U}_I}{\mathcal{U}}+\frac{K_I}{3}\right)(\bar{W}_{\bar{J}}+K_{\bar{J}}\bar{W}) \right).
\end{eqnarray}
Thus, the final fermionic masses are made by combinations of the three contributions above as follows:
\begin{eqnarray}
 m_{3/2} &=& We^{K/2},\nonumber\\
  m_{IJ}^{(g)} &=& m_{IJ}^{(0)}+m_{IJ}^{FI}+m_{IJ}^{(\nu)} \nonumber\\
&=& e^{K/2}(W_{IJ} + K_{IJ}W+K_JW_I+K_IW_J + K_IK_JW)
\nonumber\\
&& -e^{K/2} G^{K\bar{L}}\partial_I G_{J\bar{L}}(W_K + K_KW) -\frac{2}{3}  (W_I+K_IW)(W_J+K_JW),\nonumber\\
m_{IA}^{(g)} &=& m_{IA}^{(0)}+m_{IA}^{FI}+m_{IA}^{(\nu)} \nonumber\\
&=&
 i\sqrt{2}[\partial_I\mathcal{P}_A-\frac{1}{4}f_{ABI}(\textrm{Re}f)^{-1~BC}\mathcal{P}_C]-i\frac{2}{3\sqrt{2}W}(W_I+K_IW) \mathcal{P}_A\nonumber
\\
m_{AB}^{(g)} &=&m_{IA}^{(0)}+m_{IA}^{FI}+m_{IA}^{(\nu)}  \nonumber\\
&=&  -\frac{1}{2}e^{K/2}f_{ABI} G^{I\bar{J}}(\bar{W}_{\bar{J}}+K_{\bar{J}}\bar{W}) + \frac{1}{3e^{K/2}W} \mathcal{P}_A\mathcal{P}_B\nonumber\\
m_{I\lambda}^{(g)} &=& m_{I\lambda}^{(0)}+m_{I\lambda}^{FI}+m_{I\lambda}^{(\nu)} \nonumber\\
&=& -\frac{i}{\sqrt{2}}\frac{\mathcal{U}_I}{\mathcal{U}}-\frac{i\sqrt{2}}{3W}(W_I+K_IW) \mathcal{U} = m_{\lambda I}^{(g)},\nonumber\\
m_{\lambda\lambda}^{(g)} &=& m_{\lambda\lambda}^{(0)}+m_{\lambda\lambda}^{FI}+m_{\lambda\lambda}^{(\nu)} \nonumber\\
&=& -e^{K/2} \left( \bar{W} + 4G^{I\bar{J}}\left(\frac{\mathcal{U}_I}{\mathcal{U}}+\frac{K_I}{3}\right)(\bar{W}_{\bar{J}}+K_{\bar{J}}\bar{W}) \right) +\frac{\mathcal{U}^2}{3e^{K/2}W}. \label{General_Fermion_Masses}
\end{eqnarray}

%% file: chapters/alpha_cal.tex
The selection rules we defined are
\begin{eqnarray}
\textrm{For  } (d,f,a),
&& \Tilde{\mathcal{C}}:\quad (0,0,0) \implies \Gamma(\Tilde{\mathcal{C}})=0 \qquad \delta_{add} = 2,4,\nonumber\\
&& \Tilde{\mathcal{Z}}:\quad {\color{red}(0, 1,0)}\implies \Gamma(\Tilde{\mathcal{Z}})=1\qquad \delta_{add} =5/2,~9/2,\nonumber\\
&& \Tilde{\mathcal{B}_a}:\quad {\color{orange} (1,0,0)},{\color{blue}(0,2,0)}\implies \Gamma(\Tilde{\mathcal{B}_a})=1,2\qquad \delta_{add} =3,\nonumber\\
&& \Tilde{\Lambda}:\quad {\color{red}(1,1,0)},{\color{red}(1,1,1)},{\color{blue}(0,3,0)}\implies \Gamma(\Tilde{\Lambda})=2,3\qquad \delta_{add} =3/2,\nonumber\\
&& \Tilde{\mathcal{D}}:\quad (0,0,0),{\color{orange}(2,0,0)},{\color{blue}(0,2,0)},{\color{orange}(0,0,2)},\nonumber\\
&& \qquad \qquad {\color{blue}(0,2,1)},{\color{blue}(1,2,0)},{\color{blue}(0,4,0)}\implies \Gamma(\Tilde{\mathcal{D}})=0,2,3,4\qquad \delta_{add} =0. \nonumber
\end{eqnarray}

\section{Case (0,0,0) coupled to $\tilde{\mathcal{C}}$ with $\delta_{add}=2,4$}

\begin{itemize}
    \item Case of $b=0$ (with no derivatives):
    \begin{eqnarray}
     (N -u_2)_{b=0}&=& 2 ,\\
    (\delta_{tot}-4)_{b=0} &=&2+ 6_{4\lambda} +\delta_{add}-4
    \end{eqnarray}
\end{itemize}
which gives
\begin{eqnarray}
\alpha = \frac{2}{4+\delta_{add}} = \frac{1}{2},\frac{1}{3},\frac{1}{4}.
\end{eqnarray}

\section{Case (0,1,0) coupled to $\tilde{\mathcal{Z}}$ with $\delta_{add}=5/2,9/2$}

\begin{itemize}
     \item Case of $b=0$ (with derivatives):
    \begin{eqnarray}
     (N -u_2)_{b=0}&=& 2 ,\\
    (\delta_{tot}-4)_{b=0} &=&6_{4\lambda} +\frac{3}{2}f_{s1}
    +\frac{3}{2}f_{z1} +\frac{3}{2}f_{T1}+\delta_{add}-4
    \end{eqnarray}
    which gives
    \begin{eqnarray}
    \alpha = \frac{2}{7/2+\delta_{add}} = \frac{1}{3}, \frac{1}{4}.
    \end{eqnarray}
    \item  Case of $b=1$:
    \begin{eqnarray}
     (N -u_2)_{b=1=f_{W1}^{\lambda}}&=& 2  ,\\
     (\delta_{tot}-4)_{b=1=f_{W1}^{\lambda}} &=&3_{2\lambda}+\delta_{add}-\frac{1}{2},
    \end{eqnarray}
    which gives
    \begin{eqnarray}
    \alpha = \frac{2}{5/2+\delta_{add}} = \frac{2}{5}, \frac{2}{7}
    \end{eqnarray}
    \begin{eqnarray}
  (N -u_2)_{b=1=f_{W2}^{\lambda}}&=& 1  ,\\
  (\delta_{tot}-4)_{b=1=f_{W2}^{\lambda}} &=& 3_{2\lambda}+\delta_{add}-\frac{5}{2},
    \end{eqnarray}
    which gives
    \begin{eqnarray}
    \alpha = \frac{1}{1/2+\delta_{add}} = \frac{1}{3}, \frac{1}{5}.
    \end{eqnarray}
\end{itemize}

\section{Case (1,0,0) coupled to $\tilde{\mathcal{B}}$ with $\delta_{add}=3$}

\begin{itemize}
     \item Case of $b=0$ (with derivatives):
    \begin{eqnarray}
     (N -u_2)_{b=0}&=& 2 ,\\
    (\delta_{tot}-4)_{b=0} &=&6_{4\lambda} +2d_{s1}+d_{s2}+4d_{s3}^{\psi}+2d_{z1}
   +4d_{z2}^{\psi}+\delta_{add}-4
    \end{eqnarray}
    which gives
    \begin{eqnarray}
    \alpha = \frac{2}{5+\{1,2,4\}} = \frac{1}{3},\frac{2}{7}, \frac{2}{9}.
    \end{eqnarray}
\end{itemize}

\section{Case (0,2,0) coupled to $\tilde{\mathcal{B}}$ with $\delta_{add}=3$}

\begin{itemize}
     \item Case of $b=0$ (with derivatives):
    \begin{eqnarray}
     (N -u_2)_{b=0}&=& 2 ,\\
    (\delta_{tot}-4)_{b=0} &=&6_{4\lambda} +\frac{3}{2}f_{s1}
   +\frac{3}{2}f_{z1} +\frac{3}{2}f_{T1}-1
    \end{eqnarray}
    which gives $ \alpha = \frac{1}{4}$. 
    \item  Case of $b=1$:
    \begin{eqnarray}
     (N -u_2)_{b=1=f_{W1}^{\lambda}}&=& 2 + f_{T2}^{\lambda} + 2f_{T3}^{\lambda} ,\\
     (\delta_{tot}-4)_{b=1=f_{W1}^{\lambda}} &=&3_{2\lambda} +\frac{3}{2}f_{s1}+\frac{3}{2}f_{z1} +\frac{3}{2}f_{T1}+\frac{5}{2}f_{T2}^{\lambda}+\frac{11}{2}f_{T3}^{\lambda}+\frac{5}{2},
    \end{eqnarray}
    which gives 
    \begin{eqnarray}
    \alpha = \frac{2}{11/2+3/2} = \frac{2}{7}, \frac{3}{11/2+5/2} =\frac{3}{8}, \frac{4}{11/2+11/2} = \frac{4}{11}.
    \end{eqnarray}
    \begin{eqnarray}
  (N -u_2)_{b=1=f_{W2}^{\lambda}}&=& 1  + f_{T2}^{\lambda} + 2f_{T3}^{\lambda} ,\\
  (\delta_{tot}-4)_{b=1=f_{W2}^{\lambda}} &=& 3_{2\lambda} +\frac{3}{2}f_{s1}+\frac{3}{2}f_{z1} +\frac{3}{2}f_{T1}+\frac{5}{2}f_{T2}^{\lambda}+\frac{11}{2}f_{T3}^{\lambda} +\frac{1}{2},
    \end{eqnarray}
     which gives 
    \begin{eqnarray}
    \alpha = \frac{1}{7/2+3/2} = \frac{1}{5}, \frac{2}{7/2+5/2} =\frac{1}{3}, \frac{3}{7/2+11/2} = \frac{1}{3}.
    \end{eqnarray}
    \item Case of $b=2$:
    \begin{eqnarray}
    (N -u_2)_{b=2=f_{W1}^{\lambda}} &=& 2    ,\\
    (\delta_{tot}-4)_{b=2=f_{W1}^{\lambda}} &=& 6,
    \end{eqnarray}
    which gives $\alpha = 1/3$.
    \begin{eqnarray}
    (N -u_2)_{b=2=f_{W2}^{\lambda}} &=& 0,\\
    (\delta_{tot}-4)_{b=2=f_{W2}^{\lambda}} &=& 
   2,
    \end{eqnarray}
     which gives $\alpha = 0$.
    \begin{eqnarray}
    (N -u_2)_{b=2,f_{W1}^{\lambda}=f_{W2}^{\lambda}=1} &=&1  ,\\
    (\delta_{tot}-4)_{b=2,f_{W1}^{\lambda}=f_{W2}^{\lambda}=1} &=& 4. 
    \end{eqnarray}
     which gives $\alpha = 1/4$.
\end{itemize}

\section{Case (1,1,0) coupled to $\tilde{\Lambda}$ with $\delta_{add}=3/2$}

\begin{itemize}
     \item Case of $b=0$ (with derivatives):
    \begin{eqnarray}
     (N -u_2)_{b=0}&=& 2 ,\\
    (\delta_{tot}-4)_{b=0} &=&6_{4\lambda} +2d_{s1}+d_{s2}+4d_{s3}^{\psi}+\frac{3}{2}f_{s1}+2d_{z1}
    +4d_{z2}^{\psi}+\frac{3}{2}f_{z1} +\frac{3}{2}f_{T1}-5/2
    \end{eqnarray}
    The most largest one is given by $\alpha = 1/3$ when $d_{s2}=f_{s1}=1$.
    \item  Case of $b=1$:
    \begin{eqnarray}
     (N -u_2)_{b=1=f_{W1}^{\lambda}}&=& 2 + d_{T3}^{\lambda} + 2d_{T4}^{\lambda} ,\\
     (\delta_{tot}-4)_{b=1=f_{W1}^{\lambda}} &=&3_{2\lambda} +2d_{s1}+d_{s2}+4d_{s3}^{\psi}+2d_{z1}+4d_{z2}^{\psi} +5d_{T3}^{\lambda}+7d_{T4}^{\lambda} +1,
    \end{eqnarray}
    which gives $\alpha = 2/5, 1/3, 4/11$.
    \begin{eqnarray}
  (N -u_2)_{b=1=f_{W2}^{\lambda}}&=& 1  + d_{T3}^{\lambda} + 2d_{T4}^{\lambda}  ,\\
  (\delta_{tot}-4)_{b=1=f_{W2}^{\lambda}} &=& 3_{2\lambda} +2d_{s1}+d_{s2}+4d_{s3}^{\psi}+2d_{z1}+4d_{z2}^{\psi} +5d_{T3}^{\lambda}+7d_{T4}^{\lambda} -1,
    \end{eqnarray}
    which gives $\alpha = 1/3, 2/7$. 
\end{itemize}

\section{Case (1,1,1) coupled to $\tilde{\Lambda}$ with $\delta_{add}=3/2$}

\begin{itemize}
     \item Case of $b=0$ (with derivatives):
    \begin{eqnarray}
     (N -u_2)_{b=0}&=& 2-a_{s1} -a_{z1} - a_{T1} ,\\
    (\delta_{tot}-4)_{b=0} &=&6_{4\lambda} +2d_{s1}+d_{s2}+4d_{s3}^{\psi}+\frac{3}{2}f_{s1}+2d_{z1}
    \nonumber\\
    && +4d_{z2}^{\psi}+\frac{3}{2}f_{z1} +\frac{3}{2}f_{T1}+3 a_{T2}+3/2-4
    \end{eqnarray}
    which gives $\alpha = 1/6, 2/9$.
    \item  Case of $b=1$:
    \begin{eqnarray}
     (N -u_2)_{b=1=f_{W1}^{\lambda}}&=& 2 + d_{T3}^{\lambda} + 2d_{T4}^{\lambda} -a_{s1} -a_{z1} - a_{T1},\\
     (\delta_{tot}-4)_{b=1=f_{W1}^{\lambda}} &=&3_{2\lambda} +2d_{s1}+d_{s2}+4d_{s3}^{\psi}+2d_{z1}+4d_{z2}^{\psi} \nonumber\\
 &&+5d_{T3}^{\lambda}+7d_{T4}^{\lambda}+3 a_{T2} +1,
    \end{eqnarray}
    which gives $\alpha = 1/4,1/5,2/7,2/9,3/11$.
    \begin{eqnarray}
  (N -u_2)_{b=1=f_{W2}^{\lambda}}&=& 1  + d_{T3}^{\lambda} + 2d_{T4}^{\lambda} -a_{s1} -a_{z1} - a_{T1} ,\\
  (\delta_{tot}-4)_{b=1=f_{W2}^{\lambda}} &=& 3_{2\lambda} +2d_{s1}+d_{s2}+4d_{s3}^{\psi}+2d_{z1}+4d_{z2}^{\psi} \nonumber\\
 &&+5d_{T3}^{\lambda}+7d_{T4}^{\lambda}+3 a_{T2} -1,
    \end{eqnarray}
    which gives $\alpha = 1/7, 1/6, 1/5, 1/4, 2/9$.
\end{itemize}

\section{Case (0,3,0) coupled to $\tilde{\Lambda}$ with $\delta_{add}=3/2$}

\begin{itemize}
     \item Case of $b=0$ (with derivatives):
    \begin{eqnarray}
     (N -u_2)_{b=0}&=& 2 ,\\
    (\delta_{tot}-4)_{b=0} &=&6_{4\lambda}+\frac{3}{2}f_{s1}
    +\frac{3}{2}f_{z1} +\frac{3}{2}f_{T1}+3/2-4
    \end{eqnarray}
    which gives $\alpha = 1/4$.
    \item  Case of $b=1$:
    \begin{eqnarray}
     (N -u_2)_{b=1=f_{W1}^{\lambda}}&=& 2  + f_{T2}^{\lambda} + 2f_{T3}^{\lambda} ,\\
     (\delta_{tot}-4)_{b=1=f_{W1}^{\lambda}} &=&3_{2\lambda} +\frac{3}{2}f_{s1}+\frac{3}{2}f_{z1} +\frac{3}{2}f_{T1}+\frac{5}{2}f_{T2}^{\lambda}+\frac{11}{2}f_{T3}^{\lambda} +3/2-\frac{1}{2},
    \end{eqnarray}
    which gives $\alpha= 2/7, 3/8, 4/9, 5/12, 6/15$.
    \begin{eqnarray}
  (N -u_2)_{b=1=f_{W2}^{\lambda}}&=& 1 + f_{T2}^{\lambda} + 2f_{T3}^{\lambda} ,\\
  (\delta_{tot}-4)_{b=1=f_{W2}^{\lambda}} &=& 3_{2\lambda} +\frac{3}{2}f_{s1}+\frac{3}{2}f_{z1} +\frac{3}{2}f_{T1}+\frac{5}{2}f_{T2}^{\lambda}+\frac{11}{2}f_{T3}^{\lambda} +3/2-\frac{5}{2},
    \end{eqnarray}
    which gives $\alpha= 1/5, 1/3, 3/7, 2/5, 5/13$.
    \item Case of $b=2$:
    \begin{eqnarray}
    (N -u_2)_{b=2=f_{W1}^{\lambda}} &=& 2  + f_{T2}^{\lambda} + 2f_{T3}^{\lambda}  ,\\
    (\delta_{tot}-4)_{b=2=f_{W1}^{\lambda}} &=& 
    \frac{3}{2}f_{s1}+\frac{3}{2}f_{z1} +\frac{3}{2}f_{T1}+\frac{5}{2}f_{T2}^{\lambda}+\frac{11}{2}f_{T3}^{\lambda} +3/2+3,
    \end{eqnarray}
    which gives $\alpha = 1/3, 3/7, 2/5$.
    \begin{eqnarray}
    (N -u_2)_{b=2=f_{W2}^{\lambda}} &=&  f_{T2}^{\lambda} + 2f_{T3}^{\lambda} ,\\
    (\delta_{tot}-4)_{b=2=f_{W2}^{\lambda}} &=& 
   \frac{3}{2}f_{s1}+\frac{3}{2}f_{z1} +\frac{3}{2}f_{T1}+\frac{5}{2}f_{T2}^{\lambda}+\frac{11}{2}f_{T3}^{\lambda} +3/2-1,
    \end{eqnarray}
    which gives $\alpha = 0, 1/3$.
    \begin{eqnarray}
    (N -u_2)_{b=2,f_{W1}^{\lambda}=f_{W2}^{\lambda}=1} &=&1   + f_{T2}^{\lambda} + 2f_{T3}^{\lambda} ,\\
    (\delta_{tot}-4)_{b=2,f_{W1}^{\lambda}=f_{W2}^{\lambda}=1} &=& 
   \frac{3}{2}f_{s1}+\frac{3}{2}f_{z1} +\frac{3}{2}f_{T1}+\frac{5}{2}f_{T2}^{\lambda}+\frac{11}{2}f_{T3}^{\lambda}+3/2+1. 
    \end{eqnarray}
    which gives $\alpha = 1/4, 2/5, 3/8$.
\end{itemize}

\section{Case (0,0,0) coupled to $\tilde{\mathcal{D}}$ with $\delta_{add}=0$}

\begin{itemize}
    \item Case of $b=0$ (with no derivatives):
    \begin{eqnarray}
     (N -u_2)_{b=0}&=& 2 ,\\
    (\delta_{tot}-4)_{b=0} &=&2+ 6_{4\lambda} -4
    \end{eqnarray}
\end{itemize}
which gives $\alpha = 1/2$.

\section{Case (2,0,0) coupled to $\tilde{\mathcal{D}}$ with $\delta_{add}=0$}

\begin{itemize}
     \item Case of $b=0$ (with derivatives):
    \begin{eqnarray}
     (N -u_2)_{b=0}&=& 2 ,\\
    (\delta_{tot}-4)_{b=0} &=&6_{4\lambda} +2d_{s1}+d_{s2}+4d_{s3}^{\psi}+2d_{z1}
  +4d_{z2}^{\psi}-4
    \end{eqnarray}
\end{itemize}
The largest one is given by $\alpha = 1/2$.

\section{Case (0,2,0) coupled to $\tilde{\mathcal{D}}$ with $\delta_{add}=0$}

\begin{itemize}
     \item Case of $b=0$ (with derivatives):
    \begin{eqnarray}
     (N -u_2)_{b=0}&=& 2 ,\\
    (\delta_{tot}-4)_{b=0} &=&6_{4\lambda} +\frac{3}{2}f_{s1}
  +\frac{3}{2}f_{z1} +\frac{3}{2}f_{T1}-4
    \end{eqnarray}
    which gives $\alpha = 2/5$.
    \item  Case of $b=1$:
    \begin{eqnarray}
     (N -u_2)_{b=1=f_{W1}^{\lambda}}&=& 2  + f_{T2}^{\lambda} + 2f_{T3}^{\lambda} ,\\
     (\delta_{tot}-4)_{b=1=f_{W1}^{\lambda}} &=&3_{2\lambda} +\frac{3}{2}f_{s1}+\frac{3}{2}f_{z1} +\frac{3}{2}f_{T1}+\frac{5}{2}f_{T2}^{\lambda}+\frac{11}{2}f_{T3}^{\lambda} -\frac{1}{2},
    \end{eqnarray}
    which gives $\alpha = 4/11, 3/5, 1/2$. {\color{red} In particular, $\alpha=3/5$ is obtained when $f_{T2}^{\lambda}=1$}.
    \begin{eqnarray}
  (N -u_2)_{b=1=f_{W2}^{\lambda}}&=& 1 + f_{T2}^{\lambda} + 2f_{T3}^{\lambda} ,\\
  (\delta_{tot}-4)_{b=1=f_{W2}^{\lambda}} &=& 3_{2\lambda} +\frac{3}{2}f_{s1}+\frac{3}{2}f_{z1} +\frac{3}{2}f_{T1}+\frac{5}{2}f_{T2}^{\lambda}+\frac{11}{2}f_{T3}^{\lambda} -\frac{5}{2},
    \end{eqnarray}
    which gives $\alpha = 1/2, 2/3$. {\color{red} In particular, $\alpha=2/3$ is obtained when $f_{T2}^{\lambda}=1$, which is identified as the largest power of $H$. Notice that this arises via the kinetic mixing term of the fermions of $W$ and $T$ multiplets in the $\tilde{\mathcal{D}}$ term action of the new FI term, i.e. $\mathcal{L}_{newFI}e^{-1}\supset \frac{D}{4}f_{W\bar{T}} \bar{\Omega}^W\cancel{\mathcal{D}}\Omega^{\bar{T}}\sim D^{-2}\mathcal{O}^{(7)}_{4\lambda}$}.
    \item Case of $b=2$: This case corresponds to the fermionic kinetic term in the $\tilde{\mathcal{D}}$ of the supergravity action. Thus, all the terms from this case vanish on-shell (i.e. $\mathcal{D}_a F^{ab} =0, \cancel{\mathcal{D}}\lambda=0$ and using Bianchi identities $\mathcal{D}_{[a} F_{b]c}=0$) because $\cancel{\mathcal{D}}\Omega^W=\cancel{\mathcal{D}}\Big(\sqrt{2}P_L(-\frac{1}{2}\gamma\cdot \hat{F} + iD)\lambda\Big)=0$ on-shell and $\cancel{\mathcal{D}}D \sim \mathcal{U}^{(1)}D \sim 0$ along the potential minima. 
\end{itemize}

\section{Case (0,0,2) coupled to $\tilde{\mathcal{D}}$ with $\delta_{add}=0$}

\begin{itemize}
     \item Case of $b=0$ (with derivatives):
    \begin{eqnarray}
     (N -u_2)_{b=0}&=& 2 -a_{s1} -a_{z1} - a_{T1},\\
    (\delta_{tot}-4)_{b=0} &=&6_{4\lambda} +3a_{T2}-4 .
    \end{eqnarray}
    which gives $\alpha=0,1/5,1/4$.
\end{itemize}

\section{Case (0,2,1) coupled to $\tilde{\mathcal{D}}$ with $\delta_{add}=0$}

\begin{itemize}
     \item Case of $b=0$ (with derivatives):
    \begin{eqnarray}
     (N -u_2)_{b=0}&=& 2-a_{s1} -a_{z1} - a_{T1} ,\\
    (\delta_{tot}-4)_{b=0} &=&6_{4\lambda} +\frac{3}{2}f_{s1}+2d_{z1}
 +\frac{3}{2}f_{z1} +\frac{3}{2}f_{T1}+3 a_{T2}-4
    \end{eqnarray}
    which gives $\alpha = 1/5, 1/4$.
    \item  Case of $b=1$:
    \begin{eqnarray}
     (N -u_2)_{b=1=f_{W1}^{\lambda}}&=& 2 + f_{T2}^{\lambda} + 2f_{T3}^{\lambda}-a_{s1} -a_{z1} - a_{T1} ,\\
     (\delta_{tot}-4)_{b=1=f_{W1}^{\lambda}} &=&3_{2\lambda} +\frac{3}{2}f_{s1}+\frac{3}{2}f_{z1} +\frac{3}{2}f_{T1}+\frac{5}{2}f_{T2}^{\lambda}+\frac{11}{2}f_{T3}^{\lambda}+3 a_{T2} -\frac{1}{2},
    \end{eqnarray}
    which gives $\alpha = 1/4, 2/7, 2/5, 3/8, 4/11$.
    \begin{eqnarray}
  (N -u_2)_{b=1=f_{W2}^{\lambda}}&=& 1  + f_{T2}^{\lambda} + 2f_{T3}^{\lambda} -a_{s1} -a_{z1} - a_{T1},\\
  (\delta_{tot}-4)_{b=1=f_{W2}^{\lambda}} &=& 3_{2\lambda} +\frac{3}{2}f_{s1}+\frac{3}{2}f_{z1} +\frac{3}{2}f_{T1}+\frac{5}{2}f_{T2}^{\lambda}+\frac{11}{2}f_{T3}^{\lambda}+3 a_{T2} -\frac{5}{2},
    \end{eqnarray}
        which gives $\alpha = 0, 1/5, 1/3$.
    \item Case of $b=2$:
    \begin{eqnarray}
    (N -u_2)_{b=2=f_{W1}^{\lambda}} &=& 2   -a_{s1} -a_{z1} - a_{T1} ,\\
    (\delta_{tot}-4)_{b=2=f_{W1}^{\lambda}} &=& 
   3 a_{T2} +3,
    \end{eqnarray}
    which gives $\alpha = 1/3$.
    \begin{eqnarray}
    (N -u_2)_{b=2=f_{W2}^{\lambda}} &=&   -a_{s1} -a_{z1} - a_{T1},\\
    (\delta_{tot}-4)_{b=2=f_{W2}^{\lambda}} &=& 
  3 a_{T2} -1,
    \end{eqnarray}
    which gives $\alpha = 0$ and renormalizable terms.
    \begin{eqnarray}
    (N -u_2)_{b=2,f_{W1}^{\lambda}=f_{W2}^{\lambda}=1} &=&1 -a_{s1} -a_{z1} - a_{T1} ,\\
    (\delta_{tot}-4)_{b=2,f_{W1}^{\lambda}=f_{W2}^{\lambda}=1} &=& 
  3 a_{T2} +1. 
    \end{eqnarray}
    which gives $\alpha =0, 1/4$.
\end{itemize}

\section{Case (1,2,0) coupled to $\tilde{\mathcal{D}}$ with $\delta_{add}=0$}

\begin{itemize}
     \item Case of $b=0$ (with derivatives):
    \begin{eqnarray}
     (N -u_2)_{b=0}&=& 2 ,\\
    (\delta_{tot}-4)_{b=0} &=&6_{4\lambda} +2d_{s1}+d_{s2}+4d_{s3}^{\psi}+\frac{3}{2}f_{s1}+2d_{z1}
 +4d_{z2}^{\psi}+\frac{3}{2}f_{z1} +\frac{3}{2}f_{T1}-4
    \end{eqnarray}
    which gives $\alpha = 1/3$ as the largest one.
    \item  Case of $b=1$:
    \begin{eqnarray}
     (N -u_2)_{b=1=f_{W1}^{\lambda}}&=& 2 + d_{T3}^{\lambda} + 2d_{T4}^{\lambda} + f_{T2}^{\lambda} + 2f_{T3}^{\lambda} ,\\
     (\delta_{tot}-4)_{b=1=f_{W1}^{\lambda}} &=&3_{2\lambda} +2d_{s1}+d_{s2}+4d_{s3}^{\psi}+\frac{3}{2}f_{s1}+2d_{z1}+4d_{z2}^{\psi}+\frac{3}{2}f_{z1} \nonumber\\
 &&+5d_{T3}^{\lambda}+7d_{T4}^{\lambda}+\frac{3}{2}f_{T1}+\frac{5}{2}f_{T2}^{\lambda}+\frac{11}{2}f_{T3}^{\lambda} -\frac{1}{2},
    \end{eqnarray}
    which gives $\alpha= 2/5, 1/3, 4/11, 1/2, 4/9$.
    \begin{eqnarray}
  (N -u_2)_{b=1=f_{W2}^{\lambda}}&=& 1  + d_{T3}^{\lambda} + 2d_{T4}^{\lambda} + f_{T2}^{\lambda} + 2f_{T3}^{\lambda} ,\\
  (\delta_{tot}-4)_{b=1=f_{W2}^{\lambda}} &=& 3_{2\lambda} +2d_{s1}+d_{s2}+4d_{s3}^{\psi}+\frac{3}{2}f_{s1}+2d_{z1}+4d_{z2}^{\psi}+\frac{3}{2}f_{z1} \nonumber\\
 &&+5d_{T3}^{\lambda}+7d_{T4}^{\lambda}+\frac{3}{2}f_{T1}+\frac{5}{2}f_{T2}^{\lambda}+\frac{11}{2}f_{T3}^{\lambda} -\frac{5}{2},
    \end{eqnarray}
        which gives $\alpha= 1/3, 2/7, 1/2, 3/7$.
    \item Case of $b=2$:
    \begin{eqnarray}
    (N -u_2)_{b=2=f_{W1}^{\lambda}} &=& 2  + d_{T3}^{\lambda} + 2d_{T4}^{\lambda}   ,\\
    (\delta_{tot}-4)_{b=2=f_{W1}^{\lambda}} &=& 
    2d_{s1}+d_{s2}+4d_{s3}^{\psi}+2d_{z1}+4d_{z2}^{\psi} +5d_{T3}^{\lambda}+7d_{T4}^{\lambda}+3,
    \end{eqnarray}
    which gives $\alpha=1/2,3/8,2/5$.
    \begin{eqnarray}
    (N -u_2)_{b=2=f_{W2}^{\lambda}} &=&  d_{T3}^{\lambda} + 2d_{T4}^{\lambda} ,\\
    (\delta_{tot}-4)_{b=2=f_{W2}^{\lambda}} &=& 
    2d_{s1}+d_{s2}+4d_{s3}^{\psi}+2d_{z1}+4d_{z2}^{\psi} +5d_{T3}^{\lambda}+7d_{T4}^{\lambda} -1,
    \end{eqnarray}
    which gives $\alpha=0,1/4,1/3$.
    \begin{eqnarray}
    (N -u_2)_{b=2,f_{W1}^{\lambda}=f_{W2}^{\lambda}=1} &=&1  + d_{T3}^{\lambda} + 2d_{T4}^{\lambda} ,\\
    (\delta_{tot}-4)_{b=2,f_{W1}^{\lambda}=f_{W2}^{\lambda}=1} &=& 
   2d_{s1}+d_{s2}+4d_{s3}^{\psi}+2d_{z1}+4d_{z2}^{\psi} +5d_{T3}^{\lambda}+7d_{T4}^{\lambda} +1. 
    \end{eqnarray}
    which gives $\alpha = 1/2, 1/3, 3/8$.
\end{itemize}

\section{Case (0,4,0) coupled to $\tilde{\mathcal{D}}$ with $\delta_{add}=0$}

\begin{itemize}
     \item Case of $b=0$ (with derivatives):
    \begin{eqnarray}
     (N -u_2)_{b=0}&=& 2 ,\\
    (\delta_{tot}-4)_{b=0} &=&6_{4\lambda} +\frac{3}{2}f_{s1}+\frac{3}{2}f_{z1} +\frac{3}{2}f_{T1}-4
    \end{eqnarray}
    which gives $\alpha = 1/4$ since $f=4$.
    \item  Case of $b=1$:
    \begin{eqnarray}
     (N -u_2)_{b=1=f_{W1}^{\lambda}}&=& 2  + f_{T2}^{\lambda} + 2f_{T3}^{\lambda} ,\\
     (\delta_{tot}-4)_{b=1=f_{W1}^{\lambda}} &=&3_{2\lambda} +\frac{3}{2}f_{s1}+\frac{3}{2}f_{z1} +\frac{3}{2}f_{T1}+\frac{5}{2}f_{T2}^{\lambda}+\frac{11}{2}f_{T3}^{\lambda} -\frac{1}{2},
    \end{eqnarray}
    which gives $\alpha= 2/7, 4/9, 5/12, 2/5$.
    \begin{eqnarray}
  (N -u_2)_{b=1=f_{W2}^{\lambda}}&=& 1   + f_{T2}^{\lambda} + 2f_{T3}^{\lambda} ,\\
  (\delta_{tot}-4)_{b=1=f_{W2}^{\lambda}} &=& 3_{2\lambda} +\frac{3}{2}f_{s1}+\frac{3}{2}f_{z1} +\frac{3}{2}f_{T1}+\frac{5}{2}f_{T2}^{\lambda}+\frac{11}{2}f_{T3}^{\lambda} -\frac{5}{2},
    \end{eqnarray}
    which gives $\alpha=1/5, 2/7, 2/5, 5/13$.
    \item Case of $b=2$:
    \begin{eqnarray}
    (N -u_2)_{b=2=f_{W1}^{\lambda}} &=& 2   + f_{T2}^{\lambda} + 2f_{T3}^{\lambda}  ,\\
    (\delta_{tot}-4)_{b=2=f_{W1}^{\lambda}} &=& 
    \frac{3}{2}f_{s1}+\frac{3}{2}f_{z1}  +\frac{3}{2}f_{T1}+\frac{5}{2}f_{T2}^{\lambda}+\frac{11}{2}f_{T3}^{\lambda} +3,
    \end{eqnarray}
    which gives $\alpha= 1/3, 3/7, 1/2, 5/11, 3/7$.
    \begin{eqnarray}
    (N -u_2)_{b=2=f_{W2}^{\lambda}} &=&   f_{T2}^{\lambda} + 2f_{T3}^{\lambda} ,\\
    (\delta_{tot}-4)_{b=2=f_{W2}^{\lambda}} &=& 
    \frac{3}{2}f_{s1}+\frac{3}{2}f_{z1} +\frac{3}{2}f_{T1}+\frac{5}{2}f_{T2}^{\lambda}+\frac{11}{2}f_{T3}^{\lambda} -1,
    \end{eqnarray}
    which gives $\alpha= 0, 1/2, 3/7, 2/5$.
    \begin{eqnarray}
    (N -u_2)_{b=2,f_{W1}^{\lambda}=f_{W2}^{\lambda}=1} &=&1   + f_{T2}^{\lambda} + 2f_{T3}^{\lambda} ,\\
    (\delta_{tot}-4)_{b=2,f_{W1}^{\lambda}=f_{W2}^{\lambda}=1} &=& 
   \frac{3}{2}f_{s1}+\frac{3}{2}f_{z1}+\frac{3}{2}f_{T1}+\frac{5}{2}f_{T2}^{\lambda}+\frac{11}{2}f_{T3}^{\lambda} +1. 
    \end{eqnarray}
    which gives $\alpha=1/4, 2/5, 1/2, 4/9, 5/12$.
\end{itemize}

%% file: NYU PhD Dissertation (Hun Jang)/thesis.bib
@article{FPR_Integrating_out_functional_method,
    author = "Fuentes-Martin, Javier and Portoles, Jorge and Ruiz-Femenia, Pedro",
    title = "{Integrating out heavy particles with functional methods: a simplified framework}",
    eprint = "1607.02142",
    archivePrefix = "arXiv",
    primaryClass = "hep-ph",
    reportNumber = "IFIC-16-28, TUM-HEP-1047-16",
    doi = "10.1007/JHEP09(2016)156",
    journal = "JHEP",
    volume = "09",
    pages = "156",
    year = "2016"
}

@article{Dittmaier,
    author = "Dittmaier, Stefan and Schuhmacher, Sebastian and Stahlhofen, Maximilian",
    title = "{Integrating out heavy fields in the path integral using the background-field method: general formalism}",
    eprint = "2102.12020",
    archivePrefix = "arXiv",
    primaryClass = "hep-ph",
    reportNumber = "FR-PHENO-2020-010",
    doi = "10.1140/epjc/s10052-021-09587-7",
    journal = "Eur. Phys. J. C",
    volume = "81",
    number = "9",
    pages = "826",
    year = "2021"
}

@article{Planck2018_review,
    author = "Aghanim, N. and others",
    collaboration = "Planck",
    title = "{Planck 2018 results. I. Overview and the cosmological legacy of Planck}",
    eprint = "1807.06205",
    archivePrefix = "arXiv",
    primaryClass = "astro-ph.CO",
    doi = "10.1051/0004-6361/201833880",
    journal = "Astron. Astrophys.",
    volume = "641",
    pages = "A1",
    year = "2020"
}

@article{Planck2018_cosmo_para,
    author = "Aghanim, N. and others",
    collaboration = "Planck",
    title = "{Planck 2018 results. VI. Cosmological parameters}",
    eprint = "1807.06209",
    archivePrefix = "arXiv",
    primaryClass = "astro-ph.CO",
    doi = "10.1051/0004-6361/201833910",
    journal = "Astron. Astrophys.",
    volume = "641",
    pages = "A6",
    year = "2020",
    note = "[Erratum: Astron.Astrophys. 652, C4 (2021)]"
}

@article{Planck2018_infl,
    author = "Akrami, Y. and others",
    collaboration = "Planck",
    title = "{Planck 2018 results. X. Constraints on inflation}",
    eprint = "1807.06211",
    archivePrefix = "arXiv",
    primaryClass = "astro-ph.CO",
    doi = "10.1051/0004-6361/201833887",
    journal = "Astron. Astrophys.",
    volume = "641",
    pages = "A10",
    year = "2020"
}

@article{Baumann_cosmo,
    author = "Baumann, Daniel",
    title = "{Primordial Cosmology}",
    eprint = "1807.03098",
    archivePrefix = "arXiv",
    primaryClass = "hep-th",
    doi = "10.22323/1.305.0009",
    journal = "PoS",
    volume = "TASI2017",
    pages = "009",
    year = "2018"
}

@online{Cosmology,
  author = {Daniel Baumann},
  title = {Cosmology},
  year = unidentified,
  url = {https://cmb.wintherscoming.no/pdfs/baumann.pdf},
  urldate = {2022-05-04}
}

@article{Higgs_discover,
    author = "Aad, Georges and others",
    collaboration = "ATLAS",
    title = "{Observation of a new particle in the search for the Standard Model Higgs boson with the ATLAS detector at the LHC}",
    eprint = "1207.7214",
    archivePrefix = "arXiv",
    primaryClass = "hep-ex",
    reportNumber = "CERN-PH-EP-2012-218",
    doi = "10.1016/j.physletb.2012.08.020",
    journal = "Phys. Lett. B",
    volume = "716",
    pages = "1--29",
    year = "2012"
}

@article{BSM_SUSY,
    author = "Vempati, S. K.",
    editor = "Mulders, M. and Yuan, C. Z.",
    title = "{Physics beyond the Standard Model (Mostly Supersymmetry)}",
    doi = "10.23730/CYRSP-2018-002.87",
    journal = "CERN Yellow Rep. School Proc.",
    volume = "2",
    pages = "87--128",
    year = "2018"
}

@article{Quevedo_SUSY,
    author = "Quevedo, Fernando and Krippendorf, Sven and Schlotterer, Oliver",
    title = "{Cambridge Lectures on Supersymmetry and Extra Dimensions}",
    eprint = "1011.1491",
    archivePrefix = "arXiv",
    primaryClass = "hep-th",
    reportNumber = "DAMTP-2010-90, MPP-2010-143",
    month = "11",
    year = "2010"
}

@book{Superconformal_Freedman,
    author = "Freedman, Daniel Z. and Van Proeyen, Antoine",
    title = "{Supergravity}",
    isbn = "978-1-139-36806-3, 978-0-521-19401-3",
    publisher = "Cambridge Univ. Press",
    address = "Cambridge, UK",
    month = "5",
    year = "2012"
}

@book{Superspace_DallAgata,
    author = "Dall\textquoteright{}Agata, Gianguido and Zagermann, Marco",
    title = "{Supergravity: From First Principles to Modern Applications}",
    doi = "10.1007/978-3-662-63980-1",
    isbn = "978-3-662-63978-8, 978-3-662-63980-1",
    series = "Lecture Notes in Physics",
    volume = "991",
    month = "7",
    year = "2021"
}

@book{SUGRAprimer,
    author = "Rausch de Traubenberg, Michel and Valenzuela, Mauricio",
    title = "{A Supergravity Primer}: {From Geometrical Principles to the Final Lagrangian}",
    doi = "10.1142/11557",
    isbn = "978-981-12-1051-8, 978-981-12-1053-2",
    publisher = "WSP",
    address = "Singapur",
    year = "2020"
}

@article{No_SUSY,
    author = "Aad, Georges and others",
    collaboration = "ATLAS",
    title = "{Search for R-parity-violating supersymmetry in a final state containing leptons and many jets with the ATLAS experiment using $\sqrt{s} = 13 { TeV}$ proton\textendash{}proton collision data}",
    eprint = "2106.09609",
    archivePrefix = "arXiv",
    primaryClass = "hep-ex",
    reportNumber = "CERN-EP-2021-066",
    doi = "10.1140/epjc/s10052-021-09761-x",
    journal = "Eur. Phys. J. C",
    volume = "81",
    number = "11",
    pages = "1023",
    year = "2021"
}

@article{One_loop,
    author = "Gaillard, Mary K. and Nelson, Brent D.",
    title = "{On quadratic divergences in supergravity, vacuum energy and the supersymmetric flavor problem}",
    eprint = "hep-ph/0511234",
    archivePrefix = "arXiv",
    reportNumber = "LBNL-59069, UCB-PTH-05-38, UPR-1136-T",
    doi = "10.1016/j.nuclphysb.2006.05.035",
    journal = "Nucl. Phys. B",
    volume = "751",
    pages = "75--107",
    year = "2006"
}

@article{Taylor,
    author = "Taylor, Washington",
    title = "{TASI Lectures on Supergravity and String Vacua in Various Dimensions}",
    eprint = "1104.2051",
    archivePrefix = "arXiv",
    primaryClass = "hep-th",
    reportNumber = "MIT-CTP-4227",
    month = "4",
    year = "2011"
}

@inproceedings{MSSMWorkingGroup,
    author = "Djouadi, A. and others",
    collaboration = "MSSM Working Group",
    title = "{The Minimal supersymmetric standard model: Group summary report}",
    booktitle = "{GDR (Groupement De Recherche) - Supersymetrie}",
    eprint = "hep-ph/9901246",
    archivePrefix = "arXiv",
    reportNumber = "PM-98-45",
    month = "12",
    year = "1998"
}

@article{Ferrara_SUGRA40,
    author = "Ferrara, S. and Sagnotti, A.",
    editor = "Grzadkowski, Bohdan and Kalinowski, Jan and Krawczyk, Maria",
    title = "{Supergravity at 40: Reflections and Perspectives}",
    eprint = "1702.00743",
    archivePrefix = "arXiv",
    primaryClass = "hep-th",
    reportNumber = "CERN-TH-2017-005",
    doi = "10.1393/ncr/i2017-10136-6",
    journal = "Riv. Nuovo Cim.",
    volume = "40",
    number = "6",
    pages = "279--295",
    year = "2017"
}

@article{Yamaguchi,
    author = "Yamaguchi, Masahide",
    title = "{Supergravity based inflation models: a review}",
    eprint = "1101.2488",
    archivePrefix = "arXiv",
    primaryClass = "astro-ph.CO",
    doi = "10.1088/0264-9381/28/10/103001",
    journal = "Class. Quant. Grav.",
    volume = "28",
    pages = "103001",
    year = "2011"
}

@article{eta_problem,
    author = "Easson, Damien A. and Gregory, Ruth",
    title = "{Circumventing the eta problem}",
    eprint = "0902.1798",
    archivePrefix = "arXiv",
    primaryClass = "hep-th",
    reportNumber = "IPMU-09-0009, DCPT-09-09",
    doi = "10.1103/PhysRevD.80.083518",
    journal = "Phys. Rev. D",
    volume = "80",
    pages = "083518",
    year = "2009"
}

@article{Ellis_no_scale,
    author = "Ellis, John",
    title = "{No-scale supergravity inflation: A bridge between string theory and particle physics?}",
    doi = "10.1142/S0218271816300275",
    journal = "Int. J. Mod. Phys. D",
    volume = "25",
    number = "14",
    pages = "1630027",
    year = "2016"
}

@article{KKLT,
    author = "Kachru, Shamit and Kallosh, Renata and Linde, Andrei D. and Trivedi, Sandip P.",
    title = "{De Sitter vacua in string theory}",
    eprint = "hep-th/0301240",
    archivePrefix = "arXiv",
    reportNumber = "SLAC-PUB-9630, SU-ITP-03-01, TIFR-TH-03-03",
    doi = "10.1103/PhysRevD.68.046005",
    journal = "Phys. Rev. D",
    volume = "68",
    pages = "046005",
    year = "2003"
}

@article{KKLMMT,
    author = "Kachru, Shamit and Kallosh, Renata and Linde, Andrei D. and Maldacena, Juan Martin and McAllister, Liam P. and Trivedi, Sandip P.",
    title = "{Towards inflation in string theory}",
    eprint = "hep-th/0308055",
    archivePrefix = "arXiv",
    reportNumber = "SLAC-PUB-9669, SU-ITP-03-18, TIFR-TH-03-06",
    doi = "10.1088/1475-7516/2003/10/013",
    journal = "JCAP",
    volume = "10",
    pages = "013",
    year = "2003"
}

@inproceedings{Baumann,
    author = "Baumann, Daniel",
    title = "{Inflation}",
    booktitle = "{Theoretical Advanced Study Institute in Elementary Particle Physics}: {Physics of the Large and the Small}",
    eprint = "0907.5424",
    archivePrefix = "arXiv",
    primaryClass = "hep-th",
    reportNumber = "TASI-2009",
    doi = "10.1142/9789814327183_0010",
    pages = "523--686",
    year = "2011"
}

@online{EFT,
  author = {Adam Falkowski},
  title = {Lectures on Effective Field Theories},
  year = 2020,
  url = {https://indico.in2p3.fr/event/22195/contributions/86017/attachments/59873/81148/eftlectures.pdf},
  urldate = {2022-04-15}
}

@online{Marko,
  author = {Marko Vojinovi\'{c}},
  title = {Renormalization in QFT},
  year = 2014,
  url = {http://www.markovojinovic.com/professional/pdf/2014-Lisbon-TQFTclub-Renormalization-Lecture.pdf},
  urldate = {2022-05-06}
}

@book{Wess,
    author = "Wess, J. and Bagger, J.",
    title = "{Supersymmetry and supergravity}",
    isbn = "978-0-691-02530-8",
    publisher = "Princeton University Press",
    address = "Princeton, NJ, USA",
    year = "1992"
}

@book{Kirsten,
    author = "Muller-Kirsten, H. J. W. and Wiedemann, A.",
    title = "{SUPERSYMMETRY: AN INTRODUCTION WITH CONCEPTUAL AND CALCULATIONAL DETAILS}",
    reportNumber = "PRINT-86-0955, PRINT-86-0955 (KAISERSLAUTERN)",
    month = "7",
    year = "1986"
}

@unpublished{seahra,
title = {The Classical and Quantum Mechanics of Systems with Constraints},
author = {Sanjeev S. Seahra},
url = {https://sanjeev.seahra.ca/wp-content/uploads/2022/01/constrained_quantization.pdf},
year = {2002},
date = {2002-05-23},
urldate = {2002-05-23}
}

@article{Jon_Allen,
    author = "Allen, Jon and Matzner, Richard A.",
    title = "{Gauge fixing and constrained dynamics}",
    eprint = "2007.06641",
    archivePrefix = "arXiv",
    primaryClass = "math-ph",
    reportNumber = "WTTG 06-2020",
    doi = "10.1140/epjp/s13360-021-02258-2",
    journal = "Eur. Phys. J. Plus",
    volume = "137",
    number = "1",
    pages = "41",
    year = "2022"
}

@book{Henneaux,
    author = "Henneaux, M. and Teitelboim, C.",
    title = "{Quantization of gauge systems}",
    isbn = "978-0-691-03769-1",
    year = "1992"
}

@article{Banados,
    author = "Ba\~nados, M\'aximo and Reyes, Ignacio A.",
    title = "{A short review on Noether\textquoteright{}s theorems, gauge symmetries and boundary terms}",
    eprint = "1601.03616",
    archivePrefix = "arXiv",
    primaryClass = "hep-th",
    doi = "10.1142/S0218271816300214",
    journal = "Int. J. Mod. Phys. D",
    volume = "25",
    number = "10",
    pages = "1630021",
    year = "2016"
}

@article{Xavier,
    author = "Bekaert, Xavier and Park, Jeong-Hyuck",
    title = "{Symmetries and dynamics in constrained systems}",
    eprint = "0902.4754",
    archivePrefix = "arXiv",
    primaryClass = "math-ph",
    doi = "10.1140/epjc/s10052-009-0973-7",
    journal = "Eur. Phys. J. C",
    volume = "61",
    pages = "141--183",
    year = "2009"
}

@article{Brown,
    author = "Brown, J. David",
    title = "{Singular Lagrangians, Constrained Hamiltonian Systems and Gauge Invariance: An Example of the Dirac\textendash{}Bergmann Algorithm}",
    eprint = "2201.06558",
    archivePrefix = "arXiv",
    primaryClass = "gr-qc",
    doi = "10.3390/universe8030171",
    journal = "Universe",
    volume = "8",
    number = "3",
    pages = "171",
    year = "2022"
}

@phdthesis{Yamada,
    author = "Yamada, Yusuke",
    title = "{Construction of higher-derivative supergravity models via superconformal formulation}",
    school = "Waseda U.",
    month = "2",
    year = "2016"
}

@article{Brandt,
    author = "Brandt, Friedemann",
    title = "{Lectures on supergravity}",
    eprint = "hep-th/0204035",
    archivePrefix = "arXiv",
    reportNumber = "MIS-LECTURE-NOTES-SERIES-13-2002",
    doi = "10.1002/1521-3978(200210)50:10/11<1126::AID-PROP1126>3.0.CO;2-B",
    journal = "Fortsch. Phys.",
    volume = "50",
    pages = "1126--1172",
    year = "2002"
}

@article{Stueckelberg,
    author = "Ruegg, Henri and Ruiz-Altaba, Marti",
    title = "{The Stueckelberg field}",
    eprint = "hep-th/0304245",
    archivePrefix = "arXiv",
    reportNumber = "UGVA-DPT-2003-04-1106, UGVA-DPT-04-1106",
    doi = "10.1142/S0217751X04019755",
    journal = "Int. J. Mod. Phys. A",
    volume = "19",
    pages = "3265--3348",
    year = "2004"
}

@inproceedings{SUPER_CONFORMAL_TENSORCAL,
    author = "Van Proeyen, Antoine",
    title = "{Superconformal tensor calculus in $\mathcal{N}=1$ and $\mathcal{N}=2$ supergravity}",
    booktitle = "{19th Winter School and Workshop on Theoretical Physics: Supersymmetry and Supergravity}",
    reportNumber = "CERN-TH-3579",
    month = "4",
    year = "1983"
}

@book{Nakahara,
    author = "Nakahara, M.",
    title = "{Geometry, topology and physics}",
    year = "2003"
}

@article{Lewis_cov_deriv,
	doi = {10.1119/1.3153503},
	url = {https://doi.org/10.1119%2F1.3153503},
	year = 2009,
	month = {Sep},
	publisher = {American Association of Physics Teachers ({AAPT})},
	volume = {77},
	number = {9},
	pages = {839--843},
	author = {Clinton L. Lewis},
	title = {Explicit gauge covariant Euler{\textendash}Lagrange equation},
	journal = {American Journal of Physics}
}

@article{DS17,
    author = "Domcke, Valerie and Schmitz, Kai",
    title = "{Unified model of D-term inflation}",
    eprint = "1702.02173",
    archivePrefix = "arXiv",
    primaryClass = "hep-ph",
    doi = "10.1103/PhysRevD.95.075020",
    journal = "Phys. Rev. D",
    volume = "95",
    number = "7",
    pages = "075020",
    year = "2017"
}

@article{DS17_2,
    author = "Domcke, Valerie and Schmitz, Kai",
    title = "{Inflation from High-Scale Supersymmetry Breaking}",
    eprint = "1712.08121",
    archivePrefix = "arXiv",
    primaryClass = "hep-ph",
    reportNumber = "DESY-17-223",
    doi = "10.1103/PhysRevD.97.115025",
    journal = "Phys. Rev. D",
    volume = "97",
    number = "11",
    pages = "115025",
    year = "2018"
}

@article{InflationMSSM_Chakravarty,
    author = "Chakravarty, Girish Kumar and Gupta, Gaveshna and Lambiase, Gaetano and Mohanty, Subhendra",
    title = "{Plateau Inflation in SUGRA-MSSM}",
    eprint = "1604.02556",
    archivePrefix = "arXiv",
    primaryClass = "hep-ph",
    doi = "10.1016/j.physletb.2016.06.053",
    journal = "Phys. Lett. B",
    volume = "760",
    pages = "263--266",
    year = "2016"
}

@article{InflationMSSM_Pallis,
    author = "Pallis, C.",
    title = "{Linking Starobinsky-Type Inflation in no-Scale Supergravity to MSSM}",
    eprint = "1312.3623",
    archivePrefix = "arXiv",
    primaryClass = "hep-ph",
    doi = "10.1088/1475-7516/2014/04/024",
    journal = "JCAP",
    volume = "04",
    pages = "024",
    year = "2014",
    note = "[Erratum: JCAP 07, E01 (2017)]"
}

@article{InflationNMSSM_FKLMV,
    author = "Ferrara, Sergio and Kallosh, Renata and Linde, Andrei and Marrani, Alessio and Van Proeyen, Antoine",
    title = "{Jordan Frame Supergravity and Inflation in NMSSM}",
    eprint = "1004.0712",
    archivePrefix = "arXiv",
    primaryClass = "hep-th",
    reportNumber = "CERN-PH-TH-2010-070, SU-ITP-2010-12",
    doi = "10.1103/PhysRevD.82.045003",
    journal = "Phys. Rev. D",
    volume = "82",
    pages = "045003",
    year = "2010"
}

@article{InflationMSSM_KP,
    author = "Kaminska, Anna and Pacholek, Pawel",
    title = "{Inflation and Preheating in Supergravity with MSSM Flat Directions}",
    eprint = "0901.0478",
    archivePrefix = "arXiv",
    primaryClass = "hep-ph",
    doi = "10.1088/1475-7516/2009/06/024",
    journal = "JCAP",
    volume = "06",
    pages = "024",
    year = "2009"
}

@article{InflationMSSM_EMN,
    author = "Enqvist, Kari and Mether, Lotta and Nurmi, Sami",
    title = "{Supergravity origin of the MSSM inflation}",
    eprint = "0706.2355",
    archivePrefix = "arXiv",
    primaryClass = "hep-th",
    reportNumber = "HIP-2007-33-TH",
    doi = "10.1088/1475-7516/2007/11/014",
    journal = "JCAP",
    volume = "11",
    pages = "014",
    year = "2007"
}

@article{MSSMinflation,
    author = "Allahverdi, Rouzbeh and Enqvist, Kari and Garcia-Bellido, Juan and Mazumdar, Anupam",
    title = "{Gauge invariant MSSM inflaton}",
    eprint = "hep-ph/0605035",
    archivePrefix = "arXiv",
    reportNumber = "NORDITA-2006-13, IFT-UAM-CSIC-06-18, HIP-2006-22-TH",
    doi = "10.1103/PhysRevLett.97.191304",
    journal = "Phys. Rev. Lett.",
    volume = "97",
    pages = "191304",
    year = "2006"
}

@article{MSSMinflation_2,
    author = "Allahverdi, Rouzbeh and Enqvist, Kari and Garcia-Bellido, Juan and Jokinen, Asko and Mazumdar, Anupam",
    title = "{MSSM flat direction inflation: Slow roll, stability, fine tunning and reheating}",
    eprint = "hep-ph/0610134",
    archivePrefix = "arXiv",
    reportNumber = "IFT-UAM-CSIC-06-47",
    doi = "10.1088/1475-7516/2007/06/019",
    journal = "JCAP",
    volume = "06",
    pages = "019",
    year = "2007"
}

@article{lln1,
    author = "Li, Tianjun and Li, Zhijin and Nanopoulos, Dimitri V.",
    title = "{Chaotic Inflation in No-Scale Supergravity with String Inspired Moduli Stabilization}",
    eprint = "1405.0197",
    archivePrefix = "arXiv",
    primaryClass = "hep-th",
    reportNumber = "ACT-4-14, MIFPA-14-15",
    doi = "10.1140/epjc/s10052-015-3291-2",
    journal = "Eur. Phys. J. C",
    volume = "75",
    number = "2",
    pages = "55",
    year = "2015"
}

@article{lln2,
    author = "Li, Tianjun and Li, Zhijin and Nanopoulos, Dimitri V.",
    title = "{Natural Inflation with Natural Trans-Planckian Axion Decay Constant from Anomalous $U(1)_X$}",
    eprint = "1405.1804",
    archivePrefix = "arXiv",
    primaryClass = "hep-th",
    doi = "10.1007/JHEP07(2014)052",
    journal = "JHEP",
    volume = "07",
    pages = "052",
    year = "2014"
}

@article{lln3,
    author = "Li, Tianjun and Li, Zhijin and Nanopoulos, Dimitri V.",
    title = "{Aligned Natural Inflation and Moduli Stabilization from Anomalous $U(1)$ Gauge Symmetries}",
    eprint = "1407.1819",
    archivePrefix = "arXiv",
    primaryClass = "hep-th",
    reportNumber = "ACT-8-14, MIFPA-14-22",
    doi = "10.1007/JHEP11(2014)012",
    journal = "JHEP",
    volume = "11",
    pages = "012",
    year = "2014"
}

@article{LightScalar,
    author = "Vennin, Vincent and Koyama, Kazuya and Wands, David",
    title = "{Inflation with an extra light scalar field after Planck}",
    eprint = "1512.03403",
    archivePrefix = "arXiv",
    primaryClass = "astro-ph.CO",
    doi = "10.1088/1475-7516/2016/03/024",
    journal = "JCAP",
    volume = "03",
    pages = "024",
    year = "2016"
}

@article{acik,
    author = "Antoniadis, Ignatios and Chatrabhuti, Auttakit and Isono, Hiroshi and Knoops, Rob",
    title = "{The cosmological constant in Supergravity}",
    eprint = "1805.00852",
    archivePrefix = "arXiv",
    primaryClass = "hep-th",
    doi = "10.1140/epjc/s10052-018-6175-4",
    journal = "Eur. Phys. J. C",
    volume = "78",
    number = "9",
    pages = "718",
    year = "2018"
}

@article{CFTV,
    author = "Cribiori, Niccol\`o and Farakos, Fotis and Tournoy, Magnus and van Proeyen, Antoine",
    title = "{Fayet-Iliopoulos terms in supergravity without gauged R-symmetry}",
    eprint = "1712.08601",
    archivePrefix = "arXiv",
    primaryClass = "hep-th",
    doi = "10.1007/JHEP04(2018)032",
    journal = "JHEP",
    volume = "04",
    pages = "032",
    year = "2018"
}

@article{oldACIK,
    author = "Antoniadis, Ignatios and Chatrabhuti, Auttakit and Isono, Hiroshi and Knoops, Rob",
    title = "{Fayet\textendash{}Iliopoulos terms in supergravity and D-term inflation}",
    eprint = "1803.03817",
    archivePrefix = "arXiv",
    primaryClass = "hep-th",
    doi = "10.1140/epjc/s10052-018-5861-6",
    journal = "Eur. Phys. J. C",
    volume = "78",
    number = "5",
    pages = "366",
    year = "2018"
}

@article{Kuzenko,
    author = "Kuzenko, Sergei M.",
    title = "{Taking a vector supermultiplet apart: Alternative Fayet\textendash{}Iliopoulos-type terms}",
    eprint = "1801.04794",
    archivePrefix = "arXiv",
    primaryClass = "hep-th",
    doi = "10.1016/j.physletb.2018.04.051",
    journal = "Phys. Lett. B",
    volume = "781",
    pages = "723--727",
    year = "2018"
}

@article{ar,
    author = "Antoniadis, Ignatios and Rondeau, Fran\c{c}ois",
    title = {{New K\"ahler invariant Fayet\textendash{}Iliopoulos terms in supergravity and cosmological applications}},
    eprint = "1912.08117",
    archivePrefix = "arXiv",
    primaryClass = "hep-th",
    doi = "10.1140/epjc/s10052-020-7912-z",
    journal = "Eur. Phys. J. C",
    volume = "80",
    number = "4",
    pages = "346",
    year = "2020"
}

@article{AKK,
    author = "Aldabergenov, Yermek and Ketov, Sergei V. and Knoops, Rob",
    title = "{General couplings of a vector multiplet in $N=1$ supergravity with new FI terms}",
    eprint = "1806.04290",
    archivePrefix = "arXiv",
    primaryClass = "hep-th",
    reportNumber = "IPMU18-0104",
    doi = "10.1016/j.physletb.2018.07.072",
    journal = "Phys. Lett. B",
    volume = "785",
    pages = "284--287",
    year = "2018"
}

@article{jp2,
    author = "Jang, Hun and Porrati, Massimo",
    title = {{Inflation, gravity mediated supersymmetry breaking, and de Sitter vacua in supergravity with a K\"ahler-invariant Fayet-Iliopoulos term}},
    eprint = "2102.11358",
    archivePrefix = "arXiv",
    primaryClass = "hep-th",
    doi = "10.1103/PhysRevD.103.105006",
    journal = "Phys. Rev. D",
    volume = "103",
    number = "10",
    pages = "105006",
    year = "2021"
}

@article{R_symm,
    author = "Freedman, Daniel Z.",
    title = "{Supergravity with Axial Gauge Invariance}",
    reportNumber = "ITP-SB-76-50",
    doi = "10.1103/PhysRevD.15.1173",
    journal = "Phys. Rev. D",
    volume = "15",
    pages = "1173",
    year = "1977"
}

@article{R_symm_2,
    author = "Das, A. and Fischler, M. and Rocek, M.",
    title = "{SuperHiggs Effect in a New Class of Scalar Models and a Model of Super QED}",
    doi = "10.1103/PhysRevD.16.3427",
    journal = "Phys. Rev. D",
    volume = "16",
    pages = "3427--3436",
    year = "1977"
}

@article{R_symm_3,
    author = "de Wit, B. and van Nieuwenhuizen, P.",
    title = "{The Auxiliary Field Structure in Chirally Extended Supergravity}",
    reportNumber = "CERN-TH-2478",
    doi = "10.1016/0550-3213(78)90188-8",
    journal = "Nucl. Phys. B",
    volume = "139",
    pages = "216--220",
    year = "1978"
}

@article{barb,
    author = "Barbieri, Riccardo and Ferrara, S. and Nanopoulos, Dimitri V. and Stelle, K. S.",
    title = "{Supergravity, R Invariance and Spontaneous Supersymmetry Breaking}",
    reportNumber = "CERN-TH-3243",
    doi = "10.1016/0370-2693(82)90825-5",
    journal = "Phys. Lett. B",
    volume = "113",
    pages = "219--222",
    year = "1982"
}

@article{Nathan,
    author = "Seiberg, Nathan",
    title = "{Modifying the Sum Over Topological Sectors and Constraints on Supergravity}",
    eprint = "1005.0002",
    archivePrefix = "arXiv",
    primaryClass = "hep-th",
    doi = "10.1007/JHEP07(2010)070",
    journal = "JHEP",
    volume = "07",
    pages = "070",
    year = "2010"
}

@article{jp1,
    author = "Jang, Hun and Porrati, Massimo",
    title = "{Constraining Liberated Supergravity}",
    eprint = "2010.06789",
    archivePrefix = "arXiv",
    primaryClass = "hep-th",
    doi = "10.1103/PhysRevD.103.025008",
    journal = "Phys. Rev. D",
    volume = "103",
    number = "2",
    pages = "025008",
    year = "2021"
}

@article{jp3,
    author = "Jang, Hun and Porrati, Massimo",
    title = "{Component actions of liberated $ \mathcal{N} $ = 1 supergravity and new Fayet-Iliopoulos terms in superconformal tensor calculus}",
    eprint = "2108.04469",
    archivePrefix = "arXiv",
    primaryClass = "hep-th",
    doi = "10.1007/JHEP11(2021)075",
    journal = "JHEP",
    volume = "11",
    pages = "075",
    year = "2021"
}

@article{cfgvnv,
    author = "Cremmer, E. and Julia, B. and Scherk, Joel and Ferrara, S. and Girardello, L. and van Nieuwenhuizen, P.",
    title = "{Spontaneous Symmetry Breaking and Higgs Effect in Supergravity Without Cosmological Constant}",
    reportNumber = "CERN-TH-2554",
    doi = "10.1016/0550-3213(79)90417-6",
    journal = "Nucl. Phys. B",
    volume = "147",
    pages = "105",
    year = "1979"
}

@article{cfgvnv_2,
    author = "Cremmer, E. and Ferrara, S. and Girardello, L. and Van Proeyen, Antoine",
    editor = "Salam, A. and Sezgin, E.",
    title = "{Yang-Mills Theories with Local Supersymmetry: Lagrangian, Transformation Laws and SuperHiggs Effect}",
    reportNumber = "CERN-TH-3348",
    doi = "10.1016/0550-3213(83)90679-X",
    journal = "Nucl. Phys. B",
    volume = "212",
    pages = "413",
    year = "1983"
}

@article{Linear,
    author = "Ferrara, Sergio and Kallosh, Renata and Van Proeyen, Antoine and Wrase, Timm",
    title = "{Linear Versus Non-linear Supersymmetry, in General}",
    eprint = "1603.02653",
    archivePrefix = "arXiv",
    primaryClass = "hep-th",
    reportNumber = "CERN-TH-2016-052, TUW-16-05",
    doi = "10.1007/JHEP04(2016)065",
    journal = "JHEP",
    volume = "04",
    pages = "065",
    year = "2016"
}

@article{fkr,
    author = "Farakos, Fotis and Kehagias, Alex and Riotto, Antonio",
    title = "{Liberated $ \mathcal{N} $ = 1 supergravity}",
    eprint = "1805.01877",
    archivePrefix = "arXiv",
    primaryClass = "hep-th",
    doi = "10.1007/JHEP06(2018)011",
    journal = "JHEP",
    volume = "06",
    pages = "011",
    year = "2018"
}

@article{kyy,
    author = "Kugo, Taichiro and Yokokura, Ryo and Yoshioka, Koichi",
    title = "{Component versus superspace approaches to D = 4, N = 1 conformal supergravity}",
    eprint = "1602.04441",
    archivePrefix = "arXiv",
    primaryClass = "hep-th",
    doi = "10.1093/ptep/ptw090",
    journal = "PTEP",
    volume = "2016",
    number = "7",
    pages = "073",
    year = "2016"
}

@article{jp4,
    author = "Jang, Hun",
    title = "{Relaxed Supergravity}",
    eprint = "2112.02464",
    archivePrefix = "arXiv",
    primaryClass = "hep-th",
    month = "12",
    year = "2021"
}

@article{FP,
    author = "Ferrara, Sergio and Porrati, Massimo",
    title = "{Minimal $R+R^2$ Supergravity Models of Inflation Coupled to Matter}",
    eprint = "1407.6164",
    archivePrefix = "arXiv",
    primaryClass = "hep-th",
    reportNumber = "CERN-PH-TH-2014-133",
    doi = "10.1016/j.physletb.2014.08.050",
    journal = "Phys. Lett. B",
    volume = "737",
    pages = "135--138",
    year = "2014"
}

@article{HOC,
    author = "Ferrara, Sergio and Kallosh, Renata and Linde, Andrei and Porrati, Massimo",
    title = "{Higher Order Corrections in Minimal Supergravity Models of Inflation}",
    eprint = "1309.1085",
    archivePrefix = "arXiv",
    primaryClass = "hep-th",
    reportNumber = "CERN-PH-TH-2013-214",
    doi = "10.1088/1475-7516/2013/11/046",
    journal = "JCAP",
    volume = "11",
    pages = "046",
    year = "2013"
}

@article{High_SUSY,
    author = "Ellis, Sebastian A. R. and Wells, James D.",
    title = "{High-scale supersymmetry, the Higgs boson mass, and gauge unification}",
    eprint = "1706.00013",
    archivePrefix = "arXiv",
    primaryClass = "hep-ph",
    reportNumber = "MCTP-17-06",
    doi = "10.1103/PhysRevD.96.055024",
    journal = "Phys. Rev. D",
    volume = "96",
    number = "5",
    pages = "055024",
    year = "2017"
}

@article{StringRealization,
    author = "Cribiori, Niccol\`o and Farakos, Fotis and Tournoy, Magnus",
    title = "{Supersymmetric Born-Infeld actions and new Fayet-Iliopoulos terms}",
    eprint = "1811.08424",
    archivePrefix = "arXiv",
    primaryClass = "hep-th",
    doi = "10.1007/JHEP03(2019)050",
    journal = "JHEP",
    volume = "03",
    pages = "050",
    year = "2019"
}

@article{GM,
    author = "Nilles, Hans Peter",
    title = "{Supersymmetry, Supergravity and Particle Physics}",
    reportNumber = "UGVA-DPT-1983-12-412",
    doi = "10.1016/0370-1573(84)90008-5",
    journal = "Phys. Rept.",
    volume = "110",
    pages = "1--162",
    year = "1984"
}

@article{CFT,
    author = "Cribiori, Niccol\`o and Farakos, Fotis and Tournoy, Magnus",
    title = "{Supersymmetric Born-Infeld actions and new Fayet-Iliopoulos terms}",
    eprint = "1811.08424",
    archivePrefix = "arXiv",
    primaryClass = "hep-th",
    doi = "10.1007/JHEP03(2019)050",
    journal = "JHEP",
    volume = "03",
    pages = "050",
    year = "2019"
}

@article{EeVGravitino,
    author = "Dudas, Emilian and Mambrini, Yann and Olive, Keith",
    title = "{Case for an EeV Gravitino}",
    eprint = "1704.03008",
    archivePrefix = "arXiv",
    primaryClass = "hep-ph",
    doi = "10.1103/PhysRevLett.119.051801",
    journal = "Phys. Rev. Lett.",
    volume = "119",
    number = "5",
    pages = "051801",
    year = "2017"
}

@article{Inf_HighSUSY_EeVGrav,
    author = "Dudas, Emilian and Gherghetta, Tony and Mambrini, Yann and Olive, Keith A.",
    title = "{Inflation and High-Scale Supersymmetry with an EeV Gravitino}",
    eprint = "1710.07341",
    archivePrefix = "arXiv",
    primaryClass = "hep-ph",
    reportNumber = "CPHT-RR053.102017, LPT-ORSAY-17-41, UMN-TH-3704-17, FTPI-MINN-17-20, LPT--ORSAY-17-41, UMN--TH--3704-17, FTPI--MINN--17-20",
    doi = "10.1103/PhysRevD.96.115032",
    journal = "Phys. Rev. D",
    volume = "96",
    number = "11",
    pages = "115032",
    year = "2017"
}

@article{HeavyGravitino,
    author = "Meissner, Krzysztof A. and Nicolai, Hermann",
    title = "{Standard Model Fermions and Infinite-Dime-nsional R-Symmetries}",
    eprint = "1804.09606",
    archivePrefix = "arXiv",
    primaryClass = "hep-th",
    doi = "10.1103/PhysRevLett.121.091601",
    journal = "Phys. Rev. Lett.",
    volume = "121",
    number = "9",
    pages = "091601",
    year = "2018"
}

@article{GDM,
    author = "Meissner, Krzysztof A. and Nicolai, Hermann",
    title = "{Planck Mass Charged Gravitino Dark Matter}",
    eprint = "1809.01441",
    archivePrefix = "arXiv",
    primaryClass = "hep-ph",
    doi = "10.1103/PhysRevD.100.035001",
    journal = "Phys. Rev. D",
    volume = "100",
    number = "3",
    pages = "035001",
    year = "2019"
}

@article{HDM,
    author = "Clark, Michael and Depoian, Amanda and Elshimy, Bahaa and Kopec, Abigail and Lang, Rafael F. and Li, Shengchao and Qin, Juehang",
    title = "{Direct Detection Limits on Heavy Dark Matter}",
    eprint = "2009.07909",
    archivePrefix = "arXiv",
    primaryClass = "hep-ph",
    doi = "10.1103/PhysRevD.102.123026",
    journal = "Phys. Rev. D",
    volume = "102",
    number = "12",
    pages = "123026",
    year = "2020"
}

@article{softSUSYbreaking,
    author = "Girardello, L. and Grisaru, Marcus T.",
    title = "{Soft Breaking of Supersymmetry}",
    reportNumber = "Print-81-0592 (HARVARD)",
    doi = "10.1016/0550-3213(82)90512-0",
    journal = "Nucl. Phys. B",
    volume = "194",
    pages = "65",
    year = "1982"
}

@book{WeinbergSUSY,
    author = "Weinberg, Steven",
    title = "{The quantum theory of fields. Vol. 3: Supersymmetry}",
    isbn = "978-0-521-67055-5, 978-1-139-63263-8, 978-0-521-67055-5",
    publisher = "Cambridge University Press",
    month = "6",
    year = "2013"
}

@article{HighSUSYGUT,
    author = "Ellis, Sebastian A. R. and Wells, James D.",
    title = "{High-scale supersymmetry, the Higgs boson mass, and gauge unification}",
    eprint = "1706.00013",
    archivePrefix = "arXiv",
    primaryClass = "hep-ph",
    reportNumber = "MCTP-17-06",
    doi = "10.1103/PhysRevD.96.055024",
    journal = "Phys. Rev. D",
    volume = "96",
    number = "5",
    pages = "055024",
    year = "2017"
}

@unpublished{jp5,
    author = "Jang, Hun and Porrati, Massimo",
    %title = "{Inflation and MSSM from Supergravity with
%New Fayet-Iliopoulos Terms}",
    note = "In preparation for publication"
}

@article{RPP,
    author = "Zyla, P.A. and others",
    collaboration = "Particle Data Group",
    title = "{Review of Particle Physics}",
    doi = "10.1093/ptep/ptaa104",
    journal = "PTEP",
    volume = "2020",
    number = "8",
    pages = "083C01",
    year = "2020",
    note = "and 2021 update"
}

@article{Fermion_mass_matrix,
    author = "Fuks, Benjamin and Klasen, Michael and Schmiemann, Saskia and Sunder, Marthijn",
    title = "{Realistic simplified gaugino-higgsino models in the MSSM}",
    eprint = "1710.09941",
    archivePrefix = "arXiv",
    primaryClass = "hep-ph",
    reportNumber = "MS-TP-17-14",
    doi = "10.1140/epjc/s10052-018-5695-2",
    journal = "Eur. Phys. J. C",
    volume = "78",
    number = "3",
    pages = "209",
    year = "2018"
}

@article{WIMPZILLA1,
    author = "Ziaeepour, Houri",
    editor = "Khlopov, M. Yu. and Prokhorov, M. E. and Starobinsky, A. A.",
    title = "{A Decaying ultra heavy dark matter (WIMPZILLA): Review of recent progress}",
    eprint = "astro-ph/0005299",
    archivePrefix = "arXiv",
    journal = "Grav. Cosmol. Suppl.",
    volume = "6",
    pages = "128--133",
    year = "2000"
}

@article{WIMPZILLA2,
    author = "Park, Jong-Chul and Park, Seong Chan",
    title = "{A testable scenario of WIMPZILLA with Dark Radiation}",
    eprint = "1305.5013",
    archivePrefix = "arXiv",
    primaryClass = "hep-ph",
    doi = "10.1016/j.physletb.2013.11.027",
    journal = "Phys. Lett. B",
    volume = "728",
    pages = "41--44",
    year = "2014"
}

@article{WIMPZILLA3,
    author = "Farzinnia, Arsham and Kouwn, Seyen",
    title = "{Classically scale invariant inflation, supermassive WIMPs, and adimensional gravity}",
    eprint = "1512.05890",
    archivePrefix = "arXiv",
    primaryClass = "hep-ph",
    reportNumber = "CTPU-15-25",
    doi = "10.1103/PhysRevD.93.063528",
    journal = "Phys. Rev. D",
    volume = "93",
    number = "6",
    pages = "063528",
    year = "2016"
}

@article{WIMPZILLA4,
    author = "Kolb, Edward W. and Long, Andrew J.",
    title = "{Superheavy dark matter through Higgs portal operators}",
    eprint = "1708.04293",
    archivePrefix = "arXiv",
    primaryClass = "astro-ph.CO",
    doi = "10.1103/PhysRevD.96.103540",
    journal = "Phys. Rev. D",
    volume = "96",
    number = "10",
    pages = "103540",
    year = "2017"
}

@article{SuperheavyDM,
    author = "Alcantara, Esteban and Anchordoqui, Luis A. and Soriano, Jorge F.",
    title = "{Hunting for superheavy dark matter with the highest-energy cosmic rays}",
    eprint = "1903.05429",
    archivePrefix = "arXiv",
    primaryClass = "hep-ph",
    doi = "10.1103/PhysRevD.99.103016",
    journal = "Phys. Rev. D",
    volume = "99",
    number = "10",
    pages = "103016",
    year = "2019"
}

@article{Swampland,
    author = "Vafa, Cumrun",
    title = "{The String landscape and the swampland}",
    eprint = "hep-th/0509212",
    archivePrefix = "arXiv",
    reportNumber = "HUTP-05-A043",
    month = "9",
    year = "2005"
}

@article{Recent_Review_Swampland1,
    author = "Brennan, T. Daniel and Carta, Federico and Vafa, Cumrun",
    title = "{The String Landscape, the Swampland, and the Missing Corner}",
    eprint = "1711.00864",
    archivePrefix = "arXiv",
    primaryClass = "hep-th",
    reportNumber = "IFT-UAM-CSIC-17-105",
    doi = "10.22323/1.305.0015",
    journal = "PoS",
    volume = "TASI2017",
    pages = "015",
    year = "2017"
}

@article{Recent_Review_Swampland2,
    author = "Palti, Eran",
    title = "{The Swampland: Introduction and Review}",
    eprint = "1903.06239",
    archivePrefix = "arXiv",
    primaryClass = "hep-th",
    reportNumber = "MPP-2019-53",
    doi = "10.1002/prop.201900037",
    journal = "Fortsch. Phys.",
    volume = "67",
    number = "6",
    pages = "1900037",
    year = "2019"
}

@article{de_Sitter_Conjecture,
    author = "Obied, Georges and Ooguri, Hirosi and Spodyneiko, Lev and Vafa, Cumrun",
    title = "{De Sitter Space and the Swampland}",
    eprint = "1806.08362",
    archivePrefix = "arXiv",
    primaryClass = "hep-th",
    reportNumber = "CALT-TH-2018-020, IPMU18-0100",
    month = "6",
    year = "2018"
}

@article{MWI,
    author = "Berghaus, Kim V. and Graham, Peter W. and Kaplan, David E.",
    title = "{Minimal Warm Inflation}",
    eprint = "1910.07525",
    archivePrefix = "arXiv",
    primaryClass = "hep-ph",
    doi = "10.1088/1475-7516/2020/03/034",
    journal = "JCAP",
    volume = "03",
    pages = "034",
    year = "2020"
}

@article{MWI_dSC,
    author = "Das, Suratna and Goswami, Gaurav and Krishnan, Chethan",
    title = "{Swampland, axions, and minimal warm inflation}",
    eprint = "1911.00323",
    archivePrefix = "arXiv",
    primaryClass = "hep-th",
    doi = "10.1103/PhysRevD.101.103529",
    journal = "Phys. Rev. D",
    volume = "101",
    number = "10",
    pages = "103529",
    year = "2020"
}

@article{MWI_TCC,
    author = "Kamali, Vahid and Moshafi, Hossein and Ebrahimi, Saeid",
    title = "{Minimal warm inflation and TCC}",
    eprint = "2111.11436",
    archivePrefix = "arXiv",
    primaryClass = "gr-qc",
    month = "11",
    year = "2021"
}
